\documentclass[11pt,reqno]{amsart}
\usepackage{enumitem}
\usepackage{url}

\usepackage[utf8]{inputenc}

\usepackage{tikz}

\usepackage[colorlinks]{hyperref}
\usepackage{dcolumn}
\usepackage{enumitem}
\usepackage{bm}

\usepackage{amsfonts,amsthm,amsmath,amssymb}
\usepackage{chngcntr}
\usepackage{float}  
\usepackage[ruled,vlined]{algorithm2e}
\usepackage{algpseudocode}

\newcommand{\Mod}[1]{\ (\mathrm{mod}\ #1)}  

\theoremstyle{plain}
\newtheorem{theorem}{Theorem}
\newtheorem{lemma}{Lemma}
\newtheorem{conjecture}{Conjecture}

\definecolor{bermuda}{RGB}{141, 204, 191}
\definecolor{cancan}{RGB}{209, 135, 160}

\calclayout

\begin{document}

\title{Infinitely growing configurations in Emil Post's tag system problem}

\author{Nikita V. Kurilenko}

\begin{abstract}
Emil Post's tag system problem is the question of whether or not a tag system $\{N=3, P(0)=00, P(1)=1101\}$ has a configuration, simulation of which will never halt or end up in a loop.
For the past decades, there were several attempts to find an answer to this question, including a recent study, during which the first $2^{84}$ initial configurations were checked \cite{WolframS2021}.
This paper presents a family of configurations of this type in a form of strings $a^{n} b c^{m}$, that evolve to $a^{n+1} b c^{m+1}$ after a finite amount of steps.
The proof of this behavior for all non-negative $n$ and $m$ is described further in a paper as a finite verification procedure, which is computationally bounded by 20000 iterations of tag.
All corresponding code can be found at \url{https://github.com/nikitakurilenko/post-tag-infinitely-growing-configurations}.
\end{abstract}

\maketitle

\section{Introduction}

A tag system is a computational model introduced by Emil Post, capable of universal computations \cite{CockeJMinskyM1964}.
Every tag system is a single string manipulation rule that transforms one string to another by adding symbols to the right side and then removing symbols from the left side.
Added symbols depend only on the first symbol of an input string, and the amount of deleted symbols is fixed.
Therefore any tag system can be represented by a positive integer $N$, called \textit{deletion number}, and a symbol-to-string function $P: A \rightarrow A^{*}$ (where A is an alphabet of a string), called \textit{production rules}.
In his paper about canonical systems \cite{PostE1943}, Emil Post mentioned the tag system $\{N=3, P(0)=00, P(1)=1101\}$ as being "intractable".
Since then this tag system appeared in literature as an example of a simple dynamic system that produces chaotic behavior, similar to the behavior of Syracuse function from the famous Collatz conjecture \cite{LagariasJC2010}, which is still an open problem.
\textit{Emil Post's tag system problem} asks if there are strings that will show unbounded growth during manipulations of the system $\{N=3, P(0)=00, P(1)=1101\}$.

As already mentioned in \cite{WolframS2021} there is no extensive literature on this topic.
The main experiments were conducted by Watabale \cite{WatanabeS1963}, Hayes \cite{HayesB1986}, Asveld \cite{AsveldPRJ1992}, Shearer \cite{ShearerJB1996}, De Mol \cite{DeMolL2011}, and, recently, Wolfram \cite{WolframS2021}.
In each study, a large set of initial configurations was checked, and in several of them, new families of periodic evolutions were discovered.
The most noticeable works on this problem are \cite{DeMolL2011} and \cite{WolframS2021}, as \cite{DeMolL2011} studied a large set of similar tag systems, and \cite{WolframS2021} checked a huge amount of possible candidates and presented insightful visualizations of chaotic behavior.
Since the quality of computational tools is getting better every year, the problem, if we assume the existence of configurations with unbounded growth, is getting closer to resolution.
In this paper, we present a family of configurations of this type and study their behavior in detail.

\section{Main statement}

Consider the following strings $a$, $b$, $c$ (for more compact visualization see Appendix \ref{appendixa}):
\footnotesize

\begin{equation*}
a = 000011011101110100\end{equation*}

\begin{equation*}
b = \begin{aligned}
{} & 00000011011101000000110100110100001101110111011101110111010000001101110111 \\
   & 01000000110111011101000000110111011101000000110111011101001101110111011101 \\
   & 11010011011101110111011101000000000011011101110100001101110100000011011101 \\
   & 11010000001101110111010000001101110111010000000011011101110111010011011101 \\
   & 11010000110111010011011101110111011101001101001101110111010000001101110111 \\
   & 01000000110111011101000000110111011101110111010011010000110111010011010000 \\
   & 11011101001101000011011101001101000011011101001101000000000011010011010000 \\
   & 00110111011101000000000000001101000000000000001101110111010000001101110111 \\
   & 01000000110111011101110111010011010011010011010011011101110100000011011101 \\
   & 11010011011101110111011101001101110111011101110100110111011101110111010011 \\
   & 01110111011101110100000000000000110111011101000000110111011101001101001101 \\
   & 00000011011101110100000000110111010000000011011101000000001101110100000000 \\
   & 00000000000000001101110100110100001101110100110100110111011101001101110111 \\
   & 01110111010011010000110111010011010011010011011101110100000011011101110100 \\
   & 00000011011101001101000011011101001101001101110111010011011101110111011101 \\
   & 00110100001101110100110100001101110100110100001101110100110100001101110100 \\
   & 11011101110100000000001101110111010000001101110111010000001101110111010011 \\
   & 01110111010000110111010000000011011101001101110111010000001101110111011101 \\
   & 11010011011101110100110111011101110111010000001101110111010000001101110111 \\
   & 01000000110111011101110111011101110111011101001101000011011101001101000000 \\
   & 11010000000000110111011101000000110111011101000000110111011101001101110111 \\
   & 01001101001101000000000011011101110111011101110111010011010011011101110100 \\
   & 00001101110111010011011101110100110100110100110111011101000000110111011101 \\
   & 00110100110100000011010011010011010011010000001101110111011101110100110111 \\
   & 01110100000011011101110100110100001101110100110100000011011101110111011101 \\
   & 00110111011101110111010011010000110111010011010000001101000000110111011101 \\
   & 11011101000000110111011101000000110111011101000000000011011101110100000011 \\
   & 01110111011101110100110111011101001101110111010011010000110111010011010000 \\
   & 11011101001101001101110111010000000000110111011101000000110111011101000000 \\
   & 11011101110100000011011101110100000000001101000000000000000000000011011101 \\
   & 11011101110100110100110100110100001101110111011101000000000000001101001101 \\
   & 00001101110111011101000000001101110111011101001101000011011101001101000011 \\
   & 0111010011010000000000110111011101 \end{aligned}\end{equation*}

\begin{equation*}c = 000000110111011101000000110111011101000000110111011101\end{equation*}

\normalsize
These strings have an important property that makes them suitable for the theorem related to Post's tag system problem.

\begin{theorem}\label{maintheorem}
For Post's tag system there are strings $a$, $b$ and $c$ such that every string $a^{n} b c^{m}$ will eventually evolve to string $a^{n+1} b c^{m+1}$.
\end{theorem}

While the lengths of $a$ and $c$ are small, the length of $b$ is 2402, which makes this example hard to find.
Note that in particular, $b$ evolves to $abc$ (after 10444 iterations), so $b$ is a zero element of a constructed set.
It is not known yet whether or not there is a shorter example of $b$, but it is possible.
Larger examples are very likely to exist.

After $a$, $b$, $c$ candidates are given it is easy to check, that the statement is valid for all $n$ and $m$ less than 100.
The next step is to formally prove Theorem \ref{maintheorem} for every other value of $n$ and $m$.
We will do this in the next section by using properties of $a$, $b$, and $c$ and by constructing a verification procedure to fully describe the evolution of words $ a^{n} b c^{m} $.

\section{Proof of the main statement}

We use lower indexing with square brackets to indicate string cutting (from the beginning of a string); for example $1010001_{[2]} = 10001$, $101_{[0]} = 101$.
Cutting indices are written in modulo 3, so for example if $n = 1$ then $101100_{[-n]} = 1100$ and $0111000_{[n+n+n+n]} = 111000$.

To avoid corner cases it is assumed that every string in every statement has a large enough length (at least four symbols), so statements like $(a b)_{[x]} = a_{[x]} b$ always hold.

Similarly to string cutting a function $l$ is defined as a length of string modulo 3.
As with low indexing addition and subtraction of values of $l$ conducted through modular arithmetic: if $l(a) = 2$ and $l(b) = 1$ then $l(a b) = 0$ and $l(a a b) = 2$.
By definition the following statement is true:
\begin{lemma}\label{lemmagcuttinginds}
$l(a_{[n]}) = l(a) - n$
\end{lemma} 
\begin{proof}
Each symbol deletion decreases length by 1. $n$ deletions decreases length by $n$, and $l(a)$ decreases by n modulo 3.
\end{proof}

To simplify and prove statements about the behavior of Post's tag system, we define three supporting functions $g$, $f$, and $F$, that break down evolutions of strings $ a^{n} b c^{m} $ into analyzable groups of operations.

A function $g$ takes every third element of an input string (discards other elements) and replaces it with $1101$ if an element is $1$, else replaces it with $00$.
For example, $g(1) = 1101$, $g(000100) = 001101$.
The following formula is useful to operate $g$:

\begin{lemma}\label{lemmag}
$g( (ab)_{[x]} ) = g( a_{[x]} ) g( b_{[x-l(a)]} )$
\end{lemma} 
\begin{proof}

If $G_{s}$ is a set that includes every third element of a string $s$, then it is easy to see that if the length of $a$ is divisible by 3, then $G_{ab} = G_{a} \cup G_{b}$ (hence $g(ab) = g(a)g(b)$).

If $l(a) = 2$ (length of $a$ equals $3 \times k + 2$ for some non-negative $k$), $g(ab)$ in particular processes every third element of $b$ starting with the second symbol, instead of the first symbol, because "remainder" symbols of $a$ will shift a set of processed symbols of $b$.
This can be summarized in a formula
\begin{equation*} l(a) = 2 \implies G_{ab} = G_{a} \cup G_{b_{[1]}}. \end{equation*}
Similarly,
\begin{equation*} l(a) = 1 \implies G_{ab} = G_{a} \cup G_{b_{[2]}}, \end{equation*}
which concludes formula
\begin{equation*} G_{ab} = G_{a} \cup G_{b_{[-l(a)]}}, \end{equation*}
hence proving
\begin{equation*} g( ab ) = g ( a ) g ( b_{[-l(a)]} ).\end{equation*}

This formula can split $g( (ab)_{[x]} )$ into two parts:
\begin{equation*} g( (ab)_{[x]} ) = g( a_{[x]} b ) = g ( a_{[x]} ) g ( b_{[-l(a_{[x]})]}).\end{equation*}
By Lemma \ref{lemmagcuttinginds}:
\begin{equation*} g ( a_{[x]} ) g ( b_{[-l(a_{[x]})]}) = g ( a_{[x]} ) g ( b_{[-(l(a)-x)]}) = g ( a_{[x]} ) g ( b_{[x-l(a)]}).\end{equation*}
\end{proof}

We define $f$ as a single transformation of the tag system (adding $1101$ or $00$ and deleting 3 symbols) and $F$ as a function that applies $f$ to an input string until all symbols of the initial string are removed.
By definition, the following statement is correct:
\begin{lemma}\label{lemmaF}
$F(a) = g(a)_{[-l(a)]}$
\end{lemma}

The verification procedure related to Theorem \ref{maintheorem} is the following:

\begin{lemma}\label{lemmaverification}
For binary strings $a$, $b$, $c$, $d$, $e$, $f$ (of length at least 3) and numbers $x, y \in \{ 0, 1, 2\}$:
\begin{equation*}x - l(b) = y \Mod{3}, l(a) = l(c) = 0, g(a_{[x]}) = d, g(b_{[x]}) = e,g(c_{[y]}) = f \implies \end{equation*}
\begin{equation*} \implies \forall n, m \in \mathbb{N}_{0} : F( ( a^{n} b c^{m} )_{[x]}) = (d^{n} e f^{m})_{[y]}\end{equation*}
\end{lemma} 
\begin{proof}

By Lemma \ref{lemmaF}:
\begin{equation*}F( ( a^{n} b c^{m} )_{[x]}) = g( ( a^{n} b c^{m} )_{[x]} )_{[-l( ( a^{n} b c^{m} )_{[x]})]}.\end{equation*}

Lemma \ref{lemmagcuttinginds}, a property $l(st) = l(s) + l(t)$, conditions $l(a) = l(c) = 0$ and $x - l(b) = y$ can simplify cutting index:
\begin{equation}-l( ( a^{n} b c^{m} )_{[x]}) = - (l( a^{n} b c^{m}) - x) = x - l( a^{n} b c^{m} ) = x - (l(a) \times n + l (b) + l(c) \times m) = x - l(b) = y.\end{equation}

To simplify $g( ( a^{n} b c^{m} )_{[x]} )$, first, we need to iteratively apply Lemma \ref{lemmag} and Lemma \ref{lemmagcuttinginds} to the left part of string and use condition $l(a) = 0$:
\begin{equation*}g( ( a^{n} b c^{m} )_{[x]} ) = g( a_{[x]} ) g( ( a^{n-1} b c^{m} )_{[x - l(a)]} ) = g( a_{[x]} ) g( ( a^{n-1} b c^{m} )_{[x]} ),\end{equation*}
\begin{equation*}g( ( a^{n} b c^{m} )_{[x]} ) = g( a_{[x]} )^{n} g( (b c^{m} )_{[x]} ).\end{equation*}

Then we need to apply Lemma \ref{lemmag} to $b$ part and use condition $x - l(b) = y$:
\begin{equation*}g( a_{[x]} )^{n} g( (b c^{m} )_{[x]} ) = g( a_{[x]} )^{n} g( b_{[x]} ) g( (c^{m} )_{[x-l(b)]}) = g( a_{[x]} )^{n} g( b_{[x]} ) g( (c^{m} )_{[y]})\end{equation*}

Finally, we need to simplify the right side of a string the same way the left side was simplified:
\begin{equation*}g( a_{[x]} )^{n} g( b_{[x]} ) g( (c^{m} )_{[y]}) = g( a_{[x]} )^{n} g( b_{[x]} ) g(c_{[y]}) g( (c^{m-1} )_{[y-l(c)]}) = g( a_{x} )^{n} g( b_{[x]} ) g(c_{[y]}) g( (c^{m-1} )_{[y]})\end{equation*}
\begin{equation}g( a_{[x]} )^{n} g( b_{[x]} ) g( (c^{m} )_{[y]}) = g( a_{[x]} )^{n} g( b_{[x]} ) g(c_{[y]})^{m}\end{equation}

Combining (1) and (2) with remaining conditions produces the desired equation:
\begin{equation*}F( ( a^{n} b c^{m} )_{[x]}) = g( ( a^{n} b c^{m} )_{[x]} )_{[-l( ( a^{n} b c^{m} )_{[x]})]} = (g( a_{[x]} )^{n} g( b_{[x]} ) g(c_{[y]})^{m})_{[y]} = (d^{n} e f^{m})_{[y]}\end{equation*}

\end{proof}

Lemma \ref{lemmaverification} essentially builds a procedure to verify that a slightly truncated string in a form of $ a^{n} b c^{m} $ goes to a slightly truncated string in a form of $ d^{n} e f^{m} $ in a finite amount of steps for every possible $ n $ and $ m $.

To prove Theorem \ref{maintheorem} we need to create a sequence of quadruplets $(a_{i}, b_{i}, c_{i}, x_{i})$ (triples of strings and corresponding cutting indices) that starts with quadruplet $(a_{1}, b_{1}, c_{1}, x_{1})$ and ends with quadruplet $(a_{n+1}, b_{n+1}, c_{n+1}, x_{n+1}) = (a_{1}, a_{1}b_{1}c_{1}, c_{1}, x_{1})$, then iteratively apply Lemma \ref{lemmaverification} to each pair of adjacent quadruplets.
Such sequence $\omega$ of 14 quadruplets is presented in Appendix \ref{appendixa} of this paper.
Using Lemma \ref{lemmaverification} we can verify the following lemma for $\omega$:

\begin{lemma}\label{lemmaw} The following properties of $\omega$ are valid:
\begin{equation*}\textnormal{\textit{(i)}} \hspace{10.0pt} \forall i \in \{1, \dotsb , 13\}, \forall n, m \in \mathbb{N}_{0}: F( \omega_{i,1}^{n} \omega_{i,2} \omega_{i,3}^{m} )_{[\omega_{i,4}]} = ( \omega_{i+1,1}^{n} \omega_{i+1,2} \omega_{i+1,3}^{m} )_{[\omega_{i+1,4}]}\end{equation*}
\begin{equation*}\textnormal{\textit{(ii)}} \hspace{10.0pt} (\omega_{14,1}, \omega_{14,2}, \omega_{14,3}, \omega_{14,4}) = (\omega_{1,1}, \omega_{1,1}\omega_{1,2}\omega_{1,3}, \omega_{1,3}, \omega_{1,4})\end{equation*}
\end{lemma} 

Since conditions of Lemma \ref{lemmaverification} are equations in a form $l(s) = k$ or $g({s_1}_{[k_1]}) = {s_2}_{[k_2]}$, they can be easily verified by a computer (see the Github repository).

\section{Notion of building blocks}

In this section we briefly shift focus of this paper to heuristics that were used to find strings $a$, $b$ and $c$.
We introduce the notion of building blocks and make an important claim about the properties of these objects.
These structures were implemented along with simple heuristics and programming tricks, but since they have no scientific value, they wouldn't be mentioned.

Suppose $B = \{\varepsilon, v,vv\}\{0,1\}\{uu0,uu1\}^{*} \{\varepsilon,w,ww\}$.
A \textit{converting set} $B(a)$ for string $a$ from $\{v,u,w,0,1\}^{*}$ is defined as following:
\begin{equation*}
\begin{aligned}
B(a) = \{b \in B | {} & b \textrm{ derives from } a \textrm{ by replacing some of } w \textrm{ symbols with } u, \\
                      & \textrm{ some of } 0 \textrm{ and } 1 \textrm{ symbols with } v, u \textrm{ or } w\}
\end{aligned}
\end{equation*}

\vspace{\baselineskip}

\vspace{\baselineskip}

\vspace{\baselineskip}

\vspace{\baselineskip}

\vspace{\baselineskip}

\vspace{\baselineskip}

\vspace{\baselineskip}

\vspace{\baselineskip}

\vspace{\baselineskip}

For example:
\begin{figure}[H]
\begin{equation*}\hspace{23.1pt}B(1uu11100) = \{1uu1uu0w\},\end{equation*}
\begin{equation*}\hspace{22.1pt}B(v1w0) = \{v1ww\},\end{equation*}
\begin{equation*}\hspace{83.2pt}B(0000) = \{0uu0, v0ww, vv0w\},\end{equation*}
\begin{equation*}\hspace{70.5pt}B(111) = \{1ww, v1w, vv1\},\end{equation*}
\begin{equation*}\hspace{95.6pt}B(10101) = \{1uu0w, v0uu1, vv1ww\},\end{equation*}
\begin{equation*}\hspace{0.1pt}B(w1v) = \{\}.\end{equation*}
\end{figure}

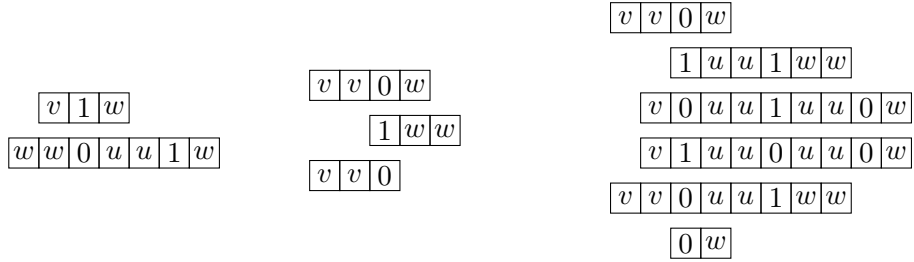
\begin{figure}[h]
\begin{tikzpicture}

\draw (0.40,-1.20) rectangle ++ (.4,-.4); \node[text centered] at (0.60, -1.40) {$v$};
\draw (0.80,-1.20) rectangle ++ (.4,-.4); \node[text centered] at (1.00, -1.40) {$1$};
\draw (1.20,-1.20) rectangle ++ (.4,-.4); \node[text centered] at (1.40, -1.40) {$w$};
\draw (0.00,-1.80) rectangle ++ (.4,-.4); \node[text centered] at (0.20, -2.00) {$w$};
\draw (0.40,-1.80) rectangle ++ (.4,-.4); \node[text centered] at (0.60, -2.00) {$w$};
\draw (0.80,-1.80) rectangle ++ (.4,-.4); \node[text centered] at (1.00, -2.00) {$0$};
\draw (1.20,-1.80) rectangle ++ (.4,-.4); \node[text centered] at (1.40, -2.00) {$u$};
\draw (1.60,-1.80) rectangle ++ (.4,-.4); \node[text centered] at (1.80, -2.00) {$u$};
\draw (2.00,-1.80) rectangle ++ (.4,-.4); \node[text centered] at (2.20, -2.00) {$1$};
\draw (2.40,-1.80) rectangle ++ (.4,-.4); \node[text centered] at (2.60, -2.00) {$w$};
\draw (4.00,-0.90) rectangle ++ (.4,-.4); \node[text centered] at (4.20, -1.10) {$v$};
\draw (4.40,-0.90) rectangle ++ (.4,-.4); \node[text centered] at (4.60, -1.10) {$v$};
\draw (4.80,-0.90) rectangle ++ (.4,-.4); \node[text centered] at (5.00, -1.10) {$0$};
\draw (5.20,-0.90) rectangle ++ (.4,-.4); \node[text centered] at (5.40, -1.10) {$w$};
\draw (4.80,-1.50) rectangle ++ (.4,-.4); \node[text centered] at (5.00, -1.70) {$1$};
\draw (5.20,-1.50) rectangle ++ (.4,-.4); \node[text centered] at (5.40, -1.70) {$w$};
\draw (5.60,-1.50) rectangle ++ (.4,-.4); \node[text centered] at (5.80, -1.70) {$w$};
\draw (4.00,-2.10) rectangle ++ (.4,-.4); \node[text centered] at (4.20, -2.30) {$v$};
\draw (4.40,-2.10) rectangle ++ (.4,-.4); \node[text centered] at (4.60, -2.30) {$v$};
\draw (4.80,-2.10) rectangle ++ (.4,-.4); \node[text centered] at (5.00, -2.30) {$0$};
\draw (8.00,0.00) rectangle ++ (.4,-.4); \node[text centered] at (8.20, -0.20) {$v$};
\draw (8.40,0.00) rectangle ++ (.4,-.4); \node[text centered] at (8.60, -0.20) {$v$};
\draw (8.80,0.00) rectangle ++ (.4,-.4); \node[text centered] at (9.00, -0.20) {$0$};
\draw (9.20,0.00) rectangle ++ (.4,-.4); \node[text centered] at (9.40, -0.20) {$w$};
\draw (8.80,-0.60) rectangle ++ (.4,-.4); \node[text centered] at (9.00, -0.80) {$1$};
\draw (9.20,-0.60) rectangle ++ (.4,-.4); \node[text centered] at (9.40, -0.80) {$u$};
\draw (9.60,-0.60) rectangle ++ (.4,-.4); \node[text centered] at (9.80, -0.80) {$u$};
\draw (10.00,-0.60) rectangle ++ (.4,-.4); \node[text centered] at (10.20, -0.80) {$1$};
\draw (10.40,-0.60) rectangle ++ (.4,-.4); \node[text centered] at (10.60, -0.80) {$w$};
\draw (10.80,-0.60) rectangle ++ (.4,-.4); \node[text centered] at (11.00, -0.80) {$w$};
\draw (8.40,-1.20) rectangle ++ (.4,-.4); \node[text centered] at (8.60, -1.40) {$v$};
\draw (8.80,-1.20) rectangle ++ (.4,-.4); \node[text centered] at (9.00, -1.40) {$0$};
\draw (9.20,-1.20) rectangle ++ (.4,-.4); \node[text centered] at (9.40, -1.40) {$u$};
\draw (9.60,-1.20) rectangle ++ (.4,-.4); \node[text centered] at (9.80, -1.40) {$u$};
\draw (10.00,-1.20) rectangle ++ (.4,-.4); \node[text centered] at (10.20, -1.40) {$1$};
\draw (10.40,-1.20) rectangle ++ (.4,-.4); \node[text centered] at (10.60, -1.40) {$u$};
\draw (10.80,-1.20) rectangle ++ (.4,-.4); \node[text centered] at (11.00, -1.40) {$u$};
\draw (11.20,-1.20) rectangle ++ (.4,-.4); \node[text centered] at (11.40, -1.40) {$0$};
\draw (11.60,-1.20) rectangle ++ (.4,-.4); \node[text centered] at (11.80, -1.40) {$w$};
\draw (8.40,-1.80) rectangle ++ (.4,-.4); \node[text centered] at (8.60, -2.00) {$v$};
\draw (8.80,-1.80) rectangle ++ (.4,-.4); \node[text centered] at (9.00, -2.00) {$1$};
\draw (9.20,-1.80) rectangle ++ (.4,-.4); \node[text centered] at (9.40, -2.00) {$u$};
\draw (9.60,-1.80) rectangle ++ (.4,-.4); \node[text centered] at (9.80, -2.00) {$u$};
\draw (10.00,-1.80) rectangle ++ (.4,-.4); \node[text centered] at (10.20, -2.00) {$0$};
\draw (10.40,-1.80) rectangle ++ (.4,-.4); \node[text centered] at (10.60, -2.00) {$u$};
\draw (10.80,-1.80) rectangle ++ (.4,-.4); \node[text centered] at (11.00, -2.00) {$u$};
\draw (11.20,-1.80) rectangle ++ (.4,-.4); \node[text centered] at (11.40, -2.00) {$0$};
\draw (11.60,-1.80) rectangle ++ (.4,-.4); \node[text centered] at (11.80, -2.00) {$w$};
\draw (8.00,-2.40) rectangle ++ (.4,-.4); \node[text centered] at (8.20, -2.60) {$v$};
\draw (8.40,-2.40) rectangle ++ (.4,-.4); \node[text centered] at (8.60, -2.60) {$v$};
\draw (8.80,-2.40) rectangle ++ (.4,-.4); \node[text centered] at (9.00, -2.60) {$0$};
\draw (9.20,-2.40) rectangle ++ (.4,-.4); \node[text centered] at (9.40, -2.60) {$u$};
\draw (9.60,-2.40) rectangle ++ (.4,-.4); \node[text centered] at (9.80, -2.60) {$u$};
\draw (10.00,-2.40) rectangle ++ (.4,-.4); \node[text centered] at (10.20, -2.60) {$1$};
\draw (10.40,-2.40) rectangle ++ (.4,-.4); \node[text centered] at (10.60, -2.60) {$w$};
\draw (10.80,-2.40) rectangle ++ (.4,-.4); \node[text centered] at (11.00, -2.60) {$w$};
\draw (8.80,-3.00) rectangle ++ (.4,-.4); \node[text centered] at (9.00, -3.20) {$0$};
\draw (9.20,-3.00) rectangle ++ (.4,-.4); \node[text centered] at (9.40, -3.20) {$w$};

\end{tikzpicture}
\caption{Illustration of building blocks.} \label{figillustrateblocks}
\end{figure}

We call $n$ strings from $B$ a \textit{building block}. We illustrate them as aligned arrays (See Fig. \ref{figillustrateblocks}).

A function $c(a, x): A^{*} \times A \rightarrow \mathbb{N}_{0}$ defined as a function that counts a number of symbols $x$ in string $a$.

\begin{figure}[h]
\begin{tikzpicture}

\draw (0.40,0.00) rectangle ++ (.4,-.4); \node[text centered] at (0.60, -0.20) {$v$};
\draw (0.80,0.00) rectangle ++ (.4,-.4); \node[text centered] at (1.00, -0.20) {$1$};
\draw (1.20,0.00) rectangle ++ (.4,-.4); \node[text centered] at (1.40, -0.20) {$u$};
\draw (1.60,0.00) rectangle ++ (.4,-.4); \node[text centered] at (1.80, -0.20) {$u$};
\draw (2.00,0.00) rectangle ++ (.4,-.4); \node[text centered] at (2.20, -0.20) {$0$};
\draw (2.40,0.00) rectangle ++ (.4,-.4); \node[text centered] at (2.60, -0.20) {$w$};
\draw (2.80,0.00) rectangle ++ (.4,-.4); \node[text centered] at (3.00, -0.20) {$w$};
\draw (0.40,-1.00) rectangle ++ (.4,-.4); \node[text centered] at (0.60, -1.20) {$v$};
\draw (0.80,-1.00) rectangle ++ (.4,-.4); \node[text centered] at (1.00, -1.20) {$1$};
\draw (1.20,-1.00) rectangle ++ (.4,-.4); \node[text centered] at (1.40, -1.20) {$u$};
\draw (1.60,-1.00) rectangle ++ (.4,-.4); \node[text centered] at (1.80, -1.20) {$u$};
\draw (2.00,-1.00) rectangle ++ (.4,-.4); \node[text centered] at (2.20, -1.20) {$1$};
\draw (2.40,-1.00) rectangle ++ (.4,-.4); \node[text centered] at (2.60, -1.20) {$u$};
\draw (2.80,-1.00) rectangle ++ (.4,-.4); \node[text centered] at (3.00, -1.20) {$u$};
\draw (3.20,-1.00) rectangle ++ (.4,-.4); \node[text centered] at (3.40, -1.20) {$0$};
\draw (3.60,-1.00) rectangle ++ (.4,-.4); \node[text centered] at (3.80, -1.20) {$w$};
\draw (4.00,-1.00) rectangle ++ (.4,-.4); \node[text centered] at (4.20, -1.20) {$w$};
\draw (0.00,-2.00) rectangle ++ (.4,-.4); \node[text centered] at (0.20, -2.20) {$v$};
\draw (0.40,-2.00) rectangle ++ (.4,-.4); \node[text centered] at (0.60, -2.20) {$v$};
\draw (0.80,-2.00) rectangle ++ (.4,-.4); \node[text centered] at (1.00, -2.20) {$0$};
\draw (1.20,-2.00) rectangle ++ (.4,-.4); \node[text centered] at (1.40, -2.20) {$u$};
\draw (1.60,-2.00) rectangle ++ (.4,-.4); \node[text centered] at (1.80, -2.20) {$u$};
\draw (2.00,-2.00) rectangle ++ (.4,-.4); \node[text centered] at (2.20, -2.20) {$0$};
\draw (2.40,-2.00) rectangle ++ (.4,-.4); \node[text centered] at (2.60, -2.20) {$u$};
\draw (2.80,-2.00) rectangle ++ (.4,-.4); \node[text centered] at (3.00, -2.20) {$u$};
\draw (3.20,-2.00) rectangle ++ (.4,-.4); \node[text centered] at (3.40, -2.20) {$0$};
\draw (0.80,-3.00) rectangle ++ (.4,-.4); \node[text centered] at (1.00, -3.20) {$0$};
\draw (1.20,-3.00) rectangle ++ (.4,-.4); \node[text centered] at (1.40, -3.20) {$u$};
\draw (1.60,-3.00) rectangle ++ (.4,-.4); \node[text centered] at (1.80, -3.20) {$u$};
\draw (2.00,-3.00) rectangle ++ (.4,-.4); \node[text centered] at (2.20, -3.20) {$1$};
\draw (2.40,-3.00) rectangle ++ (.4,-.4); \node[text centered] at (2.60, -3.20) {$u$};
\draw (2.80,-3.00) rectangle ++ (.4,-.4); \node[text centered] at (3.00, -3.20) {$u$};
\draw (3.20,-3.00) rectangle ++ (.4,-.4); \node[text centered] at (3.40, -3.20) {$0$};
\draw (10.40,0.00) rectangle ++ (.4,-.4); \node[text centered] at (10.60, -0.20) {$v$};
\draw (10.80,0.00) rectangle ++ (.4,-.4); \node[text centered] at (11.00, -0.20) {$1$};
\draw (11.20,0.00) rectangle ++ (.4,-.4); \node[text centered] at (11.40, -0.20) {$u$};
\draw (11.60,0.00) rectangle ++ (.4,-.4); \node[text centered] at (11.80, -0.20) {$u$};
\draw (12.00,0.00) rectangle ++ (.4,-.4); \node[text centered] at (12.20, -0.20) {$0$};
\draw (12.40,0.00) rectangle ++ (.4,-.4); \node[text centered] at (12.60, -0.20) {$w$};
\draw (12.80,0.00) rectangle ++ (.4,-.4); \node[text centered] at (13.00, -0.20) {$w$};
\draw (13.20,0.00) rectangle ++ (.4,-.4); \node[text centered] at (13.40, -0.20) {$1$};
\draw (13.60,0.00) rectangle ++ (.4,-.4); \node[text centered] at (13.80, -0.20) {$1$};
\draw (10.40,-1.00) rectangle ++ (.4,-.4); \node[text centered] at (10.60, -1.20) {$v$};
\draw (10.80,-1.00) rectangle ++ (.4,-.4); \node[text centered] at (11.00, -1.20) {$1$};
\draw (11.20,-1.00) rectangle ++ (.4,-.4); \node[text centered] at (11.40, -1.20) {$u$};
\draw (11.60,-1.00) rectangle ++ (.4,-.4); \node[text centered] at (11.80, -1.20) {$u$};
\draw (12.00,-1.00) rectangle ++ (.4,-.4); \node[text centered] at (12.20, -1.20) {$1$};
\draw (12.40,-1.00) rectangle ++ (.4,-.4); \node[text centered] at (12.60, -1.20) {$u$};
\draw (12.80,-1.00) rectangle ++ (.4,-.4); \node[text centered] at (13.00, -1.20) {$u$};
\draw (13.20,-1.00) rectangle ++ (.4,-.4); \node[text centered] at (13.40, -1.20) {$0$};
\draw (13.60,-1.00) rectangle ++ (.4,-.4); \node[text centered] at (13.80, -1.20) {$w$};
\draw (14.00,-1.00) rectangle ++ (.4,-.4); \node[text centered] at (14.20, -1.20) {$w$};
\draw (14.40,-1.00) rectangle ++ (.4,-.4); \node[text centered] at (14.60, -1.20) {$1$};
\draw (14.80,-1.00) rectangle ++ (.4,-.4); \node[text centered] at (15.00, -1.20) {$1$};
\draw (15.20,-1.00) rectangle ++ (.4,-.4); \node[text centered] at (15.40, -1.20) {$0$};
\draw (15.60,-1.00) rectangle ++ (.4,-.4); \node[text centered] at (15.80, -1.20) {$1$};
\draw (10.00,-2.00) rectangle ++ (.4,-.4); \node[text centered] at (10.20, -2.20) {$v$};
\draw (10.40,-2.00) rectangle ++ (.4,-.4); \node[text centered] at (10.60, -2.20) {$v$};
\draw (10.80,-2.00) rectangle ++ (.4,-.4); \node[text centered] at (11.00, -2.20) {$0$};
\draw (11.20,-2.00) rectangle ++ (.4,-.4); \node[text centered] at (11.40, -2.20) {$u$};
\draw (11.60,-2.00) rectangle ++ (.4,-.4); \node[text centered] at (11.80, -2.20) {$u$};
\draw (12.00,-2.00) rectangle ++ (.4,-.4); \node[text centered] at (12.20, -2.20) {$0$};
\draw (12.40,-2.00) rectangle ++ (.4,-.4); \node[text centered] at (12.60, -2.20) {$u$};
\draw (12.80,-2.00) rectangle ++ (.4,-.4); \node[text centered] at (13.00, -2.20) {$u$};
\draw (13.20,-2.00) rectangle ++ (.4,-.4); \node[text centered] at (13.40, -2.20) {$0$};
\draw (13.60,-2.00) rectangle ++ (.4,-.4); \node[text centered] at (13.80, -2.20) {$1$};
\draw (14.00,-2.00) rectangle ++ (.4,-.4); \node[text centered] at (14.20, -2.20) {$1$};
\draw (14.40,-2.00) rectangle ++ (.4,-.4); \node[text centered] at (14.60, -2.20) {$0$};
\draw (14.80,-2.00) rectangle ++ (.4,-.4); \node[text centered] at (15.00, -2.20) {$1$};
\draw (15.20,-2.00) rectangle ++ (.4,-.4); \node[text centered] at (15.40, -2.20) {$1$};
\draw (15.60,-2.00) rectangle ++ (.4,-.4); \node[text centered] at (15.80, -2.20) {$1$};
\draw (16.00,-2.00) rectangle ++ (.4,-.4); \node[text centered] at (16.20, -2.20) {$0$};
\draw (16.40,-2.00) rectangle ++ (.4,-.4); \node[text centered] at (16.60, -2.20) {$1$};
\draw (10.80,-3.00) rectangle ++ (.4,-.4); \node[text centered] at (11.00, -3.20) {$0$};
\draw (11.20,-3.00) rectangle ++ (.4,-.4); \node[text centered] at (11.40, -3.20) {$u$};
\draw (11.60,-3.00) rectangle ++ (.4,-.4); \node[text centered] at (11.80, -3.20) {$u$};
\draw (12.00,-3.00) rectangle ++ (.4,-.4); \node[text centered] at (12.20, -3.20) {$1$};
\draw (12.40,-3.00) rectangle ++ (.4,-.4); \node[text centered] at (12.60, -3.20) {$u$};
\draw (12.80,-3.00) rectangle ++ (.4,-.4); \node[text centered] at (13.00, -3.20) {$u$};
\draw (13.20,-3.00) rectangle ++ (.4,-.4); \node[text centered] at (13.40, -3.20) {$0$};
\draw (13.60,-3.00) rectangle ++ (.4,-.4); \node[text centered] at (13.80, -3.20) {$0$};
\draw (14.00,-3.00) rectangle ++ (.4,-.4); \node[text centered] at (14.20, -3.20) {$0$};
\draw (14.40,-3.00) rectangle ++ (.4,-.4); \node[text centered] at (14.60, -3.20) {$1$};
\draw (14.80,-3.00) rectangle ++ (.4,-.4); \node[text centered] at (15.00, -3.20) {$1$};
\draw (15.20,-3.00) rectangle ++ (.4,-.4); \node[text centered] at (15.40, -3.20) {$0$};
\draw (15.60,-3.00) rectangle ++ (.4,-.4); \node[text centered] at (15.80, -3.20) {$1$};
\draw [rounded corners=2pt] (14.4,-0.95) -- (13.20,-0.45) -- (13.60,-0.45) -- (16.00,-0.95);
\draw [rounded corners=2pt] (13.6,-1.95) -- (14.40,-1.45) -- (14.80,-1.45) -- (15.20,-1.95);
\draw [rounded corners=2pt] (15.2,-1.95) -- (15.60,-1.45) -- (16.00,-1.45) -- (16.80,-1.95);
\draw [rounded corners=2pt] (13.6,-2.95) -- (14.40,-2.45) -- (14.80,-2.45) -- (14.40,-2.95);
\draw [rounded corners=2pt] (14.4,-2.95) -- (15.60,-2.45) -- (16.00,-2.45) -- (16.00,-2.95);
\draw (4.40,-4.00) rectangle ++ (.4,-.4); \node[text centered] at (4.60, -4.20) {$v$};
\draw (4.80,-4.00) rectangle ++ (.4,-.4); \node[text centered] at (5.00, -4.20) {$1$};
\draw (5.20,-4.00) rectangle ++ (.4,-.4); \node[text centered] at (5.40, -4.20) {$u$};
\draw (5.60,-4.00) rectangle ++ (.4,-.4); \node[text centered] at (5.80, -4.20) {$u$};
\draw (6.00,-4.00) rectangle ++ (.4,-.4); \node[text centered] at (6.20, -4.20) {$0$};
\draw (6.40,-4.00) rectangle ++ (.4,-.4); \node[text centered] at (6.60, -4.20) {$u$};
\draw (6.80,-4.00) rectangle ++ (.4,-.4); \node[text centered] at (7.00, -4.20) {$u$};
\draw (7.20,-4.00) rectangle ++ (.4,-.4); \node[text centered] at (7.40, -4.20) {$1$};
\draw (7.60,-4.00) rectangle ++ (.4,-.4); \node[text centered] at (7.80, -4.20) {$w$};
\draw (4.40,-5.00) rectangle ++ (.4,-.4); \node[text centered] at (4.60, -5.20) {$v$};
\draw (4.80,-5.00) rectangle ++ (.4,-.4); \node[text centered] at (5.00, -5.20) {$1$};
\draw (5.20,-5.00) rectangle ++ (.4,-.4); \node[text centered] at (5.40, -5.20) {$u$};
\draw (5.60,-5.00) rectangle ++ (.4,-.4); \node[text centered] at (5.80, -5.20) {$u$};
\draw (6.00,-5.00) rectangle ++ (.4,-.4); \node[text centered] at (6.20, -5.20) {$1$};
\draw (6.40,-5.00) rectangle ++ (.4,-.4); \node[text centered] at (6.60, -5.20) {$u$};
\draw (6.80,-5.00) rectangle ++ (.4,-.4); \node[text centered] at (7.00, -5.20) {$u$};
\draw (7.20,-5.00) rectangle ++ (.4,-.4); \node[text centered] at (7.40, -5.20) {$0$};
\draw (7.60,-5.00) rectangle ++ (.4,-.4); \node[text centered] at (7.80, -5.20) {$u$};
\draw (8.00,-5.00) rectangle ++ (.4,-.4); \node[text centered] at (8.20, -5.20) {$u$};
\draw (8.40,-5.00) rectangle ++ (.4,-.4); \node[text centered] at (8.60, -5.20) {$1$};
\draw (8.80,-5.00) rectangle ++ (.4,-.4); \node[text centered] at (9.00, -5.20) {$u$};
\draw (9.20,-5.00) rectangle ++ (.4,-.4); \node[text centered] at (9.40, -5.20) {$u$};
\draw (9.60,-5.00) rectangle ++ (.4,-.4); \node[text centered] at (9.80, -5.20) {$1$};
\draw (4.00,-6.00) rectangle ++ (.4,-.4); \node[text centered] at (4.20, -6.20) {$v$};
\draw (4.40,-6.00) rectangle ++ (.4,-.4); \node[text centered] at (4.60, -6.20) {$v$};
\draw (4.80,-6.00) rectangle ++ (.4,-.4); \node[text centered] at (5.00, -6.20) {$0$};
\draw (5.20,-6.00) rectangle ++ (.4,-.4); \node[text centered] at (5.40, -6.20) {$u$};
\draw (5.60,-6.00) rectangle ++ (.4,-.4); \node[text centered] at (5.80, -6.20) {$u$};
\draw (6.00,-6.00) rectangle ++ (.4,-.4); \node[text centered] at (6.20, -6.20) {$0$};
\draw (6.40,-6.00) rectangle ++ (.4,-.4); \node[text centered] at (6.60, -6.20) {$u$};
\draw (6.80,-6.00) rectangle ++ (.4,-.4); \node[text centered] at (7.00, -6.20) {$u$};
\draw (7.20,-6.00) rectangle ++ (.4,-.4); \node[text centered] at (7.40, -6.20) {$0$};
\draw (7.60,-6.00) rectangle ++ (.4,-.4); \node[text centered] at (7.80, -6.20) {$u$};
\draw (8.00,-6.00) rectangle ++ (.4,-.4); \node[text centered] at (8.20, -6.20) {$u$};
\draw (8.40,-6.00) rectangle ++ (.4,-.4); \node[text centered] at (8.60, -6.20) {$0$};
\draw (8.80,-6.00) rectangle ++ (.4,-.4); \node[text centered] at (9.00, -6.20) {$u$};
\draw (9.20,-6.00) rectangle ++ (.4,-.4); \node[text centered] at (9.40, -6.20) {$u$};
\draw (9.60,-6.00) rectangle ++ (.4,-.4); \node[text centered] at (9.80, -6.20) {$1$};
\draw (10.00,-6.00) rectangle ++ (.4,-.4); \node[text centered] at (10.20, -6.20) {$w$};
\draw (10.40,-6.00) rectangle ++ (.4,-.4); \node[text centered] at (10.60, -6.20) {$w$};
\draw (4.80,-7.00) rectangle ++ (.4,-.4); \node[text centered] at (5.00, -7.20) {$0$};
\draw (5.20,-7.00) rectangle ++ (.4,-.4); \node[text centered] at (5.40, -7.20) {$u$};
\draw (5.60,-7.00) rectangle ++ (.4,-.4); \node[text centered] at (5.80, -7.20) {$u$};
\draw (6.00,-7.00) rectangle ++ (.4,-.4); \node[text centered] at (6.20, -7.20) {$1$};
\draw (6.40,-7.00) rectangle ++ (.4,-.4); \node[text centered] at (6.60, -7.20) {$u$};
\draw (6.80,-7.00) rectangle ++ (.4,-.4); \node[text centered] at (7.00, -7.20) {$u$};
\draw (7.20,-7.00) rectangle ++ (.4,-.4); \node[text centered] at (7.40, -7.20) {$0$};
\draw (7.60,-7.00) rectangle ++ (.4,-.4); \node[text centered] at (7.80, -7.20) {$u$};
\draw (8.00,-7.00) rectangle ++ (.4,-.4); \node[text centered] at (8.20, -7.20) {$u$};
\draw (8.40,-7.00) rectangle ++ (.4,-.4); \node[text centered] at (8.60, -7.20) {$1$};
\draw (8.80,-7.00) rectangle ++ (.4,-.4); \node[text centered] at (9.00, -7.20) {$u$};
\draw (9.20,-7.00) rectangle ++ (.4,-.4); \node[text centered] at (9.40, -7.20) {$u$};
\draw (9.60,-7.00) rectangle ++ (.4,-.4); \node[text centered] at (9.80, -7.20) {$1$};

\draw [->] (5.2, -1.7) -- (9.1, -1.7);

\draw [->] (9.6, -3.3) -- (9.0, -3.9);

\end{tikzpicture}
\caption{Illustration of Algorithm \ref{alg1}.} \label{figillustratealg1}
\end{figure}
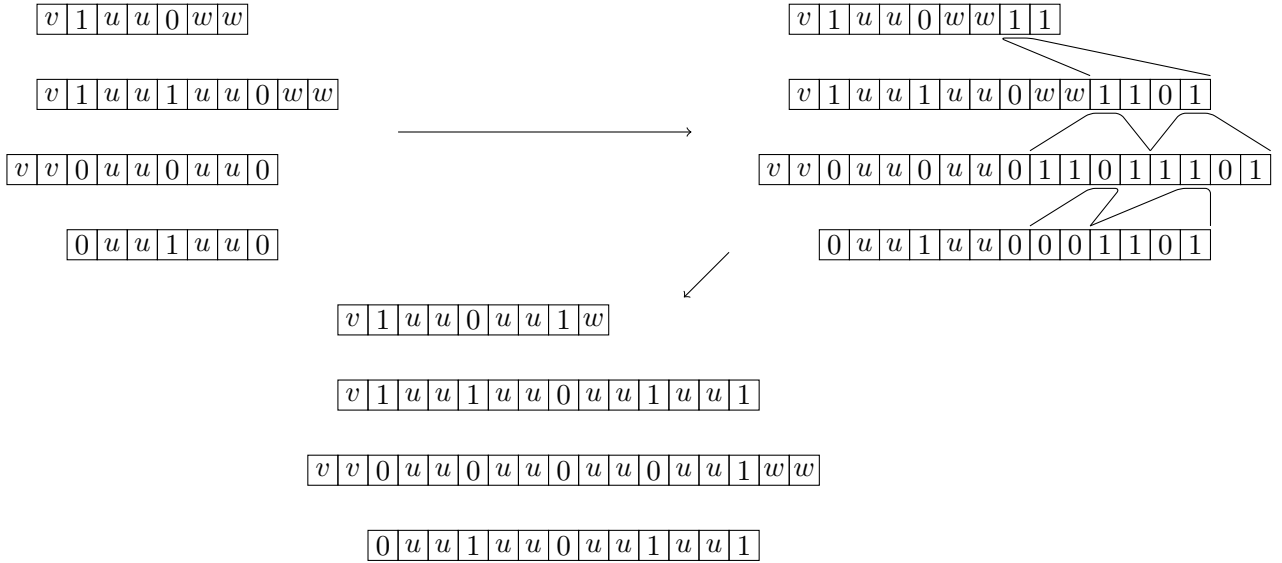

An important procedure that helps to work with building blocks is block extension.
The procedure (for the right) is presented as Algorithm \ref{alg1} and illustrated in Fig. \ref{figillustratealg1}.
The extension to the left side works the same way, except we add $\hat{b}_{i}$ to the left side and converting set needs to be redefined.

\begin{algorithm}[h]
\caption{Building block extension from the right side}  \label{alg1}
\textbf{Input:} Building block $(b_{1}, \dots, b_{n+1})$\\
\textbf{Output:} None \Comment \textit{in-place algorithm} \\
\texttt{\\}
Create a string $\hat{b}_{1}$ that $B(b_{1} \hat{b}_{1}) = \{b\}$ and $c(b, 0) + c(b, 1) - c(b_{1}, 0) - c(b_{1}, 1) = 1$\\
\For{$i$ in $\{1, \dotsb, n\}$}{
   add $\hat{b}_{i}$ to the right side of $b_{i}$\\
   replace $b_{i} \hat{b}_{i}$ with element from $B({b}_{i}\hat{b}_{i})$\\
   \eIf{all symbols of $\hat{b}_{i}$ in $b_{i} \hat{b}_{i}$ are changed during replacement}{
       \textbf{return}
   }
   {
       form a string $\hat{b}_{i+1}$ by taking unchanged $0$'s and $1$'s of $\hat{b}_{i}$ and changing $0$'s to $00$'s and$1$'s to $1101$'s
   }
}
\textbf{return}
\end{algorithm}

\begin{figure}[h]
\begin{tikzpicture}

\draw (0.4,0.0) rectangle ++ (.4,-.4); \node[text centered] at (0.6, -0.2) {$v$};
\draw (0.8,0.0) rectangle ++ (.4,-.4); \node[text centered] at (1.0, -0.2) {$1$};
\draw (1.2,0.0) rectangle ++ (.4,-.4); \node[text centered] at (1.4, -0.2) {$w$};
\draw (0.8,-1.0) rectangle ++ (.4,-.4); \node[text centered] at (1.0, -1.2) {$1$};
\draw (1.2,-1.0) rectangle ++ (.4,-.4); \node[text centered] at (1.4, -1.2) {$1$};
\draw (1.6,-1.0) rectangle ++ (.4,-.4); \node[text centered] at (1.8, -1.2) {$0$};
\draw (2.0,-1.0) rectangle ++ (.4,-.4); \node[text centered] at (2.2, -1.2) {$1$};
\draw (0.0,-2.0) rectangle ++ (.4,-.4); \node[text centered] at (0.2, -2.2) {$1$};
\draw (0.4,-2.0) rectangle ++ (.4,-.4); \node[text centered] at (0.6, -2.2) {$1$};
\draw (0.8,-2.0) rectangle ++ (.4,-.4); \node[text centered] at (1.0, -2.2) {$0$};
\draw (1.2,-2.0) rectangle ++ (.4,-.4); \node[text centered] at (1.4, -2.2) {$1$};
\draw (1.6,-2.0) rectangle ++ (.4,-.4); \node[text centered] at (1.8, -2.2) {$1$};
\draw (2.0,-2.0) rectangle ++ (.4,-.4); \node[text centered] at (2.2, -2.2) {$1$};
\draw (2.4,-2.0) rectangle ++ (.4,-.4); \node[text centered] at (2.6, -2.2) {$0$};
\draw (2.8,-2.0) rectangle ++ (.4,-.4); \node[text centered] at (3.0, -2.2) {$1$};
\draw (0.4,-3.0) rectangle ++ (.4,-.4); \node[text centered] at (0.6, -3.2) {$0$};
\draw (0.8,-3.0) rectangle ++ (.4,-.4); \node[text centered] at (1.0, -3.2) {$0$};
\draw (1.2,-3.0) rectangle ++ (.4,-.4); \node[text centered] at (1.4, -3.2) {$1$};
\draw (1.6,-3.0) rectangle ++ (.4,-.4); \node[text centered] at (1.8, -3.2) {$1$};
\draw (2.0,-3.0) rectangle ++ (.4,-.4); \node[text centered] at (2.2, -3.2) {$0$};
\draw (2.4,-3.0) rectangle ++ (.4,-.4); \node[text centered] at (2.6, -3.2) {$1$};
\draw (0.4,-4.0) rectangle ++ (.4,-.4); \node[text centered] at (0.6, -4.2) {$0$};
\draw (0.8,-4.0) rectangle ++ (.4,-.4); \node[text centered] at (1.0, -4.2) {$0$};
\draw (1.2,-4.0) rectangle ++ (.4,-.4); \node[text centered] at (1.4, -4.2) {$0$};
\draw (1.6,-4.0) rectangle ++ (.4,-.4); \node[text centered] at (1.8, -4.2) {$0$};
\draw (0.8,-5.0) rectangle ++ (.4,-.4); \node[text centered] at (1.0, -5.2) {$0$};
\draw (1.2,-5.0) rectangle ++ (.4,-.4); \node[text centered] at (1.4, -5.2) {$0$};
\draw [rounded corners=2pt] (0.8,-0.95) -- (0.8,-0.45) -- (1.2,-0.45) -- (2.4,-0.95);
\draw [rounded corners=2pt] (0.0,-1.95) -- (0.8,-1.45) -- (1.2,-1.45) -- (1.6,-1.95);
\draw [rounded corners=2pt] (1.6,-1.95) -- (2.0,-1.45) -- (2.4,-1.45) -- (3.2,-1.95);
\draw [rounded corners=2pt] (0.4,-2.95) -- (0.8,-2.45) -- (1.2,-2.45) -- (1.2,-2.95);
\draw [rounded corners=2pt] (1.2,-2.95) -- (2.0,-2.45) -- (2.4,-2.45) -- (2.8,-2.95);
\draw [rounded corners=2pt] (0.4,-3.95) -- (0.8,-3.45) -- (1.2,-3.45) -- (1.2,-3.95);
\draw [rounded corners=2pt] (1.2,-3.95) -- (2.0,-3.45) -- (2.4,-3.45) -- (2.0,-3.95);
\draw [rounded corners=2pt] (0.8,-4.95) -- (0.8,-4.45) -- (1.2,-4.45) -- (1.6,-4.95);
\draw (5.4,0.0) rectangle ++ (.4,-.4); \node[text centered] at (5.6, -0.2) {$v$};
\draw (5.8,0.0) rectangle ++ (.4,-.4); \node[text centered] at (6.0, -0.2) {$1$};
\draw (6.2,0.0) rectangle ++ (.4,-.4); \node[text centered] at (6.4, -0.2) {$w$};
\draw (5.8,-1.0) rectangle ++ (.4,-.4); \node[text centered] at (6.0, -1.2) {$1$};
\draw (6.2,-1.0) rectangle ++ (.4,-.4); \node[text centered] at (6.4, -1.2) {$u$};
\draw (6.6,-1.0) rectangle ++ (.4,-.4); \node[text centered] at (6.8, -1.2) {$u$};
\draw (7.0,-1.0) rectangle ++ (.4,-.4); \node[text centered] at (7.2, -1.2) {$1$};
\draw (5.0,-2.0) rectangle ++ (.4,-.4); \node[text centered] at (5.2, -2.2) {$v$};
\draw (5.4,-2.0) rectangle ++ (.4,-.4); \node[text centered] at (5.6, -2.2) {$v$};
\draw (5.8,-2.0) rectangle ++ (.4,-.4); \node[text centered] at (6.0, -2.2) {$0$};
\draw (6.2,-2.0) rectangle ++ (.4,-.4); \node[text centered] at (6.4, -2.2) {$u$};
\draw (6.6,-2.0) rectangle ++ (.4,-.4); \node[text centered] at (6.8, -2.2) {$u$};
\draw (7.0,-2.0) rectangle ++ (.4,-.4); \node[text centered] at (7.2, -2.2) {$1$};
\draw (7.4,-2.0) rectangle ++ (.4,-.4); \node[text centered] at (7.6, -2.2) {$w$};
\draw (7.8,-2.0) rectangle ++ (.4,-.4); \node[text centered] at (8.0, -2.2) {$w$};
\draw (5.4,-3.0) rectangle ++ (.4,-.4); \node[text centered] at (5.6, -3.2) {$v$};
\draw (5.8,-3.0) rectangle ++ (.4,-.4); \node[text centered] at (6.0, -3.2) {$0$};
\draw (6.2,-3.0) rectangle ++ (.4,-.4); \node[text centered] at (6.4, -3.2) {$u$};
\draw (6.6,-3.0) rectangle ++ (.4,-.4); \node[text centered] at (6.8, -3.2) {$u$};
\draw (7.0,-3.0) rectangle ++ (.4,-.4); \node[text centered] at (7.2, -3.2) {$0$};
\draw (7.4,-3.0) rectangle ++ (.4,-.4); \node[text centered] at (7.6, -3.2) {$w$};
\draw (5.4,-4.0) rectangle ++ (.4,-.4); \node[text centered] at (5.6, -4.2) {$v$};
\draw (5.8,-4.0) rectangle ++ (.4,-.4); \node[text centered] at (6.0, -4.2) {$0$};
\draw (6.2,-4.0) rectangle ++ (.4,-.4); \node[text centered] at (6.4, -4.2) {$w$};
\draw (6.6,-4.0) rectangle ++ (.4,-.4); \node[text centered] at (6.8, -4.2) {$w$};
\draw (5.8,-5.0) rectangle ++ (.4,-.4); \node[text centered] at (6.0, -5.2) {$0$};
\draw (6.2,-5.0) rectangle ++ (.4,-.4); \node[text centered] at (6.4, -5.2) {$w$};
\draw [->] (3.6, -2.7) -- (4.7, -2.7);

\end{tikzpicture}
\caption{Illustration of Algorithm \ref{alg2}.} \label{figillustratealg2}
\end{figure}
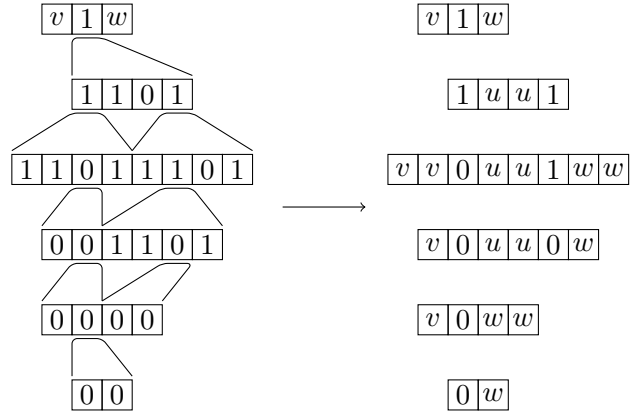

We call a building block an \textit{initial building block} if it is constructed with another procedure, defined as Algorithm \ref{alg2} and illustrated in Fig \ref{figillustratealg2}

\begin{algorithm}[h]
\caption{Creation of initial block}  \label{alg2}
\textbf{Input:} None\\
\textbf{Output:} Building block $(b_{1}, \dots, b_{n+1})$ or None\\
\texttt{\\}
Choose $\hat{b}_{1}$ from $\{\epsilon, v, vv\}\{0, 1\}\{\epsilon, w, ww\}$\\
\For{$i$ in $\{1, \dotsb, n\}$}{
   replace $\hat{b}_{i}$ with element $b_{i}$ from $B(\hat{b}_{i})$\\
   \eIf{$c(b_{i}, 0) + c(b_{i}, 1) = 0 $}{
       \textbf{return} None
   }
   {
       form a string $\hat{b}_{i+1}$ by taking $0$'s and $1$'s of ${b}_{i}$ and changing $0$'s to $00$'s and $1$'s to $1101$'s
   }
}
replace $\hat{b}_{i}$ with element $b_{i}$ from $B(\hat{b}_{i})$\\
\textbf{return} $(b_{1}, \dots, b_{n+1})$
\end{algorithm}

\begin{conjecture}\label{conjecture1}
If $(b_{1}, \dots, b_{n+1})$ is a building block and:
\begin{enumerate}[label=(\roman*)]
  \item \label{cond1} $(b_{1}, \dots, b_{n+1})$ is constructed from initial building block by several extensions\\
  \item \label{cond2} $b_{1} = b_{n+1}$\\
  \item \label{cond3} $\forall i \in \{1, \dotsb , n\}: c(b_{i}, w) + c(b_{i+1}, v) = 2$\\
  \item \label{cond4} $\exists i \in \{1, \dotsb , n\}: c(b_{i}, v) = 0$\end{enumerate}
then $(b_{1}, \dots, b_{n+1})$ can be reconstructed to periodic evolution of Post's system of tag.
\end{conjecture}

This conjecture has several important implications.
Mainly it creates a new method for finding periodic evolutions for Post's system of tag. 
Since for a fixed $n$ the amount of possible initial configurations and block extensions is limited, the search becomes much easier.
In particular, most extensions can not be performed one after another, and we can graph connections between them to use this graph to speed up computations.
The search goal now becomes to start from any initial building block and extend it using constructed graph so conditions \ref{cond2}, \ref{cond3} and \ref{cond4} are met.
This method allows to check (in implicit way) configurations with extremely large lengths, since now we iterate over loop lengths and not word lengths, hence producing more interesting results.

Conjecture \ref{conjecture1} concerns periodic evolutions of Post's system of tag, but it can be used for non-periodic evolutions as well - the only difference is a slightly more complicated condition \ref{cond2}.
In general, switching from "vertical search" to "horizontal search" is a useful tool to analyze behavior of any system of tag.
This tool is not well suited for Collatz conjecture, as the length of any building block will be too big to use extensions graph, but we hope that maybe in the future these structures will help to find insights for this problem too.

\section{Acknowledgements}

I want to thank my company management, notably Eugene I. Khorev, for providing me with a flexible work schedule and computational power to complete this project.
I also thank Kirill V. Sapunov for reading the initial draft of this paper.
I thank prof. Alexander P. Ryjov for supporting me during publication process.

\appendix  
\section{Sequence $\omega$}\label{appendixa}

Strings $a$, $b$ and $c$ can be rewritten more compactly since all of them consist of substrings $00$ ( \tikz{\filldraw[fill=cancan, draw=black] (0, 0) rectangle (0.25, 0.25);} ) and $1101$ ( 
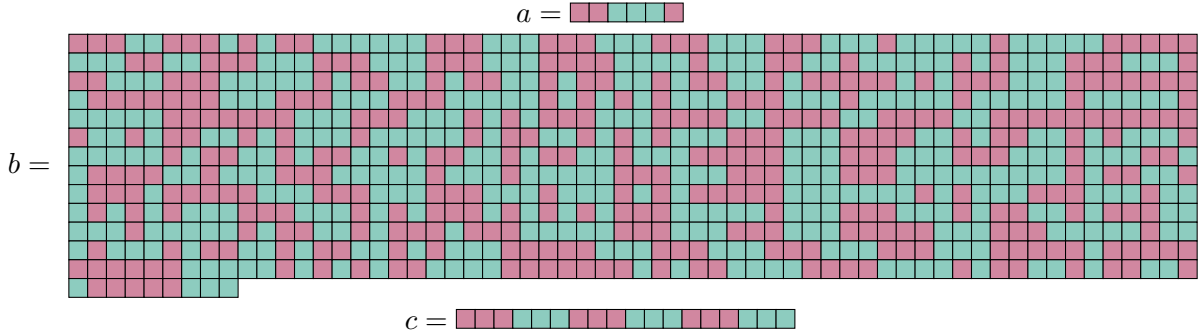
\tikz{\filldraw[fill=bermuda, draw=black] (0, 0) rectangle (0.25, 0.25);} ):
\begin{equation*}a = \tikz{\filldraw[fill=cancan, draw=black] (0.00, 0.00) rectangle (0.25, 0.25);
                           \filldraw[fill=cancan, draw=black] (0.25, 0.00) rectangle (0.50, 0.25);
                           \filldraw[fill=bermuda, draw=black] (0.50, 0.00) rectangle (0.75, 0.25);
                           \filldraw[fill=bermuda, draw=black] (0.75, 0.00) rectangle (1.00, 0.25);
                           \filldraw[fill=bermuda, draw=black] (1.00, 0.00) rectangle (1.25, 0.25);
                           \filldraw[fill=cancan, draw=black] (1.25, 0.00) rectangle (1.50, 0.25);}\end{equation*}
\begin{equation*}
b = \vcenter{\hbox{ \tikz{\filldraw[fill=cancan, draw=black] (0.00, 0.00) rectangle (0.25, -0.25);
                          \filldraw[fill=cancan, draw=black] (0.25, 0.00) rectangle (0.50, -0.25);
                          \filldraw[fill=cancan, draw=black] (0.50, 0.00) rectangle (0.75, -0.25);
                          \filldraw[fill=bermuda, draw=black] (0.75, 0.00) rectangle (1.00, -0.25);
                          \filldraw[fill=bermuda, draw=black] (1.00, 0.00) rectangle (1.25, -0.25);
                          \filldraw[fill=cancan, draw=black] (1.25, 0.00) rectangle (1.50, -0.25);
                          \filldraw[fill=cancan, draw=black] (1.50, 0.00) rectangle (1.75, -0.25);
                          \filldraw[fill=cancan, draw=black] (1.75, 0.00) rectangle (2.00, -0.25);
                          \filldraw[fill=bermuda, draw=black] (2.00, 0.00) rectangle (2.25, -0.25);
                          \filldraw[fill=cancan, draw=black] (2.25, 0.00) rectangle (2.50, -0.25);
                          \filldraw[fill=bermuda, draw=black] (2.50, 0.00) rectangle (2.75, -0.25);
                          \filldraw[fill=cancan, draw=black] (2.75, 0.00) rectangle (3.00, -0.25);
                          \filldraw[fill=cancan, draw=black] (3.00, 0.00) rectangle (3.25, -0.25);
                          \filldraw[fill=bermuda, draw=black] (3.25, 0.00) rectangle (3.50, -0.25);
                          \filldraw[fill=bermuda, draw=black] (3.50, 0.00) rectangle (3.75, -0.25);
                          \filldraw[fill=bermuda, draw=black] (3.75, 0.00) rectangle (4.00, -0.25);
                          \filldraw[fill=bermuda, draw=black] (4.00, 0.00) rectangle (4.25, -0.25);
                          \filldraw[fill=bermuda, draw=black] (4.25, 0.00) rectangle (4.50, -0.25);
                          \filldraw[fill=bermuda, draw=black] (4.50, 0.00) rectangle (4.75, -0.25);
                          \filldraw[fill=cancan, draw=black] (4.75, 0.00) rectangle (5.00, -0.25);
                          \filldraw[fill=cancan, draw=black] (5.00, 0.00) rectangle (5.25, -0.25);
                          \filldraw[fill=cancan, draw=black] (5.25, 0.00) rectangle (5.50, -0.25);
                          \filldraw[fill=bermuda, draw=black] (5.50, 0.00) rectangle (5.75, -0.25);
                          \filldraw[fill=bermuda, draw=black] (5.75, 0.00) rectangle (6.00, -0.25);
                          \filldraw[fill=bermuda, draw=black] (6.00, 0.00) rectangle (6.25, -0.25);
                          \filldraw[fill=cancan, draw=black] (6.25, 0.00) rectangle (6.50, -0.25);
                          \filldraw[fill=cancan, draw=black] (6.50, 0.00) rectangle (6.75, -0.25);
                          \filldraw[fill=cancan, draw=black] (6.75, 0.00) rectangle (7.00, -0.25);
                          \filldraw[fill=bermuda, draw=black] (7.00, 0.00) rectangle (7.25, -0.25);
                          \filldraw[fill=bermuda, draw=black] (7.25, 0.00) rectangle (7.50, -0.25);
                          \filldraw[fill=bermuda, draw=black] (7.50, 0.00) rectangle (7.75, -0.25);
                          \filldraw[fill=cancan, draw=black] (7.75, 0.00) rectangle (8.00, -0.25);
                          \filldraw[fill=cancan, draw=black] (8.00, 0.00) rectangle (8.25, -0.25);
                          \filldraw[fill=cancan, draw=black] (8.25, 0.00) rectangle (8.50, -0.25);
                          \filldraw[fill=bermuda, draw=black] (8.50, 0.00) rectangle (8.75, -0.25);
                          \filldraw[fill=bermuda, draw=black] (8.75, 0.00) rectangle (9.00, -0.25);
                          \filldraw[fill=bermuda, draw=black] (9.00, 0.00) rectangle (9.25, -0.25);
                          \filldraw[fill=cancan, draw=black] (9.25, 0.00) rectangle (9.50, -0.25);
                          \filldraw[fill=cancan, draw=black] (9.50, 0.00) rectangle (9.75, -0.25);
                          \filldraw[fill=cancan, draw=black] (9.75, 0.00) rectangle (10.00, -0.25);
                          \filldraw[fill=bermuda, draw=black] (10.00, 0.00) rectangle (10.25, -0.25);
                          \filldraw[fill=bermuda, draw=black] (10.25, 0.00) rectangle (10.50, -0.25);
                          \filldraw[fill=bermuda, draw=black] (10.50, 0.00) rectangle (10.75, -0.25);
                          \filldraw[fill=cancan, draw=black] (10.75, 0.00) rectangle (11.00, -0.25);
                          \filldraw[fill=bermuda, draw=black] (11.00, 0.00) rectangle (11.25, -0.25);
                          \filldraw[fill=bermuda, draw=black] (11.25, 0.00) rectangle (11.50, -0.25);
                          \filldraw[fill=bermuda, draw=black] (11.50, 0.00) rectangle (11.75, -0.25);
                          \filldraw[fill=bermuda, draw=black] (11.75, 0.00) rectangle (12.00, -0.25);
                          \filldraw[fill=bermuda, draw=black] (12.00, 0.00) rectangle (12.25, -0.25);
                          \filldraw[fill=cancan, draw=black] (12.25, 0.00) rectangle (12.50, -0.25);
                          \filldraw[fill=bermuda, draw=black] (12.50, 0.00) rectangle (12.75, -0.25);
                          \filldraw[fill=bermuda, draw=black] (12.75, 0.00) rectangle (13.00, -0.25);
                          \filldraw[fill=bermuda, draw=black] (13.00, 0.00) rectangle (13.25, -0.25);
                          \filldraw[fill=bermuda, draw=black] (13.25, 0.00) rectangle (13.50, -0.25);
                          \filldraw[fill=bermuda, draw=black] (13.50, 0.00) rectangle (13.75, -0.25);
                          \filldraw[fill=cancan, draw=black] (13.75, 0.00) rectangle (14.00, -0.25);
                          \filldraw[fill=cancan, draw=black] (14.00, 0.00) rectangle (14.25, -0.25);
                          \filldraw[fill=cancan, draw=black] (14.25, 0.00) rectangle (14.50, -0.25);
                          \filldraw[fill=cancan, draw=black] (14.50, 0.00) rectangle (14.75, -0.25);
                          \filldraw[fill=cancan, draw=black] (14.75, 0.00) rectangle (15.00, -0.25);
                          \filldraw[fill=bermuda, draw=black] (0.00, -0.25) rectangle (0.25, -0.50);
                          \filldraw[fill=bermuda, draw=black] (0.25, -0.25) rectangle (0.50, -0.50);
                          \filldraw[fill=bermuda, draw=black] (0.50, -0.25) rectangle (0.75, -0.50);
                          \filldraw[fill=cancan, draw=black] (0.75, -0.25) rectangle (1.00, -0.50);
                          \filldraw[fill=cancan, draw=black] (1.00, -0.25) rectangle (1.25, -0.50);
                          \filldraw[fill=bermuda, draw=black] (1.25, -0.25) rectangle (1.50, -0.50);
                          \filldraw[fill=bermuda, draw=black] (1.50, -0.25) rectangle (1.75, -0.50);
                          \filldraw[fill=cancan, draw=black] (1.75, -0.25) rectangle (2.00, -0.50);
                          \filldraw[fill=cancan, draw=black] (2.00, -0.25) rectangle (2.25, -0.50);
                          \filldraw[fill=cancan, draw=black] (2.25, -0.25) rectangle (2.50, -0.50);
                          \filldraw[fill=bermuda, draw=black] (2.50, -0.25) rectangle (2.75, -0.50);
                          \filldraw[fill=bermuda, draw=black] (2.75, -0.25) rectangle (3.00, -0.50);
                          \filldraw[fill=bermuda, draw=black] (3.00, -0.25) rectangle (3.25, -0.50);
                          \filldraw[fill=cancan, draw=black] (3.25, -0.25) rectangle (3.50, -0.50);
                          \filldraw[fill=cancan, draw=black] (3.50, -0.25) rectangle (3.75, -0.50);
                          \filldraw[fill=cancan, draw=black] (3.75, -0.25) rectangle (4.00, -0.50);
                          \filldraw[fill=bermuda, draw=black] (4.00, -0.25) rectangle (4.25, -0.50);
                          \filldraw[fill=bermuda, draw=black] (4.25, -0.25) rectangle (4.50, -0.50);
                          \filldraw[fill=bermuda, draw=black] (4.50, -0.25) rectangle (4.75, -0.50);
                          \filldraw[fill=cancan, draw=black] (4.75, -0.25) rectangle (5.00, -0.50);
                          \filldraw[fill=cancan, draw=black] (5.00, -0.25) rectangle (5.25, -0.50);
                          \filldraw[fill=cancan, draw=black] (5.25, -0.25) rectangle (5.50, -0.50);
                          \filldraw[fill=bermuda, draw=black] (5.50, -0.25) rectangle (5.75, -0.50);
                          \filldraw[fill=bermuda, draw=black] (5.75, -0.25) rectangle (6.00, -0.50);
                          \filldraw[fill=bermuda, draw=black] (6.00, -0.25) rectangle (6.25, -0.50);
                          \filldraw[fill=cancan, draw=black] (6.25, -0.25) rectangle (6.50, -0.50);
                          \filldraw[fill=cancan, draw=black] (6.50, -0.25) rectangle (6.75, -0.50);
                          \filldraw[fill=cancan, draw=black] (6.75, -0.25) rectangle (7.00, -0.50);
                          \filldraw[fill=cancan, draw=black] (7.00, -0.25) rectangle (7.25, -0.50);
                          \filldraw[fill=bermuda, draw=black] (7.25, -0.25) rectangle (7.50, -0.50);
                          \filldraw[fill=bermuda, draw=black] (7.50, -0.25) rectangle (7.75, -0.50);
                          \filldraw[fill=bermuda, draw=black] (7.75, -0.25) rectangle (8.00, -0.50);
                          \filldraw[fill=bermuda, draw=black] (8.00, -0.25) rectangle (8.25, -0.50);
                          \filldraw[fill=cancan, draw=black] (8.25, -0.25) rectangle (8.50, -0.50);
                          \filldraw[fill=bermuda, draw=black] (8.50, -0.25) rectangle (8.75, -0.50);
                          \filldraw[fill=bermuda, draw=black] (8.75, -0.25) rectangle (9.00, -0.50);
                          \filldraw[fill=bermuda, draw=black] (9.00, -0.25) rectangle (9.25, -0.50);
                          \filldraw[fill=cancan, draw=black] (9.25, -0.25) rectangle (9.50, -0.50);
                          \filldraw[fill=cancan, draw=black] (9.50, -0.25) rectangle (9.75, -0.50);
                          \filldraw[fill=bermuda, draw=black] (9.75, -0.25) rectangle (10.00, -0.50);
                          \filldraw[fill=bermuda, draw=black] (10.00, -0.25) rectangle (10.25, -0.50);
                          \filldraw[fill=cancan, draw=black] (10.25, -0.25) rectangle (10.50, -0.50);
                          \filldraw[fill=bermuda, draw=black] (10.50, -0.25) rectangle (10.75, -0.50);
                          \filldraw[fill=bermuda, draw=black] (10.75, -0.25) rectangle (11.00, -0.50);
                          \filldraw[fill=bermuda, draw=black] (11.00, -0.25) rectangle (11.25, -0.50);
                          \filldraw[fill=bermuda, draw=black] (11.25, -0.25) rectangle (11.50, -0.50);
                          \filldraw[fill=bermuda, draw=black] (11.50, -0.25) rectangle (11.75, -0.50);
                          \filldraw[fill=cancan, draw=black] (11.75, -0.25) rectangle (12.00, -0.50);
                          \filldraw[fill=bermuda, draw=black] (12.00, -0.25) rectangle (12.25, -0.50);
                          \filldraw[fill=cancan, draw=black] (12.25, -0.25) rectangle (12.50, -0.50);
                          \filldraw[fill=bermuda, draw=black] (12.50, -0.25) rectangle (12.75, -0.50);
                          \filldraw[fill=bermuda, draw=black] (12.75, -0.25) rectangle (13.00, -0.50);
                          \filldraw[fill=bermuda, draw=black] (13.00, -0.25) rectangle (13.25, -0.50);
                          \filldraw[fill=cancan, draw=black] (13.25, -0.25) rectangle (13.50, -0.50);
                          \filldraw[fill=cancan, draw=black] (13.50, -0.25) rectangle (13.75, -0.50);
                          \filldraw[fill=cancan, draw=black] (13.75, -0.25) rectangle (14.00, -0.50);
                          \filldraw[fill=bermuda, draw=black] (14.00, -0.25) rectangle (14.25, -0.50);
                          \filldraw[fill=bermuda, draw=black] (14.25, -0.25) rectangle (14.50, -0.50);
                          \filldraw[fill=bermuda, draw=black] (14.50, -0.25) rectangle (14.75, -0.50);
                          \filldraw[fill=cancan, draw=black] (14.75, -0.25) rectangle (15.00, -0.50);
                          \filldraw[fill=cancan, draw=black] (0.00, -0.50) rectangle (0.25, -0.75);
                          \filldraw[fill=cancan, draw=black] (0.25, -0.50) rectangle (0.50, -0.75);
                          \filldraw[fill=bermuda, draw=black] (0.50, -0.50) rectangle (0.75, -0.75);
                          \filldraw[fill=bermuda, draw=black] (0.75, -0.50) rectangle (1.00, -0.75);
                          \filldraw[fill=bermuda, draw=black] (1.00, -0.50) rectangle (1.25, -0.75);
                          \filldraw[fill=cancan, draw=black] (1.25, -0.50) rectangle (1.50, -0.75);
                          \filldraw[fill=cancan, draw=black] (1.50, -0.50) rectangle (1.75, -0.75);
                          \filldraw[fill=cancan, draw=black] (1.75, -0.50) rectangle (2.00, -0.75);
                          \filldraw[fill=bermuda, draw=black] (2.00, -0.50) rectangle (2.25, -0.75);
                          \filldraw[fill=bermuda, draw=black] (2.25, -0.50) rectangle (2.50, -0.75);
                          \filldraw[fill=bermuda, draw=black] (2.50, -0.50) rectangle (2.75, -0.75);
                          \filldraw[fill=bermuda, draw=black] (2.75, -0.50) rectangle (3.00, -0.75);
                          \filldraw[fill=bermuda, draw=black] (3.00, -0.50) rectangle (3.25, -0.75);
                          \filldraw[fill=cancan, draw=black] (3.25, -0.50) rectangle (3.50, -0.75);
                          \filldraw[fill=bermuda, draw=black] (3.50, -0.50) rectangle (3.75, -0.75);
                          \filldraw[fill=cancan, draw=black] (3.75, -0.50) rectangle (4.00, -0.75);
                          \filldraw[fill=cancan, draw=black] (4.00, -0.50) rectangle (4.25, -0.75);
                          \filldraw[fill=bermuda, draw=black] (4.25, -0.50) rectangle (4.50, -0.75);
                          \filldraw[fill=bermuda, draw=black] (4.50, -0.50) rectangle (4.75, -0.75);
                          \filldraw[fill=cancan, draw=black] (4.75, -0.50) rectangle (5.00, -0.75);
                          \filldraw[fill=bermuda, draw=black] (5.00, -0.50) rectangle (5.25, -0.75);
                          \filldraw[fill=cancan, draw=black] (5.25, -0.50) rectangle (5.50, -0.75);
                          \filldraw[fill=cancan, draw=black] (5.50, -0.50) rectangle (5.75, -0.75);
                          \filldraw[fill=bermuda, draw=black] (5.75, -0.50) rectangle (6.00, -0.75);
                          \filldraw[fill=bermuda, draw=black] (6.00, -0.50) rectangle (6.25, -0.75);
                          \filldraw[fill=cancan, draw=black] (6.25, -0.50) rectangle (6.50, -0.75);
                          \filldraw[fill=bermuda, draw=black] (6.50, -0.50) rectangle (6.75, -0.75);
                          \filldraw[fill=cancan, draw=black] (6.75, -0.50) rectangle (7.00, -0.75);
                          \filldraw[fill=cancan, draw=black] (7.00, -0.50) rectangle (7.25, -0.75);
                          \filldraw[fill=bermuda, draw=black] (7.25, -0.50) rectangle (7.50, -0.75);
                          \filldraw[fill=bermuda, draw=black] (7.50, -0.50) rectangle (7.75, -0.75);
                          \filldraw[fill=cancan, draw=black] (7.75, -0.50) rectangle (8.00, -0.75);
                          \filldraw[fill=bermuda, draw=black] (8.00, -0.50) rectangle (8.25, -0.75);
                          \filldraw[fill=cancan, draw=black] (8.25, -0.50) rectangle (8.50, -0.75);
                          \filldraw[fill=cancan, draw=black] (8.50, -0.50) rectangle (8.75, -0.75);
                          \filldraw[fill=bermuda, draw=black] (8.75, -0.50) rectangle (9.00, -0.75);
                          \filldraw[fill=bermuda, draw=black] (9.00, -0.50) rectangle (9.25, -0.75);
                          \filldraw[fill=cancan, draw=black] (9.25, -0.50) rectangle (9.50, -0.75);
                          \filldraw[fill=bermuda, draw=black] (9.50, -0.50) rectangle (9.75, -0.75);
                          \filldraw[fill=cancan, draw=black] (9.75, -0.50) rectangle (10.00, -0.75);
                          \filldraw[fill=cancan, draw=black] (10.00, -0.50) rectangle (10.25, -0.75);
                          \filldraw[fill=cancan, draw=black] (10.25, -0.50) rectangle (10.50, -0.75);
                          \filldraw[fill=cancan, draw=black] (10.50, -0.50) rectangle (10.75, -0.75);
                          \filldraw[fill=cancan, draw=black] (10.75, -0.50) rectangle (11.00, -0.75);
                          \filldraw[fill=bermuda, draw=black] (11.00, -0.50) rectangle (11.25, -0.75);
                          \filldraw[fill=cancan, draw=black] (11.25, -0.50) rectangle (11.50, -0.75);
                          \filldraw[fill=bermuda, draw=black] (11.50, -0.50) rectangle (11.75, -0.75);
                          \filldraw[fill=cancan, draw=black] (11.75, -0.50) rectangle (12.00, -0.75);
                          \filldraw[fill=cancan, draw=black] (12.00, -0.50) rectangle (12.25, -0.75);
                          \filldraw[fill=cancan, draw=black] (12.25, -0.50) rectangle (12.50, -0.75);
                          \filldraw[fill=bermuda, draw=black] (12.50, -0.50) rectangle (12.75, -0.75);
                          \filldraw[fill=bermuda, draw=black] (12.75, -0.50) rectangle (13.00, -0.75);
                          \filldraw[fill=bermuda, draw=black] (13.00, -0.50) rectangle (13.25, -0.75);
                          \filldraw[fill=cancan, draw=black] (13.25, -0.50) rectangle (13.50, -0.75);
                          \filldraw[fill=cancan, draw=black] (13.50, -0.50) rectangle (13.75, -0.75);
                          \filldraw[fill=cancan, draw=black] (13.75, -0.50) rectangle (14.00, -0.75);
                          \filldraw[fill=cancan, draw=black] (14.00, -0.50) rectangle (14.25, -0.75);
                          \filldraw[fill=cancan, draw=black] (14.25, -0.50) rectangle (14.50, -0.75);
                          \filldraw[fill=cancan, draw=black] (14.50, -0.50) rectangle (14.75, -0.75);
                          \filldraw[fill=cancan, draw=black] (14.75, -0.50) rectangle (15.00, -0.75);
                          \filldraw[fill=bermuda, draw=black] (0.00, -0.75) rectangle (0.25, -1.00);
                          \filldraw[fill=cancan, draw=black] (0.25, -0.75) rectangle (0.50, -1.00);
                          \filldraw[fill=cancan, draw=black] (0.50, -0.75) rectangle (0.75, -1.00);
                          \filldraw[fill=cancan, draw=black] (0.75, -0.75) rectangle (1.00, -1.00);
                          \filldraw[fill=cancan, draw=black] (1.00, -0.75) rectangle (1.25, -1.00);
                          \filldraw[fill=cancan, draw=black] (1.25, -0.75) rectangle (1.50, -1.00);
                          \filldraw[fill=cancan, draw=black] (1.50, -0.75) rectangle (1.75, -1.00);
                          \filldraw[fill=cancan, draw=black] (1.75, -0.75) rectangle (2.00, -1.00);
                          \filldraw[fill=bermuda, draw=black] (2.00, -0.75) rectangle (2.25, -1.00);
                          \filldraw[fill=bermuda, draw=black] (2.25, -0.75) rectangle (2.50, -1.00);
                          \filldraw[fill=bermuda, draw=black] (2.50, -0.75) rectangle (2.75, -1.00);
                          \filldraw[fill=cancan, draw=black] (2.75, -0.75) rectangle (3.00, -1.00);
                          \filldraw[fill=cancan, draw=black] (3.00, -0.75) rectangle (3.25, -1.00);
                          \filldraw[fill=cancan, draw=black] (3.25, -0.75) rectangle (3.50, -1.00);
                          \filldraw[fill=bermuda, draw=black] (3.50, -0.75) rectangle (3.75, -1.00);
                          \filldraw[fill=bermuda, draw=black] (3.75, -0.75) rectangle (4.00, -1.00);
                          \filldraw[fill=bermuda, draw=black] (4.00, -0.75) rectangle (4.25, -1.00);
                          \filldraw[fill=cancan, draw=black] (4.25, -0.75) rectangle (4.50, -1.00);
                          \filldraw[fill=cancan, draw=black] (4.50, -0.75) rectangle (4.75, -1.00);
                          \filldraw[fill=cancan, draw=black] (4.75, -0.75) rectangle (5.00, -1.00);
                          \filldraw[fill=bermuda, draw=black] (5.00, -0.75) rectangle (5.25, -1.00);
                          \filldraw[fill=bermuda, draw=black] (5.25, -0.75) rectangle (5.50, -1.00);
                          \filldraw[fill=bermuda, draw=black] (5.50, -0.75) rectangle (5.75, -1.00);
                          \filldraw[fill=bermuda, draw=black] (5.75, -0.75) rectangle (6.00, -1.00);
                          \filldraw[fill=bermuda, draw=black] (6.00, -0.75) rectangle (6.25, -1.00);
                          \filldraw[fill=cancan, draw=black] (6.25, -0.75) rectangle (6.50, -1.00);
                          \filldraw[fill=bermuda, draw=black] (6.50, -0.75) rectangle (6.75, -1.00);
                          \filldraw[fill=cancan, draw=black] (6.75, -0.75) rectangle (7.00, -1.00);
                          \filldraw[fill=bermuda, draw=black] (7.00, -0.75) rectangle (7.25, -1.00);
                          \filldraw[fill=cancan, draw=black] (7.25, -0.75) rectangle (7.50, -1.00);
                          \filldraw[fill=bermuda, draw=black] (7.50, -0.75) rectangle (7.75, -1.00);
                          \filldraw[fill=cancan, draw=black] (7.75, -0.75) rectangle (8.00, -1.00);
                          \filldraw[fill=bermuda, draw=black] (8.00, -0.75) rectangle (8.25, -1.00);
                          \filldraw[fill=bermuda, draw=black] (8.25, -0.75) rectangle (8.50, -1.00);
                          \filldraw[fill=bermuda, draw=black] (8.50, -0.75) rectangle (8.75, -1.00);
                          \filldraw[fill=cancan, draw=black] (8.75, -0.75) rectangle (9.00, -1.00);
                          \filldraw[fill=cancan, draw=black] (9.00, -0.75) rectangle (9.25, -1.00);
                          \filldraw[fill=cancan, draw=black] (9.25, -0.75) rectangle (9.50, -1.00);
                          \filldraw[fill=bermuda, draw=black] (9.50, -0.75) rectangle (9.75, -1.00);
                          \filldraw[fill=bermuda, draw=black] (9.75, -0.75) rectangle (10.00, -1.00);
                          \filldraw[fill=bermuda, draw=black] (10.00, -0.75) rectangle (10.25, -1.00);
                          \filldraw[fill=cancan, draw=black] (10.25, -0.75) rectangle (10.50, -1.00);
                          \filldraw[fill=bermuda, draw=black] (10.50, -0.75) rectangle (10.75, -1.00);
                          \filldraw[fill=bermuda, draw=black] (10.75, -0.75) rectangle (11.00, -1.00);
                          \filldraw[fill=bermuda, draw=black] (11.00, -0.75) rectangle (11.25, -1.00);
                          \filldraw[fill=bermuda, draw=black] (11.25, -0.75) rectangle (11.50, -1.00);
                          \filldraw[fill=bermuda, draw=black] (11.50, -0.75) rectangle (11.75, -1.00);
                          \filldraw[fill=cancan, draw=black] (11.75, -0.75) rectangle (12.00, -1.00);
                          \filldraw[fill=bermuda, draw=black] (12.00, -0.75) rectangle (12.25, -1.00);
                          \filldraw[fill=bermuda, draw=black] (12.25, -0.75) rectangle (12.50, -1.00);
                          \filldraw[fill=bermuda, draw=black] (12.50, -0.75) rectangle (12.75, -1.00);
                          \filldraw[fill=bermuda, draw=black] (12.75, -0.75) rectangle (13.00, -1.00);
                          \filldraw[fill=bermuda, draw=black] (13.00, -0.75) rectangle (13.25, -1.00);
                          \filldraw[fill=cancan, draw=black] (13.25, -0.75) rectangle (13.50, -1.00);
                          \filldraw[fill=bermuda, draw=black] (13.50, -0.75) rectangle (13.75, -1.00);
                          \filldraw[fill=bermuda, draw=black] (13.75, -0.75) rectangle (14.00, -1.00);
                          \filldraw[fill=bermuda, draw=black] (14.00, -0.75) rectangle (14.25, -1.00);
                          \filldraw[fill=bermuda, draw=black] (14.25, -0.75) rectangle (14.50, -1.00);
                          \filldraw[fill=bermuda, draw=black] (14.50, -0.75) rectangle (14.75, -1.00);
                          \filldraw[fill=cancan, draw=black] (14.75, -0.75) rectangle (15.00, -1.00);
                          \filldraw[fill=bermuda, draw=black] (0.00, -1.00) rectangle (0.25, -1.25);
                          \filldraw[fill=bermuda, draw=black] (0.25, -1.00) rectangle (0.50, -1.25);
                          \filldraw[fill=bermuda, draw=black] (0.50, -1.00) rectangle (0.75, -1.25);
                          \filldraw[fill=bermuda, draw=black] (0.75, -1.00) rectangle (1.00, -1.25);
                          \filldraw[fill=bermuda, draw=black] (1.00, -1.00) rectangle (1.25, -1.25);
                          \filldraw[fill=cancan, draw=black] (1.25, -1.00) rectangle (1.50, -1.25);
                          \filldraw[fill=cancan, draw=black] (1.50, -1.00) rectangle (1.75, -1.25);
                          \filldraw[fill=cancan, draw=black] (1.75, -1.00) rectangle (2.00, -1.25);
                          \filldraw[fill=cancan, draw=black] (2.00, -1.00) rectangle (2.25, -1.25);
                          \filldraw[fill=cancan, draw=black] (2.25, -1.00) rectangle (2.50, -1.25);
                          \filldraw[fill=cancan, draw=black] (2.50, -1.00) rectangle (2.75, -1.25);
                          \filldraw[fill=cancan, draw=black] (2.75, -1.00) rectangle (3.00, -1.25);
                          \filldraw[fill=bermuda, draw=black] (3.00, -1.00) rectangle (3.25, -1.25);
                          \filldraw[fill=bermuda, draw=black] (3.25, -1.00) rectangle (3.50, -1.25);
                          \filldraw[fill=bermuda, draw=black] (3.50, -1.00) rectangle (3.75, -1.25);
                          \filldraw[fill=cancan, draw=black] (3.75, -1.00) rectangle (4.00, -1.25);
                          \filldraw[fill=cancan, draw=black] (4.00, -1.00) rectangle (4.25, -1.25);
                          \filldraw[fill=cancan, draw=black] (4.25, -1.00) rectangle (4.50, -1.25);
                          \filldraw[fill=bermuda, draw=black] (4.50, -1.00) rectangle (4.75, -1.25);
                          \filldraw[fill=bermuda, draw=black] (4.75, -1.00) rectangle (5.00, -1.25);
                          \filldraw[fill=bermuda, draw=black] (5.00, -1.00) rectangle (5.25, -1.25);
                          \filldraw[fill=cancan, draw=black] (5.25, -1.00) rectangle (5.50, -1.25);
                          \filldraw[fill=bermuda, draw=black] (5.50, -1.00) rectangle (5.75, -1.25);
                          \filldraw[fill=cancan, draw=black] (5.75, -1.00) rectangle (6.00, -1.25);
                          \filldraw[fill=bermuda, draw=black] (6.00, -1.00) rectangle (6.25, -1.25);
                          \filldraw[fill=cancan, draw=black] (6.25, -1.00) rectangle (6.50, -1.25);
                          \filldraw[fill=cancan, draw=black] (6.50, -1.00) rectangle (6.75, -1.25);
                          \filldraw[fill=cancan, draw=black] (6.75, -1.00) rectangle (7.00, -1.25);
                          \filldraw[fill=bermuda, draw=black] (7.00, -1.00) rectangle (7.25, -1.25);
                          \filldraw[fill=bermuda, draw=black] (7.25, -1.00) rectangle (7.50, -1.25);
                          \filldraw[fill=bermuda, draw=black] (7.50, -1.00) rectangle (7.75, -1.25);
                          \filldraw[fill=cancan, draw=black] (7.75, -1.00) rectangle (8.00, -1.25);
                          \filldraw[fill=cancan, draw=black] (8.00, -1.00) rectangle (8.25, -1.25);
                          \filldraw[fill=cancan, draw=black] (8.25, -1.00) rectangle (8.50, -1.25);
                          \filldraw[fill=cancan, draw=black] (8.50, -1.00) rectangle (8.75, -1.25);
                          \filldraw[fill=bermuda, draw=black] (8.75, -1.00) rectangle (9.00, -1.25);
                          \filldraw[fill=bermuda, draw=black] (9.00, -1.00) rectangle (9.25, -1.25);
                          \filldraw[fill=cancan, draw=black] (9.25, -1.00) rectangle (9.50, -1.25);
                          \filldraw[fill=cancan, draw=black] (9.50, -1.00) rectangle (9.75, -1.25);
                          \filldraw[fill=cancan, draw=black] (9.75, -1.00) rectangle (10.00, -1.25);
                          \filldraw[fill=cancan, draw=black] (10.00, -1.00) rectangle (10.25, -1.25);
                          \filldraw[fill=bermuda, draw=black] (10.25, -1.00) rectangle (10.50, -1.25);
                          \filldraw[fill=bermuda, draw=black] (10.50, -1.00) rectangle (10.75, -1.25);
                          \filldraw[fill=cancan, draw=black] (10.75, -1.00) rectangle (11.00, -1.25);
                          \filldraw[fill=cancan, draw=black] (11.00, -1.00) rectangle (11.25, -1.25);
                          \filldraw[fill=cancan, draw=black] (11.25, -1.00) rectangle (11.50, -1.25);
                          \filldraw[fill=cancan, draw=black] (11.50, -1.00) rectangle (11.75, -1.25);
                          \filldraw[fill=bermuda, draw=black] (11.75, -1.00) rectangle (12.00, -1.25);
                          \filldraw[fill=bermuda, draw=black] (12.00, -1.00) rectangle (12.25, -1.25);
                          \filldraw[fill=cancan, draw=black] (12.25, -1.00) rectangle (12.50, -1.25);
                          \filldraw[fill=cancan, draw=black] (12.50, -1.00) rectangle (12.75, -1.25);
                          \filldraw[fill=cancan, draw=black] (12.75, -1.00) rectangle (13.00, -1.25);
                          \filldraw[fill=cancan, draw=black] (13.00, -1.00) rectangle (13.25, -1.25);
                          \filldraw[fill=cancan, draw=black] (13.25, -1.00) rectangle (13.50, -1.25);
                          \filldraw[fill=cancan, draw=black] (13.50, -1.00) rectangle (13.75, -1.25);
                          \filldraw[fill=cancan, draw=black] (13.75, -1.00) rectangle (14.00, -1.25);
                          \filldraw[fill=cancan, draw=black] (14.00, -1.00) rectangle (14.25, -1.25);
                          \filldraw[fill=cancan, draw=black] (14.25, -1.00) rectangle (14.50, -1.25);
                          \filldraw[fill=cancan, draw=black] (14.50, -1.00) rectangle (14.75, -1.25);
                          \filldraw[fill=cancan, draw=black] (14.75, -1.00) rectangle (15.00, -1.25);
                          \filldraw[fill=cancan, draw=black] (0.00, -1.25) rectangle (0.25, -1.50);
                          \filldraw[fill=bermuda, draw=black] (0.25, -1.25) rectangle (0.50, -1.50);
                          \filldraw[fill=bermuda, draw=black] (0.50, -1.25) rectangle (0.75, -1.50);
                          \filldraw[fill=cancan, draw=black] (0.75, -1.25) rectangle (1.00, -1.50);
                          \filldraw[fill=bermuda, draw=black] (1.00, -1.25) rectangle (1.25, -1.50);
                          \filldraw[fill=cancan, draw=black] (1.25, -1.25) rectangle (1.50, -1.50);
                          \filldraw[fill=cancan, draw=black] (1.50, -1.25) rectangle (1.75, -1.50);
                          \filldraw[fill=bermuda, draw=black] (1.75, -1.25) rectangle (2.00, -1.50);
                          \filldraw[fill=bermuda, draw=black] (2.00, -1.25) rectangle (2.25, -1.50);
                          \filldraw[fill=cancan, draw=black] (2.25, -1.25) rectangle (2.50, -1.50);
                          \filldraw[fill=bermuda, draw=black] (2.50, -1.25) rectangle (2.75, -1.50);
                          \filldraw[fill=cancan, draw=black] (2.75, -1.25) rectangle (3.00, -1.50);
                          \filldraw[fill=bermuda, draw=black] (3.00, -1.25) rectangle (3.25, -1.50);
                          \filldraw[fill=bermuda, draw=black] (3.25, -1.25) rectangle (3.50, -1.50);
                          \filldraw[fill=bermuda, draw=black] (3.50, -1.25) rectangle (3.75, -1.50);
                          \filldraw[fill=cancan, draw=black] (3.75, -1.25) rectangle (4.00, -1.50);
                          \filldraw[fill=bermuda, draw=black] (4.00, -1.25) rectangle (4.25, -1.50);
                          \filldraw[fill=bermuda, draw=black] (4.25, -1.25) rectangle (4.50, -1.50);
                          \filldraw[fill=bermuda, draw=black] (4.50, -1.25) rectangle (4.75, -1.50);
                          \filldraw[fill=bermuda, draw=black] (4.75, -1.25) rectangle (5.00, -1.50);
                          \filldraw[fill=bermuda, draw=black] (5.00, -1.25) rectangle (5.25, -1.50);
                          \filldraw[fill=cancan, draw=black] (5.25, -1.25) rectangle (5.50, -1.50);
                          \filldraw[fill=bermuda, draw=black] (5.50, -1.25) rectangle (5.75, -1.50);
                          \filldraw[fill=cancan, draw=black] (5.75, -1.25) rectangle (6.00, -1.50);
                          \filldraw[fill=cancan, draw=black] (6.00, -1.25) rectangle (6.25, -1.50);
                          \filldraw[fill=bermuda, draw=black] (6.25, -1.25) rectangle (6.50, -1.50);
                          \filldraw[fill=bermuda, draw=black] (6.50, -1.25) rectangle (6.75, -1.50);
                          \filldraw[fill=cancan, draw=black] (6.75, -1.25) rectangle (7.00, -1.50);
                          \filldraw[fill=bermuda, draw=black] (7.00, -1.25) rectangle (7.25, -1.50);
                          \filldraw[fill=cancan, draw=black] (7.25, -1.25) rectangle (7.50, -1.50);
                          \filldraw[fill=bermuda, draw=black] (7.50, -1.25) rectangle (7.75, -1.50);
                          \filldraw[fill=cancan, draw=black] (7.75, -1.25) rectangle (8.00, -1.50);
                          \filldraw[fill=bermuda, draw=black] (8.00, -1.25) rectangle (8.25, -1.50);
                          \filldraw[fill=bermuda, draw=black] (8.25, -1.25) rectangle (8.50, -1.50);
                          \filldraw[fill=bermuda, draw=black] (8.50, -1.25) rectangle (8.75, -1.50);
                          \filldraw[fill=cancan, draw=black] (8.75, -1.25) rectangle (9.00, -1.50);
                          \filldraw[fill=cancan, draw=black] (9.00, -1.25) rectangle (9.25, -1.50);
                          \filldraw[fill=cancan, draw=black] (9.25, -1.25) rectangle (9.50, -1.50);
                          \filldraw[fill=bermuda, draw=black] (9.50, -1.25) rectangle (9.75, -1.50);
                          \filldraw[fill=bermuda, draw=black] (9.75, -1.25) rectangle (10.00, -1.50);
                          \filldraw[fill=bermuda, draw=black] (10.00, -1.25) rectangle (10.25, -1.50);
                          \filldraw[fill=cancan, draw=black] (10.25, -1.25) rectangle (10.50, -1.50);
                          \filldraw[fill=cancan, draw=black] (10.50, -1.25) rectangle (10.75, -1.50);
                          \filldraw[fill=cancan, draw=black] (10.75, -1.25) rectangle (11.00, -1.50);
                          \filldraw[fill=cancan, draw=black] (11.00, -1.25) rectangle (11.25, -1.50);
                          \filldraw[fill=bermuda, draw=black] (11.25, -1.25) rectangle (11.50, -1.50);
                          \filldraw[fill=bermuda, draw=black] (11.50, -1.25) rectangle (11.75, -1.50);
                          \filldraw[fill=cancan, draw=black] (11.75, -1.25) rectangle (12.00, -1.50);
                          \filldraw[fill=bermuda, draw=black] (12.00, -1.25) rectangle (12.25, -1.50);
                          \filldraw[fill=cancan, draw=black] (12.25, -1.25) rectangle (12.50, -1.50);
                          \filldraw[fill=cancan, draw=black] (12.50, -1.25) rectangle (12.75, -1.50);
                          \filldraw[fill=bermuda, draw=black] (12.75, -1.25) rectangle (13.00, -1.50);
                          \filldraw[fill=bermuda, draw=black] (13.00, -1.25) rectangle (13.25, -1.50);
                          \filldraw[fill=cancan, draw=black] (13.25, -1.25) rectangle (13.50, -1.50);
                          \filldraw[fill=bermuda, draw=black] (13.50, -1.25) rectangle (13.75, -1.50);
                          \filldraw[fill=cancan, draw=black] (13.75, -1.25) rectangle (14.00, -1.50);
                          \filldraw[fill=bermuda, draw=black] (14.00, -1.25) rectangle (14.25, -1.50);
                          \filldraw[fill=bermuda, draw=black] (14.25, -1.25) rectangle (14.50, -1.50);
                          \filldraw[fill=bermuda, draw=black] (14.50, -1.25) rectangle (14.75, -1.50);
                          \filldraw[fill=cancan, draw=black] (14.75, -1.25) rectangle (15.00, -1.50);
                          \filldraw[fill=bermuda, draw=black] (0.00, -1.50) rectangle (0.25, -1.75);
                          \filldraw[fill=bermuda, draw=black] (0.25, -1.50) rectangle (0.50, -1.75);
                          \filldraw[fill=bermuda, draw=black] (0.50, -1.50) rectangle (0.75, -1.75);
                          \filldraw[fill=bermuda, draw=black] (0.75, -1.50) rectangle (1.00, -1.75);
                          \filldraw[fill=bermuda, draw=black] (1.00, -1.50) rectangle (1.25, -1.75);
                          \filldraw[fill=cancan, draw=black] (1.25, -1.50) rectangle (1.50, -1.75);
                          \filldraw[fill=bermuda, draw=black] (1.50, -1.50) rectangle (1.75, -1.75);
                          \filldraw[fill=cancan, draw=black] (1.75, -1.50) rectangle (2.00, -1.75);
                          \filldraw[fill=cancan, draw=black] (2.00, -1.50) rectangle (2.25, -1.75);
                          \filldraw[fill=bermuda, draw=black] (2.25, -1.50) rectangle (2.50, -1.75);
                          \filldraw[fill=bermuda, draw=black] (2.50, -1.50) rectangle (2.75, -1.75);
                          \filldraw[fill=cancan, draw=black] (2.75, -1.50) rectangle (3.00, -1.75);
                          \filldraw[fill=bermuda, draw=black] (3.00, -1.50) rectangle (3.25, -1.75);
                          \filldraw[fill=cancan, draw=black] (3.25, -1.50) rectangle (3.50, -1.75);
                          \filldraw[fill=cancan, draw=black] (3.50, -1.50) rectangle (3.75, -1.75);
                          \filldraw[fill=bermuda, draw=black] (3.75, -1.50) rectangle (4.00, -1.75);
                          \filldraw[fill=bermuda, draw=black] (4.00, -1.50) rectangle (4.25, -1.75);
                          \filldraw[fill=cancan, draw=black] (4.25, -1.50) rectangle (4.50, -1.75);
                          \filldraw[fill=bermuda, draw=black] (4.50, -1.50) rectangle (4.75, -1.75);
                          \filldraw[fill=cancan, draw=black] (4.75, -1.50) rectangle (5.00, -1.75);
                          \filldraw[fill=cancan, draw=black] (5.00, -1.50) rectangle (5.25, -1.75);
                          \filldraw[fill=bermuda, draw=black] (5.25, -1.50) rectangle (5.50, -1.75);
                          \filldraw[fill=bermuda, draw=black] (5.50, -1.50) rectangle (5.75, -1.75);
                          \filldraw[fill=cancan, draw=black] (5.75, -1.50) rectangle (6.00, -1.75);
                          \filldraw[fill=bermuda, draw=black] (6.00, -1.50) rectangle (6.25, -1.75);
                          \filldraw[fill=cancan, draw=black] (6.25, -1.50) rectangle (6.50, -1.75);
                          \filldraw[fill=cancan, draw=black] (6.50, -1.50) rectangle (6.75, -1.75);
                          \filldraw[fill=bermuda, draw=black] (6.75, -1.50) rectangle (7.00, -1.75);
                          \filldraw[fill=bermuda, draw=black] (7.00, -1.50) rectangle (7.25, -1.75);
                          \filldraw[fill=cancan, draw=black] (7.25, -1.50) rectangle (7.50, -1.75);
                          \filldraw[fill=bermuda, draw=black] (7.50, -1.50) rectangle (7.75, -1.75);
                          \filldraw[fill=bermuda, draw=black] (7.75, -1.50) rectangle (8.00, -1.75);
                          \filldraw[fill=bermuda, draw=black] (8.00, -1.50) rectangle (8.25, -1.75);
                          \filldraw[fill=cancan, draw=black] (8.25, -1.50) rectangle (8.50, -1.75);
                          \filldraw[fill=cancan, draw=black] (8.50, -1.50) rectangle (8.75, -1.75);
                          \filldraw[fill=cancan, draw=black] (8.75, -1.50) rectangle (9.00, -1.75);
                          \filldraw[fill=cancan, draw=black] (9.00, -1.50) rectangle (9.25, -1.75);
                          \filldraw[fill=cancan, draw=black] (9.25, -1.50) rectangle (9.50, -1.75);
                          \filldraw[fill=bermuda, draw=black] (9.50, -1.50) rectangle (9.75, -1.75);
                          \filldraw[fill=bermuda, draw=black] (9.75, -1.50) rectangle (10.00, -1.75);
                          \filldraw[fill=bermuda, draw=black] (10.00, -1.50) rectangle (10.25, -1.75);
                          \filldraw[fill=cancan, draw=black] (10.25, -1.50) rectangle (10.50, -1.75);
                          \filldraw[fill=cancan, draw=black] (10.50, -1.50) rectangle (10.75, -1.75);
                          \filldraw[fill=cancan, draw=black] (10.75, -1.50) rectangle (11.00, -1.75);
                          \filldraw[fill=bermuda, draw=black] (11.00, -1.50) rectangle (11.25, -1.75);
                          \filldraw[fill=bermuda, draw=black] (11.25, -1.50) rectangle (11.50, -1.75);
                          \filldraw[fill=bermuda, draw=black] (11.50, -1.50) rectangle (11.75, -1.75);
                          \filldraw[fill=cancan, draw=black] (11.75, -1.50) rectangle (12.00, -1.75);
                          \filldraw[fill=cancan, draw=black] (12.00, -1.50) rectangle (12.25, -1.75);
                          \filldraw[fill=cancan, draw=black] (12.25, -1.50) rectangle (12.50, -1.75);
                          \filldraw[fill=bermuda, draw=black] (12.50, -1.50) rectangle (12.75, -1.75);
                          \filldraw[fill=bermuda, draw=black] (12.75, -1.50) rectangle (13.00, -1.75);
                          \filldraw[fill=bermuda, draw=black] (13.00, -1.50) rectangle (13.25, -1.75);
                          \filldraw[fill=cancan, draw=black] (13.25, -1.50) rectangle (13.50, -1.75);
                          \filldraw[fill=bermuda, draw=black] (13.50, -1.50) rectangle (13.75, -1.75);
                          \filldraw[fill=bermuda, draw=black] (13.75, -1.50) rectangle (14.00, -1.75);
                          \filldraw[fill=bermuda, draw=black] (14.00, -1.50) rectangle (14.25, -1.75);
                          \filldraw[fill=cancan, draw=black] (14.25, -1.50) rectangle (14.50, -1.75);
                          \filldraw[fill=cancan, draw=black] (14.50, -1.50) rectangle (14.75, -1.75);
                          \filldraw[fill=bermuda, draw=black] (14.75, -1.50) rectangle (15.00, -1.75);
                          \filldraw[fill=bermuda, draw=black] (0.00, -1.75) rectangle (0.25, -2.00);
                          \filldraw[fill=cancan, draw=black] (0.25, -1.75) rectangle (0.50, -2.00);
                          \filldraw[fill=cancan, draw=black] (0.50, -1.75) rectangle (0.75, -2.00);
                          \filldraw[fill=cancan, draw=black] (0.75, -1.75) rectangle (1.00, -2.00);
                          \filldraw[fill=cancan, draw=black] (1.00, -1.75) rectangle (1.25, -2.00);
                          \filldraw[fill=bermuda, draw=black] (1.25, -1.75) rectangle (1.50, -2.00);
                          \filldraw[fill=bermuda, draw=black] (1.50, -1.75) rectangle (1.75, -2.00);
                          \filldraw[fill=cancan, draw=black] (1.75, -1.75) rectangle (2.00, -2.00);
                          \filldraw[fill=bermuda, draw=black] (2.00, -1.75) rectangle (2.25, -2.00);
                          \filldraw[fill=bermuda, draw=black] (2.25, -1.75) rectangle (2.50, -2.00);
                          \filldraw[fill=bermuda, draw=black] (2.50, -1.75) rectangle (2.75, -2.00);
                          \filldraw[fill=cancan, draw=black] (2.75, -1.75) rectangle (3.00, -2.00);
                          \filldraw[fill=cancan, draw=black] (3.00, -1.75) rectangle (3.25, -2.00);
                          \filldraw[fill=cancan, draw=black] (3.25, -1.75) rectangle (3.50, -2.00);
                          \filldraw[fill=bermuda, draw=black] (3.50, -1.75) rectangle (3.75, -2.00);
                          \filldraw[fill=bermuda, draw=black] (3.75, -1.75) rectangle (4.00, -2.00);
                          \filldraw[fill=bermuda, draw=black] (4.00, -1.75) rectangle (4.25, -2.00);
                          \filldraw[fill=bermuda, draw=black] (4.25, -1.75) rectangle (4.50, -2.00);
                          \filldraw[fill=bermuda, draw=black] (4.50, -1.75) rectangle (4.75, -2.00);
                          \filldraw[fill=cancan, draw=black] (4.75, -1.75) rectangle (5.00, -2.00);
                          \filldraw[fill=bermuda, draw=black] (5.00, -1.75) rectangle (5.25, -2.00);
                          \filldraw[fill=bermuda, draw=black] (5.25, -1.75) rectangle (5.50, -2.00);
                          \filldraw[fill=bermuda, draw=black] (5.50, -1.75) rectangle (5.75, -2.00);
                          \filldraw[fill=cancan, draw=black] (5.75, -1.75) rectangle (6.00, -2.00);
                          \filldraw[fill=bermuda, draw=black] (6.00, -1.75) rectangle (6.25, -2.00);
                          \filldraw[fill=bermuda, draw=black] (6.25, -1.75) rectangle (6.50, -2.00);
                          \filldraw[fill=bermuda, draw=black] (6.50, -1.75) rectangle (6.75, -2.00);
                          \filldraw[fill=bermuda, draw=black] (6.75, -1.75) rectangle (7.00, -2.00);
                          \filldraw[fill=bermuda, draw=black] (7.00, -1.75) rectangle (7.25, -2.00);
                          \filldraw[fill=cancan, draw=black] (7.25, -1.75) rectangle (7.50, -2.00);
                          \filldraw[fill=cancan, draw=black] (7.50, -1.75) rectangle (7.75, -2.00);
                          \filldraw[fill=cancan, draw=black] (7.75, -1.75) rectangle (8.00, -2.00);
                          \filldraw[fill=bermuda, draw=black] (8.00, -1.75) rectangle (8.25, -2.00);
                          \filldraw[fill=bermuda, draw=black] (8.25, -1.75) rectangle (8.50, -2.00);
                          \filldraw[fill=bermuda, draw=black] (8.50, -1.75) rectangle (8.75, -2.00);
                          \filldraw[fill=cancan, draw=black] (8.75, -1.75) rectangle (9.00, -2.00);
                          \filldraw[fill=cancan, draw=black] (9.00, -1.75) rectangle (9.25, -2.00);
                          \filldraw[fill=cancan, draw=black] (9.25, -1.75) rectangle (9.50, -2.00);
                          \filldraw[fill=bermuda, draw=black] (9.50, -1.75) rectangle (9.75, -2.00);
                          \filldraw[fill=bermuda, draw=black] (9.75, -1.75) rectangle (10.00, -2.00);
                          \filldraw[fill=bermuda, draw=black] (10.00, -1.75) rectangle (10.25, -2.00);
                          \filldraw[fill=cancan, draw=black] (10.25, -1.75) rectangle (10.50, -2.00);
                          \filldraw[fill=cancan, draw=black] (10.50, -1.75) rectangle (10.75, -2.00);
                          \filldraw[fill=cancan, draw=black] (10.75, -1.75) rectangle (11.00, -2.00);
                          \filldraw[fill=bermuda, draw=black] (11.00, -1.75) rectangle (11.25, -2.00);
                          \filldraw[fill=bermuda, draw=black] (11.25, -1.75) rectangle (11.50, -2.00);
                          \filldraw[fill=bermuda, draw=black] (11.50, -1.75) rectangle (11.75, -2.00);
                          \filldraw[fill=bermuda, draw=black] (11.75, -1.75) rectangle (12.00, -2.00);
                          \filldraw[fill=bermuda, draw=black] (12.00, -1.75) rectangle (12.25, -2.00);
                          \filldraw[fill=bermuda, draw=black] (12.25, -1.75) rectangle (12.50, -2.00);
                          \filldraw[fill=bermuda, draw=black] (12.50, -1.75) rectangle (12.75, -2.00);
                          \filldraw[fill=bermuda, draw=black] (12.75, -1.75) rectangle (13.00, -2.00);
                          \filldraw[fill=bermuda, draw=black] (13.00, -1.75) rectangle (13.25, -2.00);
                          \filldraw[fill=cancan, draw=black] (13.25, -1.75) rectangle (13.50, -2.00);
                          \filldraw[fill=bermuda, draw=black] (13.50, -1.75) rectangle (13.75, -2.00);
                          \filldraw[fill=cancan, draw=black] (13.75, -1.75) rectangle (14.00, -2.00);
                          \filldraw[fill=cancan, draw=black] (14.00, -1.75) rectangle (14.25, -2.00);
                          \filldraw[fill=bermuda, draw=black] (14.25, -1.75) rectangle (14.50, -2.00);
                          \filldraw[fill=bermuda, draw=black] (14.50, -1.75) rectangle (14.75, -2.00);
                          \filldraw[fill=cancan, draw=black] (14.75, -1.75) rectangle (15.00, -2.00);
                          \filldraw[fill=bermuda, draw=black] (0.00, -2.00) rectangle (0.25, -2.25);
                          \filldraw[fill=cancan, draw=black] (0.25, -2.00) rectangle (0.50, -2.25);
                          \filldraw[fill=cancan, draw=black] (0.50, -2.00) rectangle (0.75, -2.25);
                          \filldraw[fill=cancan, draw=black] (0.75, -2.00) rectangle (1.00, -2.25);
                          \filldraw[fill=bermuda, draw=black] (1.00, -2.00) rectangle (1.25, -2.25);
                          \filldraw[fill=cancan, draw=black] (1.25, -2.00) rectangle (1.50, -2.25);
                          \filldraw[fill=cancan, draw=black] (1.50, -2.00) rectangle (1.75, -2.25);
                          \filldraw[fill=cancan, draw=black] (1.75, -2.00) rectangle (2.00, -2.25);
                          \filldraw[fill=cancan, draw=black] (2.00, -2.00) rectangle (2.25, -2.25);
                          \filldraw[fill=cancan, draw=black] (2.25, -2.00) rectangle (2.50, -2.25);
                          \filldraw[fill=bermuda, draw=black] (2.50, -2.00) rectangle (2.75, -2.25);
                          \filldraw[fill=bermuda, draw=black] (2.75, -2.00) rectangle (3.00, -2.25);
                          \filldraw[fill=bermuda, draw=black] (3.00, -2.00) rectangle (3.25, -2.25);
                          \filldraw[fill=cancan, draw=black] (3.25, -2.00) rectangle (3.50, -2.25);
                          \filldraw[fill=cancan, draw=black] (3.50, -2.00) rectangle (3.75, -2.25);
                          \filldraw[fill=cancan, draw=black] (3.75, -2.00) rectangle (4.00, -2.25);
                          \filldraw[fill=bermuda, draw=black] (4.00, -2.00) rectangle (4.25, -2.25);
                          \filldraw[fill=bermuda, draw=black] (4.25, -2.00) rectangle (4.50, -2.25);
                          \filldraw[fill=bermuda, draw=black] (4.50, -2.00) rectangle (4.75, -2.25);
                          \filldraw[fill=cancan, draw=black] (4.75, -2.00) rectangle (5.00, -2.25);
                          \filldraw[fill=cancan, draw=black] (5.00, -2.00) rectangle (5.25, -2.25);
                          \filldraw[fill=cancan, draw=black] (5.25, -2.00) rectangle (5.50, -2.25);
                          \filldraw[fill=bermuda, draw=black] (5.50, -2.00) rectangle (5.75, -2.25);
                          \filldraw[fill=bermuda, draw=black] (5.75, -2.00) rectangle (6.00, -2.25);
                          \filldraw[fill=bermuda, draw=black] (6.00, -2.00) rectangle (6.25, -2.25);
                          \filldraw[fill=cancan, draw=black] (6.25, -2.00) rectangle (6.50, -2.25);
                          \filldraw[fill=bermuda, draw=black] (6.50, -2.00) rectangle (6.75, -2.25);
                          \filldraw[fill=bermuda, draw=black] (6.75, -2.00) rectangle (7.00, -2.25);
                          \filldraw[fill=bermuda, draw=black] (7.00, -2.00) rectangle (7.25, -2.25);
                          \filldraw[fill=cancan, draw=black] (7.25, -2.00) rectangle (7.50, -2.25);
                          \filldraw[fill=bermuda, draw=black] (7.50, -2.00) rectangle (7.75, -2.25);
                          \filldraw[fill=cancan, draw=black] (7.75, -2.00) rectangle (8.00, -2.25);
                          \filldraw[fill=bermuda, draw=black] (8.00, -2.00) rectangle (8.25, -2.25);
                          \filldraw[fill=cancan, draw=black] (8.25, -2.00) rectangle (8.50, -2.25);
                          \filldraw[fill=cancan, draw=black] (8.50, -2.00) rectangle (8.75, -2.25);
                          \filldraw[fill=cancan, draw=black] (8.75, -2.00) rectangle (9.00, -2.25);
                          \filldraw[fill=cancan, draw=black] (9.00, -2.00) rectangle (9.25, -2.25);
                          \filldraw[fill=cancan, draw=black] (9.25, -2.00) rectangle (9.50, -2.25);
                          \filldraw[fill=bermuda, draw=black] (9.50, -2.00) rectangle (9.75, -2.25);
                          \filldraw[fill=bermuda, draw=black] (9.75, -2.00) rectangle (10.00, -2.25);
                          \filldraw[fill=bermuda, draw=black] (10.00, -2.00) rectangle (10.25, -2.25);
                          \filldraw[fill=bermuda, draw=black] (10.25, -2.00) rectangle (10.50, -2.25);
                          \filldraw[fill=bermuda, draw=black] (10.50, -2.00) rectangle (10.75, -2.25);
                          \filldraw[fill=bermuda, draw=black] (10.75, -2.00) rectangle (11.00, -2.25);
                          \filldraw[fill=bermuda, draw=black] (11.00, -2.00) rectangle (11.25, -2.25);
                          \filldraw[fill=cancan, draw=black] (11.25, -2.00) rectangle (11.50, -2.25);
                          \filldraw[fill=bermuda, draw=black] (11.50, -2.00) rectangle (11.75, -2.25);
                          \filldraw[fill=cancan, draw=black] (11.75, -2.00) rectangle (12.00, -2.25);
                          \filldraw[fill=bermuda, draw=black] (12.00, -2.00) rectangle (12.25, -2.25);
                          \filldraw[fill=bermuda, draw=black] (12.25, -2.00) rectangle (12.50, -2.25);
                          \filldraw[fill=bermuda, draw=black] (12.50, -2.00) rectangle (12.75, -2.25);
                          \filldraw[fill=cancan, draw=black] (12.75, -2.00) rectangle (13.00, -2.25);
                          \filldraw[fill=cancan, draw=black] (13.00, -2.00) rectangle (13.25, -2.25);
                          \filldraw[fill=cancan, draw=black] (13.25, -2.00) rectangle (13.50, -2.25);
                          \filldraw[fill=bermuda, draw=black] (13.50, -2.00) rectangle (13.75, -2.25);
                          \filldraw[fill=bermuda, draw=black] (13.75, -2.00) rectangle (14.00, -2.25);
                          \filldraw[fill=bermuda, draw=black] (14.00, -2.00) rectangle (14.25, -2.25);
                          \filldraw[fill=cancan, draw=black] (14.25, -2.00) rectangle (14.50, -2.25);
                          \filldraw[fill=bermuda, draw=black] (14.50, -2.00) rectangle (14.75, -2.25);
                          \filldraw[fill=bermuda, draw=black] (14.75, -2.00) rectangle (15.00, -2.25);
                          \filldraw[fill=bermuda, draw=black] (0.00, -2.25) rectangle (0.25, -2.50);
                          \filldraw[fill=cancan, draw=black] (0.25, -2.25) rectangle (0.50, -2.50);
                          \filldraw[fill=bermuda, draw=black] (0.50, -2.25) rectangle (0.75, -2.50);
                          \filldraw[fill=cancan, draw=black] (0.75, -2.25) rectangle (1.00, -2.50);
                          \filldraw[fill=bermuda, draw=black] (1.00, -2.25) rectangle (1.25, -2.50);
                          \filldraw[fill=cancan, draw=black] (1.25, -2.25) rectangle (1.50, -2.50);
                          \filldraw[fill=bermuda, draw=black] (1.50, -2.25) rectangle (1.75, -2.50);
                          \filldraw[fill=bermuda, draw=black] (1.75, -2.25) rectangle (2.00, -2.50);
                          \filldraw[fill=bermuda, draw=black] (2.00, -2.25) rectangle (2.25, -2.50);
                          \filldraw[fill=cancan, draw=black] (2.25, -2.25) rectangle (2.50, -2.50);
                          \filldraw[fill=cancan, draw=black] (2.50, -2.25) rectangle (2.75, -2.50);
                          \filldraw[fill=cancan, draw=black] (2.75, -2.25) rectangle (3.00, -2.50);
                          \filldraw[fill=bermuda, draw=black] (3.00, -2.25) rectangle (3.25, -2.50);
                          \filldraw[fill=bermuda, draw=black] (3.25, -2.25) rectangle (3.50, -2.50);
                          \filldraw[fill=bermuda, draw=black] (3.50, -2.25) rectangle (3.75, -2.50);
                          \filldraw[fill=cancan, draw=black] (3.75, -2.25) rectangle (4.00, -2.50);
                          \filldraw[fill=bermuda, draw=black] (4.00, -2.25) rectangle (4.25, -2.50);
                          \filldraw[fill=cancan, draw=black] (4.25, -2.25) rectangle (4.50, -2.50);
                          \filldraw[fill=bermuda, draw=black] (4.50, -2.25) rectangle (4.75, -2.50);
                          \filldraw[fill=cancan, draw=black] (4.75, -2.25) rectangle (5.00, -2.50);
                          \filldraw[fill=cancan, draw=black] (5.00, -2.25) rectangle (5.25, -2.50);
                          \filldraw[fill=cancan, draw=black] (5.25, -2.25) rectangle (5.50, -2.50);
                          \filldraw[fill=bermuda, draw=black] (5.50, -2.25) rectangle (5.75, -2.50);
                          \filldraw[fill=cancan, draw=black] (5.75, -2.25) rectangle (6.00, -2.50);
                          \filldraw[fill=bermuda, draw=black] (6.00, -2.25) rectangle (6.25, -2.50);
                          \filldraw[fill=cancan, draw=black] (6.25, -2.25) rectangle (6.50, -2.50);
                          \filldraw[fill=bermuda, draw=black] (6.50, -2.25) rectangle (6.75, -2.50);
                          \filldraw[fill=cancan, draw=black] (6.75, -2.25) rectangle (7.00, -2.50);
                          \filldraw[fill=bermuda, draw=black] (7.00, -2.25) rectangle (7.25, -2.50);
                          \filldraw[fill=cancan, draw=black] (7.25, -2.25) rectangle (7.50, -2.50);
                          \filldraw[fill=cancan, draw=black] (7.50, -2.25) rectangle (7.75, -2.50);
                          \filldraw[fill=cancan, draw=black] (7.75, -2.25) rectangle (8.00, -2.50);
                          \filldraw[fill=bermuda, draw=black] (8.00, -2.25) rectangle (8.25, -2.50);
                          \filldraw[fill=bermuda, draw=black] (8.25, -2.25) rectangle (8.50, -2.50);
                          \filldraw[fill=bermuda, draw=black] (8.50, -2.25) rectangle (8.75, -2.50);
                          \filldraw[fill=bermuda, draw=black] (8.75, -2.25) rectangle (9.00, -2.50);
                          \filldraw[fill=bermuda, draw=black] (9.00, -2.25) rectangle (9.25, -2.50);
                          \filldraw[fill=cancan, draw=black] (9.25, -2.25) rectangle (9.50, -2.50);
                          \filldraw[fill=bermuda, draw=black] (9.50, -2.25) rectangle (9.75, -2.50);
                          \filldraw[fill=bermuda, draw=black] (9.75, -2.25) rectangle (10.00, -2.50);
                          \filldraw[fill=bermuda, draw=black] (10.00, -2.25) rectangle (10.25, -2.50);
                          \filldraw[fill=cancan, draw=black] (10.25, -2.25) rectangle (10.50, -2.50);
                          \filldraw[fill=cancan, draw=black] (10.50, -2.25) rectangle (10.75, -2.50);
                          \filldraw[fill=cancan, draw=black] (10.75, -2.25) rectangle (11.00, -2.50);
                          \filldraw[fill=bermuda, draw=black] (11.00, -2.25) rectangle (11.25, -2.50);
                          \filldraw[fill=bermuda, draw=black] (11.25, -2.25) rectangle (11.50, -2.50);
                          \filldraw[fill=bermuda, draw=black] (11.50, -2.25) rectangle (11.75, -2.50);
                          \filldraw[fill=cancan, draw=black] (11.75, -2.25) rectangle (12.00, -2.50);
                          \filldraw[fill=bermuda, draw=black] (12.00, -2.25) rectangle (12.25, -2.50);
                          \filldraw[fill=cancan, draw=black] (12.25, -2.25) rectangle (12.50, -2.50);
                          \filldraw[fill=cancan, draw=black] (12.50, -2.25) rectangle (12.75, -2.50);
                          \filldraw[fill=bermuda, draw=black] (12.75, -2.25) rectangle (13.00, -2.50);
                          \filldraw[fill=bermuda, draw=black] (13.00, -2.25) rectangle (13.25, -2.50);
                          \filldraw[fill=cancan, draw=black] (13.25, -2.25) rectangle (13.50, -2.50);
                          \filldraw[fill=bermuda, draw=black] (13.50, -2.25) rectangle (13.75, -2.50);
                          \filldraw[fill=cancan, draw=black] (13.75, -2.25) rectangle (14.00, -2.50);
                          \filldraw[fill=cancan, draw=black] (14.00, -2.25) rectangle (14.25, -2.50);
                          \filldraw[fill=cancan, draw=black] (14.25, -2.25) rectangle (14.50, -2.50);
                          \filldraw[fill=bermuda, draw=black] (14.50, -2.25) rectangle (14.75, -2.50);
                          \filldraw[fill=bermuda, draw=black] (14.75, -2.25) rectangle (15.00, -2.50);
                          \filldraw[fill=bermuda, draw=black] (0.00, -2.50) rectangle (0.25, -2.75);
                          \filldraw[fill=bermuda, draw=black] (0.25, -2.50) rectangle (0.50, -2.75);
                          \filldraw[fill=bermuda, draw=black] (0.50, -2.50) rectangle (0.75, -2.75);
                          \filldraw[fill=cancan, draw=black] (0.75, -2.50) rectangle (1.00, -2.75);
                          \filldraw[fill=bermuda, draw=black] (1.00, -2.50) rectangle (1.25, -2.75);
                          \filldraw[fill=bermuda, draw=black] (1.25, -2.50) rectangle (1.50, -2.75);
                          \filldraw[fill=bermuda, draw=black] (1.50, -2.50) rectangle (1.75, -2.75);
                          \filldraw[fill=bermuda, draw=black] (1.75, -2.50) rectangle (2.00, -2.75);
                          \filldraw[fill=bermuda, draw=black] (2.00, -2.50) rectangle (2.25, -2.75);
                          \filldraw[fill=cancan, draw=black] (2.25, -2.50) rectangle (2.50, -2.75);
                          \filldraw[fill=bermuda, draw=black] (2.50, -2.50) rectangle (2.75, -2.75);
                          \filldraw[fill=cancan, draw=black] (2.75, -2.50) rectangle (3.00, -2.75);
                          \filldraw[fill=cancan, draw=black] (3.00, -2.50) rectangle (3.25, -2.75);
                          \filldraw[fill=bermuda, draw=black] (3.25, -2.50) rectangle (3.50, -2.75);
                          \filldraw[fill=bermuda, draw=black] (3.50, -2.50) rectangle (3.75, -2.75);
                          \filldraw[fill=cancan, draw=black] (3.75, -2.50) rectangle (4.00, -2.75);
                          \filldraw[fill=bermuda, draw=black] (4.00, -2.50) rectangle (4.25, -2.75);
                          \filldraw[fill=cancan, draw=black] (4.25, -2.50) rectangle (4.50, -2.75);
                          \filldraw[fill=cancan, draw=black] (4.50, -2.50) rectangle (4.75, -2.75);
                          \filldraw[fill=cancan, draw=black] (4.75, -2.50) rectangle (5.00, -2.75);
                          \filldraw[fill=bermuda, draw=black] (5.00, -2.50) rectangle (5.25, -2.75);
                          \filldraw[fill=cancan, draw=black] (5.25, -2.50) rectangle (5.50, -2.75);
                          \filldraw[fill=cancan, draw=black] (5.50, -2.50) rectangle (5.75, -2.75);
                          \filldraw[fill=cancan, draw=black] (5.75, -2.50) rectangle (6.00, -2.75);
                          \filldraw[fill=bermuda, draw=black] (6.00, -2.50) rectangle (6.25, -2.75);
                          \filldraw[fill=bermuda, draw=black] (6.25, -2.50) rectangle (6.50, -2.75);
                          \filldraw[fill=bermuda, draw=black] (6.50, -2.50) rectangle (6.75, -2.75);
                          \filldraw[fill=bermuda, draw=black] (6.75, -2.50) rectangle (7.00, -2.75);
                          \filldraw[fill=bermuda, draw=black] (7.00, -2.50) rectangle (7.25, -2.75);
                          \filldraw[fill=cancan, draw=black] (7.25, -2.50) rectangle (7.50, -2.75);
                          \filldraw[fill=cancan, draw=black] (7.50, -2.50) rectangle (7.75, -2.75);
                          \filldraw[fill=cancan, draw=black] (7.75, -2.50) rectangle (8.00, -2.75);
                          \filldraw[fill=bermuda, draw=black] (8.00, -2.50) rectangle (8.25, -2.75);
                          \filldraw[fill=bermuda, draw=black] (8.25, -2.50) rectangle (8.50, -2.75);
                          \filldraw[fill=bermuda, draw=black] (8.50, -2.50) rectangle (8.75, -2.75);
                          \filldraw[fill=cancan, draw=black] (8.75, -2.50) rectangle (9.00, -2.75);
                          \filldraw[fill=cancan, draw=black] (9.00, -2.50) rectangle (9.25, -2.75);
                          \filldraw[fill=cancan, draw=black] (9.25, -2.50) rectangle (9.50, -2.75);
                          \filldraw[fill=bermuda, draw=black] (9.50, -2.50) rectangle (9.75, -2.75);
                          \filldraw[fill=bermuda, draw=black] (9.75, -2.50) rectangle (10.00, -2.75);
                          \filldraw[fill=bermuda, draw=black] (10.00, -2.50) rectangle (10.25, -2.75);
                          \filldraw[fill=cancan, draw=black] (10.25, -2.50) rectangle (10.50, -2.75);
                          \filldraw[fill=cancan, draw=black] (10.50, -2.50) rectangle (10.75, -2.75);
                          \filldraw[fill=cancan, draw=black] (10.75, -2.50) rectangle (11.00, -2.75);
                          \filldraw[fill=cancan, draw=black] (11.00, -2.50) rectangle (11.25, -2.75);
                          \filldraw[fill=cancan, draw=black] (11.25, -2.50) rectangle (11.50, -2.75);
                          \filldraw[fill=bermuda, draw=black] (11.50, -2.50) rectangle (11.75, -2.75);
                          \filldraw[fill=bermuda, draw=black] (11.75, -2.50) rectangle (12.00, -2.75);
                          \filldraw[fill=bermuda, draw=black] (12.00, -2.50) rectangle (12.25, -2.75);
                          \filldraw[fill=cancan, draw=black] (12.25, -2.50) rectangle (12.50, -2.75);
                          \filldraw[fill=cancan, draw=black] (12.50, -2.50) rectangle (12.75, -2.75);
                          \filldraw[fill=cancan, draw=black] (12.75, -2.50) rectangle (13.00, -2.75);
                          \filldraw[fill=bermuda, draw=black] (13.00, -2.50) rectangle (13.25, -2.75);
                          \filldraw[fill=bermuda, draw=black] (13.25, -2.50) rectangle (13.50, -2.75);
                          \filldraw[fill=bermuda, draw=black] (13.50, -2.50) rectangle (13.75, -2.75);
                          \filldraw[fill=bermuda, draw=black] (13.75, -2.50) rectangle (14.00, -2.75);
                          \filldraw[fill=bermuda, draw=black] (14.00, -2.50) rectangle (14.25, -2.75);
                          \filldraw[fill=cancan, draw=black] (14.25, -2.50) rectangle (14.50, -2.75);
                          \filldraw[fill=bermuda, draw=black] (14.50, -2.50) rectangle (14.75, -2.75);
                          \filldraw[fill=bermuda, draw=black] (14.75, -2.50) rectangle (15.00, -2.75);
                          \filldraw[fill=bermuda, draw=black] (0.00, -2.75) rectangle (0.25, -3.00);
                          \filldraw[fill=cancan, draw=black] (0.25, -2.75) rectangle (0.50, -3.00);
                          \filldraw[fill=bermuda, draw=black] (0.50, -2.75) rectangle (0.75, -3.00);
                          \filldraw[fill=bermuda, draw=black] (0.75, -2.75) rectangle (1.00, -3.00);
                          \filldraw[fill=bermuda, draw=black] (1.00, -2.75) rectangle (1.25, -3.00);
                          \filldraw[fill=cancan, draw=black] (1.25, -2.75) rectangle (1.50, -3.00);
                          \filldraw[fill=bermuda, draw=black] (1.50, -2.75) rectangle (1.75, -3.00);
                          \filldraw[fill=cancan, draw=black] (1.75, -2.75) rectangle (2.00, -3.00);
                          \filldraw[fill=cancan, draw=black] (2.00, -2.75) rectangle (2.25, -3.00);
                          \filldraw[fill=bermuda, draw=black] (2.25, -2.75) rectangle (2.50, -3.00);
                          \filldraw[fill=bermuda, draw=black] (2.50, -2.75) rectangle (2.75, -3.00);
                          \filldraw[fill=cancan, draw=black] (2.75, -2.75) rectangle (3.00, -3.00);
                          \filldraw[fill=bermuda, draw=black] (3.00, -2.75) rectangle (3.25, -3.00);
                          \filldraw[fill=cancan, draw=black] (3.25, -2.75) rectangle (3.50, -3.00);
                          \filldraw[fill=cancan, draw=black] (3.50, -2.75) rectangle (3.75, -3.00);
                          \filldraw[fill=bermuda, draw=black] (3.75, -2.75) rectangle (4.00, -3.00);
                          \filldraw[fill=bermuda, draw=black] (4.00, -2.75) rectangle (4.25, -3.00);
                          \filldraw[fill=cancan, draw=black] (4.25, -2.75) rectangle (4.50, -3.00);
                          \filldraw[fill=bermuda, draw=black] (4.50, -2.75) rectangle (4.75, -3.00);
                          \filldraw[fill=cancan, draw=black] (4.75, -2.75) rectangle (5.00, -3.00);
                          \filldraw[fill=bermuda, draw=black] (5.00, -2.75) rectangle (5.25, -3.00);
                          \filldraw[fill=bermuda, draw=black] (5.25, -2.75) rectangle (5.50, -3.00);
                          \filldraw[fill=bermuda, draw=black] (5.50, -2.75) rectangle (5.75, -3.00);
                          \filldraw[fill=cancan, draw=black] (5.75, -2.75) rectangle (6.00, -3.00);
                          \filldraw[fill=cancan, draw=black] (6.00, -2.75) rectangle (6.25, -3.00);
                          \filldraw[fill=cancan, draw=black] (6.25, -2.75) rectangle (6.50, -3.00);
                          \filldraw[fill=cancan, draw=black] (6.50, -2.75) rectangle (6.75, -3.00);
                          \filldraw[fill=cancan, draw=black] (6.75, -2.75) rectangle (7.00, -3.00);
                          \filldraw[fill=bermuda, draw=black] (7.00, -2.75) rectangle (7.25, -3.00);
                          \filldraw[fill=bermuda, draw=black] (7.25, -2.75) rectangle (7.50, -3.00);
                          \filldraw[fill=bermuda, draw=black] (7.50, -2.75) rectangle (7.75, -3.00);
                          \filldraw[fill=cancan, draw=black] (7.75, -2.75) rectangle (8.00, -3.00);
                          \filldraw[fill=cancan, draw=black] (8.00, -2.75) rectangle (8.25, -3.00);
                          \filldraw[fill=cancan, draw=black] (8.25, -2.75) rectangle (8.50, -3.00);
                          \filldraw[fill=bermuda, draw=black] (8.50, -2.75) rectangle (8.75, -3.00);
                          \filldraw[fill=bermuda, draw=black] (8.75, -2.75) rectangle (9.00, -3.00);
                          \filldraw[fill=bermuda, draw=black] (9.00, -2.75) rectangle (9.25, -3.00);
                          \filldraw[fill=cancan, draw=black] (9.25, -2.75) rectangle (9.50, -3.00);
                          \filldraw[fill=cancan, draw=black] (9.50, -2.75) rectangle (9.75, -3.00);
                          \filldraw[fill=cancan, draw=black] (9.75, -2.75) rectangle (10.00, -3.00);
                          \filldraw[fill=bermuda, draw=black] (10.00, -2.75) rectangle (10.25, -3.00);
                          \filldraw[fill=bermuda, draw=black] (10.25, -2.75) rectangle (10.50, -3.00);
                          \filldraw[fill=bermuda, draw=black] (10.50, -2.75) rectangle (10.75, -3.00);
                          \filldraw[fill=cancan, draw=black] (10.75, -2.75) rectangle (11.00, -3.00);
                          \filldraw[fill=cancan, draw=black] (11.00, -2.75) rectangle (11.25, -3.00);
                          \filldraw[fill=cancan, draw=black] (11.25, -2.75) rectangle (11.50, -3.00);
                          \filldraw[fill=bermuda, draw=black] (11.50, -2.75) rectangle (11.75, -3.00);
                          \filldraw[fill=bermuda, draw=black] (11.75, -2.75) rectangle (12.00, -3.00);
                          \filldraw[fill=bermuda, draw=black] (12.00, -2.75) rectangle (12.25, -3.00);
                          \filldraw[fill=cancan, draw=black] (12.25, -2.75) rectangle (12.50, -3.00);
                          \filldraw[fill=cancan, draw=black] (12.50, -2.75) rectangle (12.75, -3.00);
                          \filldraw[fill=cancan, draw=black] (12.75, -2.75) rectangle (13.00, -3.00);
                          \filldraw[fill=cancan, draw=black] (13.00, -2.75) rectangle (13.25, -3.00);
                          \filldraw[fill=cancan, draw=black] (13.25, -2.75) rectangle (13.50, -3.00);
                          \filldraw[fill=bermuda, draw=black] (13.50, -2.75) rectangle (13.75, -3.00);
                          \filldraw[fill=cancan, draw=black] (13.75, -2.75) rectangle (14.00, -3.00);
                          \filldraw[fill=cancan, draw=black] (14.00, -2.75) rectangle (14.25, -3.00);
                          \filldraw[fill=cancan, draw=black] (14.25, -2.75) rectangle (14.50, -3.00);
                          \filldraw[fill=cancan, draw=black] (14.50, -2.75) rectangle (14.75, -3.00);
                          \filldraw[fill=cancan, draw=black] (14.75, -2.75) rectangle (15.00, -3.00);
                          \filldraw[fill=cancan, draw=black] (0.00, -3.00) rectangle (0.25, -3.25);
                          \filldraw[fill=cancan, draw=black] (0.25, -3.00) rectangle (0.50, -3.25);
                          \filldraw[fill=cancan, draw=black] (0.50, -3.00) rectangle (0.75, -3.25);
                          \filldraw[fill=cancan, draw=black] (0.75, -3.00) rectangle (1.00, -3.25);
                          \filldraw[fill=cancan, draw=black] (1.00, -3.00) rectangle (1.25, -3.25);
                          \filldraw[fill=cancan, draw=black] (1.25, -3.00) rectangle (1.50, -3.25);
                          \filldraw[fill=bermuda, draw=black] (1.50, -3.00) rectangle (1.75, -3.25);
                          \filldraw[fill=bermuda, draw=black] (1.75, -3.00) rectangle (2.00, -3.25);
                          \filldraw[fill=bermuda, draw=black] (2.00, -3.00) rectangle (2.25, -3.25);
                          \filldraw[fill=bermuda, draw=black] (2.25, -3.00) rectangle (2.50, -3.25);
                          \filldraw[fill=bermuda, draw=black] (2.50, -3.00) rectangle (2.75, -3.25);
                          \filldraw[fill=cancan, draw=black] (2.75, -3.00) rectangle (3.00, -3.25);
                          \filldraw[fill=bermuda, draw=black] (3.00, -3.00) rectangle (3.25, -3.25);
                          \filldraw[fill=cancan, draw=black] (3.25, -3.00) rectangle (3.50, -3.25);
                          \filldraw[fill=bermuda, draw=black] (3.50, -3.00) rectangle (3.75, -3.25);
                          \filldraw[fill=cancan, draw=black] (3.75, -3.00) rectangle (4.00, -3.25);
                          \filldraw[fill=bermuda, draw=black] (4.00, -3.00) rectangle (4.25, -3.25);
                          \filldraw[fill=cancan, draw=black] (4.25, -3.00) rectangle (4.50, -3.25);
                          \filldraw[fill=cancan, draw=black] (4.50, -3.00) rectangle (4.75, -3.25);
                          \filldraw[fill=bermuda, draw=black] (4.75, -3.00) rectangle (5.00, -3.25);
                          \filldraw[fill=bermuda, draw=black] (5.00, -3.00) rectangle (5.25, -3.25);
                          \filldraw[fill=bermuda, draw=black] (5.25, -3.00) rectangle (5.50, -3.25);
                          \filldraw[fill=bermuda, draw=black] (5.50, -3.00) rectangle (5.75, -3.25);
                          \filldraw[fill=cancan, draw=black] (5.75, -3.00) rectangle (6.00, -3.25);
                          \filldraw[fill=cancan, draw=black] (6.00, -3.00) rectangle (6.25, -3.25);
                          \filldraw[fill=cancan, draw=black] (6.25, -3.00) rectangle (6.50, -3.25);
                          \filldraw[fill=cancan, draw=black] (6.50, -3.00) rectangle (6.75, -3.25);
                          \filldraw[fill=cancan, draw=black] (6.75, -3.00) rectangle (7.00, -3.25);
                          \filldraw[fill=cancan, draw=black] (7.00, -3.00) rectangle (7.25, -3.25);
                          \filldraw[fill=cancan, draw=black] (7.25, -3.00) rectangle (7.50, -3.25);
                          \filldraw[fill=bermuda, draw=black] (7.50, -3.00) rectangle (7.75, -3.25);
                          \filldraw[fill=cancan, draw=black] (7.75, -3.00) rectangle (8.00, -3.25);
                          \filldraw[fill=bermuda, draw=black] (8.00, -3.00) rectangle (8.25, -3.25);
                          \filldraw[fill=cancan, draw=black] (8.25, -3.00) rectangle (8.50, -3.25);
                          \filldraw[fill=cancan, draw=black] (8.50, -3.00) rectangle (8.75, -3.25);
                          \filldraw[fill=bermuda, draw=black] (8.75, -3.00) rectangle (9.00, -3.25);
                          \filldraw[fill=bermuda, draw=black] (9.00, -3.00) rectangle (9.25, -3.25);
                          \filldraw[fill=bermuda, draw=black] (9.25, -3.00) rectangle (9.50, -3.25);
                          \filldraw[fill=bermuda, draw=black] (9.50, -3.00) rectangle (9.75, -3.25);
                          \filldraw[fill=cancan, draw=black] (9.75, -3.00) rectangle (10.00, -3.25);
                          \filldraw[fill=cancan, draw=black] (10.00, -3.00) rectangle (10.25, -3.25);
                          \filldraw[fill=cancan, draw=black] (10.25, -3.00) rectangle (10.50, -3.25);
                          \filldraw[fill=cancan, draw=black] (10.50, -3.00) rectangle (10.75, -3.25);
                          \filldraw[fill=bermuda, draw=black] (10.75, -3.00) rectangle (11.00, -3.25);
                          \filldraw[fill=bermuda, draw=black] (11.00, -3.00) rectangle (11.25, -3.25);
                          \filldraw[fill=bermuda, draw=black] (11.25, -3.00) rectangle (11.50, -3.25);
                          \filldraw[fill=bermuda, draw=black] (11.50, -3.00) rectangle (11.75, -3.25);
                          \filldraw[fill=cancan, draw=black] (11.75, -3.00) rectangle (12.00, -3.25);
                          \filldraw[fill=bermuda, draw=black] (12.00, -3.00) rectangle (12.25, -3.25);
                          \filldraw[fill=cancan, draw=black] (12.25, -3.00) rectangle (12.50, -3.25);
                          \filldraw[fill=cancan, draw=black] (12.50, -3.00) rectangle (12.75, -3.25);
                          \filldraw[fill=bermuda, draw=black] (12.75, -3.00) rectangle (13.00, -3.25);
                          \filldraw[fill=bermuda, draw=black] (13.00, -3.00) rectangle (13.25, -3.25);
                          \filldraw[fill=cancan, draw=black] (13.25, -3.00) rectangle (13.50, -3.25);
                          \filldraw[fill=bermuda, draw=black] (13.50, -3.00) rectangle (13.75, -3.25);
                          \filldraw[fill=cancan, draw=black] (13.75, -3.00) rectangle (14.00, -3.25);
                          \filldraw[fill=cancan, draw=black] (14.00, -3.00) rectangle (14.25, -3.25);
                          \filldraw[fill=bermuda, draw=black] (14.25, -3.00) rectangle (14.50, -3.25);
                          \filldraw[fill=bermuda, draw=black] (14.50, -3.00) rectangle (14.75, -3.25);
                          \filldraw[fill=cancan, draw=black] (14.75, -3.00) rectangle (15.00, -3.25);
                          \filldraw[fill=bermuda, draw=black] (0.00, -3.25) rectangle (0.25, -3.50);
                          \filldraw[fill=cancan, draw=black] (0.25, -3.25) rectangle (0.50, -3.50);
                          \filldraw[fill=cancan, draw=black] (0.50, -3.25) rectangle (0.75, -3.50);
                          \filldraw[fill=cancan, draw=black] (0.75, -3.25) rectangle (1.00, -3.50);
                          \filldraw[fill=cancan, draw=black] (1.00, -3.25) rectangle (1.25, -3.50);
                          \filldraw[fill=cancan, draw=black] (1.25, -3.25) rectangle (1.50, -3.50);
                          \filldraw[fill=bermuda, draw=black] (1.50, -3.25) rectangle (1.75, -3.50);
                          \filldraw[fill=bermuda, draw=black] (1.75, -3.25) rectangle (2.00, -3.50);
                          \filldraw[fill=bermuda, draw=black] (2.00, -3.25) rectangle (2.25, -3.50);} }}\end{equation*}
\begin{equation*}c = \tikz{\filldraw[fill=cancan, draw=black] (0.00, 0.00) rectangle (0.25, 0.25);
                           \filldraw[fill=cancan, draw=black] (0.25, 0.00) rectangle (0.50, 0.25);
                           \filldraw[fill=cancan, draw=black] (0.50, 0.00) rectangle (0.75, 0.25);
                           \filldraw[fill=bermuda, draw=black] (0.75, 0.00) rectangle (1.00, 0.25);
                           \filldraw[fill=bermuda, draw=black] (1.00, 0.00) rectangle (1.25, 0.25);
                           \filldraw[fill=bermuda, draw=black] (1.25, 0.00) rectangle (1.50, 0.25);
                           \filldraw[fill=cancan, draw=black] (1.50, 0.00) rectangle (1.75, 0.25);
                           \filldraw[fill=cancan, draw=black] (1.75, 0.00) rectangle (2.00, 0.25);
                           \filldraw[fill=cancan, draw=black] (2.00, 0.00) rectangle (2.25, 0.25);
                           \filldraw[fill=bermuda, draw=black] (2.25, 0.00) rectangle (2.50, 0.25);
                           \filldraw[fill=bermuda, draw=black] (2.50, 0.00) rectangle (2.75, 0.25);
                           \filldraw[fill=bermuda, draw=black] (2.75, 0.00) rectangle (3.00, 0.25);
                           \filldraw[fill=cancan, draw=black] (3.00, 0.00) rectangle (3.25, 0.25);
                           \filldraw[fill=cancan, draw=black] (3.25, 0.00) rectangle (3.50, 0.25);
                           \filldraw[fill=cancan, draw=black] (3.50, 0.00) rectangle (3.75, 0.25);
                           \filldraw[fill=bermuda, draw=black] (3.75, 0.00) rectangle (4.00, 0.25);
                           \filldraw[fill=bermuda, draw=black] (4.00, 0.00) rectangle (4.25, 0.25);
                           \filldraw[fill=bermuda, draw=black] (4.25, 0.00) rectangle (4.50, 0.25);}\end{equation*}
All strings from sequence $\omega$ can be rewritten the same way too. As described above the sequence contains 14 quadruplets $(a_{1}, b_{1}, c_{1}, x_{1})$, $\dotsb$, $(a_{14}, b_{14}, c_{14}, x_{14})$ to represent strings $( a_{1}^{n} b_{1} c_{1}^{m} )_{[x_{1}]}$, $\dotsb$, $( a_{14}^{n} b_{14} c_{14}^{m} )_{[x_{14}]}$.
The values of subarrays $a_{1, \dotsb, 14}$, $c_{1, \dotsb, 14}$ and $x_{1, \dotsb, 14}$ are the following:
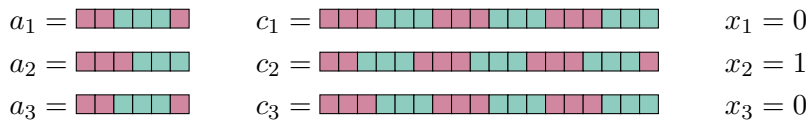
\begin{equation*}\end{equation*}  
\begin{equation*}\hspace{4pt} a_{1} = \tikz{\filldraw[fill=cancan, draw=black] (0.00, 0.00) rectangle (0.25, 0.25);
                                              \filldraw[fill=cancan, draw=black] (0.25, 0.00) rectangle (0.50, 0.25);
                                              \filldraw[fill=bermuda, draw=black] (0.50, 0.00) rectangle (0.75, 0.25);
                                              \filldraw[fill=bermuda, draw=black] (0.75, 0.00) rectangle (1.00, 0.25);
                                              \filldraw[fill=bermuda, draw=black] (1.00, 0.00) rectangle (1.25, 0.25);
                                              \filldraw[fill=cancan, draw=black] (1.25, 0.00) rectangle (1.50, 0.25);}
\hspace{25.0pt} c_{1} = \tikz{\filldraw[fill=cancan, draw=black] (0.00, 0.00) rectangle (0.25, 0.25);
                              \filldraw[fill=cancan, draw=black] (0.25, 0.00) rectangle (0.50, 0.25);
                              \filldraw[fill=cancan, draw=black] (0.50, 0.00) rectangle (0.75, 0.25);
                              \filldraw[fill=bermuda, draw=black] (0.75, 0.00) rectangle (1.00, 0.25);
                              \filldraw[fill=bermuda, draw=black] (1.00, 0.00) rectangle (1.25, 0.25);
                              \filldraw[fill=bermuda, draw=black] (1.25, 0.00) rectangle (1.50, 0.25);
                              \filldraw[fill=cancan, draw=black] (1.50, 0.00) rectangle (1.75, 0.25);
                              \filldraw[fill=cancan, draw=black] (1.75, 0.00) rectangle (2.00, 0.25);
                              \filldraw[fill=cancan, draw=black] (2.00, 0.00) rectangle (2.25, 0.25);
                              \filldraw[fill=bermuda, draw=black] (2.25, 0.00) rectangle (2.50, 0.25);
                              \filldraw[fill=bermuda, draw=black] (2.50, 0.00) rectangle (2.75, 0.25);
                              \filldraw[fill=bermuda, draw=black] (2.75, 0.00) rectangle (3.00, 0.25);
                              \filldraw[fill=cancan, draw=black] (3.00, 0.00) rectangle (3.25, 0.25);
                              \filldraw[fill=cancan, draw=black] (3.25, 0.00) rectangle (3.50, 0.25);
                              \filldraw[fill=cancan, draw=black] (3.50, 0.00) rectangle (3.75, 0.25);
                              \filldraw[fill=bermuda, draw=black] (3.75, 0.00) rectangle (4.00, 0.25);
                              \filldraw[fill=bermuda, draw=black] (4.00, 0.00) rectangle (4.25, 0.25);
                              \filldraw[fill=bermuda, draw=black] (4.25, 0.00) rectangle (4.50, 0.25);}
\hspace{25.0pt} x_{1} = 0\end{equation*}
\begin{equation*}\hspace{4pt} a_{2} = \tikz{\filldraw[fill=cancan, draw=black] (0.00, 0.00) rectangle (0.25, 0.25);
                                              \filldraw[fill=cancan, draw=black] (0.25, 0.00) rectangle (0.50, 0.25);
                                              \filldraw[fill=cancan, draw=black] (0.50, 0.00) rectangle (0.75, 0.25);
                                              \filldraw[fill=bermuda, draw=black] (0.75, 0.00) rectangle (1.00, 0.25);
                                              \filldraw[fill=bermuda, draw=black] (1.00, 0.00) rectangle (1.25, 0.25);
                                              \filldraw[fill=bermuda, draw=black] (1.25, 0.00) rectangle (1.50, 0.25);}
\hspace{25.0pt} c_{2} = \tikz{\filldraw[fill=cancan, draw=black] (0.00, 0.00) rectangle (0.25, 0.25);
                              \filldraw[fill=cancan, draw=black] (0.25, 0.00) rectangle (0.50, 0.25);
                              \filldraw[fill=bermuda, draw=black] (0.50, 0.00) rectangle (0.75, 0.25);
                              \filldraw[fill=bermuda, draw=black] (0.75, 0.00) rectangle (1.00, 0.25);
                              \filldraw[fill=bermuda, draw=black] (1.00, 0.00) rectangle (1.25, 0.25);
                              \filldraw[fill=cancan, draw=black] (1.25, 0.00) rectangle (1.50, 0.25);
                              \filldraw[fill=cancan, draw=black] (1.50, 0.00) rectangle (1.75, 0.25);
                              \filldraw[fill=cancan, draw=black] (1.75, 0.00) rectangle (2.00, 0.25);
                              \filldraw[fill=bermuda, draw=black] (2.00, 0.00) rectangle (2.25, 0.25);
                              \filldraw[fill=bermuda, draw=black] (2.25, 0.00) rectangle (2.50, 0.25);
                              \filldraw[fill=bermuda, draw=black] (2.50, 0.00) rectangle (2.75, 0.25);
                              \filldraw[fill=cancan, draw=black] (2.75, 0.00) rectangle (3.00, 0.25);
                              \filldraw[fill=cancan, draw=black] (3.00, 0.00) rectangle (3.25, 0.25);
                              \filldraw[fill=cancan, draw=black] (3.25, 0.00) rectangle (3.50, 0.25);
                              \filldraw[fill=bermuda, draw=black] (3.50, 0.00) rectangle (3.75, 0.25);
                              \filldraw[fill=bermuda, draw=black] (3.75, 0.00) rectangle (4.00, 0.25);
                              \filldraw[fill=bermuda, draw=black] (4.00, 0.00) rectangle (4.25, 0.25);
                              \filldraw[fill=cancan, draw=black] (4.25, 0.00) rectangle (4.50, 0.25);}
\hspace{25.0pt} x_{2} = 1\end{equation*}
\begin{equation*}\hspace{4pt} a_{3} = \tikz{\filldraw[fill=cancan, draw=black] (0.00, 0.00) rectangle (0.25, 0.25);
                                              \filldraw[fill=cancan, draw=black] (0.25, 0.00) rectangle (0.50, 0.25);
                                              \filldraw[fill=bermuda, draw=black] (0.50, 0.00) rectangle (0.75, 0.25);
                                              \filldraw[fill=bermuda, draw=black] (0.75, 0.00) rectangle (1.00, 0.25);
                                              \filldraw[fill=bermuda, draw=black] (1.00, 0.00) rectangle (1.25, 0.25);
                                              \filldraw[fill=cancan, draw=black] (1.25, 0.00) rectangle (1.50, 0.25);}
\hspace{25.0pt} c_{3} = \tikz{\filldraw[fill=cancan, draw=black] (0.00, 0.00) rectangle (0.25, 0.25);
                              \filldraw[fill=cancan, draw=black] (0.25, 0.00) rectangle (0.50, 0.25);
                              \filldraw[fill=cancan, draw=black] (0.50, 0.00) rectangle (0.75, 0.25);
                              \filldraw[fill=bermuda, draw=black] (0.75, 0.00) rectangle (1.00, 0.25);
                              \filldraw[fill=bermuda, draw=black] (1.00, 0.00) rectangle (1.25, 0.25);
                              \filldraw[fill=bermuda, draw=black] (1.25, 0.00) rectangle (1.50, 0.25);
                              \filldraw[fill=cancan, draw=black] (1.50, 0.00) rectangle (1.75, 0.25);
                              \filldraw[fill=cancan, draw=black] (1.75, 0.00) rectangle (2.00, 0.25);
                              \filldraw[fill=cancan, draw=black] (2.00, 0.00) rectangle (2.25, 0.25);
                              \filldraw[fill=bermuda, draw=black] (2.25, 0.00) rectangle (2.50, 0.25);
                              \filldraw[fill=bermuda, draw=black] (2.50, 0.00) rectangle (2.75, 0.25);
                              \filldraw[fill=bermuda, draw=black] (2.75, 0.00) rectangle (3.00, 0.25);
                              \filldraw[fill=cancan, draw=black] (3.00, 0.00) rectangle (3.25, 0.25);
                              \filldraw[fill=cancan, draw=black] (3.25, 0.00) rectangle (3.50, 0.25);
                              \filldraw[fill=cancan, draw=black] (3.50, 0.00) rectangle (3.75, 0.25);
                              \filldraw[fill=bermuda, draw=black] (3.75, 0.00) rectangle (4.00, 0.25);
                              \filldraw[fill=bermuda, draw=black] (4.00, 0.00) rectangle (4.25, 0.25);
                              \filldraw[fill=bermuda, draw=black] (4.25, 0.00) rectangle (4.50, 0.25);}
\hspace{25.0pt} x_{3} = 0\end{equation*}
\begin{equation*}\hspace{4pt} a_{4} = \tikz{\filldraw[fill=cancan, draw=black] (0.00, 0.00) rectangle (0.25, 0.25);
                                              \filldraw[fill=cancan, draw=black] (0.25, 0.00) rectangle (0.50, 0.25);
                                              \filldraw[fill=cancan, draw=black] (0.50, 0.00) rectangle (0.75, 0.25);
                                              \filldraw[fill=bermuda, draw=black] (0.75, 0.00) rectangle (1.00, 0.25);
                                              \filldraw[fill=bermuda, draw=black] (1.00, 0.00) rectangle (1.25, 0.25);
                                              \filldraw[fill=bermuda, draw=black] (1.25, 0.00) rectangle (1.50, 0.25);}
\hspace{25.0pt} c_{4} = \tikz{\filldraw[fill=cancan, draw=black] (0.00, 0.00) rectangle (0.25, 0.25);
                              \filldraw[fill=cancan, draw=black] (0.25, 0.00) rectangle (0.50, 0.25);
                              \filldraw[fill=cancan, draw=black] (0.50, 0.00) rectangle (0.75, 0.25);
                              \filldraw[fill=bermuda, draw=black] (0.75, 0.00) rectangle (1.00, 0.25);
                              \filldraw[fill=bermuda, draw=black] (1.00, 0.00) rectangle (1.25, 0.25);
                              \filldraw[fill=bermuda, draw=black] (1.25, 0.00) rectangle (1.50, 0.25);
                              \filldraw[fill=cancan, draw=black] (1.50, 0.00) rectangle (1.75, 0.25);
                              \filldraw[fill=cancan, draw=black] (1.75, 0.00) rectangle (2.00, 0.25);
                              \filldraw[fill=cancan, draw=black] (2.00, 0.00) rectangle (2.25, 0.25);
                              \filldraw[fill=bermuda, draw=black] (2.25, 0.00) rectangle (2.50, 0.25);
                              \filldraw[fill=bermuda, draw=black] (2.50, 0.00) rectangle (2.75, 0.25);
                              \filldraw[fill=bermuda, draw=black] (2.75, 0.00) rectangle (3.00, 0.25);
                              \filldraw[fill=cancan, draw=black] (3.00, 0.00) rectangle (3.25, 0.25);
                              \filldraw[fill=cancan, draw=black] (3.25, 0.00) rectangle (3.50, 0.25);
                              \filldraw[fill=cancan, draw=black] (3.50, 0.00) rectangle (3.75, 0.25);
                              \filldraw[fill=bermuda, draw=black] (3.75, 0.00) rectangle (4.00, 0.25);
                              \filldraw[fill=bermuda, draw=black] (4.00, 0.00) rectangle (4.25, 0.25);
                              \filldraw[fill=bermuda, draw=black] (4.25, 0.00) rectangle (4.50, 0.25);}
\hspace{25.0pt} x_{4} = 2\end{equation*}
\begin{equation*}\hspace{4pt} a_{5} = \tikz{\filldraw[fill=cancan, draw=black] (0.00, 0.00) rectangle (0.25, 0.25);
                                              \filldraw[fill=cancan, draw=black] (0.25, 0.00) rectangle (0.50, 0.25);
                                              \filldraw[fill=cancan, draw=black] (0.50, 0.00) rectangle (0.75, 0.25);
                                              \filldraw[fill=bermuda, draw=black] (0.75, 0.00) rectangle (1.00, 0.25);
                                              \filldraw[fill=bermuda, draw=black] (1.00, 0.00) rectangle (1.25, 0.25);
                                              \filldraw[fill=bermuda, draw=black] (1.25, 0.00) rectangle (1.50, 0.25);}
\hspace{25.0pt} c_{5} = \tikz{\filldraw[fill=cancan, draw=black] (0.00, 0.00) rectangle (0.25, 0.25);
                              \filldraw[fill=cancan, draw=black] (0.25, 0.00) rectangle (0.50, 0.25);
                              \filldraw[fill=bermuda, draw=black] (0.50, 0.00) rectangle (0.75, 0.25);
                              \filldraw[fill=bermuda, draw=black] (0.75, 0.00) rectangle (1.00, 0.25);
                              \filldraw[fill=bermuda, draw=black] (1.00, 0.00) rectangle (1.25, 0.25);
                              \filldraw[fill=cancan, draw=black] (1.25, 0.00) rectangle (1.50, 0.25);
                              \filldraw[fill=cancan, draw=black] (1.50, 0.00) rectangle (1.75, 0.25);
                              \filldraw[fill=cancan, draw=black] (1.75, 0.00) rectangle (2.00, 0.25);
                              \filldraw[fill=bermuda, draw=black] (2.00, 0.00) rectangle (2.25, 0.25);
                              \filldraw[fill=bermuda, draw=black] (2.25, 0.00) rectangle (2.50, 0.25);
                              \filldraw[fill=bermuda, draw=black] (2.50, 0.00) rectangle (2.75, 0.25);
                              \filldraw[fill=cancan, draw=black] (2.75, 0.00) rectangle (3.00, 0.25);
                              \filldraw[fill=cancan, draw=black] (3.00, 0.00) rectangle (3.25, 0.25);
                              \filldraw[fill=cancan, draw=black] (3.25, 0.00) rectangle (3.50, 0.25);
                              \filldraw[fill=bermuda, draw=black] (3.50, 0.00) rectangle (3.75, 0.25);
                              \filldraw[fill=bermuda, draw=black] (3.75, 0.00) rectangle (4.00, 0.25);
                              \filldraw[fill=bermuda, draw=black] (4.00, 0.00) rectangle (4.25, 0.25);
                              \filldraw[fill=cancan, draw=black] (4.25, 0.00) rectangle (4.50, 0.25);}
\hspace{25.0pt} x_{5} = 1\end{equation*}
\begin{equation*}\hspace{4pt} a_{6} = \tikz{\filldraw[fill=cancan, draw=black] (0.00, 0.00) rectangle (0.25, 0.25);
                                              \filldraw[fill=cancan, draw=black] (0.25, 0.00) rectangle (0.50, 0.25);
                                              \filldraw[fill=bermuda, draw=black] (0.50, 0.00) rectangle (0.75, 0.25);
                                              \filldraw[fill=bermuda, draw=black] (0.75, 0.00) rectangle (1.00, 0.25);
                                              \filldraw[fill=bermuda, draw=black] (1.00, 0.00) rectangle (1.25, 0.25);
                                              \filldraw[fill=cancan, draw=black] (1.25, 0.00) rectangle (1.50, 0.25);}
\hspace{25.0pt} c_{6} = \tikz{\filldraw[fill=cancan, draw=black] (0.00, 0.00) rectangle (0.25, 0.25);
                              \filldraw[fill=cancan, draw=black] (0.25, 0.00) rectangle (0.50, 0.25);
                              \filldraw[fill=cancan, draw=black] (0.50, 0.00) rectangle (0.75, 0.25);
                              \filldraw[fill=bermuda, draw=black] (0.75, 0.00) rectangle (1.00, 0.25);
                              \filldraw[fill=bermuda, draw=black] (1.00, 0.00) rectangle (1.25, 0.25);
                              \filldraw[fill=bermuda, draw=black] (1.25, 0.00) rectangle (1.50, 0.25);
                              \filldraw[fill=cancan, draw=black] (1.50, 0.00) rectangle (1.75, 0.25);
                              \filldraw[fill=cancan, draw=black] (1.75, 0.00) rectangle (2.00, 0.25);
                              \filldraw[fill=cancan, draw=black] (2.00, 0.00) rectangle (2.25, 0.25);
                              \filldraw[fill=bermuda, draw=black] (2.25, 0.00) rectangle (2.50, 0.25);
                              \filldraw[fill=bermuda, draw=black] (2.50, 0.00) rectangle (2.75, 0.25);
                              \filldraw[fill=bermuda, draw=black] (2.75, 0.00) rectangle (3.00, 0.25);
                              \filldraw[fill=cancan, draw=black] (3.00, 0.00) rectangle (3.25, 0.25);
                              \filldraw[fill=cancan, draw=black] (3.25, 0.00) rectangle (3.50, 0.25);
                              \filldraw[fill=cancan, draw=black] (3.50, 0.00) rectangle (3.75, 0.25);
                              \filldraw[fill=bermuda, draw=black] (3.75, 0.00) rectangle (4.00, 0.25);
                              \filldraw[fill=bermuda, draw=black] (4.00, 0.00) rectangle (4.25, 0.25);
                              \filldraw[fill=bermuda, draw=black] (4.25, 0.00) rectangle (4.50, 0.25);}
\hspace{25.0pt} x_{6} = 0\end{equation*}
\begin{equation*}\hspace{4pt} a_{7} = \tikz{\filldraw[fill=cancan, draw=black] (0.00, 0.00) rectangle (0.25, 0.25);
                                              \filldraw[fill=cancan, draw=black] (0.25, 0.00) rectangle (0.50, 0.25);
                                              \filldraw[fill=cancan, draw=black] (0.50, 0.00) rectangle (0.75, 0.25);
                                              \filldraw[fill=bermuda, draw=black] (0.75, 0.00) rectangle (1.00, 0.25);
                                              \filldraw[fill=bermuda, draw=black] (1.00, 0.00) rectangle (1.25, 0.25);
                                              \filldraw[fill=bermuda, draw=black] (1.25, 0.00) rectangle (1.50, 0.25);}
\hspace{25.0pt} c_{7} = \tikz{\filldraw[fill=cancan, draw=black] (0.00, 0.00) rectangle (0.25, 0.25);
                              \filldraw[fill=cancan, draw=black] (0.25, 0.00) rectangle (0.50, 0.25);
                              \filldraw[fill=bermuda, draw=black] (0.50, 0.00) rectangle (0.75, 0.25);
                              \filldraw[fill=bermuda, draw=black] (0.75, 0.00) rectangle (1.00, 0.25);
                              \filldraw[fill=bermuda, draw=black] (1.00, 0.00) rectangle (1.25, 0.25);
                              \filldraw[fill=cancan, draw=black] (1.25, 0.00) rectangle (1.50, 0.25);
                              \filldraw[fill=cancan, draw=black] (1.50, 0.00) rectangle (1.75, 0.25);
                              \filldraw[fill=cancan, draw=black] (1.75, 0.00) rectangle (2.00, 0.25);
                              \filldraw[fill=bermuda, draw=black] (2.00, 0.00) rectangle (2.25, 0.25);
                              \filldraw[fill=bermuda, draw=black] (2.25, 0.00) rectangle (2.50, 0.25);
                              \filldraw[fill=bermuda, draw=black] (2.50, 0.00) rectangle (2.75, 0.25);
                              \filldraw[fill=cancan, draw=black] (2.75, 0.00) rectangle (3.00, 0.25);
                              \filldraw[fill=cancan, draw=black] (3.00, 0.00) rectangle (3.25, 0.25);
                              \filldraw[fill=cancan, draw=black] (3.25, 0.00) rectangle (3.50, 0.25);
                              \filldraw[fill=bermuda, draw=black] (3.50, 0.00) rectangle (3.75, 0.25);
                              \filldraw[fill=bermuda, draw=black] (3.75, 0.00) rectangle (4.00, 0.25);
                              \filldraw[fill=bermuda, draw=black] (4.00, 0.00) rectangle (4.25, 0.25);
                              \filldraw[fill=cancan, draw=black] (4.25, 0.00) rectangle (4.50, 0.25);}
\hspace{25.0pt} x_{7} = 1\end{equation*}
\begin{equation*}\hspace{4pt} a_{8} = \tikz{\filldraw[fill=cancan, draw=black] (0.00, 0.00) rectangle (0.25, 0.25);
                                              \filldraw[fill=cancan, draw=black] (0.25, 0.00) rectangle (0.50, 0.25);
                                              \filldraw[fill=bermuda, draw=black] (0.50, 0.00) rectangle (0.75, 0.25);
                                              \filldraw[fill=bermuda, draw=black] (0.75, 0.00) rectangle (1.00, 0.25);
                                              \filldraw[fill=bermuda, draw=black] (1.00, 0.00) rectangle (1.25, 0.25);
                                              \filldraw[fill=cancan, draw=black] (1.25, 0.00) rectangle (1.50, 0.25);}
\hspace{25.0pt} c_{8} = \tikz{\filldraw[fill=cancan, draw=black] (0.00, 0.00) rectangle (0.25, 0.25);
                              \filldraw[fill=cancan, draw=black] (0.25, 0.00) rectangle (0.50, 0.25);
                              \filldraw[fill=cancan, draw=black] (0.50, 0.00) rectangle (0.75, 0.25);
                              \filldraw[fill=bermuda, draw=black] (0.75, 0.00) rectangle (1.00, 0.25);
                              \filldraw[fill=bermuda, draw=black] (1.00, 0.00) rectangle (1.25, 0.25);
                              \filldraw[fill=bermuda, draw=black] (1.25, 0.00) rectangle (1.50, 0.25);
                              \filldraw[fill=cancan, draw=black] (1.50, 0.00) rectangle (1.75, 0.25);
                              \filldraw[fill=cancan, draw=black] (1.75, 0.00) rectangle (2.00, 0.25);
                              \filldraw[fill=cancan, draw=black] (2.00, 0.00) rectangle (2.25, 0.25);
                              \filldraw[fill=bermuda, draw=black] (2.25, 0.00) rectangle (2.50, 0.25);
                              \filldraw[fill=bermuda, draw=black] (2.50, 0.00) rectangle (2.75, 0.25);
                              \filldraw[fill=bermuda, draw=black] (2.75, 0.00) rectangle (3.00, 0.25);
                              \filldraw[fill=cancan, draw=black] (3.00, 0.00) rectangle (3.25, 0.25);
                              \filldraw[fill=cancan, draw=black] (3.25, 0.00) rectangle (3.50, 0.25);
                              \filldraw[fill=cancan, draw=black] (3.50, 0.00) rectangle (3.75, 0.25);
                              \filldraw[fill=bermuda, draw=black] (3.75, 0.00) rectangle (4.00, 0.25);
                              \filldraw[fill=bermuda, draw=black] (4.00, 0.00) rectangle (4.25, 0.25);
                              \filldraw[fill=bermuda, draw=black] (4.25, 0.00) rectangle (4.50, 0.25);}
\hspace{25.0pt} x_{8} = 0\end{equation*}
\begin{equation*}\hspace{4pt} a_{9} = \tikz{\filldraw[fill=cancan, draw=black] (0.00, 0.00) rectangle (0.25, 0.25);
                                              \filldraw[fill=cancan, draw=black] (0.25, 0.00) rectangle (0.50, 0.25);
                                              \filldraw[fill=cancan, draw=black] (0.50, 0.00) rectangle (0.75, 0.25);
                                              \filldraw[fill=bermuda, draw=black] (0.75, 0.00) rectangle (1.00, 0.25);
                                              \filldraw[fill=bermuda, draw=black] (1.00, 0.00) rectangle (1.25, 0.25);
                                              \filldraw[fill=bermuda, draw=black] (1.25, 0.00) rectangle (1.50, 0.25);}
\hspace{25.0pt} c_{9} = \tikz{\filldraw[fill=cancan, draw=black] (0.00, 0.00) rectangle (0.25, 0.25);
                              \filldraw[fill=cancan, draw=black] (0.25, 0.00) rectangle (0.50, 0.25);
                              \filldraw[fill=bermuda, draw=black] (0.50, 0.00) rectangle (0.75, 0.25);
                              \filldraw[fill=bermuda, draw=black] (0.75, 0.00) rectangle (1.00, 0.25);
                              \filldraw[fill=bermuda, draw=black] (1.00, 0.00) rectangle (1.25, 0.25);
                              \filldraw[fill=cancan, draw=black] (1.25, 0.00) rectangle (1.50, 0.25);
                              \filldraw[fill=cancan, draw=black] (1.50, 0.00) rectangle (1.75, 0.25);
                              \filldraw[fill=cancan, draw=black] (1.75, 0.00) rectangle (2.00, 0.25);
                              \filldraw[fill=bermuda, draw=black] (2.00, 0.00) rectangle (2.25, 0.25);
                              \filldraw[fill=bermuda, draw=black] (2.25, 0.00) rectangle (2.50, 0.25);
                              \filldraw[fill=bermuda, draw=black] (2.50, 0.00) rectangle (2.75, 0.25);
                              \filldraw[fill=cancan, draw=black] (2.75, 0.00) rectangle (3.00, 0.25);
                              \filldraw[fill=cancan, draw=black] (3.00, 0.00) rectangle (3.25, 0.25);
                              \filldraw[fill=cancan, draw=black] (3.25, 0.00) rectangle (3.50, 0.25);
                              \filldraw[fill=bermuda, draw=black] (3.50, 0.00) rectangle (3.75, 0.25);
                              \filldraw[fill=bermuda, draw=black] (3.75, 0.00) rectangle (4.00, 0.25);
                              \filldraw[fill=bermuda, draw=black] (4.00, 0.00) rectangle (4.25, 0.25);
                              \filldraw[fill=cancan, draw=black] (4.25, 0.00) rectangle (4.50, 0.25);}
\hspace{25.0pt} x_{9} = 1\end{equation*}
\begin{equation*}\hspace{0.3pt} a_{10} = \tikz{\filldraw[fill=cancan, draw=black] (0.00, 0.00) rectangle (0.25, 0.25);
                                              \filldraw[fill=cancan, draw=black] (0.25, 0.00) rectangle (0.50, 0.25);
                                              \filldraw[fill=bermuda, draw=black] (0.50, 0.00) rectangle (0.75, 0.25);
                                              \filldraw[fill=bermuda, draw=black] (0.75, 0.00) rectangle (1.00, 0.25);
                                              \filldraw[fill=bermuda, draw=black] (1.00, 0.00) rectangle (1.25, 0.25);
                                              \filldraw[fill=cancan, draw=black] (1.25, 0.00) rectangle (1.50, 0.25);}
\hspace{21.0pt} c_{10} = \tikz{\filldraw[fill=cancan, draw=black] (0.00, 0.00) rectangle (0.25, 0.25);
                              \filldraw[fill=bermuda, draw=black] (0.25, 0.00) rectangle (0.50, 0.25);
                              \filldraw[fill=bermuda, draw=black] (0.50, 0.00) rectangle (0.75, 0.25);
                              \filldraw[fill=bermuda, draw=black] (0.75, 0.00) rectangle (1.00, 0.25);
                              \filldraw[fill=cancan, draw=black] (1.00, 0.00) rectangle (1.25, 0.25);
                              \filldraw[fill=cancan, draw=black] (1.25, 0.00) rectangle (1.50, 0.25);
                              \filldraw[fill=cancan, draw=black] (1.50, 0.00) rectangle (1.75, 0.25);
                              \filldraw[fill=bermuda, draw=black] (1.75, 0.00) rectangle (2.00, 0.25);
                              \filldraw[fill=bermuda, draw=black] (2.00, 0.00) rectangle (2.25, 0.25);
                              \filldraw[fill=bermuda, draw=black] (2.25, 0.00) rectangle (2.50, 0.25);
                              \filldraw[fill=cancan, draw=black] (2.50, 0.00) rectangle (2.75, 0.25);
                              \filldraw[fill=cancan, draw=black] (2.75, 0.00) rectangle (3.00, 0.25);
                              \filldraw[fill=cancan, draw=black] (3.00, 0.00) rectangle (3.25, 0.25);
                              \filldraw[fill=bermuda, draw=black] (3.25, 0.00) rectangle (3.50, 0.25);
                              \filldraw[fill=bermuda, draw=black] (3.50, 0.00) rectangle (3.75, 0.25);
                              \filldraw[fill=bermuda, draw=black] (3.75, 0.00) rectangle (4.00, 0.25);
                              \filldraw[fill=cancan, draw=black] (4.00, 0.00) rectangle (4.25, 0.25);
                              \filldraw[fill=cancan, draw=black] (4.25, 0.00) rectangle (4.50, 0.25);}
\hspace{21.0pt} x_{10} = 2\end{equation*}
\begin{equation*}\hspace{0.3pt} a_{11} = \tikz{\filldraw[fill=cancan, draw=black] (0.00, 0.00) rectangle (0.25, 0.25);
                                              \filldraw[fill=bermuda, draw=black] (0.25, 0.00) rectangle (0.50, 0.25);
                                              \filldraw[fill=bermuda, draw=black] (0.50, 0.00) rectangle (0.75, 0.25);
                                              \filldraw[fill=bermuda, draw=black] (0.75, 0.00) rectangle (1.00, 0.25);
                                              \filldraw[fill=cancan, draw=black] (1.00, 0.00) rectangle (1.25, 0.25);
                                              \filldraw[fill=cancan, draw=black] (1.25, 0.00) rectangle (1.50, 0.25);}
\hspace{21.0pt} c_{11} = \tikz{\filldraw[fill=cancan, draw=black] (0.00, 0.00) rectangle (0.25, 0.25);
                              \filldraw[fill=bermuda, draw=black] (0.25, 0.00) rectangle (0.50, 0.25);
                              \filldraw[fill=bermuda, draw=black] (0.50, 0.00) rectangle (0.75, 0.25);
                              \filldraw[fill=bermuda, draw=black] (0.75, 0.00) rectangle (1.00, 0.25);
                              \filldraw[fill=cancan, draw=black] (1.00, 0.00) rectangle (1.25, 0.25);
                              \filldraw[fill=cancan, draw=black] (1.25, 0.00) rectangle (1.50, 0.25);
                              \filldraw[fill=cancan, draw=black] (1.50, 0.00) rectangle (1.75, 0.25);
                              \filldraw[fill=bermuda, draw=black] (1.75, 0.00) rectangle (2.00, 0.25);
                              \filldraw[fill=bermuda, draw=black] (2.00, 0.00) rectangle (2.25, 0.25);
                              \filldraw[fill=bermuda, draw=black] (2.25, 0.00) rectangle (2.50, 0.25);
                              \filldraw[fill=cancan, draw=black] (2.50, 0.00) rectangle (2.75, 0.25);
                              \filldraw[fill=cancan, draw=black] (2.75, 0.00) rectangle (3.00, 0.25);
                              \filldraw[fill=cancan, draw=black] (3.00, 0.00) rectangle (3.25, 0.25);
                              \filldraw[fill=bermuda, draw=black] (3.25, 0.00) rectangle (3.50, 0.25);
                              \filldraw[fill=bermuda, draw=black] (3.50, 0.00) rectangle (3.75, 0.25);
                              \filldraw[fill=bermuda, draw=black] (3.75, 0.00) rectangle (4.00, 0.25);
                              \filldraw[fill=cancan, draw=black] (4.00, 0.00) rectangle (4.25, 0.25);
                              \filldraw[fill=cancan, draw=black] (4.25, 0.00) rectangle (4.50, 0.25);}
\hspace{21.0pt} x_{11} = 0\end{equation*}
\begin{equation*}\hspace{0.3pt} a_{12} = \tikz{\filldraw[fill=cancan, draw=black] (0.00, 0.00) rectangle (0.25, 0.25);
                                              \filldraw[fill=bermuda, draw=black] (0.25, 0.00) rectangle (0.50, 0.25);
                                              \filldraw[fill=bermuda, draw=black] (0.50, 0.00) rectangle (0.75, 0.25);
                                              \filldraw[fill=bermuda, draw=black] (0.75, 0.00) rectangle (1.00, 0.25);
                                              \filldraw[fill=cancan, draw=black] (1.00, 0.00) rectangle (1.25, 0.25);
                                              \filldraw[fill=cancan, draw=black] (1.25, 0.00) rectangle (1.50, 0.25);}
\hspace{21.0pt} c_{12} = \tikz{\filldraw[fill=cancan, draw=black] (0.00, 0.00) rectangle (0.25, 0.25);
                              \filldraw[fill=bermuda, draw=black] (0.25, 0.00) rectangle (0.50, 0.25);
                              \filldraw[fill=bermuda, draw=black] (0.50, 0.00) rectangle (0.75, 0.25);
                              \filldraw[fill=bermuda, draw=black] (0.75, 0.00) rectangle (1.00, 0.25);
                              \filldraw[fill=cancan, draw=black] (1.00, 0.00) rectangle (1.25, 0.25);
                              \filldraw[fill=cancan, draw=black] (1.25, 0.00) rectangle (1.50, 0.25);
                              \filldraw[fill=cancan, draw=black] (1.50, 0.00) rectangle (1.75, 0.25);
                              \filldraw[fill=bermuda, draw=black] (1.75, 0.00) rectangle (2.00, 0.25);
                              \filldraw[fill=bermuda, draw=black] (2.00, 0.00) rectangle (2.25, 0.25);
                              \filldraw[fill=bermuda, draw=black] (2.25, 0.00) rectangle (2.50, 0.25);
                              \filldraw[fill=cancan, draw=black] (2.50, 0.00) rectangle (2.75, 0.25);
                              \filldraw[fill=cancan, draw=black] (2.75, 0.00) rectangle (3.00, 0.25);
                              \filldraw[fill=cancan, draw=black] (3.00, 0.00) rectangle (3.25, 0.25);
                              \filldraw[fill=bermuda, draw=black] (3.25, 0.00) rectangle (3.50, 0.25);
                              \filldraw[fill=bermuda, draw=black] (3.50, 0.00) rectangle (3.75, 0.25);
                              \filldraw[fill=bermuda, draw=black] (3.75, 0.00) rectangle (4.00, 0.25);
                              \filldraw[fill=cancan, draw=black] (4.00, 0.00) rectangle (4.25, 0.25);
                              \filldraw[fill=cancan, draw=black] (4.25, 0.00) rectangle (4.50, 0.25);}
\hspace{21.0pt} x_{12} = 0\end{equation*}
\begin{equation*}\hspace{0.3pt} a_{13} = \tikz{\filldraw[fill=cancan, draw=black] (0.00, 0.00) rectangle (0.25, 0.25);
                                              \filldraw[fill=bermuda, draw=black] (0.25, 0.00) rectangle (0.50, 0.25);
                                              \filldraw[fill=bermuda, draw=black] (0.50, 0.00) rectangle (0.75, 0.25);
                                              \filldraw[fill=bermuda, draw=black] (0.75, 0.00) rectangle (1.00, 0.25);
                                              \filldraw[fill=cancan, draw=black] (1.00, 0.00) rectangle (1.25, 0.25);
                                              \filldraw[fill=cancan, draw=black] (1.25, 0.00) rectangle (1.50, 0.25);}
\hspace{21.0pt} c_{13} = \tikz{\filldraw[fill=cancan, draw=black] (0.00, 0.00) rectangle (0.25, 0.25);
                              \filldraw[fill=cancan, draw=black] (0.25, 0.00) rectangle (0.50, 0.25);
                              \filldraw[fill=bermuda, draw=black] (0.50, 0.00) rectangle (0.75, 0.25);
                              \filldraw[fill=bermuda, draw=black] (0.75, 0.00) rectangle (1.00, 0.25);
                              \filldraw[fill=bermuda, draw=black] (1.00, 0.00) rectangle (1.25, 0.25);
                              \filldraw[fill=cancan, draw=black] (1.25, 0.00) rectangle (1.50, 0.25);
                              \filldraw[fill=cancan, draw=black] (1.50, 0.00) rectangle (1.75, 0.25);
                              \filldraw[fill=cancan, draw=black] (1.75, 0.00) rectangle (2.00, 0.25);
                              \filldraw[fill=bermuda, draw=black] (2.00, 0.00) rectangle (2.25, 0.25);
                              \filldraw[fill=bermuda, draw=black] (2.25, 0.00) rectangle (2.50, 0.25);
                              \filldraw[fill=bermuda, draw=black] (2.50, 0.00) rectangle (2.75, 0.25);
                              \filldraw[fill=cancan, draw=black] (2.75, 0.00) rectangle (3.00, 0.25);
                              \filldraw[fill=cancan, draw=black] (3.00, 0.00) rectangle (3.25, 0.25);
                              \filldraw[fill=cancan, draw=black] (3.25, 0.00) rectangle (3.50, 0.25);
                              \filldraw[fill=bermuda, draw=black] (3.50, 0.00) rectangle (3.75, 0.25);
                              \filldraw[fill=bermuda, draw=black] (3.75, 0.00) rectangle (4.00, 0.25);
                              \filldraw[fill=bermuda, draw=black] (4.00, 0.00) rectangle (4.25, 0.25);
                              \filldraw[fill=cancan, draw=black] (4.25, 0.00) rectangle (4.50, 0.25);}
\hspace{21.0pt} x_{13} = 1\end{equation*}
\begin{equation*}\hspace{0.3pt} a_{14} = \tikz{\filldraw[fill=cancan, draw=black] (0.00, 0.00) rectangle (0.25, 0.25);
                                              \filldraw[fill=cancan, draw=black] (0.25, 0.00) rectangle (0.50, 0.25);
                                              \filldraw[fill=bermuda, draw=black] (0.50, 0.00) rectangle (0.75, 0.25);
                                              \filldraw[fill=bermuda, draw=black] (0.75, 0.00) rectangle (1.00, 0.25);
                                              \filldraw[fill=bermuda, draw=black] (1.00, 0.00) rectangle (1.25, 0.25);
                                              \filldraw[fill=cancan, draw=black] (1.25, 0.00) rectangle (1.50, 0.25);}
\hspace{21.0pt} c_{14} = \tikz{\filldraw[fill=cancan, draw=black] (0.00, 0.00) rectangle (0.25, 0.25);
                              \filldraw[fill=cancan, draw=black] (0.25, 0.00) rectangle (0.50, 0.25);
                              \filldraw[fill=cancan, draw=black] (0.50, 0.00) rectangle (0.75, 0.25);
                              \filldraw[fill=bermuda, draw=black] (0.75, 0.00) rectangle (1.00, 0.25);
                              \filldraw[fill=bermuda, draw=black] (1.00, 0.00) rectangle (1.25, 0.25);
                              \filldraw[fill=bermuda, draw=black] (1.25, 0.00) rectangle (1.50, 0.25);
                              \filldraw[fill=cancan, draw=black] (1.50, 0.00) rectangle (1.75, 0.25);
                              \filldraw[fill=cancan, draw=black] (1.75, 0.00) rectangle (2.00, 0.25);
                              \filldraw[fill=cancan, draw=black] (2.00, 0.00) rectangle (2.25, 0.25);
                              \filldraw[fill=bermuda, draw=black] (2.25, 0.00) rectangle (2.50, 0.25);
                              \filldraw[fill=bermuda, draw=black] (2.50, 0.00) rectangle (2.75, 0.25);
                              \filldraw[fill=bermuda, draw=black] (2.75, 0.00) rectangle (3.00, 0.25);
                              \filldraw[fill=cancan, draw=black] (3.00, 0.00) rectangle (3.25, 0.25);
                              \filldraw[fill=cancan, draw=black] (3.25, 0.00) rectangle (3.50, 0.25);
                              \filldraw[fill=cancan, draw=black] (3.50, 0.00) rectangle (3.75, 0.25);
                              \filldraw[fill=bermuda, draw=black] (3.75, 0.00) rectangle (4.00, 0.25);
                              \filldraw[fill=bermuda, draw=black] (4.00, 0.00) rectangle (4.25, 0.25);
                              \filldraw[fill=bermuda, draw=black] (4.25, 0.00) rectangle (4.50, 0.25);}
\hspace{21.0pt} x_{14} = 0\end{equation*}
\begin{equation*}\end{equation*}  

Strings $b_{1, \dotsb, 14}$ are significantly longer and will be presented separately:

\begin{equation*}
\hspace{4.6pt} b_{1} = \vcenter{\hbox{ \tikz{
\filldraw[fill=cancan, draw=black] (0.00, 0.00) rectangle (0.25, -0.25);
\filldraw[fill=cancan, draw=black] (0.25, 0.00) rectangle (0.50, -0.25);
\filldraw[fill=cancan, draw=black] (0.50, 0.00) rectangle (0.75, -0.25);
\filldraw[fill=bermuda, draw=black] (0.75, 0.00) rectangle (1.00, -0.25);
\filldraw[fill=bermuda, draw=black] (1.00, 0.00) rectangle (1.25, -0.25);
\filldraw[fill=cancan, draw=black] (1.25, 0.00) rectangle (1.50, -0.25);
\filldraw[fill=cancan, draw=black] (1.50, 0.00) rectangle (1.75, -0.25);
\filldraw[fill=cancan, draw=black] (1.75, 0.00) rectangle (2.00, -0.25);
\filldraw[fill=bermuda, draw=black] (2.00, 0.00) rectangle (2.25, -0.25);
\filldraw[fill=cancan, draw=black] (2.25, 0.00) rectangle (2.50, -0.25);
\filldraw[fill=bermuda, draw=black] (2.50, 0.00) rectangle (2.75, -0.25);
\filldraw[fill=cancan, draw=black] (2.75, 0.00) rectangle (3.00, -0.25);
\filldraw[fill=cancan, draw=black] (3.00, 0.00) rectangle (3.25, -0.25);
\filldraw[fill=bermuda, draw=black] (3.25, 0.00) rectangle (3.50, -0.25);
\filldraw[fill=bermuda, draw=black] (3.50, 0.00) rectangle (3.75, -0.25);
\filldraw[fill=bermuda, draw=black] (3.75, 0.00) rectangle (4.00, -0.25);
\filldraw[fill=bermuda, draw=black] (4.00, 0.00) rectangle (4.25, -0.25);
\filldraw[fill=bermuda, draw=black] (4.25, 0.00) rectangle (4.50, -0.25);
\filldraw[fill=bermuda, draw=black] (4.50, 0.00) rectangle (4.75, -0.25);
\filldraw[fill=cancan, draw=black] (4.75, 0.00) rectangle (5.00, -0.25);
\filldraw[fill=cancan, draw=black] (5.00, 0.00) rectangle (5.25, -0.25);
\filldraw[fill=cancan, draw=black] (5.25, 0.00) rectangle (5.50, -0.25);
\filldraw[fill=bermuda, draw=black] (5.50, 0.00) rectangle (5.75, -0.25);
\filldraw[fill=bermuda, draw=black] (5.75, 0.00) rectangle (6.00, -0.25);
\filldraw[fill=bermuda, draw=black] (6.00, 0.00) rectangle (6.25, -0.25);
\filldraw[fill=cancan, draw=black] (6.25, 0.00) rectangle (6.50, -0.25);
\filldraw[fill=cancan, draw=black] (6.50, 0.00) rectangle (6.75, -0.25);
\filldraw[fill=cancan, draw=black] (6.75, 0.00) rectangle (7.00, -0.25);
\filldraw[fill=bermuda, draw=black] (7.00, 0.00) rectangle (7.25, -0.25);
\filldraw[fill=bermuda, draw=black] (7.25, 0.00) rectangle (7.50, -0.25);
\filldraw[fill=bermuda, draw=black] (7.50, 0.00) rectangle (7.75, -0.25);
\filldraw[fill=cancan, draw=black] (7.75, 0.00) rectangle (8.00, -0.25);
\filldraw[fill=cancan, draw=black] (8.00, 0.00) rectangle (8.25, -0.25);
\filldraw[fill=cancan, draw=black] (8.25, 0.00) rectangle (8.50, -0.25);
\filldraw[fill=bermuda, draw=black] (8.50, 0.00) rectangle (8.75, -0.25);
\filldraw[fill=bermuda, draw=black] (8.75, 0.00) rectangle (9.00, -0.25);
\filldraw[fill=bermuda, draw=black] (9.00, 0.00) rectangle (9.25, -0.25);
\filldraw[fill=cancan, draw=black] (9.25, 0.00) rectangle (9.50, -0.25);
\filldraw[fill=cancan, draw=black] (9.50, 0.00) rectangle (9.75, -0.25);
\filldraw[fill=cancan, draw=black] (9.75, 0.00) rectangle (10.00, -0.25);
\filldraw[fill=bermuda, draw=black] (10.00, 0.00) rectangle (10.25, -0.25);
\filldraw[fill=bermuda, draw=black] (10.25, 0.00) rectangle (10.50, -0.25);
\filldraw[fill=bermuda, draw=black] (10.50, 0.00) rectangle (10.75, -0.25);
\filldraw[fill=cancan, draw=black] (10.75, 0.00) rectangle (11.00, -0.25);
\filldraw[fill=bermuda, draw=black] (11.00, 0.00) rectangle (11.25, -0.25);
\filldraw[fill=bermuda, draw=black] (11.25, 0.00) rectangle (11.50, -0.25);
\filldraw[fill=bermuda, draw=black] (11.50, 0.00) rectangle (11.75, -0.25);
\filldraw[fill=bermuda, draw=black] (11.75, 0.00) rectangle (12.00, -0.25);
\filldraw[fill=bermuda, draw=black] (12.00, 0.00) rectangle (12.25, -0.25);
\filldraw[fill=cancan, draw=black] (12.25, 0.00) rectangle (12.50, -0.25);
\filldraw[fill=bermuda, draw=black] (12.50, 0.00) rectangle (12.75, -0.25);
\filldraw[fill=bermuda, draw=black] (12.75, 0.00) rectangle (13.00, -0.25);
\filldraw[fill=bermuda, draw=black] (13.00, 0.00) rectangle (13.25, -0.25);
\filldraw[fill=bermuda, draw=black] (13.25, 0.00) rectangle (13.50, -0.25);
\filldraw[fill=bermuda, draw=black] (13.50, 0.00) rectangle (13.75, -0.25);
\filldraw[fill=cancan, draw=black] (13.75, 0.00) rectangle (14.00, -0.25);
\filldraw[fill=cancan, draw=black] (14.00, 0.00) rectangle (14.25, -0.25);
\filldraw[fill=cancan, draw=black] (14.25, 0.00) rectangle (14.50, -0.25);
\filldraw[fill=cancan, draw=black] (14.50, 0.00) rectangle (14.75, -0.25);
\filldraw[fill=cancan, draw=black] (14.75, 0.00) rectangle (15.00, -0.25);
\filldraw[fill=bermuda, draw=black] (0.00, -0.25) rectangle (0.25, -0.50);
\filldraw[fill=bermuda, draw=black] (0.25, -0.25) rectangle (0.50, -0.50);
\filldraw[fill=bermuda, draw=black] (0.50, -0.25) rectangle (0.75, -0.50);
\filldraw[fill=cancan, draw=black] (0.75, -0.25) rectangle (1.00, -0.50);
\filldraw[fill=cancan, draw=black] (1.00, -0.25) rectangle (1.25, -0.50);
\filldraw[fill=bermuda, draw=black] (1.25, -0.25) rectangle (1.50, -0.50);
\filldraw[fill=bermuda, draw=black] (1.50, -0.25) rectangle (1.75, -0.50);
\filldraw[fill=cancan, draw=black] (1.75, -0.25) rectangle (2.00, -0.50);
\filldraw[fill=cancan, draw=black] (2.00, -0.25) rectangle (2.25, -0.50);
\filldraw[fill=cancan, draw=black] (2.25, -0.25) rectangle (2.50, -0.50);
\filldraw[fill=bermuda, draw=black] (2.50, -0.25) rectangle (2.75, -0.50);
\filldraw[fill=bermuda, draw=black] (2.75, -0.25) rectangle (3.00, -0.50);
\filldraw[fill=bermuda, draw=black] (3.00, -0.25) rectangle (3.25, -0.50);
\filldraw[fill=cancan, draw=black] (3.25, -0.25) rectangle (3.50, -0.50);
\filldraw[fill=cancan, draw=black] (3.50, -0.25) rectangle (3.75, -0.50);
\filldraw[fill=cancan, draw=black] (3.75, -0.25) rectangle (4.00, -0.50);
\filldraw[fill=bermuda, draw=black] (4.00, -0.25) rectangle (4.25, -0.50);
\filldraw[fill=bermuda, draw=black] (4.25, -0.25) rectangle (4.50, -0.50);
\filldraw[fill=bermuda, draw=black] (4.50, -0.25) rectangle (4.75, -0.50);
\filldraw[fill=cancan, draw=black] (4.75, -0.25) rectangle (5.00, -0.50);
\filldraw[fill=cancan, draw=black] (5.00, -0.25) rectangle (5.25, -0.50);
\filldraw[fill=cancan, draw=black] (5.25, -0.25) rectangle (5.50, -0.50);
\filldraw[fill=bermuda, draw=black] (5.50, -0.25) rectangle (5.75, -0.50);
\filldraw[fill=bermuda, draw=black] (5.75, -0.25) rectangle (6.00, -0.50);
\filldraw[fill=bermuda, draw=black] (6.00, -0.25) rectangle (6.25, -0.50);
\filldraw[fill=cancan, draw=black] (6.25, -0.25) rectangle (6.50, -0.50);
\filldraw[fill=cancan, draw=black] (6.50, -0.25) rectangle (6.75, -0.50);
\filldraw[fill=cancan, draw=black] (6.75, -0.25) rectangle (7.00, -0.50);
\filldraw[fill=cancan, draw=black] (7.00, -0.25) rectangle (7.25, -0.50);
\filldraw[fill=bermuda, draw=black] (7.25, -0.25) rectangle (7.50, -0.50);
\filldraw[fill=bermuda, draw=black] (7.50, -0.25) rectangle (7.75, -0.50);
\filldraw[fill=bermuda, draw=black] (7.75, -0.25) rectangle (8.00, -0.50);
\filldraw[fill=bermuda, draw=black] (8.00, -0.25) rectangle (8.25, -0.50);
\filldraw[fill=cancan, draw=black] (8.25, -0.25) rectangle (8.50, -0.50);
\filldraw[fill=bermuda, draw=black] (8.50, -0.25) rectangle (8.75, -0.50);
\filldraw[fill=bermuda, draw=black] (8.75, -0.25) rectangle (9.00, -0.50);
\filldraw[fill=bermuda, draw=black] (9.00, -0.25) rectangle (9.25, -0.50);
\filldraw[fill=cancan, draw=black] (9.25, -0.25) rectangle (9.50, -0.50);
\filldraw[fill=cancan, draw=black] (9.50, -0.25) rectangle (9.75, -0.50);
\filldraw[fill=bermuda, draw=black] (9.75, -0.25) rectangle (10.00, -0.50);
\filldraw[fill=bermuda, draw=black] (10.00, -0.25) rectangle (10.25, -0.50);
\filldraw[fill=cancan, draw=black] (10.25, -0.25) rectangle (10.50, -0.50);
\filldraw[fill=bermuda, draw=black] (10.50, -0.25) rectangle (10.75, -0.50);
\filldraw[fill=bermuda, draw=black] (10.75, -0.25) rectangle (11.00, -0.50);
\filldraw[fill=bermuda, draw=black] (11.00, -0.25) rectangle (11.25, -0.50);
\filldraw[fill=bermuda, draw=black] (11.25, -0.25) rectangle (11.50, -0.50);
\filldraw[fill=bermuda, draw=black] (11.50, -0.25) rectangle (11.75, -0.50);
\filldraw[fill=cancan, draw=black] (11.75, -0.25) rectangle (12.00, -0.50);
\filldraw[fill=bermuda, draw=black] (12.00, -0.25) rectangle (12.25, -0.50);
\filldraw[fill=cancan, draw=black] (12.25, -0.25) rectangle (12.50, -0.50);
\filldraw[fill=bermuda, draw=black] (12.50, -0.25) rectangle (12.75, -0.50);
\filldraw[fill=bermuda, draw=black] (12.75, -0.25) rectangle (13.00, -0.50);
\filldraw[fill=bermuda, draw=black] (13.00, -0.25) rectangle (13.25, -0.50);
\filldraw[fill=cancan, draw=black] (13.25, -0.25) rectangle (13.50, -0.50);
\filldraw[fill=cancan, draw=black] (13.50, -0.25) rectangle (13.75, -0.50);
\filldraw[fill=cancan, draw=black] (13.75, -0.25) rectangle (14.00, -0.50);
\filldraw[fill=bermuda, draw=black] (14.00, -0.25) rectangle (14.25, -0.50);
\filldraw[fill=bermuda, draw=black] (14.25, -0.25) rectangle (14.50, -0.50);
\filldraw[fill=bermuda, draw=black] (14.50, -0.25) rectangle (14.75, -0.50);
\filldraw[fill=cancan, draw=black] (14.75, -0.25) rectangle (15.00, -0.50);
\filldraw[fill=cancan, draw=black] (0.00, -0.50) rectangle (0.25, -0.75);
\filldraw[fill=cancan, draw=black] (0.25, -0.50) rectangle (0.50, -0.75);
\filldraw[fill=bermuda, draw=black] (0.50, -0.50) rectangle (0.75, -0.75);
\filldraw[fill=bermuda, draw=black] (0.75, -0.50) rectangle (1.00, -0.75);
\filldraw[fill=bermuda, draw=black] (1.00, -0.50) rectangle (1.25, -0.75);
\filldraw[fill=cancan, draw=black] (1.25, -0.50) rectangle (1.50, -0.75);
\filldraw[fill=cancan, draw=black] (1.50, -0.50) rectangle (1.75, -0.75);
\filldraw[fill=cancan, draw=black] (1.75, -0.50) rectangle (2.00, -0.75);
\filldraw[fill=bermuda, draw=black] (2.00, -0.50) rectangle (2.25, -0.75);
\filldraw[fill=bermuda, draw=black] (2.25, -0.50) rectangle (2.50, -0.75);
\filldraw[fill=bermuda, draw=black] (2.50, -0.50) rectangle (2.75, -0.75);
\filldraw[fill=bermuda, draw=black] (2.75, -0.50) rectangle (3.00, -0.75);
\filldraw[fill=bermuda, draw=black] (3.00, -0.50) rectangle (3.25, -0.75);
\filldraw[fill=cancan, draw=black] (3.25, -0.50) rectangle (3.50, -0.75);
\filldraw[fill=bermuda, draw=black] (3.50, -0.50) rectangle (3.75, -0.75);
\filldraw[fill=cancan, draw=black] (3.75, -0.50) rectangle (4.00, -0.75);
\filldraw[fill=cancan, draw=black] (4.00, -0.50) rectangle (4.25, -0.75);
\filldraw[fill=bermuda, draw=black] (4.25, -0.50) rectangle (4.50, -0.75);
\filldraw[fill=bermuda, draw=black] (4.50, -0.50) rectangle (4.75, -0.75);
\filldraw[fill=cancan, draw=black] (4.75, -0.50) rectangle (5.00, -0.75);
\filldraw[fill=bermuda, draw=black] (5.00, -0.50) rectangle (5.25, -0.75);
\filldraw[fill=cancan, draw=black] (5.25, -0.50) rectangle (5.50, -0.75);
\filldraw[fill=cancan, draw=black] (5.50, -0.50) rectangle (5.75, -0.75);
\filldraw[fill=bermuda, draw=black] (5.75, -0.50) rectangle (6.00, -0.75);
\filldraw[fill=bermuda, draw=black] (6.00, -0.50) rectangle (6.25, -0.75);
\filldraw[fill=cancan, draw=black] (6.25, -0.50) rectangle (6.50, -0.75);
\filldraw[fill=bermuda, draw=black] (6.50, -0.50) rectangle (6.75, -0.75);
\filldraw[fill=cancan, draw=black] (6.75, -0.50) rectangle (7.00, -0.75);
\filldraw[fill=cancan, draw=black] (7.00, -0.50) rectangle (7.25, -0.75);
\filldraw[fill=bermuda, draw=black] (7.25, -0.50) rectangle (7.50, -0.75);
\filldraw[fill=bermuda, draw=black] (7.50, -0.50) rectangle (7.75, -0.75);
\filldraw[fill=cancan, draw=black] (7.75, -0.50) rectangle (8.00, -0.75);
\filldraw[fill=bermuda, draw=black] (8.00, -0.50) rectangle (8.25, -0.75);
\filldraw[fill=cancan, draw=black] (8.25, -0.50) rectangle (8.50, -0.75);
\filldraw[fill=cancan, draw=black] (8.50, -0.50) rectangle (8.75, -0.75);
\filldraw[fill=bermuda, draw=black] (8.75, -0.50) rectangle (9.00, -0.75);
\filldraw[fill=bermuda, draw=black] (9.00, -0.50) rectangle (9.25, -0.75);
\filldraw[fill=cancan, draw=black] (9.25, -0.50) rectangle (9.50, -0.75);
\filldraw[fill=bermuda, draw=black] (9.50, -0.50) rectangle (9.75, -0.75);
\filldraw[fill=cancan, draw=black] (9.75, -0.50) rectangle (10.00, -0.75);
\filldraw[fill=cancan, draw=black] (10.00, -0.50) rectangle (10.25, -0.75);
\filldraw[fill=cancan, draw=black] (10.25, -0.50) rectangle (10.50, -0.75);
\filldraw[fill=cancan, draw=black] (10.50, -0.50) rectangle (10.75, -0.75);
\filldraw[fill=cancan, draw=black] (10.75, -0.50) rectangle (11.00, -0.75);
\filldraw[fill=bermuda, draw=black] (11.00, -0.50) rectangle (11.25, -0.75);
\filldraw[fill=cancan, draw=black] (11.25, -0.50) rectangle (11.50, -0.75);
\filldraw[fill=bermuda, draw=black] (11.50, -0.50) rectangle (11.75, -0.75);
\filldraw[fill=cancan, draw=black] (11.75, -0.50) rectangle (12.00, -0.75);
\filldraw[fill=cancan, draw=black] (12.00, -0.50) rectangle (12.25, -0.75);
\filldraw[fill=cancan, draw=black] (12.25, -0.50) rectangle (12.50, -0.75);
\filldraw[fill=bermuda, draw=black] (12.50, -0.50) rectangle (12.75, -0.75);
\filldraw[fill=bermuda, draw=black] (12.75, -0.50) rectangle (13.00, -0.75);
\filldraw[fill=bermuda, draw=black] (13.00, -0.50) rectangle (13.25, -0.75);
\filldraw[fill=cancan, draw=black] (13.25, -0.50) rectangle (13.50, -0.75);
\filldraw[fill=cancan, draw=black] (13.50, -0.50) rectangle (13.75, -0.75);
\filldraw[fill=cancan, draw=black] (13.75, -0.50) rectangle (14.00, -0.75);
\filldraw[fill=cancan, draw=black] (14.00, -0.50) rectangle (14.25, -0.75);
\filldraw[fill=cancan, draw=black] (14.25, -0.50) rectangle (14.50, -0.75);
\filldraw[fill=cancan, draw=black] (14.50, -0.50) rectangle (14.75, -0.75);
\filldraw[fill=cancan, draw=black] (14.75, -0.50) rectangle (15.00, -0.75);
\filldraw[fill=bermuda, draw=black] (0.00, -0.75) rectangle (0.25, -1.00);
\filldraw[fill=cancan, draw=black] (0.25, -0.75) rectangle (0.50, -1.00);
\filldraw[fill=cancan, draw=black] (0.50, -0.75) rectangle (0.75, -1.00);
\filldraw[fill=cancan, draw=black] (0.75, -0.75) rectangle (1.00, -1.00);
\filldraw[fill=cancan, draw=black] (1.00, -0.75) rectangle (1.25, -1.00);
\filldraw[fill=cancan, draw=black] (1.25, -0.75) rectangle (1.50, -1.00);
\filldraw[fill=cancan, draw=black] (1.50, -0.75) rectangle (1.75, -1.00);
\filldraw[fill=cancan, draw=black] (1.75, -0.75) rectangle (2.00, -1.00);
\filldraw[fill=bermuda, draw=black] (2.00, -0.75) rectangle (2.25, -1.00);
\filldraw[fill=bermuda, draw=black] (2.25, -0.75) rectangle (2.50, -1.00);
\filldraw[fill=bermuda, draw=black] (2.50, -0.75) rectangle (2.75, -1.00);
\filldraw[fill=cancan, draw=black] (2.75, -0.75) rectangle (3.00, -1.00);
\filldraw[fill=cancan, draw=black] (3.00, -0.75) rectangle (3.25, -1.00);
\filldraw[fill=cancan, draw=black] (3.25, -0.75) rectangle (3.50, -1.00);
\filldraw[fill=bermuda, draw=black] (3.50, -0.75) rectangle (3.75, -1.00);
\filldraw[fill=bermuda, draw=black] (3.75, -0.75) rectangle (4.00, -1.00);
\filldraw[fill=bermuda, draw=black] (4.00, -0.75) rectangle (4.25, -1.00);
\filldraw[fill=cancan, draw=black] (4.25, -0.75) rectangle (4.50, -1.00);
\filldraw[fill=cancan, draw=black] (4.50, -0.75) rectangle (4.75, -1.00);
\filldraw[fill=cancan, draw=black] (4.75, -0.75) rectangle (5.00, -1.00);
\filldraw[fill=bermuda, draw=black] (5.00, -0.75) rectangle (5.25, -1.00);
\filldraw[fill=bermuda, draw=black] (5.25, -0.75) rectangle (5.50, -1.00);
\filldraw[fill=bermuda, draw=black] (5.50, -0.75) rectangle (5.75, -1.00);
\filldraw[fill=bermuda, draw=black] (5.75, -0.75) rectangle (6.00, -1.00);
\filldraw[fill=bermuda, draw=black] (6.00, -0.75) rectangle (6.25, -1.00);
\filldraw[fill=cancan, draw=black] (6.25, -0.75) rectangle (6.50, -1.00);
\filldraw[fill=bermuda, draw=black] (6.50, -0.75) rectangle (6.75, -1.00);
\filldraw[fill=cancan, draw=black] (6.75, -0.75) rectangle (7.00, -1.00);
\filldraw[fill=bermuda, draw=black] (7.00, -0.75) rectangle (7.25, -1.00);
\filldraw[fill=cancan, draw=black] (7.25, -0.75) rectangle (7.50, -1.00);
\filldraw[fill=bermuda, draw=black] (7.50, -0.75) rectangle (7.75, -1.00);
\filldraw[fill=cancan, draw=black] (7.75, -0.75) rectangle (8.00, -1.00);
\filldraw[fill=bermuda, draw=black] (8.00, -0.75) rectangle (8.25, -1.00);
\filldraw[fill=bermuda, draw=black] (8.25, -0.75) rectangle (8.50, -1.00);
\filldraw[fill=bermuda, draw=black] (8.50, -0.75) rectangle (8.75, -1.00);
\filldraw[fill=cancan, draw=black] (8.75, -0.75) rectangle (9.00, -1.00);
\filldraw[fill=cancan, draw=black] (9.00, -0.75) rectangle (9.25, -1.00);
\filldraw[fill=cancan, draw=black] (9.25, -0.75) rectangle (9.50, -1.00);
\filldraw[fill=bermuda, draw=black] (9.50, -0.75) rectangle (9.75, -1.00);
\filldraw[fill=bermuda, draw=black] (9.75, -0.75) rectangle (10.00, -1.00);
\filldraw[fill=bermuda, draw=black] (10.00, -0.75) rectangle (10.25, -1.00);
\filldraw[fill=cancan, draw=black] (10.25, -0.75) rectangle (10.50, -1.00);
\filldraw[fill=bermuda, draw=black] (10.50, -0.75) rectangle (10.75, -1.00);
\filldraw[fill=bermuda, draw=black] (10.75, -0.75) rectangle (11.00, -1.00);
\filldraw[fill=bermuda, draw=black] (11.00, -0.75) rectangle (11.25, -1.00);
\filldraw[fill=bermuda, draw=black] (11.25, -0.75) rectangle (11.50, -1.00);
\filldraw[fill=bermuda, draw=black] (11.50, -0.75) rectangle (11.75, -1.00);
\filldraw[fill=cancan, draw=black] (11.75, -0.75) rectangle (12.00, -1.00);
\filldraw[fill=bermuda, draw=black] (12.00, -0.75) rectangle (12.25, -1.00);
\filldraw[fill=bermuda, draw=black] (12.25, -0.75) rectangle (12.50, -1.00);
\filldraw[fill=bermuda, draw=black] (12.50, -0.75) rectangle (12.75, -1.00);
\filldraw[fill=bermuda, draw=black] (12.75, -0.75) rectangle (13.00, -1.00);
\filldraw[fill=bermuda, draw=black] (13.00, -0.75) rectangle (13.25, -1.00);
\filldraw[fill=cancan, draw=black] (13.25, -0.75) rectangle (13.50, -1.00);
\filldraw[fill=bermuda, draw=black] (13.50, -0.75) rectangle (13.75, -1.00);
\filldraw[fill=bermuda, draw=black] (13.75, -0.75) rectangle (14.00, -1.00);
\filldraw[fill=bermuda, draw=black] (14.00, -0.75) rectangle (14.25, -1.00);
\filldraw[fill=bermuda, draw=black] (14.25, -0.75) rectangle (14.50, -1.00);
\filldraw[fill=bermuda, draw=black] (14.50, -0.75) rectangle (14.75, -1.00);
\filldraw[fill=cancan, draw=black] (14.75, -0.75) rectangle (15.00, -1.00);
\filldraw[fill=bermuda, draw=black] (0.00, -1.00) rectangle (0.25, -1.25);
\filldraw[fill=bermuda, draw=black] (0.25, -1.00) rectangle (0.50, -1.25);
\filldraw[fill=bermuda, draw=black] (0.50, -1.00) rectangle (0.75, -1.25);
\filldraw[fill=bermuda, draw=black] (0.75, -1.00) rectangle (1.00, -1.25);
\filldraw[fill=bermuda, draw=black] (1.00, -1.00) rectangle (1.25, -1.25);
\filldraw[fill=cancan, draw=black] (1.25, -1.00) rectangle (1.50, -1.25);
\filldraw[fill=cancan, draw=black] (1.50, -1.00) rectangle (1.75, -1.25);
\filldraw[fill=cancan, draw=black] (1.75, -1.00) rectangle (2.00, -1.25);
\filldraw[fill=cancan, draw=black] (2.00, -1.00) rectangle (2.25, -1.25);
\filldraw[fill=cancan, draw=black] (2.25, -1.00) rectangle (2.50, -1.25);
\filldraw[fill=cancan, draw=black] (2.50, -1.00) rectangle (2.75, -1.25);
\filldraw[fill=cancan, draw=black] (2.75, -1.00) rectangle (3.00, -1.25);
\filldraw[fill=bermuda, draw=black] (3.00, -1.00) rectangle (3.25, -1.25);
\filldraw[fill=bermuda, draw=black] (3.25, -1.00) rectangle (3.50, -1.25);
\filldraw[fill=bermuda, draw=black] (3.50, -1.00) rectangle (3.75, -1.25);
\filldraw[fill=cancan, draw=black] (3.75, -1.00) rectangle (4.00, -1.25);
\filldraw[fill=cancan, draw=black] (4.00, -1.00) rectangle (4.25, -1.25);
\filldraw[fill=cancan, draw=black] (4.25, -1.00) rectangle (4.50, -1.25);
\filldraw[fill=bermuda, draw=black] (4.50, -1.00) rectangle (4.75, -1.25);
\filldraw[fill=bermuda, draw=black] (4.75, -1.00) rectangle (5.00, -1.25);
\filldraw[fill=bermuda, draw=black] (5.00, -1.00) rectangle (5.25, -1.25);
\filldraw[fill=cancan, draw=black] (5.25, -1.00) rectangle (5.50, -1.25);
\filldraw[fill=bermuda, draw=black] (5.50, -1.00) rectangle (5.75, -1.25);
\filldraw[fill=cancan, draw=black] (5.75, -1.00) rectangle (6.00, -1.25);
\filldraw[fill=bermuda, draw=black] (6.00, -1.00) rectangle (6.25, -1.25);
\filldraw[fill=cancan, draw=black] (6.25, -1.00) rectangle (6.50, -1.25);
\filldraw[fill=cancan, draw=black] (6.50, -1.00) rectangle (6.75, -1.25);
\filldraw[fill=cancan, draw=black] (6.75, -1.00) rectangle (7.00, -1.25);
\filldraw[fill=bermuda, draw=black] (7.00, -1.00) rectangle (7.25, -1.25);
\filldraw[fill=bermuda, draw=black] (7.25, -1.00) rectangle (7.50, -1.25);
\filldraw[fill=bermuda, draw=black] (7.50, -1.00) rectangle (7.75, -1.25);
\filldraw[fill=cancan, draw=black] (7.75, -1.00) rectangle (8.00, -1.25);
\filldraw[fill=cancan, draw=black] (8.00, -1.00) rectangle (8.25, -1.25);
\filldraw[fill=cancan, draw=black] (8.25, -1.00) rectangle (8.50, -1.25);
\filldraw[fill=cancan, draw=black] (8.50, -1.00) rectangle (8.75, -1.25);
\filldraw[fill=bermuda, draw=black] (8.75, -1.00) rectangle (9.00, -1.25);
\filldraw[fill=bermuda, draw=black] (9.00, -1.00) rectangle (9.25, -1.25);
\filldraw[fill=cancan, draw=black] (9.25, -1.00) rectangle (9.50, -1.25);
\filldraw[fill=cancan, draw=black] (9.50, -1.00) rectangle (9.75, -1.25);
\filldraw[fill=cancan, draw=black] (9.75, -1.00) rectangle (10.00, -1.25);
\filldraw[fill=cancan, draw=black] (10.00, -1.00) rectangle (10.25, -1.25);
\filldraw[fill=bermuda, draw=black] (10.25, -1.00) rectangle (10.50, -1.25);
\filldraw[fill=bermuda, draw=black] (10.50, -1.00) rectangle (10.75, -1.25);
\filldraw[fill=cancan, draw=black] (10.75, -1.00) rectangle (11.00, -1.25);
\filldraw[fill=cancan, draw=black] (11.00, -1.00) rectangle (11.25, -1.25);
\filldraw[fill=cancan, draw=black] (11.25, -1.00) rectangle (11.50, -1.25);
\filldraw[fill=cancan, draw=black] (11.50, -1.00) rectangle (11.75, -1.25);
\filldraw[fill=bermuda, draw=black] (11.75, -1.00) rectangle (12.00, -1.25);
\filldraw[fill=bermuda, draw=black] (12.00, -1.00) rectangle (12.25, -1.25);
\filldraw[fill=cancan, draw=black] (12.25, -1.00) rectangle (12.50, -1.25);
\filldraw[fill=cancan, draw=black] (12.50, -1.00) rectangle (12.75, -1.25);
\filldraw[fill=cancan, draw=black] (12.75, -1.00) rectangle (13.00, -1.25);
\filldraw[fill=cancan, draw=black] (13.00, -1.00) rectangle (13.25, -1.25);
\filldraw[fill=cancan, draw=black] (13.25, -1.00) rectangle (13.50, -1.25);
\filldraw[fill=cancan, draw=black] (13.50, -1.00) rectangle (13.75, -1.25);
\filldraw[fill=cancan, draw=black] (13.75, -1.00) rectangle (14.00, -1.25);
\filldraw[fill=cancan, draw=black] (14.00, -1.00) rectangle (14.25, -1.25);
\filldraw[fill=cancan, draw=black] (14.25, -1.00) rectangle (14.50, -1.25);
\filldraw[fill=cancan, draw=black] (14.50, -1.00) rectangle (14.75, -1.25);
\filldraw[fill=cancan, draw=black] (14.75, -1.00) rectangle (15.00, -1.25);
\filldraw[fill=cancan, draw=black] (0.00, -1.25) rectangle (0.25, -1.50);
\filldraw[fill=bermuda, draw=black] (0.25, -1.25) rectangle (0.50, -1.50);
\filldraw[fill=bermuda, draw=black] (0.50, -1.25) rectangle (0.75, -1.50);
\filldraw[fill=cancan, draw=black] (0.75, -1.25) rectangle (1.00, -1.50);
\filldraw[fill=bermuda, draw=black] (1.00, -1.25) rectangle (1.25, -1.50);
\filldraw[fill=cancan, draw=black] (1.25, -1.25) rectangle (1.50, -1.50);
\filldraw[fill=cancan, draw=black] (1.50, -1.25) rectangle (1.75, -1.50);
\filldraw[fill=bermuda, draw=black] (1.75, -1.25) rectangle (2.00, -1.50);
\filldraw[fill=bermuda, draw=black] (2.00, -1.25) rectangle (2.25, -1.50);
\filldraw[fill=cancan, draw=black] (2.25, -1.25) rectangle (2.50, -1.50);
\filldraw[fill=bermuda, draw=black] (2.50, -1.25) rectangle (2.75, -1.50);
\filldraw[fill=cancan, draw=black] (2.75, -1.25) rectangle (3.00, -1.50);
\filldraw[fill=bermuda, draw=black] (3.00, -1.25) rectangle (3.25, -1.50);
\filldraw[fill=bermuda, draw=black] (3.25, -1.25) rectangle (3.50, -1.50);
\filldraw[fill=bermuda, draw=black] (3.50, -1.25) rectangle (3.75, -1.50);
\filldraw[fill=cancan, draw=black] (3.75, -1.25) rectangle (4.00, -1.50);
\filldraw[fill=bermuda, draw=black] (4.00, -1.25) rectangle (4.25, -1.50);
\filldraw[fill=bermuda, draw=black] (4.25, -1.25) rectangle (4.50, -1.50);
\filldraw[fill=bermuda, draw=black] (4.50, -1.25) rectangle (4.75, -1.50);
\filldraw[fill=bermuda, draw=black] (4.75, -1.25) rectangle (5.00, -1.50);
\filldraw[fill=bermuda, draw=black] (5.00, -1.25) rectangle (5.25, -1.50);
\filldraw[fill=cancan, draw=black] (5.25, -1.25) rectangle (5.50, -1.50);
\filldraw[fill=bermuda, draw=black] (5.50, -1.25) rectangle (5.75, -1.50);
\filldraw[fill=cancan, draw=black] (5.75, -1.25) rectangle (6.00, -1.50);
\filldraw[fill=cancan, draw=black] (6.00, -1.25) rectangle (6.25, -1.50);
\filldraw[fill=bermuda, draw=black] (6.25, -1.25) rectangle (6.50, -1.50);
\filldraw[fill=bermuda, draw=black] (6.50, -1.25) rectangle (6.75, -1.50);
\filldraw[fill=cancan, draw=black] (6.75, -1.25) rectangle (7.00, -1.50);
\filldraw[fill=bermuda, draw=black] (7.00, -1.25) rectangle (7.25, -1.50);
\filldraw[fill=cancan, draw=black] (7.25, -1.25) rectangle (7.50, -1.50);
\filldraw[fill=bermuda, draw=black] (7.50, -1.25) rectangle (7.75, -1.50);
\filldraw[fill=cancan, draw=black] (7.75, -1.25) rectangle (8.00, -1.50);
\filldraw[fill=bermuda, draw=black] (8.00, -1.25) rectangle (8.25, -1.50);
\filldraw[fill=bermuda, draw=black] (8.25, -1.25) rectangle (8.50, -1.50);
\filldraw[fill=bermuda, draw=black] (8.50, -1.25) rectangle (8.75, -1.50);
\filldraw[fill=cancan, draw=black] (8.75, -1.25) rectangle (9.00, -1.50);
\filldraw[fill=cancan, draw=black] (9.00, -1.25) rectangle (9.25, -1.50);
\filldraw[fill=cancan, draw=black] (9.25, -1.25) rectangle (9.50, -1.50);
\filldraw[fill=bermuda, draw=black] (9.50, -1.25) rectangle (9.75, -1.50);
\filldraw[fill=bermuda, draw=black] (9.75, -1.25) rectangle (10.00, -1.50);
\filldraw[fill=bermuda, draw=black] (10.00, -1.25) rectangle (10.25, -1.50);
\filldraw[fill=cancan, draw=black] (10.25, -1.25) rectangle (10.50, -1.50);
\filldraw[fill=cancan, draw=black] (10.50, -1.25) rectangle (10.75, -1.50);
\filldraw[fill=cancan, draw=black] (10.75, -1.25) rectangle (11.00, -1.50);
\filldraw[fill=cancan, draw=black] (11.00, -1.25) rectangle (11.25, -1.50);
\filldraw[fill=bermuda, draw=black] (11.25, -1.25) rectangle (11.50, -1.50);
\filldraw[fill=bermuda, draw=black] (11.50, -1.25) rectangle (11.75, -1.50);
\filldraw[fill=cancan, draw=black] (11.75, -1.25) rectangle (12.00, -1.50);
\filldraw[fill=bermuda, draw=black] (12.00, -1.25) rectangle (12.25, -1.50);
\filldraw[fill=cancan, draw=black] (12.25, -1.25) rectangle (12.50, -1.50);
\filldraw[fill=cancan, draw=black] (12.50, -1.25) rectangle (12.75, -1.50);
\filldraw[fill=bermuda, draw=black] (12.75, -1.25) rectangle (13.00, -1.50);
\filldraw[fill=bermuda, draw=black] (13.00, -1.25) rectangle (13.25, -1.50);
\filldraw[fill=cancan, draw=black] (13.25, -1.25) rectangle (13.50, -1.50);
\filldraw[fill=bermuda, draw=black] (13.50, -1.25) rectangle (13.75, -1.50);
\filldraw[fill=cancan, draw=black] (13.75, -1.25) rectangle (14.00, -1.50);
\filldraw[fill=bermuda, draw=black] (14.00, -1.25) rectangle (14.25, -1.50);
\filldraw[fill=bermuda, draw=black] (14.25, -1.25) rectangle (14.50, -1.50);
\filldraw[fill=bermuda, draw=black] (14.50, -1.25) rectangle (14.75, -1.50);
\filldraw[fill=cancan, draw=black] (14.75, -1.25) rectangle (15.00, -1.50);
\filldraw[fill=bermuda, draw=black] (0.00, -1.50) rectangle (0.25, -1.75);
\filldraw[fill=bermuda, draw=black] (0.25, -1.50) rectangle (0.50, -1.75);
\filldraw[fill=bermuda, draw=black] (0.50, -1.50) rectangle (0.75, -1.75);
\filldraw[fill=bermuda, draw=black] (0.75, -1.50) rectangle (1.00, -1.75);
\filldraw[fill=bermuda, draw=black] (1.00, -1.50) rectangle (1.25, -1.75);
\filldraw[fill=cancan, draw=black] (1.25, -1.50) rectangle (1.50, -1.75);
\filldraw[fill=bermuda, draw=black] (1.50, -1.50) rectangle (1.75, -1.75);
\filldraw[fill=cancan, draw=black] (1.75, -1.50) rectangle (2.00, -1.75);
\filldraw[fill=cancan, draw=black] (2.00, -1.50) rectangle (2.25, -1.75);
\filldraw[fill=bermuda, draw=black] (2.25, -1.50) rectangle (2.50, -1.75);
\filldraw[fill=bermuda, draw=black] (2.50, -1.50) rectangle (2.75, -1.75);
\filldraw[fill=cancan, draw=black] (2.75, -1.50) rectangle (3.00, -1.75);
\filldraw[fill=bermuda, draw=black] (3.00, -1.50) rectangle (3.25, -1.75);
\filldraw[fill=cancan, draw=black] (3.25, -1.50) rectangle (3.50, -1.75);
\filldraw[fill=cancan, draw=black] (3.50, -1.50) rectangle (3.75, -1.75);
\filldraw[fill=bermuda, draw=black] (3.75, -1.50) rectangle (4.00, -1.75);
\filldraw[fill=bermuda, draw=black] (4.00, -1.50) rectangle (4.25, -1.75);
\filldraw[fill=cancan, draw=black] (4.25, -1.50) rectangle (4.50, -1.75);
\filldraw[fill=bermuda, draw=black] (4.50, -1.50) rectangle (4.75, -1.75);
\filldraw[fill=cancan, draw=black] (4.75, -1.50) rectangle (5.00, -1.75);
\filldraw[fill=cancan, draw=black] (5.00, -1.50) rectangle (5.25, -1.75);
\filldraw[fill=bermuda, draw=black] (5.25, -1.50) rectangle (5.50, -1.75);
\filldraw[fill=bermuda, draw=black] (5.50, -1.50) rectangle (5.75, -1.75);
\filldraw[fill=cancan, draw=black] (5.75, -1.50) rectangle (6.00, -1.75);
\filldraw[fill=bermuda, draw=black] (6.00, -1.50) rectangle (6.25, -1.75);
\filldraw[fill=cancan, draw=black] (6.25, -1.50) rectangle (6.50, -1.75);
\filldraw[fill=cancan, draw=black] (6.50, -1.50) rectangle (6.75, -1.75);
\filldraw[fill=bermuda, draw=black] (6.75, -1.50) rectangle (7.00, -1.75);
\filldraw[fill=bermuda, draw=black] (7.00, -1.50) rectangle (7.25, -1.75);
\filldraw[fill=cancan, draw=black] (7.25, -1.50) rectangle (7.50, -1.75);
\filldraw[fill=bermuda, draw=black] (7.50, -1.50) rectangle (7.75, -1.75);
\filldraw[fill=bermuda, draw=black] (7.75, -1.50) rectangle (8.00, -1.75);
\filldraw[fill=bermuda, draw=black] (8.00, -1.50) rectangle (8.25, -1.75);
\filldraw[fill=cancan, draw=black] (8.25, -1.50) rectangle (8.50, -1.75);
\filldraw[fill=cancan, draw=black] (8.50, -1.50) rectangle (8.75, -1.75);
\filldraw[fill=cancan, draw=black] (8.75, -1.50) rectangle (9.00, -1.75);
\filldraw[fill=cancan, draw=black] (9.00, -1.50) rectangle (9.25, -1.75);
\filldraw[fill=cancan, draw=black] (9.25, -1.50) rectangle (9.50, -1.75);
\filldraw[fill=bermuda, draw=black] (9.50, -1.50) rectangle (9.75, -1.75);
\filldraw[fill=bermuda, draw=black] (9.75, -1.50) rectangle (10.00, -1.75);
\filldraw[fill=bermuda, draw=black] (10.00, -1.50) rectangle (10.25, -1.75);
\filldraw[fill=cancan, draw=black] (10.25, -1.50) rectangle (10.50, -1.75);
\filldraw[fill=cancan, draw=black] (10.50, -1.50) rectangle (10.75, -1.75);
\filldraw[fill=cancan, draw=black] (10.75, -1.50) rectangle (11.00, -1.75);
\filldraw[fill=bermuda, draw=black] (11.00, -1.50) rectangle (11.25, -1.75);
\filldraw[fill=bermuda, draw=black] (11.25, -1.50) rectangle (11.50, -1.75);
\filldraw[fill=bermuda, draw=black] (11.50, -1.50) rectangle (11.75, -1.75);
\filldraw[fill=cancan, draw=black] (11.75, -1.50) rectangle (12.00, -1.75);
\filldraw[fill=cancan, draw=black] (12.00, -1.50) rectangle (12.25, -1.75);
\filldraw[fill=cancan, draw=black] (12.25, -1.50) rectangle (12.50, -1.75);
\filldraw[fill=bermuda, draw=black] (12.50, -1.50) rectangle (12.75, -1.75);
\filldraw[fill=bermuda, draw=black] (12.75, -1.50) rectangle (13.00, -1.75);
\filldraw[fill=bermuda, draw=black] (13.00, -1.50) rectangle (13.25, -1.75);
\filldraw[fill=cancan, draw=black] (13.25, -1.50) rectangle (13.50, -1.75);
\filldraw[fill=bermuda, draw=black] (13.50, -1.50) rectangle (13.75, -1.75);
\filldraw[fill=bermuda, draw=black] (13.75, -1.50) rectangle (14.00, -1.75);
\filldraw[fill=bermuda, draw=black] (14.00, -1.50) rectangle (14.25, -1.75);
\filldraw[fill=cancan, draw=black] (14.25, -1.50) rectangle (14.50, -1.75);
\filldraw[fill=cancan, draw=black] (14.50, -1.50) rectangle (14.75, -1.75);
\filldraw[fill=bermuda, draw=black] (14.75, -1.50) rectangle (15.00, -1.75);
\filldraw[fill=bermuda, draw=black] (0.00, -1.75) rectangle (0.25, -2.00);
\filldraw[fill=cancan, draw=black] (0.25, -1.75) rectangle (0.50, -2.00);
\filldraw[fill=cancan, draw=black] (0.50, -1.75) rectangle (0.75, -2.00);
\filldraw[fill=cancan, draw=black] (0.75, -1.75) rectangle (1.00, -2.00);
\filldraw[fill=cancan, draw=black] (1.00, -1.75) rectangle (1.25, -2.00);
\filldraw[fill=bermuda, draw=black] (1.25, -1.75) rectangle (1.50, -2.00);
\filldraw[fill=bermuda, draw=black] (1.50, -1.75) rectangle (1.75, -2.00);
\filldraw[fill=cancan, draw=black] (1.75, -1.75) rectangle (2.00, -2.00);
\filldraw[fill=bermuda, draw=black] (2.00, -1.75) rectangle (2.25, -2.00);
\filldraw[fill=bermuda, draw=black] (2.25, -1.75) rectangle (2.50, -2.00);
\filldraw[fill=bermuda, draw=black] (2.50, -1.75) rectangle (2.75, -2.00);
\filldraw[fill=cancan, draw=black] (2.75, -1.75) rectangle (3.00, -2.00);
\filldraw[fill=cancan, draw=black] (3.00, -1.75) rectangle (3.25, -2.00);
\filldraw[fill=cancan, draw=black] (3.25, -1.75) rectangle (3.50, -2.00);
\filldraw[fill=bermuda, draw=black] (3.50, -1.75) rectangle (3.75, -2.00);
\filldraw[fill=bermuda, draw=black] (3.75, -1.75) rectangle (4.00, -2.00);
\filldraw[fill=bermuda, draw=black] (4.00, -1.75) rectangle (4.25, -2.00);
\filldraw[fill=bermuda, draw=black] (4.25, -1.75) rectangle (4.50, -2.00);
\filldraw[fill=bermuda, draw=black] (4.50, -1.75) rectangle (4.75, -2.00);
\filldraw[fill=cancan, draw=black] (4.75, -1.75) rectangle (5.00, -2.00);
\filldraw[fill=bermuda, draw=black] (5.00, -1.75) rectangle (5.25, -2.00);
\filldraw[fill=bermuda, draw=black] (5.25, -1.75) rectangle (5.50, -2.00);
\filldraw[fill=bermuda, draw=black] (5.50, -1.75) rectangle (5.75, -2.00);
\filldraw[fill=cancan, draw=black] (5.75, -1.75) rectangle (6.00, -2.00);
\filldraw[fill=bermuda, draw=black] (6.00, -1.75) rectangle (6.25, -2.00);
\filldraw[fill=bermuda, draw=black] (6.25, -1.75) rectangle (6.50, -2.00);
\filldraw[fill=bermuda, draw=black] (6.50, -1.75) rectangle (6.75, -2.00);
\filldraw[fill=bermuda, draw=black] (6.75, -1.75) rectangle (7.00, -2.00);
\filldraw[fill=bermuda, draw=black] (7.00, -1.75) rectangle (7.25, -2.00);
\filldraw[fill=cancan, draw=black] (7.25, -1.75) rectangle (7.50, -2.00);
\filldraw[fill=cancan, draw=black] (7.50, -1.75) rectangle (7.75, -2.00);
\filldraw[fill=cancan, draw=black] (7.75, -1.75) rectangle (8.00, -2.00);
\filldraw[fill=bermuda, draw=black] (8.00, -1.75) rectangle (8.25, -2.00);
\filldraw[fill=bermuda, draw=black] (8.25, -1.75) rectangle (8.50, -2.00);
\filldraw[fill=bermuda, draw=black] (8.50, -1.75) rectangle (8.75, -2.00);
\filldraw[fill=cancan, draw=black] (8.75, -1.75) rectangle (9.00, -2.00);
\filldraw[fill=cancan, draw=black] (9.00, -1.75) rectangle (9.25, -2.00);
\filldraw[fill=cancan, draw=black] (9.25, -1.75) rectangle (9.50, -2.00);
\filldraw[fill=bermuda, draw=black] (9.50, -1.75) rectangle (9.75, -2.00);
\filldraw[fill=bermuda, draw=black] (9.75, -1.75) rectangle (10.00, -2.00);
\filldraw[fill=bermuda, draw=black] (10.00, -1.75) rectangle (10.25, -2.00);
\filldraw[fill=cancan, draw=black] (10.25, -1.75) rectangle (10.50, -2.00);
\filldraw[fill=cancan, draw=black] (10.50, -1.75) rectangle (10.75, -2.00);
\filldraw[fill=cancan, draw=black] (10.75, -1.75) rectangle (11.00, -2.00);
\filldraw[fill=bermuda, draw=black] (11.00, -1.75) rectangle (11.25, -2.00);
\filldraw[fill=bermuda, draw=black] (11.25, -1.75) rectangle (11.50, -2.00);
\filldraw[fill=bermuda, draw=black] (11.50, -1.75) rectangle (11.75, -2.00);
\filldraw[fill=bermuda, draw=black] (11.75, -1.75) rectangle (12.00, -2.00);
\filldraw[fill=bermuda, draw=black] (12.00, -1.75) rectangle (12.25, -2.00);
\filldraw[fill=bermuda, draw=black] (12.25, -1.75) rectangle (12.50, -2.00);
\filldraw[fill=bermuda, draw=black] (12.50, -1.75) rectangle (12.75, -2.00);
\filldraw[fill=bermuda, draw=black] (12.75, -1.75) rectangle (13.00, -2.00);
\filldraw[fill=bermuda, draw=black] (13.00, -1.75) rectangle (13.25, -2.00);
\filldraw[fill=cancan, draw=black] (13.25, -1.75) rectangle (13.50, -2.00);
\filldraw[fill=bermuda, draw=black] (13.50, -1.75) rectangle (13.75, -2.00);
\filldraw[fill=cancan, draw=black] (13.75, -1.75) rectangle (14.00, -2.00);
\filldraw[fill=cancan, draw=black] (14.00, -1.75) rectangle (14.25, -2.00);
\filldraw[fill=bermuda, draw=black] (14.25, -1.75) rectangle (14.50, -2.00);
\filldraw[fill=bermuda, draw=black] (14.50, -1.75) rectangle (14.75, -2.00);
\filldraw[fill=cancan, draw=black] (14.75, -1.75) rectangle (15.00, -2.00);
\filldraw[fill=bermuda, draw=black] (0.00, -2.00) rectangle (0.25, -2.25);
\filldraw[fill=cancan, draw=black] (0.25, -2.00) rectangle (0.50, -2.25);
\filldraw[fill=cancan, draw=black] (0.50, -2.00) rectangle (0.75, -2.25);
\filldraw[fill=cancan, draw=black] (0.75, -2.00) rectangle (1.00, -2.25);
\filldraw[fill=bermuda, draw=black] (1.00, -2.00) rectangle (1.25, -2.25);
\filldraw[fill=cancan, draw=black] (1.25, -2.00) rectangle (1.50, -2.25);
\filldraw[fill=cancan, draw=black] (1.50, -2.00) rectangle (1.75, -2.25);
\filldraw[fill=cancan, draw=black] (1.75, -2.00) rectangle (2.00, -2.25);
\filldraw[fill=cancan, draw=black] (2.00, -2.00) rectangle (2.25, -2.25);
\filldraw[fill=cancan, draw=black] (2.25, -2.00) rectangle (2.50, -2.25);
\filldraw[fill=bermuda, draw=black] (2.50, -2.00) rectangle (2.75, -2.25);
\filldraw[fill=bermuda, draw=black] (2.75, -2.00) rectangle (3.00, -2.25);
\filldraw[fill=bermuda, draw=black] (3.00, -2.00) rectangle (3.25, -2.25);
\filldraw[fill=cancan, draw=black] (3.25, -2.00) rectangle (3.50, -2.25);
\filldraw[fill=cancan, draw=black] (3.50, -2.00) rectangle (3.75, -2.25);
\filldraw[fill=cancan, draw=black] (3.75, -2.00) rectangle (4.00, -2.25);
\filldraw[fill=bermuda, draw=black] (4.00, -2.00) rectangle (4.25, -2.25);
\filldraw[fill=bermuda, draw=black] (4.25, -2.00) rectangle (4.50, -2.25);
\filldraw[fill=bermuda, draw=black] (4.50, -2.00) rectangle (4.75, -2.25);
\filldraw[fill=cancan, draw=black] (4.75, -2.00) rectangle (5.00, -2.25);
\filldraw[fill=cancan, draw=black] (5.00, -2.00) rectangle (5.25, -2.25);
\filldraw[fill=cancan, draw=black] (5.25, -2.00) rectangle (5.50, -2.25);
\filldraw[fill=bermuda, draw=black] (5.50, -2.00) rectangle (5.75, -2.25);
\filldraw[fill=bermuda, draw=black] (5.75, -2.00) rectangle (6.00, -2.25);
\filldraw[fill=bermuda, draw=black] (6.00, -2.00) rectangle (6.25, -2.25);
\filldraw[fill=cancan, draw=black] (6.25, -2.00) rectangle (6.50, -2.25);
\filldraw[fill=bermuda, draw=black] (6.50, -2.00) rectangle (6.75, -2.25);
\filldraw[fill=bermuda, draw=black] (6.75, -2.00) rectangle (7.00, -2.25);
\filldraw[fill=bermuda, draw=black] (7.00, -2.00) rectangle (7.25, -2.25);
\filldraw[fill=cancan, draw=black] (7.25, -2.00) rectangle (7.50, -2.25);
\filldraw[fill=bermuda, draw=black] (7.50, -2.00) rectangle (7.75, -2.25);
\filldraw[fill=cancan, draw=black] (7.75, -2.00) rectangle (8.00, -2.25);
\filldraw[fill=bermuda, draw=black] (8.00, -2.00) rectangle (8.25, -2.25);
\filldraw[fill=cancan, draw=black] (8.25, -2.00) rectangle (8.50, -2.25);
\filldraw[fill=cancan, draw=black] (8.50, -2.00) rectangle (8.75, -2.25);
\filldraw[fill=cancan, draw=black] (8.75, -2.00) rectangle (9.00, -2.25);
\filldraw[fill=cancan, draw=black] (9.00, -2.00) rectangle (9.25, -2.25);
\filldraw[fill=cancan, draw=black] (9.25, -2.00) rectangle (9.50, -2.25);
\filldraw[fill=bermuda, draw=black] (9.50, -2.00) rectangle (9.75, -2.25);
\filldraw[fill=bermuda, draw=black] (9.75, -2.00) rectangle (10.00, -2.25);
\filldraw[fill=bermuda, draw=black] (10.00, -2.00) rectangle (10.25, -2.25);
\filldraw[fill=bermuda, draw=black] (10.25, -2.00) rectangle (10.50, -2.25);
\filldraw[fill=bermuda, draw=black] (10.50, -2.00) rectangle (10.75, -2.25);
\filldraw[fill=bermuda, draw=black] (10.75, -2.00) rectangle (11.00, -2.25);
\filldraw[fill=bermuda, draw=black] (11.00, -2.00) rectangle (11.25, -2.25);
\filldraw[fill=cancan, draw=black] (11.25, -2.00) rectangle (11.50, -2.25);
\filldraw[fill=bermuda, draw=black] (11.50, -2.00) rectangle (11.75, -2.25);
\filldraw[fill=cancan, draw=black] (11.75, -2.00) rectangle (12.00, -2.25);
\filldraw[fill=bermuda, draw=black] (12.00, -2.00) rectangle (12.25, -2.25);
\filldraw[fill=bermuda, draw=black] (12.25, -2.00) rectangle (12.50, -2.25);
\filldraw[fill=bermuda, draw=black] (12.50, -2.00) rectangle (12.75, -2.25);
\filldraw[fill=cancan, draw=black] (12.75, -2.00) rectangle (13.00, -2.25);
\filldraw[fill=cancan, draw=black] (13.00, -2.00) rectangle (13.25, -2.25);
\filldraw[fill=cancan, draw=black] (13.25, -2.00) rectangle (13.50, -2.25);
\filldraw[fill=bermuda, draw=black] (13.50, -2.00) rectangle (13.75, -2.25);
\filldraw[fill=bermuda, draw=black] (13.75, -2.00) rectangle (14.00, -2.25);
\filldraw[fill=bermuda, draw=black] (14.00, -2.00) rectangle (14.25, -2.25);
\filldraw[fill=cancan, draw=black] (14.25, -2.00) rectangle (14.50, -2.25);
\filldraw[fill=bermuda, draw=black] (14.50, -2.00) rectangle (14.75, -2.25);
\filldraw[fill=bermuda, draw=black] (14.75, -2.00) rectangle (15.00, -2.25);
\filldraw[fill=bermuda, draw=black] (0.00, -2.25) rectangle (0.25, -2.50);
\filldraw[fill=cancan, draw=black] (0.25, -2.25) rectangle (0.50, -2.50);
\filldraw[fill=bermuda, draw=black] (0.50, -2.25) rectangle (0.75, -2.50);
\filldraw[fill=cancan, draw=black] (0.75, -2.25) rectangle (1.00, -2.50);
\filldraw[fill=bermuda, draw=black] (1.00, -2.25) rectangle (1.25, -2.50);
\filldraw[fill=cancan, draw=black] (1.25, -2.25) rectangle (1.50, -2.50);
\filldraw[fill=bermuda, draw=black] (1.50, -2.25) rectangle (1.75, -2.50);
\filldraw[fill=bermuda, draw=black] (1.75, -2.25) rectangle (2.00, -2.50);
\filldraw[fill=bermuda, draw=black] (2.00, -2.25) rectangle (2.25, -2.50);
\filldraw[fill=cancan, draw=black] (2.25, -2.25) rectangle (2.50, -2.50);
\filldraw[fill=cancan, draw=black] (2.50, -2.25) rectangle (2.75, -2.50);
\filldraw[fill=cancan, draw=black] (2.75, -2.25) rectangle (3.00, -2.50);
\filldraw[fill=bermuda, draw=black] (3.00, -2.25) rectangle (3.25, -2.50);
\filldraw[fill=bermuda, draw=black] (3.25, -2.25) rectangle (3.50, -2.50);
\filldraw[fill=bermuda, draw=black] (3.50, -2.25) rectangle (3.75, -2.50);
\filldraw[fill=cancan, draw=black] (3.75, -2.25) rectangle (4.00, -2.50);
\filldraw[fill=bermuda, draw=black] (4.00, -2.25) rectangle (4.25, -2.50);
\filldraw[fill=cancan, draw=black] (4.25, -2.25) rectangle (4.50, -2.50);
\filldraw[fill=bermuda, draw=black] (4.50, -2.25) rectangle (4.75, -2.50);
\filldraw[fill=cancan, draw=black] (4.75, -2.25) rectangle (5.00, -2.50);
\filldraw[fill=cancan, draw=black] (5.00, -2.25) rectangle (5.25, -2.50);
\filldraw[fill=cancan, draw=black] (5.25, -2.25) rectangle (5.50, -2.50);
\filldraw[fill=bermuda, draw=black] (5.50, -2.25) rectangle (5.75, -2.50);
\filldraw[fill=cancan, draw=black] (5.75, -2.25) rectangle (6.00, -2.50);
\filldraw[fill=bermuda, draw=black] (6.00, -2.25) rectangle (6.25, -2.50);
\filldraw[fill=cancan, draw=black] (6.25, -2.25) rectangle (6.50, -2.50);
\filldraw[fill=bermuda, draw=black] (6.50, -2.25) rectangle (6.75, -2.50);
\filldraw[fill=cancan, draw=black] (6.75, -2.25) rectangle (7.00, -2.50);
\filldraw[fill=bermuda, draw=black] (7.00, -2.25) rectangle (7.25, -2.50);
\filldraw[fill=cancan, draw=black] (7.25, -2.25) rectangle (7.50, -2.50);
\filldraw[fill=cancan, draw=black] (7.50, -2.25) rectangle (7.75, -2.50);
\filldraw[fill=cancan, draw=black] (7.75, -2.25) rectangle (8.00, -2.50);
\filldraw[fill=bermuda, draw=black] (8.00, -2.25) rectangle (8.25, -2.50);
\filldraw[fill=bermuda, draw=black] (8.25, -2.25) rectangle (8.50, -2.50);
\filldraw[fill=bermuda, draw=black] (8.50, -2.25) rectangle (8.75, -2.50);
\filldraw[fill=bermuda, draw=black] (8.75, -2.25) rectangle (9.00, -2.50);
\filldraw[fill=bermuda, draw=black] (9.00, -2.25) rectangle (9.25, -2.50);
\filldraw[fill=cancan, draw=black] (9.25, -2.25) rectangle (9.50, -2.50);
\filldraw[fill=bermuda, draw=black] (9.50, -2.25) rectangle (9.75, -2.50);
\filldraw[fill=bermuda, draw=black] (9.75, -2.25) rectangle (10.00, -2.50);
\filldraw[fill=bermuda, draw=black] (10.00, -2.25) rectangle (10.25, -2.50);
\filldraw[fill=cancan, draw=black] (10.25, -2.25) rectangle (10.50, -2.50);
\filldraw[fill=cancan, draw=black] (10.50, -2.25) rectangle (10.75, -2.50);
\filldraw[fill=cancan, draw=black] (10.75, -2.25) rectangle (11.00, -2.50);
\filldraw[fill=bermuda, draw=black] (11.00, -2.25) rectangle (11.25, -2.50);
\filldraw[fill=bermuda, draw=black] (11.25, -2.25) rectangle (11.50, -2.50);
\filldraw[fill=bermuda, draw=black] (11.50, -2.25) rectangle (11.75, -2.50);
\filldraw[fill=cancan, draw=black] (11.75, -2.25) rectangle (12.00, -2.50);
\filldraw[fill=bermuda, draw=black] (12.00, -2.25) rectangle (12.25, -2.50);
\filldraw[fill=cancan, draw=black] (12.25, -2.25) rectangle (12.50, -2.50);
\filldraw[fill=cancan, draw=black] (12.50, -2.25) rectangle (12.75, -2.50);
\filldraw[fill=bermuda, draw=black] (12.75, -2.25) rectangle (13.00, -2.50);
\filldraw[fill=bermuda, draw=black] (13.00, -2.25) rectangle (13.25, -2.50);
\filldraw[fill=cancan, draw=black] (13.25, -2.25) rectangle (13.50, -2.50);
\filldraw[fill=bermuda, draw=black] (13.50, -2.25) rectangle (13.75, -2.50);
\filldraw[fill=cancan, draw=black] (13.75, -2.25) rectangle (14.00, -2.50);
\filldraw[fill=cancan, draw=black] (14.00, -2.25) rectangle (14.25, -2.50);
\filldraw[fill=cancan, draw=black] (14.25, -2.25) rectangle (14.50, -2.50);
\filldraw[fill=bermuda, draw=black] (14.50, -2.25) rectangle (14.75, -2.50);
\filldraw[fill=bermuda, draw=black] (14.75, -2.25) rectangle (15.00, -2.50);
\filldraw[fill=bermuda, draw=black] (0.00, -2.50) rectangle (0.25, -2.75);
\filldraw[fill=bermuda, draw=black] (0.25, -2.50) rectangle (0.50, -2.75);
\filldraw[fill=bermuda, draw=black] (0.50, -2.50) rectangle (0.75, -2.75);
\filldraw[fill=cancan, draw=black] (0.75, -2.50) rectangle (1.00, -2.75);
\filldraw[fill=bermuda, draw=black] (1.00, -2.50) rectangle (1.25, -2.75);
\filldraw[fill=bermuda, draw=black] (1.25, -2.50) rectangle (1.50, -2.75);
\filldraw[fill=bermuda, draw=black] (1.50, -2.50) rectangle (1.75, -2.75);
\filldraw[fill=bermuda, draw=black] (1.75, -2.50) rectangle (2.00, -2.75);
\filldraw[fill=bermuda, draw=black] (2.00, -2.50) rectangle (2.25, -2.75);
\filldraw[fill=cancan, draw=black] (2.25, -2.50) rectangle (2.50, -2.75);
\filldraw[fill=bermuda, draw=black] (2.50, -2.50) rectangle (2.75, -2.75);
\filldraw[fill=cancan, draw=black] (2.75, -2.50) rectangle (3.00, -2.75);
\filldraw[fill=cancan, draw=black] (3.00, -2.50) rectangle (3.25, -2.75);
\filldraw[fill=bermuda, draw=black] (3.25, -2.50) rectangle (3.50, -2.75);
\filldraw[fill=bermuda, draw=black] (3.50, -2.50) rectangle (3.75, -2.75);
\filldraw[fill=cancan, draw=black] (3.75, -2.50) rectangle (4.00, -2.75);
\filldraw[fill=bermuda, draw=black] (4.00, -2.50) rectangle (4.25, -2.75);
\filldraw[fill=cancan, draw=black] (4.25, -2.50) rectangle (4.50, -2.75);
\filldraw[fill=cancan, draw=black] (4.50, -2.50) rectangle (4.75, -2.75);
\filldraw[fill=cancan, draw=black] (4.75, -2.50) rectangle (5.00, -2.75);
\filldraw[fill=bermuda, draw=black] (5.00, -2.50) rectangle (5.25, -2.75);
\filldraw[fill=cancan, draw=black] (5.25, -2.50) rectangle (5.50, -2.75);
\filldraw[fill=cancan, draw=black] (5.50, -2.50) rectangle (5.75, -2.75);
\filldraw[fill=cancan, draw=black] (5.75, -2.50) rectangle (6.00, -2.75);
\filldraw[fill=bermuda, draw=black] (6.00, -2.50) rectangle (6.25, -2.75);
\filldraw[fill=bermuda, draw=black] (6.25, -2.50) rectangle (6.50, -2.75);
\filldraw[fill=bermuda, draw=black] (6.50, -2.50) rectangle (6.75, -2.75);
\filldraw[fill=bermuda, draw=black] (6.75, -2.50) rectangle (7.00, -2.75);
\filldraw[fill=bermuda, draw=black] (7.00, -2.50) rectangle (7.25, -2.75);
\filldraw[fill=cancan, draw=black] (7.25, -2.50) rectangle (7.50, -2.75);
\filldraw[fill=cancan, draw=black] (7.50, -2.50) rectangle (7.75, -2.75);
\filldraw[fill=cancan, draw=black] (7.75, -2.50) rectangle (8.00, -2.75);
\filldraw[fill=bermuda, draw=black] (8.00, -2.50) rectangle (8.25, -2.75);
\filldraw[fill=bermuda, draw=black] (8.25, -2.50) rectangle (8.50, -2.75);
\filldraw[fill=bermuda, draw=black] (8.50, -2.50) rectangle (8.75, -2.75);
\filldraw[fill=cancan, draw=black] (8.75, -2.50) rectangle (9.00, -2.75);
\filldraw[fill=cancan, draw=black] (9.00, -2.50) rectangle (9.25, -2.75);
\filldraw[fill=cancan, draw=black] (9.25, -2.50) rectangle (9.50, -2.75);
\filldraw[fill=bermuda, draw=black] (9.50, -2.50) rectangle (9.75, -2.75);
\filldraw[fill=bermuda, draw=black] (9.75, -2.50) rectangle (10.00, -2.75);
\filldraw[fill=bermuda, draw=black] (10.00, -2.50) rectangle (10.25, -2.75);
\filldraw[fill=cancan, draw=black] (10.25, -2.50) rectangle (10.50, -2.75);
\filldraw[fill=cancan, draw=black] (10.50, -2.50) rectangle (10.75, -2.75);
\filldraw[fill=cancan, draw=black] (10.75, -2.50) rectangle (11.00, -2.75);
\filldraw[fill=cancan, draw=black] (11.00, -2.50) rectangle (11.25, -2.75);
\filldraw[fill=cancan, draw=black] (11.25, -2.50) rectangle (11.50, -2.75);
\filldraw[fill=bermuda, draw=black] (11.50, -2.50) rectangle (11.75, -2.75);
\filldraw[fill=bermuda, draw=black] (11.75, -2.50) rectangle (12.00, -2.75);
\filldraw[fill=bermuda, draw=black] (12.00, -2.50) rectangle (12.25, -2.75);
\filldraw[fill=cancan, draw=black] (12.25, -2.50) rectangle (12.50, -2.75);
\filldraw[fill=cancan, draw=black] (12.50, -2.50) rectangle (12.75, -2.75);
\filldraw[fill=cancan, draw=black] (12.75, -2.50) rectangle (13.00, -2.75);
\filldraw[fill=bermuda, draw=black] (13.00, -2.50) rectangle (13.25, -2.75);
\filldraw[fill=bermuda, draw=black] (13.25, -2.50) rectangle (13.50, -2.75);
\filldraw[fill=bermuda, draw=black] (13.50, -2.50) rectangle (13.75, -2.75);
\filldraw[fill=bermuda, draw=black] (13.75, -2.50) rectangle (14.00, -2.75);
\filldraw[fill=bermuda, draw=black] (14.00, -2.50) rectangle (14.25, -2.75);
\filldraw[fill=cancan, draw=black] (14.25, -2.50) rectangle (14.50, -2.75);
\filldraw[fill=bermuda, draw=black] (14.50, -2.50) rectangle (14.75, -2.75);
\filldraw[fill=bermuda, draw=black] (14.75, -2.50) rectangle (15.00, -2.75);
\filldraw[fill=bermuda, draw=black] (0.00, -2.75) rectangle (0.25, -3.00);
\filldraw[fill=cancan, draw=black] (0.25, -2.75) rectangle (0.50, -3.00);
\filldraw[fill=bermuda, draw=black] (0.50, -2.75) rectangle (0.75, -3.00);
\filldraw[fill=bermuda, draw=black] (0.75, -2.75) rectangle (1.00, -3.00);
\filldraw[fill=bermuda, draw=black] (1.00, -2.75) rectangle (1.25, -3.00);
\filldraw[fill=cancan, draw=black] (1.25, -2.75) rectangle (1.50, -3.00);
\filldraw[fill=bermuda, draw=black] (1.50, -2.75) rectangle (1.75, -3.00);
\filldraw[fill=cancan, draw=black] (1.75, -2.75) rectangle (2.00, -3.00);
\filldraw[fill=cancan, draw=black] (2.00, -2.75) rectangle (2.25, -3.00);
\filldraw[fill=bermuda, draw=black] (2.25, -2.75) rectangle (2.50, -3.00);
\filldraw[fill=bermuda, draw=black] (2.50, -2.75) rectangle (2.75, -3.00);
\filldraw[fill=cancan, draw=black] (2.75, -2.75) rectangle (3.00, -3.00);
\filldraw[fill=bermuda, draw=black] (3.00, -2.75) rectangle (3.25, -3.00);
\filldraw[fill=cancan, draw=black] (3.25, -2.75) rectangle (3.50, -3.00);
\filldraw[fill=cancan, draw=black] (3.50, -2.75) rectangle (3.75, -3.00);
\filldraw[fill=bermuda, draw=black] (3.75, -2.75) rectangle (4.00, -3.00);
\filldraw[fill=bermuda, draw=black] (4.00, -2.75) rectangle (4.25, -3.00);
\filldraw[fill=cancan, draw=black] (4.25, -2.75) rectangle (4.50, -3.00);
\filldraw[fill=bermuda, draw=black] (4.50, -2.75) rectangle (4.75, -3.00);
\filldraw[fill=cancan, draw=black] (4.75, -2.75) rectangle (5.00, -3.00);
\filldraw[fill=bermuda, draw=black] (5.00, -2.75) rectangle (5.25, -3.00);
\filldraw[fill=bermuda, draw=black] (5.25, -2.75) rectangle (5.50, -3.00);
\filldraw[fill=bermuda, draw=black] (5.50, -2.75) rectangle (5.75, -3.00);
\filldraw[fill=cancan, draw=black] (5.75, -2.75) rectangle (6.00, -3.00);
\filldraw[fill=cancan, draw=black] (6.00, -2.75) rectangle (6.25, -3.00);
\filldraw[fill=cancan, draw=black] (6.25, -2.75) rectangle (6.50, -3.00);
\filldraw[fill=cancan, draw=black] (6.50, -2.75) rectangle (6.75, -3.00);
\filldraw[fill=cancan, draw=black] (6.75, -2.75) rectangle (7.00, -3.00);
\filldraw[fill=bermuda, draw=black] (7.00, -2.75) rectangle (7.25, -3.00);
\filldraw[fill=bermuda, draw=black] (7.25, -2.75) rectangle (7.50, -3.00);
\filldraw[fill=bermuda, draw=black] (7.50, -2.75) rectangle (7.75, -3.00);
\filldraw[fill=cancan, draw=black] (7.75, -2.75) rectangle (8.00, -3.00);
\filldraw[fill=cancan, draw=black] (8.00, -2.75) rectangle (8.25, -3.00);
\filldraw[fill=cancan, draw=black] (8.25, -2.75) rectangle (8.50, -3.00);
\filldraw[fill=bermuda, draw=black] (8.50, -2.75) rectangle (8.75, -3.00);
\filldraw[fill=bermuda, draw=black] (8.75, -2.75) rectangle (9.00, -3.00);
\filldraw[fill=bermuda, draw=black] (9.00, -2.75) rectangle (9.25, -3.00);
\filldraw[fill=cancan, draw=black] (9.25, -2.75) rectangle (9.50, -3.00);
\filldraw[fill=cancan, draw=black] (9.50, -2.75) rectangle (9.75, -3.00);
\filldraw[fill=cancan, draw=black] (9.75, -2.75) rectangle (10.00, -3.00);
\filldraw[fill=bermuda, draw=black] (10.00, -2.75) rectangle (10.25, -3.00);
\filldraw[fill=bermuda, draw=black] (10.25, -2.75) rectangle (10.50, -3.00);
\filldraw[fill=bermuda, draw=black] (10.50, -2.75) rectangle (10.75, -3.00);
\filldraw[fill=cancan, draw=black] (10.75, -2.75) rectangle (11.00, -3.00);
\filldraw[fill=cancan, draw=black] (11.00, -2.75) rectangle (11.25, -3.00);
\filldraw[fill=cancan, draw=black] (11.25, -2.75) rectangle (11.50, -3.00);
\filldraw[fill=bermuda, draw=black] (11.50, -2.75) rectangle (11.75, -3.00);
\filldraw[fill=bermuda, draw=black] (11.75, -2.75) rectangle (12.00, -3.00);
\filldraw[fill=bermuda, draw=black] (12.00, -2.75) rectangle (12.25, -3.00);
\filldraw[fill=cancan, draw=black] (12.25, -2.75) rectangle (12.50, -3.00);
\filldraw[fill=cancan, draw=black] (12.50, -2.75) rectangle (12.75, -3.00);
\filldraw[fill=cancan, draw=black] (12.75, -2.75) rectangle (13.00, -3.00);
\filldraw[fill=cancan, draw=black] (13.00, -2.75) rectangle (13.25, -3.00);
\filldraw[fill=cancan, draw=black] (13.25, -2.75) rectangle (13.50, -3.00);
\filldraw[fill=bermuda, draw=black] (13.50, -2.75) rectangle (13.75, -3.00);
\filldraw[fill=cancan, draw=black] (13.75, -2.75) rectangle (14.00, -3.00);
\filldraw[fill=cancan, draw=black] (14.00, -2.75) rectangle (14.25, -3.00);
\filldraw[fill=cancan, draw=black] (14.25, -2.75) rectangle (14.50, -3.00);
\filldraw[fill=cancan, draw=black] (14.50, -2.75) rectangle (14.75, -3.00);
\filldraw[fill=cancan, draw=black] (14.75, -2.75) rectangle (15.00, -3.00);
\filldraw[fill=cancan, draw=black] (0.00, -3.00) rectangle (0.25, -3.25);
\filldraw[fill=cancan, draw=black] (0.25, -3.00) rectangle (0.50, -3.25);
\filldraw[fill=cancan, draw=black] (0.50, -3.00) rectangle (0.75, -3.25);
\filldraw[fill=cancan, draw=black] (0.75, -3.00) rectangle (1.00, -3.25);
\filldraw[fill=cancan, draw=black] (1.00, -3.00) rectangle (1.25, -3.25);
\filldraw[fill=cancan, draw=black] (1.25, -3.00) rectangle (1.50, -3.25);
\filldraw[fill=bermuda, draw=black] (1.50, -3.00) rectangle (1.75, -3.25);
\filldraw[fill=bermuda, draw=black] (1.75, -3.00) rectangle (2.00, -3.25);
\filldraw[fill=bermuda, draw=black] (2.00, -3.00) rectangle (2.25, -3.25);
\filldraw[fill=bermuda, draw=black] (2.25, -3.00) rectangle (2.50, -3.25);
\filldraw[fill=bermuda, draw=black] (2.50, -3.00) rectangle (2.75, -3.25);
\filldraw[fill=cancan, draw=black] (2.75, -3.00) rectangle (3.00, -3.25);
\filldraw[fill=bermuda, draw=black] (3.00, -3.00) rectangle (3.25, -3.25);
\filldraw[fill=cancan, draw=black] (3.25, -3.00) rectangle (3.50, -3.25);
\filldraw[fill=bermuda, draw=black] (3.50, -3.00) rectangle (3.75, -3.25);
\filldraw[fill=cancan, draw=black] (3.75, -3.00) rectangle (4.00, -3.25);
\filldraw[fill=bermuda, draw=black] (4.00, -3.00) rectangle (4.25, -3.25);
\filldraw[fill=cancan, draw=black] (4.25, -3.00) rectangle (4.50, -3.25);
\filldraw[fill=cancan, draw=black] (4.50, -3.00) rectangle (4.75, -3.25);
\filldraw[fill=bermuda, draw=black] (4.75, -3.00) rectangle (5.00, -3.25);
\filldraw[fill=bermuda, draw=black] (5.00, -3.00) rectangle (5.25, -3.25);
\filldraw[fill=bermuda, draw=black] (5.25, -3.00) rectangle (5.50, -3.25);
\filldraw[fill=bermuda, draw=black] (5.50, -3.00) rectangle (5.75, -3.25);
\filldraw[fill=cancan, draw=black] (5.75, -3.00) rectangle (6.00, -3.25);
\filldraw[fill=cancan, draw=black] (6.00, -3.00) rectangle (6.25, -3.25);
\filldraw[fill=cancan, draw=black] (6.25, -3.00) rectangle (6.50, -3.25);
\filldraw[fill=cancan, draw=black] (6.50, -3.00) rectangle (6.75, -3.25);
\filldraw[fill=cancan, draw=black] (6.75, -3.00) rectangle (7.00, -3.25);
\filldraw[fill=cancan, draw=black] (7.00, -3.00) rectangle (7.25, -3.25);
\filldraw[fill=cancan, draw=black] (7.25, -3.00) rectangle (7.50, -3.25);
\filldraw[fill=bermuda, draw=black] (7.50, -3.00) rectangle (7.75, -3.25);
\filldraw[fill=cancan, draw=black] (7.75, -3.00) rectangle (8.00, -3.25);
\filldraw[fill=bermuda, draw=black] (8.00, -3.00) rectangle (8.25, -3.25);
\filldraw[fill=cancan, draw=black] (8.25, -3.00) rectangle (8.50, -3.25);
\filldraw[fill=cancan, draw=black] (8.50, -3.00) rectangle (8.75, -3.25);
\filldraw[fill=bermuda, draw=black] (8.75, -3.00) rectangle (9.00, -3.25);
\filldraw[fill=bermuda, draw=black] (9.00, -3.00) rectangle (9.25, -3.25);
\filldraw[fill=bermuda, draw=black] (9.25, -3.00) rectangle (9.50, -3.25);
\filldraw[fill=bermuda, draw=black] (9.50, -3.00) rectangle (9.75, -3.25);
\filldraw[fill=cancan, draw=black] (9.75, -3.00) rectangle (10.00, -3.25);
\filldraw[fill=cancan, draw=black] (10.00, -3.00) rectangle (10.25, -3.25);
\filldraw[fill=cancan, draw=black] (10.25, -3.00) rectangle (10.50, -3.25);
\filldraw[fill=cancan, draw=black] (10.50, -3.00) rectangle (10.75, -3.25);
\filldraw[fill=bermuda, draw=black] (10.75, -3.00) rectangle (11.00, -3.25);
\filldraw[fill=bermuda, draw=black] (11.00, -3.00) rectangle (11.25, -3.25);
\filldraw[fill=bermuda, draw=black] (11.25, -3.00) rectangle (11.50, -3.25);
\filldraw[fill=bermuda, draw=black] (11.50, -3.00) rectangle (11.75, -3.25);
\filldraw[fill=cancan, draw=black] (11.75, -3.00) rectangle (12.00, -3.25);
\filldraw[fill=bermuda, draw=black] (12.00, -3.00) rectangle (12.25, -3.25);
\filldraw[fill=cancan, draw=black] (12.25, -3.00) rectangle (12.50, -3.25);
\filldraw[fill=cancan, draw=black] (12.50, -3.00) rectangle (12.75, -3.25);
\filldraw[fill=bermuda, draw=black] (12.75, -3.00) rectangle (13.00, -3.25);
\filldraw[fill=bermuda, draw=black] (13.00, -3.00) rectangle (13.25, -3.25);
\filldraw[fill=cancan, draw=black] (13.25, -3.00) rectangle (13.50, -3.25);
\filldraw[fill=bermuda, draw=black] (13.50, -3.00) rectangle (13.75, -3.25);
\filldraw[fill=cancan, draw=black] (13.75, -3.00) rectangle (14.00, -3.25);
\filldraw[fill=cancan, draw=black] (14.00, -3.00) rectangle (14.25, -3.25);
\filldraw[fill=bermuda, draw=black] (14.25, -3.00) rectangle (14.50, -3.25);
\filldraw[fill=bermuda, draw=black] (14.50, -3.00) rectangle (14.75, -3.25);
\filldraw[fill=cancan, draw=black] (14.75, -3.00) rectangle (15.00, -3.25);
\filldraw[fill=bermuda, draw=black] (0.00, -3.25) rectangle (0.25, -3.50);
\filldraw[fill=cancan, draw=black] (0.25, -3.25) rectangle (0.50, -3.50);
\filldraw[fill=cancan, draw=black] (0.50, -3.25) rectangle (0.75, -3.50);
\filldraw[fill=cancan, draw=black] (0.75, -3.25) rectangle (1.00, -3.50);
\filldraw[fill=cancan, draw=black] (1.00, -3.25) rectangle (1.25, -3.50);
\filldraw[fill=cancan, draw=black] (1.25, -3.25) rectangle (1.50, -3.50);
\filldraw[fill=bermuda, draw=black] (1.50, -3.25) rectangle (1.75, -3.50);
\filldraw[fill=bermuda, draw=black] (1.75, -3.25) rectangle (2.00, -3.50);
\filldraw[fill=bermuda, draw=black] (2.00, -3.25) rectangle (2.25, -3.50);
} } }\end{equation*}
\begin{equation*}
\hspace{4.6pt} b_{2} = \vcenter{\hbox{ \tikz{
\filldraw[fill=cancan, draw=black] (0.00, 0.00) rectangle (0.25, -0.25);
\filldraw[fill=cancan, draw=black] (0.25, 0.00) rectangle (0.50, -0.25);
\filldraw[fill=bermuda, draw=black] (0.50, 0.00) rectangle (0.75, -0.25);
\filldraw[fill=bermuda, draw=black] (0.75, 0.00) rectangle (1.00, -0.25);
\filldraw[fill=cancan, draw=black] (1.00, 0.00) rectangle (1.25, -0.25);
\filldraw[fill=cancan, draw=black] (1.25, 0.00) rectangle (1.50, -0.25);
\filldraw[fill=cancan, draw=black] (1.50, 0.00) rectangle (1.75, -0.25);
\filldraw[fill=bermuda, draw=black] (1.75, 0.00) rectangle (2.00, -0.25);
\filldraw[fill=cancan, draw=black] (2.00, 0.00) rectangle (2.25, -0.25);
\filldraw[fill=bermuda, draw=black] (2.25, 0.00) rectangle (2.50, -0.25);
\filldraw[fill=cancan, draw=black] (2.50, 0.00) rectangle (2.75, -0.25);
\filldraw[fill=cancan, draw=black] (2.75, 0.00) rectangle (3.00, -0.25);
\filldraw[fill=cancan, draw=black] (3.00, 0.00) rectangle (3.25, -0.25);
\filldraw[fill=bermuda, draw=black] (3.25, 0.00) rectangle (3.50, -0.25);
\filldraw[fill=bermuda, draw=black] (3.50, 0.00) rectangle (3.75, -0.25);
\filldraw[fill=bermuda, draw=black] (3.75, 0.00) rectangle (4.00, -0.25);
\filldraw[fill=cancan, draw=black] (4.00, 0.00) rectangle (4.25, -0.25);
\filldraw[fill=bermuda, draw=black] (4.25, 0.00) rectangle (4.50, -0.25);
\filldraw[fill=bermuda, draw=black] (4.50, 0.00) rectangle (4.75, -0.25);
\filldraw[fill=bermuda, draw=black] (4.75, 0.00) rectangle (5.00, -0.25);
\filldraw[fill=cancan, draw=black] (5.00, 0.00) rectangle (5.25, -0.25);
\filldraw[fill=cancan, draw=black] (5.25, 0.00) rectangle (5.50, -0.25);
\filldraw[fill=cancan, draw=black] (5.50, 0.00) rectangle (5.75, -0.25);
\filldraw[fill=bermuda, draw=black] (5.75, 0.00) rectangle (6.00, -0.25);
\filldraw[fill=bermuda, draw=black] (6.00, 0.00) rectangle (6.25, -0.25);
\filldraw[fill=bermuda, draw=black] (6.25, 0.00) rectangle (6.50, -0.25);
\filldraw[fill=cancan, draw=black] (6.50, 0.00) rectangle (6.75, -0.25);
\filldraw[fill=cancan, draw=black] (6.75, 0.00) rectangle (7.00, -0.25);
\filldraw[fill=cancan, draw=black] (7.00, 0.00) rectangle (7.25, -0.25);
\filldraw[fill=bermuda, draw=black] (7.25, 0.00) rectangle (7.50, -0.25);
\filldraw[fill=bermuda, draw=black] (7.50, 0.00) rectangle (7.75, -0.25);
\filldraw[fill=bermuda, draw=black] (7.75, 0.00) rectangle (8.00, -0.25);
\filldraw[fill=cancan, draw=black] (8.00, 0.00) rectangle (8.25, -0.25);
\filldraw[fill=cancan, draw=black] (8.25, 0.00) rectangle (8.50, -0.25);
\filldraw[fill=cancan, draw=black] (8.50, 0.00) rectangle (8.75, -0.25);
\filldraw[fill=bermuda, draw=black] (8.75, 0.00) rectangle (9.00, -0.25);
\filldraw[fill=bermuda, draw=black] (9.00, 0.00) rectangle (9.25, -0.25);
\filldraw[fill=bermuda, draw=black] (9.25, 0.00) rectangle (9.50, -0.25);
\filldraw[fill=cancan, draw=black] (9.50, 0.00) rectangle (9.75, -0.25);
\filldraw[fill=cancan, draw=black] (9.75, 0.00) rectangle (10.00, -0.25);
\filldraw[fill=cancan, draw=black] (10.00, 0.00) rectangle (10.25, -0.25);
\filldraw[fill=bermuda, draw=black] (10.25, 0.00) rectangle (10.50, -0.25);
\filldraw[fill=bermuda, draw=black] (10.50, 0.00) rectangle (10.75, -0.25);
\filldraw[fill=bermuda, draw=black] (10.75, 0.00) rectangle (11.00, -0.25);
\filldraw[fill=bermuda, draw=black] (11.00, 0.00) rectangle (11.25, -0.25);
\filldraw[fill=bermuda, draw=black] (11.25, 0.00) rectangle (11.50, -0.25);
\filldraw[fill=cancan, draw=black] (11.50, 0.00) rectangle (11.75, -0.25);
\filldraw[fill=bermuda, draw=black] (11.75, 0.00) rectangle (12.00, -0.25);
\filldraw[fill=bermuda, draw=black] (12.00, 0.00) rectangle (12.25, -0.25);
\filldraw[fill=bermuda, draw=black] (12.25, 0.00) rectangle (12.50, -0.25);
\filldraw[fill=cancan, draw=black] (12.50, 0.00) rectangle (12.75, -0.25);
\filldraw[fill=cancan, draw=black] (12.75, 0.00) rectangle (13.00, -0.25);
\filldraw[fill=cancan, draw=black] (13.00, 0.00) rectangle (13.25, -0.25);
\filldraw[fill=bermuda, draw=black] (13.25, 0.00) rectangle (13.50, -0.25);
\filldraw[fill=bermuda, draw=black] (13.50, 0.00) rectangle (13.75, -0.25);
\filldraw[fill=bermuda, draw=black] (13.75, 0.00) rectangle (14.00, -0.25);
\filldraw[fill=cancan, draw=black] (14.00, 0.00) rectangle (14.25, -0.25);
\filldraw[fill=bermuda, draw=black] (14.25, 0.00) rectangle (14.50, -0.25);
\filldraw[fill=cancan, draw=black] (14.50, 0.00) rectangle (14.75, -0.25);
\filldraw[fill=cancan, draw=black] (14.75, 0.00) rectangle (15.00, -0.25);
\filldraw[fill=cancan, draw=black] (0.00, -0.25) rectangle (0.25, -0.50);
\filldraw[fill=cancan, draw=black] (0.25, -0.25) rectangle (0.50, -0.50);
\filldraw[fill=cancan, draw=black] (0.50, -0.25) rectangle (0.75, -0.50);
\filldraw[fill=bermuda, draw=black] (0.75, -0.25) rectangle (1.00, -0.50);
\filldraw[fill=bermuda, draw=black] (1.00, -0.25) rectangle (1.25, -0.50);
\filldraw[fill=bermuda, draw=black] (1.25, -0.25) rectangle (1.50, -0.50);
\filldraw[fill=cancan, draw=black] (1.50, -0.25) rectangle (1.75, -0.50);
\filldraw[fill=bermuda, draw=black] (1.75, -0.25) rectangle (2.00, -0.50);
\filldraw[fill=bermuda, draw=black] (2.00, -0.25) rectangle (2.25, -0.50);
\filldraw[fill=bermuda, draw=black] (2.25, -0.25) rectangle (2.50, -0.50);
\filldraw[fill=cancan, draw=black] (2.50, -0.25) rectangle (2.75, -0.50);
\filldraw[fill=cancan, draw=black] (2.75, -0.25) rectangle (3.00, -0.50);
\filldraw[fill=cancan, draw=black] (3.00, -0.25) rectangle (3.25, -0.50);
\filldraw[fill=bermuda, draw=black] (3.25, -0.25) rectangle (3.50, -0.50);
\filldraw[fill=bermuda, draw=black] (3.50, -0.25) rectangle (3.75, -0.50);
\filldraw[fill=bermuda, draw=black] (3.75, -0.25) rectangle (4.00, -0.50);
\filldraw[fill=cancan, draw=black] (4.00, -0.25) rectangle (4.25, -0.50);
\filldraw[fill=cancan, draw=black] (4.25, -0.25) rectangle (4.50, -0.50);
\filldraw[fill=cancan, draw=black] (4.50, -0.25) rectangle (4.75, -0.50);
\filldraw[fill=bermuda, draw=black] (4.75, -0.25) rectangle (5.00, -0.50);
\filldraw[fill=bermuda, draw=black] (5.00, -0.25) rectangle (5.25, -0.50);
\filldraw[fill=bermuda, draw=black] (5.25, -0.25) rectangle (5.50, -0.50);
\filldraw[fill=cancan, draw=black] (5.50, -0.25) rectangle (5.75, -0.50);
\filldraw[fill=cancan, draw=black] (5.75, -0.25) rectangle (6.00, -0.50);
\filldraw[fill=cancan, draw=black] (6.00, -0.25) rectangle (6.25, -0.50);
\filldraw[fill=bermuda, draw=black] (6.25, -0.25) rectangle (6.50, -0.50);
\filldraw[fill=bermuda, draw=black] (6.50, -0.25) rectangle (6.75, -0.50);
\filldraw[fill=bermuda, draw=black] (6.75, -0.25) rectangle (7.00, -0.50);
\filldraw[fill=cancan, draw=black] (7.00, -0.25) rectangle (7.25, -0.50);
\filldraw[fill=cancan, draw=black] (7.25, -0.25) rectangle (7.50, -0.50);
\filldraw[fill=bermuda, draw=black] (7.50, -0.25) rectangle (7.75, -0.50);
\filldraw[fill=bermuda, draw=black] (7.75, -0.25) rectangle (8.00, -0.50);
\filldraw[fill=cancan, draw=black] (8.00, -0.25) rectangle (8.25, -0.50);
\filldraw[fill=bermuda, draw=black] (8.25, -0.25) rectangle (8.50, -0.50);
\filldraw[fill=bermuda, draw=black] (8.50, -0.25) rectangle (8.75, -0.50);
\filldraw[fill=bermuda, draw=black] (8.75, -0.25) rectangle (9.00, -0.50);
\filldraw[fill=bermuda, draw=black] (9.00, -0.25) rectangle (9.25, -0.50);
\filldraw[fill=bermuda, draw=black] (9.25, -0.25) rectangle (9.50, -0.50);
\filldraw[fill=cancan, draw=black] (9.50, -0.25) rectangle (9.75, -0.50);
\filldraw[fill=bermuda, draw=black] (9.75, -0.25) rectangle (10.00, -0.50);
\filldraw[fill=cancan, draw=black] (10.00, -0.25) rectangle (10.25, -0.50);
\filldraw[fill=cancan, draw=black] (10.25, -0.25) rectangle (10.50, -0.50);
\filldraw[fill=cancan, draw=black] (10.50, -0.25) rectangle (10.75, -0.50);
\filldraw[fill=bermuda, draw=black] (10.75, -0.25) rectangle (11.00, -0.50);
\filldraw[fill=cancan, draw=black] (11.00, -0.25) rectangle (11.25, -0.50);
\filldraw[fill=bermuda, draw=black] (11.25, -0.25) rectangle (11.50, -0.50);
\filldraw[fill=bermuda, draw=black] (11.50, -0.25) rectangle (11.75, -0.50);
\filldraw[fill=bermuda, draw=black] (11.75, -0.25) rectangle (12.00, -0.50);
\filldraw[fill=cancan, draw=black] (12.00, -0.25) rectangle (12.25, -0.50);
\filldraw[fill=bermuda, draw=black] (12.25, -0.25) rectangle (12.50, -0.50);
\filldraw[fill=bermuda, draw=black] (12.50, -0.25) rectangle (12.75, -0.50);
\filldraw[fill=bermuda, draw=black] (12.75, -0.25) rectangle (13.00, -0.50);
\filldraw[fill=bermuda, draw=black] (13.00, -0.25) rectangle (13.25, -0.50);
\filldraw[fill=bermuda, draw=black] (13.25, -0.25) rectangle (13.50, -0.50);
\filldraw[fill=bermuda, draw=black] (13.50, -0.25) rectangle (13.75, -0.50);
\filldraw[fill=bermuda, draw=black] (13.75, -0.25) rectangle (14.00, -0.50);
\filldraw[fill=cancan, draw=black] (14.00, -0.25) rectangle (14.25, -0.50);
\filldraw[fill=bermuda, draw=black] (14.25, -0.25) rectangle (14.50, -0.50);
\filldraw[fill=cancan, draw=black] (14.50, -0.25) rectangle (14.75, -0.50);
\filldraw[fill=cancan, draw=black] (14.75, -0.25) rectangle (15.00, -0.50);
\filldraw[fill=bermuda, draw=black] (0.00, -0.50) rectangle (0.25, -0.75);
\filldraw[fill=bermuda, draw=black] (0.25, -0.50) rectangle (0.50, -0.75);
\filldraw[fill=cancan, draw=black] (0.50, -0.50) rectangle (0.75, -0.75);
\filldraw[fill=bermuda, draw=black] (0.75, -0.50) rectangle (1.00, -0.75);
\filldraw[fill=cancan, draw=black] (1.00, -0.50) rectangle (1.25, -0.75);
\filldraw[fill=cancan, draw=black] (1.25, -0.50) rectangle (1.50, -0.75);
\filldraw[fill=bermuda, draw=black] (1.50, -0.50) rectangle (1.75, -0.75);
\filldraw[fill=bermuda, draw=black] (1.75, -0.50) rectangle (2.00, -0.75);
\filldraw[fill=cancan, draw=black] (2.00, -0.50) rectangle (2.25, -0.75);
\filldraw[fill=bermuda, draw=black] (2.25, -0.50) rectangle (2.50, -0.75);
\filldraw[fill=cancan, draw=black] (2.50, -0.50) rectangle (2.75, -0.75);
\filldraw[fill=cancan, draw=black] (2.75, -0.50) rectangle (3.00, -0.75);
\filldraw[fill=bermuda, draw=black] (3.00, -0.50) rectangle (3.25, -0.75);
\filldraw[fill=bermuda, draw=black] (3.25, -0.50) rectangle (3.50, -0.75);
\filldraw[fill=cancan, draw=black] (3.50, -0.50) rectangle (3.75, -0.75);
\filldraw[fill=bermuda, draw=black] (3.75, -0.50) rectangle (4.00, -0.75);
\filldraw[fill=bermuda, draw=black] (4.00, -0.50) rectangle (4.25, -0.75);
\filldraw[fill=bermuda, draw=black] (4.25, -0.50) rectangle (4.50, -0.75);
\filldraw[fill=cancan, draw=black] (4.50, -0.50) rectangle (4.75, -0.75);
\filldraw[fill=cancan, draw=black] (4.75, -0.50) rectangle (5.00, -0.75);
\filldraw[fill=cancan, draw=black] (5.00, -0.50) rectangle (5.25, -0.75);
\filldraw[fill=cancan, draw=black] (5.25, -0.50) rectangle (5.50, -0.75);
\filldraw[fill=bermuda, draw=black] (5.50, -0.50) rectangle (5.75, -0.75);
\filldraw[fill=bermuda, draw=black] (5.75, -0.50) rectangle (6.00, -0.75);
\filldraw[fill=cancan, draw=black] (6.00, -0.50) rectangle (6.25, -0.75);
\filldraw[fill=cancan, draw=black] (6.25, -0.50) rectangle (6.50, -0.75);
\filldraw[fill=cancan, draw=black] (6.50, -0.50) rectangle (6.75, -0.75);
\filldraw[fill=cancan, draw=black] (6.75, -0.50) rectangle (7.00, -0.75);
\filldraw[fill=bermuda, draw=black] (7.00, -0.50) rectangle (7.25, -0.75);
\filldraw[fill=bermuda, draw=black] (7.25, -0.50) rectangle (7.50, -0.75);
\filldraw[fill=cancan, draw=black] (7.50, -0.50) rectangle (7.75, -0.75);
\filldraw[fill=cancan, draw=black] (7.75, -0.50) rectangle (8.00, -0.75);
\filldraw[fill=cancan, draw=black] (8.00, -0.50) rectangle (8.25, -0.75);
\filldraw[fill=cancan, draw=black] (8.25, -0.50) rectangle (8.50, -0.75);
\filldraw[fill=bermuda, draw=black] (8.50, -0.50) rectangle (8.75, -0.75);
\filldraw[fill=bermuda, draw=black] (8.75, -0.50) rectangle (9.00, -0.75);
\filldraw[fill=cancan, draw=black] (9.00, -0.50) rectangle (9.25, -0.75);
\filldraw[fill=cancan, draw=black] (9.25, -0.50) rectangle (9.50, -0.75);
\filldraw[fill=cancan, draw=black] (9.50, -0.50) rectangle (9.75, -0.75);
\filldraw[fill=cancan, draw=black] (9.75, -0.50) rectangle (10.00, -0.75);
\filldraw[fill=bermuda, draw=black] (10.00, -0.50) rectangle (10.25, -0.75);
\filldraw[fill=bermuda, draw=black] (10.25, -0.50) rectangle (10.50, -0.75);
\filldraw[fill=cancan, draw=black] (10.50, -0.50) rectangle (10.75, -0.75);
\filldraw[fill=cancan, draw=black] (10.75, -0.50) rectangle (11.00, -0.75);
\filldraw[fill=cancan, draw=black] (11.00, -0.50) rectangle (11.25, -0.75);
\filldraw[fill=cancan, draw=black] (11.25, -0.50) rectangle (11.50, -0.75);
\filldraw[fill=cancan, draw=black] (11.50, -0.50) rectangle (11.75, -0.75);
\filldraw[fill=cancan, draw=black] (11.75, -0.50) rectangle (12.00, -0.75);
\filldraw[fill=bermuda, draw=black] (12.00, -0.50) rectangle (12.25, -0.75);
\filldraw[fill=bermuda, draw=black] (12.25, -0.50) rectangle (12.50, -0.75);
\filldraw[fill=bermuda, draw=black] (12.50, -0.50) rectangle (12.75, -0.75);
\filldraw[fill=bermuda, draw=black] (12.75, -0.50) rectangle (13.00, -0.75);
\filldraw[fill=cancan, draw=black] (13.00, -0.50) rectangle (13.25, -0.75);
\filldraw[fill=cancan, draw=black] (13.25, -0.50) rectangle (13.50, -0.75);
\filldraw[fill=cancan, draw=black] (13.50, -0.50) rectangle (13.75, -0.75);
\filldraw[fill=bermuda, draw=black] (13.75, -0.50) rectangle (14.00, -0.75);
\filldraw[fill=bermuda, draw=black] (14.00, -0.50) rectangle (14.25, -0.75);
\filldraw[fill=bermuda, draw=black] (14.25, -0.50) rectangle (14.50, -0.75);
\filldraw[fill=cancan, draw=black] (14.50, -0.50) rectangle (14.75, -0.75);
\filldraw[fill=cancan, draw=black] (14.75, -0.50) rectangle (15.00, -0.75);
\filldraw[fill=cancan, draw=black] (0.00, -0.75) rectangle (0.25, -1.00);
\filldraw[fill=cancan, draw=black] (0.25, -0.75) rectangle (0.50, -1.00);
\filldraw[fill=bermuda, draw=black] (0.50, -0.75) rectangle (0.75, -1.00);
\filldraw[fill=bermuda, draw=black] (0.75, -0.75) rectangle (1.00, -1.00);
\filldraw[fill=cancan, draw=black] (1.00, -0.75) rectangle (1.25, -1.00);
\filldraw[fill=cancan, draw=black] (1.25, -0.75) rectangle (1.50, -1.00);
\filldraw[fill=cancan, draw=black] (1.50, -0.75) rectangle (1.75, -1.00);
\filldraw[fill=cancan, draw=black] (1.75, -0.75) rectangle (2.00, -1.00);
\filldraw[fill=bermuda, draw=black] (2.00, -0.75) rectangle (2.25, -1.00);
\filldraw[fill=bermuda, draw=black] (2.25, -0.75) rectangle (2.50, -1.00);
\filldraw[fill=cancan, draw=black] (2.50, -0.75) rectangle (2.75, -1.00);
\filldraw[fill=bermuda, draw=black] (2.75, -0.75) rectangle (3.00, -1.00);
\filldraw[fill=cancan, draw=black] (3.00, -0.75) rectangle (3.25, -1.00);
\filldraw[fill=cancan, draw=black] (3.25, -0.75) rectangle (3.50, -1.00);
\filldraw[fill=bermuda, draw=black] (3.50, -0.75) rectangle (3.75, -1.00);
\filldraw[fill=bermuda, draw=black] (3.75, -0.75) rectangle (4.00, -1.00);
\filldraw[fill=cancan, draw=black] (4.00, -0.75) rectangle (4.25, -1.00);
\filldraw[fill=bermuda, draw=black] (4.25, -0.75) rectangle (4.50, -1.00);
\filldraw[fill=cancan, draw=black] (4.50, -0.75) rectangle (4.75, -1.00);
\filldraw[fill=cancan, draw=black] (4.75, -0.75) rectangle (5.00, -1.00);
\filldraw[fill=bermuda, draw=black] (5.00, -0.75) rectangle (5.25, -1.00);
\filldraw[fill=bermuda, draw=black] (5.25, -0.75) rectangle (5.50, -1.00);
\filldraw[fill=cancan, draw=black] (5.50, -0.75) rectangle (5.75, -1.00);
\filldraw[fill=bermuda, draw=black] (5.75, -0.75) rectangle (6.00, -1.00);
\filldraw[fill=bermuda, draw=black] (6.00, -0.75) rectangle (6.25, -1.00);
\filldraw[fill=bermuda, draw=black] (6.25, -0.75) rectangle (6.50, -1.00);
\filldraw[fill=cancan, draw=black] (6.50, -0.75) rectangle (6.75, -1.00);
\filldraw[fill=cancan, draw=black] (6.75, -0.75) rectangle (7.00, -1.00);
\filldraw[fill=cancan, draw=black] (7.00, -0.75) rectangle (7.25, -1.00);
\filldraw[fill=cancan, draw=black] (7.25, -0.75) rectangle (7.50, -1.00);
\filldraw[fill=cancan, draw=black] (7.50, -0.75) rectangle (7.75, -1.00);
\filldraw[fill=cancan, draw=black] (7.75, -0.75) rectangle (8.00, -1.00);
\filldraw[fill=cancan, draw=black] (8.00, -0.75) rectangle (8.25, -1.00);
\filldraw[fill=cancan, draw=black] (8.25, -0.75) rectangle (8.50, -1.00);
\filldraw[fill=cancan, draw=black] (8.50, -0.75) rectangle (8.75, -1.00);
\filldraw[fill=bermuda, draw=black] (8.75, -0.75) rectangle (9.00, -1.00);
\filldraw[fill=bermuda, draw=black] (9.00, -0.75) rectangle (9.25, -1.00);
\filldraw[fill=bermuda, draw=black] (9.25, -0.75) rectangle (9.50, -1.00);
\filldraw[fill=cancan, draw=black] (9.50, -0.75) rectangle (9.75, -1.00);
\filldraw[fill=cancan, draw=black] (9.75, -0.75) rectangle (10.00, -1.00);
\filldraw[fill=cancan, draw=black] (10.00, -0.75) rectangle (10.25, -1.00);
\filldraw[fill=bermuda, draw=black] (10.25, -0.75) rectangle (10.50, -1.00);
\filldraw[fill=bermuda, draw=black] (10.50, -0.75) rectangle (10.75, -1.00);
\filldraw[fill=bermuda, draw=black] (10.75, -0.75) rectangle (11.00, -1.00);
\filldraw[fill=bermuda, draw=black] (11.00, -0.75) rectangle (11.25, -1.00);
\filldraw[fill=bermuda, draw=black] (11.25, -0.75) rectangle (11.50, -1.00);
\filldraw[fill=cancan, draw=black] (11.50, -0.75) rectangle (11.75, -1.00);
\filldraw[fill=bermuda, draw=black] (11.75, -0.75) rectangle (12.00, -1.00);
\filldraw[fill=bermuda, draw=black] (12.00, -0.75) rectangle (12.25, -1.00);
\filldraw[fill=bermuda, draw=black] (12.25, -0.75) rectangle (12.50, -1.00);
\filldraw[fill=cancan, draw=black] (12.50, -0.75) rectangle (12.75, -1.00);
\filldraw[fill=cancan, draw=black] (12.75, -0.75) rectangle (13.00, -1.00);
\filldraw[fill=cancan, draw=black] (13.00, -0.75) rectangle (13.25, -1.00);
\filldraw[fill=bermuda, draw=black] (13.25, -0.75) rectangle (13.50, -1.00);
\filldraw[fill=bermuda, draw=black] (13.50, -0.75) rectangle (13.75, -1.00);
\filldraw[fill=bermuda, draw=black] (13.75, -0.75) rectangle (14.00, -1.00);
\filldraw[fill=cancan, draw=black] (14.00, -0.75) rectangle (14.25, -1.00);
\filldraw[fill=bermuda, draw=black] (14.25, -0.75) rectangle (14.50, -1.00);
\filldraw[fill=cancan, draw=black] (14.50, -0.75) rectangle (14.75, -1.00);
\filldraw[fill=bermuda, draw=black] (14.75, -0.75) rectangle (15.00, -1.00);
\filldraw[fill=bermuda, draw=black] (0.00, -1.00) rectangle (0.25, -1.25);
\filldraw[fill=bermuda, draw=black] (0.25, -1.00) rectangle (0.50, -1.25);
\filldraw[fill=cancan, draw=black] (0.50, -1.00) rectangle (0.75, -1.25);
\filldraw[fill=bermuda, draw=black] (0.75, -1.00) rectangle (1.00, -1.25);
\filldraw[fill=bermuda, draw=black] (1.00, -1.00) rectangle (1.25, -1.25);
\filldraw[fill=bermuda, draw=black] (1.25, -1.00) rectangle (1.50, -1.25);
\filldraw[fill=bermuda, draw=black] (1.50, -1.00) rectangle (1.75, -1.25);
\filldraw[fill=bermuda, draw=black] (1.75, -1.00) rectangle (2.00, -1.25);
\filldraw[fill=cancan, draw=black] (2.00, -1.00) rectangle (2.25, -1.25);
\filldraw[fill=bermuda, draw=black] (2.25, -1.00) rectangle (2.50, -1.25);
\filldraw[fill=bermuda, draw=black] (2.50, -1.00) rectangle (2.75, -1.25);
\filldraw[fill=bermuda, draw=black] (2.75, -1.00) rectangle (3.00, -1.25);
\filldraw[fill=cancan, draw=black] (3.00, -1.00) rectangle (3.25, -1.25);
\filldraw[fill=cancan, draw=black] (3.25, -1.00) rectangle (3.50, -1.25);
\filldraw[fill=cancan, draw=black] (3.50, -1.00) rectangle (3.75, -1.25);
\filldraw[fill=cancan, draw=black] (3.75, -1.00) rectangle (4.00, -1.25);
\filldraw[fill=cancan, draw=black] (4.00, -1.00) rectangle (4.25, -1.25);
\filldraw[fill=cancan, draw=black] (4.25, -1.00) rectangle (4.50, -1.25);
\filldraw[fill=cancan, draw=black] (4.50, -1.00) rectangle (4.75, -1.25);
\filldraw[fill=bermuda, draw=black] (4.75, -1.00) rectangle (5.00, -1.25);
\filldraw[fill=bermuda, draw=black] (5.00, -1.00) rectangle (5.25, -1.25);
\filldraw[fill=bermuda, draw=black] (5.25, -1.00) rectangle (5.50, -1.25);
\filldraw[fill=cancan, draw=black] (5.50, -1.00) rectangle (5.75, -1.25);
\filldraw[fill=cancan, draw=black] (5.75, -1.00) rectangle (6.00, -1.25);
\filldraw[fill=cancan, draw=black] (6.00, -1.00) rectangle (6.25, -1.25);
\filldraw[fill=bermuda, draw=black] (6.25, -1.00) rectangle (6.50, -1.25);
\filldraw[fill=bermuda, draw=black] (6.50, -1.00) rectangle (6.75, -1.25);
\filldraw[fill=bermuda, draw=black] (6.75, -1.00) rectangle (7.00, -1.25);
\filldraw[fill=bermuda, draw=black] (7.00, -1.00) rectangle (7.25, -1.25);
\filldraw[fill=bermuda, draw=black] (7.25, -1.00) rectangle (7.50, -1.25);
\filldraw[fill=bermuda, draw=black] (7.50, -1.00) rectangle (7.75, -1.25);
\filldraw[fill=bermuda, draw=black] (7.75, -1.00) rectangle (8.00, -1.25);
\filldraw[fill=cancan, draw=black] (8.00, -1.00) rectangle (8.25, -1.25);
\filldraw[fill=cancan, draw=black] (8.25, -1.00) rectangle (8.50, -1.25);
\filldraw[fill=cancan, draw=black] (8.50, -1.00) rectangle (8.75, -1.25);
\filldraw[fill=bermuda, draw=black] (8.75, -1.00) rectangle (9.00, -1.25);
\filldraw[fill=bermuda, draw=black] (9.00, -1.00) rectangle (9.25, -1.25);
\filldraw[fill=bermuda, draw=black] (9.25, -1.00) rectangle (9.50, -1.25);
\filldraw[fill=cancan, draw=black] (9.50, -1.00) rectangle (9.75, -1.25);
\filldraw[fill=cancan, draw=black] (9.75, -1.00) rectangle (10.00, -1.25);
\filldraw[fill=bermuda, draw=black] (10.00, -1.00) rectangle (10.25, -1.25);
\filldraw[fill=bermuda, draw=black] (10.25, -1.00) rectangle (10.50, -1.25);
\filldraw[fill=cancan, draw=black] (10.50, -1.00) rectangle (10.75, -1.25);
\filldraw[fill=cancan, draw=black] (10.75, -1.00) rectangle (11.00, -1.25);
\filldraw[fill=cancan, draw=black] (11.00, -1.00) rectangle (11.25, -1.25);
\filldraw[fill=cancan, draw=black] (11.25, -1.00) rectangle (11.50, -1.25);
\filldraw[fill=cancan, draw=black] (11.50, -1.00) rectangle (11.75, -1.25);
\filldraw[fill=bermuda, draw=black] (11.75, -1.00) rectangle (12.00, -1.25);
\filldraw[fill=cancan, draw=black] (12.00, -1.00) rectangle (12.25, -1.25);
\filldraw[fill=cancan, draw=black] (12.25, -1.00) rectangle (12.50, -1.25);
\filldraw[fill=cancan, draw=black] (12.50, -1.00) rectangle (12.75, -1.25);
\filldraw[fill=bermuda, draw=black] (12.75, -1.00) rectangle (13.00, -1.25);
\filldraw[fill=bermuda, draw=black] (13.00, -1.00) rectangle (13.25, -1.25);
\filldraw[fill=bermuda, draw=black] (13.25, -1.00) rectangle (13.50, -1.25);
\filldraw[fill=cancan, draw=black] (13.50, -1.00) rectangle (13.75, -1.25);
\filldraw[fill=cancan, draw=black] (13.75, -1.00) rectangle (14.00, -1.25);
\filldraw[fill=cancan, draw=black] (14.00, -1.00) rectangle (14.25, -1.25);
\filldraw[fill=cancan, draw=black] (14.25, -1.00) rectangle (14.50, -1.25);
\filldraw[fill=cancan, draw=black] (14.50, -1.00) rectangle (14.75, -1.25);
\filldraw[fill=cancan, draw=black] (14.75, -1.00) rectangle (15.00, -1.25);
\filldraw[fill=cancan, draw=black] (0.00, -1.25) rectangle (0.25, -1.50);
\filldraw[fill=cancan, draw=black] (0.25, -1.25) rectangle (0.50, -1.50);
\filldraw[fill=cancan, draw=black] (0.50, -1.25) rectangle (0.75, -1.50);
\filldraw[fill=bermuda, draw=black] (0.75, -1.25) rectangle (1.00, -1.50);
\filldraw[fill=cancan, draw=black] (1.00, -1.25) rectangle (1.25, -1.50);
\filldraw[fill=bermuda, draw=black] (1.25, -1.25) rectangle (1.50, -1.50);
\filldraw[fill=cancan, draw=black] (1.50, -1.25) rectangle (1.75, -1.50);
\filldraw[fill=cancan, draw=black] (1.75, -1.25) rectangle (2.00, -1.50);
\filldraw[fill=cancan, draw=black] (2.00, -1.25) rectangle (2.25, -1.50);
\filldraw[fill=bermuda, draw=black] (2.25, -1.25) rectangle (2.50, -1.50);
\filldraw[fill=cancan, draw=black] (2.50, -1.25) rectangle (2.75, -1.50);
\filldraw[fill=bermuda, draw=black] (2.75, -1.25) rectangle (3.00, -1.50);
\filldraw[fill=cancan, draw=black] (3.00, -1.25) rectangle (3.25, -1.50);
\filldraw[fill=bermuda, draw=black] (3.25, -1.25) rectangle (3.50, -1.50);
\filldraw[fill=bermuda, draw=black] (3.50, -1.25) rectangle (3.75, -1.50);
\filldraw[fill=bermuda, draw=black] (3.75, -1.25) rectangle (4.00, -1.50);
\filldraw[fill=cancan, draw=black] (4.00, -1.25) rectangle (4.25, -1.50);
\filldraw[fill=cancan, draw=black] (4.25, -1.25) rectangle (4.50, -1.50);
\filldraw[fill=cancan, draw=black] (4.50, -1.25) rectangle (4.75, -1.50);
\filldraw[fill=bermuda, draw=black] (4.75, -1.25) rectangle (5.00, -1.50);
\filldraw[fill=bermuda, draw=black] (5.00, -1.25) rectangle (5.25, -1.50);
\filldraw[fill=bermuda, draw=black] (5.25, -1.25) rectangle (5.50, -1.50);
\filldraw[fill=cancan, draw=black] (5.50, -1.25) rectangle (5.75, -1.50);
\filldraw[fill=bermuda, draw=black] (5.75, -1.25) rectangle (6.00, -1.50);
\filldraw[fill=cancan, draw=black] (6.00, -1.25) rectangle (6.25, -1.50);
\filldraw[fill=bermuda, draw=black] (6.25, -1.25) rectangle (6.50, -1.50);
\filldraw[fill=cancan, draw=black] (6.50, -1.25) rectangle (6.75, -1.50);
\filldraw[fill=cancan, draw=black] (6.75, -1.25) rectangle (7.00, -1.50);
\filldraw[fill=cancan, draw=black] (7.00, -1.25) rectangle (7.25, -1.50);
\filldraw[fill=bermuda, draw=black] (7.25, -1.25) rectangle (7.50, -1.50);
\filldraw[fill=cancan, draw=black] (7.50, -1.25) rectangle (7.75, -1.50);
\filldraw[fill=bermuda, draw=black] (7.75, -1.25) rectangle (8.00, -1.50);
\filldraw[fill=cancan, draw=black] (8.00, -1.25) rectangle (8.25, -1.50);
\filldraw[fill=bermuda, draw=black] (8.25, -1.25) rectangle (8.50, -1.50);
\filldraw[fill=cancan, draw=black] (8.50, -1.25) rectangle (8.75, -1.50);
\filldraw[fill=bermuda, draw=black] (8.75, -1.25) rectangle (9.00, -1.50);
\filldraw[fill=bermuda, draw=black] (9.00, -1.25) rectangle (9.25, -1.50);
\filldraw[fill=bermuda, draw=black] (9.25, -1.25) rectangle (9.50, -1.50);
\filldraw[fill=cancan, draw=black] (9.50, -1.25) rectangle (9.75, -1.50);
\filldraw[fill=cancan, draw=black] (9.75, -1.25) rectangle (10.00, -1.50);
\filldraw[fill=cancan, draw=black] (10.00, -1.25) rectangle (10.25, -1.50);
\filldraw[fill=bermuda, draw=black] (10.25, -1.25) rectangle (10.50, -1.50);
\filldraw[fill=bermuda, draw=black] (10.50, -1.25) rectangle (10.75, -1.50);
\filldraw[fill=bermuda, draw=black] (10.75, -1.25) rectangle (11.00, -1.50);
\filldraw[fill=cancan, draw=black] (11.00, -1.25) rectangle (11.25, -1.50);
\filldraw[fill=cancan, draw=black] (11.25, -1.25) rectangle (11.50, -1.50);
\filldraw[fill=cancan, draw=black] (11.50, -1.25) rectangle (11.75, -1.50);
\filldraw[fill=cancan, draw=black] (11.75, -1.25) rectangle (12.00, -1.50);
\filldraw[fill=cancan, draw=black] (12.00, -1.25) rectangle (12.25, -1.50);
\filldraw[fill=bermuda, draw=black] (12.25, -1.25) rectangle (12.50, -1.50);
\filldraw[fill=cancan, draw=black] (12.50, -1.25) rectangle (12.75, -1.50);
\filldraw[fill=bermuda, draw=black] (12.75, -1.25) rectangle (13.00, -1.50);
\filldraw[fill=cancan, draw=black] (13.00, -1.25) rectangle (13.25, -1.50);
\filldraw[fill=cancan, draw=black] (13.25, -1.25) rectangle (13.50, -1.50);
\filldraw[fill=cancan, draw=black] (13.50, -1.25) rectangle (13.75, -1.50);
\filldraw[fill=bermuda, draw=black] (13.75, -1.25) rectangle (14.00, -1.50);
\filldraw[fill=cancan, draw=black] (14.00, -1.25) rectangle (14.25, -1.50);
\filldraw[fill=bermuda, draw=black] (14.25, -1.25) rectangle (14.50, -1.50);
\filldraw[fill=cancan, draw=black] (14.50, -1.25) rectangle (14.75, -1.50);
\filldraw[fill=bermuda, draw=black] (14.75, -1.25) rectangle (15.00, -1.50);
\filldraw[fill=bermuda, draw=black] (0.00, -1.50) rectangle (0.25, -1.75);
\filldraw[fill=bermuda, draw=black] (0.25, -1.50) rectangle (0.50, -1.75);
\filldraw[fill=cancan, draw=black] (0.50, -1.50) rectangle (0.75, -1.75);
\filldraw[fill=cancan, draw=black] (0.75, -1.50) rectangle (1.00, -1.75);
\filldraw[fill=cancan, draw=black] (1.00, -1.50) rectangle (1.25, -1.75);
\filldraw[fill=bermuda, draw=black] (1.25, -1.50) rectangle (1.50, -1.75);
\filldraw[fill=bermuda, draw=black] (1.50, -1.50) rectangle (1.75, -1.75);
\filldraw[fill=bermuda, draw=black] (1.75, -1.50) rectangle (2.00, -1.75);
\filldraw[fill=cancan, draw=black] (2.00, -1.50) rectangle (2.25, -1.75);
\filldraw[fill=bermuda, draw=black] (2.25, -1.50) rectangle (2.50, -1.75);
\filldraw[fill=cancan, draw=black] (2.50, -1.50) rectangle (2.75, -1.75);
\filldraw[fill=bermuda, draw=black] (2.75, -1.50) rectangle (3.00, -1.75);
\filldraw[fill=cancan, draw=black] (3.00, -1.50) rectangle (3.25, -1.75);
\filldraw[fill=cancan, draw=black] (3.25, -1.50) rectangle (3.50, -1.75);
\filldraw[fill=cancan, draw=black] (3.50, -1.50) rectangle (3.75, -1.75);
\filldraw[fill=bermuda, draw=black] (3.75, -1.50) rectangle (4.00, -1.75);
\filldraw[fill=cancan, draw=black] (4.00, -1.50) rectangle (4.25, -1.75);
\filldraw[fill=bermuda, draw=black] (4.25, -1.50) rectangle (4.50, -1.75);
\filldraw[fill=cancan, draw=black] (4.50, -1.50) rectangle (4.75, -1.75);
\filldraw[fill=cancan, draw=black] (4.75, -1.50) rectangle (5.00, -1.75);
\filldraw[fill=cancan, draw=black] (5.00, -1.50) rectangle (5.25, -1.75);
\filldraw[fill=bermuda, draw=black] (5.25, -1.50) rectangle (5.50, -1.75);
\filldraw[fill=cancan, draw=black] (5.50, -1.50) rectangle (5.75, -1.75);
\filldraw[fill=bermuda, draw=black] (5.75, -1.50) rectangle (6.00, -1.75);
\filldraw[fill=cancan, draw=black] (6.00, -1.50) rectangle (6.25, -1.75);
\filldraw[fill=cancan, draw=black] (6.25, -1.50) rectangle (6.50, -1.75);
\filldraw[fill=cancan, draw=black] (6.50, -1.50) rectangle (6.75, -1.75);
\filldraw[fill=bermuda, draw=black] (6.75, -1.50) rectangle (7.00, -1.75);
\filldraw[fill=cancan, draw=black] (7.00, -1.50) rectangle (7.25, -1.75);
\filldraw[fill=bermuda, draw=black] (7.25, -1.50) rectangle (7.50, -1.75);
\filldraw[fill=cancan, draw=black] (7.50, -1.50) rectangle (7.75, -1.75);
\filldraw[fill=cancan, draw=black] (7.75, -1.50) rectangle (8.00, -1.75);
\filldraw[fill=cancan, draw=black] (8.00, -1.50) rectangle (8.25, -1.75);
\filldraw[fill=bermuda, draw=black] (8.25, -1.50) rectangle (8.50, -1.75);
\filldraw[fill=cancan, draw=black] (8.50, -1.50) rectangle (8.75, -1.75);
\filldraw[fill=bermuda, draw=black] (8.75, -1.50) rectangle (9.00, -1.75);
\filldraw[fill=bermuda, draw=black] (9.00, -1.50) rectangle (9.25, -1.75);
\filldraw[fill=bermuda, draw=black] (9.25, -1.50) rectangle (9.50, -1.75);
\filldraw[fill=cancan, draw=black] (9.50, -1.50) rectangle (9.75, -1.75);
\filldraw[fill=cancan, draw=black] (9.75, -1.50) rectangle (10.00, -1.75);
\filldraw[fill=cancan, draw=black] (10.00, -1.50) rectangle (10.25, -1.75);
\filldraw[fill=cancan, draw=black] (10.25, -1.50) rectangle (10.50, -1.75);
\filldraw[fill=bermuda, draw=black] (10.50, -1.50) rectangle (10.75, -1.75);
\filldraw[fill=bermuda, draw=black] (10.75, -1.50) rectangle (11.00, -1.75);
\filldraw[fill=cancan, draw=black] (11.00, -1.50) rectangle (11.25, -1.75);
\filldraw[fill=bermuda, draw=black] (11.25, -1.50) rectangle (11.50, -1.75);
\filldraw[fill=cancan, draw=black] (11.50, -1.50) rectangle (11.75, -1.75);
\filldraw[fill=cancan, draw=black] (11.75, -1.50) rectangle (12.00, -1.75);
\filldraw[fill=bermuda, draw=black] (12.00, -1.50) rectangle (12.25, -1.75);
\filldraw[fill=bermuda, draw=black] (12.25, -1.50) rectangle (12.50, -1.75);
\filldraw[fill=cancan, draw=black] (12.50, -1.50) rectangle (12.75, -1.75);
\filldraw[fill=bermuda, draw=black] (12.75, -1.50) rectangle (13.00, -1.75);
\filldraw[fill=cancan, draw=black] (13.00, -1.50) rectangle (13.25, -1.75);
\filldraw[fill=cancan, draw=black] (13.25, -1.50) rectangle (13.50, -1.75);
\filldraw[fill=bermuda, draw=black] (13.50, -1.50) rectangle (13.75, -1.75);
\filldraw[fill=bermuda, draw=black] (13.75, -1.50) rectangle (14.00, -1.75);
\filldraw[fill=cancan, draw=black] (14.00, -1.50) rectangle (14.25, -1.75);
\filldraw[fill=bermuda, draw=black] (14.25, -1.50) rectangle (14.50, -1.75);
\filldraw[fill=cancan, draw=black] (14.50, -1.50) rectangle (14.75, -1.75);
\filldraw[fill=bermuda, draw=black] (14.75, -1.50) rectangle (15.00, -1.75);
\filldraw[fill=bermuda, draw=black] (0.00, -1.75) rectangle (0.25, -2.00);
\filldraw[fill=bermuda, draw=black] (0.25, -1.75) rectangle (0.50, -2.00);
\filldraw[fill=cancan, draw=black] (0.50, -1.75) rectangle (0.75, -2.00);
\filldraw[fill=cancan, draw=black] (0.75, -1.75) rectangle (1.00, -2.00);
\filldraw[fill=bermuda, draw=black] (1.00, -1.75) rectangle (1.25, -2.00);
\filldraw[fill=bermuda, draw=black] (1.25, -1.75) rectangle (1.50, -2.00);
\filldraw[fill=cancan, draw=black] (1.50, -1.75) rectangle (1.75, -2.00);
\filldraw[fill=cancan, draw=black] (1.75, -1.75) rectangle (2.00, -2.00);
\filldraw[fill=cancan, draw=black] (2.00, -1.75) rectangle (2.25, -2.00);
\filldraw[fill=cancan, draw=black] (2.25, -1.75) rectangle (2.50, -2.00);
\filldraw[fill=cancan, draw=black] (2.50, -1.75) rectangle (2.75, -2.00);
\filldraw[fill=bermuda, draw=black] (2.75, -1.75) rectangle (3.00, -2.00);
\filldraw[fill=cancan, draw=black] (3.00, -1.75) rectangle (3.25, -2.00);
\filldraw[fill=bermuda, draw=black] (3.25, -1.75) rectangle (3.50, -2.00);
\filldraw[fill=bermuda, draw=black] (3.50, -1.75) rectangle (3.75, -2.00);
\filldraw[fill=bermuda, draw=black] (3.75, -1.75) rectangle (4.00, -2.00);
\filldraw[fill=cancan, draw=black] (4.00, -1.75) rectangle (4.25, -2.00);
\filldraw[fill=cancan, draw=black] (4.25, -1.75) rectangle (4.50, -2.00);
\filldraw[fill=cancan, draw=black] (4.50, -1.75) rectangle (4.75, -2.00);
\filldraw[fill=bermuda, draw=black] (4.75, -1.75) rectangle (5.00, -2.00);
\filldraw[fill=bermuda, draw=black] (5.00, -1.75) rectangle (5.25, -2.00);
\filldraw[fill=bermuda, draw=black] (5.25, -1.75) rectangle (5.50, -2.00);
\filldraw[fill=cancan, draw=black] (5.50, -1.75) rectangle (5.75, -2.00);
\filldraw[fill=bermuda, draw=black] (5.75, -1.75) rectangle (6.00, -2.00);
\filldraw[fill=bermuda, draw=black] (6.00, -1.75) rectangle (6.25, -2.00);
\filldraw[fill=bermuda, draw=black] (6.25, -1.75) rectangle (6.50, -2.00);
\filldraw[fill=bermuda, draw=black] (6.50, -1.75) rectangle (6.75, -2.00);
\filldraw[fill=bermuda, draw=black] (6.75, -1.75) rectangle (7.00, -2.00);
\filldraw[fill=cancan, draw=black] (7.00, -1.75) rectangle (7.25, -2.00);
\filldraw[fill=bermuda, draw=black] (7.25, -1.75) rectangle (7.50, -2.00);
\filldraw[fill=cancan, draw=black] (7.50, -1.75) rectangle (7.75, -2.00);
\filldraw[fill=bermuda, draw=black] (7.75, -1.75) rectangle (8.00, -2.00);
\filldraw[fill=bermuda, draw=black] (8.00, -1.75) rectangle (8.25, -2.00);
\filldraw[fill=bermuda, draw=black] (8.25, -1.75) rectangle (8.50, -2.00);
\filldraw[fill=cancan, draw=black] (8.50, -1.75) rectangle (8.75, -2.00);
\filldraw[fill=bermuda, draw=black] (8.75, -1.75) rectangle (9.00, -2.00);
\filldraw[fill=bermuda, draw=black] (9.00, -1.75) rectangle (9.25, -2.00);
\filldraw[fill=bermuda, draw=black] (9.25, -1.75) rectangle (9.50, -2.00);
\filldraw[fill=cancan, draw=black] (9.50, -1.75) rectangle (9.75, -2.00);
\filldraw[fill=cancan, draw=black] (9.75, -1.75) rectangle (10.00, -2.00);
\filldraw[fill=cancan, draw=black] (10.00, -1.75) rectangle (10.25, -2.00);
\filldraw[fill=bermuda, draw=black] (10.25, -1.75) rectangle (10.50, -2.00);
\filldraw[fill=bermuda, draw=black] (10.50, -1.75) rectangle (10.75, -2.00);
\filldraw[fill=bermuda, draw=black] (10.75, -1.75) rectangle (11.00, -2.00);
\filldraw[fill=cancan, draw=black] (11.00, -1.75) rectangle (11.25, -2.00);
\filldraw[fill=cancan, draw=black] (11.25, -1.75) rectangle (11.50, -2.00);
\filldraw[fill=cancan, draw=black] (11.50, -1.75) rectangle (11.75, -2.00);
\filldraw[fill=bermuda, draw=black] (11.75, -1.75) rectangle (12.00, -2.00);
\filldraw[fill=bermuda, draw=black] (12.00, -1.75) rectangle (12.25, -2.00);
\filldraw[fill=bermuda, draw=black] (12.25, -1.75) rectangle (12.50, -2.00);
\filldraw[fill=cancan, draw=black] (12.50, -1.75) rectangle (12.75, -2.00);
\filldraw[fill=cancan, draw=black] (12.75, -1.75) rectangle (13.00, -2.00);
\filldraw[fill=cancan, draw=black] (13.00, -1.75) rectangle (13.25, -2.00);
\filldraw[fill=bermuda, draw=black] (13.25, -1.75) rectangle (13.50, -2.00);
\filldraw[fill=bermuda, draw=black] (13.50, -1.75) rectangle (13.75, -2.00);
\filldraw[fill=bermuda, draw=black] (13.75, -1.75) rectangle (14.00, -2.00);
\filldraw[fill=cancan, draw=black] (14.00, -1.75) rectangle (14.25, -2.00);
\filldraw[fill=bermuda, draw=black] (14.25, -1.75) rectangle (14.50, -2.00);
\filldraw[fill=bermuda, draw=black] (14.50, -1.75) rectangle (14.75, -2.00);
\filldraw[fill=bermuda, draw=black] (14.75, -1.75) rectangle (15.00, -2.00);
\filldraw[fill=cancan, draw=black] (0.00, -2.00) rectangle (0.25, -2.25);
\filldraw[fill=bermuda, draw=black] (0.25, -2.00) rectangle (0.50, -2.25);
\filldraw[fill=bermuda, draw=black] (0.50, -2.00) rectangle (0.75, -2.25);
\filldraw[fill=bermuda, draw=black] (0.75, -2.00) rectangle (1.00, -2.25);
\filldraw[fill=bermuda, draw=black] (1.00, -2.00) rectangle (1.25, -2.25);
\filldraw[fill=bermuda, draw=black] (1.25, -2.00) rectangle (1.50, -2.25);
\filldraw[fill=cancan, draw=black] (1.50, -2.00) rectangle (1.75, -2.25);
\filldraw[fill=bermuda, draw=black] (1.75, -2.00) rectangle (2.00, -2.25);
\filldraw[fill=bermuda, draw=black] (2.00, -2.00) rectangle (2.25, -2.25);
\filldraw[fill=bermuda, draw=black] (2.25, -2.00) rectangle (2.50, -2.25);
\filldraw[fill=bermuda, draw=black] (2.50, -2.00) rectangle (2.75, -2.25);
\filldraw[fill=bermuda, draw=black] (2.75, -2.00) rectangle (3.00, -2.25);
\filldraw[fill=cancan, draw=black] (3.00, -2.00) rectangle (3.25, -2.25);
\filldraw[fill=cancan, draw=black] (3.25, -2.00) rectangle (3.50, -2.25);
\filldraw[fill=cancan, draw=black] (3.50, -2.00) rectangle (3.75, -2.25);
\filldraw[fill=cancan, draw=black] (3.75, -2.00) rectangle (4.00, -2.25);
\filldraw[fill=cancan, draw=black] (4.00, -2.00) rectangle (4.25, -2.25);
\filldraw[fill=cancan, draw=black] (4.25, -2.00) rectangle (4.50, -2.25);
\filldraw[fill=bermuda, draw=black] (4.50, -2.00) rectangle (4.75, -2.25);
\filldraw[fill=bermuda, draw=black] (4.75, -2.00) rectangle (5.00, -2.25);
\filldraw[fill=cancan, draw=black] (5.00, -2.00) rectangle (5.25, -2.25);
\filldraw[fill=bermuda, draw=black] (5.25, -2.00) rectangle (5.50, -2.25);
\filldraw[fill=cancan, draw=black] (5.50, -2.00) rectangle (5.75, -2.25);
\filldraw[fill=cancan, draw=black] (5.75, -2.00) rectangle (6.00, -2.25);
\filldraw[fill=bermuda, draw=black] (6.00, -2.00) rectangle (6.25, -2.25);
\filldraw[fill=bermuda, draw=black] (6.25, -2.00) rectangle (6.50, -2.25);
\filldraw[fill=cancan, draw=black] (6.50, -2.00) rectangle (6.75, -2.25);
\filldraw[fill=bermuda, draw=black] (6.75, -2.00) rectangle (7.00, -2.25);
\filldraw[fill=cancan, draw=black] (7.00, -2.00) rectangle (7.25, -2.25);
\filldraw[fill=cancan, draw=black] (7.25, -2.00) rectangle (7.50, -2.25);
\filldraw[fill=bermuda, draw=black] (7.50, -2.00) rectangle (7.75, -2.25);
\filldraw[fill=bermuda, draw=black] (7.75, -2.00) rectangle (8.00, -2.25);
\filldraw[fill=cancan, draw=black] (8.00, -2.00) rectangle (8.25, -2.25);
\filldraw[fill=bermuda, draw=black] (8.25, -2.00) rectangle (8.50, -2.25);
\filldraw[fill=cancan, draw=black] (8.50, -2.00) rectangle (8.75, -2.25);
\filldraw[fill=bermuda, draw=black] (8.75, -2.00) rectangle (9.00, -2.25);
\filldraw[fill=bermuda, draw=black] (9.00, -2.00) rectangle (9.25, -2.25);
\filldraw[fill=bermuda, draw=black] (9.25, -2.00) rectangle (9.50, -2.25);
\filldraw[fill=cancan, draw=black] (9.50, -2.00) rectangle (9.75, -2.25);
\filldraw[fill=cancan, draw=black] (9.75, -2.00) rectangle (10.00, -2.25);
\filldraw[fill=cancan, draw=black] (10.00, -2.00) rectangle (10.25, -2.25);
\filldraw[fill=cancan, draw=black] (10.25, -2.00) rectangle (10.50, -2.25);
\filldraw[fill=cancan, draw=black] (10.50, -2.00) rectangle (10.75, -2.25);
\filldraw[fill=cancan, draw=black] (10.75, -2.00) rectangle (11.00, -2.25);
\filldraw[fill=cancan, draw=black] (11.00, -2.00) rectangle (11.25, -2.25);
\filldraw[fill=cancan, draw=black] (11.25, -2.00) rectangle (11.50, -2.25);
\filldraw[fill=bermuda, draw=black] (11.50, -2.00) rectangle (11.75, -2.25);
\filldraw[fill=bermuda, draw=black] (11.75, -2.00) rectangle (12.00, -2.25);
\filldraw[fill=cancan, draw=black] (12.00, -2.00) rectangle (12.25, -2.25);
\filldraw[fill=bermuda, draw=black] (12.25, -2.00) rectangle (12.50, -2.25);
\filldraw[fill=bermuda, draw=black] (12.50, -2.00) rectangle (12.75, -2.25);
\filldraw[fill=bermuda, draw=black] (12.75, -2.00) rectangle (13.00, -2.25);
\filldraw[fill=cancan, draw=black] (13.00, -2.00) rectangle (13.25, -2.25);
\filldraw[fill=bermuda, draw=black] (13.25, -2.00) rectangle (13.50, -2.25);
\filldraw[fill=bermuda, draw=black] (13.50, -2.00) rectangle (13.75, -2.25);
\filldraw[fill=bermuda, draw=black] (13.75, -2.00) rectangle (14.00, -2.25);
\filldraw[fill=bermuda, draw=black] (14.00, -2.00) rectangle (14.25, -2.25);
\filldraw[fill=bermuda, draw=black] (14.25, -2.00) rectangle (14.50, -2.25);
\filldraw[fill=bermuda, draw=black] (14.50, -2.00) rectangle (14.75, -2.25);
\filldraw[fill=bermuda, draw=black] (14.75, -2.00) rectangle (15.00, -2.25);
\filldraw[fill=cancan, draw=black] (0.00, -2.25) rectangle (0.25, -2.50);
\filldraw[fill=bermuda, draw=black] (0.25, -2.25) rectangle (0.50, -2.50);
\filldraw[fill=cancan, draw=black] (0.50, -2.25) rectangle (0.75, -2.50);
\filldraw[fill=cancan, draw=black] (0.75, -2.25) rectangle (1.00, -2.50);
\filldraw[fill=bermuda, draw=black] (1.00, -2.25) rectangle (1.25, -2.50);
\filldraw[fill=bermuda, draw=black] (1.25, -2.25) rectangle (1.50, -2.50);
\filldraw[fill=cancan, draw=black] (1.50, -2.25) rectangle (1.75, -2.50);
\filldraw[fill=bermuda, draw=black] (1.75, -2.25) rectangle (2.00, -2.50);
\filldraw[fill=cancan, draw=black] (2.00, -2.25) rectangle (2.25, -2.50);
\filldraw[fill=bermuda, draw=black] (2.25, -2.25) rectangle (2.50, -2.50);
\filldraw[fill=bermuda, draw=black] (2.50, -2.25) rectangle (2.75, -2.50);
\filldraw[fill=bermuda, draw=black] (2.75, -2.25) rectangle (3.00, -2.50);
\filldraw[fill=cancan, draw=black] (3.00, -2.25) rectangle (3.25, -2.50);
\filldraw[fill=cancan, draw=black] (3.25, -2.25) rectangle (3.50, -2.50);
\filldraw[fill=cancan, draw=black] (3.50, -2.25) rectangle (3.75, -2.50);
\filldraw[fill=cancan, draw=black] (3.75, -2.25) rectangle (4.00, -2.50);
\filldraw[fill=cancan, draw=black] (4.00, -2.25) rectangle (4.25, -2.50);
\filldraw[fill=cancan, draw=black] (4.25, -2.25) rectangle (4.50, -2.50);
\filldraw[fill=cancan, draw=black] (4.50, -2.25) rectangle (4.75, -2.50);
\filldraw[fill=bermuda, draw=black] (4.75, -2.25) rectangle (5.00, -2.50);
\filldraw[fill=bermuda, draw=black] (5.00, -2.25) rectangle (5.25, -2.50);
\filldraw[fill=bermuda, draw=black] (5.25, -2.25) rectangle (5.50, -2.50);
\filldraw[fill=cancan, draw=black] (5.50, -2.25) rectangle (5.75, -2.50);
\filldraw[fill=cancan, draw=black] (5.75, -2.25) rectangle (6.00, -2.50);
\filldraw[fill=cancan, draw=black] (6.00, -2.25) rectangle (6.25, -2.50);
\filldraw[fill=bermuda, draw=black] (6.25, -2.25) rectangle (6.50, -2.50);
\filldraw[fill=bermuda, draw=black] (6.50, -2.25) rectangle (6.75, -2.50);
\filldraw[fill=bermuda, draw=black] (6.75, -2.25) rectangle (7.00, -2.50);
\filldraw[fill=bermuda, draw=black] (7.00, -2.25) rectangle (7.25, -2.50);
\filldraw[fill=bermuda, draw=black] (7.25, -2.25) rectangle (7.50, -2.50);
\filldraw[fill=bermuda, draw=black] (7.50, -2.25) rectangle (7.75, -2.50);
\filldraw[fill=bermuda, draw=black] (7.75, -2.25) rectangle (8.00, -2.50);
\filldraw[fill=cancan, draw=black] (8.00, -2.25) rectangle (8.25, -2.50);
\filldraw[fill=cancan, draw=black] (8.25, -2.25) rectangle (8.50, -2.50);
\filldraw[fill=cancan, draw=black] (8.50, -2.25) rectangle (8.75, -2.50);
\filldraw[fill=cancan, draw=black] (8.75, -2.25) rectangle (9.00, -2.50);
\filldraw[fill=cancan, draw=black] (9.00, -2.25) rectangle (9.25, -2.50);
\filldraw[fill=cancan, draw=black] (9.25, -2.25) rectangle (9.50, -2.50);
\filldraw[fill=cancan, draw=black] (9.50, -2.25) rectangle (9.75, -2.50);
\filldraw[fill=cancan, draw=black] (9.75, -2.25) rectangle (10.00, -2.50);
\filldraw[fill=cancan, draw=black] (10.00, -2.25) rectangle (10.25, -2.50);
\filldraw[fill=cancan, draw=black] (10.25, -2.25) rectangle (10.50, -2.50);
\filldraw[fill=cancan, draw=black] (10.50, -2.25) rectangle (10.75, -2.50);
\filldraw[fill=bermuda, draw=black] (10.75, -2.25) rectangle (11.00, -2.50);
\filldraw[fill=bermuda, draw=black] (11.00, -2.25) rectangle (11.25, -2.50);
\filldraw[fill=bermuda, draw=black] (11.25, -2.25) rectangle (11.50, -2.50);
\filldraw[fill=cancan, draw=black] (11.50, -2.25) rectangle (11.75, -2.50);
\filldraw[fill=bermuda, draw=black] (11.75, -2.25) rectangle (12.00, -2.50);
\filldraw[fill=bermuda, draw=black] (12.00, -2.25) rectangle (12.25, -2.50);
\filldraw[fill=bermuda, draw=black] (12.25, -2.25) rectangle (12.50, -2.50);
\filldraw[fill=bermuda, draw=black] (12.50, -2.25) rectangle (12.75, -2.50);
\filldraw[fill=bermuda, draw=black] (12.75, -2.25) rectangle (13.00, -2.50);
\filldraw[fill=cancan, draw=black] (13.00, -2.25) rectangle (13.25, -2.50);
\filldraw[fill=bermuda, draw=black] (13.25, -2.25) rectangle (13.50, -2.50);
\filldraw[fill=cancan, draw=black] (13.50, -2.25) rectangle (13.75, -2.50);
\filldraw[fill=cancan, draw=black] (13.75, -2.25) rectangle (14.00, -2.50);
\filldraw[fill=bermuda, draw=black] (14.00, -2.25) rectangle (14.25, -2.50);
\filldraw[fill=bermuda, draw=black] (14.25, -2.25) rectangle (14.50, -2.50);
\filldraw[fill=cancan, draw=black] (14.50, -2.25) rectangle (14.75, -2.50);
\filldraw[fill=bermuda, draw=black] (14.75, -2.25) rectangle (15.00, -2.50);
\filldraw[fill=cancan, draw=black] (0.00, -2.50) rectangle (0.25, -2.75);
\filldraw[fill=bermuda, draw=black] (0.25, -2.50) rectangle (0.50, -2.75);
\filldraw[fill=cancan, draw=black] (0.50, -2.50) rectangle (0.75, -2.75);
\filldraw[fill=cancan, draw=black] (0.75, -2.50) rectangle (1.00, -2.75);
\filldraw[fill=cancan, draw=black] (1.00, -2.50) rectangle (1.25, -2.75);
\filldraw[fill=bermuda, draw=black] (1.25, -2.50) rectangle (1.50, -2.75);
\filldraw[fill=cancan, draw=black] (1.50, -2.50) rectangle (1.75, -2.75);
\filldraw[fill=bermuda, draw=black] (1.75, -2.50) rectangle (2.00, -2.75);
\filldraw[fill=cancan, draw=black] (2.00, -2.50) rectangle (2.25, -2.75);
\filldraw[fill=cancan, draw=black] (2.25, -2.50) rectangle (2.50, -2.75);
\filldraw[fill=bermuda, draw=black] (2.50, -2.50) rectangle (2.75, -2.75);
\filldraw[fill=bermuda, draw=black] (2.75, -2.50) rectangle (3.00, -2.75);
\filldraw[fill=cancan, draw=black] (3.00, -2.50) rectangle (3.25, -2.75);
\filldraw[fill=bermuda, draw=black] (3.25, -2.50) rectangle (3.50, -2.75);
\filldraw[fill=bermuda, draw=black] (3.50, -2.50) rectangle (3.75, -2.75);
\filldraw[fill=bermuda, draw=black] (3.75, -2.50) rectangle (4.00, -2.75);
\filldraw[fill=cancan, draw=black] (4.00, -2.50) rectangle (4.25, -2.75);
\filldraw[fill=cancan, draw=black] (4.25, -2.50) rectangle (4.50, -2.75);
\filldraw[fill=cancan, draw=black] (4.50, -2.50) rectangle (4.75, -2.75);
\filldraw[fill=bermuda, draw=black] (4.75, -2.50) rectangle (5.00, -2.75);
\filldraw[fill=bermuda, draw=black] (5.00, -2.50) rectangle (5.25, -2.75);
\filldraw[fill=bermuda, draw=black] (5.25, -2.50) rectangle (5.50, -2.75);
\filldraw[fill=cancan, draw=black] (5.50, -2.50) rectangle (5.75, -2.75);
\filldraw[fill=bermuda, draw=black] (5.75, -2.50) rectangle (6.00, -2.75);
\filldraw[fill=cancan, draw=black] (6.00, -2.50) rectangle (6.25, -2.75);
\filldraw[fill=bermuda, draw=black] (6.25, -2.50) rectangle (6.50, -2.75);
\filldraw[fill=cancan, draw=black] (6.50, -2.50) rectangle (6.75, -2.75);
\filldraw[fill=cancan, draw=black] (6.75, -2.50) rectangle (7.00, -2.75);
\filldraw[fill=cancan, draw=black] (7.00, -2.50) rectangle (7.25, -2.75);
\filldraw[fill=bermuda, draw=black] (7.25, -2.50) rectangle (7.50, -2.75);
\filldraw[fill=cancan, draw=black] (7.50, -2.50) rectangle (7.75, -2.75);
\filldraw[fill=bermuda, draw=black] (7.75, -2.50) rectangle (8.00, -2.75);
\filldraw[fill=cancan, draw=black] (8.00, -2.50) rectangle (8.25, -2.75);
\filldraw[fill=cancan, draw=black] (8.25, -2.50) rectangle (8.50, -2.75);
\filldraw[fill=bermuda, draw=black] (8.50, -2.50) rectangle (8.75, -2.75);
\filldraw[fill=bermuda, draw=black] (8.75, -2.50) rectangle (9.00, -2.75);
\filldraw[fill=cancan, draw=black] (9.00, -2.50) rectangle (9.25, -2.75);
\filldraw[fill=cancan, draw=black] (9.25, -2.50) rectangle (9.50, -2.75);
\filldraw[fill=cancan, draw=black] (9.50, -2.50) rectangle (9.75, -2.75);
\filldraw[fill=bermuda, draw=black] (9.75, -2.50) rectangle (10.00, -2.75);
\filldraw[fill=bermuda, draw=black] (10.00, -2.50) rectangle (10.25, -2.75);
\filldraw[fill=bermuda, draw=black] (10.25, -2.50) rectangle (10.50, -2.75);
\filldraw[fill=cancan, draw=black] (10.50, -2.50) rectangle (10.75, -2.75);
\filldraw[fill=bermuda, draw=black] (10.75, -2.50) rectangle (11.00, -2.75);
\filldraw[fill=cancan, draw=black] (11.00, -2.50) rectangle (11.25, -2.75);
\filldraw[fill=cancan, draw=black] (11.25, -2.50) rectangle (11.50, -2.75);
\filldraw[fill=bermuda, draw=black] (11.50, -2.50) rectangle (11.75, -2.75);
\filldraw[fill=bermuda, draw=black] (11.75, -2.50) rectangle (12.00, -2.75);
\filldraw[fill=cancan, draw=black] (12.00, -2.50) rectangle (12.25, -2.75);
\filldraw[fill=bermuda, draw=black] (12.25, -2.50) rectangle (12.50, -2.75);
\filldraw[fill=cancan, draw=black] (12.50, -2.50) rectangle (12.75, -2.75);
\filldraw[fill=cancan, draw=black] (12.75, -2.50) rectangle (13.00, -2.75);
\filldraw[fill=bermuda, draw=black] (13.00, -2.50) rectangle (13.25, -2.75);
\filldraw[fill=bermuda, draw=black] (13.25, -2.50) rectangle (13.50, -2.75);
\filldraw[fill=cancan, draw=black] (13.50, -2.50) rectangle (13.75, -2.75);
\filldraw[fill=bermuda, draw=black] (13.75, -2.50) rectangle (14.00, -2.75);
\filldraw[fill=cancan, draw=black] (14.00, -2.50) rectangle (14.25, -2.75);
\filldraw[fill=cancan, draw=black] (14.25, -2.50) rectangle (14.50, -2.75);
\filldraw[fill=cancan, draw=black] (14.50, -2.50) rectangle (14.75, -2.75);
\filldraw[fill=cancan, draw=black] (14.75, -2.50) rectangle (15.00, -2.75);
\filldraw[fill=cancan, draw=black] (0.00, -2.75) rectangle (0.25, -3.00);
\filldraw[fill=bermuda, draw=black] (0.25, -2.75) rectangle (0.50, -3.00);
\filldraw[fill=bermuda, draw=black] (0.50, -2.75) rectangle (0.75, -3.00);
\filldraw[fill=bermuda, draw=black] (0.75, -2.75) rectangle (1.00, -3.00);
\filldraw[fill=cancan, draw=black] (1.00, -2.75) rectangle (1.25, -3.00);
\filldraw[fill=cancan, draw=black] (1.25, -2.75) rectangle (1.50, -3.00);
\filldraw[fill=cancan, draw=black] (1.50, -2.75) rectangle (1.75, -3.00);
\filldraw[fill=bermuda, draw=black] (1.75, -2.75) rectangle (2.00, -3.00);
\filldraw[fill=bermuda, draw=black] (2.00, -2.75) rectangle (2.25, -3.00);
\filldraw[fill=bermuda, draw=black] (2.25, -2.75) rectangle (2.50, -3.00);
\filldraw[fill=cancan, draw=black] (2.50, -2.75) rectangle (2.75, -3.00);
\filldraw[fill=bermuda, draw=black] (2.75, -2.75) rectangle (3.00, -3.00);
\filldraw[fill=cancan, draw=black] (3.00, -2.75) rectangle (3.25, -3.00);
\filldraw[fill=bermuda, draw=black] (3.25, -2.75) rectangle (3.50, -3.00);
\filldraw[fill=bermuda, draw=black] (3.50, -2.75) rectangle (3.75, -3.00);
\filldraw[fill=bermuda, draw=black] (3.75, -2.75) rectangle (4.00, -3.00);
\filldraw[fill=cancan, draw=black] (4.00, -2.75) rectangle (4.25, -3.00);
\filldraw[fill=cancan, draw=black] (4.25, -2.75) rectangle (4.50, -3.00);
\filldraw[fill=cancan, draw=black] (4.50, -2.75) rectangle (4.75, -3.00);
\filldraw[fill=bermuda, draw=black] (4.75, -2.75) rectangle (5.00, -3.00);
\filldraw[fill=bermuda, draw=black] (5.00, -2.75) rectangle (5.25, -3.00);
\filldraw[fill=bermuda, draw=black] (5.25, -2.75) rectangle (5.50, -3.00);
\filldraw[fill=bermuda, draw=black] (5.50, -2.75) rectangle (5.75, -3.00);
\filldraw[fill=bermuda, draw=black] (5.75, -2.75) rectangle (6.00, -3.00);
\filldraw[fill=cancan, draw=black] (6.00, -2.75) rectangle (6.25, -3.00);
\filldraw[fill=bermuda, draw=black] (6.25, -2.75) rectangle (6.50, -3.00);
\filldraw[fill=bermuda, draw=black] (6.50, -2.75) rectangle (6.75, -3.00);
\filldraw[fill=bermuda, draw=black] (6.75, -2.75) rectangle (7.00, -3.00);
\filldraw[fill=bermuda, draw=black] (7.00, -2.75) rectangle (7.25, -3.00);
\filldraw[fill=bermuda, draw=black] (7.25, -2.75) rectangle (7.50, -3.00);
\filldraw[fill=cancan, draw=black] (7.50, -2.75) rectangle (7.75, -3.00);
\filldraw[fill=bermuda, draw=black] (7.75, -2.75) rectangle (8.00, -3.00);
\filldraw[fill=bermuda, draw=black] (8.00, -2.75) rectangle (8.25, -3.00);
\filldraw[fill=bermuda, draw=black] (8.25, -2.75) rectangle (8.50, -3.00);
\filldraw[fill=bermuda, draw=black] (8.50, -2.75) rectangle (8.75, -3.00);
\filldraw[fill=bermuda, draw=black] (8.75, -2.75) rectangle (9.00, -3.00);
\filldraw[fill=bermuda, draw=black] (9.00, -2.75) rectangle (9.25, -3.00);
\filldraw[fill=bermuda, draw=black] (9.25, -2.75) rectangle (9.50, -3.00);
\filldraw[fill=cancan, draw=black] (9.50, -2.75) rectangle (9.75, -3.00);
\filldraw[fill=bermuda, draw=black] (9.75, -2.75) rectangle (10.00, -3.00);
\filldraw[fill=cancan, draw=black] (10.00, -2.75) rectangle (10.25, -3.00);
\filldraw[fill=cancan, draw=black] (10.25, -2.75) rectangle (10.50, -3.00);
\filldraw[fill=cancan, draw=black] (10.50, -2.75) rectangle (10.75, -3.00);
\filldraw[fill=cancan, draw=black] (10.75, -2.75) rectangle (11.00, -3.00);
\filldraw[fill=cancan, draw=black] (11.00, -2.75) rectangle (11.25, -3.00);
\filldraw[fill=bermuda, draw=black] (11.25, -2.75) rectangle (11.50, -3.00);
\filldraw[fill=bermuda, draw=black] (11.50, -2.75) rectangle (11.75, -3.00);
\filldraw[fill=bermuda, draw=black] (11.75, -2.75) rectangle (12.00, -3.00);
\filldraw[fill=cancan, draw=black] (12.00, -2.75) rectangle (12.25, -3.00);
\filldraw[fill=cancan, draw=black] (12.25, -2.75) rectangle (12.50, -3.00);
\filldraw[fill=cancan, draw=black] (12.50, -2.75) rectangle (12.75, -3.00);
\filldraw[fill=bermuda, draw=black] (12.75, -2.75) rectangle (13.00, -3.00);
\filldraw[fill=bermuda, draw=black] (13.00, -2.75) rectangle (13.25, -3.00);
\filldraw[fill=bermuda, draw=black] (13.25, -2.75) rectangle (13.50, -3.00);
\filldraw[fill=cancan, draw=black] (13.50, -2.75) rectangle (13.75, -3.00);
\filldraw[fill=cancan, draw=black] (13.75, -2.75) rectangle (14.00, -3.00);
\filldraw[fill=cancan, draw=black] (14.00, -2.75) rectangle (14.25, -3.00);
\filldraw[fill=bermuda, draw=black] (14.25, -2.75) rectangle (14.50, -3.00);
\filldraw[fill=bermuda, draw=black] (14.50, -2.75) rectangle (14.75, -3.00);
\filldraw[fill=bermuda, draw=black] (14.75, -2.75) rectangle (15.00, -3.00);
\filldraw[fill=cancan, draw=black] (0.00, -3.00) rectangle (0.25, -3.25);
\filldraw[fill=cancan, draw=black] (0.25, -3.00) rectangle (0.50, -3.25);
\filldraw[fill=cancan, draw=black] (0.50, -3.00) rectangle (0.75, -3.25);
\filldraw[fill=bermuda, draw=black] (0.75, -3.00) rectangle (1.00, -3.25);
\filldraw[fill=bermuda, draw=black] (1.00, -3.00) rectangle (1.25, -3.25);
\filldraw[fill=bermuda, draw=black] (1.25, -3.00) rectangle (1.50, -3.25);
\filldraw[fill=cancan, draw=black] (1.50, -3.00) rectangle (1.75, -3.25);
\filldraw[fill=cancan, draw=black] (1.75, -3.00) rectangle (2.00, -3.25);
\filldraw[fill=cancan, draw=black] (2.00, -3.00) rectangle (2.25, -3.25);
\filldraw[fill=bermuda, draw=black] (2.25, -3.00) rectangle (2.50, -3.25);
\filldraw[fill=cancan, draw=black] (2.50, -3.00) rectangle (2.75, -3.25);
\filldraw[fill=cancan, draw=black] (2.75, -3.00) rectangle (3.00, -3.25);
\filldraw[fill=cancan, draw=black] (3.00, -3.00) rectangle (3.25, -3.25);
\filldraw[fill=cancan, draw=black] (3.25, -3.00) rectangle (3.50, -3.25);
\filldraw[fill=cancan, draw=black] (3.50, -3.00) rectangle (3.75, -3.25);
\filldraw[fill=cancan, draw=black] (3.75, -3.00) rectangle (4.00, -3.25);
\filldraw[fill=cancan, draw=black] (4.00, -3.00) rectangle (4.25, -3.25);
\filldraw[fill=cancan, draw=black] (4.25, -3.00) rectangle (4.50, -3.25);
\filldraw[fill=cancan, draw=black] (4.50, -3.00) rectangle (4.75, -3.25);
\filldraw[fill=bermuda, draw=black] (4.75, -3.00) rectangle (5.00, -3.25);
\filldraw[fill=bermuda, draw=black] (5.00, -3.00) rectangle (5.25, -3.25);
\filldraw[fill=bermuda, draw=black] (5.25, -3.00) rectangle (5.50, -3.25);
\filldraw[fill=cancan, draw=black] (5.50, -3.00) rectangle (5.75, -3.25);
\filldraw[fill=bermuda, draw=black] (5.75, -3.00) rectangle (6.00, -3.25);
\filldraw[fill=cancan, draw=black] (6.00, -3.00) rectangle (6.25, -3.25);
\filldraw[fill=bermuda, draw=black] (6.25, -3.00) rectangle (6.50, -3.25);
\filldraw[fill=cancan, draw=black] (6.50, -3.00) rectangle (6.75, -3.25);
\filldraw[fill=bermuda, draw=black] (6.75, -3.00) rectangle (7.00, -3.25);
\filldraw[fill=cancan, draw=black] (7.00, -3.00) rectangle (7.25, -3.25);
\filldraw[fill=bermuda, draw=black] (7.25, -3.00) rectangle (7.50, -3.25);
\filldraw[fill=cancan, draw=black] (7.50, -3.00) rectangle (7.75, -3.25);
\filldraw[fill=cancan, draw=black] (7.75, -3.00) rectangle (8.00, -3.25);
\filldraw[fill=cancan, draw=black] (8.00, -3.00) rectangle (8.25, -3.25);
\filldraw[fill=bermuda, draw=black] (8.25, -3.00) rectangle (8.50, -3.25);
\filldraw[fill=bermuda, draw=black] (8.50, -3.00) rectangle (8.75, -3.25);
\filldraw[fill=bermuda, draw=black] (8.75, -3.00) rectangle (9.00, -3.25);
\filldraw[fill=cancan, draw=black] (9.00, -3.00) rectangle (9.25, -3.25);
\filldraw[fill=cancan, draw=black] (9.25, -3.00) rectangle (9.50, -3.25);
\filldraw[fill=cancan, draw=black] (9.50, -3.00) rectangle (9.75, -3.25);
\filldraw[fill=cancan, draw=black] (9.75, -3.00) rectangle (10.00, -3.25);
\filldraw[fill=cancan, draw=black] (10.00, -3.00) rectangle (10.25, -3.25);
\filldraw[fill=cancan, draw=black] (10.25, -3.00) rectangle (10.50, -3.25);
\filldraw[fill=cancan, draw=black] (10.50, -3.00) rectangle (10.75, -3.25);
\filldraw[fill=cancan, draw=black] (10.75, -3.00) rectangle (11.00, -3.25);
\filldraw[fill=cancan, draw=black] (11.00, -3.00) rectangle (11.25, -3.25);
\filldraw[fill=cancan, draw=black] (11.25, -3.00) rectangle (11.50, -3.25);
\filldraw[fill=bermuda, draw=black] (11.50, -3.00) rectangle (11.75, -3.25);
\filldraw[fill=bermuda, draw=black] (11.75, -3.00) rectangle (12.00, -3.25);
\filldraw[fill=cancan, draw=black] (12.00, -3.00) rectangle (12.25, -3.25);
\filldraw[fill=bermuda, draw=black] (12.25, -3.00) rectangle (12.50, -3.25);
\filldraw[fill=bermuda, draw=black] (12.50, -3.00) rectangle (12.75, -3.25);
\filldraw[fill=bermuda, draw=black] (12.75, -3.00) rectangle (13.00, -3.25);
\filldraw[fill=cancan, draw=black] (13.00, -3.00) rectangle (13.25, -3.25);
\filldraw[fill=cancan, draw=black] (13.25, -3.00) rectangle (13.50, -3.25);
\filldraw[fill=bermuda, draw=black] (13.50, -3.00) rectangle (13.75, -3.25);
\filldraw[fill=bermuda, draw=black] (13.75, -3.00) rectangle (14.00, -3.25);
\filldraw[fill=cancan, draw=black] (14.00, -3.00) rectangle (14.25, -3.25);
\filldraw[fill=bermuda, draw=black] (14.25, -3.00) rectangle (14.50, -3.25);
\filldraw[fill=bermuda, draw=black] (14.50, -3.00) rectangle (14.75, -3.25);
\filldraw[fill=bermuda, draw=black] (14.75, -3.00) rectangle (15.00, -3.25);
\filldraw[fill=bermuda, draw=black] (0.00, -3.25) rectangle (0.25, -3.50);
\filldraw[fill=bermuda, draw=black] (0.25, -3.25) rectangle (0.50, -3.50);
\filldraw[fill=cancan, draw=black] (0.50, -3.25) rectangle (0.75, -3.50);
\filldraw[fill=bermuda, draw=black] (0.75, -3.25) rectangle (1.00, -3.50);
\filldraw[fill=bermuda, draw=black] (1.00, -3.25) rectangle (1.25, -3.50);
\filldraw[fill=bermuda, draw=black] (1.25, -3.25) rectangle (1.50, -3.50);
\filldraw[fill=bermuda, draw=black] (1.50, -3.25) rectangle (1.75, -3.50);
\filldraw[fill=bermuda, draw=black] (1.75, -3.25) rectangle (2.00, -3.50);
\filldraw[fill=cancan, draw=black] (2.00, -3.25) rectangle (2.25, -3.50);
\filldraw[fill=bermuda, draw=black] (2.25, -3.25) rectangle (2.50, -3.50);
\filldraw[fill=bermuda, draw=black] (2.50, -3.25) rectangle (2.75, -3.50);
\filldraw[fill=bermuda, draw=black] (2.75, -3.25) rectangle (3.00, -3.50);
\filldraw[fill=bermuda, draw=black] (3.00, -3.25) rectangle (3.25, -3.50);
\filldraw[fill=bermuda, draw=black] (3.25, -3.25) rectangle (3.50, -3.50);
\filldraw[fill=cancan, draw=black] (3.50, -3.25) rectangle (3.75, -3.50);
\filldraw[fill=cancan, draw=black] (3.75, -3.25) rectangle (4.00, -3.50);
\filldraw[fill=cancan, draw=black] (4.00, -3.25) rectangle (4.25, -3.50);
\filldraw[fill=bermuda, draw=black] (4.25, -3.25) rectangle (4.50, -3.50);
\filldraw[fill=bermuda, draw=black] (4.50, -3.25) rectangle (4.75, -3.50);
\filldraw[fill=bermuda, draw=black] (4.75, -3.25) rectangle (5.00, -3.50);
\filldraw[fill=cancan, draw=black] (5.00, -3.25) rectangle (5.25, -3.50);
} } }\end{equation*}
\begin{equation*}
\hspace{4.6pt} b_{3} = \vcenter{\hbox{ \tikz{
\filldraw[fill=cancan, draw=black] (0.00, 0.00) rectangle (0.25, -0.25);
\filldraw[fill=bermuda, draw=black] (0.25, 0.00) rectangle (0.50, -0.25);
\filldraw[fill=bermuda, draw=black] (0.50, 0.00) rectangle (0.75, -0.25);
\filldraw[fill=cancan, draw=black] (0.75, 0.00) rectangle (1.00, -0.25);
\filldraw[fill=cancan, draw=black] (1.00, 0.00) rectangle (1.25, -0.25);
\filldraw[fill=cancan, draw=black] (1.25, 0.00) rectangle (1.50, -0.25);
\filldraw[fill=bermuda, draw=black] (1.50, 0.00) rectangle (1.75, -0.25);
\filldraw[fill=cancan, draw=black] (1.75, 0.00) rectangle (2.00, -0.25);
\filldraw[fill=bermuda, draw=black] (2.00, 0.00) rectangle (2.25, -0.25);
\filldraw[fill=cancan, draw=black] (2.25, 0.00) rectangle (2.50, -0.25);
\filldraw[fill=cancan, draw=black] (2.50, 0.00) rectangle (2.75, -0.25);
\filldraw[fill=bermuda, draw=black] (2.75, 0.00) rectangle (3.00, -0.25);
\filldraw[fill=bermuda, draw=black] (3.00, 0.00) rectangle (3.25, -0.25);
\filldraw[fill=cancan, draw=black] (3.25, 0.00) rectangle (3.50, -0.25);
\filldraw[fill=bermuda, draw=black] (3.50, 0.00) rectangle (3.75, -0.25);
\filldraw[fill=cancan, draw=black] (3.75, 0.00) rectangle (4.00, -0.25);
\filldraw[fill=bermuda, draw=black] (4.00, 0.00) rectangle (4.25, -0.25);
\filldraw[fill=bermuda, draw=black] (4.25, 0.00) rectangle (4.50, -0.25);
\filldraw[fill=bermuda, draw=black] (4.50, 0.00) rectangle (4.75, -0.25);
\filldraw[fill=cancan, draw=black] (4.75, 0.00) rectangle (5.00, -0.25);
\filldraw[fill=cancan, draw=black] (5.00, 0.00) rectangle (5.25, -0.25);
\filldraw[fill=cancan, draw=black] (5.25, 0.00) rectangle (5.50, -0.25);
\filldraw[fill=bermuda, draw=black] (5.50, 0.00) rectangle (5.75, -0.25);
\filldraw[fill=bermuda, draw=black] (5.75, 0.00) rectangle (6.00, -0.25);
\filldraw[fill=bermuda, draw=black] (6.00, 0.00) rectangle (6.25, -0.25);
\filldraw[fill=cancan, draw=black] (6.25, 0.00) rectangle (6.50, -0.25);
\filldraw[fill=cancan, draw=black] (6.50, 0.00) rectangle (6.75, -0.25);
\filldraw[fill=cancan, draw=black] (6.75, 0.00) rectangle (7.00, -0.25);
\filldraw[fill=bermuda, draw=black] (7.00, 0.00) rectangle (7.25, -0.25);
\filldraw[fill=bermuda, draw=black] (7.25, 0.00) rectangle (7.50, -0.25);
\filldraw[fill=bermuda, draw=black] (7.50, 0.00) rectangle (7.75, -0.25);
\filldraw[fill=cancan, draw=black] (7.75, 0.00) rectangle (8.00, -0.25);
\filldraw[fill=cancan, draw=black] (8.00, 0.00) rectangle (8.25, -0.25);
\filldraw[fill=cancan, draw=black] (8.25, 0.00) rectangle (8.50, -0.25);
\filldraw[fill=bermuda, draw=black] (8.50, 0.00) rectangle (8.75, -0.25);
\filldraw[fill=bermuda, draw=black] (8.75, 0.00) rectangle (9.00, -0.25);
\filldraw[fill=bermuda, draw=black] (9.00, 0.00) rectangle (9.25, -0.25);
\filldraw[fill=cancan, draw=black] (9.25, 0.00) rectangle (9.50, -0.25);
\filldraw[fill=cancan, draw=black] (9.50, 0.00) rectangle (9.75, -0.25);
\filldraw[fill=cancan, draw=black] (9.75, 0.00) rectangle (10.00, -0.25);
\filldraw[fill=bermuda, draw=black] (10.00, 0.00) rectangle (10.25, -0.25);
\filldraw[fill=bermuda, draw=black] (10.25, 0.00) rectangle (10.50, -0.25);
\filldraw[fill=bermuda, draw=black] (10.50, 0.00) rectangle (10.75, -0.25);
\filldraw[fill=cancan, draw=black] (10.75, 0.00) rectangle (11.00, -0.25);
\filldraw[fill=bermuda, draw=black] (11.00, 0.00) rectangle (11.25, -0.25);
\filldraw[fill=bermuda, draw=black] (11.25, 0.00) rectangle (11.50, -0.25);
\filldraw[fill=bermuda, draw=black] (11.50, 0.00) rectangle (11.75, -0.25);
\filldraw[fill=bermuda, draw=black] (11.75, 0.00) rectangle (12.00, -0.25);
\filldraw[fill=bermuda, draw=black] (12.00, 0.00) rectangle (12.25, -0.25);
\filldraw[fill=cancan, draw=black] (12.25, 0.00) rectangle (12.50, -0.25);
\filldraw[fill=bermuda, draw=black] (12.50, 0.00) rectangle (12.75, -0.25);
\filldraw[fill=cancan, draw=black] (12.75, 0.00) rectangle (13.00, -0.25);
\filldraw[fill=cancan, draw=black] (13.00, 0.00) rectangle (13.25, -0.25);
\filldraw[fill=bermuda, draw=black] (13.25, 0.00) rectangle (13.50, -0.25);
\filldraw[fill=bermuda, draw=black] (13.50, 0.00) rectangle (13.75, -0.25);
\filldraw[fill=cancan, draw=black] (13.75, 0.00) rectangle (14.00, -0.25);
\filldraw[fill=bermuda, draw=black] (14.00, 0.00) rectangle (14.25, -0.25);
\filldraw[fill=cancan, draw=black] (14.25, 0.00) rectangle (14.50, -0.25);
\filldraw[fill=bermuda, draw=black] (14.50, 0.00) rectangle (14.75, -0.25);
\filldraw[fill=cancan, draw=black] (14.75, 0.00) rectangle (15.00, -0.25);
\filldraw[fill=cancan, draw=black] (0.00, -0.25) rectangle (0.25, -0.50);
\filldraw[fill=cancan, draw=black] (0.25, -0.25) rectangle (0.50, -0.50);
\filldraw[fill=cancan, draw=black] (0.50, -0.25) rectangle (0.75, -0.50);
\filldraw[fill=cancan, draw=black] (0.75, -0.25) rectangle (1.00, -0.50);
\filldraw[fill=bermuda, draw=black] (1.00, -0.25) rectangle (1.25, -0.50);
\filldraw[fill=bermuda, draw=black] (1.25, -0.25) rectangle (1.50, -0.50);
\filldraw[fill=bermuda, draw=black] (1.50, -0.25) rectangle (1.75, -0.50);
\filldraw[fill=bermuda, draw=black] (1.75, -0.25) rectangle (2.00, -0.50);
\filldraw[fill=bermuda, draw=black] (2.00, -0.25) rectangle (2.25, -0.50);
\filldraw[fill=cancan, draw=black] (2.25, -0.25) rectangle (2.50, -0.50);
\filldraw[fill=bermuda, draw=black] (2.50, -0.25) rectangle (2.75, -0.50);
\filldraw[fill=cancan, draw=black] (2.75, -0.25) rectangle (3.00, -0.50);
\filldraw[fill=cancan, draw=black] (3.00, -0.25) rectangle (3.25, -0.50);
\filldraw[fill=bermuda, draw=black] (3.25, -0.25) rectangle (3.50, -0.50);
\filldraw[fill=bermuda, draw=black] (3.50, -0.25) rectangle (3.75, -0.50);
\filldraw[fill=cancan, draw=black] (3.75, -0.25) rectangle (4.00, -0.50);
\filldraw[fill=bermuda, draw=black] (4.00, -0.25) rectangle (4.25, -0.50);
\filldraw[fill=cancan, draw=black] (4.25, -0.25) rectangle (4.50, -0.50);
\filldraw[fill=cancan, draw=black] (4.50, -0.25) rectangle (4.75, -0.50);
\filldraw[fill=bermuda, draw=black] (4.75, -0.25) rectangle (5.00, -0.50);
\filldraw[fill=bermuda, draw=black] (5.00, -0.25) rectangle (5.25, -0.50);
\filldraw[fill=cancan, draw=black] (5.25, -0.25) rectangle (5.50, -0.50);
\filldraw[fill=bermuda, draw=black] (5.50, -0.25) rectangle (5.75, -0.50);
\filldraw[fill=cancan, draw=black] (5.75, -0.25) rectangle (6.00, -0.50);
\filldraw[fill=cancan, draw=black] (6.00, -0.25) rectangle (6.25, -0.50);
\filldraw[fill=bermuda, draw=black] (6.25, -0.25) rectangle (6.50, -0.50);
\filldraw[fill=bermuda, draw=black] (6.50, -0.25) rectangle (6.75, -0.50);
\filldraw[fill=cancan, draw=black] (6.75, -0.25) rectangle (7.00, -0.50);
\filldraw[fill=bermuda, draw=black] (7.00, -0.25) rectangle (7.25, -0.50);
\filldraw[fill=cancan, draw=black] (7.25, -0.25) rectangle (7.50, -0.50);
\filldraw[fill=cancan, draw=black] (7.50, -0.25) rectangle (7.75, -0.50);
\filldraw[fill=cancan, draw=black] (7.75, -0.25) rectangle (8.00, -0.50);
\filldraw[fill=bermuda, draw=black] (8.00, -0.25) rectangle (8.25, -0.50);
\filldraw[fill=cancan, draw=black] (8.25, -0.25) rectangle (8.50, -0.50);
\filldraw[fill=bermuda, draw=black] (8.50, -0.25) rectangle (8.75, -0.50);
\filldraw[fill=bermuda, draw=black] (8.75, -0.25) rectangle (9.00, -0.50);
\filldraw[fill=bermuda, draw=black] (9.00, -0.25) rectangle (9.25, -0.50);
\filldraw[fill=cancan, draw=black] (9.25, -0.25) rectangle (9.50, -0.50);
\filldraw[fill=bermuda, draw=black] (9.50, -0.25) rectangle (9.75, -0.50);
\filldraw[fill=bermuda, draw=black] (9.75, -0.25) rectangle (10.00, -0.50);
\filldraw[fill=bermuda, draw=black] (10.00, -0.25) rectangle (10.25, -0.50);
\filldraw[fill=bermuda, draw=black] (10.25, -0.25) rectangle (10.50, -0.50);
\filldraw[fill=bermuda, draw=black] (10.50, -0.25) rectangle (10.75, -0.50);
\filldraw[fill=cancan, draw=black] (10.75, -0.25) rectangle (11.00, -0.50);
\filldraw[fill=cancan, draw=black] (11.00, -0.25) rectangle (11.25, -0.50);
\filldraw[fill=cancan, draw=black] (11.25, -0.25) rectangle (11.50, -0.50);
\filldraw[fill=cancan, draw=black] (11.50, -0.25) rectangle (11.75, -0.50);
\filldraw[fill=cancan, draw=black] (11.75, -0.25) rectangle (12.00, -0.50);
\filldraw[fill=bermuda, draw=black] (12.00, -0.25) rectangle (12.25, -0.50);
\filldraw[fill=bermuda, draw=black] (12.25, -0.25) rectangle (12.50, -0.50);
\filldraw[fill=bermuda, draw=black] (12.50, -0.25) rectangle (12.75, -0.50);
\filldraw[fill=bermuda, draw=black] (12.75, -0.25) rectangle (13.00, -0.50);
\filldraw[fill=bermuda, draw=black] (13.00, -0.25) rectangle (13.25, -0.50);
\filldraw[fill=cancan, draw=black] (13.25, -0.25) rectangle (13.50, -0.50);
\filldraw[fill=bermuda, draw=black] (13.50, -0.25) rectangle (13.75, -0.50);
\filldraw[fill=bermuda, draw=black] (13.75, -0.25) rectangle (14.00, -0.50);
\filldraw[fill=bermuda, draw=black] (14.00, -0.25) rectangle (14.25, -0.50);
\filldraw[fill=cancan, draw=black] (14.25, -0.25) rectangle (14.50, -0.50);
\filldraw[fill=bermuda, draw=black] (14.50, -0.25) rectangle (14.75, -0.50);
\filldraw[fill=bermuda, draw=black] (14.75, -0.25) rectangle (15.00, -0.50);
\filldraw[fill=bermuda, draw=black] (0.00, -0.50) rectangle (0.25, -0.75);
\filldraw[fill=bermuda, draw=black] (0.25, -0.50) rectangle (0.50, -0.75);
\filldraw[fill=bermuda, draw=black] (0.50, -0.50) rectangle (0.75, -0.75);
\filldraw[fill=cancan, draw=black] (0.75, -0.50) rectangle (1.00, -0.75);
\filldraw[fill=bermuda, draw=black] (1.00, -0.50) rectangle (1.25, -0.75);
\filldraw[fill=bermuda, draw=black] (1.25, -0.50) rectangle (1.50, -0.75);
\filldraw[fill=bermuda, draw=black] (1.50, -0.50) rectangle (1.75, -0.75);
\filldraw[fill=bermuda, draw=black] (1.75, -0.50) rectangle (2.00, -0.75);
\filldraw[fill=bermuda, draw=black] (2.00, -0.50) rectangle (2.25, -0.75);
\filldraw[fill=cancan, draw=black] (2.25, -0.50) rectangle (2.50, -0.75);
\filldraw[fill=bermuda, draw=black] (2.50, -0.50) rectangle (2.75, -0.75);
\filldraw[fill=bermuda, draw=black] (2.75, -0.50) rectangle (3.00, -0.75);
\filldraw[fill=bermuda, draw=black] (3.00, -0.50) rectangle (3.25, -0.75);
\filldraw[fill=bermuda, draw=black] (3.25, -0.50) rectangle (3.50, -0.75);
\filldraw[fill=bermuda, draw=black] (3.50, -0.50) rectangle (3.75, -0.75);
\filldraw[fill=cancan, draw=black] (3.75, -0.50) rectangle (4.00, -0.75);
\filldraw[fill=bermuda, draw=black] (4.00, -0.50) rectangle (4.25, -0.75);
\filldraw[fill=bermuda, draw=black] (4.25, -0.50) rectangle (4.50, -0.75);
\filldraw[fill=bermuda, draw=black] (4.50, -0.50) rectangle (4.75, -0.75);
\filldraw[fill=bermuda, draw=black] (4.75, -0.50) rectangle (5.00, -0.75);
\filldraw[fill=bermuda, draw=black] (5.00, -0.50) rectangle (5.25, -0.75);
\filldraw[fill=cancan, draw=black] (5.25, -0.50) rectangle (5.50, -0.75);
\filldraw[fill=bermuda, draw=black] (5.50, -0.50) rectangle (5.75, -0.75);
\filldraw[fill=cancan, draw=black] (5.75, -0.50) rectangle (6.00, -0.75);
\filldraw[fill=cancan, draw=black] (6.00, -0.50) rectangle (6.25, -0.75);
\filldraw[fill=cancan, draw=black] (6.25, -0.50) rectangle (6.50, -0.75);
\filldraw[fill=bermuda, draw=black] (6.50, -0.50) rectangle (6.75, -0.75);
\filldraw[fill=bermuda, draw=black] (6.75, -0.50) rectangle (7.00, -0.75);
\filldraw[fill=bermuda, draw=black] (7.00, -0.50) rectangle (7.25, -0.75);
\filldraw[fill=cancan, draw=black] (7.25, -0.50) rectangle (7.50, -0.75);
\filldraw[fill=cancan, draw=black] (7.50, -0.50) rectangle (7.75, -0.75);
\filldraw[fill=bermuda, draw=black] (7.75, -0.50) rectangle (8.00, -0.75);
\filldraw[fill=bermuda, draw=black] (8.00, -0.50) rectangle (8.25, -0.75);
\filldraw[fill=cancan, draw=black] (8.25, -0.50) rectangle (8.50, -0.75);
\filldraw[fill=cancan, draw=black] (8.50, -0.50) rectangle (8.75, -0.75);
\filldraw[fill=cancan, draw=black] (8.75, -0.50) rectangle (9.00, -0.75);
\filldraw[fill=cancan, draw=black] (9.00, -0.50) rectangle (9.25, -0.75);
\filldraw[fill=cancan, draw=black] (9.25, -0.50) rectangle (9.50, -0.75);
\filldraw[fill=bermuda, draw=black] (9.50, -0.50) rectangle (9.75, -0.75);
\filldraw[fill=cancan, draw=black] (9.75, -0.50) rectangle (10.00, -0.75);
\filldraw[fill=cancan, draw=black] (10.00, -0.50) rectangle (10.25, -0.75);
\filldraw[fill=cancan, draw=black] (10.25, -0.50) rectangle (10.50, -0.75);
\filldraw[fill=bermuda, draw=black] (10.50, -0.50) rectangle (10.75, -0.75);
\filldraw[fill=bermuda, draw=black] (10.75, -0.50) rectangle (11.00, -0.75);
\filldraw[fill=bermuda, draw=black] (11.00, -0.50) rectangle (11.25, -0.75);
\filldraw[fill=cancan, draw=black] (11.25, -0.50) rectangle (11.50, -0.75);
\filldraw[fill=cancan, draw=black] (11.50, -0.50) rectangle (11.75, -0.75);
\filldraw[fill=cancan, draw=black] (11.75, -0.50) rectangle (12.00, -0.75);
\filldraw[fill=cancan, draw=black] (12.00, -0.50) rectangle (12.25, -0.75);
\filldraw[fill=cancan, draw=black] (12.25, -0.50) rectangle (12.50, -0.75);
\filldraw[fill=bermuda, draw=black] (12.50, -0.50) rectangle (12.75, -0.75);
\filldraw[fill=bermuda, draw=black] (12.75, -0.50) rectangle (13.00, -0.75);
\filldraw[fill=bermuda, draw=black] (13.00, -0.50) rectangle (13.25, -0.75);
\filldraw[fill=cancan, draw=black] (13.25, -0.50) rectangle (13.50, -0.75);
\filldraw[fill=cancan, draw=black] (13.50, -0.50) rectangle (13.75, -0.75);
\filldraw[fill=cancan, draw=black] (13.75, -0.50) rectangle (14.00, -0.75);
\filldraw[fill=bermuda, draw=black] (14.00, -0.50) rectangle (14.25, -0.75);
\filldraw[fill=bermuda, draw=black] (14.25, -0.50) rectangle (14.50, -0.75);
\filldraw[fill=bermuda, draw=black] (14.50, -0.50) rectangle (14.75, -0.75);
\filldraw[fill=cancan, draw=black] (14.75, -0.50) rectangle (15.00, -0.75);
\filldraw[fill=cancan, draw=black] (0.00, -0.75) rectangle (0.25, -1.00);
\filldraw[fill=cancan, draw=black] (0.25, -0.75) rectangle (0.50, -1.00);
\filldraw[fill=cancan, draw=black] (0.50, -0.75) rectangle (0.75, -1.00);
\filldraw[fill=cancan, draw=black] (0.75, -0.75) rectangle (1.00, -1.00);
\filldraw[fill=bermuda, draw=black] (1.00, -0.75) rectangle (1.25, -1.00);
\filldraw[fill=cancan, draw=black] (1.25, -0.75) rectangle (1.50, -1.00);
\filldraw[fill=cancan, draw=black] (1.50, -0.75) rectangle (1.75, -1.00);
\filldraw[fill=cancan, draw=black] (1.75, -0.75) rectangle (2.00, -1.00);
\filldraw[fill=bermuda, draw=black] (2.00, -0.75) rectangle (2.25, -1.00);
\filldraw[fill=bermuda, draw=black] (2.25, -0.75) rectangle (2.50, -1.00);
\filldraw[fill=bermuda, draw=black] (2.50, -0.75) rectangle (2.75, -1.00);
\filldraw[fill=bermuda, draw=black] (2.75, -0.75) rectangle (3.00, -1.00);
\filldraw[fill=bermuda, draw=black] (3.00, -0.75) rectangle (3.25, -1.00);
\filldraw[fill=cancan, draw=black] (3.25, -0.75) rectangle (3.50, -1.00);
\filldraw[fill=bermuda, draw=black] (3.50, -0.75) rectangle (3.75, -1.00);
\filldraw[fill=bermuda, draw=black] (3.75, -0.75) rectangle (4.00, -1.00);
\filldraw[fill=bermuda, draw=black] (4.00, -0.75) rectangle (4.25, -1.00);
\filldraw[fill=bermuda, draw=black] (4.25, -0.75) rectangle (4.50, -1.00);
\filldraw[fill=bermuda, draw=black] (4.50, -0.75) rectangle (4.75, -1.00);
\filldraw[fill=cancan, draw=black] (4.75, -0.75) rectangle (5.00, -1.00);
\filldraw[fill=bermuda, draw=black] (5.00, -0.75) rectangle (5.25, -1.00);
\filldraw[fill=bermuda, draw=black] (5.25, -0.75) rectangle (5.50, -1.00);
\filldraw[fill=bermuda, draw=black] (5.50, -0.75) rectangle (5.75, -1.00);
\filldraw[fill=bermuda, draw=black] (5.75, -0.75) rectangle (6.00, -1.00);
\filldraw[fill=bermuda, draw=black] (6.00, -0.75) rectangle (6.25, -1.00);
\filldraw[fill=cancan, draw=black] (6.25, -0.75) rectangle (6.50, -1.00);
\filldraw[fill=bermuda, draw=black] (6.50, -0.75) rectangle (6.75, -1.00);
\filldraw[fill=cancan, draw=black] (6.75, -0.75) rectangle (7.00, -1.00);
\filldraw[fill=cancan, draw=black] (7.00, -0.75) rectangle (7.25, -1.00);
\filldraw[fill=cancan, draw=black] (7.25, -0.75) rectangle (7.50, -1.00);
\filldraw[fill=cancan, draw=black] (7.50, -0.75) rectangle (7.75, -1.00);
\filldraw[fill=cancan, draw=black] (7.75, -0.75) rectangle (8.00, -1.00);
\filldraw[fill=cancan, draw=black] (8.00, -0.75) rectangle (8.25, -1.00);
\filldraw[fill=bermuda, draw=black] (8.25, -0.75) rectangle (8.50, -1.00);
\filldraw[fill=bermuda, draw=black] (8.50, -0.75) rectangle (8.75, -1.00);
\filldraw[fill=cancan, draw=black] (8.75, -0.75) rectangle (9.00, -1.00);
\filldraw[fill=bermuda, draw=black] (9.00, -0.75) rectangle (9.25, -1.00);
\filldraw[fill=cancan, draw=black] (9.25, -0.75) rectangle (9.50, -1.00);
\filldraw[fill=cancan, draw=black] (9.50, -0.75) rectangle (9.75, -1.00);
\filldraw[fill=bermuda, draw=black] (9.75, -0.75) rectangle (10.00, -1.00);
\filldraw[fill=bermuda, draw=black] (10.00, -0.75) rectangle (10.25, -1.00);
\filldraw[fill=cancan, draw=black] (10.25, -0.75) rectangle (10.50, -1.00);
\filldraw[fill=bermuda, draw=black] (10.50, -0.75) rectangle (10.75, -1.00);
\filldraw[fill=bermuda, draw=black] (10.75, -0.75) rectangle (11.00, -1.00);
\filldraw[fill=bermuda, draw=black] (11.00, -0.75) rectangle (11.25, -1.00);
\filldraw[fill=cancan, draw=black] (11.25, -0.75) rectangle (11.50, -1.00);
\filldraw[fill=cancan, draw=black] (11.50, -0.75) rectangle (11.75, -1.00);
\filldraw[fill=cancan, draw=black] (11.75, -0.75) rectangle (12.00, -1.00);
\filldraw[fill=bermuda, draw=black] (12.00, -0.75) rectangle (12.25, -1.00);
\filldraw[fill=bermuda, draw=black] (12.25, -0.75) rectangle (12.50, -1.00);
\filldraw[fill=bermuda, draw=black] (12.50, -0.75) rectangle (12.75, -1.00);
\filldraw[fill=cancan, draw=black] (12.75, -0.75) rectangle (13.00, -1.00);
\filldraw[fill=cancan, draw=black] (13.00, -0.75) rectangle (13.25, -1.00);
\filldraw[fill=cancan, draw=black] (13.25, -0.75) rectangle (13.50, -1.00);
\filldraw[fill=bermuda, draw=black] (13.50, -0.75) rectangle (13.75, -1.00);
\filldraw[fill=bermuda, draw=black] (13.75, -0.75) rectangle (14.00, -1.00);
\filldraw[fill=bermuda, draw=black] (14.00, -0.75) rectangle (14.25, -1.00);
\filldraw[fill=bermuda, draw=black] (14.25, -0.75) rectangle (14.50, -1.00);
\filldraw[fill=bermuda, draw=black] (14.50, -0.75) rectangle (14.75, -1.00);
\filldraw[fill=bermuda, draw=black] (14.75, -0.75) rectangle (15.00, -1.00);
\filldraw[fill=bermuda, draw=black] (0.00, -1.00) rectangle (0.25, -1.25);
\filldraw[fill=cancan, draw=black] (0.25, -1.00) rectangle (0.50, -1.25);
\filldraw[fill=bermuda, draw=black] (0.50, -1.00) rectangle (0.75, -1.25);
\filldraw[fill=cancan, draw=black] (0.75, -1.00) rectangle (1.00, -1.25);
\filldraw[fill=bermuda, draw=black] (1.00, -1.00) rectangle (1.25, -1.25);
\filldraw[fill=bermuda, draw=black] (1.25, -1.00) rectangle (1.50, -1.25);
\filldraw[fill=bermuda, draw=black] (1.50, -1.00) rectangle (1.75, -1.25);
\filldraw[fill=cancan, draw=black] (1.75, -1.00) rectangle (2.00, -1.25);
\filldraw[fill=bermuda, draw=black] (2.00, -1.00) rectangle (2.25, -1.25);
\filldraw[fill=bermuda, draw=black] (2.25, -1.00) rectangle (2.50, -1.25);
\filldraw[fill=bermuda, draw=black] (2.50, -1.00) rectangle (2.75, -1.25);
\filldraw[fill=bermuda, draw=black] (2.75, -1.00) rectangle (3.00, -1.25);
\filldraw[fill=bermuda, draw=black] (3.00, -1.00) rectangle (3.25, -1.25);
\filldraw[fill=cancan, draw=black] (3.25, -1.00) rectangle (3.50, -1.25);
\filldraw[fill=bermuda, draw=black] (3.50, -1.00) rectangle (3.75, -1.25);
\filldraw[fill=cancan, draw=black] (3.75, -1.00) rectangle (4.00, -1.25);
\filldraw[fill=cancan, draw=black] (4.00, -1.00) rectangle (4.25, -1.25);
\filldraw[fill=cancan, draw=black] (4.25, -1.00) rectangle (4.50, -1.25);
\filldraw[fill=cancan, draw=black] (4.50, -1.00) rectangle (4.75, -1.25);
\filldraw[fill=cancan, draw=black] (4.75, -1.00) rectangle (5.00, -1.25);
\filldraw[fill=bermuda, draw=black] (5.00, -1.00) rectangle (5.25, -1.25);
\filldraw[fill=bermuda, draw=black] (5.25, -1.00) rectangle (5.50, -1.25);
\filldraw[fill=bermuda, draw=black] (5.50, -1.00) rectangle (5.75, -1.25);
\filldraw[fill=cancan, draw=black] (5.75, -1.00) rectangle (6.00, -1.25);
\filldraw[fill=cancan, draw=black] (6.00, -1.00) rectangle (6.25, -1.25);
\filldraw[fill=cancan, draw=black] (6.25, -1.00) rectangle (6.50, -1.25);
\filldraw[fill=bermuda, draw=black] (6.50, -1.00) rectangle (6.75, -1.25);
\filldraw[fill=bermuda, draw=black] (6.75, -1.00) rectangle (7.00, -1.25);
\filldraw[fill=bermuda, draw=black] (7.00, -1.00) rectangle (7.25, -1.25);
\filldraw[fill=cancan, draw=black] (7.25, -1.00) rectangle (7.50, -1.25);
\filldraw[fill=bermuda, draw=black] (7.50, -1.00) rectangle (7.75, -1.25);
\filldraw[fill=bermuda, draw=black] (7.75, -1.00) rectangle (8.00, -1.25);
\filldraw[fill=bermuda, draw=black] (8.00, -1.00) rectangle (8.25, -1.25);
\filldraw[fill=cancan, draw=black] (8.25, -1.00) rectangle (8.50, -1.25);
\filldraw[fill=bermuda, draw=black] (8.50, -1.00) rectangle (8.75, -1.25);
\filldraw[fill=cancan, draw=black] (8.75, -1.00) rectangle (9.00, -1.25);
\filldraw[fill=cancan, draw=black] (9.00, -1.00) rectangle (9.25, -1.25);
\filldraw[fill=bermuda, draw=black] (9.25, -1.00) rectangle (9.50, -1.25);
\filldraw[fill=bermuda, draw=black] (9.50, -1.00) rectangle (9.75, -1.25);
\filldraw[fill=cancan, draw=black] (9.75, -1.00) rectangle (10.00, -1.25);
\filldraw[fill=bermuda, draw=black] (10.00, -1.00) rectangle (10.25, -1.25);
\filldraw[fill=cancan, draw=black] (10.25, -1.00) rectangle (10.50, -1.25);
\filldraw[fill=cancan, draw=black] (10.50, -1.00) rectangle (10.75, -1.25);
\filldraw[fill=cancan, draw=black] (10.75, -1.00) rectangle (11.00, -1.25);
\filldraw[fill=bermuda, draw=black] (11.00, -1.00) rectangle (11.25, -1.25);
\filldraw[fill=cancan, draw=black] (11.25, -1.00) rectangle (11.50, -1.25);
\filldraw[fill=cancan, draw=black] (11.50, -1.00) rectangle (11.75, -1.25);
\filldraw[fill=cancan, draw=black] (11.75, -1.00) rectangle (12.00, -1.25);
\filldraw[fill=cancan, draw=black] (12.00, -1.00) rectangle (12.25, -1.25);
\filldraw[fill=cancan, draw=black] (12.25, -1.00) rectangle (12.50, -1.25);
\filldraw[fill=cancan, draw=black] (12.50, -1.00) rectangle (12.75, -1.25);
\filldraw[fill=cancan, draw=black] (12.75, -1.00) rectangle (13.00, -1.25);
\filldraw[fill=bermuda, draw=black] (13.00, -1.00) rectangle (13.25, -1.25);
\filldraw[fill=bermuda, draw=black] (13.25, -1.00) rectangle (13.50, -1.25);
\filldraw[fill=bermuda, draw=black] (13.50, -1.00) rectangle (13.75, -1.25);
\filldraw[fill=cancan, draw=black] (13.75, -1.00) rectangle (14.00, -1.25);
\filldraw[fill=cancan, draw=black] (14.00, -1.00) rectangle (14.25, -1.25);
\filldraw[fill=cancan, draw=black] (14.25, -1.00) rectangle (14.50, -1.25);
\filldraw[fill=cancan, draw=black] (14.50, -1.00) rectangle (14.75, -1.25);
\filldraw[fill=cancan, draw=black] (14.75, -1.00) rectangle (15.00, -1.25);
\filldraw[fill=cancan, draw=black] (0.00, -1.25) rectangle (0.25, -1.50);
\filldraw[fill=cancan, draw=black] (0.25, -1.25) rectangle (0.50, -1.50);
\filldraw[fill=bermuda, draw=black] (0.50, -1.25) rectangle (0.75, -1.50);
\filldraw[fill=cancan, draw=black] (0.75, -1.25) rectangle (1.00, -1.50);
\filldraw[fill=bermuda, draw=black] (1.00, -1.25) rectangle (1.25, -1.50);
\filldraw[fill=cancan, draw=black] (1.25, -1.25) rectangle (1.50, -1.50);
\filldraw[fill=cancan, draw=black] (1.50, -1.25) rectangle (1.75, -1.50);
\filldraw[fill=bermuda, draw=black] (1.75, -1.25) rectangle (2.00, -1.50);
\filldraw[fill=bermuda, draw=black] (2.00, -1.25) rectangle (2.25, -1.50);
\filldraw[fill=bermuda, draw=black] (2.25, -1.25) rectangle (2.50, -1.50);
\filldraw[fill=bermuda, draw=black] (2.50, -1.25) rectangle (2.75, -1.50);
\filldraw[fill=bermuda, draw=black] (2.75, -1.25) rectangle (3.00, -1.50);
\filldraw[fill=bermuda, draw=black] (3.00, -1.25) rectangle (3.25, -1.50);
\filldraw[fill=cancan, draw=black] (3.25, -1.25) rectangle (3.50, -1.50);
\filldraw[fill=bermuda, draw=black] (3.50, -1.25) rectangle (3.75, -1.50);
\filldraw[fill=cancan, draw=black] (3.75, -1.25) rectangle (4.00, -1.50);
\filldraw[fill=cancan, draw=black] (4.00, -1.25) rectangle (4.25, -1.50);
\filldraw[fill=bermuda, draw=black] (4.25, -1.25) rectangle (4.50, -1.50);
\filldraw[fill=bermuda, draw=black] (4.50, -1.25) rectangle (4.75, -1.50);
\filldraw[fill=cancan, draw=black] (4.75, -1.25) rectangle (5.00, -1.50);
\filldraw[fill=bermuda, draw=black] (5.00, -1.25) rectangle (5.25, -1.50);
\filldraw[fill=cancan, draw=black] (5.25, -1.25) rectangle (5.50, -1.50);
\filldraw[fill=bermuda, draw=black] (5.50, -1.25) rectangle (5.75, -1.50);
\filldraw[fill=cancan, draw=black] (5.75, -1.25) rectangle (6.00, -1.50);
\filldraw[fill=bermuda, draw=black] (6.00, -1.25) rectangle (6.25, -1.50);
\filldraw[fill=cancan, draw=black] (6.25, -1.25) rectangle (6.50, -1.50);
\filldraw[fill=cancan, draw=black] (6.50, -1.25) rectangle (6.75, -1.50);
\filldraw[fill=bermuda, draw=black] (6.75, -1.25) rectangle (7.00, -1.50);
\filldraw[fill=bermuda, draw=black] (7.00, -1.25) rectangle (7.25, -1.50);
\filldraw[fill=bermuda, draw=black] (7.25, -1.25) rectangle (7.50, -1.50);
\filldraw[fill=bermuda, draw=black] (7.50, -1.25) rectangle (7.75, -1.50);
\filldraw[fill=bermuda, draw=black] (7.75, -1.25) rectangle (8.00, -1.50);
\filldraw[fill=bermuda, draw=black] (8.00, -1.25) rectangle (8.25, -1.50);
\filldraw[fill=bermuda, draw=black] (8.25, -1.25) rectangle (8.50, -1.50);
\filldraw[fill=bermuda, draw=black] (8.50, -1.25) rectangle (8.75, -1.50);
\filldraw[fill=cancan, draw=black] (8.75, -1.25) rectangle (9.00, -1.50);
\filldraw[fill=bermuda, draw=black] (9.00, -1.25) rectangle (9.25, -1.50);
\filldraw[fill=cancan, draw=black] (9.25, -1.25) rectangle (9.50, -1.50);
\filldraw[fill=cancan, draw=black] (9.50, -1.25) rectangle (9.75, -1.50);
\filldraw[fill=bermuda, draw=black] (9.75, -1.25) rectangle (10.00, -1.50);
\filldraw[fill=bermuda, draw=black] (10.00, -1.25) rectangle (10.25, -1.50);
\filldraw[fill=cancan, draw=black] (10.25, -1.25) rectangle (10.50, -1.50);
\filldraw[fill=bermuda, draw=black] (10.50, -1.25) rectangle (10.75, -1.50);
\filldraw[fill=cancan, draw=black] (10.75, -1.25) rectangle (11.00, -1.50);
\filldraw[fill=cancan, draw=black] (11.00, -1.25) rectangle (11.25, -1.50);
\filldraw[fill=cancan, draw=black] (11.25, -1.25) rectangle (11.50, -1.50);
\filldraw[fill=cancan, draw=black] (11.50, -1.25) rectangle (11.75, -1.50);
\filldraw[fill=cancan, draw=black] (11.75, -1.25) rectangle (12.00, -1.50);
\filldraw[fill=cancan, draw=black] (12.00, -1.25) rectangle (12.25, -1.50);
\filldraw[fill=cancan, draw=black] (12.25, -1.25) rectangle (12.50, -1.50);
\filldraw[fill=cancan, draw=black] (12.50, -1.25) rectangle (12.75, -1.50);
\filldraw[fill=cancan, draw=black] (12.75, -1.25) rectangle (13.00, -1.50);
\filldraw[fill=bermuda, draw=black] (13.00, -1.25) rectangle (13.25, -1.50);
\filldraw[fill=cancan, draw=black] (13.25, -1.25) rectangle (13.50, -1.50);
\filldraw[fill=bermuda, draw=black] (13.50, -1.25) rectangle (13.75, -1.50);
\filldraw[fill=cancan, draw=black] (13.75, -1.25) rectangle (14.00, -1.50);
\filldraw[fill=bermuda, draw=black] (14.00, -1.25) rectangle (14.25, -1.50);
\filldraw[fill=bermuda, draw=black] (14.25, -1.25) rectangle (14.50, -1.50);
\filldraw[fill=bermuda, draw=black] (14.50, -1.25) rectangle (14.75, -1.50);
\filldraw[fill=cancan, draw=black] (14.75, -1.25) rectangle (15.00, -1.50);
\filldraw[fill=cancan, draw=black] (0.00, -1.50) rectangle (0.25, -1.75);
\filldraw[fill=cancan, draw=black] (0.25, -1.50) rectangle (0.50, -1.75);
\filldraw[fill=bermuda, draw=black] (0.50, -1.50) rectangle (0.75, -1.75);
\filldraw[fill=bermuda, draw=black] (0.75, -1.50) rectangle (1.00, -1.75);
\filldraw[fill=bermuda, draw=black] (1.00, -1.50) rectangle (1.25, -1.75);
\filldraw[fill=cancan, draw=black] (1.25, -1.50) rectangle (1.50, -1.75);
\filldraw[fill=cancan, draw=black] (1.50, -1.50) rectangle (1.75, -1.75);
\filldraw[fill=cancan, draw=black] (1.75, -1.50) rectangle (2.00, -1.75);
\filldraw[fill=cancan, draw=black] (2.00, -1.50) rectangle (2.25, -1.75);
\filldraw[fill=cancan, draw=black] (2.25, -1.50) rectangle (2.50, -1.75);
\filldraw[fill=cancan, draw=black] (2.50, -1.50) rectangle (2.75, -1.75);
\filldraw[fill=cancan, draw=black] (2.75, -1.50) rectangle (3.00, -1.75);
\filldraw[fill=bermuda, draw=black] (3.00, -1.50) rectangle (3.25, -1.75);
\filldraw[fill=cancan, draw=black] (3.25, -1.50) rectangle (3.50, -1.75);
\filldraw[fill=bermuda, draw=black] (3.50, -1.50) rectangle (3.75, -1.75);
\filldraw[fill=cancan, draw=black] (3.75, -1.50) rectangle (4.00, -1.75);
\filldraw[fill=cancan, draw=black] (4.00, -1.50) rectangle (4.25, -1.75);
\filldraw[fill=bermuda, draw=black] (4.25, -1.50) rectangle (4.50, -1.75);
\filldraw[fill=bermuda, draw=black] (4.50, -1.50) rectangle (4.75, -1.75);
\filldraw[fill=bermuda, draw=black] (4.75, -1.50) rectangle (5.00, -1.75);
\filldraw[fill=bermuda, draw=black] (5.00, -1.50) rectangle (5.25, -1.75);
\filldraw[fill=cancan, draw=black] (5.25, -1.50) rectangle (5.50, -1.75);
\filldraw[fill=cancan, draw=black] (5.50, -1.50) rectangle (5.75, -1.75);
\filldraw[fill=cancan, draw=black] (5.75, -1.50) rectangle (6.00, -1.75);
\filldraw[fill=cancan, draw=black] (6.00, -1.50) rectangle (6.25, -1.75);
\filldraw[fill=cancan, draw=black] (6.25, -1.50) rectangle (6.50, -1.75);
\filldraw[fill=cancan, draw=black] (6.50, -1.50) rectangle (6.75, -1.75);
\filldraw[fill=cancan, draw=black] (6.75, -1.50) rectangle (7.00, -1.75);
\filldraw[fill=bermuda, draw=black] (7.00, -1.50) rectangle (7.25, -1.75);
\filldraw[fill=cancan, draw=black] (7.25, -1.50) rectangle (7.50, -1.75);
\filldraw[fill=bermuda, draw=black] (7.50, -1.50) rectangle (7.75, -1.75);
\filldraw[fill=bermuda, draw=black] (7.75, -1.50) rectangle (8.00, -1.75);
\filldraw[fill=bermuda, draw=black] (8.00, -1.50) rectangle (8.25, -1.75);
\filldraw[fill=cancan, draw=black] (8.25, -1.50) rectangle (8.50, -1.75);
\filldraw[fill=cancan, draw=black] (8.50, -1.50) rectangle (8.75, -1.75);
\filldraw[fill=cancan, draw=black] (8.75, -1.50) rectangle (9.00, -1.75);
\filldraw[fill=cancan, draw=black] (9.00, -1.50) rectangle (9.25, -1.75);
\filldraw[fill=cancan, draw=black] (9.25, -1.50) rectangle (9.50, -1.75);
\filldraw[fill=bermuda, draw=black] (9.50, -1.50) rectangle (9.75, -1.75);
\filldraw[fill=cancan, draw=black] (9.75, -1.50) rectangle (10.00, -1.75);
\filldraw[fill=bermuda, draw=black] (10.00, -1.50) rectangle (10.25, -1.75);
\filldraw[fill=cancan, draw=black] (10.25, -1.50) rectangle (10.50, -1.75);
\filldraw[fill=cancan, draw=black] (10.50, -1.50) rectangle (10.75, -1.75);
\filldraw[fill=cancan, draw=black] (10.75, -1.50) rectangle (11.00, -1.75);
\filldraw[fill=bermuda, draw=black] (11.00, -1.50) rectangle (11.25, -1.75);
\filldraw[fill=cancan, draw=black] (11.25, -1.50) rectangle (11.50, -1.75);
\filldraw[fill=bermuda, draw=black] (11.50, -1.50) rectangle (11.75, -1.75);
\filldraw[fill=cancan, draw=black] (11.75, -1.50) rectangle (12.00, -1.75);
\filldraw[fill=cancan, draw=black] (12.00, -1.50) rectangle (12.25, -1.75);
\filldraw[fill=cancan, draw=black] (12.25, -1.50) rectangle (12.50, -1.75);
\filldraw[fill=bermuda, draw=black] (12.50, -1.50) rectangle (12.75, -1.75);
\filldraw[fill=cancan, draw=black] (12.75, -1.50) rectangle (13.00, -1.75);
\filldraw[fill=bermuda, draw=black] (13.00, -1.50) rectangle (13.25, -1.75);
\filldraw[fill=cancan, draw=black] (13.25, -1.50) rectangle (13.50, -1.75);
\filldraw[fill=bermuda, draw=black] (13.50, -1.50) rectangle (13.75, -1.75);
\filldraw[fill=bermuda, draw=black] (13.75, -1.50) rectangle (14.00, -1.75);
\filldraw[fill=bermuda, draw=black] (14.00, -1.50) rectangle (14.25, -1.75);
\filldraw[fill=cancan, draw=black] (14.25, -1.50) rectangle (14.50, -1.75);
\filldraw[fill=cancan, draw=black] (14.50, -1.50) rectangle (14.75, -1.75);
\filldraw[fill=bermuda, draw=black] (14.75, -1.50) rectangle (15.00, -1.75);
\filldraw[fill=bermuda, draw=black] (0.00, -1.75) rectangle (0.25, -2.00);
\filldraw[fill=cancan, draw=black] (0.25, -1.75) rectangle (0.50, -2.00);
\filldraw[fill=cancan, draw=black] (0.50, -1.75) rectangle (0.75, -2.00);
\filldraw[fill=cancan, draw=black] (0.75, -1.75) rectangle (1.00, -2.00);
\filldraw[fill=cancan, draw=black] (1.00, -1.75) rectangle (1.25, -2.00);
\filldraw[fill=bermuda, draw=black] (1.25, -1.75) rectangle (1.50, -2.00);
\filldraw[fill=bermuda, draw=black] (1.50, -1.75) rectangle (1.75, -2.00);
\filldraw[fill=bermuda, draw=black] (1.75, -1.75) rectangle (2.00, -2.00);
\filldraw[fill=bermuda, draw=black] (2.00, -1.75) rectangle (2.25, -2.00);
\filldraw[fill=cancan, draw=black] (2.25, -1.75) rectangle (2.50, -2.00);
\filldraw[fill=bermuda, draw=black] (2.50, -1.75) rectangle (2.75, -2.00);
\filldraw[fill=cancan, draw=black] (2.75, -1.75) rectangle (3.00, -2.00);
\filldraw[fill=cancan, draw=black] (3.00, -1.75) rectangle (3.25, -2.00);
\filldraw[fill=bermuda, draw=black] (3.25, -1.75) rectangle (3.50, -2.00);
\filldraw[fill=bermuda, draw=black] (3.50, -1.75) rectangle (3.75, -2.00);
\filldraw[fill=cancan, draw=black] (3.75, -1.75) rectangle (4.00, -2.00);
\filldraw[fill=bermuda, draw=black] (4.00, -1.75) rectangle (4.25, -2.00);
\filldraw[fill=cancan, draw=black] (4.25, -1.75) rectangle (4.50, -2.00);
\filldraw[fill=bermuda, draw=black] (4.50, -1.75) rectangle (4.75, -2.00);
\filldraw[fill=bermuda, draw=black] (4.75, -1.75) rectangle (5.00, -2.00);
\filldraw[fill=bermuda, draw=black] (5.00, -1.75) rectangle (5.25, -2.00);
\filldraw[fill=cancan, draw=black] (5.25, -1.75) rectangle (5.50, -2.00);
\filldraw[fill=bermuda, draw=black] (5.50, -1.75) rectangle (5.75, -2.00);
\filldraw[fill=bermuda, draw=black] (5.75, -1.75) rectangle (6.00, -2.00);
\filldraw[fill=bermuda, draw=black] (6.00, -1.75) rectangle (6.25, -2.00);
\filldraw[fill=bermuda, draw=black] (6.25, -1.75) rectangle (6.50, -2.00);
\filldraw[fill=bermuda, draw=black] (6.50, -1.75) rectangle (6.75, -2.00);
\filldraw[fill=bermuda, draw=black] (6.75, -1.75) rectangle (7.00, -2.00);
\filldraw[fill=bermuda, draw=black] (7.00, -1.75) rectangle (7.25, -2.00);
\filldraw[fill=cancan, draw=black] (7.25, -1.75) rectangle (7.50, -2.00);
\filldraw[fill=bermuda, draw=black] (7.50, -1.75) rectangle (7.75, -2.00);
\filldraw[fill=cancan, draw=black] (7.75, -1.75) rectangle (8.00, -2.00);
\filldraw[fill=bermuda, draw=black] (8.00, -1.75) rectangle (8.25, -2.00);
\filldraw[fill=bermuda, draw=black] (8.25, -1.75) rectangle (8.50, -2.00);
\filldraw[fill=bermuda, draw=black] (8.50, -1.75) rectangle (8.75, -2.00);
\filldraw[fill=cancan, draw=black] (8.75, -1.75) rectangle (9.00, -2.00);
\filldraw[fill=cancan, draw=black] (9.00, -1.75) rectangle (9.25, -2.00);
\filldraw[fill=cancan, draw=black] (9.25, -1.75) rectangle (9.50, -2.00);
\filldraw[fill=bermuda, draw=black] (9.50, -1.75) rectangle (9.75, -2.00);
\filldraw[fill=bermuda, draw=black] (9.75, -1.75) rectangle (10.00, -2.00);
\filldraw[fill=bermuda, draw=black] (10.00, -1.75) rectangle (10.25, -2.00);
\filldraw[fill=cancan, draw=black] (10.25, -1.75) rectangle (10.50, -2.00);
\filldraw[fill=cancan, draw=black] (10.50, -1.75) rectangle (10.75, -2.00);
\filldraw[fill=cancan, draw=black] (10.75, -1.75) rectangle (11.00, -2.00);
\filldraw[fill=bermuda, draw=black] (11.00, -1.75) rectangle (11.25, -2.00);
\filldraw[fill=bermuda, draw=black] (11.25, -1.75) rectangle (11.50, -2.00);
\filldraw[fill=bermuda, draw=black] (11.50, -1.75) rectangle (11.75, -2.00);
\filldraw[fill=cancan, draw=black] (11.75, -1.75) rectangle (12.00, -2.00);
\filldraw[fill=cancan, draw=black] (12.00, -1.75) rectangle (12.25, -2.00);
\filldraw[fill=cancan, draw=black] (12.25, -1.75) rectangle (12.50, -2.00);
\filldraw[fill=bermuda, draw=black] (12.50, -1.75) rectangle (12.75, -2.00);
\filldraw[fill=bermuda, draw=black] (12.75, -1.75) rectangle (13.00, -2.00);
\filldraw[fill=bermuda, draw=black] (13.00, -1.75) rectangle (13.25, -2.00);
\filldraw[fill=cancan, draw=black] (13.25, -1.75) rectangle (13.50, -2.00);
\filldraw[fill=cancan, draw=black] (13.50, -1.75) rectangle (13.75, -2.00);
\filldraw[fill=cancan, draw=black] (13.75, -1.75) rectangle (14.00, -2.00);
\filldraw[fill=bermuda, draw=black] (14.00, -1.75) rectangle (14.25, -2.00);
\filldraw[fill=bermuda, draw=black] (14.25, -1.75) rectangle (14.50, -2.00);
\filldraw[fill=bermuda, draw=black] (14.50, -1.75) rectangle (14.75, -2.00);
\filldraw[fill=bermuda, draw=black] (14.75, -1.75) rectangle (15.00, -2.00);
\filldraw[fill=bermuda, draw=black] (0.00, -2.00) rectangle (0.25, -2.25);
\filldraw[fill=cancan, draw=black] (0.25, -2.00) rectangle (0.50, -2.25);
\filldraw[fill=bermuda, draw=black] (0.50, -2.00) rectangle (0.75, -2.25);
\filldraw[fill=bermuda, draw=black] (0.75, -2.00) rectangle (1.00, -2.25);
\filldraw[fill=bermuda, draw=black] (1.00, -2.00) rectangle (1.25, -2.25);
\filldraw[fill=cancan, draw=black] (1.25, -2.00) rectangle (1.50, -2.25);
\filldraw[fill=cancan, draw=black] (1.50, -2.00) rectangle (1.75, -2.25);
\filldraw[fill=cancan, draw=black] (1.75, -2.00) rectangle (2.00, -2.25);
\filldraw[fill=bermuda, draw=black] (2.00, -2.00) rectangle (2.25, -2.25);
\filldraw[fill=bermuda, draw=black] (2.25, -2.00) rectangle (2.50, -2.25);
\filldraw[fill=bermuda, draw=black] (2.50, -2.00) rectangle (2.75, -2.25);
\filldraw[fill=cancan, draw=black] (2.75, -2.00) rectangle (3.00, -2.25);
\filldraw[fill=bermuda, draw=black] (3.00, -2.00) rectangle (3.25, -2.25);
\filldraw[fill=cancan, draw=black] (3.25, -2.00) rectangle (3.50, -2.25);
\filldraw[fill=cancan, draw=black] (3.50, -2.00) rectangle (3.75, -2.25);
\filldraw[fill=cancan, draw=black] (3.75, -2.00) rectangle (4.00, -2.25);
\filldraw[fill=cancan, draw=black] (4.00, -2.00) rectangle (4.25, -2.25);
\filldraw[fill=bermuda, draw=black] (4.25, -2.00) rectangle (4.50, -2.25);
\filldraw[fill=bermuda, draw=black] (4.50, -2.00) rectangle (4.75, -2.25);
\filldraw[fill=cancan, draw=black] (4.75, -2.00) rectangle (5.00, -2.25);
\filldraw[fill=cancan, draw=black] (5.00, -2.00) rectangle (5.25, -2.25);
\filldraw[fill=cancan, draw=black] (5.25, -2.00) rectangle (5.50, -2.25);
\filldraw[fill=cancan, draw=black] (5.50, -2.00) rectangle (5.75, -2.25);
\filldraw[fill=bermuda, draw=black] (5.75, -2.00) rectangle (6.00, -2.25);
\filldraw[fill=bermuda, draw=black] (6.00, -2.00) rectangle (6.25, -2.25);
\filldraw[fill=cancan, draw=black] (6.25, -2.00) rectangle (6.50, -2.25);
\filldraw[fill=cancan, draw=black] (6.50, -2.00) rectangle (6.75, -2.25);
\filldraw[fill=cancan, draw=black] (6.75, -2.00) rectangle (7.00, -2.25);
\filldraw[fill=cancan, draw=black] (7.00, -2.00) rectangle (7.25, -2.25);
\filldraw[fill=bermuda, draw=black] (7.25, -2.00) rectangle (7.50, -2.25);
\filldraw[fill=bermuda, draw=black] (7.50, -2.00) rectangle (7.75, -2.25);
\filldraw[fill=cancan, draw=black] (7.75, -2.00) rectangle (8.00, -2.25);
\filldraw[fill=cancan, draw=black] (8.00, -2.00) rectangle (8.25, -2.25);
\filldraw[fill=cancan, draw=black] (8.25, -2.00) rectangle (8.50, -2.25);
\filldraw[fill=cancan, draw=black] (8.50, -2.00) rectangle (8.75, -2.25);
\filldraw[fill=cancan, draw=black] (8.75, -2.00) rectangle (9.00, -2.25);
\filldraw[fill=bermuda, draw=black] (9.00, -2.00) rectangle (9.25, -2.25);
\filldraw[fill=bermuda, draw=black] (9.25, -2.00) rectangle (9.50, -2.25);
\filldraw[fill=bermuda, draw=black] (9.50, -2.00) rectangle (9.75, -2.25);
\filldraw[fill=cancan, draw=black] (9.75, -2.00) rectangle (10.00, -2.25);
\filldraw[fill=cancan, draw=black] (10.00, -2.00) rectangle (10.25, -2.25);
\filldraw[fill=cancan, draw=black] (10.25, -2.00) rectangle (10.50, -2.25);
\filldraw[fill=cancan, draw=black] (10.50, -2.00) rectangle (10.75, -2.25);
\filldraw[fill=cancan, draw=black] (10.75, -2.00) rectangle (11.00, -2.25);
\filldraw[fill=bermuda, draw=black] (11.00, -2.00) rectangle (11.25, -2.25);
\filldraw[fill=bermuda, draw=black] (11.25, -2.00) rectangle (11.50, -2.25);
\filldraw[fill=bermuda, draw=black] (11.50, -2.00) rectangle (11.75, -2.25);
\filldraw[fill=bermuda, draw=black] (11.75, -2.00) rectangle (12.00, -2.25);
\filldraw[fill=bermuda, draw=black] (12.00, -2.00) rectangle (12.25, -2.25);
\filldraw[fill=cancan, draw=black] (12.25, -2.00) rectangle (12.50, -2.25);
\filldraw[fill=bermuda, draw=black] (12.50, -2.00) rectangle (12.75, -2.25);
\filldraw[fill=cancan, draw=black] (12.75, -2.00) rectangle (13.00, -2.25);
\filldraw[fill=bermuda, draw=black] (13.00, -2.00) rectangle (13.25, -2.25);
\filldraw[fill=bermuda, draw=black] (13.25, -2.00) rectangle (13.50, -2.25);
\filldraw[fill=bermuda, draw=black] (13.50, -2.00) rectangle (13.75, -2.25);
\filldraw[fill=cancan, draw=black] (13.75, -2.00) rectangle (14.00, -2.25);
\filldraw[fill=bermuda, draw=black] (14.00, -2.00) rectangle (14.25, -2.25);
\filldraw[fill=bermuda, draw=black] (14.25, -2.00) rectangle (14.50, -2.25);
\filldraw[fill=bermuda, draw=black] (14.50, -2.00) rectangle (14.75, -2.25);
\filldraw[fill=cancan, draw=black] (14.75, -2.00) rectangle (15.00, -2.25);
\filldraw[fill=bermuda, draw=black] (0.00, -2.25) rectangle (0.25, -2.50);
\filldraw[fill=cancan, draw=black] (0.25, -2.25) rectangle (0.50, -2.50);
\filldraw[fill=bermuda, draw=black] (0.50, -2.25) rectangle (0.75, -2.50);
\filldraw[fill=cancan, draw=black] (0.75, -2.25) rectangle (1.00, -2.50);
\filldraw[fill=cancan, draw=black] (1.00, -2.25) rectangle (1.25, -2.50);
\filldraw[fill=cancan, draw=black] (1.25, -2.25) rectangle (1.50, -2.50);
\filldraw[fill=bermuda, draw=black] (1.50, -2.25) rectangle (1.75, -2.50);
\filldraw[fill=cancan, draw=black] (1.75, -2.25) rectangle (2.00, -2.50);
\filldraw[fill=bermuda, draw=black] (2.00, -2.25) rectangle (2.25, -2.50);
\filldraw[fill=cancan, draw=black] (2.25, -2.25) rectangle (2.50, -2.50);
\filldraw[fill=bermuda, draw=black] (2.50, -2.25) rectangle (2.75, -2.50);
\filldraw[fill=bermuda, draw=black] (2.75, -2.25) rectangle (3.00, -2.50);
\filldraw[fill=bermuda, draw=black] (3.00, -2.25) rectangle (3.25, -2.50);
\filldraw[fill=cancan, draw=black] (3.25, -2.25) rectangle (3.50, -2.50);
\filldraw[fill=cancan, draw=black] (3.50, -2.25) rectangle (3.75, -2.50);
\filldraw[fill=cancan, draw=black] (3.75, -2.25) rectangle (4.00, -2.50);
\filldraw[fill=cancan, draw=black] (4.00, -2.25) rectangle (4.25, -2.50);
\filldraw[fill=cancan, draw=black] (4.25, -2.25) rectangle (4.50, -2.50);
\filldraw[fill=cancan, draw=black] (4.50, -2.25) rectangle (4.75, -2.50);
\filldraw[fill=cancan, draw=black] (4.75, -2.25) rectangle (5.00, -2.50);
\filldraw[fill=bermuda, draw=black] (5.00, -2.25) rectangle (5.25, -2.50);
\filldraw[fill=bermuda, draw=black] (5.25, -2.25) rectangle (5.50, -2.50);
\filldraw[fill=bermuda, draw=black] (5.50, -2.25) rectangle (5.75, -2.50);
\filldraw[fill=cancan, draw=black] (5.75, -2.25) rectangle (6.00, -2.50);
\filldraw[fill=cancan, draw=black] (6.00, -2.25) rectangle (6.25, -2.50);
\filldraw[fill=cancan, draw=black] (6.25, -2.25) rectangle (6.50, -2.50);
\filldraw[fill=bermuda, draw=black] (6.50, -2.25) rectangle (6.75, -2.50);
\filldraw[fill=bermuda, draw=black] (6.75, -2.25) rectangle (7.00, -2.50);
\filldraw[fill=bermuda, draw=black] (7.00, -2.25) rectangle (7.25, -2.50);
\filldraw[fill=cancan, draw=black] (7.25, -2.25) rectangle (7.50, -2.50);
\filldraw[fill=bermuda, draw=black] (7.50, -2.25) rectangle (7.75, -2.50);
\filldraw[fill=bermuda, draw=black] (7.75, -2.25) rectangle (8.00, -2.50);
\filldraw[fill=bermuda, draw=black] (8.00, -2.25) rectangle (8.25, -2.50);
\filldraw[fill=cancan, draw=black] (8.25, -2.25) rectangle (8.50, -2.50);
\filldraw[fill=cancan, draw=black] (8.50, -2.25) rectangle (8.75, -2.50);
\filldraw[fill=cancan, draw=black] (8.75, -2.25) rectangle (9.00, -2.50);
\filldraw[fill=cancan, draw=black] (9.00, -2.25) rectangle (9.25, -2.50);
\filldraw[fill=cancan, draw=black] (9.25, -2.25) rectangle (9.50, -2.50);
\filldraw[fill=cancan, draw=black] (9.50, -2.25) rectangle (9.75, -2.50);
\filldraw[fill=cancan, draw=black] (9.75, -2.25) rectangle (10.00, -2.50);
\filldraw[fill=cancan, draw=black] (10.00, -2.25) rectangle (10.25, -2.50);
\filldraw[fill=bermuda, draw=black] (10.25, -2.25) rectangle (10.50, -2.50);
\filldraw[fill=bermuda, draw=black] (10.50, -2.25) rectangle (10.75, -2.50);
\filldraw[fill=cancan, draw=black] (10.75, -2.25) rectangle (11.00, -2.50);
\filldraw[fill=bermuda, draw=black] (11.00, -2.25) rectangle (11.25, -2.50);
\filldraw[fill=cancan, draw=black] (11.25, -2.25) rectangle (11.50, -2.50);
\filldraw[fill=bermuda, draw=black] (11.50, -2.25) rectangle (11.75, -2.50);
\filldraw[fill=bermuda, draw=black] (11.75, -2.25) rectangle (12.00, -2.50);
\filldraw[fill=bermuda, draw=black] (12.00, -2.25) rectangle (12.25, -2.50);
\filldraw[fill=cancan, draw=black] (12.25, -2.25) rectangle (12.50, -2.50);
\filldraw[fill=bermuda, draw=black] (12.50, -2.25) rectangle (12.75, -2.50);
\filldraw[fill=bermuda, draw=black] (12.75, -2.25) rectangle (13.00, -2.50);
\filldraw[fill=bermuda, draw=black] (13.00, -2.25) rectangle (13.25, -2.50);
\filldraw[fill=bermuda, draw=black] (13.25, -2.25) rectangle (13.50, -2.50);
\filldraw[fill=bermuda, draw=black] (13.50, -2.25) rectangle (13.75, -2.50);
\filldraw[fill=cancan, draw=black] (13.75, -2.25) rectangle (14.00, -2.50);
\filldraw[fill=bermuda, draw=black] (14.00, -2.25) rectangle (14.25, -2.50);
\filldraw[fill=bermuda, draw=black] (14.25, -2.25) rectangle (14.50, -2.50);
\filldraw[fill=bermuda, draw=black] (14.50, -2.25) rectangle (14.75, -2.50);
\filldraw[fill=bermuda, draw=black] (14.75, -2.25) rectangle (15.00, -2.50);
\filldraw[fill=bermuda, draw=black] (0.00, -2.50) rectangle (0.25, -2.75);
\filldraw[fill=bermuda, draw=black] (0.25, -2.50) rectangle (0.50, -2.75);
\filldraw[fill=bermuda, draw=black] (0.50, -2.50) rectangle (0.75, -2.75);
\filldraw[fill=cancan, draw=black] (0.75, -2.50) rectangle (1.00, -2.75);
\filldraw[fill=cancan, draw=black] (1.00, -2.50) rectangle (1.25, -2.75);
\filldraw[fill=cancan, draw=black] (1.25, -2.50) rectangle (1.50, -2.75);
\filldraw[fill=cancan, draw=black] (1.50, -2.50) rectangle (1.75, -2.75);
\filldraw[fill=cancan, draw=black] (1.75, -2.50) rectangle (2.00, -2.75);
\filldraw[fill=cancan, draw=black] (2.00, -2.50) rectangle (2.25, -2.75);
\filldraw[fill=bermuda, draw=black] (2.25, -2.50) rectangle (2.50, -2.75);
\filldraw[fill=bermuda, draw=black] (2.50, -2.50) rectangle (2.75, -2.75);
\filldraw[fill=cancan, draw=black] (2.75, -2.50) rectangle (3.00, -2.75);
\filldraw[fill=cancan, draw=black] (3.00, -2.50) rectangle (3.25, -2.75);
\filldraw[fill=cancan, draw=black] (3.25, -2.50) rectangle (3.50, -2.75);
\filldraw[fill=bermuda, draw=black] (3.50, -2.50) rectangle (3.75, -2.75);
\filldraw[fill=bermuda, draw=black] (3.75, -2.50) rectangle (4.00, -2.75);
\filldraw[fill=bermuda, draw=black] (4.00, -2.50) rectangle (4.25, -2.75);
\filldraw[fill=cancan, draw=black] (4.25, -2.50) rectangle (4.50, -2.75);
\filldraw[fill=cancan, draw=black] (4.50, -2.50) rectangle (4.75, -2.75);
\filldraw[fill=cancan, draw=black] (4.75, -2.50) rectangle (5.00, -2.75);
\filldraw[fill=bermuda, draw=black] (5.00, -2.50) rectangle (5.25, -2.75);
\filldraw[fill=bermuda, draw=black] (5.25, -2.50) rectangle (5.50, -2.75);
\filldraw[fill=bermuda, draw=black] (5.50, -2.50) rectangle (5.75, -2.75);
\filldraw[fill=bermuda, draw=black] (5.75, -2.50) rectangle (6.00, -2.75);
\filldraw[fill=bermuda, draw=black] (6.00, -2.50) rectangle (6.25, -2.75);
\filldraw[fill=bermuda, draw=black] (6.25, -2.50) rectangle (6.50, -2.75);
\filldraw[fill=bermuda, draw=black] (6.50, -2.50) rectangle (6.75, -2.75);
\filldraw[fill=cancan, draw=black] (6.75, -2.50) rectangle (7.00, -2.75);
\filldraw[fill=cancan, draw=black] (7.00, -2.50) rectangle (7.25, -2.75);
\filldraw[fill=cancan, draw=black] (7.25, -2.50) rectangle (7.50, -2.75);
\filldraw[fill=cancan, draw=black] (7.50, -2.50) rectangle (7.75, -2.75);
\filldraw[fill=cancan, draw=black] (7.75, -2.50) rectangle (8.00, -2.75);
\filldraw[fill=cancan, draw=black] (8.00, -2.50) rectangle (8.25, -2.75);
\filldraw[fill=bermuda, draw=black] (8.25, -2.50) rectangle (8.50, -2.75);
\filldraw[fill=bermuda, draw=black] (8.50, -2.50) rectangle (8.75, -2.75);
\filldraw[fill=cancan, draw=black] (8.75, -2.50) rectangle (9.00, -2.75);
\filldraw[fill=cancan, draw=black] (9.00, -2.50) rectangle (9.25, -2.75);
\filldraw[fill=cancan, draw=black] (9.25, -2.50) rectangle (9.50, -2.75);
\filldraw[fill=bermuda, draw=black] (9.50, -2.50) rectangle (9.75, -2.75);
\filldraw[fill=bermuda, draw=black] (9.75, -2.50) rectangle (10.00, -2.75);
\filldraw[fill=bermuda, draw=black] (10.00, -2.50) rectangle (10.25, -2.75);
\filldraw[fill=cancan, draw=black] (10.25, -2.50) rectangle (10.50, -2.75);
\filldraw[fill=cancan, draw=black] (10.50, -2.50) rectangle (10.75, -2.75);
\filldraw[fill=cancan, draw=black] (10.75, -2.50) rectangle (11.00, -2.75);
\filldraw[fill=cancan, draw=black] (11.00, -2.50) rectangle (11.25, -2.75);
\filldraw[fill=bermuda, draw=black] (11.25, -2.50) rectangle (11.50, -2.75);
\filldraw[fill=bermuda, draw=black] (11.50, -2.50) rectangle (11.75, -2.75);
\filldraw[fill=cancan, draw=black] (11.75, -2.50) rectangle (12.00, -2.75);
\filldraw[fill=cancan, draw=black] (12.00, -2.50) rectangle (12.25, -2.75);
\filldraw[fill=cancan, draw=black] (12.25, -2.50) rectangle (12.50, -2.75);
\filldraw[fill=cancan, draw=black] (12.50, -2.50) rectangle (12.75, -2.75);
\filldraw[fill=bermuda, draw=black] (12.75, -2.50) rectangle (13.00, -2.75);
\filldraw[fill=bermuda, draw=black] (13.00, -2.50) rectangle (13.25, -2.75);
\filldraw[fill=cancan, draw=black] (13.25, -2.50) rectangle (13.50, -2.75);
\filldraw[fill=cancan, draw=black] (13.50, -2.50) rectangle (13.75, -2.75);
\filldraw[fill=cancan, draw=black] (13.75, -2.50) rectangle (14.00, -2.75);
\filldraw[fill=cancan, draw=black] (14.00, -2.50) rectangle (14.25, -2.75);
\filldraw[fill=cancan, draw=black] (14.25, -2.50) rectangle (14.50, -2.75);
\filldraw[fill=cancan, draw=black] (14.50, -2.50) rectangle (14.75, -2.75);
\filldraw[fill=bermuda, draw=black] (14.75, -2.50) rectangle (15.00, -2.75);
\filldraw[fill=bermuda, draw=black] (0.00, -2.75) rectangle (0.25, -3.00);
\filldraw[fill=cancan, draw=black] (0.25, -2.75) rectangle (0.50, -3.00);
\filldraw[fill=bermuda, draw=black] (0.50, -2.75) rectangle (0.75, -3.00);
\filldraw[fill=cancan, draw=black] (0.75, -2.75) rectangle (1.00, -3.00);
\filldraw[fill=cancan, draw=black] (1.00, -2.75) rectangle (1.25, -3.00);
\filldraw[fill=bermuda, draw=black] (1.25, -2.75) rectangle (1.50, -3.00);
\filldraw[fill=bermuda, draw=black] (1.50, -2.75) rectangle (1.75, -3.00);
\filldraw[fill=cancan, draw=black] (1.75, -2.75) rectangle (2.00, -3.00);
\filldraw[fill=bermuda, draw=black] (2.00, -2.75) rectangle (2.25, -3.00);
\filldraw[fill=cancan, draw=black] (2.25, -2.75) rectangle (2.50, -3.00);
\filldraw[fill=bermuda, draw=black] (2.50, -2.75) rectangle (2.75, -3.00);
\filldraw[fill=cancan, draw=black] (2.75, -2.75) rectangle (3.00, -3.00);
\filldraw[fill=bermuda, draw=black] (3.00, -2.75) rectangle (3.25, -3.00);
\filldraw[fill=bermuda, draw=black] (3.25, -2.75) rectangle (3.50, -3.00);
\filldraw[fill=bermuda, draw=black] (3.50, -2.75) rectangle (3.75, -3.00);
\filldraw[fill=cancan, draw=black] (3.75, -2.75) rectangle (4.00, -3.00);
\filldraw[fill=cancan, draw=black] (4.00, -2.75) rectangle (4.25, -3.00);
\filldraw[fill=cancan, draw=black] (4.25, -2.75) rectangle (4.50, -3.00);
\filldraw[fill=bermuda, draw=black] (4.50, -2.75) rectangle (4.75, -3.00);
\filldraw[fill=bermuda, draw=black] (4.75, -2.75) rectangle (5.00, -3.00);
\filldraw[fill=bermuda, draw=black] (5.00, -2.75) rectangle (5.25, -3.00);
\filldraw[fill=cancan, draw=black] (5.25, -2.75) rectangle (5.50, -3.00);
\filldraw[fill=bermuda, draw=black] (5.50, -2.75) rectangle (5.75, -3.00);
\filldraw[fill=bermuda, draw=black] (5.75, -2.75) rectangle (6.00, -3.00);
\filldraw[fill=bermuda, draw=black] (6.00, -2.75) rectangle (6.25, -3.00);
\filldraw[fill=bermuda, draw=black] (6.25, -2.75) rectangle (6.50, -3.00);
\filldraw[fill=bermuda, draw=black] (6.50, -2.75) rectangle (6.75, -3.00);
\filldraw[fill=cancan, draw=black] (6.75, -2.75) rectangle (7.00, -3.00);
\filldraw[fill=bermuda, draw=black] (7.00, -2.75) rectangle (7.25, -3.00);
\filldraw[fill=bermuda, draw=black] (7.25, -2.75) rectangle (7.50, -3.00);
\filldraw[fill=bermuda, draw=black] (7.50, -2.75) rectangle (7.75, -3.00);
\filldraw[fill=cancan, draw=black] (7.75, -2.75) rectangle (8.00, -3.00);
\filldraw[fill=cancan, draw=black] (8.00, -2.75) rectangle (8.25, -3.00);
\filldraw[fill=cancan, draw=black] (8.25, -2.75) rectangle (8.50, -3.00);
\filldraw[fill=bermuda, draw=black] (8.50, -2.75) rectangle (8.75, -3.00);
\filldraw[fill=bermuda, draw=black] (8.75, -2.75) rectangle (9.00, -3.00);
\filldraw[fill=bermuda, draw=black] (9.00, -2.75) rectangle (9.25, -3.00);
\filldraw[fill=cancan, draw=black] (9.25, -2.75) rectangle (9.50, -3.00);
\filldraw[fill=bermuda, draw=black] (9.50, -2.75) rectangle (9.75, -3.00);
\filldraw[fill=bermuda, draw=black] (9.75, -2.75) rectangle (10.00, -3.00);
\filldraw[fill=bermuda, draw=black] (10.00, -2.75) rectangle (10.25, -3.00);
\filldraw[fill=cancan, draw=black] (10.25, -2.75) rectangle (10.50, -3.00);
\filldraw[fill=cancan, draw=black] (10.50, -2.75) rectangle (10.75, -3.00);
\filldraw[fill=cancan, draw=black] (10.75, -2.75) rectangle (11.00, -3.00);
\filldraw[fill=cancan, draw=black] (11.00, -2.75) rectangle (11.25, -3.00);
\filldraw[fill=cancan, draw=black] (11.25, -2.75) rectangle (11.50, -3.00);
\filldraw[fill=cancan, draw=black] (11.50, -2.75) rectangle (11.75, -3.00);
\filldraw[fill=bermuda, draw=black] (11.75, -2.75) rectangle (12.00, -3.00);
\filldraw[fill=bermuda, draw=black] (12.00, -2.75) rectangle (12.25, -3.00);
\filldraw[fill=cancan, draw=black] (12.25, -2.75) rectangle (12.50, -3.00);
\filldraw[fill=bermuda, draw=black] (12.50, -2.75) rectangle (12.75, -3.00);
\filldraw[fill=cancan, draw=black] (12.75, -2.75) rectangle (13.00, -3.00);
\filldraw[fill=cancan, draw=black] (13.00, -2.75) rectangle (13.25, -3.00);
\filldraw[fill=bermuda, draw=black] (13.25, -2.75) rectangle (13.50, -3.00);
\filldraw[fill=bermuda, draw=black] (13.50, -2.75) rectangle (13.75, -3.00);
\filldraw[fill=cancan, draw=black] (13.75, -2.75) rectangle (14.00, -3.00);
\filldraw[fill=bermuda, draw=black] (14.00, -2.75) rectangle (14.25, -3.00);
\filldraw[fill=cancan, draw=black] (14.25, -2.75) rectangle (14.50, -3.00);
\filldraw[fill=cancan, draw=black] (14.50, -2.75) rectangle (14.75, -3.00);
\filldraw[fill=bermuda, draw=black] (14.75, -2.75) rectangle (15.00, -3.00);
\filldraw[fill=bermuda, draw=black] (0.00, -3.00) rectangle (0.25, -3.25);
\filldraw[fill=cancan, draw=black] (0.25, -3.00) rectangle (0.50, -3.25);
\filldraw[fill=bermuda, draw=black] (0.50, -3.00) rectangle (0.75, -3.25);
\filldraw[fill=cancan, draw=black] (0.75, -3.00) rectangle (1.00, -3.25);
\filldraw[fill=cancan, draw=black] (1.00, -3.00) rectangle (1.25, -3.25);
\filldraw[fill=bermuda, draw=black] (1.25, -3.00) rectangle (1.50, -3.25);
\filldraw[fill=bermuda, draw=black] (1.50, -3.00) rectangle (1.75, -3.25);
\filldraw[fill=cancan, draw=black] (1.75, -3.00) rectangle (2.00, -3.25);
\filldraw[fill=bermuda, draw=black] (2.00, -3.00) rectangle (2.25, -3.25);
\filldraw[fill=cancan, draw=black] (2.25, -3.00) rectangle (2.50, -3.25);
\filldraw[fill=cancan, draw=black] (2.50, -3.00) rectangle (2.75, -3.25);
\filldraw[fill=bermuda, draw=black] (2.75, -3.00) rectangle (3.00, -3.25);
\filldraw[fill=bermuda, draw=black] (3.00, -3.00) rectangle (3.25, -3.25);
\filldraw[fill=cancan, draw=black] (3.25, -3.00) rectangle (3.50, -3.25);
\filldraw[fill=cancan, draw=black] (3.50, -3.00) rectangle (3.75, -3.25);
\filldraw[fill=cancan, draw=black] (3.75, -3.00) rectangle (4.00, -3.25);
\filldraw[fill=cancan, draw=black] (4.00, -3.00) rectangle (4.25, -3.25);
\filldraw[fill=cancan, draw=black] (4.25, -3.00) rectangle (4.50, -3.25);
\filldraw[fill=cancan, draw=black] (4.50, -3.00) rectangle (4.75, -3.25);
\filldraw[fill=cancan, draw=black] (4.75, -3.00) rectangle (5.00, -3.25);
\filldraw[fill=bermuda, draw=black] (5.00, -3.00) rectangle (5.25, -3.25);
\filldraw[fill=bermuda, draw=black] (5.25, -3.00) rectangle (5.50, -3.25);
\filldraw[fill=bermuda, draw=black] (5.50, -3.00) rectangle (5.75, -3.25);
\filldraw[fill=bermuda, draw=black] (5.75, -3.00) rectangle (6.00, -3.25);
\filldraw[fill=bermuda, draw=black] (6.00, -3.00) rectangle (6.25, -3.25);
\filldraw[fill=bermuda, draw=black] (6.25, -3.00) rectangle (6.50, -3.25);
\filldraw[fill=bermuda, draw=black] (6.50, -3.00) rectangle (6.75, -3.25);
\filldraw[fill=bermuda, draw=black] (6.75, -3.00) rectangle (7.00, -3.25);
\filldraw[fill=bermuda, draw=black] (7.00, -3.00) rectangle (7.25, -3.25);
\filldraw[fill=bermuda, draw=black] (7.25, -3.00) rectangle (7.50, -3.25);
\filldraw[fill=bermuda, draw=black] (7.50, -3.00) rectangle (7.75, -3.25);
\filldraw[fill=cancan, draw=black] (7.75, -3.00) rectangle (8.00, -3.25);
\filldraw[fill=cancan, draw=black] (8.00, -3.00) rectangle (8.25, -3.25);
\filldraw[fill=cancan, draw=black] (8.25, -3.00) rectangle (8.50, -3.25);
\filldraw[fill=bermuda, draw=black] (8.50, -3.00) rectangle (8.75, -3.25);
\filldraw[fill=bermuda, draw=black] (8.75, -3.00) rectangle (9.00, -3.25);
\filldraw[fill=bermuda, draw=black] (9.00, -3.00) rectangle (9.25, -3.25);
\filldraw[fill=cancan, draw=black] (9.25, -3.00) rectangle (9.50, -3.25);
\filldraw[fill=cancan, draw=black] (9.50, -3.00) rectangle (9.75, -3.25);
\filldraw[fill=cancan, draw=black] (9.75, -3.00) rectangle (10.00, -3.25);
\filldraw[fill=cancan, draw=black] (10.00, -3.00) rectangle (10.25, -3.25);
\filldraw[fill=cancan, draw=black] (10.25, -3.00) rectangle (10.50, -3.25);
\filldraw[fill=cancan, draw=black] (10.50, -3.00) rectangle (10.75, -3.25);
\filldraw[fill=bermuda, draw=black] (10.75, -3.00) rectangle (11.00, -3.25);
\filldraw[fill=bermuda, draw=black] (11.00, -3.00) rectangle (11.25, -3.25);
\filldraw[fill=cancan, draw=black] (11.25, -3.00) rectangle (11.50, -3.25);
\filldraw[fill=cancan, draw=black] (11.50, -3.00) rectangle (11.75, -3.25);
\filldraw[fill=cancan, draw=black] (11.75, -3.00) rectangle (12.00, -3.25);
\filldraw[fill=bermuda, draw=black] (12.00, -3.00) rectangle (12.25, -3.25);
\filldraw[fill=bermuda, draw=black] (12.25, -3.00) rectangle (12.50, -3.25);
\filldraw[fill=bermuda, draw=black] (12.50, -3.00) rectangle (12.75, -3.25);
\filldraw[fill=cancan, draw=black] (12.75, -3.00) rectangle (13.00, -3.25);
\filldraw[fill=bermuda, draw=black] (13.00, -3.00) rectangle (13.25, -3.25);
\filldraw[fill=bermuda, draw=black] (13.25, -3.00) rectangle (13.50, -3.25);
\filldraw[fill=bermuda, draw=black] (13.50, -3.00) rectangle (13.75, -3.25);
\filldraw[fill=bermuda, draw=black] (13.75, -3.00) rectangle (14.00, -3.25);
\filldraw[fill=bermuda, draw=black] (14.00, -3.00) rectangle (14.25, -3.25);
\filldraw[fill=cancan, draw=black] (14.25, -3.00) rectangle (14.50, -3.25);
\filldraw[fill=bermuda, draw=black] (14.50, -3.00) rectangle (14.75, -3.25);
\filldraw[fill=bermuda, draw=black] (14.75, -3.00) rectangle (15.00, -3.25);
\filldraw[fill=bermuda, draw=black] (0.00, -3.25) rectangle (0.25, -3.50);
\filldraw[fill=cancan, draw=black] (0.25, -3.25) rectangle (0.50, -3.50);
\filldraw[fill=cancan, draw=black] (0.50, -3.25) rectangle (0.75, -3.50);
\filldraw[fill=cancan, draw=black] (0.75, -3.25) rectangle (1.00, -3.50);
\filldraw[fill=bermuda, draw=black] (1.00, -3.25) rectangle (1.25, -3.50);
\filldraw[fill=bermuda, draw=black] (1.25, -3.25) rectangle (1.50, -3.50);
\filldraw[fill=bermuda, draw=black] (1.50, -3.25) rectangle (1.75, -3.50);
\filldraw[fill=cancan, draw=black] (1.75, -3.25) rectangle (2.00, -3.50);
\filldraw[fill=bermuda, draw=black] (2.00, -3.25) rectangle (2.25, -3.50);
\filldraw[fill=cancan, draw=black] (2.25, -3.25) rectangle (2.50, -3.50);
\filldraw[fill=bermuda, draw=black] (2.50, -3.25) rectangle (2.75, -3.50);
\filldraw[fill=bermuda, draw=black] (2.75, -3.25) rectangle (3.00, -3.50);
\filldraw[fill=bermuda, draw=black] (3.00, -3.25) rectangle (3.25, -3.50);
\filldraw[fill=cancan, draw=black] (3.25, -3.25) rectangle (3.50, -3.50);
\filldraw[fill=bermuda, draw=black] (3.50, -3.25) rectangle (3.75, -3.50);
\filldraw[fill=bermuda, draw=black] (3.75, -3.25) rectangle (4.00, -3.50);
\filldraw[fill=bermuda, draw=black] (4.00, -3.25) rectangle (4.25, -3.50);
\filldraw[fill=cancan, draw=black] (4.25, -3.25) rectangle (4.50, -3.50);
\filldraw[fill=cancan, draw=black] (4.50, -3.25) rectangle (4.75, -3.50);
\filldraw[fill=cancan, draw=black] (4.75, -3.25) rectangle (5.00, -3.50);
\filldraw[fill=bermuda, draw=black] (5.00, -3.25) rectangle (5.25, -3.50);
\filldraw[fill=bermuda, draw=black] (5.25, -3.25) rectangle (5.50, -3.50);
\filldraw[fill=bermuda, draw=black] (5.50, -3.25) rectangle (5.75, -3.50);
} } }\end{equation*}
\begin{equation*}
\hspace{4.6pt} b_{4} = \vcenter{\hbox{ \tikz{
\filldraw[fill=cancan, draw=black] (0.00, 0.00) rectangle (0.25, -0.25);
\filldraw[fill=bermuda, draw=black] (0.25, 0.00) rectangle (0.50, -0.25);
\filldraw[fill=bermuda, draw=black] (0.50, 0.00) rectangle (0.75, -0.25);
\filldraw[fill=bermuda, draw=black] (0.75, 0.00) rectangle (1.00, -0.25);
\filldraw[fill=cancan, draw=black] (1.00, 0.00) rectangle (1.25, -0.25);
\filldraw[fill=cancan, draw=black] (1.25, 0.00) rectangle (1.50, -0.25);
\filldraw[fill=cancan, draw=black] (1.50, 0.00) rectangle (1.75, -0.25);
\filldraw[fill=cancan, draw=black] (1.75, 0.00) rectangle (2.00, -0.25);
\filldraw[fill=cancan, draw=black] (2.00, 0.00) rectangle (2.25, -0.25);
\filldraw[fill=cancan, draw=black] (2.25, 0.00) rectangle (2.50, -0.25);
\filldraw[fill=bermuda, draw=black] (2.50, 0.00) rectangle (2.75, -0.25);
\filldraw[fill=bermuda, draw=black] (2.75, 0.00) rectangle (3.00, -0.25);
\filldraw[fill=cancan, draw=black] (3.00, 0.00) rectangle (3.25, -0.25);
\filldraw[fill=cancan, draw=black] (3.25, 0.00) rectangle (3.50, -0.25);
\filldraw[fill=cancan, draw=black] (3.50, 0.00) rectangle (3.75, -0.25);
\filldraw[fill=cancan, draw=black] (3.75, 0.00) rectangle (4.00, -0.25);
\filldraw[fill=cancan, draw=black] (4.00, 0.00) rectangle (4.25, -0.25);
\filldraw[fill=bermuda, draw=black] (4.25, 0.00) rectangle (4.50, -0.25);
\filldraw[fill=bermuda, draw=black] (4.50, 0.00) rectangle (4.75, -0.25);
\filldraw[fill=bermuda, draw=black] (4.75, 0.00) rectangle (5.00, -0.25);
\filldraw[fill=cancan, draw=black] (5.00, 0.00) rectangle (5.25, -0.25);
\filldraw[fill=cancan, draw=black] (5.25, 0.00) rectangle (5.50, -0.25);
\filldraw[fill=cancan, draw=black] (5.50, 0.00) rectangle (5.75, -0.25);
\filldraw[fill=bermuda, draw=black] (5.75, 0.00) rectangle (6.00, -0.25);
\filldraw[fill=bermuda, draw=black] (6.00, 0.00) rectangle (6.25, -0.25);
\filldraw[fill=bermuda, draw=black] (6.25, 0.00) rectangle (6.50, -0.25);
\filldraw[fill=cancan, draw=black] (6.50, 0.00) rectangle (6.75, -0.25);
\filldraw[fill=cancan, draw=black] (6.75, 0.00) rectangle (7.00, -0.25);
\filldraw[fill=cancan, draw=black] (7.00, 0.00) rectangle (7.25, -0.25);
\filldraw[fill=bermuda, draw=black] (7.25, 0.00) rectangle (7.50, -0.25);
\filldraw[fill=bermuda, draw=black] (7.50, 0.00) rectangle (7.75, -0.25);
\filldraw[fill=bermuda, draw=black] (7.75, 0.00) rectangle (8.00, -0.25);
\filldraw[fill=cancan, draw=black] (8.00, 0.00) rectangle (8.25, -0.25);
\filldraw[fill=cancan, draw=black] (8.25, 0.00) rectangle (8.50, -0.25);
\filldraw[fill=cancan, draw=black] (8.50, 0.00) rectangle (8.75, -0.25);
\filldraw[fill=bermuda, draw=black] (8.75, 0.00) rectangle (9.00, -0.25);
\filldraw[fill=bermuda, draw=black] (9.00, 0.00) rectangle (9.25, -0.25);
\filldraw[fill=bermuda, draw=black] (9.25, 0.00) rectangle (9.50, -0.25);
\filldraw[fill=cancan, draw=black] (9.50, 0.00) rectangle (9.75, -0.25);
\filldraw[fill=cancan, draw=black] (9.75, 0.00) rectangle (10.00, -0.25);
\filldraw[fill=cancan, draw=black] (10.00, 0.00) rectangle (10.25, -0.25);
\filldraw[fill=bermuda, draw=black] (10.25, 0.00) rectangle (10.50, -0.25);
\filldraw[fill=bermuda, draw=black] (10.50, 0.00) rectangle (10.75, -0.25);
\filldraw[fill=bermuda, draw=black] (10.75, 0.00) rectangle (11.00, -0.25);
\filldraw[fill=bermuda, draw=black] (11.00, 0.00) rectangle (11.25, -0.25);
\filldraw[fill=bermuda, draw=black] (11.25, 0.00) rectangle (11.50, -0.25);
\filldraw[fill=cancan, draw=black] (11.50, 0.00) rectangle (11.75, -0.25);
\filldraw[fill=bermuda, draw=black] (11.75, 0.00) rectangle (12.00, -0.25);
\filldraw[fill=bermuda, draw=black] (12.00, 0.00) rectangle (12.25, -0.25);
\filldraw[fill=bermuda, draw=black] (12.25, 0.00) rectangle (12.50, -0.25);
\filldraw[fill=cancan, draw=black] (12.50, 0.00) rectangle (12.75, -0.25);
\filldraw[fill=cancan, draw=black] (12.75, 0.00) rectangle (13.00, -0.25);
\filldraw[fill=cancan, draw=black] (13.00, 0.00) rectangle (13.25, -0.25);
\filldraw[fill=cancan, draw=black] (13.25, 0.00) rectangle (13.50, -0.25);
\filldraw[fill=bermuda, draw=black] (13.50, 0.00) rectangle (13.75, -0.25);
\filldraw[fill=bermuda, draw=black] (13.75, 0.00) rectangle (14.00, -0.25);
\filldraw[fill=cancan, draw=black] (14.00, 0.00) rectangle (14.25, -0.25);
\filldraw[fill=cancan, draw=black] (14.25, 0.00) rectangle (14.50, -0.25);
\filldraw[fill=cancan, draw=black] (14.50, 0.00) rectangle (14.75, -0.25);
\filldraw[fill=cancan, draw=black] (14.75, 0.00) rectangle (15.00, -0.25);
\filldraw[fill=cancan, draw=black] (0.00, -0.25) rectangle (0.25, -0.50);
\filldraw[fill=cancan, draw=black] (0.25, -0.25) rectangle (0.50, -0.50);
\filldraw[fill=cancan, draw=black] (0.50, -0.25) rectangle (0.75, -0.50);
\filldraw[fill=cancan, draw=black] (0.75, -0.25) rectangle (1.00, -0.50);
\filldraw[fill=bermuda, draw=black] (1.00, -0.25) rectangle (1.25, -0.50);
\filldraw[fill=bermuda, draw=black] (1.25, -0.25) rectangle (1.50, -0.50);
\filldraw[fill=cancan, draw=black] (1.50, -0.25) rectangle (1.75, -0.50);
\filldraw[fill=bermuda, draw=black] (1.75, -0.25) rectangle (2.00, -0.50);
\filldraw[fill=bermuda, draw=black] (2.00, -0.25) rectangle (2.25, -0.50);
\filldraw[fill=bermuda, draw=black] (2.25, -0.25) rectangle (2.50, -0.50);
\filldraw[fill=cancan, draw=black] (2.50, -0.25) rectangle (2.75, -0.50);
\filldraw[fill=cancan, draw=black] (2.75, -0.25) rectangle (3.00, -0.50);
\filldraw[fill=cancan, draw=black] (3.00, -0.25) rectangle (3.25, -0.50);
\filldraw[fill=cancan, draw=black] (3.25, -0.25) rectangle (3.50, -0.50);
\filldraw[fill=bermuda, draw=black] (3.50, -0.25) rectangle (3.75, -0.50);
\filldraw[fill=bermuda, draw=black] (3.75, -0.25) rectangle (4.00, -0.50);
\filldraw[fill=cancan, draw=black] (4.00, -0.25) rectangle (4.25, -0.50);
\filldraw[fill=cancan, draw=black] (4.25, -0.25) rectangle (4.50, -0.50);
\filldraw[fill=cancan, draw=black] (4.50, -0.25) rectangle (4.75, -0.50);
\filldraw[fill=cancan, draw=black] (4.75, -0.25) rectangle (5.00, -0.50);
\filldraw[fill=bermuda, draw=black] (5.00, -0.25) rectangle (5.25, -0.50);
\filldraw[fill=bermuda, draw=black] (5.25, -0.25) rectangle (5.50, -0.50);
\filldraw[fill=cancan, draw=black] (5.50, -0.25) rectangle (5.75, -0.50);
\filldraw[fill=cancan, draw=black] (5.75, -0.25) rectangle (6.00, -0.50);
\filldraw[fill=cancan, draw=black] (6.00, -0.25) rectangle (6.25, -0.50);
\filldraw[fill=cancan, draw=black] (6.25, -0.25) rectangle (6.50, -0.50);
\filldraw[fill=bermuda, draw=black] (6.50, -0.25) rectangle (6.75, -0.50);
\filldraw[fill=bermuda, draw=black] (6.75, -0.25) rectangle (7.00, -0.50);
\filldraw[fill=cancan, draw=black] (7.00, -0.25) rectangle (7.25, -0.50);
\filldraw[fill=cancan, draw=black] (7.25, -0.25) rectangle (7.50, -0.50);
\filldraw[fill=cancan, draw=black] (7.50, -0.25) rectangle (7.75, -0.50);
\filldraw[fill=cancan, draw=black] (7.75, -0.25) rectangle (8.00, -0.50);
\filldraw[fill=cancan, draw=black] (8.00, -0.25) rectangle (8.25, -0.50);
\filldraw[fill=bermuda, draw=black] (8.25, -0.25) rectangle (8.50, -0.50);
\filldraw[fill=cancan, draw=black] (8.50, -0.25) rectangle (8.75, -0.50);
\filldraw[fill=bermuda, draw=black] (8.75, -0.25) rectangle (9.00, -0.50);
\filldraw[fill=bermuda, draw=black] (9.00, -0.25) rectangle (9.25, -0.50);
\filldraw[fill=bermuda, draw=black] (9.25, -0.25) rectangle (9.50, -0.50);
\filldraw[fill=cancan, draw=black] (9.50, -0.25) rectangle (9.75, -0.50);
\filldraw[fill=cancan, draw=black] (9.75, -0.25) rectangle (10.00, -0.50);
\filldraw[fill=cancan, draw=black] (10.00, -0.25) rectangle (10.25, -0.50);
\filldraw[fill=bermuda, draw=black] (10.25, -0.25) rectangle (10.50, -0.50);
\filldraw[fill=bermuda, draw=black] (10.50, -0.25) rectangle (10.75, -0.50);
\filldraw[fill=bermuda, draw=black] (10.75, -0.25) rectangle (11.00, -0.50);
\filldraw[fill=cancan, draw=black] (11.00, -0.25) rectangle (11.25, -0.50);
\filldraw[fill=bermuda, draw=black] (11.25, -0.25) rectangle (11.50, -0.50);
\filldraw[fill=cancan, draw=black] (11.50, -0.25) rectangle (11.75, -0.50);
\filldraw[fill=cancan, draw=black] (11.75, -0.25) rectangle (12.00, -0.50);
\filldraw[fill=cancan, draw=black] (12.00, -0.25) rectangle (12.25, -0.50);
\filldraw[fill=cancan, draw=black] (12.25, -0.25) rectangle (12.50, -0.50);
\filldraw[fill=cancan, draw=black] (12.50, -0.25) rectangle (12.75, -0.50);
\filldraw[fill=bermuda, draw=black] (12.75, -0.25) rectangle (13.00, -0.50);
\filldraw[fill=bermuda, draw=black] (13.00, -0.25) rectangle (13.25, -0.50);
\filldraw[fill=bermuda, draw=black] (13.25, -0.25) rectangle (13.50, -0.50);
\filldraw[fill=cancan, draw=black] (13.50, -0.25) rectangle (13.75, -0.50);
\filldraw[fill=bermuda, draw=black] (13.75, -0.25) rectangle (14.00, -0.50);
\filldraw[fill=cancan, draw=black] (14.00, -0.25) rectangle (14.25, -0.50);
\filldraw[fill=bermuda, draw=black] (14.25, -0.25) rectangle (14.50, -0.50);
\filldraw[fill=bermuda, draw=black] (14.50, -0.25) rectangle (14.75, -0.50);
\filldraw[fill=bermuda, draw=black] (14.75, -0.25) rectangle (15.00, -0.50);
\filldraw[fill=cancan, draw=black] (0.00, -0.50) rectangle (0.25, -0.75);
\filldraw[fill=cancan, draw=black] (0.25, -0.50) rectangle (0.50, -0.75);
\filldraw[fill=cancan, draw=black] (0.50, -0.50) rectangle (0.75, -0.75);
\filldraw[fill=bermuda, draw=black] (0.75, -0.50) rectangle (1.00, -0.75);
\filldraw[fill=bermuda, draw=black] (1.00, -0.50) rectangle (1.25, -0.75);
\filldraw[fill=bermuda, draw=black] (1.25, -0.50) rectangle (1.50, -0.75);
\filldraw[fill=cancan, draw=black] (1.50, -0.50) rectangle (1.75, -0.75);
\filldraw[fill=bermuda, draw=black] (1.75, -0.50) rectangle (2.00, -0.75);
\filldraw[fill=cancan, draw=black] (2.00, -0.50) rectangle (2.25, -0.75);
\filldraw[fill=bermuda, draw=black] (2.25, -0.50) rectangle (2.50, -0.75);
\filldraw[fill=bermuda, draw=black] (2.50, -0.50) rectangle (2.75, -0.75);
\filldraw[fill=bermuda, draw=black] (2.75, -0.50) rectangle (3.00, -0.75);
\filldraw[fill=cancan, draw=black] (3.00, -0.50) rectangle (3.25, -0.75);
\filldraw[fill=bermuda, draw=black] (3.25, -0.50) rectangle (3.50, -0.75);
\filldraw[fill=bermuda, draw=black] (3.50, -0.50) rectangle (3.75, -0.75);
\filldraw[fill=bermuda, draw=black] (3.75, -0.50) rectangle (4.00, -0.75);
\filldraw[fill=bermuda, draw=black] (4.00, -0.50) rectangle (4.25, -0.75);
\filldraw[fill=bermuda, draw=black] (4.25, -0.50) rectangle (4.50, -0.75);
\filldraw[fill=cancan, draw=black] (4.50, -0.50) rectangle (4.75, -0.75);
\filldraw[fill=bermuda, draw=black] (4.75, -0.50) rectangle (5.00, -0.75);
\filldraw[fill=bermuda, draw=black] (5.00, -0.50) rectangle (5.25, -0.75);
\filldraw[fill=bermuda, draw=black] (5.25, -0.50) rectangle (5.50, -0.75);
\filldraw[fill=cancan, draw=black] (5.50, -0.50) rectangle (5.75, -0.75);
\filldraw[fill=cancan, draw=black] (5.75, -0.50) rectangle (6.00, -0.75);
\filldraw[fill=cancan, draw=black] (6.00, -0.50) rectangle (6.25, -0.75);
\filldraw[fill=bermuda, draw=black] (6.25, -0.50) rectangle (6.50, -0.75);
\filldraw[fill=bermuda, draw=black] (6.50, -0.50) rectangle (6.75, -0.75);
\filldraw[fill=bermuda, draw=black] (6.75, -0.50) rectangle (7.00, -0.75);
\filldraw[fill=cancan, draw=black] (7.00, -0.50) rectangle (7.25, -0.75);
\filldraw[fill=bermuda, draw=black] (7.25, -0.50) rectangle (7.50, -0.75);
\filldraw[fill=cancan, draw=black] (7.50, -0.50) rectangle (7.75, -0.75);
\filldraw[fill=bermuda, draw=black] (7.75, -0.50) rectangle (8.00, -0.75);
\filldraw[fill=cancan, draw=black] (8.00, -0.50) rectangle (8.25, -0.75);
\filldraw[fill=cancan, draw=black] (8.25, -0.50) rectangle (8.50, -0.75);
\filldraw[fill=bermuda, draw=black] (8.50, -0.50) rectangle (8.75, -0.75);
\filldraw[fill=bermuda, draw=black] (8.75, -0.50) rectangle (9.00, -0.75);
\filldraw[fill=cancan, draw=black] (9.00, -0.50) rectangle (9.25, -0.75);
\filldraw[fill=bermuda, draw=black] (9.25, -0.50) rectangle (9.50, -0.75);
\filldraw[fill=cancan, draw=black] (9.50, -0.50) rectangle (9.75, -0.75);
\filldraw[fill=cancan, draw=black] (9.75, -0.50) rectangle (10.00, -0.75);
\filldraw[fill=cancan, draw=black] (10.00, -0.50) rectangle (10.25, -0.75);
\filldraw[fill=bermuda, draw=black] (10.25, -0.50) rectangle (10.50, -0.75);
\filldraw[fill=cancan, draw=black] (10.50, -0.50) rectangle (10.75, -0.75);
\filldraw[fill=cancan, draw=black] (10.75, -0.50) rectangle (11.00, -0.75);
\filldraw[fill=cancan, draw=black] (11.00, -0.50) rectangle (11.25, -0.75);
\filldraw[fill=cancan, draw=black] (11.25, -0.50) rectangle (11.50, -0.75);
\filldraw[fill=cancan, draw=black] (11.50, -0.50) rectangle (11.75, -0.75);
\filldraw[fill=cancan, draw=black] (11.75, -0.50) rectangle (12.00, -0.75);
\filldraw[fill=cancan, draw=black] (12.00, -0.50) rectangle (12.25, -0.75);
\filldraw[fill=bermuda, draw=black] (12.25, -0.50) rectangle (12.50, -0.75);
\filldraw[fill=bermuda, draw=black] (12.50, -0.50) rectangle (12.75, -0.75);
\filldraw[fill=bermuda, draw=black] (12.75, -0.50) rectangle (13.00, -0.75);
\filldraw[fill=cancan, draw=black] (13.00, -0.50) rectangle (13.25, -0.75);
\filldraw[fill=cancan, draw=black] (13.25, -0.50) rectangle (13.50, -0.75);
\filldraw[fill=cancan, draw=black] (13.50, -0.50) rectangle (13.75, -0.75);
\filldraw[fill=cancan, draw=black] (13.75, -0.50) rectangle (14.00, -0.75);
\filldraw[fill=bermuda, draw=black] (14.00, -0.50) rectangle (14.25, -0.75);
\filldraw[fill=bermuda, draw=black] (14.25, -0.50) rectangle (14.50, -0.75);
\filldraw[fill=cancan, draw=black] (14.50, -0.50) rectangle (14.75, -0.75);
\filldraw[fill=bermuda, draw=black] (14.75, -0.50) rectangle (15.00, -0.75);
\filldraw[fill=cancan, draw=black] (0.00, -0.75) rectangle (0.25, -1.00);
\filldraw[fill=cancan, draw=black] (0.25, -0.75) rectangle (0.50, -1.00);
\filldraw[fill=bermuda, draw=black] (0.50, -0.75) rectangle (0.75, -1.00);
\filldraw[fill=bermuda, draw=black] (0.75, -0.75) rectangle (1.00, -1.00);
\filldraw[fill=cancan, draw=black] (1.00, -0.75) rectangle (1.25, -1.00);
\filldraw[fill=bermuda, draw=black] (1.25, -0.75) rectangle (1.50, -1.00);
\filldraw[fill=cancan, draw=black] (1.50, -0.75) rectangle (1.75, -1.00);
\filldraw[fill=cancan, draw=black] (1.75, -0.75) rectangle (2.00, -1.00);
\filldraw[fill=cancan, draw=black] (2.00, -0.75) rectangle (2.25, -1.00);
\filldraw[fill=cancan, draw=black] (2.25, -0.75) rectangle (2.50, -1.00);
\filldraw[fill=cancan, draw=black] (2.50, -0.75) rectangle (2.75, -1.00);
\filldraw[fill=cancan, draw=black] (2.75, -0.75) rectangle (3.00, -1.00);
\filldraw[fill=cancan, draw=black] (3.00, -0.75) rectangle (3.25, -1.00);
\filldraw[fill=bermuda, draw=black] (3.25, -0.75) rectangle (3.50, -1.00);
\filldraw[fill=bermuda, draw=black] (3.50, -0.75) rectangle (3.75, -1.00);
\filldraw[fill=bermuda, draw=black] (3.75, -0.75) rectangle (4.00, -1.00);
\filldraw[fill=cancan, draw=black] (4.00, -0.75) rectangle (4.25, -1.00);
\filldraw[fill=bermuda, draw=black] (4.25, -0.75) rectangle (4.50, -1.00);
\filldraw[fill=bermuda, draw=black] (4.50, -0.75) rectangle (4.75, -1.00);
\filldraw[fill=bermuda, draw=black] (4.75, -0.75) rectangle (5.00, -1.00);
\filldraw[fill=bermuda, draw=black] (5.00, -0.75) rectangle (5.25, -1.00);
\filldraw[fill=bermuda, draw=black] (5.25, -0.75) rectangle (5.50, -1.00);
\filldraw[fill=cancan, draw=black] (5.50, -0.75) rectangle (5.75, -1.00);
\filldraw[fill=bermuda, draw=black] (5.75, -0.75) rectangle (6.00, -1.00);
\filldraw[fill=bermuda, draw=black] (6.00, -0.75) rectangle (6.25, -1.00);
\filldraw[fill=bermuda, draw=black] (6.25, -0.75) rectangle (6.50, -1.00);
\filldraw[fill=cancan, draw=black] (6.50, -0.75) rectangle (6.75, -1.00);
\filldraw[fill=cancan, draw=black] (6.75, -0.75) rectangle (7.00, -1.00);
\filldraw[fill=cancan, draw=black] (7.00, -0.75) rectangle (7.25, -1.00);
\filldraw[fill=bermuda, draw=black] (7.25, -0.75) rectangle (7.50, -1.00);
\filldraw[fill=bermuda, draw=black] (7.50, -0.75) rectangle (7.75, -1.00);
\filldraw[fill=bermuda, draw=black] (7.75, -0.75) rectangle (8.00, -1.00);
\filldraw[fill=cancan, draw=black] (8.00, -0.75) rectangle (8.25, -1.00);
\filldraw[fill=bermuda, draw=black] (8.25, -0.75) rectangle (8.50, -1.00);
\filldraw[fill=cancan, draw=black] (8.50, -0.75) rectangle (8.75, -1.00);
\filldraw[fill=bermuda, draw=black] (8.75, -0.75) rectangle (9.00, -1.00);
\filldraw[fill=cancan, draw=black] (9.00, -0.75) rectangle (9.25, -1.00);
\filldraw[fill=cancan, draw=black] (9.25, -0.75) rectangle (9.50, -1.00);
\filldraw[fill=cancan, draw=black] (9.50, -0.75) rectangle (9.75, -1.00);
\filldraw[fill=cancan, draw=black] (9.75, -0.75) rectangle (10.00, -1.00);
\filldraw[fill=bermuda, draw=black] (10.00, -0.75) rectangle (10.25, -1.00);
\filldraw[fill=bermuda, draw=black] (10.25, -0.75) rectangle (10.50, -1.00);
\filldraw[fill=cancan, draw=black] (10.50, -0.75) rectangle (10.75, -1.00);
\filldraw[fill=cancan, draw=black] (10.75, -0.75) rectangle (11.00, -1.00);
\filldraw[fill=cancan, draw=black] (11.00, -0.75) rectangle (11.25, -1.00);
\filldraw[fill=cancan, draw=black] (11.25, -0.75) rectangle (11.50, -1.00);
\filldraw[fill=bermuda, draw=black] (11.50, -0.75) rectangle (11.75, -1.00);
\filldraw[fill=bermuda, draw=black] (11.75, -0.75) rectangle (12.00, -1.00);
\filldraw[fill=cancan, draw=black] (12.00, -0.75) rectangle (12.25, -1.00);
\filldraw[fill=cancan, draw=black] (12.25, -0.75) rectangle (12.50, -1.00);
\filldraw[fill=cancan, draw=black] (12.50, -0.75) rectangle (12.75, -1.00);
\filldraw[fill=bermuda, draw=black] (12.75, -0.75) rectangle (13.00, -1.00);
\filldraw[fill=bermuda, draw=black] (13.00, -0.75) rectangle (13.25, -1.00);
\filldraw[fill=bermuda, draw=black] (13.25, -0.75) rectangle (13.50, -1.00);
\filldraw[fill=cancan, draw=black] (13.50, -0.75) rectangle (13.75, -1.00);
\filldraw[fill=cancan, draw=black] (13.75, -0.75) rectangle (14.00, -1.00);
\filldraw[fill=cancan, draw=black] (14.00, -0.75) rectangle (14.25, -1.00);
\filldraw[fill=bermuda, draw=black] (14.25, -0.75) rectangle (14.50, -1.00);
\filldraw[fill=bermuda, draw=black] (14.50, -0.75) rectangle (14.75, -1.00);
\filldraw[fill=bermuda, draw=black] (14.75, -0.75) rectangle (15.00, -1.00);
\filldraw[fill=cancan, draw=black] (0.00, -1.00) rectangle (0.25, -1.25);
\filldraw[fill=cancan, draw=black] (0.25, -1.00) rectangle (0.50, -1.25);
\filldraw[fill=cancan, draw=black] (0.50, -1.00) rectangle (0.75, -1.25);
\filldraw[fill=bermuda, draw=black] (0.75, -1.00) rectangle (1.00, -1.25);
\filldraw[fill=bermuda, draw=black] (1.00, -1.00) rectangle (1.25, -1.25);
\filldraw[fill=bermuda, draw=black] (1.25, -1.00) rectangle (1.50, -1.25);
\filldraw[fill=cancan, draw=black] (1.50, -1.00) rectangle (1.75, -1.25);
\filldraw[fill=bermuda, draw=black] (1.75, -1.00) rectangle (2.00, -1.25);
\filldraw[fill=bermuda, draw=black] (2.00, -1.00) rectangle (2.25, -1.25);
\filldraw[fill=bermuda, draw=black] (2.25, -1.00) rectangle (2.50, -1.25);
\filldraw[fill=cancan, draw=black] (2.50, -1.00) rectangle (2.75, -1.25);
\filldraw[fill=cancan, draw=black] (2.75, -1.00) rectangle (3.00, -1.25);
\filldraw[fill=cancan, draw=black] (3.00, -1.00) rectangle (3.25, -1.25);
\filldraw[fill=cancan, draw=black] (3.25, -1.00) rectangle (3.50, -1.25);
\filldraw[fill=cancan, draw=black] (3.50, -1.00) rectangle (3.75, -1.25);
\filldraw[fill=bermuda, draw=black] (3.75, -1.00) rectangle (4.00, -1.25);
\filldraw[fill=bermuda, draw=black] (4.00, -1.00) rectangle (4.25, -1.25);
\filldraw[fill=bermuda, draw=black] (4.25, -1.00) rectangle (4.50, -1.25);
\filldraw[fill=bermuda, draw=black] (4.50, -1.00) rectangle (4.75, -1.25);
\filldraw[fill=bermuda, draw=black] (4.75, -1.00) rectangle (5.00, -1.25);
\filldraw[fill=cancan, draw=black] (5.00, -1.00) rectangle (5.25, -1.25);
\filldraw[fill=bermuda, draw=black] (5.25, -1.00) rectangle (5.50, -1.25);
\filldraw[fill=bermuda, draw=black] (5.50, -1.00) rectangle (5.75, -1.25);
\filldraw[fill=bermuda, draw=black] (5.75, -1.00) rectangle (6.00, -1.25);
\filldraw[fill=cancan, draw=black] (6.00, -1.00) rectangle (6.25, -1.25);
\filldraw[fill=cancan, draw=black] (6.25, -1.00) rectangle (6.50, -1.25);
\filldraw[fill=cancan, draw=black] (6.50, -1.00) rectangle (6.75, -1.25);
\filldraw[fill=cancan, draw=black] (6.75, -1.00) rectangle (7.00, -1.25);
\filldraw[fill=cancan, draw=black] (7.00, -1.00) rectangle (7.25, -1.25);
\filldraw[fill=cancan, draw=black] (7.25, -1.00) rectangle (7.50, -1.25);
\filldraw[fill=bermuda, draw=black] (7.50, -1.00) rectangle (7.75, -1.25);
\filldraw[fill=bermuda, draw=black] (7.75, -1.00) rectangle (8.00, -1.25);
\filldraw[fill=cancan, draw=black] (8.00, -1.00) rectangle (8.25, -1.25);
\filldraw[fill=bermuda, draw=black] (8.25, -1.00) rectangle (8.50, -1.25);
\filldraw[fill=cancan, draw=black] (8.50, -1.00) rectangle (8.75, -1.25);
\filldraw[fill=cancan, draw=black] (8.75, -1.00) rectangle (9.00, -1.25);
\filldraw[fill=bermuda, draw=black] (9.00, -1.00) rectangle (9.25, -1.25);
\filldraw[fill=bermuda, draw=black] (9.25, -1.00) rectangle (9.50, -1.25);
\filldraw[fill=cancan, draw=black] (9.50, -1.00) rectangle (9.75, -1.25);
\filldraw[fill=bermuda, draw=black] (9.75, -1.00) rectangle (10.00, -1.25);
\filldraw[fill=cancan, draw=black] (10.00, -1.00) rectangle (10.25, -1.25);
\filldraw[fill=bermuda, draw=black] (10.25, -1.00) rectangle (10.50, -1.25);
\filldraw[fill=bermuda, draw=black] (10.50, -1.00) rectangle (10.75, -1.25);
\filldraw[fill=bermuda, draw=black] (10.75, -1.00) rectangle (11.00, -1.25);
\filldraw[fill=cancan, draw=black] (11.00, -1.00) rectangle (11.25, -1.25);
\filldraw[fill=cancan, draw=black] (11.25, -1.00) rectangle (11.50, -1.25);
\filldraw[fill=cancan, draw=black] (11.50, -1.00) rectangle (11.75, -1.25);
\filldraw[fill=cancan, draw=black] (11.75, -1.00) rectangle (12.00, -1.25);
\filldraw[fill=bermuda, draw=black] (12.00, -1.00) rectangle (12.25, -1.25);
\filldraw[fill=bermuda, draw=black] (12.25, -1.00) rectangle (12.50, -1.25);
\filldraw[fill=cancan, draw=black] (12.50, -1.00) rectangle (12.75, -1.25);
\filldraw[fill=cancan, draw=black] (12.75, -1.00) rectangle (13.00, -1.25);
\filldraw[fill=cancan, draw=black] (13.00, -1.00) rectangle (13.25, -1.25);
\filldraw[fill=cancan, draw=black] (13.25, -1.00) rectangle (13.50, -1.25);
\filldraw[fill=cancan, draw=black] (13.50, -1.00) rectangle (13.75, -1.25);
\filldraw[fill=bermuda, draw=black] (13.75, -1.00) rectangle (14.00, -1.25);
\filldraw[fill=cancan, draw=black] (14.00, -1.00) rectangle (14.25, -1.25);
\filldraw[fill=cancan, draw=black] (14.25, -1.00) rectangle (14.50, -1.25);
\filldraw[fill=cancan, draw=black] (14.50, -1.00) rectangle (14.75, -1.25);
\filldraw[fill=cancan, draw=black] (14.75, -1.00) rectangle (15.00, -1.25);
\filldraw[fill=cancan, draw=black] (0.00, -1.25) rectangle (0.25, -1.50);
\filldraw[fill=bermuda, draw=black] (0.25, -1.25) rectangle (0.50, -1.50);
\filldraw[fill=bermuda, draw=black] (0.50, -1.25) rectangle (0.75, -1.50);
\filldraw[fill=bermuda, draw=black] (0.75, -1.25) rectangle (1.00, -1.50);
\filldraw[fill=cancan, draw=black] (1.00, -1.25) rectangle (1.25, -1.50);
\filldraw[fill=cancan, draw=black] (1.25, -1.25) rectangle (1.50, -1.50);
\filldraw[fill=cancan, draw=black] (1.50, -1.25) rectangle (1.75, -1.50);
\filldraw[fill=cancan, draw=black] (1.75, -1.25) rectangle (2.00, -1.50);
\filldraw[fill=cancan, draw=black] (2.00, -1.25) rectangle (2.25, -1.50);
\filldraw[fill=cancan, draw=black] (2.25, -1.25) rectangle (2.50, -1.50);
\filldraw[fill=cancan, draw=black] (2.50, -1.25) rectangle (2.75, -1.50);
\filldraw[fill=cancan, draw=black] (2.75, -1.25) rectangle (3.00, -1.50);
\filldraw[fill=cancan, draw=black] (3.00, -1.25) rectangle (3.25, -1.50);
\filldraw[fill=cancan, draw=black] (3.25, -1.25) rectangle (3.50, -1.50);
\filldraw[fill=bermuda, draw=black] (3.50, -1.25) rectangle (3.75, -1.50);
\filldraw[fill=bermuda, draw=black] (3.75, -1.25) rectangle (4.00, -1.50);
\filldraw[fill=cancan, draw=black] (4.00, -1.25) rectangle (4.25, -1.50);
\filldraw[fill=bermuda, draw=black] (4.25, -1.25) rectangle (4.50, -1.50);
\filldraw[fill=bermuda, draw=black] (4.50, -1.25) rectangle (4.75, -1.50);
\filldraw[fill=bermuda, draw=black] (4.75, -1.25) rectangle (5.00, -1.50);
\filldraw[fill=cancan, draw=black] (5.00, -1.25) rectangle (5.25, -1.50);
\filldraw[fill=bermuda, draw=black] (5.25, -1.25) rectangle (5.50, -1.50);
\filldraw[fill=cancan, draw=black] (5.50, -1.25) rectangle (5.75, -1.50);
\filldraw[fill=bermuda, draw=black] (5.75, -1.25) rectangle (6.00, -1.50);
\filldraw[fill=cancan, draw=black] (6.00, -1.25) rectangle (6.25, -1.50);
\filldraw[fill=cancan, draw=black] (6.25, -1.25) rectangle (6.50, -1.50);
\filldraw[fill=cancan, draw=black] (6.50, -1.25) rectangle (6.75, -1.50);
\filldraw[fill=bermuda, draw=black] (6.75, -1.25) rectangle (7.00, -1.50);
\filldraw[fill=cancan, draw=black] (7.00, -1.25) rectangle (7.25, -1.50);
\filldraw[fill=bermuda, draw=black] (7.25, -1.25) rectangle (7.50, -1.50);
\filldraw[fill=cancan, draw=black] (7.50, -1.25) rectangle (7.75, -1.50);
\filldraw[fill=bermuda, draw=black] (7.75, -1.25) rectangle (8.00, -1.50);
\filldraw[fill=cancan, draw=black] (8.00, -1.25) rectangle (8.25, -1.50);
\filldraw[fill=bermuda, draw=black] (8.25, -1.25) rectangle (8.50, -1.50);
\filldraw[fill=cancan, draw=black] (8.50, -1.25) rectangle (8.75, -1.50);
\filldraw[fill=cancan, draw=black] (8.75, -1.25) rectangle (9.00, -1.50);
\filldraw[fill=cancan, draw=black] (9.00, -1.25) rectangle (9.25, -1.50);
\filldraw[fill=bermuda, draw=black] (9.25, -1.25) rectangle (9.50, -1.50);
\filldraw[fill=bermuda, draw=black] (9.50, -1.25) rectangle (9.75, -1.50);
\filldraw[fill=bermuda, draw=black] (9.75, -1.25) rectangle (10.00, -1.50);
\filldraw[fill=cancan, draw=black] (10.00, -1.25) rectangle (10.25, -1.50);
\filldraw[fill=bermuda, draw=black] (10.25, -1.25) rectangle (10.50, -1.50);
\filldraw[fill=bermuda, draw=black] (10.50, -1.25) rectangle (10.75, -1.50);
\filldraw[fill=bermuda, draw=black] (10.75, -1.25) rectangle (11.00, -1.50);
\filldraw[fill=cancan, draw=black] (11.00, -1.25) rectangle (11.25, -1.50);
\filldraw[fill=bermuda, draw=black] (11.25, -1.25) rectangle (11.50, -1.50);
\filldraw[fill=cancan, draw=black] (11.50, -1.25) rectangle (11.75, -1.50);
\filldraw[fill=bermuda, draw=black] (11.75, -1.25) rectangle (12.00, -1.50);
\filldraw[fill=cancan, draw=black] (12.00, -1.25) rectangle (12.25, -1.50);
\filldraw[fill=cancan, draw=black] (12.25, -1.25) rectangle (12.50, -1.50);
\filldraw[fill=cancan, draw=black] (12.50, -1.25) rectangle (12.75, -1.50);
\filldraw[fill=bermuda, draw=black] (12.75, -1.25) rectangle (13.00, -1.50);
\filldraw[fill=cancan, draw=black] (13.00, -1.25) rectangle (13.25, -1.50);
\filldraw[fill=bermuda, draw=black] (13.25, -1.25) rectangle (13.50, -1.50);
\filldraw[fill=cancan, draw=black] (13.50, -1.25) rectangle (13.75, -1.50);
\filldraw[fill=cancan, draw=black] (13.75, -1.25) rectangle (14.00, -1.50);
\filldraw[fill=cancan, draw=black] (14.00, -1.25) rectangle (14.25, -1.50);
\filldraw[fill=cancan, draw=black] (14.25, -1.25) rectangle (14.50, -1.50);
\filldraw[fill=cancan, draw=black] (14.50, -1.25) rectangle (14.75, -1.50);
\filldraw[fill=cancan, draw=black] (14.75, -1.25) rectangle (15.00, -1.50);
\filldraw[fill=bermuda, draw=black] (0.00, -1.50) rectangle (0.25, -1.75);
\filldraw[fill=bermuda, draw=black] (0.25, -1.50) rectangle (0.50, -1.75);
\filldraw[fill=bermuda, draw=black] (0.50, -1.50) rectangle (0.75, -1.75);
\filldraw[fill=bermuda, draw=black] (0.75, -1.50) rectangle (1.00, -1.75);
\filldraw[fill=bermuda, draw=black] (1.00, -1.50) rectangle (1.25, -1.75);
\filldraw[fill=bermuda, draw=black] (1.25, -1.50) rectangle (1.50, -1.75);
\filldraw[fill=cancan, draw=black] (1.50, -1.50) rectangle (1.75, -1.75);
\filldraw[fill=bermuda, draw=black] (1.75, -1.50) rectangle (2.00, -1.75);
\filldraw[fill=cancan, draw=black] (2.00, -1.50) rectangle (2.25, -1.75);
\filldraw[fill=cancan, draw=black] (2.25, -1.50) rectangle (2.50, -1.75);
\filldraw[fill=bermuda, draw=black] (2.50, -1.50) rectangle (2.75, -1.75);
\filldraw[fill=bermuda, draw=black] (2.75, -1.50) rectangle (3.00, -1.75);
\filldraw[fill=cancan, draw=black] (3.00, -1.50) rectangle (3.25, -1.75);
\filldraw[fill=bermuda, draw=black] (3.25, -1.50) rectangle (3.50, -1.75);
\filldraw[fill=cancan, draw=black] (3.50, -1.50) rectangle (3.75, -1.75);
\filldraw[fill=cancan, draw=black] (3.75, -1.50) rectangle (4.00, -1.75);
\filldraw[fill=cancan, draw=black] (4.00, -1.50) rectangle (4.25, -1.75);
\filldraw[fill=cancan, draw=black] (4.25, -1.50) rectangle (4.50, -1.75);
\filldraw[fill=cancan, draw=black] (4.50, -1.50) rectangle (4.75, -1.75);
\filldraw[fill=bermuda, draw=black] (4.75, -1.50) rectangle (5.00, -1.75);
\filldraw[fill=cancan, draw=black] (5.00, -1.50) rectangle (5.25, -1.75);
\filldraw[fill=bermuda, draw=black] (5.25, -1.50) rectangle (5.50, -1.75);
\filldraw[fill=cancan, draw=black] (5.50, -1.50) rectangle (5.75, -1.75);
\filldraw[fill=cancan, draw=black] (5.75, -1.50) rectangle (6.00, -1.75);
\filldraw[fill=cancan, draw=black] (6.00, -1.50) rectangle (6.25, -1.75);
\filldraw[fill=bermuda, draw=black] (6.25, -1.50) rectangle (6.50, -1.75);
\filldraw[fill=bermuda, draw=black] (6.50, -1.50) rectangle (6.75, -1.75);
\filldraw[fill=bermuda, draw=black] (6.75, -1.50) rectangle (7.00, -1.75);
\filldraw[fill=cancan, draw=black] (7.00, -1.50) rectangle (7.25, -1.75);
\filldraw[fill=cancan, draw=black] (7.25, -1.50) rectangle (7.50, -1.75);
\filldraw[fill=cancan, draw=black] (7.50, -1.50) rectangle (7.75, -1.75);
\filldraw[fill=cancan, draw=black] (7.75, -1.50) rectangle (8.00, -1.75);
\filldraw[fill=cancan, draw=black] (8.00, -1.50) rectangle (8.25, -1.75);
\filldraw[fill=cancan, draw=black] (8.25, -1.50) rectangle (8.50, -1.75);
\filldraw[fill=cancan, draw=black] (8.50, -1.50) rectangle (8.75, -1.75);
\filldraw[fill=cancan, draw=black] (8.75, -1.50) rectangle (9.00, -1.75);
\filldraw[fill=cancan, draw=black] (9.00, -1.50) rectangle (9.25, -1.75);
\filldraw[fill=bermuda, draw=black] (9.25, -1.50) rectangle (9.50, -1.75);
\filldraw[fill=bermuda, draw=black] (9.50, -1.50) rectangle (9.75, -1.75);
\filldraw[fill=bermuda, draw=black] (9.75, -1.50) rectangle (10.00, -1.75);
\filldraw[fill=cancan, draw=black] (10.00, -1.50) rectangle (10.25, -1.75);
\filldraw[fill=cancan, draw=black] (10.25, -1.50) rectangle (10.50, -1.75);
\filldraw[fill=cancan, draw=black] (10.50, -1.50) rectangle (10.75, -1.75);
\filldraw[fill=bermuda, draw=black] (10.75, -1.50) rectangle (11.00, -1.75);
\filldraw[fill=cancan, draw=black] (11.00, -1.50) rectangle (11.25, -1.75);
\filldraw[fill=bermuda, draw=black] (11.25, -1.50) rectangle (11.50, -1.75);
\filldraw[fill=cancan, draw=black] (11.50, -1.50) rectangle (11.75, -1.75);
\filldraw[fill=cancan, draw=black] (11.75, -1.50) rectangle (12.00, -1.75);
\filldraw[fill=bermuda, draw=black] (12.00, -1.50) rectangle (12.25, -1.75);
\filldraw[fill=bermuda, draw=black] (12.25, -1.50) rectangle (12.50, -1.75);
\filldraw[fill=bermuda, draw=black] (12.50, -1.50) rectangle (12.75, -1.75);
\filldraw[fill=bermuda, draw=black] (12.75, -1.50) rectangle (13.00, -1.75);
\filldraw[fill=cancan, draw=black] (13.00, -1.50) rectangle (13.25, -1.75);
\filldraw[fill=cancan, draw=black] (13.25, -1.50) rectangle (13.50, -1.75);
\filldraw[fill=cancan, draw=black] (13.50, -1.50) rectangle (13.75, -1.75);
\filldraw[fill=cancan, draw=black] (13.75, -1.50) rectangle (14.00, -1.75);
\filldraw[fill=cancan, draw=black] (14.00, -1.50) rectangle (14.25, -1.75);
\filldraw[fill=cancan, draw=black] (14.25, -1.50) rectangle (14.50, -1.75);
\filldraw[fill=cancan, draw=black] (14.50, -1.50) rectangle (14.75, -1.75);
\filldraw[fill=bermuda, draw=black] (14.75, -1.50) rectangle (15.00, -1.75);
\filldraw[fill=bermuda, draw=black] (0.00, -1.75) rectangle (0.25, -2.00);
\filldraw[fill=bermuda, draw=black] (0.25, -1.75) rectangle (0.50, -2.00);
\filldraw[fill=cancan, draw=black] (0.50, -1.75) rectangle (0.75, -2.00);
\filldraw[fill=bermuda, draw=black] (0.75, -1.75) rectangle (1.00, -2.00);
\filldraw[fill=bermuda, draw=black] (1.00, -1.75) rectangle (1.25, -2.00);
\filldraw[fill=bermuda, draw=black] (1.25, -1.75) rectangle (1.50, -2.00);
\filldraw[fill=cancan, draw=black] (1.50, -1.75) rectangle (1.75, -2.00);
\filldraw[fill=cancan, draw=black] (1.75, -1.75) rectangle (2.00, -2.00);
\filldraw[fill=bermuda, draw=black] (2.00, -1.75) rectangle (2.25, -2.00);
\filldraw[fill=bermuda, draw=black] (2.25, -1.75) rectangle (2.50, -2.00);
\filldraw[fill=cancan, draw=black] (2.50, -1.75) rectangle (2.75, -2.00);
\filldraw[fill=bermuda, draw=black] (2.75, -1.75) rectangle (3.00, -2.00);
\filldraw[fill=bermuda, draw=black] (3.00, -1.75) rectangle (3.25, -2.00);
\filldraw[fill=bermuda, draw=black] (3.25, -1.75) rectangle (3.50, -2.00);
\filldraw[fill=bermuda, draw=black] (3.50, -1.75) rectangle (3.75, -2.00);
\filldraw[fill=bermuda, draw=black] (3.75, -1.75) rectangle (4.00, -2.00);
\filldraw[fill=cancan, draw=black] (4.00, -1.75) rectangle (4.25, -2.00);
\filldraw[fill=bermuda, draw=black] (4.25, -1.75) rectangle (4.50, -2.00);
\filldraw[fill=bermuda, draw=black] (4.50, -1.75) rectangle (4.75, -2.00);
\filldraw[fill=bermuda, draw=black] (4.75, -1.75) rectangle (5.00, -2.00);
\filldraw[fill=bermuda, draw=black] (5.00, -1.75) rectangle (5.25, -2.00);
\filldraw[fill=bermuda, draw=black] (5.25, -1.75) rectangle (5.50, -2.00);
\filldraw[fill=bermuda, draw=black] (5.50, -1.75) rectangle (5.75, -2.00);
\filldraw[fill=bermuda, draw=black] (5.75, -1.75) rectangle (6.00, -2.00);
\filldraw[fill=cancan, draw=black] (6.00, -1.75) rectangle (6.25, -2.00);
\filldraw[fill=bermuda, draw=black] (6.25, -1.75) rectangle (6.50, -2.00);
\filldraw[fill=cancan, draw=black] (6.50, -1.75) rectangle (6.75, -2.00);
\filldraw[fill=bermuda, draw=black] (6.75, -1.75) rectangle (7.00, -2.00);
\filldraw[fill=bermuda, draw=black] (7.00, -1.75) rectangle (7.25, -2.00);
\filldraw[fill=bermuda, draw=black] (7.25, -1.75) rectangle (7.50, -2.00);
\filldraw[fill=cancan, draw=black] (7.50, -1.75) rectangle (7.75, -2.00);
\filldraw[fill=bermuda, draw=black] (7.75, -1.75) rectangle (8.00, -2.00);
\filldraw[fill=bermuda, draw=black] (8.00, -1.75) rectangle (8.25, -2.00);
\filldraw[fill=bermuda, draw=black] (8.25, -1.75) rectangle (8.50, -2.00);
\filldraw[fill=cancan, draw=black] (8.50, -1.75) rectangle (8.75, -2.00);
\filldraw[fill=bermuda, draw=black] (8.75, -1.75) rectangle (9.00, -2.00);
\filldraw[fill=cancan, draw=black] (9.00, -1.75) rectangle (9.25, -2.00);
\filldraw[fill=bermuda, draw=black] (9.25, -1.75) rectangle (9.50, -2.00);
\filldraw[fill=cancan, draw=black] (9.50, -1.75) rectangle (9.75, -2.00);
\filldraw[fill=bermuda, draw=black] (9.75, -1.75) rectangle (10.00, -2.00);
\filldraw[fill=bermuda, draw=black] (10.00, -1.75) rectangle (10.25, -2.00);
\filldraw[fill=bermuda, draw=black] (10.25, -1.75) rectangle (10.50, -2.00);
\filldraw[fill=cancan, draw=black] (10.50, -1.75) rectangle (10.75, -2.00);
\filldraw[fill=cancan, draw=black] (10.75, -1.75) rectangle (11.00, -2.00);
\filldraw[fill=cancan, draw=black] (11.00, -1.75) rectangle (11.25, -2.00);
\filldraw[fill=bermuda, draw=black] (11.25, -1.75) rectangle (11.50, -2.00);
\filldraw[fill=bermuda, draw=black] (11.50, -1.75) rectangle (11.75, -2.00);
\filldraw[fill=bermuda, draw=black] (11.75, -1.75) rectangle (12.00, -2.00);
\filldraw[fill=cancan, draw=black] (12.00, -1.75) rectangle (12.25, -2.00);
\filldraw[fill=cancan, draw=black] (12.25, -1.75) rectangle (12.50, -2.00);
\filldraw[fill=cancan, draw=black] (12.50, -1.75) rectangle (12.75, -2.00);
\filldraw[fill=bermuda, draw=black] (12.75, -1.75) rectangle (13.00, -2.00);
\filldraw[fill=bermuda, draw=black] (13.00, -1.75) rectangle (13.25, -2.00);
\filldraw[fill=bermuda, draw=black] (13.25, -1.75) rectangle (13.50, -2.00);
\filldraw[fill=cancan, draw=black] (13.50, -1.75) rectangle (13.75, -2.00);
\filldraw[fill=cancan, draw=black] (13.75, -1.75) rectangle (14.00, -2.00);
\filldraw[fill=cancan, draw=black] (14.00, -1.75) rectangle (14.25, -2.00);
\filldraw[fill=bermuda, draw=black] (14.25, -1.75) rectangle (14.50, -2.00);
\filldraw[fill=bermuda, draw=black] (14.50, -1.75) rectangle (14.75, -2.00);
\filldraw[fill=bermuda, draw=black] (14.75, -1.75) rectangle (15.00, -2.00);
\filldraw[fill=cancan, draw=black] (0.00, -2.00) rectangle (0.25, -2.25);
\filldraw[fill=cancan, draw=black] (0.25, -2.00) rectangle (0.50, -2.25);
\filldraw[fill=cancan, draw=black] (0.50, -2.00) rectangle (0.75, -2.25);
\filldraw[fill=bermuda, draw=black] (0.75, -2.00) rectangle (1.00, -2.25);
\filldraw[fill=bermuda, draw=black] (1.00, -2.00) rectangle (1.25, -2.25);
\filldraw[fill=bermuda, draw=black] (1.25, -2.00) rectangle (1.50, -2.25);
\filldraw[fill=cancan, draw=black] (1.50, -2.00) rectangle (1.75, -2.25);
\filldraw[fill=bermuda, draw=black] (1.75, -2.00) rectangle (2.00, -2.25);
\filldraw[fill=bermuda, draw=black] (2.00, -2.00) rectangle (2.25, -2.25);
\filldraw[fill=bermuda, draw=black] (2.25, -2.00) rectangle (2.50, -2.25);
\filldraw[fill=bermuda, draw=black] (2.50, -2.00) rectangle (2.75, -2.25);
\filldraw[fill=bermuda, draw=black] (2.75, -2.00) rectangle (3.00, -2.25);
\filldraw[fill=cancan, draw=black] (3.00, -2.00) rectangle (3.25, -2.25);
\filldraw[fill=bermuda, draw=black] (3.25, -2.00) rectangle (3.50, -2.25);
\filldraw[fill=cancan, draw=black] (3.50, -2.00) rectangle (3.75, -2.25);
\filldraw[fill=cancan, draw=black] (3.75, -2.00) rectangle (4.00, -2.25);
\filldraw[fill=bermuda, draw=black] (4.00, -2.00) rectangle (4.25, -2.25);
\filldraw[fill=bermuda, draw=black] (4.25, -2.00) rectangle (4.50, -2.25);
\filldraw[fill=cancan, draw=black] (4.50, -2.00) rectangle (4.75, -2.25);
\filldraw[fill=bermuda, draw=black] (4.75, -2.00) rectangle (5.00, -2.25);
\filldraw[fill=cancan, draw=black] (5.00, -2.00) rectangle (5.25, -2.25);
\filldraw[fill=bermuda, draw=black] (5.25, -2.00) rectangle (5.50, -2.25);
\filldraw[fill=cancan, draw=black] (5.50, -2.00) rectangle (5.75, -2.25);
\filldraw[fill=cancan, draw=black] (5.75, -2.00) rectangle (6.00, -2.25);
\filldraw[fill=cancan, draw=black] (6.00, -2.00) rectangle (6.25, -2.25);
\filldraw[fill=bermuda, draw=black] (6.25, -2.00) rectangle (6.50, -2.25);
\filldraw[fill=bermuda, draw=black] (6.50, -2.00) rectangle (6.75, -2.25);
\filldraw[fill=bermuda, draw=black] (6.75, -2.00) rectangle (7.00, -2.25);
\filldraw[fill=cancan, draw=black] (7.00, -2.00) rectangle (7.25, -2.25);
\filldraw[fill=cancan, draw=black] (7.25, -2.00) rectangle (7.50, -2.25);
\filldraw[fill=bermuda, draw=black] (7.50, -2.00) rectangle (7.75, -2.25);
\filldraw[fill=bermuda, draw=black] (7.75, -2.00) rectangle (8.00, -2.25);
\filldraw[fill=cancan, draw=black] (8.00, -2.00) rectangle (8.25, -2.25);
\filldraw[fill=cancan, draw=black] (8.25, -2.00) rectangle (8.50, -2.25);
\filldraw[fill=cancan, draw=black] (8.50, -2.00) rectangle (8.75, -2.25);
\filldraw[fill=cancan, draw=black] (8.75, -2.00) rectangle (9.00, -2.25);
\filldraw[fill=cancan, draw=black] (9.00, -2.00) rectangle (9.25, -2.25);
\filldraw[fill=bermuda, draw=black] (9.25, -2.00) rectangle (9.50, -2.25);
\filldraw[fill=cancan, draw=black] (9.50, -2.00) rectangle (9.75, -2.25);
\filldraw[fill=cancan, draw=black] (9.75, -2.00) rectangle (10.00, -2.25);
\filldraw[fill=cancan, draw=black] (10.00, -2.00) rectangle (10.25, -2.25);
\filldraw[fill=cancan, draw=black] (10.25, -2.00) rectangle (10.50, -2.25);
\filldraw[fill=cancan, draw=black] (10.50, -2.00) rectangle (10.75, -2.25);
\filldraw[fill=bermuda, draw=black] (10.75, -2.00) rectangle (11.00, -2.25);
\filldraw[fill=bermuda, draw=black] (11.00, -2.00) rectangle (11.25, -2.25);
\filldraw[fill=bermuda, draw=black] (11.25, -2.00) rectangle (11.50, -2.25);
\filldraw[fill=cancan, draw=black] (11.50, -2.00) rectangle (11.75, -2.25);
\filldraw[fill=cancan, draw=black] (11.75, -2.00) rectangle (12.00, -2.25);
\filldraw[fill=cancan, draw=black] (12.00, -2.00) rectangle (12.25, -2.25);
\filldraw[fill=bermuda, draw=black] (12.25, -2.00) rectangle (12.50, -2.25);
\filldraw[fill=bermuda, draw=black] (12.50, -2.00) rectangle (12.75, -2.25);
\filldraw[fill=bermuda, draw=black] (12.75, -2.00) rectangle (13.00, -2.25);
\filldraw[fill=cancan, draw=black] (13.00, -2.00) rectangle (13.25, -2.25);
\filldraw[fill=bermuda, draw=black] (13.25, -2.00) rectangle (13.50, -2.25);
\filldraw[fill=bermuda, draw=black] (13.50, -2.00) rectangle (13.75, -2.25);
\filldraw[fill=bermuda, draw=black] (13.75, -2.00) rectangle (14.00, -2.25);
\filldraw[fill=bermuda, draw=black] (14.00, -2.00) rectangle (14.25, -2.25);
\filldraw[fill=bermuda, draw=black] (14.25, -2.00) rectangle (14.50, -2.25);
\filldraw[fill=bermuda, draw=black] (14.50, -2.00) rectangle (14.75, -2.25);
\filldraw[fill=bermuda, draw=black] (14.75, -2.00) rectangle (15.00, -2.25);
\filldraw[fill=cancan, draw=black] (0.00, -2.25) rectangle (0.25, -2.50);
\filldraw[fill=bermuda, draw=black] (0.25, -2.25) rectangle (0.50, -2.50);
\filldraw[fill=cancan, draw=black] (0.50, -2.25) rectangle (0.75, -2.50);
\filldraw[fill=bermuda, draw=black] (0.75, -2.25) rectangle (1.00, -2.50);
\filldraw[fill=bermuda, draw=black] (1.00, -2.25) rectangle (1.25, -2.50);
\filldraw[fill=bermuda, draw=black] (1.25, -2.25) rectangle (1.50, -2.50);
\filldraw[fill=cancan, draw=black] (1.50, -2.25) rectangle (1.75, -2.50);
\filldraw[fill=cancan, draw=black] (1.75, -2.25) rectangle (2.00, -2.50);
\filldraw[fill=cancan, draw=black] (2.00, -2.25) rectangle (2.25, -2.50);
\filldraw[fill=cancan, draw=black] (2.25, -2.25) rectangle (2.50, -2.50);
\filldraw[fill=cancan, draw=black] (2.50, -2.25) rectangle (2.75, -2.50);
\filldraw[fill=cancan, draw=black] (2.75, -2.25) rectangle (3.00, -2.50);
\filldraw[fill=cancan, draw=black] (3.00, -2.25) rectangle (3.25, -2.50);
\filldraw[fill=bermuda, draw=black] (3.25, -2.25) rectangle (3.50, -2.50);
\filldraw[fill=cancan, draw=black] (3.50, -2.25) rectangle (3.75, -2.50);
\filldraw[fill=bermuda, draw=black] (3.75, -2.25) rectangle (4.00, -2.50);
\filldraw[fill=cancan, draw=black] (4.00, -2.25) rectangle (4.25, -2.50);
\filldraw[fill=bermuda, draw=black] (4.25, -2.25) rectangle (4.50, -2.50);
\filldraw[fill=bermuda, draw=black] (4.50, -2.25) rectangle (4.75, -2.50);
\filldraw[fill=bermuda, draw=black] (4.75, -2.25) rectangle (5.00, -2.50);
\filldraw[fill=cancan, draw=black] (5.00, -2.25) rectangle (5.25, -2.50);
\filldraw[fill=cancan, draw=black] (5.25, -2.25) rectangle (5.50, -2.50);
\filldraw[fill=cancan, draw=black] (5.50, -2.25) rectangle (5.75, -2.50);
\filldraw[fill=cancan, draw=black] (5.75, -2.25) rectangle (6.00, -2.50);
\filldraw[fill=cancan, draw=black] (6.00, -2.25) rectangle (6.25, -2.50);
\filldraw[fill=cancan, draw=black] (6.25, -2.25) rectangle (6.50, -2.50);
\filldraw[fill=cancan, draw=black] (6.50, -2.25) rectangle (6.75, -2.50);
\filldraw[fill=bermuda, draw=black] (6.75, -2.25) rectangle (7.00, -2.50);
\filldraw[fill=bermuda, draw=black] (7.00, -2.25) rectangle (7.25, -2.50);
\filldraw[fill=bermuda, draw=black] (7.25, -2.25) rectangle (7.50, -2.50);
\filldraw[fill=cancan, draw=black] (7.50, -2.25) rectangle (7.75, -2.50);
\filldraw[fill=cancan, draw=black] (7.75, -2.25) rectangle (8.00, -2.50);
\filldraw[fill=cancan, draw=black] (8.00, -2.25) rectangle (8.25, -2.50);
\filldraw[fill=bermuda, draw=black] (8.25, -2.25) rectangle (8.50, -2.50);
\filldraw[fill=bermuda, draw=black] (8.50, -2.25) rectangle (8.75, -2.50);
\filldraw[fill=bermuda, draw=black] (8.75, -2.25) rectangle (9.00, -2.50);
\filldraw[fill=bermuda, draw=black] (9.00, -2.25) rectangle (9.25, -2.50);
\filldraw[fill=bermuda, draw=black] (9.25, -2.25) rectangle (9.50, -2.50);
\filldraw[fill=cancan, draw=black] (9.50, -2.25) rectangle (9.75, -2.50);
\filldraw[fill=bermuda, draw=black] (9.75, -2.25) rectangle (10.00, -2.50);
\filldraw[fill=cancan, draw=black] (10.00, -2.25) rectangle (10.25, -2.50);
\filldraw[fill=cancan, draw=black] (10.25, -2.25) rectangle (10.50, -2.50);
\filldraw[fill=cancan, draw=black] (10.50, -2.25) rectangle (10.75, -2.50);
\filldraw[fill=cancan, draw=black] (10.75, -2.25) rectangle (11.00, -2.50);
\filldraw[fill=cancan, draw=black] (11.00, -2.25) rectangle (11.25, -2.50);
\filldraw[fill=cancan, draw=black] (11.25, -2.25) rectangle (11.50, -2.50);
\filldraw[fill=cancan, draw=black] (11.50, -2.25) rectangle (11.75, -2.50);
\filldraw[fill=bermuda, draw=black] (11.75, -2.25) rectangle (12.00, -2.50);
\filldraw[fill=cancan, draw=black] (12.00, -2.25) rectangle (12.25, -2.50);
\filldraw[fill=bermuda, draw=black] (12.25, -2.25) rectangle (12.50, -2.50);
\filldraw[fill=cancan, draw=black] (12.50, -2.25) rectangle (12.75, -2.50);
\filldraw[fill=bermuda, draw=black] (12.75, -2.25) rectangle (13.00, -2.50);
\filldraw[fill=bermuda, draw=black] (13.00, -2.25) rectangle (13.25, -2.50);
\filldraw[fill=bermuda, draw=black] (13.25, -2.25) rectangle (13.50, -2.50);
\filldraw[fill=cancan, draw=black] (13.50, -2.25) rectangle (13.75, -2.50);
\filldraw[fill=cancan, draw=black] (13.75, -2.25) rectangle (14.00, -2.50);
\filldraw[fill=cancan, draw=black] (14.00, -2.25) rectangle (14.25, -2.50);
\filldraw[fill=bermuda, draw=black] (14.25, -2.25) rectangle (14.50, -2.50);
\filldraw[fill=bermuda, draw=black] (14.50, -2.25) rectangle (14.75, -2.50);
\filldraw[fill=bermuda, draw=black] (14.75, -2.25) rectangle (15.00, -2.50);
\filldraw[fill=cancan, draw=black] (0.00, -2.50) rectangle (0.25, -2.75);
\filldraw[fill=bermuda, draw=black] (0.25, -2.50) rectangle (0.50, -2.75);
\filldraw[fill=cancan, draw=black] (0.50, -2.50) rectangle (0.75, -2.75);
\filldraw[fill=bermuda, draw=black] (0.75, -2.50) rectangle (1.00, -2.75);
\filldraw[fill=bermuda, draw=black] (1.00, -2.50) rectangle (1.25, -2.75);
\filldraw[fill=bermuda, draw=black] (1.25, -2.50) rectangle (1.50, -2.75);
\filldraw[fill=cancan, draw=black] (1.50, -2.50) rectangle (1.75, -2.75);
\filldraw[fill=bermuda, draw=black] (1.75, -2.50) rectangle (2.00, -2.75);
\filldraw[fill=bermuda, draw=black] (2.00, -2.50) rectangle (2.25, -2.75);
\filldraw[fill=bermuda, draw=black] (2.25, -2.50) rectangle (2.50, -2.75);
\filldraw[fill=cancan, draw=black] (2.50, -2.50) rectangle (2.75, -2.75);
\filldraw[fill=bermuda, draw=black] (2.75, -2.50) rectangle (3.00, -2.75);
\filldraw[fill=cancan, draw=black] (3.00, -2.50) rectangle (3.25, -2.75);
\filldraw[fill=cancan, draw=black] (3.25, -2.50) rectangle (3.50, -2.75);
\filldraw[fill=cancan, draw=black] (3.50, -2.50) rectangle (3.75, -2.75);
\filldraw[fill=cancan, draw=black] (3.75, -2.50) rectangle (4.00, -2.75);
\filldraw[fill=bermuda, draw=black] (4.00, -2.50) rectangle (4.25, -2.75);
\filldraw[fill=bermuda, draw=black] (4.25, -2.50) rectangle (4.50, -2.75);
\filldraw[fill=cancan, draw=black] (4.50, -2.50) rectangle (4.75, -2.75);
\filldraw[fill=cancan, draw=black] (4.75, -2.50) rectangle (5.00, -2.75);
\filldraw[fill=cancan, draw=black] (5.00, -2.50) rectangle (5.25, -2.75);
\filldraw[fill=bermuda, draw=black] (5.25, -2.50) rectangle (5.50, -2.75);
\filldraw[fill=bermuda, draw=black] (5.50, -2.50) rectangle (5.75, -2.75);
\filldraw[fill=bermuda, draw=black] (5.75, -2.50) rectangle (6.00, -2.75);
\filldraw[fill=cancan, draw=black] (6.00, -2.50) rectangle (6.25, -2.75);
\filldraw[fill=cancan, draw=black] (6.25, -2.50) rectangle (6.50, -2.75);
\filldraw[fill=cancan, draw=black] (6.50, -2.50) rectangle (6.75, -2.75);
\filldraw[fill=bermuda, draw=black] (6.75, -2.50) rectangle (7.00, -2.75);
\filldraw[fill=bermuda, draw=black] (7.00, -2.50) rectangle (7.25, -2.75);
\filldraw[fill=bermuda, draw=black] (7.25, -2.50) rectangle (7.50, -2.75);
\filldraw[fill=cancan, draw=black] (7.50, -2.50) rectangle (7.75, -2.75);
\filldraw[fill=bermuda, draw=black] (7.75, -2.50) rectangle (8.00, -2.75);
\filldraw[fill=bermuda, draw=black] (8.00, -2.50) rectangle (8.25, -2.75);
\filldraw[fill=bermuda, draw=black] (8.25, -2.50) rectangle (8.50, -2.75);
\filldraw[fill=cancan, draw=black] (8.50, -2.50) rectangle (8.75, -2.75);
\filldraw[fill=bermuda, draw=black] (8.75, -2.50) rectangle (9.00, -2.75);
\filldraw[fill=cancan, draw=black] (9.00, -2.50) rectangle (9.25, -2.75);
\filldraw[fill=cancan, draw=black] (9.25, -2.50) rectangle (9.50, -2.75);
\filldraw[fill=cancan, draw=black] (9.50, -2.50) rectangle (9.75, -2.75);
\filldraw[fill=cancan, draw=black] (9.75, -2.50) rectangle (10.00, -2.75);
\filldraw[fill=bermuda, draw=black] (10.00, -2.50) rectangle (10.25, -2.75);
\filldraw[fill=bermuda, draw=black] (10.25, -2.50) rectangle (10.50, -2.75);
\filldraw[fill=cancan, draw=black] (10.50, -2.50) rectangle (10.75, -2.75);
\filldraw[fill=cancan, draw=black] (10.75, -2.50) rectangle (11.00, -2.75);
\filldraw[fill=cancan, draw=black] (11.00, -2.50) rectangle (11.25, -2.75);
\filldraw[fill=bermuda, draw=black] (11.25, -2.50) rectangle (11.50, -2.75);
\filldraw[fill=bermuda, draw=black] (11.50, -2.50) rectangle (11.75, -2.75);
\filldraw[fill=bermuda, draw=black] (11.75, -2.50) rectangle (12.00, -2.75);
\filldraw[fill=cancan, draw=black] (12.00, -2.50) rectangle (12.25, -2.75);
\filldraw[fill=cancan, draw=black] (12.25, -2.50) rectangle (12.50, -2.75);
\filldraw[fill=cancan, draw=black] (12.50, -2.50) rectangle (12.75, -2.75);
\filldraw[fill=cancan, draw=black] (12.75, -2.50) rectangle (13.00, -2.75);
\filldraw[fill=cancan, draw=black] (13.00, -2.50) rectangle (13.25, -2.75);
\filldraw[fill=bermuda, draw=black] (13.25, -2.50) rectangle (13.50, -2.75);
\filldraw[fill=cancan, draw=black] (13.50, -2.50) rectangle (13.75, -2.75);
\filldraw[fill=cancan, draw=black] (13.75, -2.50) rectangle (14.00, -2.75);
\filldraw[fill=cancan, draw=black] (14.00, -2.50) rectangle (14.25, -2.75);
\filldraw[fill=bermuda, draw=black] (14.25, -2.50) rectangle (14.50, -2.75);
\filldraw[fill=bermuda, draw=black] (14.50, -2.50) rectangle (14.75, -2.75);
\filldraw[fill=bermuda, draw=black] (14.75, -2.50) rectangle (15.00, -2.75);
\filldraw[fill=cancan, draw=black] (0.00, -2.75) rectangle (0.25, -3.00);
\filldraw[fill=cancan, draw=black] (0.25, -2.75) rectangle (0.50, -3.00);
\filldraw[fill=cancan, draw=black] (0.50, -2.75) rectangle (0.75, -3.00);
\filldraw[fill=cancan, draw=black] (0.75, -2.75) rectangle (1.00, -3.00);
\filldraw[fill=cancan, draw=black] (1.00, -2.75) rectangle (1.25, -3.00);
\filldraw[fill=bermuda, draw=black] (1.25, -2.75) rectangle (1.50, -3.00);
\filldraw[fill=cancan, draw=black] (1.50, -2.75) rectangle (1.75, -3.00);
\filldraw[fill=bermuda, draw=black] (1.75, -2.75) rectangle (2.00, -3.00);
\filldraw[fill=cancan, draw=black] (2.00, -2.75) rectangle (2.25, -3.00);
\filldraw[fill=cancan, draw=black] (2.25, -2.75) rectangle (2.50, -3.00);
\filldraw[fill=cancan, draw=black] (2.50, -2.75) rectangle (2.75, -3.00);
\filldraw[fill=bermuda, draw=black] (2.75, -2.75) rectangle (3.00, -3.00);
\filldraw[fill=cancan, draw=black] (3.00, -2.75) rectangle (3.25, -3.00);
\filldraw[fill=bermuda, draw=black] (3.25, -2.75) rectangle (3.50, -3.00);
\filldraw[fill=cancan, draw=black] (3.50, -2.75) rectangle (3.75, -3.00);
\filldraw[fill=bermuda, draw=black] (3.75, -2.75) rectangle (4.00, -3.00);
\filldraw[fill=cancan, draw=black] (4.00, -2.75) rectangle (4.25, -3.00);
\filldraw[fill=bermuda, draw=black] (4.25, -2.75) rectangle (4.50, -3.00);
\filldraw[fill=bermuda, draw=black] (4.50, -2.75) rectangle (4.75, -3.00);
\filldraw[fill=bermuda, draw=black] (4.75, -2.75) rectangle (5.00, -3.00);
\filldraw[fill=cancan, draw=black] (5.00, -2.75) rectangle (5.25, -3.00);
\filldraw[fill=cancan, draw=black] (5.25, -2.75) rectangle (5.50, -3.00);
\filldraw[fill=cancan, draw=black] (5.50, -2.75) rectangle (5.75, -3.00);
\filldraw[fill=bermuda, draw=black] (5.75, -2.75) rectangle (6.00, -3.00);
\filldraw[fill=bermuda, draw=black] (6.00, -2.75) rectangle (6.25, -3.00);
\filldraw[fill=bermuda, draw=black] (6.25, -2.75) rectangle (6.50, -3.00);
\filldraw[fill=cancan, draw=black] (6.50, -2.75) rectangle (6.75, -3.00);
\filldraw[fill=cancan, draw=black] (6.75, -2.75) rectangle (7.00, -3.00);
\filldraw[fill=cancan, draw=black] (7.00, -2.75) rectangle (7.25, -3.00);
\filldraw[fill=bermuda, draw=black] (7.25, -2.75) rectangle (7.50, -3.00);
\filldraw[fill=bermuda, draw=black] (7.50, -2.75) rectangle (7.75, -3.00);
\filldraw[fill=bermuda, draw=black] (7.75, -2.75) rectangle (8.00, -3.00);
\filldraw[fill=cancan, draw=black] (8.00, -2.75) rectangle (8.25, -3.00);
\filldraw[fill=bermuda, draw=black] (8.25, -2.75) rectangle (8.50, -3.00);
\filldraw[fill=cancan, draw=black] (8.50, -2.75) rectangle (8.75, -3.00);
\filldraw[fill=bermuda, draw=black] (8.75, -2.75) rectangle (9.00, -3.00);
\filldraw[fill=bermuda, draw=black] (9.00, -2.75) rectangle (9.25, -3.00);
\filldraw[fill=bermuda, draw=black] (9.25, -2.75) rectangle (9.50, -3.00);
\filldraw[fill=cancan, draw=black] (9.50, -2.75) rectangle (9.75, -3.00);
\filldraw[fill=cancan, draw=black] (9.75, -2.75) rectangle (10.00, -3.00);
\filldraw[fill=cancan, draw=black] (10.00, -2.75) rectangle (10.25, -3.00);
\filldraw[fill=bermuda, draw=black] (10.25, -2.75) rectangle (10.50, -3.00);
\filldraw[fill=bermuda, draw=black] (10.50, -2.75) rectangle (10.75, -3.00);
\filldraw[fill=bermuda, draw=black] (10.75, -2.75) rectangle (11.00, -3.00);
\filldraw[fill=cancan, draw=black] (11.00, -2.75) rectangle (11.25, -3.00);
\filldraw[fill=cancan, draw=black] (11.25, -2.75) rectangle (11.50, -3.00);
\filldraw[fill=cancan, draw=black] (11.50, -2.75) rectangle (11.75, -3.00);
\filldraw[fill=bermuda, draw=black] (11.75, -2.75) rectangle (12.00, -3.00);
\filldraw[fill=bermuda, draw=black] (12.00, -2.75) rectangle (12.25, -3.00);
\filldraw[fill=bermuda, draw=black] (12.25, -2.75) rectangle (12.50, -3.00);
\filldraw[fill=cancan, draw=black] (12.50, -2.75) rectangle (12.75, -3.00);
\filldraw[fill=cancan, draw=black] (12.75, -2.75) rectangle (13.00, -3.00);
\filldraw[fill=cancan, draw=black] (13.00, -2.75) rectangle (13.25, -3.00);
\filldraw[fill=cancan, draw=black] (13.25, -2.75) rectangle (13.50, -3.00);
\filldraw[fill=cancan, draw=black] (13.50, -2.75) rectangle (13.75, -3.00);
\filldraw[fill=bermuda, draw=black] (13.75, -2.75) rectangle (14.00, -3.00);
\filldraw[fill=cancan, draw=black] (14.00, -2.75) rectangle (14.25, -3.00);
\filldraw[fill=bermuda, draw=black] (14.25, -2.75) rectangle (14.50, -3.00);
\filldraw[fill=cancan, draw=black] (14.50, -2.75) rectangle (14.75, -3.00);
\filldraw[fill=cancan, draw=black] (14.75, -2.75) rectangle (15.00, -3.00);
\filldraw[fill=cancan, draw=black] (0.00, -3.00) rectangle (0.25, -3.25);
\filldraw[fill=bermuda, draw=black] (0.25, -3.00) rectangle (0.50, -3.25);
\filldraw[fill=cancan, draw=black] (0.50, -3.00) rectangle (0.75, -3.25);
\filldraw[fill=bermuda, draw=black] (0.75, -3.00) rectangle (1.00, -3.25);
\filldraw[fill=cancan, draw=black] (1.00, -3.00) rectangle (1.25, -3.25);
\filldraw[fill=cancan, draw=black] (1.25, -3.00) rectangle (1.50, -3.25);
\filldraw[fill=cancan, draw=black] (1.50, -3.00) rectangle (1.75, -3.25);
\filldraw[fill=bermuda, draw=black] (1.75, -3.00) rectangle (2.00, -3.25);
\filldraw[fill=cancan, draw=black] (2.00, -3.00) rectangle (2.25, -3.25);
\filldraw[fill=bermuda, draw=black] (2.25, -3.00) rectangle (2.50, -3.25);
\filldraw[fill=cancan, draw=black] (2.50, -3.00) rectangle (2.75, -3.25);
\filldraw[fill=cancan, draw=black] (2.75, -3.00) rectangle (3.00, -3.25);
\filldraw[fill=cancan, draw=black] (3.00, -3.00) rectangle (3.25, -3.25);
\filldraw[fill=bermuda, draw=black] (3.25, -3.00) rectangle (3.50, -3.25);
\filldraw[fill=cancan, draw=black] (3.50, -3.00) rectangle (3.75, -3.25);
\filldraw[fill=bermuda, draw=black] (3.75, -3.00) rectangle (4.00, -3.25);
\filldraw[fill=cancan, draw=black] (4.00, -3.00) rectangle (4.25, -3.25);
\filldraw[fill=cancan, draw=black] (4.25, -3.00) rectangle (4.50, -3.25);
\filldraw[fill=cancan, draw=black] (4.50, -3.00) rectangle (4.75, -3.25);
\filldraw[fill=bermuda, draw=black] (4.75, -3.00) rectangle (5.00, -3.25);
\filldraw[fill=cancan, draw=black] (5.00, -3.00) rectangle (5.25, -3.25);
\filldraw[fill=cancan, draw=black] (5.25, -3.00) rectangle (5.50, -3.25);
\filldraw[fill=cancan, draw=black] (5.50, -3.00) rectangle (5.75, -3.25);
\filldraw[fill=cancan, draw=black] (5.75, -3.00) rectangle (6.00, -3.25);
\filldraw[fill=cancan, draw=black] (6.00, -3.00) rectangle (6.25, -3.25);
\filldraw[fill=bermuda, draw=black] (6.25, -3.00) rectangle (6.50, -3.25);
\filldraw[fill=bermuda, draw=black] (6.50, -3.00) rectangle (6.75, -3.25);
\filldraw[fill=bermuda, draw=black] (6.75, -3.00) rectangle (7.00, -3.25);
\filldraw[fill=cancan, draw=black] (7.00, -3.00) rectangle (7.25, -3.25);
\filldraw[fill=bermuda, draw=black] (7.25, -3.00) rectangle (7.50, -3.25);
\filldraw[fill=bermuda, draw=black] (7.50, -3.00) rectangle (7.75, -3.25);
\filldraw[fill=bermuda, draw=black] (7.75, -3.00) rectangle (8.00, -3.25);
\filldraw[fill=cancan, draw=black] (8.00, -3.00) rectangle (8.25, -3.25);
\filldraw[fill=bermuda, draw=black] (8.25, -3.00) rectangle (8.50, -3.25);
\filldraw[fill=bermuda, draw=black] (8.50, -3.00) rectangle (8.75, -3.25);
\filldraw[fill=bermuda, draw=black] (8.75, -3.00) rectangle (9.00, -3.25);
\filldraw[fill=cancan, draw=black] (9.00, -3.00) rectangle (9.25, -3.25);
\filldraw[fill=bermuda, draw=black] (9.25, -3.00) rectangle (9.50, -3.25);
\filldraw[fill=bermuda, draw=black] (9.50, -3.00) rectangle (9.75, -3.25);
\filldraw[fill=bermuda, draw=black] (9.75, -3.00) rectangle (10.00, -3.25);
\filldraw[fill=cancan, draw=black] (10.00, -3.00) rectangle (10.25, -3.25);
\filldraw[fill=cancan, draw=black] (10.25, -3.00) rectangle (10.50, -3.25);
\filldraw[fill=cancan, draw=black] (10.50, -3.00) rectangle (10.75, -3.25);
\filldraw[fill=bermuda, draw=black] (10.75, -3.00) rectangle (11.00, -3.25);
\filldraw[fill=bermuda, draw=black] (11.00, -3.00) rectangle (11.25, -3.25);
\filldraw[fill=bermuda, draw=black] (11.25, -3.00) rectangle (11.50, -3.25);
\filldraw[fill=cancan, draw=black] (11.50, -3.00) rectangle (11.75, -3.25);
\filldraw[fill=cancan, draw=black] (11.75, -3.00) rectangle (12.00, -3.25);
\filldraw[fill=cancan, draw=black] (12.00, -3.00) rectangle (12.25, -3.25);
\filldraw[fill=cancan, draw=black] (12.25, -3.00) rectangle (12.50, -3.25);
\filldraw[fill=cancan, draw=black] (12.50, -3.00) rectangle (12.75, -3.25);
\filldraw[fill=bermuda, draw=black] (12.75, -3.00) rectangle (13.00, -3.25);
\filldraw[fill=cancan, draw=black] (13.00, -3.00) rectangle (13.25, -3.25);
\filldraw[fill=cancan, draw=black] (13.25, -3.00) rectangle (13.50, -3.25);
\filldraw[fill=bermuda, draw=black] (13.50, -3.00) rectangle (13.75, -3.25);
\filldraw[fill=bermuda, draw=black] (13.75, -3.00) rectangle (14.00, -3.25);
\filldraw[fill=cancan, draw=black] (14.00, -3.00) rectangle (14.25, -3.25);
\filldraw[fill=bermuda, draw=black] (14.25, -3.00) rectangle (14.50, -3.25);
\filldraw[fill=cancan, draw=black] (14.50, -3.00) rectangle (14.75, -3.25);
\filldraw[fill=bermuda, draw=black] (14.75, -3.00) rectangle (15.00, -3.25);
\filldraw[fill=bermuda, draw=black] (0.00, -3.25) rectangle (0.25, -3.50);
\filldraw[fill=bermuda, draw=black] (0.25, -3.25) rectangle (0.50, -3.50);
\filldraw[fill=cancan, draw=black] (0.50, -3.25) rectangle (0.75, -3.50);
\filldraw[fill=bermuda, draw=black] (0.75, -3.25) rectangle (1.00, -3.50);
\filldraw[fill=bermuda, draw=black] (1.00, -3.25) rectangle (1.25, -3.50);
\filldraw[fill=bermuda, draw=black] (1.25, -3.25) rectangle (1.50, -3.50);
\filldraw[fill=bermuda, draw=black] (1.50, -3.25) rectangle (1.75, -3.50);
\filldraw[fill=bermuda, draw=black] (1.75, -3.25) rectangle (2.00, -3.50);
\filldraw[fill=cancan, draw=black] (2.00, -3.25) rectangle (2.25, -3.50);
\filldraw[fill=bermuda, draw=black] (2.25, -3.25) rectangle (2.50, -3.50);
\filldraw[fill=cancan, draw=black] (2.50, -3.25) rectangle (2.75, -3.50);
\filldraw[fill=cancan, draw=black] (2.75, -3.25) rectangle (3.00, -3.50);
\filldraw[fill=bermuda, draw=black] (3.00, -3.25) rectangle (3.25, -3.50);
\filldraw[fill=bermuda, draw=black] (3.25, -3.25) rectangle (3.50, -3.50);
\filldraw[fill=cancan, draw=black] (3.50, -3.25) rectangle (3.75, -3.50);
\filldraw[fill=bermuda, draw=black] (3.75, -3.25) rectangle (4.00, -3.50);
\filldraw[fill=cancan, draw=black] (4.00, -3.25) rectangle (4.25, -3.50);
\filldraw[fill=bermuda, draw=black] (4.25, -3.25) rectangle (4.50, -3.50);
\filldraw[fill=cancan, draw=black] (4.50, -3.25) rectangle (4.75, -3.50);
\filldraw[fill=bermuda, draw=black] (4.75, -3.25) rectangle (5.00, -3.50);
\filldraw[fill=bermuda, draw=black] (5.00, -3.25) rectangle (5.25, -3.50);
\filldraw[fill=bermuda, draw=black] (5.25, -3.25) rectangle (5.50, -3.50);
\filldraw[fill=cancan, draw=black] (5.50, -3.25) rectangle (5.75, -3.50);
\filldraw[fill=cancan, draw=black] (5.75, -3.25) rectangle (6.00, -3.50);
\filldraw[fill=cancan, draw=black] (6.00, -3.25) rectangle (6.25, -3.50);
\filldraw[fill=bermuda, draw=black] (6.25, -3.25) rectangle (6.50, -3.50);
\filldraw[fill=bermuda, draw=black] (6.50, -3.25) rectangle (6.75, -3.50);
\filldraw[fill=bermuda, draw=black] (6.75, -3.25) rectangle (7.00, -3.50);
\filldraw[fill=cancan, draw=black] (7.00, -3.25) rectangle (7.25, -3.50);
\filldraw[fill=cancan, draw=black] (7.25, -3.25) rectangle (7.50, -3.50);
\filldraw[fill=cancan, draw=black] (7.50, -3.25) rectangle (7.75, -3.50);
\filldraw[fill=bermuda, draw=black] (7.75, -3.25) rectangle (8.00, -3.50);
\filldraw[fill=bermuda, draw=black] (8.00, -3.25) rectangle (8.25, -3.50);
\filldraw[fill=bermuda, draw=black] (8.25, -3.25) rectangle (8.50, -3.50);
} } }\end{equation*}
\begin{equation*}
\hspace{4.6pt} b_{5} = \vcenter{\hbox{ \tikz{
\filldraw[fill=bermuda, draw=black] (0.00, 0.00) rectangle (0.25, -0.25);
\filldraw[fill=bermuda, draw=black] (0.25, 0.00) rectangle (0.50, -0.25);
\filldraw[fill=cancan, draw=black] (0.50, 0.00) rectangle (0.75, -0.25);
\filldraw[fill=bermuda, draw=black] (0.75, 0.00) rectangle (1.00, -0.25);
\filldraw[fill=cancan, draw=black] (1.00, 0.00) rectangle (1.25, -0.25);
\filldraw[fill=cancan, draw=black] (1.25, 0.00) rectangle (1.50, -0.25);
\filldraw[fill=cancan, draw=black] (1.50, 0.00) rectangle (1.75, -0.25);
\filldraw[fill=cancan, draw=black] (1.75, 0.00) rectangle (2.00, -0.25);
\filldraw[fill=bermuda, draw=black] (2.00, 0.00) rectangle (2.25, -0.25);
\filldraw[fill=bermuda, draw=black] (2.25, 0.00) rectangle (2.50, -0.25);
\filldraw[fill=cancan, draw=black] (2.50, 0.00) rectangle (2.75, -0.25);
\filldraw[fill=cancan, draw=black] (2.75, 0.00) rectangle (3.00, -0.25);
\filldraw[fill=cancan, draw=black] (3.00, 0.00) rectangle (3.25, -0.25);
\filldraw[fill=cancan, draw=black] (3.25, 0.00) rectangle (3.50, -0.25);
\filldraw[fill=bermuda, draw=black] (3.50, 0.00) rectangle (3.75, -0.25);
\filldraw[fill=bermuda, draw=black] (3.75, 0.00) rectangle (4.00, -0.25);
\filldraw[fill=cancan, draw=black] (4.00, 0.00) rectangle (4.25, -0.25);
\filldraw[fill=bermuda, draw=black] (4.25, 0.00) rectangle (4.50, -0.25);
\filldraw[fill=cancan, draw=black] (4.50, 0.00) rectangle (4.75, -0.25);
\filldraw[fill=cancan, draw=black] (4.75, 0.00) rectangle (5.00, -0.25);
\filldraw[fill=bermuda, draw=black] (5.00, 0.00) rectangle (5.25, -0.25);
\filldraw[fill=bermuda, draw=black] (5.25, 0.00) rectangle (5.50, -0.25);
\filldraw[fill=cancan, draw=black] (5.50, 0.00) rectangle (5.75, -0.25);
\filldraw[fill=bermuda, draw=black] (5.75, 0.00) rectangle (6.00, -0.25);
\filldraw[fill=cancan, draw=black] (6.00, 0.00) rectangle (6.25, -0.25);
\filldraw[fill=cancan, draw=black] (6.25, 0.00) rectangle (6.50, -0.25);
\filldraw[fill=bermuda, draw=black] (6.50, 0.00) rectangle (6.75, -0.25);
\filldraw[fill=bermuda, draw=black] (6.75, 0.00) rectangle (7.00, -0.25);
\filldraw[fill=cancan, draw=black] (7.00, 0.00) rectangle (7.25, -0.25);
\filldraw[fill=bermuda, draw=black] (7.25, 0.00) rectangle (7.50, -0.25);
\filldraw[fill=cancan, draw=black] (7.50, 0.00) rectangle (7.75, -0.25);
\filldraw[fill=cancan, draw=black] (7.75, 0.00) rectangle (8.00, -0.25);
\filldraw[fill=bermuda, draw=black] (8.00, 0.00) rectangle (8.25, -0.25);
\filldraw[fill=bermuda, draw=black] (8.25, 0.00) rectangle (8.50, -0.25);
\filldraw[fill=cancan, draw=black] (8.50, 0.00) rectangle (8.75, -0.25);
\filldraw[fill=bermuda, draw=black] (8.75, 0.00) rectangle (9.00, -0.25);
\filldraw[fill=cancan, draw=black] (9.00, 0.00) rectangle (9.25, -0.25);
\filldraw[fill=cancan, draw=black] (9.25, 0.00) rectangle (9.50, -0.25);
\filldraw[fill=bermuda, draw=black] (9.50, 0.00) rectangle (9.75, -0.25);
\filldraw[fill=bermuda, draw=black] (9.75, 0.00) rectangle (10.00, -0.25);
\filldraw[fill=cancan, draw=black] (10.00, 0.00) rectangle (10.25, -0.25);
\filldraw[fill=bermuda, draw=black] (10.25, 0.00) rectangle (10.50, -0.25);
\filldraw[fill=bermuda, draw=black] (10.50, 0.00) rectangle (10.75, -0.25);
\filldraw[fill=bermuda, draw=black] (10.75, 0.00) rectangle (11.00, -0.25);
\filldraw[fill=cancan, draw=black] (11.00, 0.00) rectangle (11.25, -0.25);
\filldraw[fill=cancan, draw=black] (11.25, 0.00) rectangle (11.50, -0.25);
\filldraw[fill=cancan, draw=black] (11.50, 0.00) rectangle (11.75, -0.25);
\filldraw[fill=bermuda, draw=black] (11.75, 0.00) rectangle (12.00, -0.25);
\filldraw[fill=bermuda, draw=black] (12.00, 0.00) rectangle (12.25, -0.25);
\filldraw[fill=bermuda, draw=black] (12.25, 0.00) rectangle (12.50, -0.25);
\filldraw[fill=cancan, draw=black] (12.50, 0.00) rectangle (12.75, -0.25);
\filldraw[fill=cancan, draw=black] (12.75, 0.00) rectangle (13.00, -0.25);
\filldraw[fill=bermuda, draw=black] (13.00, 0.00) rectangle (13.25, -0.25);
\filldraw[fill=bermuda, draw=black] (13.25, 0.00) rectangle (13.50, -0.25);
\filldraw[fill=cancan, draw=black] (13.50, 0.00) rectangle (13.75, -0.25);
\filldraw[fill=cancan, draw=black] (13.75, 0.00) rectangle (14.00, -0.25);
\filldraw[fill=cancan, draw=black] (14.00, 0.00) rectangle (14.25, -0.25);
\filldraw[fill=cancan, draw=black] (14.25, 0.00) rectangle (14.50, -0.25);
\filldraw[fill=cancan, draw=black] (14.50, 0.00) rectangle (14.75, -0.25);
\filldraw[fill=cancan, draw=black] (14.75, 0.00) rectangle (15.00, -0.25);
\filldraw[fill=bermuda, draw=black] (0.00, -0.25) rectangle (0.25, -0.50);
\filldraw[fill=bermuda, draw=black] (0.25, -0.25) rectangle (0.50, -0.50);
\filldraw[fill=cancan, draw=black] (0.50, -0.25) rectangle (0.75, -0.50);
\filldraw[fill=cancan, draw=black] (0.75, -0.25) rectangle (1.00, -0.50);
\filldraw[fill=cancan, draw=black] (1.00, -0.25) rectangle (1.25, -0.50);
\filldraw[fill=bermuda, draw=black] (1.25, -0.25) rectangle (1.50, -0.50);
\filldraw[fill=bermuda, draw=black] (1.50, -0.25) rectangle (1.75, -0.50);
\filldraw[fill=bermuda, draw=black] (1.75, -0.25) rectangle (2.00, -0.50);
\filldraw[fill=cancan, draw=black] (2.00, -0.25) rectangle (2.25, -0.50);
\filldraw[fill=cancan, draw=black] (2.25, -0.25) rectangle (2.50, -0.50);
\filldraw[fill=bermuda, draw=black] (2.50, -0.25) rectangle (2.75, -0.50);
\filldraw[fill=bermuda, draw=black] (2.75, -0.25) rectangle (3.00, -0.50);
\filldraw[fill=cancan, draw=black] (3.00, -0.25) rectangle (3.25, -0.50);
\filldraw[fill=cancan, draw=black] (3.25, -0.25) rectangle (3.50, -0.50);
\filldraw[fill=cancan, draw=black] (3.50, -0.25) rectangle (3.75, -0.50);
\filldraw[fill=cancan, draw=black] (3.75, -0.25) rectangle (4.00, -0.50);
\filldraw[fill=cancan, draw=black] (4.00, -0.25) rectangle (4.25, -0.50);
\filldraw[fill=bermuda, draw=black] (4.25, -0.25) rectangle (4.50, -0.50);
\filldraw[fill=cancan, draw=black] (4.50, -0.25) rectangle (4.75, -0.50);
\filldraw[fill=cancan, draw=black] (4.75, -0.25) rectangle (5.00, -0.50);
\filldraw[fill=cancan, draw=black] (5.00, -0.25) rectangle (5.25, -0.50);
\filldraw[fill=bermuda, draw=black] (5.25, -0.25) rectangle (5.50, -0.50);
\filldraw[fill=bermuda, draw=black] (5.50, -0.25) rectangle (5.75, -0.50);
\filldraw[fill=bermuda, draw=black] (5.75, -0.25) rectangle (6.00, -0.50);
\filldraw[fill=cancan, draw=black] (6.00, -0.25) rectangle (6.25, -0.50);
\filldraw[fill=cancan, draw=black] (6.25, -0.25) rectangle (6.50, -0.50);
\filldraw[fill=cancan, draw=black] (6.50, -0.25) rectangle (6.75, -0.50);
\filldraw[fill=bermuda, draw=black] (6.75, -0.25) rectangle (7.00, -0.50);
\filldraw[fill=cancan, draw=black] (7.00, -0.25) rectangle (7.25, -0.50);
\filldraw[fill=bermuda, draw=black] (7.25, -0.25) rectangle (7.50, -0.50);
\filldraw[fill=bermuda, draw=black] (7.50, -0.25) rectangle (7.75, -0.50);
\filldraw[fill=bermuda, draw=black] (7.75, -0.25) rectangle (8.00, -0.50);
\filldraw[fill=cancan, draw=black] (8.00, -0.25) rectangle (8.25, -0.50);
\filldraw[fill=cancan, draw=black] (8.25, -0.25) rectangle (8.50, -0.50);
\filldraw[fill=cancan, draw=black] (8.50, -0.25) rectangle (8.75, -0.50);
\filldraw[fill=bermuda, draw=black] (8.75, -0.25) rectangle (9.00, -0.50);
\filldraw[fill=bermuda, draw=black] (9.00, -0.25) rectangle (9.25, -0.50);
\filldraw[fill=bermuda, draw=black] (9.25, -0.25) rectangle (9.50, -0.50);
\filldraw[fill=cancan, draw=black] (9.50, -0.25) rectangle (9.75, -0.50);
\filldraw[fill=cancan, draw=black] (9.75, -0.25) rectangle (10.00, -0.50);
\filldraw[fill=cancan, draw=black] (10.00, -0.25) rectangle (10.25, -0.50);
\filldraw[fill=cancan, draw=black] (10.25, -0.25) rectangle (10.50, -0.50);
\filldraw[fill=cancan, draw=black] (10.50, -0.25) rectangle (10.75, -0.50);
\filldraw[fill=cancan, draw=black] (10.75, -0.25) rectangle (11.00, -0.50);
\filldraw[fill=bermuda, draw=black] (11.00, -0.25) rectangle (11.25, -0.50);
\filldraw[fill=bermuda, draw=black] (11.25, -0.25) rectangle (11.50, -0.50);
\filldraw[fill=cancan, draw=black] (11.50, -0.25) rectangle (11.75, -0.50);
\filldraw[fill=bermuda, draw=black] (11.75, -0.25) rectangle (12.00, -0.50);
\filldraw[fill=cancan, draw=black] (12.00, -0.25) rectangle (12.25, -0.50);
\filldraw[fill=bermuda, draw=black] (12.25, -0.25) rectangle (12.50, -0.50);
\filldraw[fill=cancan, draw=black] (12.50, -0.25) rectangle (12.75, -0.50);
\filldraw[fill=bermuda, draw=black] (12.75, -0.25) rectangle (13.00, -0.50);
\filldraw[fill=bermuda, draw=black] (13.00, -0.25) rectangle (13.25, -0.50);
\filldraw[fill=bermuda, draw=black] (13.25, -0.25) rectangle (13.50, -0.50);
\filldraw[fill=cancan, draw=black] (13.50, -0.25) rectangle (13.75, -0.50);
\filldraw[fill=cancan, draw=black] (13.75, -0.25) rectangle (14.00, -0.50);
\filldraw[fill=cancan, draw=black] (14.00, -0.25) rectangle (14.25, -0.50);
\filldraw[fill=bermuda, draw=black] (14.25, -0.25) rectangle (14.50, -0.50);
\filldraw[fill=bermuda, draw=black] (14.50, -0.25) rectangle (14.75, -0.50);
\filldraw[fill=bermuda, draw=black] (14.75, -0.25) rectangle (15.00, -0.50);
\filldraw[fill=cancan, draw=black] (0.00, -0.50) rectangle (0.25, -0.75);
\filldraw[fill=cancan, draw=black] (0.25, -0.50) rectangle (0.50, -0.75);
\filldraw[fill=cancan, draw=black] (0.50, -0.50) rectangle (0.75, -0.75);
\filldraw[fill=cancan, draw=black] (0.75, -0.50) rectangle (1.00, -0.75);
\filldraw[fill=cancan, draw=black] (1.00, -0.50) rectangle (1.25, -0.75);
\filldraw[fill=bermuda, draw=black] (1.25, -0.50) rectangle (1.50, -0.75);
\filldraw[fill=bermuda, draw=black] (1.50, -0.50) rectangle (1.75, -0.75);
\filldraw[fill=bermuda, draw=black] (1.75, -0.50) rectangle (2.00, -0.75);
\filldraw[fill=bermuda, draw=black] (2.00, -0.50) rectangle (2.25, -0.75);
\filldraw[fill=bermuda, draw=black] (2.25, -0.50) rectangle (2.50, -0.75);
\filldraw[fill=cancan, draw=black] (2.50, -0.50) rectangle (2.75, -0.75);
\filldraw[fill=bermuda, draw=black] (2.75, -0.50) rectangle (3.00, -0.75);
\filldraw[fill=bermuda, draw=black] (3.00, -0.50) rectangle (3.25, -0.75);
\filldraw[fill=bermuda, draw=black] (3.25, -0.50) rectangle (3.50, -0.75);
\filldraw[fill=cancan, draw=black] (3.50, -0.50) rectangle (3.75, -0.75);
\filldraw[fill=cancan, draw=black] (3.75, -0.50) rectangle (4.00, -0.75);
\filldraw[fill=cancan, draw=black] (4.00, -0.50) rectangle (4.25, -0.75);
\filldraw[fill=bermuda, draw=black] (4.25, -0.50) rectangle (4.50, -0.75);
\filldraw[fill=bermuda, draw=black] (4.50, -0.50) rectangle (4.75, -0.75);
\filldraw[fill=bermuda, draw=black] (4.75, -0.50) rectangle (5.00, -0.75);
\filldraw[fill=cancan, draw=black] (5.00, -0.50) rectangle (5.25, -0.75);
\filldraw[fill=cancan, draw=black] (5.25, -0.50) rectangle (5.50, -0.75);
\filldraw[fill=cancan, draw=black] (5.50, -0.50) rectangle (5.75, -0.75);
\filldraw[fill=bermuda, draw=black] (5.75, -0.50) rectangle (6.00, -0.75);
\filldraw[fill=bermuda, draw=black] (6.00, -0.50) rectangle (6.25, -0.75);
\filldraw[fill=bermuda, draw=black] (6.25, -0.50) rectangle (6.50, -0.75);
\filldraw[fill=bermuda, draw=black] (6.50, -0.50) rectangle (6.75, -0.75);
\filldraw[fill=bermuda, draw=black] (6.75, -0.50) rectangle (7.00, -0.75);
\filldraw[fill=bermuda, draw=black] (7.00, -0.50) rectangle (7.25, -0.75);
\filldraw[fill=bermuda, draw=black] (7.25, -0.50) rectangle (7.50, -0.75);
\filldraw[fill=cancan, draw=black] (7.50, -0.50) rectangle (7.75, -0.75);
\filldraw[fill=bermuda, draw=black] (7.75, -0.50) rectangle (8.00, -0.75);
\filldraw[fill=bermuda, draw=black] (8.00, -0.50) rectangle (8.25, -0.75);
\filldraw[fill=bermuda, draw=black] (8.25, -0.50) rectangle (8.50, -0.75);
\filldraw[fill=bermuda, draw=black] (8.50, -0.50) rectangle (8.75, -0.75);
\filldraw[fill=bermuda, draw=black] (8.75, -0.50) rectangle (9.00, -0.75);
\filldraw[fill=cancan, draw=black] (9.00, -0.50) rectangle (9.25, -0.75);
\filldraw[fill=cancan, draw=black] (9.25, -0.50) rectangle (9.50, -0.75);
\filldraw[fill=cancan, draw=black] (9.50, -0.50) rectangle (9.75, -0.75);
\filldraw[fill=cancan, draw=black] (9.75, -0.50) rectangle (10.00, -0.75);
\filldraw[fill=cancan, draw=black] (10.00, -0.50) rectangle (10.25, -0.75);
\filldraw[fill=cancan, draw=black] (10.25, -0.50) rectangle (10.50, -0.75);
\filldraw[fill=cancan, draw=black] (10.50, -0.50) rectangle (10.75, -0.75);
\filldraw[fill=cancan, draw=black] (10.75, -0.50) rectangle (11.00, -0.75);
\filldraw[fill=cancan, draw=black] (11.00, -0.50) rectangle (11.25, -0.75);
\filldraw[fill=bermuda, draw=black] (11.25, -0.50) rectangle (11.50, -0.75);
\filldraw[fill=bermuda, draw=black] (11.50, -0.50) rectangle (11.75, -0.75);
\filldraw[fill=bermuda, draw=black] (11.75, -0.50) rectangle (12.00, -0.75);
\filldraw[fill=cancan, draw=black] (12.00, -0.50) rectangle (12.25, -0.75);
\filldraw[fill=cancan, draw=black] (12.25, -0.50) rectangle (12.50, -0.75);
\filldraw[fill=bermuda, draw=black] (12.50, -0.50) rectangle (12.75, -0.75);
\filldraw[fill=bermuda, draw=black] (12.75, -0.50) rectangle (13.00, -0.75);
\filldraw[fill=cancan, draw=black] (13.00, -0.50) rectangle (13.25, -0.75);
\filldraw[fill=cancan, draw=black] (13.25, -0.50) rectangle (13.50, -0.75);
\filldraw[fill=cancan, draw=black] (13.50, -0.50) rectangle (13.75, -0.75);
\filldraw[fill=cancan, draw=black] (13.75, -0.50) rectangle (14.00, -0.75);
\filldraw[fill=bermuda, draw=black] (14.00, -0.50) rectangle (14.25, -0.75);
\filldraw[fill=bermuda, draw=black] (14.25, -0.50) rectangle (14.50, -0.75);
\filldraw[fill=cancan, draw=black] (14.50, -0.50) rectangle (14.75, -0.75);
\filldraw[fill=cancan, draw=black] (14.75, -0.50) rectangle (15.00, -0.75);
\filldraw[fill=cancan, draw=black] (0.00, -0.75) rectangle (0.25, -1.00);
\filldraw[fill=cancan, draw=black] (0.25, -0.75) rectangle (0.50, -1.00);
\filldraw[fill=cancan, draw=black] (0.50, -0.75) rectangle (0.75, -1.00);
\filldraw[fill=cancan, draw=black] (0.75, -0.75) rectangle (1.00, -1.00);
\filldraw[fill=cancan, draw=black] (1.00, -0.75) rectangle (1.25, -1.00);
\filldraw[fill=cancan, draw=black] (1.25, -0.75) rectangle (1.50, -1.00);
\filldraw[fill=cancan, draw=black] (1.50, -0.75) rectangle (1.75, -1.00);
\filldraw[fill=bermuda, draw=black] (1.75, -0.75) rectangle (2.00, -1.00);
\filldraw[fill=bermuda, draw=black] (2.00, -0.75) rectangle (2.25, -1.00);
\filldraw[fill=bermuda, draw=black] (2.25, -0.75) rectangle (2.50, -1.00);
\filldraw[fill=bermuda, draw=black] (2.50, -0.75) rectangle (2.75, -1.00);
\filldraw[fill=bermuda, draw=black] (2.75, -0.75) rectangle (3.00, -1.00);
\filldraw[fill=cancan, draw=black] (3.00, -0.75) rectangle (3.25, -1.00);
\filldraw[fill=bermuda, draw=black] (3.25, -0.75) rectangle (3.50, -1.00);
\filldraw[fill=bermuda, draw=black] (3.50, -0.75) rectangle (3.75, -1.00);
\filldraw[fill=bermuda, draw=black] (3.75, -0.75) rectangle (4.00, -1.00);
\filldraw[fill=cancan, draw=black] (4.00, -0.75) rectangle (4.25, -1.00);
\filldraw[fill=cancan, draw=black] (4.25, -0.75) rectangle (4.50, -1.00);
\filldraw[fill=cancan, draw=black] (4.50, -0.75) rectangle (4.75, -1.00);
\filldraw[fill=bermuda, draw=black] (4.75, -0.75) rectangle (5.00, -1.00);
\filldraw[fill=bermuda, draw=black] (5.00, -0.75) rectangle (5.25, -1.00);
\filldraw[fill=bermuda, draw=black] (5.25, -0.75) rectangle (5.50, -1.00);
\filldraw[fill=cancan, draw=black] (5.50, -0.75) rectangle (5.75, -1.00);
\filldraw[fill=cancan, draw=black] (5.75, -0.75) rectangle (6.00, -1.00);
\filldraw[fill=cancan, draw=black] (6.00, -0.75) rectangle (6.25, -1.00);
\filldraw[fill=bermuda, draw=black] (6.25, -0.75) rectangle (6.50, -1.00);
\filldraw[fill=bermuda, draw=black] (6.50, -0.75) rectangle (6.75, -1.00);
\filldraw[fill=bermuda, draw=black] (6.75, -0.75) rectangle (7.00, -1.00);
\filldraw[fill=bermuda, draw=black] (7.00, -0.75) rectangle (7.25, -1.00);
\filldraw[fill=bermuda, draw=black] (7.25, -0.75) rectangle (7.50, -1.00);
\filldraw[fill=bermuda, draw=black] (7.50, -0.75) rectangle (7.75, -1.00);
\filldraw[fill=bermuda, draw=black] (7.75, -0.75) rectangle (8.00, -1.00);
\filldraw[fill=cancan, draw=black] (8.00, -0.75) rectangle (8.25, -1.00);
\filldraw[fill=cancan, draw=black] (8.25, -0.75) rectangle (8.50, -1.00);
\filldraw[fill=bermuda, draw=black] (8.50, -0.75) rectangle (8.75, -1.00);
\filldraw[fill=bermuda, draw=black] (8.75, -0.75) rectangle (9.00, -1.00);
\filldraw[fill=cancan, draw=black] (9.00, -0.75) rectangle (9.25, -1.00);
\filldraw[fill=cancan, draw=black] (9.25, -0.75) rectangle (9.50, -1.00);
\filldraw[fill=cancan, draw=black] (9.50, -0.75) rectangle (9.75, -1.00);
\filldraw[fill=cancan, draw=black] (9.75, -0.75) rectangle (10.00, -1.00);
\filldraw[fill=cancan, draw=black] (10.00, -0.75) rectangle (10.25, -1.00);
\filldraw[fill=bermuda, draw=black] (10.25, -0.75) rectangle (10.50, -1.00);
\filldraw[fill=cancan, draw=black] (10.50, -0.75) rectangle (10.75, -1.00);
\filldraw[fill=cancan, draw=black] (10.75, -0.75) rectangle (11.00, -1.00);
\filldraw[fill=bermuda, draw=black] (11.00, -0.75) rectangle (11.25, -1.00);
\filldraw[fill=bermuda, draw=black] (11.25, -0.75) rectangle (11.50, -1.00);
\filldraw[fill=cancan, draw=black] (11.50, -0.75) rectangle (11.75, -1.00);
\filldraw[fill=bermuda, draw=black] (11.75, -0.75) rectangle (12.00, -1.00);
\filldraw[fill=cancan, draw=black] (12.00, -0.75) rectangle (12.25, -1.00);
\filldraw[fill=cancan, draw=black] (12.25, -0.75) rectangle (12.50, -1.00);
\filldraw[fill=bermuda, draw=black] (12.50, -0.75) rectangle (12.75, -1.00);
\filldraw[fill=bermuda, draw=black] (12.75, -0.75) rectangle (13.00, -1.00);
\filldraw[fill=cancan, draw=black] (13.00, -0.75) rectangle (13.25, -1.00);
\filldraw[fill=bermuda, draw=black] (13.25, -0.75) rectangle (13.50, -1.00);
\filldraw[fill=cancan, draw=black] (13.50, -0.75) rectangle (13.75, -1.00);
\filldraw[fill=cancan, draw=black] (13.75, -0.75) rectangle (14.00, -1.00);
\filldraw[fill=bermuda, draw=black] (14.00, -0.75) rectangle (14.25, -1.00);
\filldraw[fill=bermuda, draw=black] (14.25, -0.75) rectangle (14.50, -1.00);
\filldraw[fill=cancan, draw=black] (14.50, -0.75) rectangle (14.75, -1.00);
\filldraw[fill=bermuda, draw=black] (14.75, -0.75) rectangle (15.00, -1.00);
\filldraw[fill=cancan, draw=black] (0.00, -1.00) rectangle (0.25, -1.25);
\filldraw[fill=bermuda, draw=black] (0.25, -1.00) rectangle (0.50, -1.25);
\filldraw[fill=bermuda, draw=black] (0.50, -1.00) rectangle (0.75, -1.25);
\filldraw[fill=bermuda, draw=black] (0.75, -1.00) rectangle (1.00, -1.25);
\filldraw[fill=cancan, draw=black] (1.00, -1.00) rectangle (1.25, -1.25);
\filldraw[fill=cancan, draw=black] (1.25, -1.00) rectangle (1.50, -1.25);
\filldraw[fill=cancan, draw=black] (1.50, -1.00) rectangle (1.75, -1.25);
\filldraw[fill=cancan, draw=black] (1.75, -1.00) rectangle (2.00, -1.25);
\filldraw[fill=bermuda, draw=black] (2.00, -1.00) rectangle (2.25, -1.25);
\filldraw[fill=bermuda, draw=black] (2.25, -1.00) rectangle (2.50, -1.25);
\filldraw[fill=cancan, draw=black] (2.50, -1.00) rectangle (2.75, -1.25);
\filldraw[fill=bermuda, draw=black] (2.75, -1.00) rectangle (3.00, -1.25);
\filldraw[fill=bermuda, draw=black] (3.00, -1.00) rectangle (3.25, -1.25);
\filldraw[fill=bermuda, draw=black] (3.25, -1.00) rectangle (3.50, -1.25);
\filldraw[fill=cancan, draw=black] (3.50, -1.00) rectangle (3.75, -1.25);
\filldraw[fill=cancan, draw=black] (3.75, -1.00) rectangle (4.00, -1.25);
\filldraw[fill=cancan, draw=black] (4.00, -1.00) rectangle (4.25, -1.25);
\filldraw[fill=bermuda, draw=black] (4.25, -1.00) rectangle (4.50, -1.25);
\filldraw[fill=bermuda, draw=black] (4.50, -1.00) rectangle (4.75, -1.25);
\filldraw[fill=bermuda, draw=black] (4.75, -1.00) rectangle (5.00, -1.25);
\filldraw[fill=cancan, draw=black] (5.00, -1.00) rectangle (5.25, -1.25);
\filldraw[fill=cancan, draw=black] (5.25, -1.00) rectangle (5.50, -1.25);
\filldraw[fill=cancan, draw=black] (5.50, -1.00) rectangle (5.75, -1.25);
\filldraw[fill=cancan, draw=black] (5.75, -1.00) rectangle (6.00, -1.25);
\filldraw[fill=cancan, draw=black] (6.00, -1.00) rectangle (6.25, -1.25);
\filldraw[fill=bermuda, draw=black] (6.25, -1.00) rectangle (6.50, -1.25);
\filldraw[fill=cancan, draw=black] (6.50, -1.00) rectangle (6.75, -1.25);
\filldraw[fill=bermuda, draw=black] (6.75, -1.00) rectangle (7.00, -1.25);
\filldraw[fill=cancan, draw=black] (7.00, -1.00) rectangle (7.25, -1.25);
\filldraw[fill=cancan, draw=black] (7.25, -1.00) rectangle (7.50, -1.25);
\filldraw[fill=cancan, draw=black] (7.50, -1.00) rectangle (7.75, -1.25);
\filldraw[fill=bermuda, draw=black] (7.75, -1.00) rectangle (8.00, -1.25);
\filldraw[fill=cancan, draw=black] (8.00, -1.00) rectangle (8.25, -1.25);
\filldraw[fill=bermuda, draw=black] (8.25, -1.00) rectangle (8.50, -1.25);
\filldraw[fill=cancan, draw=black] (8.50, -1.00) rectangle (8.75, -1.25);
\filldraw[fill=bermuda, draw=black] (8.75, -1.00) rectangle (9.00, -1.25);
\filldraw[fill=bermuda, draw=black] (9.00, -1.00) rectangle (9.25, -1.25);
\filldraw[fill=bermuda, draw=black] (9.25, -1.00) rectangle (9.50, -1.25);
\filldraw[fill=cancan, draw=black] (9.50, -1.00) rectangle (9.75, -1.25);
\filldraw[fill=cancan, draw=black] (9.75, -1.00) rectangle (10.00, -1.25);
\filldraw[fill=cancan, draw=black] (10.00, -1.00) rectangle (10.25, -1.25);
\filldraw[fill=cancan, draw=black] (10.25, -1.00) rectangle (10.50, -1.25);
\filldraw[fill=cancan, draw=black] (10.50, -1.00) rectangle (10.75, -1.25);
\filldraw[fill=bermuda, draw=black] (10.75, -1.00) rectangle (11.00, -1.25);
\filldraw[fill=cancan, draw=black] (11.00, -1.00) rectangle (11.25, -1.25);
\filldraw[fill=cancan, draw=black] (11.25, -1.00) rectangle (11.50, -1.25);
\filldraw[fill=cancan, draw=black] (11.50, -1.00) rectangle (11.75, -1.25);
\filldraw[fill=cancan, draw=black] (11.75, -1.00) rectangle (12.00, -1.25);
\filldraw[fill=cancan, draw=black] (12.00, -1.00) rectangle (12.25, -1.25);
\filldraw[fill=cancan, draw=black] (12.25, -1.00) rectangle (12.50, -1.25);
\filldraw[fill=cancan, draw=black] (12.50, -1.00) rectangle (12.75, -1.25);
\filldraw[fill=cancan, draw=black] (12.75, -1.00) rectangle (13.00, -1.25);
\filldraw[fill=bermuda, draw=black] (13.00, -1.00) rectangle (13.25, -1.25);
\filldraw[fill=bermuda, draw=black] (13.25, -1.00) rectangle (13.50, -1.25);
\filldraw[fill=cancan, draw=black] (13.50, -1.00) rectangle (13.75, -1.25);
\filldraw[fill=bermuda, draw=black] (13.75, -1.00) rectangle (14.00, -1.25);
\filldraw[fill=cancan, draw=black] (14.00, -1.00) rectangle (14.25, -1.25);
\filldraw[fill=cancan, draw=black] (14.25, -1.00) rectangle (14.50, -1.25);
\filldraw[fill=cancan, draw=black] (14.50, -1.00) rectangle (14.75, -1.25);
\filldraw[fill=cancan, draw=black] (14.75, -1.00) rectangle (15.00, -1.25);
\filldraw[fill=cancan, draw=black] (0.00, -1.25) rectangle (0.25, -1.50);
\filldraw[fill=cancan, draw=black] (0.25, -1.25) rectangle (0.50, -1.50);
\filldraw[fill=cancan, draw=black] (0.50, -1.25) rectangle (0.75, -1.50);
\filldraw[fill=bermuda, draw=black] (0.75, -1.25) rectangle (1.00, -1.50);
\filldraw[fill=bermuda, draw=black] (1.00, -1.25) rectangle (1.25, -1.50);
\filldraw[fill=bermuda, draw=black] (1.25, -1.25) rectangle (1.50, -1.50);
\filldraw[fill=bermuda, draw=black] (1.50, -1.25) rectangle (1.75, -1.50);
\filldraw[fill=bermuda, draw=black] (1.75, -1.25) rectangle (2.00, -1.50);
\filldraw[fill=cancan, draw=black] (2.00, -1.25) rectangle (2.25, -1.50);
\filldraw[fill=bermuda, draw=black] (2.25, -1.25) rectangle (2.50, -1.50);
\filldraw[fill=cancan, draw=black] (2.50, -1.25) rectangle (2.75, -1.50);
\filldraw[fill=bermuda, draw=black] (2.75, -1.25) rectangle (3.00, -1.50);
\filldraw[fill=cancan, draw=black] (3.00, -1.25) rectangle (3.25, -1.50);
\filldraw[fill=bermuda, draw=black] (3.25, -1.25) rectangle (3.50, -1.50);
\filldraw[fill=cancan, draw=black] (3.50, -1.25) rectangle (3.75, -1.50);
\filldraw[fill=cancan, draw=black] (3.75, -1.25) rectangle (4.00, -1.50);
\filldraw[fill=bermuda, draw=black] (4.00, -1.25) rectangle (4.25, -1.50);
\filldraw[fill=bermuda, draw=black] (4.25, -1.25) rectangle (4.50, -1.50);
\filldraw[fill=bermuda, draw=black] (4.50, -1.25) rectangle (4.75, -1.50);
\filldraw[fill=bermuda, draw=black] (4.75, -1.25) rectangle (5.00, -1.50);
\filldraw[fill=bermuda, draw=black] (5.00, -1.25) rectangle (5.25, -1.50);
\filldraw[fill=bermuda, draw=black] (5.25, -1.25) rectangle (5.50, -1.50);
\filldraw[fill=bermuda, draw=black] (5.50, -1.25) rectangle (5.75, -1.50);
\filldraw[fill=bermuda, draw=black] (5.75, -1.25) rectangle (6.00, -1.50);
\filldraw[fill=cancan, draw=black] (6.00, -1.25) rectangle (6.25, -1.50);
\filldraw[fill=cancan, draw=black] (6.25, -1.25) rectangle (6.50, -1.50);
\filldraw[fill=cancan, draw=black] (6.50, -1.25) rectangle (6.75, -1.50);
\filldraw[fill=bermuda, draw=black] (6.75, -1.25) rectangle (7.00, -1.50);
\filldraw[fill=bermuda, draw=black] (7.00, -1.25) rectangle (7.25, -1.50);
\filldraw[fill=bermuda, draw=black] (7.25, -1.25) rectangle (7.50, -1.50);
\filldraw[fill=bermuda, draw=black] (7.50, -1.25) rectangle (7.75, -1.50);
\filldraw[fill=bermuda, draw=black] (7.75, -1.25) rectangle (8.00, -1.50);
\filldraw[fill=cancan, draw=black] (8.00, -1.25) rectangle (8.25, -1.50);
\filldraw[fill=bermuda, draw=black] (8.25, -1.25) rectangle (8.50, -1.50);
\filldraw[fill=cancan, draw=black] (8.50, -1.25) rectangle (8.75, -1.50);
\filldraw[fill=bermuda, draw=black] (8.75, -1.25) rectangle (9.00, -1.50);
\filldraw[fill=cancan, draw=black] (9.00, -1.25) rectangle (9.25, -1.50);
\filldraw[fill=bermuda, draw=black] (9.25, -1.25) rectangle (9.50, -1.50);
\filldraw[fill=cancan, draw=black] (9.50, -1.25) rectangle (9.75, -1.50);
\filldraw[fill=cancan, draw=black] (9.75, -1.25) rectangle (10.00, -1.50);
\filldraw[fill=bermuda, draw=black] (10.00, -1.25) rectangle (10.25, -1.50);
\filldraw[fill=bermuda, draw=black] (10.25, -1.25) rectangle (10.50, -1.50);
\filldraw[fill=bermuda, draw=black] (10.50, -1.25) rectangle (10.75, -1.50);
\filldraw[fill=bermuda, draw=black] (10.75, -1.25) rectangle (11.00, -1.50);
\filldraw[fill=cancan, draw=black] (11.00, -1.25) rectangle (11.25, -1.50);
\filldraw[fill=cancan, draw=black] (11.25, -1.25) rectangle (11.50, -1.50);
\filldraw[fill=cancan, draw=black] (11.50, -1.25) rectangle (11.75, -1.50);
\filldraw[fill=cancan, draw=black] (11.75, -1.25) rectangle (12.00, -1.50);
\filldraw[fill=cancan, draw=black] (12.00, -1.25) rectangle (12.25, -1.50);
\filldraw[fill=bermuda, draw=black] (12.25, -1.25) rectangle (12.50, -1.50);
\filldraw[fill=bermuda, draw=black] (12.50, -1.25) rectangle (12.75, -1.50);
\filldraw[fill=bermuda, draw=black] (12.75, -1.25) rectangle (13.00, -1.50);
\filldraw[fill=cancan, draw=black] (13.00, -1.25) rectangle (13.25, -1.50);
\filldraw[fill=bermuda, draw=black] (13.25, -1.25) rectangle (13.50, -1.50);
\filldraw[fill=bermuda, draw=black] (13.50, -1.25) rectangle (13.75, -1.50);
\filldraw[fill=bermuda, draw=black] (13.75, -1.25) rectangle (14.00, -1.50);
\filldraw[fill=bermuda, draw=black] (14.00, -1.25) rectangle (14.25, -1.50);
\filldraw[fill=bermuda, draw=black] (14.25, -1.25) rectangle (14.50, -1.50);
\filldraw[fill=cancan, draw=black] (14.50, -1.25) rectangle (14.75, -1.50);
\filldraw[fill=bermuda, draw=black] (14.75, -1.25) rectangle (15.00, -1.50);
\filldraw[fill=bermuda, draw=black] (0.00, -1.50) rectangle (0.25, -1.75);
\filldraw[fill=bermuda, draw=black] (0.25, -1.50) rectangle (0.50, -1.75);
\filldraw[fill=bermuda, draw=black] (0.50, -1.50) rectangle (0.75, -1.75);
\filldraw[fill=bermuda, draw=black] (0.75, -1.50) rectangle (1.00, -1.75);
\filldraw[fill=cancan, draw=black] (1.00, -1.50) rectangle (1.25, -1.75);
\filldraw[fill=cancan, draw=black] (1.25, -1.50) rectangle (1.50, -1.75);
\filldraw[fill=cancan, draw=black] (1.50, -1.50) rectangle (1.75, -1.75);
\filldraw[fill=bermuda, draw=black] (1.75, -1.50) rectangle (2.00, -1.75);
\filldraw[fill=cancan, draw=black] (2.00, -1.50) rectangle (2.25, -1.75);
\filldraw[fill=bermuda, draw=black] (2.25, -1.50) rectangle (2.50, -1.75);
\filldraw[fill=cancan, draw=black] (2.50, -1.50) rectangle (2.75, -1.75);
\filldraw[fill=cancan, draw=black] (2.75, -1.50) rectangle (3.00, -1.75);
\filldraw[fill=bermuda, draw=black] (3.00, -1.50) rectangle (3.25, -1.75);
\filldraw[fill=bermuda, draw=black] (3.25, -1.50) rectangle (3.50, -1.75);
\filldraw[fill=cancan, draw=black] (3.50, -1.50) rectangle (3.75, -1.75);
\filldraw[fill=bermuda, draw=black] (3.75, -1.50) rectangle (4.00, -1.75);
\filldraw[fill=cancan, draw=black] (4.00, -1.50) rectangle (4.25, -1.75);
\filldraw[fill=cancan, draw=black] (4.25, -1.50) rectangle (4.50, -1.75);
\filldraw[fill=cancan, draw=black] (4.50, -1.50) rectangle (4.75, -1.75);
\filldraw[fill=cancan, draw=black] (4.75, -1.50) rectangle (5.00, -1.75);
\filldraw[fill=cancan, draw=black] (5.00, -1.50) rectangle (5.25, -1.75);
\filldraw[fill=cancan, draw=black] (5.25, -1.50) rectangle (5.50, -1.75);
\filldraw[fill=bermuda, draw=black] (5.50, -1.50) rectangle (5.75, -1.75);
\filldraw[fill=bermuda, draw=black] (5.75, -1.50) rectangle (6.00, -1.75);
\filldraw[fill=cancan, draw=black] (6.00, -1.50) rectangle (6.25, -1.75);
\filldraw[fill=bermuda, draw=black] (6.25, -1.50) rectangle (6.50, -1.75);
\filldraw[fill=cancan, draw=black] (6.50, -1.50) rectangle (6.75, -1.75);
\filldraw[fill=cancan, draw=black] (6.75, -1.50) rectangle (7.00, -1.75);
\filldraw[fill=bermuda, draw=black] (7.00, -1.50) rectangle (7.25, -1.75);
\filldraw[fill=bermuda, draw=black] (7.25, -1.50) rectangle (7.50, -1.75);
\filldraw[fill=bermuda, draw=black] (7.50, -1.50) rectangle (7.75, -1.75);
\filldraw[fill=bermuda, draw=black] (7.75, -1.50) rectangle (8.00, -1.75);
\filldraw[fill=cancan, draw=black] (8.00, -1.50) rectangle (8.25, -1.75);
\filldraw[fill=bermuda, draw=black] (8.25, -1.50) rectangle (8.50, -1.75);
\filldraw[fill=bermuda, draw=black] (8.50, -1.50) rectangle (8.75, -1.75);
\filldraw[fill=bermuda, draw=black] (8.75, -1.50) rectangle (9.00, -1.75);
\filldraw[fill=cancan, draw=black] (9.00, -1.50) rectangle (9.25, -1.75);
\filldraw[fill=bermuda, draw=black] (9.25, -1.50) rectangle (9.50, -1.75);
\filldraw[fill=cancan, draw=black] (9.50, -1.50) rectangle (9.75, -1.75);
\filldraw[fill=cancan, draw=black] (9.75, -1.50) rectangle (10.00, -1.75);
\filldraw[fill=cancan, draw=black] (10.00, -1.50) rectangle (10.25, -1.75);
\filldraw[fill=cancan, draw=black] (10.25, -1.50) rectangle (10.50, -1.75);
\filldraw[fill=cancan, draw=black] (10.50, -1.50) rectangle (10.75, -1.75);
\filldraw[fill=bermuda, draw=black] (10.75, -1.50) rectangle (11.00, -1.75);
\filldraw[fill=bermuda, draw=black] (11.00, -1.50) rectangle (11.25, -1.75);
\filldraw[fill=bermuda, draw=black] (11.25, -1.50) rectangle (11.50, -1.75);
\filldraw[fill=cancan, draw=black] (11.50, -1.50) rectangle (11.75, -1.75);
\filldraw[fill=cancan, draw=black] (11.75, -1.50) rectangle (12.00, -1.75);
\filldraw[fill=cancan, draw=black] (12.00, -1.50) rectangle (12.25, -1.75);
\filldraw[fill=bermuda, draw=black] (12.25, -1.50) rectangle (12.50, -1.75);
\filldraw[fill=bermuda, draw=black] (12.50, -1.50) rectangle (12.75, -1.75);
\filldraw[fill=bermuda, draw=black] (12.75, -1.50) rectangle (13.00, -1.75);
\filldraw[fill=cancan, draw=black] (13.00, -1.50) rectangle (13.25, -1.75);
\filldraw[fill=bermuda, draw=black] (13.25, -1.50) rectangle (13.50, -1.75);
\filldraw[fill=bermuda, draw=black] (13.50, -1.50) rectangle (13.75, -1.75);
\filldraw[fill=bermuda, draw=black] (13.75, -1.50) rectangle (14.00, -1.75);
\filldraw[fill=bermuda, draw=black] (14.00, -1.50) rectangle (14.25, -1.75);
\filldraw[fill=bermuda, draw=black] (14.25, -1.50) rectangle (14.50, -1.75);
\filldraw[fill=cancan, draw=black] (14.50, -1.50) rectangle (14.75, -1.75);
\filldraw[fill=bermuda, draw=black] (14.75, -1.50) rectangle (15.00, -1.75);
\filldraw[fill=bermuda, draw=black] (0.00, -1.75) rectangle (0.25, -2.00);
\filldraw[fill=bermuda, draw=black] (0.25, -1.75) rectangle (0.50, -2.00);
\filldraw[fill=cancan, draw=black] (0.50, -1.75) rectangle (0.75, -2.00);
\filldraw[fill=cancan, draw=black] (0.75, -1.75) rectangle (1.00, -2.00);
\filldraw[fill=cancan, draw=black] (1.00, -1.75) rectangle (1.25, -2.00);
\filldraw[fill=bermuda, draw=black] (1.25, -1.75) rectangle (1.50, -2.00);
\filldraw[fill=bermuda, draw=black] (1.50, -1.75) rectangle (1.75, -2.00);
\filldraw[fill=bermuda, draw=black] (1.75, -1.75) rectangle (2.00, -2.00);
\filldraw[fill=cancan, draw=black] (2.00, -1.75) rectangle (2.25, -2.00);
\filldraw[fill=bermuda, draw=black] (2.25, -1.75) rectangle (2.50, -2.00);
\filldraw[fill=bermuda, draw=black] (2.50, -1.75) rectangle (2.75, -2.00);
\filldraw[fill=bermuda, draw=black] (2.75, -1.75) rectangle (3.00, -2.00);
\filldraw[fill=cancan, draw=black] (3.00, -1.75) rectangle (3.25, -2.00);
\filldraw[fill=cancan, draw=black] (3.25, -1.75) rectangle (3.50, -2.00);
\filldraw[fill=cancan, draw=black] (3.50, -1.75) rectangle (3.75, -2.00);
\filldraw[fill=cancan, draw=black] (3.75, -1.75) rectangle (4.00, -2.00);
\filldraw[fill=cancan, draw=black] (4.00, -1.75) rectangle (4.25, -2.00);
\filldraw[fill=bermuda, draw=black] (4.25, -1.75) rectangle (4.50, -2.00);
\filldraw[fill=bermuda, draw=black] (4.50, -1.75) rectangle (4.75, -2.00);
\filldraw[fill=bermuda, draw=black] (4.75, -1.75) rectangle (5.00, -2.00);
\filldraw[fill=bermuda, draw=black] (5.00, -1.75) rectangle (5.25, -2.00);
\filldraw[fill=bermuda, draw=black] (5.25, -1.75) rectangle (5.50, -2.00);
\filldraw[fill=cancan, draw=black] (5.50, -1.75) rectangle (5.75, -2.00);
\filldraw[fill=bermuda, draw=black] (5.75, -1.75) rectangle (6.00, -2.00);
\filldraw[fill=cancan, draw=black] (6.00, -1.75) rectangle (6.25, -2.00);
\filldraw[fill=bermuda, draw=black] (6.25, -1.75) rectangle (6.50, -2.00);
\filldraw[fill=cancan, draw=black] (6.50, -1.75) rectangle (6.75, -2.00);
\filldraw[fill=bermuda, draw=black] (6.75, -1.75) rectangle (7.00, -2.00);
\filldraw[fill=cancan, draw=black] (7.00, -1.75) rectangle (7.25, -2.00);
\filldraw[fill=bermuda, draw=black] (7.25, -1.75) rectangle (7.50, -2.00);
\filldraw[fill=bermuda, draw=black] (7.50, -1.75) rectangle (7.75, -2.00);
\filldraw[fill=bermuda, draw=black] (7.75, -1.75) rectangle (8.00, -2.00);
\filldraw[fill=cancan, draw=black] (8.00, -1.75) rectangle (8.25, -2.00);
\filldraw[fill=cancan, draw=black] (8.25, -1.75) rectangle (8.50, -2.00);
\filldraw[fill=cancan, draw=black] (8.50, -1.75) rectangle (8.75, -2.00);
\filldraw[fill=bermuda, draw=black] (8.75, -1.75) rectangle (9.00, -2.00);
\filldraw[fill=bermuda, draw=black] (9.00, -1.75) rectangle (9.25, -2.00);
\filldraw[fill=bermuda, draw=black] (9.25, -1.75) rectangle (9.50, -2.00);
\filldraw[fill=cancan, draw=black] (9.50, -1.75) rectangle (9.75, -2.00);
\filldraw[fill=cancan, draw=black] (9.75, -1.75) rectangle (10.00, -2.00);
\filldraw[fill=cancan, draw=black] (10.00, -1.75) rectangle (10.25, -2.00);
\filldraw[fill=bermuda, draw=black] (10.25, -1.75) rectangle (10.50, -2.00);
\filldraw[fill=bermuda, draw=black] (10.50, -1.75) rectangle (10.75, -2.00);
\filldraw[fill=bermuda, draw=black] (10.75, -1.75) rectangle (11.00, -2.00);
\filldraw[fill=cancan, draw=black] (11.00, -1.75) rectangle (11.25, -2.00);
\filldraw[fill=cancan, draw=black] (11.25, -1.75) rectangle (11.50, -2.00);
\filldraw[fill=cancan, draw=black] (11.50, -1.75) rectangle (11.75, -2.00);
\filldraw[fill=bermuda, draw=black] (11.75, -1.75) rectangle (12.00, -2.00);
\filldraw[fill=bermuda, draw=black] (12.00, -1.75) rectangle (12.25, -2.00);
\filldraw[fill=bermuda, draw=black] (12.25, -1.75) rectangle (12.50, -2.00);
\filldraw[fill=cancan, draw=black] (12.50, -1.75) rectangle (12.75, -2.00);
\filldraw[fill=cancan, draw=black] (12.75, -1.75) rectangle (13.00, -2.00);
\filldraw[fill=cancan, draw=black] (13.00, -1.75) rectangle (13.25, -2.00);
\filldraw[fill=bermuda, draw=black] (13.25, -1.75) rectangle (13.50, -2.00);
\filldraw[fill=bermuda, draw=black] (13.50, -1.75) rectangle (13.75, -2.00);
\filldraw[fill=bermuda, draw=black] (13.75, -1.75) rectangle (14.00, -2.00);
\filldraw[fill=cancan, draw=black] (14.00, -1.75) rectangle (14.25, -2.00);
\filldraw[fill=cancan, draw=black] (14.25, -1.75) rectangle (14.50, -2.00);
\filldraw[fill=cancan, draw=black] (14.50, -1.75) rectangle (14.75, -2.00);
\filldraw[fill=bermuda, draw=black] (14.75, -1.75) rectangle (15.00, -2.00);
\filldraw[fill=bermuda, draw=black] (0.00, -2.00) rectangle (0.25, -2.25);
\filldraw[fill=bermuda, draw=black] (0.25, -2.00) rectangle (0.50, -2.25);
\filldraw[fill=cancan, draw=black] (0.50, -2.00) rectangle (0.75, -2.25);
\filldraw[fill=bermuda, draw=black] (0.75, -2.00) rectangle (1.00, -2.25);
\filldraw[fill=cancan, draw=black] (1.00, -2.00) rectangle (1.25, -2.25);
\filldraw[fill=bermuda, draw=black] (1.25, -2.00) rectangle (1.50, -2.25);
\filldraw[fill=cancan, draw=black] (1.50, -2.00) rectangle (1.75, -2.25);
\filldraw[fill=cancan, draw=black] (1.75, -2.00) rectangle (2.00, -2.25);
\filldraw[fill=cancan, draw=black] (2.00, -2.00) rectangle (2.25, -2.25);
\filldraw[fill=bermuda, draw=black] (2.25, -2.00) rectangle (2.50, -2.25);
\filldraw[fill=cancan, draw=black] (2.50, -2.00) rectangle (2.75, -2.25);
\filldraw[fill=bermuda, draw=black] (2.75, -2.00) rectangle (3.00, -2.25);
\filldraw[fill=cancan, draw=black] (3.00, -2.00) rectangle (3.25, -2.25);
\filldraw[fill=bermuda, draw=black] (3.25, -2.00) rectangle (3.50, -2.25);
\filldraw[fill=cancan, draw=black] (3.50, -2.00) rectangle (3.75, -2.25);
\filldraw[fill=cancan, draw=black] (3.75, -2.00) rectangle (4.00, -2.25);
\filldraw[fill=bermuda, draw=black] (4.00, -2.00) rectangle (4.25, -2.25);
\filldraw[fill=bermuda, draw=black] (4.25, -2.00) rectangle (4.50, -2.25);
\filldraw[fill=cancan, draw=black] (4.50, -2.00) rectangle (4.75, -2.25);
\filldraw[fill=bermuda, draw=black] (4.75, -2.00) rectangle (5.00, -2.25);
\filldraw[fill=cancan, draw=black] (5.00, -2.00) rectangle (5.25, -2.25);
\filldraw[fill=cancan, draw=black] (5.25, -2.00) rectangle (5.50, -2.25);
\filldraw[fill=cancan, draw=black] (5.50, -2.00) rectangle (5.75, -2.25);
\filldraw[fill=bermuda, draw=black] (5.75, -2.00) rectangle (6.00, -2.25);
\filldraw[fill=cancan, draw=black] (6.00, -2.00) rectangle (6.25, -2.25);
\filldraw[fill=cancan, draw=black] (6.25, -2.00) rectangle (6.50, -2.25);
\filldraw[fill=cancan, draw=black] (6.50, -2.00) rectangle (6.75, -2.25);
\filldraw[fill=cancan, draw=black] (6.75, -2.00) rectangle (7.00, -2.25);
\filldraw[fill=cancan, draw=black] (7.00, -2.00) rectangle (7.25, -2.25);
\filldraw[fill=cancan, draw=black] (7.25, -2.00) rectangle (7.50, -2.25);
\filldraw[fill=cancan, draw=black] (7.50, -2.00) rectangle (7.75, -2.25);
\filldraw[fill=cancan, draw=black] (7.75, -2.00) rectangle (8.00, -2.25);
\filldraw[fill=bermuda, draw=black] (8.00, -2.00) rectangle (8.25, -2.25);
\filldraw[fill=bermuda, draw=black] (8.25, -2.00) rectangle (8.50, -2.25);
\filldraw[fill=cancan, draw=black] (8.50, -2.00) rectangle (8.75, -2.25);
\filldraw[fill=bermuda, draw=black] (8.75, -2.00) rectangle (9.00, -2.25);
\filldraw[fill=cancan, draw=black] (9.00, -2.00) rectangle (9.25, -2.25);
\filldraw[fill=cancan, draw=black] (9.25, -2.00) rectangle (9.50, -2.25);
\filldraw[fill=bermuda, draw=black] (9.50, -2.00) rectangle (9.75, -2.25);
\filldraw[fill=bermuda, draw=black] (9.75, -2.00) rectangle (10.00, -2.25);
\filldraw[fill=cancan, draw=black] (10.00, -2.00) rectangle (10.25, -2.25);
\filldraw[fill=bermuda, draw=black] (10.25, -2.00) rectangle (10.50, -2.25);
\filldraw[fill=cancan, draw=black] (10.50, -2.00) rectangle (10.75, -2.25);
\filldraw[fill=bermuda, draw=black] (10.75, -2.00) rectangle (11.00, -2.25);
\filldraw[fill=bermuda, draw=black] (11.00, -2.00) rectangle (11.25, -2.25);
\filldraw[fill=bermuda, draw=black] (11.25, -2.00) rectangle (11.50, -2.25);
\filldraw[fill=cancan, draw=black] (11.50, -2.00) rectangle (11.75, -2.25);
\filldraw[fill=bermuda, draw=black] (11.75, -2.00) rectangle (12.00, -2.25);
\filldraw[fill=bermuda, draw=black] (12.00, -2.00) rectangle (12.25, -2.25);
\filldraw[fill=bermuda, draw=black] (12.25, -2.00) rectangle (12.50, -2.25);
\filldraw[fill=cancan, draw=black] (12.50, -2.00) rectangle (12.75, -2.25);
\filldraw[fill=bermuda, draw=black] (12.75, -2.00) rectangle (13.00, -2.25);
\filldraw[fill=cancan, draw=black] (13.00, -2.00) rectangle (13.25, -2.25);
\filldraw[fill=bermuda, draw=black] (13.25, -2.00) rectangle (13.50, -2.25);
\filldraw[fill=cancan, draw=black] (13.50, -2.00) rectangle (13.75, -2.25);
\filldraw[fill=bermuda, draw=black] (13.75, -2.00) rectangle (14.00, -2.25);
\filldraw[fill=bermuda, draw=black] (14.00, -2.00) rectangle (14.25, -2.25);
\filldraw[fill=bermuda, draw=black] (14.25, -2.00) rectangle (14.50, -2.25);
\filldraw[fill=cancan, draw=black] (14.50, -2.00) rectangle (14.75, -2.25);
\filldraw[fill=cancan, draw=black] (14.75, -2.00) rectangle (15.00, -2.25);
\filldraw[fill=cancan, draw=black] (0.00, -2.25) rectangle (0.25, -2.50);
\filldraw[fill=cancan, draw=black] (0.25, -2.25) rectangle (0.50, -2.50);
\filldraw[fill=cancan, draw=black] (0.50, -2.25) rectangle (0.75, -2.50);
\filldraw[fill=cancan, draw=black] (0.75, -2.25) rectangle (1.00, -2.50);
\filldraw[fill=cancan, draw=black] (1.00, -2.25) rectangle (1.25, -2.50);
\filldraw[fill=cancan, draw=black] (1.25, -2.25) rectangle (1.50, -2.50);
\filldraw[fill=cancan, draw=black] (1.50, -2.25) rectangle (1.75, -2.50);
\filldraw[fill=cancan, draw=black] (1.75, -2.25) rectangle (2.00, -2.50);
\filldraw[fill=cancan, draw=black] (2.00, -2.25) rectangle (2.25, -2.50);
\filldraw[fill=bermuda, draw=black] (2.25, -2.25) rectangle (2.50, -2.50);
\filldraw[fill=bermuda, draw=black] (2.50, -2.25) rectangle (2.75, -2.50);
\filldraw[fill=bermuda, draw=black] (2.75, -2.25) rectangle (3.00, -2.50);
\filldraw[fill=cancan, draw=black] (3.00, -2.25) rectangle (3.25, -2.50);
\filldraw[fill=cancan, draw=black] (3.25, -2.25) rectangle (3.50, -2.50);
\filldraw[fill=cancan, draw=black] (3.50, -2.25) rectangle (3.75, -2.50);
\filldraw[fill=cancan, draw=black] (3.75, -2.25) rectangle (4.00, -2.50);
\filldraw[fill=bermuda, draw=black] (4.00, -2.25) rectangle (4.25, -2.50);
\filldraw[fill=bermuda, draw=black] (4.25, -2.25) rectangle (4.50, -2.50);
\filldraw[fill=cancan, draw=black] (4.50, -2.25) rectangle (4.75, -2.50);
\filldraw[fill=bermuda, draw=black] (4.75, -2.25) rectangle (5.00, -2.50);
\filldraw[fill=cancan, draw=black] (5.00, -2.25) rectangle (5.25, -2.50);
\filldraw[fill=cancan, draw=black] (5.25, -2.25) rectangle (5.50, -2.50);
\filldraw[fill=bermuda, draw=black] (5.50, -2.25) rectangle (5.75, -2.50);
\filldraw[fill=bermuda, draw=black] (5.75, -2.25) rectangle (6.00, -2.50);
\filldraw[fill=cancan, draw=black] (6.00, -2.25) rectangle (6.25, -2.50);
\filldraw[fill=bermuda, draw=black] (6.25, -2.25) rectangle (6.50, -2.50);
\filldraw[fill=bermuda, draw=black] (6.50, -2.25) rectangle (6.75, -2.50);
\filldraw[fill=bermuda, draw=black] (6.75, -2.25) rectangle (7.00, -2.50);
\filldraw[fill=cancan, draw=black] (7.00, -2.25) rectangle (7.25, -2.50);
\filldraw[fill=cancan, draw=black] (7.25, -2.25) rectangle (7.50, -2.50);
\filldraw[fill=cancan, draw=black] (7.50, -2.25) rectangle (7.75, -2.50);
\filldraw[fill=cancan, draw=black] (7.75, -2.25) rectangle (8.00, -2.50);
\filldraw[fill=cancan, draw=black] (8.00, -2.25) rectangle (8.25, -2.50);
\filldraw[fill=cancan, draw=black] (8.25, -2.25) rectangle (8.50, -2.50);
\filldraw[fill=cancan, draw=black] (8.50, -2.25) rectangle (8.75, -2.50);
\filldraw[fill=cancan, draw=black] (8.75, -2.25) rectangle (9.00, -2.50);
\filldraw[fill=cancan, draw=black] (9.00, -2.25) rectangle (9.25, -2.50);
\filldraw[fill=cancan, draw=black] (9.25, -2.25) rectangle (9.50, -2.50);
\filldraw[fill=cancan, draw=black] (9.50, -2.25) rectangle (9.75, -2.50);
\filldraw[fill=cancan, draw=black] (9.75, -2.25) rectangle (10.00, -2.50);
\filldraw[fill=cancan, draw=black] (10.00, -2.25) rectangle (10.25, -2.50);
\filldraw[fill=bermuda, draw=black] (10.25, -2.25) rectangle (10.50, -2.50);
\filldraw[fill=bermuda, draw=black] (10.50, -2.25) rectangle (10.75, -2.50);
\filldraw[fill=bermuda, draw=black] (10.75, -2.25) rectangle (11.00, -2.50);
\filldraw[fill=cancan, draw=black] (11.00, -2.25) rectangle (11.25, -2.50);
\filldraw[fill=cancan, draw=black] (11.25, -2.25) rectangle (11.50, -2.50);
\filldraw[fill=cancan, draw=black] (11.50, -2.25) rectangle (11.75, -2.50);
\filldraw[fill=bermuda, draw=black] (11.75, -2.25) rectangle (12.00, -2.50);
\filldraw[fill=bermuda, draw=black] (12.00, -2.25) rectangle (12.25, -2.50);
\filldraw[fill=bermuda, draw=black] (12.25, -2.25) rectangle (12.50, -2.50);
\filldraw[fill=bermuda, draw=black] (12.50, -2.25) rectangle (12.75, -2.50);
\filldraw[fill=bermuda, draw=black] (12.75, -2.25) rectangle (13.00, -2.50);
\filldraw[fill=bermuda, draw=black] (13.00, -2.25) rectangle (13.25, -2.50);
\filldraw[fill=bermuda, draw=black] (13.25, -2.25) rectangle (13.50, -2.50);
\filldraw[fill=cancan, draw=black] (13.50, -2.25) rectangle (13.75, -2.50);
\filldraw[fill=bermuda, draw=black] (13.75, -2.25) rectangle (14.00, -2.50);
\filldraw[fill=cancan, draw=black] (14.00, -2.25) rectangle (14.25, -2.50);
\filldraw[fill=bermuda, draw=black] (14.25, -2.25) rectangle (14.50, -2.50);
\filldraw[fill=bermuda, draw=black] (14.50, -2.25) rectangle (14.75, -2.50);
\filldraw[fill=bermuda, draw=black] (14.75, -2.25) rectangle (15.00, -2.50);
\filldraw[fill=cancan, draw=black] (0.00, -2.50) rectangle (0.25, -2.75);
\filldraw[fill=cancan, draw=black] (0.25, -2.50) rectangle (0.50, -2.75);
\filldraw[fill=cancan, draw=black] (0.50, -2.50) rectangle (0.75, -2.75);
\filldraw[fill=cancan, draw=black] (0.75, -2.50) rectangle (1.00, -2.75);
\filldraw[fill=cancan, draw=black] (1.00, -2.50) rectangle (1.25, -2.75);
\filldraw[fill=cancan, draw=black] (1.25, -2.50) rectangle (1.50, -2.75);
\filldraw[fill=cancan, draw=black] (1.50, -2.50) rectangle (1.75, -2.75);
\filldraw[fill=bermuda, draw=black] (1.75, -2.50) rectangle (2.00, -2.75);
\filldraw[fill=cancan, draw=black] (2.00, -2.50) rectangle (2.25, -2.75);
\filldraw[fill=cancan, draw=black] (2.25, -2.50) rectangle (2.50, -2.75);
\filldraw[fill=bermuda, draw=black] (2.50, -2.50) rectangle (2.75, -2.75);
\filldraw[fill=bermuda, draw=black] (2.75, -2.50) rectangle (3.00, -2.75);
\filldraw[fill=cancan, draw=black] (3.00, -2.50) rectangle (3.25, -2.75);
\filldraw[fill=bermuda, draw=black] (3.25, -2.50) rectangle (3.50, -2.75);
\filldraw[fill=cancan, draw=black] (3.50, -2.50) rectangle (3.75, -2.75);
\filldraw[fill=cancan, draw=black] (3.75, -2.50) rectangle (4.00, -2.75);
\filldraw[fill=bermuda, draw=black] (4.00, -2.50) rectangle (4.25, -2.75);
\filldraw[fill=bermuda, draw=black] (4.25, -2.50) rectangle (4.50, -2.75);
\filldraw[fill=cancan, draw=black] (4.50, -2.50) rectangle (4.75, -2.75);
\filldraw[fill=bermuda, draw=black] (4.75, -2.50) rectangle (5.00, -2.75);
\filldraw[fill=cancan, draw=black] (5.00, -2.50) rectangle (5.25, -2.75);
\filldraw[fill=bermuda, draw=black] (5.25, -2.50) rectangle (5.50, -2.75);
\filldraw[fill=bermuda, draw=black] (5.50, -2.50) rectangle (5.75, -2.75);
\filldraw[fill=bermuda, draw=black] (5.75, -2.50) rectangle (6.00, -2.75);
\filldraw[fill=cancan, draw=black] (6.00, -2.50) rectangle (6.25, -2.75);
\filldraw[fill=cancan, draw=black] (6.25, -2.50) rectangle (6.50, -2.75);
\filldraw[fill=cancan, draw=black] (6.50, -2.50) rectangle (6.75, -2.75);
\filldraw[fill=cancan, draw=black] (6.75, -2.50) rectangle (7.00, -2.75);
\filldraw[fill=cancan, draw=black] (7.00, -2.50) rectangle (7.25, -2.75);
\filldraw[fill=cancan, draw=black] (7.25, -2.50) rectangle (7.50, -2.75);
\filldraw[fill=cancan, draw=black] (7.50, -2.50) rectangle (7.75, -2.75);
\filldraw[fill=bermuda, draw=black] (7.75, -2.50) rectangle (8.00, -2.75);
\filldraw[fill=cancan, draw=black] (8.00, -2.50) rectangle (8.25, -2.75);
\filldraw[fill=cancan, draw=black] (8.25, -2.50) rectangle (8.50, -2.75);
\filldraw[fill=bermuda, draw=black] (8.50, -2.50) rectangle (8.75, -2.75);
\filldraw[fill=bermuda, draw=black] (8.75, -2.50) rectangle (9.00, -2.75);
\filldraw[fill=cancan, draw=black] (9.00, -2.50) rectangle (9.25, -2.75);
\filldraw[fill=bermuda, draw=black] (9.25, -2.50) rectangle (9.50, -2.75);
\filldraw[fill=cancan, draw=black] (9.50, -2.50) rectangle (9.75, -2.75);
\filldraw[fill=cancan, draw=black] (9.75, -2.50) rectangle (10.00, -2.75);
\filldraw[fill=cancan, draw=black] (10.00, -2.50) rectangle (10.25, -2.75);
\filldraw[fill=cancan, draw=black] (10.25, -2.50) rectangle (10.50, -2.75);
\filldraw[fill=cancan, draw=black] (10.50, -2.50) rectangle (10.75, -2.75);
\filldraw[fill=cancan, draw=black] (10.75, -2.50) rectangle (11.00, -2.75);
\filldraw[fill=cancan, draw=black] (11.00, -2.50) rectangle (11.25, -2.75);
\filldraw[fill=bermuda, draw=black] (11.25, -2.50) rectangle (11.50, -2.75);
\filldraw[fill=bermuda, draw=black] (11.50, -2.50) rectangle (11.75, -2.75);
\filldraw[fill=bermuda, draw=black] (11.75, -2.50) rectangle (12.00, -2.75);
\filldraw[fill=cancan, draw=black] (12.00, -2.50) rectangle (12.25, -2.75);
\filldraw[fill=cancan, draw=black] (12.25, -2.50) rectangle (12.50, -2.75);
\filldraw[fill=cancan, draw=black] (12.50, -2.50) rectangle (12.75, -2.75);
\filldraw[fill=cancan, draw=black] (12.75, -2.50) rectangle (13.00, -2.75);
\filldraw[fill=bermuda, draw=black] (13.00, -2.50) rectangle (13.25, -2.75);
\filldraw[fill=bermuda, draw=black] (13.25, -2.50) rectangle (13.50, -2.75);
\filldraw[fill=bermuda, draw=black] (13.50, -2.50) rectangle (13.75, -2.75);
\filldraw[fill=bermuda, draw=black] (13.75, -2.50) rectangle (14.00, -2.75);
\filldraw[fill=cancan, draw=black] (14.00, -2.50) rectangle (14.25, -2.75);
\filldraw[fill=cancan, draw=black] (14.25, -2.50) rectangle (14.50, -2.75);
\filldraw[fill=cancan, draw=black] (14.50, -2.50) rectangle (14.75, -2.75);
\filldraw[fill=cancan, draw=black] (14.75, -2.50) rectangle (15.00, -2.75);
\filldraw[fill=cancan, draw=black] (0.00, -2.75) rectangle (0.25, -3.00);
\filldraw[fill=cancan, draw=black] (0.25, -2.75) rectangle (0.50, -3.00);
\filldraw[fill=cancan, draw=black] (0.50, -2.75) rectangle (0.75, -3.00);
\filldraw[fill=cancan, draw=black] (0.75, -2.75) rectangle (1.00, -3.00);
\filldraw[fill=cancan, draw=black] (1.00, -2.75) rectangle (1.25, -3.00);
\filldraw[fill=bermuda, draw=black] (1.25, -2.75) rectangle (1.50, -3.00);
\filldraw[fill=bermuda, draw=black] (1.50, -2.75) rectangle (1.75, -3.00);
\filldraw[fill=bermuda, draw=black] (1.75, -2.75) rectangle (2.00, -3.00);
\filldraw[fill=cancan, draw=black] (2.00, -2.75) rectangle (2.25, -3.00);
\filldraw[fill=cancan, draw=black] (2.25, -2.75) rectangle (2.50, -3.00);
\filldraw[fill=cancan, draw=black] (2.50, -2.75) rectangle (2.75, -3.00);
\filldraw[fill=bermuda, draw=black] (2.75, -2.75) rectangle (3.00, -3.00);
\filldraw[fill=bermuda, draw=black] (3.00, -2.75) rectangle (3.25, -3.00);
\filldraw[fill=bermuda, draw=black] (3.25, -2.75) rectangle (3.50, -3.00);
\filldraw[fill=cancan, draw=black] (3.50, -2.75) rectangle (3.75, -3.00);
\filldraw[fill=cancan, draw=black] (3.75, -2.75) rectangle (4.00, -3.00);
\filldraw[fill=cancan, draw=black] (4.00, -2.75) rectangle (4.25, -3.00);
\filldraw[fill=bermuda, draw=black] (4.25, -2.75) rectangle (4.50, -3.00);
\filldraw[fill=bermuda, draw=black] (4.50, -2.75) rectangle (4.75, -3.00);
\filldraw[fill=bermuda, draw=black] (4.75, -2.75) rectangle (5.00, -3.00);
\filldraw[fill=bermuda, draw=black] (5.00, -2.75) rectangle (5.25, -3.00);
\filldraw[fill=bermuda, draw=black] (5.25, -2.75) rectangle (5.50, -3.00);
\filldraw[fill=bermuda, draw=black] (5.50, -2.75) rectangle (5.75, -3.00);
\filldraw[fill=bermuda, draw=black] (5.75, -2.75) rectangle (6.00, -3.00);
\filldraw[fill=cancan, draw=black] (6.00, -2.75) rectangle (6.25, -3.00);
\filldraw[fill=bermuda, draw=black] (6.25, -2.75) rectangle (6.50, -3.00);
\filldraw[fill=cancan, draw=black] (6.50, -2.75) rectangle (6.75, -3.00);
\filldraw[fill=cancan, draw=black] (6.75, -2.75) rectangle (7.00, -3.00);
\filldraw[fill=bermuda, draw=black] (7.00, -2.75) rectangle (7.25, -3.00);
\filldraw[fill=bermuda, draw=black] (7.25, -2.75) rectangle (7.50, -3.00);
\filldraw[fill=cancan, draw=black] (7.50, -2.75) rectangle (7.75, -3.00);
\filldraw[fill=bermuda, draw=black] (7.75, -2.75) rectangle (8.00, -3.00);
\filldraw[fill=cancan, draw=black] (8.00, -2.75) rectangle (8.25, -3.00);
\filldraw[fill=cancan, draw=black] (8.25, -2.75) rectangle (8.50, -3.00);
\filldraw[fill=bermuda, draw=black] (8.50, -2.75) rectangle (8.75, -3.00);
\filldraw[fill=bermuda, draw=black] (8.75, -2.75) rectangle (9.00, -3.00);
\filldraw[fill=cancan, draw=black] (9.00, -2.75) rectangle (9.25, -3.00);
\filldraw[fill=bermuda, draw=black] (9.25, -2.75) rectangle (9.50, -3.00);
\filldraw[fill=cancan, draw=black] (9.50, -2.75) rectangle (9.75, -3.00);
\filldraw[fill=cancan, draw=black] (9.75, -2.75) rectangle (10.00, -3.00);
\filldraw[fill=cancan, draw=black] (10.00, -2.75) rectangle (10.25, -3.00);
\filldraw[fill=cancan, draw=black] (10.25, -2.75) rectangle (10.50, -3.00);
\filldraw[fill=cancan, draw=black] (10.50, -2.75) rectangle (10.75, -3.00);
\filldraw[fill=cancan, draw=black] (10.75, -2.75) rectangle (11.00, -3.00);
\filldraw[fill=cancan, draw=black] (11.00, -2.75) rectangle (11.25, -3.00);
\filldraw[fill=cancan, draw=black] (11.25, -2.75) rectangle (11.50, -3.00);
\filldraw[fill=cancan, draw=black] (11.50, -2.75) rectangle (11.75, -3.00);
\filldraw[fill=bermuda, draw=black] (11.75, -2.75) rectangle (12.00, -3.00);
\filldraw[fill=cancan, draw=black] (12.00, -2.75) rectangle (12.25, -3.00);
\filldraw[fill=bermuda, draw=black] (12.25, -2.75) rectangle (12.50, -3.00);
\filldraw[fill=cancan, draw=black] (12.50, -2.75) rectangle (12.75, -3.00);
\filldraw[fill=cancan, draw=black] (12.75, -2.75) rectangle (13.00, -3.00);
\filldraw[fill=bermuda, draw=black] (13.00, -2.75) rectangle (13.25, -3.00);
\filldraw[fill=bermuda, draw=black] (13.25, -2.75) rectangle (13.50, -3.00);
\filldraw[fill=bermuda, draw=black] (13.50, -2.75) rectangle (13.75, -3.00);
\filldraw[fill=bermuda, draw=black] (13.75, -2.75) rectangle (14.00, -3.00);
\filldraw[fill=cancan, draw=black] (14.00, -2.75) rectangle (14.25, -3.00);
\filldraw[fill=cancan, draw=black] (14.25, -2.75) rectangle (14.50, -3.00);
\filldraw[fill=cancan, draw=black] (14.50, -2.75) rectangle (14.75, -3.00);
\filldraw[fill=cancan, draw=black] (14.75, -2.75) rectangle (15.00, -3.00);
\filldraw[fill=cancan, draw=black] (0.00, -3.00) rectangle (0.25, -3.25);
\filldraw[fill=cancan, draw=black] (0.25, -3.00) rectangle (0.50, -3.25);
\filldraw[fill=cancan, draw=black] (0.50, -3.00) rectangle (0.75, -3.25);
\filldraw[fill=bermuda, draw=black] (0.75, -3.00) rectangle (1.00, -3.25);
\filldraw[fill=cancan, draw=black] (1.00, -3.00) rectangle (1.25, -3.25);
\filldraw[fill=cancan, draw=black] (1.25, -3.00) rectangle (1.50, -3.25);
\filldraw[fill=cancan, draw=black] (1.50, -3.00) rectangle (1.75, -3.25);
\filldraw[fill=cancan, draw=black] (1.75, -3.00) rectangle (2.00, -3.25);
\filldraw[fill=cancan, draw=black] (2.00, -3.00) rectangle (2.25, -3.25);
\filldraw[fill=bermuda, draw=black] (2.25, -3.00) rectangle (2.50, -3.25);
\filldraw[fill=bermuda, draw=black] (2.50, -3.00) rectangle (2.75, -3.25);
\filldraw[fill=bermuda, draw=black] (2.75, -3.00) rectangle (3.00, -3.25);
\filldraw[fill=bermuda, draw=black] (3.00, -3.00) rectangle (3.25, -3.25);
\filldraw[fill=bermuda, draw=black] (3.25, -3.00) rectangle (3.50, -3.25);
\filldraw[fill=cancan, draw=black] (3.50, -3.00) rectangle (3.75, -3.25);
\filldraw[fill=bermuda, draw=black] (3.75, -3.00) rectangle (4.00, -3.25);
\filldraw[fill=cancan, draw=black] (4.00, -3.00) rectangle (4.25, -3.25);
\filldraw[fill=bermuda, draw=black] (4.25, -3.00) rectangle (4.50, -3.25);
\filldraw[fill=bermuda, draw=black] (4.50, -3.00) rectangle (4.75, -3.25);
\filldraw[fill=bermuda, draw=black] (4.75, -3.00) rectangle (5.00, -3.25);
\filldraw[fill=cancan, draw=black] (5.00, -3.00) rectangle (5.25, -3.25);
\filldraw[fill=cancan, draw=black] (5.25, -3.00) rectangle (5.50, -3.25);
\filldraw[fill=cancan, draw=black] (5.50, -3.00) rectangle (5.75, -3.25);
\filldraw[fill=bermuda, draw=black] (5.75, -3.00) rectangle (6.00, -3.25);
\filldraw[fill=bermuda, draw=black] (6.00, -3.00) rectangle (6.25, -3.25);
\filldraw[fill=bermuda, draw=black] (6.25, -3.00) rectangle (6.50, -3.25);
\filldraw[fill=cancan, draw=black] (6.50, -3.00) rectangle (6.75, -3.25);
\filldraw[fill=cancan, draw=black] (6.75, -3.00) rectangle (7.00, -3.25);
\filldraw[fill=cancan, draw=black] (7.00, -3.00) rectangle (7.25, -3.25);
\filldraw[fill=bermuda, draw=black] (7.25, -3.00) rectangle (7.50, -3.25);
\filldraw[fill=bermuda, draw=black] (7.50, -3.00) rectangle (7.75, -3.25);
\filldraw[fill=bermuda, draw=black] (7.75, -3.00) rectangle (8.00, -3.25);
\filldraw[fill=cancan, draw=black] (8.00, -3.00) rectangle (8.25, -3.25);
\filldraw[fill=cancan, draw=black] (8.25, -3.00) rectangle (8.50, -3.25);
\filldraw[fill=cancan, draw=black] (8.50, -3.00) rectangle (8.75, -3.25);
\filldraw[fill=bermuda, draw=black] (8.75, -3.00) rectangle (9.00, -3.25);
\filldraw[fill=cancan, draw=black] (9.00, -3.00) rectangle (9.25, -3.25);
\filldraw[fill=cancan, draw=black] (9.25, -3.00) rectangle (9.50, -3.25);
\filldraw[fill=cancan, draw=black] (9.50, -3.00) rectangle (9.75, -3.25);
\filldraw[fill=bermuda, draw=black] (9.75, -3.00) rectangle (10.00, -3.25);
\filldraw[fill=cancan, draw=black] (10.00, -3.00) rectangle (10.25, -3.25);
\filldraw[fill=bermuda, draw=black] (10.25, -3.00) rectangle (10.50, -3.25);
\filldraw[fill=cancan, draw=black] (10.50, -3.00) rectangle (10.75, -3.25);
\filldraw[fill=bermuda, draw=black] (10.75, -3.00) rectangle (11.00, -3.25);
\filldraw[fill=bermuda, draw=black] (11.00, -3.00) rectangle (11.25, -3.25);
\filldraw[fill=bermuda, draw=black] (11.25, -3.00) rectangle (11.50, -3.25);
\filldraw[fill=cancan, draw=black] (11.50, -3.00) rectangle (11.75, -3.25);
\filldraw[fill=cancan, draw=black] (11.75, -3.00) rectangle (12.00, -3.25);
\filldraw[fill=cancan, draw=black] (12.00, -3.00) rectangle (12.25, -3.25);
\filldraw[fill=bermuda, draw=black] (12.25, -3.00) rectangle (12.50, -3.25);
\filldraw[fill=bermuda, draw=black] (12.50, -3.00) rectangle (12.75, -3.25);
\filldraw[fill=bermuda, draw=black] (12.75, -3.00) rectangle (13.00, -3.25);
\filldraw[fill=cancan, draw=black] (13.00, -3.00) rectangle (13.25, -3.25);
\filldraw[fill=bermuda, draw=black] (13.25, -3.00) rectangle (13.50, -3.25);
\filldraw[fill=cancan, draw=black] (13.50, -3.00) rectangle (13.75, -3.25);
\filldraw[fill=bermuda, draw=black] (13.75, -3.00) rectangle (14.00, -3.25);
\filldraw[fill=cancan, draw=black] (14.00, -3.00) rectangle (14.25, -3.25);
\filldraw[fill=cancan, draw=black] (14.25, -3.00) rectangle (14.50, -3.25);
\filldraw[fill=cancan, draw=black] (14.50, -3.00) rectangle (14.75, -3.25);
\filldraw[fill=bermuda, draw=black] (14.75, -3.00) rectangle (15.00, -3.25);
\filldraw[fill=cancan, draw=black] (0.00, -3.25) rectangle (0.25, -3.50);
\filldraw[fill=bermuda, draw=black] (0.25, -3.25) rectangle (0.50, -3.50);
\filldraw[fill=cancan, draw=black] (0.50, -3.25) rectangle (0.75, -3.50);
\filldraw[fill=bermuda, draw=black] (0.75, -3.25) rectangle (1.00, -3.50);
\filldraw[fill=cancan, draw=black] (1.00, -3.25) rectangle (1.25, -3.50);
\filldraw[fill=bermuda, draw=black] (1.25, -3.25) rectangle (1.50, -3.50);
\filldraw[fill=bermuda, draw=black] (1.50, -3.25) rectangle (1.75, -3.50);
\filldraw[fill=bermuda, draw=black] (1.75, -3.25) rectangle (2.00, -3.50);
\filldraw[fill=cancan, draw=black] (2.00, -3.25) rectangle (2.25, -3.50);
\filldraw[fill=cancan, draw=black] (2.25, -3.25) rectangle (2.50, -3.50);
\filldraw[fill=cancan, draw=black] (2.50, -3.25) rectangle (2.75, -3.50);
\filldraw[fill=bermuda, draw=black] (2.75, -3.25) rectangle (3.00, -3.50);
\filldraw[fill=bermuda, draw=black] (3.00, -3.25) rectangle (3.25, -3.50);
\filldraw[fill=bermuda, draw=black] (3.25, -3.25) rectangle (3.50, -3.50);
\filldraw[fill=cancan, draw=black] (3.50, -3.25) rectangle (3.75, -3.50);
\filldraw[fill=cancan, draw=black] (3.75, -3.25) rectangle (4.00, -3.50);
\filldraw[fill=cancan, draw=black] (4.00, -3.25) rectangle (4.25, -3.50);
\filldraw[fill=bermuda, draw=black] (4.25, -3.25) rectangle (4.50, -3.50);
\filldraw[fill=bermuda, draw=black] (4.50, -3.25) rectangle (4.75, -3.50);
\filldraw[fill=bermuda, draw=black] (4.75, -3.25) rectangle (5.00, -3.50);
\filldraw[fill=cancan, draw=black] (5.00, -3.25) rectangle (5.25, -3.50);
} } }\end{equation*}
\begin{equation*}
\hspace{4.6pt} b_{6} = \vcenter{\hbox{ \tikz{
\filldraw[fill=bermuda, draw=black] (0.00, 0.00) rectangle (0.25, -0.25);
\filldraw[fill=bermuda, draw=black] (0.25, 0.00) rectangle (0.50, -0.25);
\filldraw[fill=bermuda, draw=black] (0.50, 0.00) rectangle (0.75, -0.25);
\filldraw[fill=bermuda, draw=black] (0.75, 0.00) rectangle (1.00, -0.25);
\filldraw[fill=bermuda, draw=black] (1.00, 0.00) rectangle (1.25, -0.25);
\filldraw[fill=cancan, draw=black] (1.25, 0.00) rectangle (1.50, -0.25);
\filldraw[fill=cancan, draw=black] (1.50, 0.00) rectangle (1.75, -0.25);
\filldraw[fill=bermuda, draw=black] (1.75, 0.00) rectangle (2.00, -0.25);
\filldraw[fill=bermuda, draw=black] (2.00, 0.00) rectangle (2.25, -0.25);
\filldraw[fill=cancan, draw=black] (2.25, 0.00) rectangle (2.50, -0.25);
\filldraw[fill=cancan, draw=black] (2.50, 0.00) rectangle (2.75, -0.25);
\filldraw[fill=cancan, draw=black] (2.75, 0.00) rectangle (3.00, -0.25);
\filldraw[fill=cancan, draw=black] (3.00, 0.00) rectangle (3.25, -0.25);
\filldraw[fill=cancan, draw=black] (3.25, 0.00) rectangle (3.50, -0.25);
\filldraw[fill=bermuda, draw=black] (3.50, 0.00) rectangle (3.75, -0.25);
\filldraw[fill=cancan, draw=black] (3.75, 0.00) rectangle (4.00, -0.25);
\filldraw[fill=bermuda, draw=black] (4.00, 0.00) rectangle (4.25, -0.25);
\filldraw[fill=cancan, draw=black] (4.25, 0.00) rectangle (4.50, -0.25);
\filldraw[fill=cancan, draw=black] (4.50, 0.00) rectangle (4.75, -0.25);
\filldraw[fill=cancan, draw=black] (4.75, 0.00) rectangle (5.00, -0.25);
\filldraw[fill=bermuda, draw=black] (5.00, 0.00) rectangle (5.25, -0.25);
\filldraw[fill=cancan, draw=black] (5.25, 0.00) rectangle (5.50, -0.25);
\filldraw[fill=bermuda, draw=black] (5.50, 0.00) rectangle (5.75, -0.25);
\filldraw[fill=cancan, draw=black] (5.75, 0.00) rectangle (6.00, -0.25);
\filldraw[fill=cancan, draw=black] (6.00, 0.00) rectangle (6.25, -0.25);
\filldraw[fill=cancan, draw=black] (6.25, 0.00) rectangle (6.50, -0.25);
\filldraw[fill=bermuda, draw=black] (6.50, 0.00) rectangle (6.75, -0.25);
\filldraw[fill=cancan, draw=black] (6.75, 0.00) rectangle (7.00, -0.25);
\filldraw[fill=bermuda, draw=black] (7.00, 0.00) rectangle (7.25, -0.25);
\filldraw[fill=cancan, draw=black] (7.25, 0.00) rectangle (7.50, -0.25);
\filldraw[fill=cancan, draw=black] (7.50, 0.00) rectangle (7.75, -0.25);
\filldraw[fill=cancan, draw=black] (7.75, 0.00) rectangle (8.00, -0.25);
\filldraw[fill=bermuda, draw=black] (8.00, 0.00) rectangle (8.25, -0.25);
\filldraw[fill=cancan, draw=black] (8.25, 0.00) rectangle (8.50, -0.25);
\filldraw[fill=bermuda, draw=black] (8.50, 0.00) rectangle (8.75, -0.25);
\filldraw[fill=cancan, draw=black] (8.75, 0.00) rectangle (9.00, -0.25);
\filldraw[fill=cancan, draw=black] (9.00, 0.00) rectangle (9.25, -0.25);
\filldraw[fill=cancan, draw=black] (9.25, 0.00) rectangle (9.50, -0.25);
\filldraw[fill=bermuda, draw=black] (9.50, 0.00) rectangle (9.75, -0.25);
\filldraw[fill=cancan, draw=black] (9.75, 0.00) rectangle (10.00, -0.25);
\filldraw[fill=bermuda, draw=black] (10.00, 0.00) rectangle (10.25, -0.25);
\filldraw[fill=bermuda, draw=black] (10.25, 0.00) rectangle (10.50, -0.25);
\filldraw[fill=bermuda, draw=black] (10.50, 0.00) rectangle (10.75, -0.25);
\filldraw[fill=cancan, draw=black] (10.75, 0.00) rectangle (11.00, -0.25);
\filldraw[fill=cancan, draw=black] (11.00, 0.00) rectangle (11.25, -0.25);
\filldraw[fill=cancan, draw=black] (11.25, 0.00) rectangle (11.50, -0.25);
\filldraw[fill=bermuda, draw=black] (11.50, 0.00) rectangle (11.75, -0.25);
\filldraw[fill=bermuda, draw=black] (11.75, 0.00) rectangle (12.00, -0.25);
\filldraw[fill=bermuda, draw=black] (12.00, 0.00) rectangle (12.25, -0.25);
\filldraw[fill=cancan, draw=black] (12.25, 0.00) rectangle (12.50, -0.25);
\filldraw[fill=cancan, draw=black] (12.50, 0.00) rectangle (12.75, -0.25);
\filldraw[fill=bermuda, draw=black] (12.75, 0.00) rectangle (13.00, -0.25);
\filldraw[fill=bermuda, draw=black] (13.00, 0.00) rectangle (13.25, -0.25);
\filldraw[fill=cancan, draw=black] (13.25, 0.00) rectangle (13.50, -0.25);
\filldraw[fill=cancan, draw=black] (13.50, 0.00) rectangle (13.75, -0.25);
\filldraw[fill=cancan, draw=black] (13.75, 0.00) rectangle (14.00, -0.25);
\filldraw[fill=cancan, draw=black] (14.00, 0.00) rectangle (14.25, -0.25);
\filldraw[fill=cancan, draw=black] (14.25, 0.00) rectangle (14.50, -0.25);
\filldraw[fill=bermuda, draw=black] (14.50, 0.00) rectangle (14.75, -0.25);
\filldraw[fill=bermuda, draw=black] (14.75, 0.00) rectangle (15.00, -0.25);
\filldraw[fill=bermuda, draw=black] (0.00, -0.25) rectangle (0.25, -0.50);
\filldraw[fill=cancan, draw=black] (0.25, -0.25) rectangle (0.50, -0.50);
\filldraw[fill=cancan, draw=black] (0.50, -0.25) rectangle (0.75, -0.50);
\filldraw[fill=cancan, draw=black] (0.75, -0.25) rectangle (1.00, -0.50);
\filldraw[fill=bermuda, draw=black] (1.00, -0.25) rectangle (1.25, -0.50);
\filldraw[fill=bermuda, draw=black] (1.25, -0.25) rectangle (1.50, -0.50);
\filldraw[fill=bermuda, draw=black] (1.50, -0.25) rectangle (1.75, -0.50);
\filldraw[fill=cancan, draw=black] (1.75, -0.25) rectangle (2.00, -0.50);
\filldraw[fill=bermuda, draw=black] (2.00, -0.25) rectangle (2.25, -0.50);
\filldraw[fill=bermuda, draw=black] (2.25, -0.25) rectangle (2.50, -0.50);
\filldraw[fill=bermuda, draw=black] (2.50, -0.25) rectangle (2.75, -0.50);
\filldraw[fill=cancan, draw=black] (2.75, -0.25) rectangle (3.00, -0.50);
\filldraw[fill=cancan, draw=black] (3.00, -0.25) rectangle (3.25, -0.50);
\filldraw[fill=cancan, draw=black] (3.25, -0.25) rectangle (3.50, -0.50);
\filldraw[fill=bermuda, draw=black] (3.50, -0.25) rectangle (3.75, -0.50);
\filldraw[fill=cancan, draw=black] (3.75, -0.25) rectangle (4.00, -0.50);
\filldraw[fill=cancan, draw=black] (4.00, -0.25) rectangle (4.25, -0.50);
\filldraw[fill=bermuda, draw=black] (4.25, -0.25) rectangle (4.50, -0.50);
\filldraw[fill=bermuda, draw=black] (4.50, -0.25) rectangle (4.75, -0.50);
\filldraw[fill=cancan, draw=black] (4.75, -0.25) rectangle (5.00, -0.50);
\filldraw[fill=bermuda, draw=black] (5.00, -0.25) rectangle (5.25, -0.50);
\filldraw[fill=cancan, draw=black] (5.25, -0.25) rectangle (5.50, -0.50);
\filldraw[fill=cancan, draw=black] (5.50, -0.25) rectangle (5.75, -0.50);
\filldraw[fill=bermuda, draw=black] (5.75, -0.25) rectangle (6.00, -0.50);
\filldraw[fill=bermuda, draw=black] (6.00, -0.25) rectangle (6.25, -0.50);
\filldraw[fill=bermuda, draw=black] (6.25, -0.25) rectangle (6.50, -0.50);
\filldraw[fill=bermuda, draw=black] (6.50, -0.25) rectangle (6.75, -0.50);
\filldraw[fill=cancan, draw=black] (6.75, -0.25) rectangle (7.00, -0.50);
\filldraw[fill=bermuda, draw=black] (7.00, -0.25) rectangle (7.25, -0.50);
\filldraw[fill=cancan, draw=black] (7.25, -0.25) rectangle (7.50, -0.50);
\filldraw[fill=cancan, draw=black] (7.50, -0.25) rectangle (7.75, -0.50);
\filldraw[fill=bermuda, draw=black] (7.75, -0.25) rectangle (8.00, -0.50);
\filldraw[fill=bermuda, draw=black] (8.00, -0.25) rectangle (8.25, -0.50);
\filldraw[fill=cancan, draw=black] (8.25, -0.25) rectangle (8.50, -0.50);
\filldraw[fill=bermuda, draw=black] (8.50, -0.25) rectangle (8.75, -0.50);
\filldraw[fill=cancan, draw=black] (8.75, -0.25) rectangle (9.00, -0.50);
\filldraw[fill=cancan, draw=black] (9.00, -0.25) rectangle (9.25, -0.50);
\filldraw[fill=cancan, draw=black] (9.25, -0.25) rectangle (9.50, -0.50);
\filldraw[fill=cancan, draw=black] (9.50, -0.25) rectangle (9.75, -0.50);
\filldraw[fill=bermuda, draw=black] (9.75, -0.25) rectangle (10.00, -0.50);
\filldraw[fill=bermuda, draw=black] (10.00, -0.25) rectangle (10.25, -0.50);
\filldraw[fill=cancan, draw=black] (10.25, -0.25) rectangle (10.50, -0.50);
\filldraw[fill=cancan, draw=black] (10.50, -0.25) rectangle (10.75, -0.50);
\filldraw[fill=cancan, draw=black] (10.75, -0.25) rectangle (11.00, -0.50);
\filldraw[fill=cancan, draw=black] (11.00, -0.25) rectangle (11.25, -0.50);
\filldraw[fill=cancan, draw=black] (11.25, -0.25) rectangle (11.50, -0.50);
\filldraw[fill=cancan, draw=black] (11.50, -0.25) rectangle (11.75, -0.50);
\filldraw[fill=cancan, draw=black] (11.75, -0.25) rectangle (12.00, -0.50);
\filldraw[fill=bermuda, draw=black] (12.00, -0.25) rectangle (12.25, -0.50);
\filldraw[fill=bermuda, draw=black] (12.25, -0.25) rectangle (12.50, -0.50);
\filldraw[fill=bermuda, draw=black] (12.50, -0.25) rectangle (12.75, -0.50);
\filldraw[fill=cancan, draw=black] (12.75, -0.25) rectangle (13.00, -0.50);
\filldraw[fill=cancan, draw=black] (13.00, -0.25) rectangle (13.25, -0.50);
\filldraw[fill=cancan, draw=black] (13.25, -0.25) rectangle (13.50, -0.50);
\filldraw[fill=bermuda, draw=black] (13.50, -0.25) rectangle (13.75, -0.50);
\filldraw[fill=bermuda, draw=black] (13.75, -0.25) rectangle (14.00, -0.50);
\filldraw[fill=bermuda, draw=black] (14.00, -0.25) rectangle (14.25, -0.50);
\filldraw[fill=cancan, draw=black] (14.25, -0.25) rectangle (14.50, -0.50);
\filldraw[fill=cancan, draw=black] (14.50, -0.25) rectangle (14.75, -0.50);
\filldraw[fill=cancan, draw=black] (14.75, -0.25) rectangle (15.00, -0.50);
\filldraw[fill=bermuda, draw=black] (0.00, -0.50) rectangle (0.25, -0.75);
\filldraw[fill=bermuda, draw=black] (0.25, -0.50) rectangle (0.50, -0.75);
\filldraw[fill=bermuda, draw=black] (0.50, -0.50) rectangle (0.75, -0.75);
\filldraw[fill=cancan, draw=black] (0.75, -0.50) rectangle (1.00, -0.75);
\filldraw[fill=bermuda, draw=black] (1.00, -0.50) rectangle (1.25, -0.75);
\filldraw[fill=bermuda, draw=black] (1.25, -0.50) rectangle (1.50, -0.75);
\filldraw[fill=bermuda, draw=black] (1.50, -0.50) rectangle (1.75, -0.75);
\filldraw[fill=bermuda, draw=black] (1.75, -0.50) rectangle (2.00, -0.75);
\filldraw[fill=bermuda, draw=black] (2.00, -0.50) rectangle (2.25, -0.75);
\filldraw[fill=cancan, draw=black] (2.25, -0.50) rectangle (2.50, -0.75);
\filldraw[fill=bermuda, draw=black] (2.50, -0.50) rectangle (2.75, -0.75);
\filldraw[fill=cancan, draw=black] (2.75, -0.50) rectangle (3.00, -0.75);
\filldraw[fill=cancan, draw=black] (3.00, -0.50) rectangle (3.25, -0.75);
\filldraw[fill=bermuda, draw=black] (3.25, -0.50) rectangle (3.50, -0.75);
\filldraw[fill=bermuda, draw=black] (3.50, -0.50) rectangle (3.75, -0.75);
\filldraw[fill=cancan, draw=black] (3.75, -0.50) rectangle (4.00, -0.75);
\filldraw[fill=bermuda, draw=black] (4.00, -0.50) rectangle (4.25, -0.75);
\filldraw[fill=cancan, draw=black] (4.25, -0.50) rectangle (4.50, -0.75);
\filldraw[fill=cancan, draw=black] (4.50, -0.50) rectangle (4.75, -0.75);
\filldraw[fill=bermuda, draw=black] (4.75, -0.50) rectangle (5.00, -0.75);
\filldraw[fill=bermuda, draw=black] (5.00, -0.50) rectangle (5.25, -0.75);
\filldraw[fill=cancan, draw=black] (5.25, -0.50) rectangle (5.50, -0.75);
\filldraw[fill=bermuda, draw=black] (5.50, -0.50) rectangle (5.75, -0.75);
\filldraw[fill=bermuda, draw=black] (5.75, -0.50) rectangle (6.00, -0.75);
\filldraw[fill=bermuda, draw=black] (6.00, -0.50) rectangle (6.25, -0.75);
\filldraw[fill=cancan, draw=black] (6.25, -0.50) rectangle (6.50, -0.75);
\filldraw[fill=bermuda, draw=black] (6.50, -0.50) rectangle (6.75, -0.75);
\filldraw[fill=bermuda, draw=black] (6.75, -0.50) rectangle (7.00, -0.75);
\filldraw[fill=bermuda, draw=black] (7.00, -0.50) rectangle (7.25, -0.75);
\filldraw[fill=bermuda, draw=black] (7.25, -0.50) rectangle (7.50, -0.75);
\filldraw[fill=bermuda, draw=black] (7.50, -0.50) rectangle (7.75, -0.75);
\filldraw[fill=cancan, draw=black] (7.75, -0.50) rectangle (8.00, -0.75);
\filldraw[fill=bermuda, draw=black] (8.00, -0.50) rectangle (8.25, -0.75);
\filldraw[fill=bermuda, draw=black] (8.25, -0.50) rectangle (8.50, -0.75);
\filldraw[fill=bermuda, draw=black] (8.50, -0.50) rectangle (8.75, -0.75);
\filldraw[fill=cancan, draw=black] (8.75, -0.50) rectangle (9.00, -0.75);
\filldraw[fill=cancan, draw=black] (9.00, -0.50) rectangle (9.25, -0.75);
\filldraw[fill=cancan, draw=black] (9.25, -0.50) rectangle (9.50, -0.75);
\filldraw[fill=cancan, draw=black] (9.50, -0.50) rectangle (9.75, -0.75);
\filldraw[fill=cancan, draw=black] (9.75, -0.50) rectangle (10.00, -0.75);
\filldraw[fill=cancan, draw=black] (10.00, -0.50) rectangle (10.25, -0.75);
\filldraw[fill=cancan, draw=black] (10.25, -0.50) rectangle (10.50, -0.75);
\filldraw[fill=bermuda, draw=black] (10.50, -0.50) rectangle (10.75, -0.75);
\filldraw[fill=bermuda, draw=black] (10.75, -0.50) rectangle (11.00, -0.75);
\filldraw[fill=bermuda, draw=black] (11.00, -0.50) rectangle (11.25, -0.75);
\filldraw[fill=cancan, draw=black] (11.25, -0.50) rectangle (11.50, -0.75);
\filldraw[fill=cancan, draw=black] (11.50, -0.50) rectangle (11.75, -0.75);
\filldraw[fill=bermuda, draw=black] (11.75, -0.50) rectangle (12.00, -0.75);
\filldraw[fill=bermuda, draw=black] (12.00, -0.50) rectangle (12.25, -0.75);
\filldraw[fill=cancan, draw=black] (12.25, -0.50) rectangle (12.50, -0.75);
\filldraw[fill=cancan, draw=black] (12.50, -0.50) rectangle (12.75, -0.75);
\filldraw[fill=cancan, draw=black] (12.75, -0.50) rectangle (13.00, -0.75);
\filldraw[fill=cancan, draw=black] (13.00, -0.50) rectangle (13.25, -0.75);
\filldraw[fill=cancan, draw=black] (13.25, -0.50) rectangle (13.50, -0.75);
\filldraw[fill=bermuda, draw=black] (13.50, -0.50) rectangle (13.75, -0.75);
\filldraw[fill=cancan, draw=black] (13.75, -0.50) rectangle (14.00, -0.75);
\filldraw[fill=cancan, draw=black] (14.00, -0.50) rectangle (14.25, -0.75);
\filldraw[fill=cancan, draw=black] (14.25, -0.50) rectangle (14.50, -0.75);
\filldraw[fill=cancan, draw=black] (14.50, -0.50) rectangle (14.75, -0.75);
\filldraw[fill=cancan, draw=black] (14.75, -0.50) rectangle (15.00, -0.75);
\filldraw[fill=cancan, draw=black] (0.00, -0.75) rectangle (0.25, -1.00);
\filldraw[fill=bermuda, draw=black] (0.25, -0.75) rectangle (0.50, -1.00);
\filldraw[fill=bermuda, draw=black] (0.50, -0.75) rectangle (0.75, -1.00);
\filldraw[fill=cancan, draw=black] (0.75, -0.75) rectangle (1.00, -1.00);
\filldraw[fill=bermuda, draw=black] (1.00, -0.75) rectangle (1.25, -1.00);
\filldraw[fill=bermuda, draw=black] (1.25, -0.75) rectangle (1.50, -1.00);
\filldraw[fill=bermuda, draw=black] (1.50, -0.75) rectangle (1.75, -1.00);
\filldraw[fill=cancan, draw=black] (1.75, -0.75) rectangle (2.00, -1.00);
\filldraw[fill=cancan, draw=black] (2.00, -0.75) rectangle (2.25, -1.00);
\filldraw[fill=cancan, draw=black] (2.25, -0.75) rectangle (2.50, -1.00);
\filldraw[fill=bermuda, draw=black] (2.50, -0.75) rectangle (2.75, -1.00);
\filldraw[fill=bermuda, draw=black] (2.75, -0.75) rectangle (3.00, -1.00);
\filldraw[fill=bermuda, draw=black] (3.00, -0.75) rectangle (3.25, -1.00);
\filldraw[fill=cancan, draw=black] (3.25, -0.75) rectangle (3.50, -1.00);
\filldraw[fill=cancan, draw=black] (3.50, -0.75) rectangle (3.75, -1.00);
\filldraw[fill=cancan, draw=black] (3.75, -0.75) rectangle (4.00, -1.00);
\filldraw[fill=bermuda, draw=black] (4.00, -0.75) rectangle (4.25, -1.00);
\filldraw[fill=bermuda, draw=black] (4.25, -0.75) rectangle (4.50, -1.00);
\filldraw[fill=bermuda, draw=black] (4.50, -0.75) rectangle (4.75, -1.00);
\filldraw[fill=cancan, draw=black] (4.75, -0.75) rectangle (5.00, -1.00);
\filldraw[fill=cancan, draw=black] (5.00, -0.75) rectangle (5.25, -1.00);
\filldraw[fill=cancan, draw=black] (5.25, -0.75) rectangle (5.50, -1.00);
\filldraw[fill=bermuda, draw=black] (5.50, -0.75) rectangle (5.75, -1.00);
\filldraw[fill=bermuda, draw=black] (5.75, -0.75) rectangle (6.00, -1.00);
\filldraw[fill=bermuda, draw=black] (6.00, -0.75) rectangle (6.25, -1.00);
\filldraw[fill=cancan, draw=black] (6.25, -0.75) rectangle (6.50, -1.00);
\filldraw[fill=bermuda, draw=black] (6.50, -0.75) rectangle (6.75, -1.00);
\filldraw[fill=bermuda, draw=black] (6.75, -0.75) rectangle (7.00, -1.00);
\filldraw[fill=bermuda, draw=black] (7.00, -0.75) rectangle (7.25, -1.00);
\filldraw[fill=cancan, draw=black] (7.25, -0.75) rectangle (7.50, -1.00);
\filldraw[fill=cancan, draw=black] (7.50, -0.75) rectangle (7.75, -1.00);
\filldraw[fill=bermuda, draw=black] (7.75, -0.75) rectangle (8.00, -1.00);
\filldraw[fill=bermuda, draw=black] (8.00, -0.75) rectangle (8.25, -1.00);
\filldraw[fill=cancan, draw=black] (8.25, -0.75) rectangle (8.50, -1.00);
\filldraw[fill=cancan, draw=black] (8.50, -0.75) rectangle (8.75, -1.00);
\filldraw[fill=cancan, draw=black] (8.75, -0.75) rectangle (9.00, -1.00);
\filldraw[fill=cancan, draw=black] (9.00, -0.75) rectangle (9.25, -1.00);
\filldraw[fill=bermuda, draw=black] (9.25, -0.75) rectangle (9.50, -1.00);
\filldraw[fill=bermuda, draw=black] (9.50, -0.75) rectangle (9.75, -1.00);
\filldraw[fill=cancan, draw=black] (9.75, -0.75) rectangle (10.00, -1.00);
\filldraw[fill=bermuda, draw=black] (10.00, -0.75) rectangle (10.25, -1.00);
\filldraw[fill=bermuda, draw=black] (10.25, -0.75) rectangle (10.50, -1.00);
\filldraw[fill=bermuda, draw=black] (10.50, -0.75) rectangle (10.75, -1.00);
\filldraw[fill=bermuda, draw=black] (10.75, -0.75) rectangle (11.00, -1.00);
\filldraw[fill=bermuda, draw=black] (11.00, -0.75) rectangle (11.25, -1.00);
\filldraw[fill=cancan, draw=black] (11.25, -0.75) rectangle (11.50, -1.00);
\filldraw[fill=bermuda, draw=black] (11.50, -0.75) rectangle (11.75, -1.00);
\filldraw[fill=bermuda, draw=black] (11.75, -0.75) rectangle (12.00, -1.00);
\filldraw[fill=bermuda, draw=black] (12.00, -0.75) rectangle (12.25, -1.00);
\filldraw[fill=bermuda, draw=black] (12.25, -0.75) rectangle (12.50, -1.00);
\filldraw[fill=bermuda, draw=black] (12.50, -0.75) rectangle (12.75, -1.00);
\filldraw[fill=cancan, draw=black] (12.75, -0.75) rectangle (13.00, -1.00);
\filldraw[fill=bermuda, draw=black] (13.00, -0.75) rectangle (13.25, -1.00);
\filldraw[fill=bermuda, draw=black] (13.25, -0.75) rectangle (13.50, -1.00);
\filldraw[fill=bermuda, draw=black] (13.50, -0.75) rectangle (13.75, -1.00);
\filldraw[fill=bermuda, draw=black] (13.75, -0.75) rectangle (14.00, -1.00);
\filldraw[fill=bermuda, draw=black] (14.00, -0.75) rectangle (14.25, -1.00);
\filldraw[fill=bermuda, draw=black] (14.25, -0.75) rectangle (14.50, -1.00);
\filldraw[fill=bermuda, draw=black] (14.50, -0.75) rectangle (14.75, -1.00);
\filldraw[fill=cancan, draw=black] (14.75, -0.75) rectangle (15.00, -1.00);
\filldraw[fill=bermuda, draw=black] (0.00, -1.00) rectangle (0.25, -1.25);
\filldraw[fill=cancan, draw=black] (0.25, -1.00) rectangle (0.50, -1.25);
\filldraw[fill=cancan, draw=black] (0.50, -1.00) rectangle (0.75, -1.25);
\filldraw[fill=cancan, draw=black] (0.75, -1.00) rectangle (1.00, -1.25);
\filldraw[fill=bermuda, draw=black] (1.00, -1.00) rectangle (1.25, -1.25);
\filldraw[fill=bermuda, draw=black] (1.25, -1.00) rectangle (1.50, -1.25);
\filldraw[fill=bermuda, draw=black] (1.50, -1.00) rectangle (1.75, -1.25);
\filldraw[fill=bermuda, draw=black] (1.75, -1.00) rectangle (2.00, -1.25);
\filldraw[fill=bermuda, draw=black] (2.00, -1.00) rectangle (2.25, -1.25);
\filldraw[fill=cancan, draw=black] (2.25, -1.00) rectangle (2.50, -1.25);
\filldraw[fill=bermuda, draw=black] (2.50, -1.00) rectangle (2.75, -1.25);
\filldraw[fill=cancan, draw=black] (2.75, -1.00) rectangle (3.00, -1.25);
\filldraw[fill=cancan, draw=black] (3.00, -1.00) rectangle (3.25, -1.25);
\filldraw[fill=bermuda, draw=black] (3.25, -1.00) rectangle (3.50, -1.25);
\filldraw[fill=bermuda, draw=black] (3.50, -1.00) rectangle (3.75, -1.25);
\filldraw[fill=cancan, draw=black] (3.75, -1.00) rectangle (4.00, -1.25);
\filldraw[fill=bermuda, draw=black] (4.00, -1.00) rectangle (4.25, -1.25);
\filldraw[fill=cancan, draw=black] (4.25, -1.00) rectangle (4.50, -1.25);
\filldraw[fill=cancan, draw=black] (4.50, -1.00) rectangle (4.75, -1.25);
\filldraw[fill=cancan, draw=black] (4.75, -1.00) rectangle (5.00, -1.25);
\filldraw[fill=cancan, draw=black] (5.00, -1.00) rectangle (5.25, -1.25);
\filldraw[fill=cancan, draw=black] (5.25, -1.00) rectangle (5.50, -1.25);
\filldraw[fill=cancan, draw=black] (5.50, -1.00) rectangle (5.75, -1.25);
\filldraw[fill=cancan, draw=black] (5.75, -1.00) rectangle (6.00, -1.25);
\filldraw[fill=cancan, draw=black] (6.00, -1.00) rectangle (6.25, -1.25);
\filldraw[fill=cancan, draw=black] (6.25, -1.00) rectangle (6.50, -1.25);
\filldraw[fill=bermuda, draw=black] (6.50, -1.00) rectangle (6.75, -1.25);
\filldraw[fill=cancan, draw=black] (6.75, -1.00) rectangle (7.00, -1.25);
\filldraw[fill=bermuda, draw=black] (7.00, -1.00) rectangle (7.25, -1.25);
\filldraw[fill=cancan, draw=black] (7.25, -1.00) rectangle (7.50, -1.25);
\filldraw[fill=bermuda, draw=black] (7.50, -1.00) rectangle (7.75, -1.25);
\filldraw[fill=bermuda, draw=black] (7.75, -1.00) rectangle (8.00, -1.25);
\filldraw[fill=bermuda, draw=black] (8.00, -1.00) rectangle (8.25, -1.25);
\filldraw[fill=cancan, draw=black] (8.25, -1.00) rectangle (8.50, -1.25);
\filldraw[fill=cancan, draw=black] (8.50, -1.00) rectangle (8.75, -1.25);
\filldraw[fill=cancan, draw=black] (8.75, -1.00) rectangle (9.00, -1.25);
\filldraw[fill=cancan, draw=black] (9.00, -1.00) rectangle (9.25, -1.25);
\filldraw[fill=bermuda, draw=black] (9.25, -1.00) rectangle (9.50, -1.25);
\filldraw[fill=bermuda, draw=black] (9.50, -1.00) rectangle (9.75, -1.25);
\filldraw[fill=cancan, draw=black] (9.75, -1.00) rectangle (10.00, -1.25);
\filldraw[fill=cancan, draw=black] (10.00, -1.00) rectangle (10.25, -1.25);
\filldraw[fill=cancan, draw=black] (10.25, -1.00) rectangle (10.50, -1.25);
\filldraw[fill=cancan, draw=black] (10.50, -1.00) rectangle (10.75, -1.25);
\filldraw[fill=cancan, draw=black] (10.75, -1.00) rectangle (11.00, -1.25);
\filldraw[fill=bermuda, draw=black] (11.00, -1.00) rectangle (11.25, -1.25);
\filldraw[fill=bermuda, draw=black] (11.25, -1.00) rectangle (11.50, -1.25);
\filldraw[fill=bermuda, draw=black] (11.50, -1.00) rectangle (11.75, -1.25);
\filldraw[fill=bermuda, draw=black] (11.75, -1.00) rectangle (12.00, -1.25);
\filldraw[fill=bermuda, draw=black] (12.00, -1.00) rectangle (12.25, -1.25);
\filldraw[fill=cancan, draw=black] (12.25, -1.00) rectangle (12.50, -1.25);
\filldraw[fill=cancan, draw=black] (12.50, -1.00) rectangle (12.75, -1.25);
\filldraw[fill=cancan, draw=black] (12.75, -1.00) rectangle (13.00, -1.25);
\filldraw[fill=cancan, draw=black] (13.00, -1.00) rectangle (13.25, -1.25);
\filldraw[fill=bermuda, draw=black] (13.25, -1.00) rectangle (13.50, -1.25);
\filldraw[fill=bermuda, draw=black] (13.50, -1.00) rectangle (13.75, -1.25);
\filldraw[fill=cancan, draw=black] (13.75, -1.00) rectangle (14.00, -1.25);
\filldraw[fill=bermuda, draw=black] (14.00, -1.00) rectangle (14.25, -1.25);
\filldraw[fill=bermuda, draw=black] (14.25, -1.00) rectangle (14.50, -1.25);
\filldraw[fill=bermuda, draw=black] (14.50, -1.00) rectangle (14.75, -1.25);
\filldraw[fill=cancan, draw=black] (14.75, -1.00) rectangle (15.00, -1.25);
\filldraw[fill=cancan, draw=black] (0.00, -1.25) rectangle (0.25, -1.50);
\filldraw[fill=cancan, draw=black] (0.25, -1.25) rectangle (0.50, -1.50);
\filldraw[fill=cancan, draw=black] (0.50, -1.25) rectangle (0.75, -1.50);
\filldraw[fill=cancan, draw=black] (0.75, -1.25) rectangle (1.00, -1.50);
\filldraw[fill=cancan, draw=black] (1.00, -1.25) rectangle (1.25, -1.50);
\filldraw[fill=cancan, draw=black] (1.25, -1.25) rectangle (1.50, -1.50);
\filldraw[fill=cancan, draw=black] (1.50, -1.25) rectangle (1.75, -1.50);
\filldraw[fill=bermuda, draw=black] (1.75, -1.25) rectangle (2.00, -1.50);
\filldraw[fill=bermuda, draw=black] (2.00, -1.25) rectangle (2.25, -1.50);
\filldraw[fill=cancan, draw=black] (2.25, -1.25) rectangle (2.50, -1.50);
\filldraw[fill=bermuda, draw=black] (2.50, -1.25) rectangle (2.75, -1.50);
\filldraw[fill=bermuda, draw=black] (2.75, -1.25) rectangle (3.00, -1.50);
\filldraw[fill=bermuda, draw=black] (3.00, -1.25) rectangle (3.25, -1.50);
\filldraw[fill=cancan, draw=black] (3.25, -1.25) rectangle (3.50, -1.50);
\filldraw[fill=bermuda, draw=black] (3.50, -1.25) rectangle (3.75, -1.50);
\filldraw[fill=bermuda, draw=black] (3.75, -1.25) rectangle (4.00, -1.50);
\filldraw[fill=bermuda, draw=black] (4.00, -1.25) rectangle (4.25, -1.50);
\filldraw[fill=cancan, draw=black] (4.25, -1.25) rectangle (4.50, -1.50);
\filldraw[fill=cancan, draw=black] (4.50, -1.25) rectangle (4.75, -1.50);
\filldraw[fill=cancan, draw=black] (4.75, -1.25) rectangle (5.00, -1.50);
\filldraw[fill=bermuda, draw=black] (5.00, -1.25) rectangle (5.25, -1.50);
\filldraw[fill=bermuda, draw=black] (5.25, -1.25) rectangle (5.50, -1.50);
\filldraw[fill=bermuda, draw=black] (5.50, -1.25) rectangle (5.75, -1.50);
\filldraw[fill=cancan, draw=black] (5.75, -1.25) rectangle (6.00, -1.50);
\filldraw[fill=bermuda, draw=black] (6.00, -1.25) rectangle (6.25, -1.50);
\filldraw[fill=bermuda, draw=black] (6.25, -1.25) rectangle (6.50, -1.50);
\filldraw[fill=bermuda, draw=black] (6.50, -1.25) rectangle (6.75, -1.50);
\filldraw[fill=bermuda, draw=black] (6.75, -1.25) rectangle (7.00, -1.50);
\filldraw[fill=bermuda, draw=black] (7.00, -1.25) rectangle (7.25, -1.50);
\filldraw[fill=bermuda, draw=black] (7.25, -1.25) rectangle (7.50, -1.50);
\filldraw[fill=bermuda, draw=black] (7.50, -1.25) rectangle (7.75, -1.50);
\filldraw[fill=bermuda, draw=black] (7.75, -1.25) rectangle (8.00, -1.50);
\filldraw[fill=bermuda, draw=black] (8.00, -1.25) rectangle (8.25, -1.50);
\filldraw[fill=cancan, draw=black] (8.25, -1.25) rectangle (8.50, -1.50);
\filldraw[fill=bermuda, draw=black] (8.50, -1.25) rectangle (8.75, -1.50);
\filldraw[fill=bermuda, draw=black] (8.75, -1.25) rectangle (9.00, -1.50);
\filldraw[fill=bermuda, draw=black] (9.00, -1.25) rectangle (9.25, -1.50);
\filldraw[fill=cancan, draw=black] (9.25, -1.25) rectangle (9.50, -1.50);
\filldraw[fill=bermuda, draw=black] (9.50, -1.25) rectangle (9.75, -1.50);
\filldraw[fill=cancan, draw=black] (9.75, -1.25) rectangle (10.00, -1.50);
\filldraw[fill=cancan, draw=black] (10.00, -1.25) rectangle (10.25, -1.50);
\filldraw[fill=cancan, draw=black] (10.25, -1.25) rectangle (10.50, -1.50);
\filldraw[fill=cancan, draw=black] (10.50, -1.25) rectangle (10.75, -1.50);
\filldraw[fill=cancan, draw=black] (10.75, -1.25) rectangle (11.00, -1.50);
\filldraw[fill=bermuda, draw=black] (11.00, -1.25) rectangle (11.25, -1.50);
\filldraw[fill=bermuda, draw=black] (11.25, -1.25) rectangle (11.50, -1.50);
\filldraw[fill=bermuda, draw=black] (11.50, -1.25) rectangle (11.75, -1.50);
\filldraw[fill=bermuda, draw=black] (11.75, -1.25) rectangle (12.00, -1.50);
\filldraw[fill=bermuda, draw=black] (12.00, -1.25) rectangle (12.25, -1.50);
\filldraw[fill=cancan, draw=black] (12.25, -1.25) rectangle (12.50, -1.50);
\filldraw[fill=bermuda, draw=black] (12.50, -1.25) rectangle (12.75, -1.50);
\filldraw[fill=bermuda, draw=black] (12.75, -1.25) rectangle (13.00, -1.50);
\filldraw[fill=bermuda, draw=black] (13.00, -1.25) rectangle (13.25, -1.50);
\filldraw[fill=cancan, draw=black] (13.25, -1.25) rectangle (13.50, -1.50);
\filldraw[fill=cancan, draw=black] (13.50, -1.25) rectangle (13.75, -1.50);
\filldraw[fill=cancan, draw=black] (13.75, -1.25) rectangle (14.00, -1.50);
\filldraw[fill=bermuda, draw=black] (14.00, -1.25) rectangle (14.25, -1.50);
\filldraw[fill=bermuda, draw=black] (14.25, -1.25) rectangle (14.50, -1.50);
\filldraw[fill=bermuda, draw=black] (14.50, -1.25) rectangle (14.75, -1.50);
\filldraw[fill=cancan, draw=black] (14.75, -1.25) rectangle (15.00, -1.50);
\filldraw[fill=bermuda, draw=black] (0.00, -1.50) rectangle (0.25, -1.75);
\filldraw[fill=cancan, draw=black] (0.25, -1.50) rectangle (0.50, -1.75);
\filldraw[fill=cancan, draw=black] (0.50, -1.50) rectangle (0.75, -1.75);
\filldraw[fill=bermuda, draw=black] (0.75, -1.50) rectangle (1.00, -1.75);
\filldraw[fill=bermuda, draw=black] (1.00, -1.50) rectangle (1.25, -1.75);
\filldraw[fill=bermuda, draw=black] (1.25, -1.50) rectangle (1.50, -1.75);
\filldraw[fill=bermuda, draw=black] (1.50, -1.50) rectangle (1.75, -1.75);
\filldraw[fill=cancan, draw=black] (1.75, -1.50) rectangle (2.00, -1.75);
\filldraw[fill=bermuda, draw=black] (2.00, -1.50) rectangle (2.25, -1.75);
\filldraw[fill=bermuda, draw=black] (2.25, -1.50) rectangle (2.50, -1.75);
\filldraw[fill=bermuda, draw=black] (2.50, -1.50) rectangle (2.75, -1.75);
\filldraw[fill=bermuda, draw=black] (2.75, -1.50) rectangle (3.00, -1.75);
\filldraw[fill=bermuda, draw=black] (3.00, -1.50) rectangle (3.25, -1.75);
\filldraw[fill=cancan, draw=black] (3.25, -1.50) rectangle (3.50, -1.75);
\filldraw[fill=cancan, draw=black] (3.50, -1.50) rectangle (3.75, -1.75);
\filldraw[fill=cancan, draw=black] (3.75, -1.50) rectangle (4.00, -1.75);
\filldraw[fill=cancan, draw=black] (4.00, -1.50) rectangle (4.25, -1.75);
\filldraw[fill=cancan, draw=black] (4.25, -1.50) rectangle (4.50, -1.75);
\filldraw[fill=bermuda, draw=black] (4.50, -1.50) rectangle (4.75, -1.75);
\filldraw[fill=cancan, draw=black] (4.75, -1.50) rectangle (5.00, -1.75);
\filldraw[fill=bermuda, draw=black] (5.00, -1.50) rectangle (5.25, -1.75);
\filldraw[fill=cancan, draw=black] (5.25, -1.50) rectangle (5.50, -1.75);
\filldraw[fill=cancan, draw=black] (5.50, -1.50) rectangle (5.75, -1.75);
\filldraw[fill=cancan, draw=black] (5.75, -1.50) rectangle (6.00, -1.75);
\filldraw[fill=bermuda, draw=black] (6.00, -1.50) rectangle (6.25, -1.75);
\filldraw[fill=bermuda, draw=black] (6.25, -1.50) rectangle (6.50, -1.75);
\filldraw[fill=bermuda, draw=black] (6.50, -1.50) rectangle (6.75, -1.75);
\filldraw[fill=cancan, draw=black] (6.75, -1.50) rectangle (7.00, -1.75);
\filldraw[fill=cancan, draw=black] (7.00, -1.50) rectangle (7.25, -1.75);
\filldraw[fill=cancan, draw=black] (7.25, -1.50) rectangle (7.50, -1.75);
\filldraw[fill=bermuda, draw=black] (7.50, -1.50) rectangle (7.75, -1.75);
\filldraw[fill=bermuda, draw=black] (7.75, -1.50) rectangle (8.00, -1.75);
\filldraw[fill=bermuda, draw=black] (8.00, -1.50) rectangle (8.25, -1.75);
\filldraw[fill=bermuda, draw=black] (8.25, -1.50) rectangle (8.50, -1.75);
\filldraw[fill=bermuda, draw=black] (8.50, -1.50) rectangle (8.75, -1.75);
\filldraw[fill=cancan, draw=black] (8.75, -1.50) rectangle (9.00, -1.75);
\filldraw[fill=cancan, draw=black] (9.00, -1.50) rectangle (9.25, -1.75);
\filldraw[fill=cancan, draw=black] (9.25, -1.50) rectangle (9.50, -1.75);
\filldraw[fill=bermuda, draw=black] (9.50, -1.50) rectangle (9.75, -1.75);
\filldraw[fill=bermuda, draw=black] (9.75, -1.50) rectangle (10.00, -1.75);
\filldraw[fill=bermuda, draw=black] (10.00, -1.50) rectangle (10.25, -1.75);
\filldraw[fill=cancan, draw=black] (10.25, -1.50) rectangle (10.50, -1.75);
\filldraw[fill=cancan, draw=black] (10.50, -1.50) rectangle (10.75, -1.75);
\filldraw[fill=cancan, draw=black] (10.75, -1.50) rectangle (11.00, -1.75);
\filldraw[fill=bermuda, draw=black] (11.00, -1.50) rectangle (11.25, -1.75);
\filldraw[fill=bermuda, draw=black] (11.25, -1.50) rectangle (11.50, -1.75);
\filldraw[fill=bermuda, draw=black] (11.50, -1.50) rectangle (11.75, -1.75);
\filldraw[fill=cancan, draw=black] (11.75, -1.50) rectangle (12.00, -1.75);
\filldraw[fill=cancan, draw=black] (12.00, -1.50) rectangle (12.25, -1.75);
\filldraw[fill=cancan, draw=black] (12.25, -1.50) rectangle (12.50, -1.75);
\filldraw[fill=bermuda, draw=black] (12.50, -1.50) rectangle (12.75, -1.75);
\filldraw[fill=bermuda, draw=black] (12.75, -1.50) rectangle (13.00, -1.75);
\filldraw[fill=bermuda, draw=black] (13.00, -1.50) rectangle (13.25, -1.75);
\filldraw[fill=cancan, draw=black] (13.25, -1.50) rectangle (13.50, -1.75);
\filldraw[fill=bermuda, draw=black] (13.50, -1.50) rectangle (13.75, -1.75);
\filldraw[fill=cancan, draw=black] (13.75, -1.50) rectangle (14.00, -1.75);
\filldraw[fill=bermuda, draw=black] (14.00, -1.50) rectangle (14.25, -1.75);
\filldraw[fill=bermuda, draw=black] (14.25, -1.50) rectangle (14.50, -1.75);
\filldraw[fill=bermuda, draw=black] (14.50, -1.50) rectangle (14.75, -1.75);
\filldraw[fill=cancan, draw=black] (14.75, -1.50) rectangle (15.00, -1.75);
\filldraw[fill=cancan, draw=black] (0.00, -1.75) rectangle (0.25, -2.00);
\filldraw[fill=cancan, draw=black] (0.25, -1.75) rectangle (0.50, -2.00);
\filldraw[fill=bermuda, draw=black] (0.50, -1.75) rectangle (0.75, -2.00);
\filldraw[fill=bermuda, draw=black] (0.75, -1.75) rectangle (1.00, -2.00);
\filldraw[fill=bermuda, draw=black] (1.00, -1.75) rectangle (1.25, -2.00);
\filldraw[fill=cancan, draw=black] (1.25, -1.75) rectangle (1.50, -2.00);
\filldraw[fill=cancan, draw=black] (1.50, -1.75) rectangle (1.75, -2.00);
\filldraw[fill=cancan, draw=black] (1.75, -1.75) rectangle (2.00, -2.00);
\filldraw[fill=bermuda, draw=black] (2.00, -1.75) rectangle (2.25, -2.00);
\filldraw[fill=bermuda, draw=black] (2.25, -1.75) rectangle (2.50, -2.00);
\filldraw[fill=bermuda, draw=black] (2.50, -1.75) rectangle (2.75, -2.00);
\filldraw[fill=cancan, draw=black] (2.75, -1.75) rectangle (3.00, -2.00);
\filldraw[fill=cancan, draw=black] (3.00, -1.75) rectangle (3.25, -2.00);
\filldraw[fill=cancan, draw=black] (3.25, -1.75) rectangle (3.50, -2.00);
\filldraw[fill=bermuda, draw=black] (3.50, -1.75) rectangle (3.75, -2.00);
\filldraw[fill=bermuda, draw=black] (3.75, -1.75) rectangle (4.00, -2.00);
\filldraw[fill=bermuda, draw=black] (4.00, -1.75) rectangle (4.25, -2.00);
\filldraw[fill=cancan, draw=black] (4.25, -1.75) rectangle (4.50, -2.00);
\filldraw[fill=bermuda, draw=black] (4.50, -1.75) rectangle (4.75, -2.00);
\filldraw[fill=bermuda, draw=black] (4.75, -1.75) rectangle (5.00, -2.00);
\filldraw[fill=bermuda, draw=black] (5.00, -1.75) rectangle (5.25, -2.00);
\filldraw[fill=bermuda, draw=black] (5.25, -1.75) rectangle (5.50, -2.00);
\filldraw[fill=bermuda, draw=black] (5.50, -1.75) rectangle (5.75, -2.00);
\filldraw[fill=bermuda, draw=black] (5.75, -1.75) rectangle (6.00, -2.00);
\filldraw[fill=bermuda, draw=black] (6.00, -1.75) rectangle (6.25, -2.00);
\filldraw[fill=bermuda, draw=black] (6.25, -1.75) rectangle (6.50, -2.00);
\filldraw[fill=bermuda, draw=black] (6.50, -1.75) rectangle (6.75, -2.00);
\filldraw[fill=bermuda, draw=black] (6.75, -1.75) rectangle (7.00, -2.00);
\filldraw[fill=bermuda, draw=black] (7.00, -1.75) rectangle (7.25, -2.00);
\filldraw[fill=cancan, draw=black] (7.25, -1.75) rectangle (7.50, -2.00);
\filldraw[fill=bermuda, draw=black] (7.50, -1.75) rectangle (7.75, -2.00);
\filldraw[fill=cancan, draw=black] (7.75, -1.75) rectangle (8.00, -2.00);
\filldraw[fill=cancan, draw=black] (8.00, -1.75) rectangle (8.25, -2.00);
\filldraw[fill=bermuda, draw=black] (8.25, -1.75) rectangle (8.50, -2.00);
\filldraw[fill=bermuda, draw=black] (8.50, -1.75) rectangle (8.75, -2.00);
\filldraw[fill=cancan, draw=black] (8.75, -1.75) rectangle (9.00, -2.00);
\filldraw[fill=bermuda, draw=black] (9.00, -1.75) rectangle (9.25, -2.00);
\filldraw[fill=cancan, draw=black] (9.25, -1.75) rectangle (9.50, -2.00);
\filldraw[fill=cancan, draw=black] (9.50, -1.75) rectangle (9.75, -2.00);
\filldraw[fill=bermuda, draw=black] (9.75, -1.75) rectangle (10.00, -2.00);
\filldraw[fill=bermuda, draw=black] (10.00, -1.75) rectangle (10.25, -2.00);
\filldraw[fill=cancan, draw=black] (10.25, -1.75) rectangle (10.50, -2.00);
\filldraw[fill=bermuda, draw=black] (10.50, -1.75) rectangle (10.75, -2.00);
\filldraw[fill=cancan, draw=black] (10.75, -1.75) rectangle (11.00, -2.00);
\filldraw[fill=cancan, draw=black] (11.00, -1.75) rectangle (11.25, -2.00);
\filldraw[fill=bermuda, draw=black] (11.25, -1.75) rectangle (11.50, -2.00);
\filldraw[fill=bermuda, draw=black] (11.50, -1.75) rectangle (11.75, -2.00);
\filldraw[fill=cancan, draw=black] (11.75, -1.75) rectangle (12.00, -2.00);
\filldraw[fill=bermuda, draw=black] (12.00, -1.75) rectangle (12.25, -2.00);
\filldraw[fill=cancan, draw=black] (12.25, -1.75) rectangle (12.50, -2.00);
\filldraw[fill=cancan, draw=black] (12.50, -1.75) rectangle (12.75, -2.00);
\filldraw[fill=bermuda, draw=black] (12.75, -1.75) rectangle (13.00, -2.00);
\filldraw[fill=bermuda, draw=black] (13.00, -1.75) rectangle (13.25, -2.00);
\filldraw[fill=cancan, draw=black] (13.25, -1.75) rectangle (13.50, -2.00);
\filldraw[fill=bermuda, draw=black] (13.50, -1.75) rectangle (13.75, -2.00);
\filldraw[fill=cancan, draw=black] (13.75, -1.75) rectangle (14.00, -2.00);
\filldraw[fill=cancan, draw=black] (14.00, -1.75) rectangle (14.25, -2.00);
\filldraw[fill=bermuda, draw=black] (14.25, -1.75) rectangle (14.50, -2.00);
\filldraw[fill=bermuda, draw=black] (14.50, -1.75) rectangle (14.75, -2.00);
\filldraw[fill=cancan, draw=black] (14.75, -1.75) rectangle (15.00, -2.00);
\filldraw[fill=bermuda, draw=black] (0.00, -2.00) rectangle (0.25, -2.25);
\filldraw[fill=cancan, draw=black] (0.25, -2.00) rectangle (0.50, -2.25);
\filldraw[fill=bermuda, draw=black] (0.50, -2.00) rectangle (0.75, -2.25);
\filldraw[fill=cancan, draw=black] (0.75, -2.00) rectangle (1.00, -2.25);
\filldraw[fill=bermuda, draw=black] (1.00, -2.00) rectangle (1.25, -2.25);
\filldraw[fill=cancan, draw=black] (1.25, -2.00) rectangle (1.50, -2.25);
\filldraw[fill=cancan, draw=black] (1.50, -2.00) rectangle (1.75, -2.25);
\filldraw[fill=bermuda, draw=black] (1.75, -2.00) rectangle (2.00, -2.25);
\filldraw[fill=bermuda, draw=black] (2.00, -2.00) rectangle (2.25, -2.25);
\filldraw[fill=bermuda, draw=black] (2.25, -2.00) rectangle (2.50, -2.25);
\filldraw[fill=bermuda, draw=black] (2.50, -2.00) rectangle (2.75, -2.25);
\filldraw[fill=bermuda, draw=black] (2.75, -2.00) rectangle (3.00, -2.25);
\filldraw[fill=bermuda, draw=black] (3.00, -2.00) rectangle (3.25, -2.25);
\filldraw[fill=cancan, draw=black] (3.25, -2.00) rectangle (3.50, -2.25);
\filldraw[fill=bermuda, draw=black] (3.50, -2.00) rectangle (3.75, -2.25);
\filldraw[fill=bermuda, draw=black] (3.75, -2.00) rectangle (4.00, -2.25);
\filldraw[fill=bermuda, draw=black] (4.00, -2.00) rectangle (4.25, -2.25);
\filldraw[fill=bermuda, draw=black] (4.25, -2.00) rectangle (4.50, -2.25);
\filldraw[fill=bermuda, draw=black] (4.50, -2.00) rectangle (4.75, -2.25);
\filldraw[fill=cancan, draw=black] (4.75, -2.00) rectangle (5.00, -2.25);
\filldraw[fill=cancan, draw=black] (5.00, -2.00) rectangle (5.25, -2.25);
\filldraw[fill=cancan, draw=black] (5.25, -2.00) rectangle (5.50, -2.25);
\filldraw[fill=cancan, draw=black] (5.50, -2.00) rectangle (5.75, -2.25);
\filldraw[fill=cancan, draw=black] (5.75, -2.00) rectangle (6.00, -2.25);
\filldraw[fill=cancan, draw=black] (6.00, -2.00) rectangle (6.25, -2.25);
\filldraw[fill=cancan, draw=black] (6.25, -2.00) rectangle (6.50, -2.25);
\filldraw[fill=cancan, draw=black] (6.50, -2.00) rectangle (6.75, -2.25);
\filldraw[fill=bermuda, draw=black] (6.75, -2.00) rectangle (7.00, -2.25);
\filldraw[fill=bermuda, draw=black] (7.00, -2.00) rectangle (7.25, -2.25);
\filldraw[fill=cancan, draw=black] (7.25, -2.00) rectangle (7.50, -2.25);
\filldraw[fill=cancan, draw=black] (7.50, -2.00) rectangle (7.75, -2.25);
\filldraw[fill=cancan, draw=black] (7.75, -2.00) rectangle (8.00, -2.25);
\filldraw[fill=cancan, draw=black] (8.00, -2.00) rectangle (8.25, -2.25);
\filldraw[fill=bermuda, draw=black] (8.25, -2.00) rectangle (8.50, -2.25);
\filldraw[fill=bermuda, draw=black] (8.50, -2.00) rectangle (8.75, -2.25);
\filldraw[fill=cancan, draw=black] (8.75, -2.00) rectangle (9.00, -2.25);
\filldraw[fill=cancan, draw=black] (9.00, -2.00) rectangle (9.25, -2.25);
\filldraw[fill=cancan, draw=black] (9.25, -2.00) rectangle (9.50, -2.25);
\filldraw[fill=cancan, draw=black] (9.50, -2.00) rectangle (9.75, -2.25);
\filldraw[fill=cancan, draw=black] (9.75, -2.00) rectangle (10.00, -2.25);
\filldraw[fill=bermuda, draw=black] (10.00, -2.00) rectangle (10.25, -2.25);
\filldraw[fill=bermuda, draw=black] (10.25, -2.00) rectangle (10.50, -2.25);
\filldraw[fill=bermuda, draw=black] (10.50, -2.00) rectangle (10.75, -2.25);
\filldraw[fill=bermuda, draw=black] (10.75, -2.00) rectangle (11.00, -2.25);
\filldraw[fill=bermuda, draw=black] (11.00, -2.00) rectangle (11.25, -2.25);
\filldraw[fill=cancan, draw=black] (11.25, -2.00) rectangle (11.50, -2.25);
\filldraw[fill=bermuda, draw=black] (11.50, -2.00) rectangle (11.75, -2.25);
\filldraw[fill=cancan, draw=black] (11.75, -2.00) rectangle (12.00, -2.25);
\filldraw[fill=bermuda, draw=black] (12.00, -2.00) rectangle (12.25, -2.25);
\filldraw[fill=cancan, draw=black] (12.25, -2.00) rectangle (12.50, -2.25);
\filldraw[fill=bermuda, draw=black] (12.50, -2.00) rectangle (12.75, -2.25);
\filldraw[fill=cancan, draw=black] (12.75, -2.00) rectangle (13.00, -2.25);
\filldraw[fill=bermuda, draw=black] (13.00, -2.00) rectangle (13.25, -2.25);
\filldraw[fill=bermuda, draw=black] (13.25, -2.00) rectangle (13.50, -2.25);
\filldraw[fill=bermuda, draw=black] (13.50, -2.00) rectangle (13.75, -2.25);
\filldraw[fill=cancan, draw=black] (13.75, -2.00) rectangle (14.00, -2.25);
\filldraw[fill=cancan, draw=black] (14.00, -2.00) rectangle (14.25, -2.25);
\filldraw[fill=cancan, draw=black] (14.25, -2.00) rectangle (14.50, -2.25);
\filldraw[fill=cancan, draw=black] (14.50, -2.00) rectangle (14.75, -2.25);
\filldraw[fill=cancan, draw=black] (14.75, -2.00) rectangle (15.00, -2.25);
\filldraw[fill=cancan, draw=black] (0.00, -2.25) rectangle (0.25, -2.50);
\filldraw[fill=cancan, draw=black] (0.25, -2.25) rectangle (0.50, -2.50);
\filldraw[fill=cancan, draw=black] (0.50, -2.25) rectangle (0.75, -2.50);
\filldraw[fill=bermuda, draw=black] (0.75, -2.25) rectangle (1.00, -2.50);
\filldraw[fill=bermuda, draw=black] (1.00, -2.25) rectangle (1.25, -2.50);
\filldraw[fill=cancan, draw=black] (1.25, -2.25) rectangle (1.50, -2.50);
\filldraw[fill=bermuda, draw=black] (1.50, -2.25) rectangle (1.75, -2.50);
\filldraw[fill=cancan, draw=black] (1.75, -2.25) rectangle (2.00, -2.50);
\filldraw[fill=cancan, draw=black] (2.00, -2.25) rectangle (2.25, -2.50);
\filldraw[fill=cancan, draw=black] (2.25, -2.25) rectangle (2.50, -2.50);
\filldraw[fill=bermuda, draw=black] (2.50, -2.25) rectangle (2.75, -2.50);
\filldraw[fill=bermuda, draw=black] (2.75, -2.25) rectangle (3.00, -2.50);
\filldraw[fill=bermuda, draw=black] (3.00, -2.25) rectangle (3.25, -2.50);
\filldraw[fill=bermuda, draw=black] (3.25, -2.25) rectangle (3.50, -2.50);
\filldraw[fill=bermuda, draw=black] (3.50, -2.25) rectangle (3.75, -2.50);
\filldraw[fill=cancan, draw=black] (3.75, -2.25) rectangle (4.00, -2.50);
\filldraw[fill=bermuda, draw=black] (4.00, -2.25) rectangle (4.25, -2.50);
\filldraw[fill=bermuda, draw=black] (4.25, -2.25) rectangle (4.50, -2.50);
\filldraw[fill=bermuda, draw=black] (4.50, -2.25) rectangle (4.75, -2.50);
\filldraw[fill=bermuda, draw=black] (4.75, -2.25) rectangle (5.00, -2.50);
\filldraw[fill=bermuda, draw=black] (5.00, -2.25) rectangle (5.25, -2.50);
\filldraw[fill=cancan, draw=black] (5.25, -2.25) rectangle (5.50, -2.50);
\filldraw[fill=bermuda, draw=black] (5.50, -2.25) rectangle (5.75, -2.50);
\filldraw[fill=cancan, draw=black] (5.75, -2.25) rectangle (6.00, -2.50);
\filldraw[fill=cancan, draw=black] (6.00, -2.25) rectangle (6.25, -2.50);
\filldraw[fill=cancan, draw=black] (6.25, -2.25) rectangle (6.50, -2.50);
\filldraw[fill=cancan, draw=black] (6.50, -2.25) rectangle (6.75, -2.50);
\filldraw[fill=cancan, draw=black] (6.75, -2.25) rectangle (7.00, -2.50);
\filldraw[fill=cancan, draw=black] (7.00, -2.25) rectangle (7.25, -2.50);
\filldraw[fill=cancan, draw=black] (7.25, -2.25) rectangle (7.50, -2.50);
\filldraw[fill=cancan, draw=black] (7.50, -2.25) rectangle (7.75, -2.50);
\filldraw[fill=cancan, draw=black] (7.75, -2.25) rectangle (8.00, -2.50);
\filldraw[fill=bermuda, draw=black] (8.00, -2.25) rectangle (8.25, -2.50);
\filldraw[fill=bermuda, draw=black] (8.25, -2.25) rectangle (8.50, -2.50);
\filldraw[fill=bermuda, draw=black] (8.50, -2.25) rectangle (8.75, -2.50);
\filldraw[fill=cancan, draw=black] (8.75, -2.25) rectangle (9.00, -2.50);
\filldraw[fill=cancan, draw=black] (9.00, -2.25) rectangle (9.25, -2.50);
\filldraw[fill=cancan, draw=black] (9.25, -2.25) rectangle (9.50, -2.50);
\filldraw[fill=bermuda, draw=black] (9.50, -2.25) rectangle (9.75, -2.50);
\filldraw[fill=bermuda, draw=black] (9.75, -2.25) rectangle (10.00, -2.50);
\filldraw[fill=bermuda, draw=black] (10.00, -2.25) rectangle (10.25, -2.50);
\filldraw[fill=cancan, draw=black] (10.25, -2.25) rectangle (10.50, -2.50);
\filldraw[fill=bermuda, draw=black] (10.50, -2.25) rectangle (10.75, -2.50);
\filldraw[fill=bermuda, draw=black] (10.75, -2.25) rectangle (11.00, -2.50);
\filldraw[fill=bermuda, draw=black] (11.00, -2.25) rectangle (11.25, -2.50);
\filldraw[fill=cancan, draw=black] (11.25, -2.25) rectangle (11.50, -2.50);
\filldraw[fill=bermuda, draw=black] (11.50, -2.25) rectangle (11.75, -2.50);
\filldraw[fill=cancan, draw=black] (11.75, -2.25) rectangle (12.00, -2.50);
\filldraw[fill=bermuda, draw=black] (12.00, -2.25) rectangle (12.25, -2.50);
\filldraw[fill=cancan, draw=black] (12.25, -2.25) rectangle (12.50, -2.50);
\filldraw[fill=bermuda, draw=black] (12.50, -2.25) rectangle (12.75, -2.50);
\filldraw[fill=bermuda, draw=black] (12.75, -2.25) rectangle (13.00, -2.50);
\filldraw[fill=bermuda, draw=black] (13.00, -2.25) rectangle (13.25, -2.50);
\filldraw[fill=cancan, draw=black] (13.25, -2.25) rectangle (13.50, -2.50);
\filldraw[fill=cancan, draw=black] (13.50, -2.25) rectangle (13.75, -2.50);
\filldraw[fill=cancan, draw=black] (13.75, -2.25) rectangle (14.00, -2.50);
\filldraw[fill=cancan, draw=black] (14.00, -2.25) rectangle (14.25, -2.50);
\filldraw[fill=cancan, draw=black] (14.25, -2.25) rectangle (14.50, -2.50);
\filldraw[fill=cancan, draw=black] (14.50, -2.25) rectangle (14.75, -2.50);
\filldraw[fill=cancan, draw=black] (14.75, -2.25) rectangle (15.00, -2.50);
\filldraw[fill=cancan, draw=black] (0.00, -2.50) rectangle (0.25, -2.75);
\filldraw[fill=bermuda, draw=black] (0.25, -2.50) rectangle (0.50, -2.75);
\filldraw[fill=bermuda, draw=black] (0.50, -2.50) rectangle (0.75, -2.75);
\filldraw[fill=cancan, draw=black] (0.75, -2.50) rectangle (1.00, -2.75);
\filldraw[fill=cancan, draw=black] (1.00, -2.50) rectangle (1.25, -2.75);
\filldraw[fill=cancan, draw=black] (1.25, -2.50) rectangle (1.50, -2.75);
\filldraw[fill=cancan, draw=black] (1.50, -2.50) rectangle (1.75, -2.75);
\filldraw[fill=bermuda, draw=black] (1.75, -2.50) rectangle (2.00, -2.75);
\filldraw[fill=bermuda, draw=black] (2.00, -2.50) rectangle (2.25, -2.75);
\filldraw[fill=cancan, draw=black] (2.25, -2.50) rectangle (2.50, -2.75);
\filldraw[fill=cancan, draw=black] (2.50, -2.50) rectangle (2.75, -2.75);
\filldraw[fill=cancan, draw=black] (2.75, -2.50) rectangle (3.00, -2.75);
\filldraw[fill=cancan, draw=black] (3.00, -2.50) rectangle (3.25, -2.75);
\filldraw[fill=cancan, draw=black] (3.25, -2.50) rectangle (3.50, -2.75);
\filldraw[fill=bermuda, draw=black] (3.50, -2.50) rectangle (3.75, -2.75);
\filldraw[fill=bermuda, draw=black] (3.75, -2.50) rectangle (4.00, -2.75);
\filldraw[fill=bermuda, draw=black] (4.00, -2.50) rectangle (4.25, -2.75);
\filldraw[fill=cancan, draw=black] (4.25, -2.50) rectangle (4.50, -2.75);
\filldraw[fill=cancan, draw=black] (4.50, -2.50) rectangle (4.75, -2.75);
\filldraw[fill=cancan, draw=black] (4.75, -2.50) rectangle (5.00, -2.75);
\filldraw[fill=cancan, draw=black] (5.00, -2.50) rectangle (5.25, -2.75);
\filldraw[fill=bermuda, draw=black] (5.25, -2.50) rectangle (5.50, -2.75);
\filldraw[fill=bermuda, draw=black] (5.50, -2.50) rectangle (5.75, -2.75);
\filldraw[fill=cancan, draw=black] (5.75, -2.50) rectangle (6.00, -2.75);
\filldraw[fill=bermuda, draw=black] (6.00, -2.50) rectangle (6.25, -2.75);
\filldraw[fill=bermuda, draw=black] (6.25, -2.50) rectangle (6.50, -2.75);
\filldraw[fill=bermuda, draw=black] (6.50, -2.50) rectangle (6.75, -2.75);
\filldraw[fill=bermuda, draw=black] (6.75, -2.50) rectangle (7.00, -2.75);
\filldraw[fill=bermuda, draw=black] (7.00, -2.50) rectangle (7.25, -2.75);
\filldraw[fill=cancan, draw=black] (7.25, -2.50) rectangle (7.50, -2.75);
\filldraw[fill=cancan, draw=black] (7.50, -2.50) rectangle (7.75, -2.75);
\filldraw[fill=cancan, draw=black] (7.75, -2.50) rectangle (8.00, -2.75);
\filldraw[fill=cancan, draw=black] (8.00, -2.50) rectangle (8.25, -2.75);
\filldraw[fill=bermuda, draw=black] (8.25, -2.50) rectangle (8.50, -2.75);
\filldraw[fill=bermuda, draw=black] (8.50, -2.50) rectangle (8.75, -2.75);
\filldraw[fill=cancan, draw=black] (8.75, -2.50) rectangle (9.00, -2.75);
\filldraw[fill=bermuda, draw=black] (9.00, -2.50) rectangle (9.25, -2.75);
\filldraw[fill=cancan, draw=black] (9.25, -2.50) rectangle (9.50, -2.75);
\filldraw[fill=cancan, draw=black] (9.50, -2.50) rectangle (9.75, -2.75);
\filldraw[fill=cancan, draw=black] (9.75, -2.50) rectangle (10.00, -2.75);
\filldraw[fill=bermuda, draw=black] (10.00, -2.50) rectangle (10.25, -2.75);
\filldraw[fill=bermuda, draw=black] (10.25, -2.50) rectangle (10.50, -2.75);
\filldraw[fill=bermuda, draw=black] (10.50, -2.50) rectangle (10.75, -2.75);
\filldraw[fill=cancan, draw=black] (10.75, -2.50) rectangle (11.00, -2.75);
\filldraw[fill=bermuda, draw=black] (11.00, -2.50) rectangle (11.25, -2.75);
\filldraw[fill=cancan, draw=black] (11.25, -2.50) rectangle (11.50, -2.75);
\filldraw[fill=cancan, draw=black] (11.50, -2.50) rectangle (11.75, -2.75);
\filldraw[fill=cancan, draw=black] (11.75, -2.50) rectangle (12.00, -2.75);
\filldraw[fill=cancan, draw=black] (12.00, -2.50) rectangle (12.25, -2.75);
\filldraw[fill=cancan, draw=black] (12.25, -2.50) rectangle (12.50, -2.75);
\filldraw[fill=cancan, draw=black] (12.50, -2.50) rectangle (12.75, -2.75);
\filldraw[fill=bermuda, draw=black] (12.75, -2.50) rectangle (13.00, -2.75);
\filldraw[fill=bermuda, draw=black] (13.00, -2.50) rectangle (13.25, -2.75);
\filldraw[fill=cancan, draw=black] (13.25, -2.50) rectangle (13.50, -2.75);
\filldraw[fill=bermuda, draw=black] (13.50, -2.50) rectangle (13.75, -2.75);
\filldraw[fill=cancan, draw=black] (13.75, -2.50) rectangle (14.00, -2.75);
\filldraw[fill=cancan, draw=black] (14.00, -2.50) rectangle (14.25, -2.75);
\filldraw[fill=bermuda, draw=black] (14.25, -2.50) rectangle (14.50, -2.75);
\filldraw[fill=bermuda, draw=black] (14.50, -2.50) rectangle (14.75, -2.75);
\filldraw[fill=cancan, draw=black] (14.75, -2.50) rectangle (15.00, -2.75);
\filldraw[fill=bermuda, draw=black] (0.00, -2.75) rectangle (0.25, -3.00);
\filldraw[fill=cancan, draw=black] (0.25, -2.75) rectangle (0.50, -3.00);
\filldraw[fill=cancan, draw=black] (0.50, -2.75) rectangle (0.75, -3.00);
\filldraw[fill=bermuda, draw=black] (0.75, -2.75) rectangle (1.00, -3.00);
\filldraw[fill=bermuda, draw=black] (1.00, -2.75) rectangle (1.25, -3.00);
\filldraw[fill=cancan, draw=black] (1.25, -2.75) rectangle (1.50, -3.00);
\filldraw[fill=bermuda, draw=black] (1.50, -2.75) rectangle (1.75, -3.00);
\filldraw[fill=bermuda, draw=black] (1.75, -2.75) rectangle (2.00, -3.00);
\filldraw[fill=bermuda, draw=black] (2.00, -2.75) rectangle (2.25, -3.00);
\filldraw[fill=cancan, draw=black] (2.25, -2.75) rectangle (2.50, -3.00);
\filldraw[fill=bermuda, draw=black] (2.50, -2.75) rectangle (2.75, -3.00);
\filldraw[fill=bermuda, draw=black] (2.75, -2.75) rectangle (3.00, -3.00);
\filldraw[fill=bermuda, draw=black] (3.00, -2.75) rectangle (3.25, -3.00);
\filldraw[fill=bermuda, draw=black] (3.25, -2.75) rectangle (3.50, -3.00);
\filldraw[fill=bermuda, draw=black] (3.50, -2.75) rectangle (3.75, -3.00);
\filldraw[fill=cancan, draw=black] (3.75, -2.75) rectangle (4.00, -3.00);
\filldraw[fill=bermuda, draw=black] (4.00, -2.75) rectangle (4.25, -3.00);
\filldraw[fill=bermuda, draw=black] (4.25, -2.75) rectangle (4.50, -3.00);
\filldraw[fill=bermuda, draw=black] (4.50, -2.75) rectangle (4.75, -3.00);
\filldraw[fill=bermuda, draw=black] (4.75, -2.75) rectangle (5.00, -3.00);
\filldraw[fill=bermuda, draw=black] (5.00, -2.75) rectangle (5.25, -3.00);
\filldraw[fill=cancan, draw=black] (5.25, -2.75) rectangle (5.50, -3.00);
\filldraw[fill=bermuda, draw=black] (5.50, -2.75) rectangle (5.75, -3.00);
\filldraw[fill=bermuda, draw=black] (5.75, -2.75) rectangle (6.00, -3.00);
\filldraw[fill=bermuda, draw=black] (6.00, -2.75) rectangle (6.25, -3.00);
\filldraw[fill=bermuda, draw=black] (6.25, -2.75) rectangle (6.50, -3.00);
\filldraw[fill=bermuda, draw=black] (6.50, -2.75) rectangle (6.75, -3.00);
\filldraw[fill=cancan, draw=black] (6.75, -2.75) rectangle (7.00, -3.00);
\filldraw[fill=cancan, draw=black] (7.00, -2.75) rectangle (7.25, -3.00);
\filldraw[fill=cancan, draw=black] (7.25, -2.75) rectangle (7.50, -3.00);
\filldraw[fill=cancan, draw=black] (7.50, -2.75) rectangle (7.75, -3.00);
\filldraw[fill=cancan, draw=black] (7.75, -2.75) rectangle (8.00, -3.00);
\filldraw[fill=cancan, draw=black] (8.00, -2.75) rectangle (8.25, -3.00);
\filldraw[fill=cancan, draw=black] (8.25, -2.75) rectangle (8.50, -3.00);
\filldraw[fill=cancan, draw=black] (8.50, -2.75) rectangle (8.75, -3.00);
\filldraw[fill=cancan, draw=black] (8.75, -2.75) rectangle (9.00, -3.00);
\filldraw[fill=cancan, draw=black] (9.00, -2.75) rectangle (9.25, -3.00);
\filldraw[fill=bermuda, draw=black] (9.25, -2.75) rectangle (9.50, -3.00);
\filldraw[fill=bermuda, draw=black] (9.50, -2.75) rectangle (9.75, -3.00);
\filldraw[fill=cancan, draw=black] (9.75, -2.75) rectangle (10.00, -3.00);
\filldraw[fill=bermuda, draw=black] (10.00, -2.75) rectangle (10.25, -3.00);
\filldraw[fill=bermuda, draw=black] (10.25, -2.75) rectangle (10.50, -3.00);
\filldraw[fill=bermuda, draw=black] (10.50, -2.75) rectangle (10.75, -3.00);
\filldraw[fill=cancan, draw=black] (10.75, -2.75) rectangle (11.00, -3.00);
\filldraw[fill=cancan, draw=black] (11.00, -2.75) rectangle (11.25, -3.00);
\filldraw[fill=cancan, draw=black] (11.25, -2.75) rectangle (11.50, -3.00);
\filldraw[fill=cancan, draw=black] (11.50, -2.75) rectangle (11.75, -3.00);
\filldraw[fill=bermuda, draw=black] (11.75, -2.75) rectangle (12.00, -3.00);
\filldraw[fill=bermuda, draw=black] (12.00, -2.75) rectangle (12.25, -3.00);
\filldraw[fill=cancan, draw=black] (12.25, -2.75) rectangle (12.50, -3.00);
\filldraw[fill=cancan, draw=black] (12.50, -2.75) rectangle (12.75, -3.00);
\filldraw[fill=cancan, draw=black] (12.75, -2.75) rectangle (13.00, -3.00);
\filldraw[fill=bermuda, draw=black] (13.00, -2.75) rectangle (13.25, -3.00);
\filldraw[fill=bermuda, draw=black] (13.25, -2.75) rectangle (13.50, -3.00);
\filldraw[fill=bermuda, draw=black] (13.50, -2.75) rectangle (13.75, -3.00);
\filldraw[fill=cancan, draw=black] (13.75, -2.75) rectangle (14.00, -3.00);
\filldraw[fill=bermuda, draw=black] (14.00, -2.75) rectangle (14.25, -3.00);
\filldraw[fill=bermuda, draw=black] (14.25, -2.75) rectangle (14.50, -3.00);
\filldraw[fill=bermuda, draw=black] (14.50, -2.75) rectangle (14.75, -3.00);
\filldraw[fill=bermuda, draw=black] (14.75, -2.75) rectangle (15.00, -3.00);
\filldraw[fill=bermuda, draw=black] (0.00, -3.00) rectangle (0.25, -3.25);
\filldraw[fill=bermuda, draw=black] (0.25, -3.00) rectangle (0.50, -3.25);
\filldraw[fill=bermuda, draw=black] (0.50, -3.00) rectangle (0.75, -3.25);
\filldraw[fill=cancan, draw=black] (0.75, -3.00) rectangle (1.00, -3.25);
\filldraw[fill=bermuda, draw=black] (1.00, -3.00) rectangle (1.25, -3.25);
\filldraw[fill=cancan, draw=black] (1.25, -3.00) rectangle (1.50, -3.25);
\filldraw[fill=cancan, draw=black] (1.50, -3.00) rectangle (1.75, -3.25);
\filldraw[fill=bermuda, draw=black] (1.75, -3.00) rectangle (2.00, -3.25);
\filldraw[fill=bermuda, draw=black] (2.00, -3.00) rectangle (2.25, -3.25);
\filldraw[fill=cancan, draw=black] (2.25, -3.00) rectangle (2.50, -3.25);
\filldraw[fill=bermuda, draw=black] (2.50, -3.00) rectangle (2.75, -3.25);
\filldraw[fill=cancan, draw=black] (2.75, -3.00) rectangle (3.00, -3.25);
\filldraw[fill=cancan, draw=black] (3.00, -3.00) rectangle (3.25, -3.25);
\filldraw[fill=bermuda, draw=black] (3.25, -3.00) rectangle (3.50, -3.25);
\filldraw[fill=bermuda, draw=black] (3.50, -3.00) rectangle (3.75, -3.25);
\filldraw[fill=cancan, draw=black] (3.75, -3.00) rectangle (4.00, -3.25);
\filldraw[fill=bermuda, draw=black] (4.00, -3.00) rectangle (4.25, -3.25);
\filldraw[fill=cancan, draw=black] (4.25, -3.00) rectangle (4.50, -3.25);
\filldraw[fill=cancan, draw=black] (4.50, -3.00) rectangle (4.75, -3.25);
\filldraw[fill=bermuda, draw=black] (4.75, -3.00) rectangle (5.00, -3.25);
\filldraw[fill=bermuda, draw=black] (5.00, -3.00) rectangle (5.25, -3.25);
\filldraw[fill=cancan, draw=black] (5.25, -3.00) rectangle (5.50, -3.25);
\filldraw[fill=cancan, draw=black] (5.50, -3.00) rectangle (5.75, -3.25);
\filldraw[fill=cancan, draw=black] (5.75, -3.00) rectangle (6.00, -3.25);
\filldraw[fill=cancan, draw=black] (6.00, -3.00) rectangle (6.25, -3.25);
\filldraw[fill=cancan, draw=black] (6.25, -3.00) rectangle (6.50, -3.25);
\filldraw[fill=cancan, draw=black] (6.50, -3.00) rectangle (6.75, -3.25);
\filldraw[fill=cancan, draw=black] (6.75, -3.00) rectangle (7.00, -3.25);
\filldraw[fill=bermuda, draw=black] (7.00, -3.00) rectangle (7.25, -3.25);
\filldraw[fill=bermuda, draw=black] (7.25, -3.00) rectangle (7.50, -3.25);
\filldraw[fill=bermuda, draw=black] (7.50, -3.00) rectangle (7.75, -3.25);
\filldraw[fill=cancan, draw=black] (7.75, -3.00) rectangle (8.00, -3.25);
\filldraw[fill=cancan, draw=black] (8.00, -3.00) rectangle (8.25, -3.25);
\filldraw[fill=cancan, draw=black] (8.25, -3.00) rectangle (8.50, -3.25);
\filldraw[fill=bermuda, draw=black] (8.50, -3.00) rectangle (8.75, -3.25);
\filldraw[fill=bermuda, draw=black] (8.75, -3.00) rectangle (9.00, -3.25);
\filldraw[fill=bermuda, draw=black] (9.00, -3.00) rectangle (9.25, -3.25);
\filldraw[fill=bermuda, draw=black] (9.25, -3.00) rectangle (9.50, -3.25);
\filldraw[fill=bermuda, draw=black] (9.50, -3.00) rectangle (9.75, -3.25);
\filldraw[fill=bermuda, draw=black] (9.75, -3.00) rectangle (10.00, -3.25);
\filldraw[fill=bermuda, draw=black] (10.00, -3.00) rectangle (10.25, -3.25);
\filldraw[fill=cancan, draw=black] (10.25, -3.00) rectangle (10.50, -3.25);
\filldraw[fill=cancan, draw=black] (10.50, -3.00) rectangle (10.75, -3.25);
\filldraw[fill=cancan, draw=black] (10.75, -3.00) rectangle (11.00, -3.25);
\filldraw[fill=cancan, draw=black] (11.00, -3.00) rectangle (11.25, -3.25);
\filldraw[fill=cancan, draw=black] (11.25, -3.00) rectangle (11.50, -3.25);
\filldraw[fill=cancan, draw=black] (11.50, -3.00) rectangle (11.75, -3.25);
\filldraw[fill=cancan, draw=black] (11.75, -3.00) rectangle (12.00, -3.25);
\filldraw[fill=cancan, draw=black] (12.00, -3.00) rectangle (12.25, -3.25);
\filldraw[fill=cancan, draw=black] (12.25, -3.00) rectangle (12.50, -3.25);
\filldraw[fill=bermuda, draw=black] (12.50, -3.00) rectangle (12.75, -3.25);
\filldraw[fill=bermuda, draw=black] (12.75, -3.00) rectangle (13.00, -3.25);
\filldraw[fill=bermuda, draw=black] (13.00, -3.00) rectangle (13.25, -3.25);
\filldraw[fill=cancan, draw=black] (13.25, -3.00) rectangle (13.50, -3.25);
\filldraw[fill=cancan, draw=black] (13.50, -3.00) rectangle (13.75, -3.25);
\filldraw[fill=cancan, draw=black] (13.75, -3.00) rectangle (14.00, -3.25);
\filldraw[fill=bermuda, draw=black] (14.00, -3.00) rectangle (14.25, -3.25);
\filldraw[fill=bermuda, draw=black] (14.25, -3.00) rectangle (14.50, -3.25);
\filldraw[fill=bermuda, draw=black] (14.50, -3.00) rectangle (14.75, -3.25);
\filldraw[fill=cancan, draw=black] (14.75, -3.00) rectangle (15.00, -3.25);
\filldraw[fill=cancan, draw=black] (0.00, -3.25) rectangle (0.25, -3.50);
\filldraw[fill=cancan, draw=black] (0.25, -3.25) rectangle (0.50, -3.50);
\filldraw[fill=bermuda, draw=black] (0.50, -3.25) rectangle (0.75, -3.50);
\filldraw[fill=bermuda, draw=black] (0.75, -3.25) rectangle (1.00, -3.50);
\filldraw[fill=bermuda, draw=black] (1.00, -3.25) rectangle (1.25, -3.50);
} } }\end{equation*}
\begin{equation*}
\hspace{4.6pt} b_{7} = \vcenter{\hbox{ \tikz{
\filldraw[fill=bermuda, draw=black] (0.00, 0.00) rectangle (0.25, -0.25);
\filldraw[fill=bermuda, draw=black] (0.25, 0.00) rectangle (0.50, -0.25);
\filldraw[fill=cancan, draw=black] (0.50, 0.00) rectangle (0.75, -0.25);
\filldraw[fill=bermuda, draw=black] (0.75, 0.00) rectangle (1.00, -0.25);
\filldraw[fill=bermuda, draw=black] (1.00, 0.00) rectangle (1.25, -0.25);
\filldraw[fill=bermuda, draw=black] (1.25, 0.00) rectangle (1.50, -0.25);
\filldraw[fill=cancan, draw=black] (1.50, 0.00) rectangle (1.75, -0.25);
\filldraw[fill=cancan, draw=black] (1.75, 0.00) rectangle (2.00, -0.25);
\filldraw[fill=bermuda, draw=black] (2.00, 0.00) rectangle (2.25, -0.25);
\filldraw[fill=bermuda, draw=black] (2.25, 0.00) rectangle (2.50, -0.25);
\filldraw[fill=cancan, draw=black] (2.50, 0.00) rectangle (2.75, -0.25);
\filldraw[fill=cancan, draw=black] (2.75, 0.00) rectangle (3.00, -0.25);
\filldraw[fill=cancan, draw=black] (3.00, 0.00) rectangle (3.25, -0.25);
\filldraw[fill=cancan, draw=black] (3.25, 0.00) rectangle (3.50, -0.25);
\filldraw[fill=bermuda, draw=black] (3.50, 0.00) rectangle (3.75, -0.25);
\filldraw[fill=bermuda, draw=black] (3.75, 0.00) rectangle (4.00, -0.25);
\filldraw[fill=bermuda, draw=black] (4.00, 0.00) rectangle (4.25, -0.25);
\filldraw[fill=bermuda, draw=black] (4.25, 0.00) rectangle (4.50, -0.25);
\filldraw[fill=cancan, draw=black] (4.50, 0.00) rectangle (4.75, -0.25);
\filldraw[fill=cancan, draw=black] (4.75, 0.00) rectangle (5.00, -0.25);
\filldraw[fill=cancan, draw=black] (5.00, 0.00) rectangle (5.25, -0.25);
\filldraw[fill=cancan, draw=black] (5.25, 0.00) rectangle (5.50, -0.25);
\filldraw[fill=cancan, draw=black] (5.50, 0.00) rectangle (5.75, -0.25);
\filldraw[fill=cancan, draw=black] (5.75, 0.00) rectangle (6.00, -0.25);
\filldraw[fill=cancan, draw=black] (6.00, 0.00) rectangle (6.25, -0.25);
\filldraw[fill=bermuda, draw=black] (6.25, 0.00) rectangle (6.50, -0.25);
\filldraw[fill=cancan, draw=black] (6.50, 0.00) rectangle (6.75, -0.25);
\filldraw[fill=bermuda, draw=black] (6.75, 0.00) rectangle (7.00, -0.25);
\filldraw[fill=cancan, draw=black] (7.00, 0.00) rectangle (7.25, -0.25);
\filldraw[fill=cancan, draw=black] (7.25, 0.00) rectangle (7.50, -0.25);
\filldraw[fill=bermuda, draw=black] (7.50, 0.00) rectangle (7.75, -0.25);
\filldraw[fill=bermuda, draw=black] (7.75, 0.00) rectangle (8.00, -0.25);
\filldraw[fill=bermuda, draw=black] (8.00, 0.00) rectangle (8.25, -0.25);
\filldraw[fill=bermuda, draw=black] (8.25, 0.00) rectangle (8.50, -0.25);
\filldraw[fill=cancan, draw=black] (8.50, 0.00) rectangle (8.75, -0.25);
\filldraw[fill=cancan, draw=black] (8.75, 0.00) rectangle (9.00, -0.25);
\filldraw[fill=cancan, draw=black] (9.00, 0.00) rectangle (9.25, -0.25);
\filldraw[fill=cancan, draw=black] (9.25, 0.00) rectangle (9.50, -0.25);
\filldraw[fill=cancan, draw=black] (9.50, 0.00) rectangle (9.75, -0.25);
\filldraw[fill=bermuda, draw=black] (9.75, 0.00) rectangle (10.00, -0.25);
\filldraw[fill=bermuda, draw=black] (10.00, 0.00) rectangle (10.25, -0.25);
\filldraw[fill=bermuda, draw=black] (10.25, 0.00) rectangle (10.50, -0.25);
\filldraw[fill=cancan, draw=black] (10.50, 0.00) rectangle (10.75, -0.25);
\filldraw[fill=cancan, draw=black] (10.75, 0.00) rectangle (11.00, -0.25);
\filldraw[fill=cancan, draw=black] (11.00, 0.00) rectangle (11.25, -0.25);
\filldraw[fill=bermuda, draw=black] (11.25, 0.00) rectangle (11.50, -0.25);
\filldraw[fill=bermuda, draw=black] (11.50, 0.00) rectangle (11.75, -0.25);
\filldraw[fill=bermuda, draw=black] (11.75, 0.00) rectangle (12.00, -0.25);
\filldraw[fill=cancan, draw=black] (12.00, 0.00) rectangle (12.25, -0.25);
\filldraw[fill=bermuda, draw=black] (12.25, 0.00) rectangle (12.50, -0.25);
\filldraw[fill=bermuda, draw=black] (12.50, 0.00) rectangle (12.75, -0.25);
\filldraw[fill=bermuda, draw=black] (12.75, 0.00) rectangle (13.00, -0.25);
\filldraw[fill=cancan, draw=black] (13.00, 0.00) rectangle (13.25, -0.25);
\filldraw[fill=cancan, draw=black] (13.25, 0.00) rectangle (13.50, -0.25);
\filldraw[fill=cancan, draw=black] (13.50, 0.00) rectangle (13.75, -0.25);
\filldraw[fill=bermuda, draw=black] (13.75, 0.00) rectangle (14.00, -0.25);
\filldraw[fill=bermuda, draw=black] (14.00, 0.00) rectangle (14.25, -0.25);
\filldraw[fill=bermuda, draw=black] (14.25, 0.00) rectangle (14.50, -0.25);
\filldraw[fill=cancan, draw=black] (14.50, 0.00) rectangle (14.75, -0.25);
\filldraw[fill=cancan, draw=black] (14.75, 0.00) rectangle (15.00, -0.25);
\filldraw[fill=cancan, draw=black] (0.00, -0.25) rectangle (0.25, -0.50);
\filldraw[fill=bermuda, draw=black] (0.25, -0.25) rectangle (0.50, -0.50);
\filldraw[fill=bermuda, draw=black] (0.50, -0.25) rectangle (0.75, -0.50);
\filldraw[fill=bermuda, draw=black] (0.75, -0.25) rectangle (1.00, -0.50);
\filldraw[fill=cancan, draw=black] (1.00, -0.25) rectangle (1.25, -0.50);
\filldraw[fill=cancan, draw=black] (1.25, -0.25) rectangle (1.50, -0.50);
\filldraw[fill=cancan, draw=black] (1.50, -0.25) rectangle (1.75, -0.50);
\filldraw[fill=bermuda, draw=black] (1.75, -0.25) rectangle (2.00, -0.50);
\filldraw[fill=bermuda, draw=black] (2.00, -0.25) rectangle (2.25, -0.50);
\filldraw[fill=bermuda, draw=black] (2.25, -0.25) rectangle (2.50, -0.50);
\filldraw[fill=cancan, draw=black] (2.50, -0.25) rectangle (2.75, -0.50);
\filldraw[fill=cancan, draw=black] (2.75, -0.25) rectangle (3.00, -0.50);
\filldraw[fill=cancan, draw=black] (3.00, -0.25) rectangle (3.25, -0.50);
\filldraw[fill=cancan, draw=black] (3.25, -0.25) rectangle (3.50, -0.50);
\filldraw[fill=bermuda, draw=black] (3.50, -0.25) rectangle (3.75, -0.50);
\filldraw[fill=bermuda, draw=black] (3.75, -0.25) rectangle (4.00, -0.50);
\filldraw[fill=cancan, draw=black] (4.00, -0.25) rectangle (4.25, -0.50);
\filldraw[fill=cancan, draw=black] (4.25, -0.25) rectangle (4.50, -0.50);
\filldraw[fill=cancan, draw=black] (4.50, -0.25) rectangle (4.75, -0.50);
\filldraw[fill=cancan, draw=black] (4.75, -0.25) rectangle (5.00, -0.50);
\filldraw[fill=bermuda, draw=black] (5.00, -0.25) rectangle (5.25, -0.50);
\filldraw[fill=bermuda, draw=black] (5.25, -0.25) rectangle (5.50, -0.50);
\filldraw[fill=cancan, draw=black] (5.50, -0.25) rectangle (5.75, -0.50);
\filldraw[fill=bermuda, draw=black] (5.75, -0.25) rectangle (6.00, -0.50);
\filldraw[fill=bermuda, draw=black] (6.00, -0.25) rectangle (6.25, -0.50);
\filldraw[fill=bermuda, draw=black] (6.25, -0.25) rectangle (6.50, -0.50);
\filldraw[fill=bermuda, draw=black] (6.50, -0.25) rectangle (6.75, -0.50);
\filldraw[fill=bermuda, draw=black] (6.75, -0.25) rectangle (7.00, -0.50);
\filldraw[fill=cancan, draw=black] (7.00, -0.25) rectangle (7.25, -0.50);
\filldraw[fill=bermuda, draw=black] (7.25, -0.25) rectangle (7.50, -0.50);
\filldraw[fill=bermuda, draw=black] (7.50, -0.25) rectangle (7.75, -0.50);
\filldraw[fill=bermuda, draw=black] (7.75, -0.25) rectangle (8.00, -0.50);
\filldraw[fill=bermuda, draw=black] (8.00, -0.25) rectangle (8.25, -0.50);
\filldraw[fill=bermuda, draw=black] (8.25, -0.25) rectangle (8.50, -0.50);
\filldraw[fill=cancan, draw=black] (8.50, -0.25) rectangle (8.75, -0.50);
\filldraw[fill=cancan, draw=black] (8.75, -0.25) rectangle (9.00, -0.50);
\filldraw[fill=bermuda, draw=black] (9.00, -0.25) rectangle (9.25, -0.50);
\filldraw[fill=bermuda, draw=black] (9.25, -0.25) rectangle (9.50, -0.50);
\filldraw[fill=cancan, draw=black] (9.50, -0.25) rectangle (9.75, -0.50);
\filldraw[fill=cancan, draw=black] (9.75, -0.25) rectangle (10.00, -0.50);
\filldraw[fill=cancan, draw=black] (10.00, -0.25) rectangle (10.25, -0.50);
\filldraw[fill=cancan, draw=black] (10.25, -0.25) rectangle (10.50, -0.50);
\filldraw[fill=cancan, draw=black] (10.50, -0.25) rectangle (10.75, -0.50);
\filldraw[fill=cancan, draw=black] (10.75, -0.25) rectangle (11.00, -0.50);
\filldraw[fill=cancan, draw=black] (11.00, -0.25) rectangle (11.25, -0.50);
\filldraw[fill=bermuda, draw=black] (11.25, -0.25) rectangle (11.50, -0.50);
\filldraw[fill=bermuda, draw=black] (11.50, -0.25) rectangle (11.75, -0.50);
\filldraw[fill=bermuda, draw=black] (11.75, -0.25) rectangle (12.00, -0.50);
\filldraw[fill=cancan, draw=black] (12.00, -0.25) rectangle (12.25, -0.50);
\filldraw[fill=cancan, draw=black] (12.25, -0.25) rectangle (12.50, -0.50);
\filldraw[fill=cancan, draw=black] (12.50, -0.25) rectangle (12.75, -0.50);
\filldraw[fill=bermuda, draw=black] (12.75, -0.25) rectangle (13.00, -0.50);
\filldraw[fill=bermuda, draw=black] (13.00, -0.25) rectangle (13.25, -0.50);
\filldraw[fill=bermuda, draw=black] (13.25, -0.25) rectangle (13.50, -0.50);
\filldraw[fill=cancan, draw=black] (13.50, -0.25) rectangle (13.75, -0.50);
\filldraw[fill=cancan, draw=black] (13.75, -0.25) rectangle (14.00, -0.50);
\filldraw[fill=cancan, draw=black] (14.00, -0.25) rectangle (14.25, -0.50);
\filldraw[fill=bermuda, draw=black] (14.25, -0.25) rectangle (14.50, -0.50);
\filldraw[fill=bermuda, draw=black] (14.50, -0.25) rectangle (14.75, -0.50);
\filldraw[fill=bermuda, draw=black] (14.75, -0.25) rectangle (15.00, -0.50);
\filldraw[fill=bermuda, draw=black] (0.00, -0.50) rectangle (0.25, -0.75);
\filldraw[fill=bermuda, draw=black] (0.25, -0.50) rectangle (0.50, -0.75);
\filldraw[fill=cancan, draw=black] (0.50, -0.50) rectangle (0.75, -0.75);
\filldraw[fill=bermuda, draw=black] (0.75, -0.50) rectangle (1.00, -0.75);
\filldraw[fill=bermuda, draw=black] (1.00, -0.50) rectangle (1.25, -0.75);
\filldraw[fill=bermuda, draw=black] (1.25, -0.50) rectangle (1.50, -0.75);
\filldraw[fill=cancan, draw=black] (1.50, -0.50) rectangle (1.75, -0.75);
\filldraw[fill=cancan, draw=black] (1.75, -0.50) rectangle (2.00, -0.75);
\filldraw[fill=cancan, draw=black] (2.00, -0.50) rectangle (2.25, -0.75);
\filldraw[fill=cancan, draw=black] (2.25, -0.50) rectangle (2.50, -0.75);
\filldraw[fill=bermuda, draw=black] (2.50, -0.50) rectangle (2.75, -0.75);
\filldraw[fill=bermuda, draw=black] (2.75, -0.50) rectangle (3.00, -0.75);
\filldraw[fill=cancan, draw=black] (3.00, -0.50) rectangle (3.25, -0.75);
\filldraw[fill=cancan, draw=black] (3.25, -0.50) rectangle (3.50, -0.75);
\filldraw[fill=cancan, draw=black] (3.50, -0.50) rectangle (3.75, -0.75);
\filldraw[fill=cancan, draw=black] (3.75, -0.50) rectangle (4.00, -0.75);
\filldraw[fill=bermuda, draw=black] (4.00, -0.50) rectangle (4.25, -0.75);
\filldraw[fill=bermuda, draw=black] (4.25, -0.50) rectangle (4.50, -0.75);
\filldraw[fill=cancan, draw=black] (4.50, -0.50) rectangle (4.75, -0.75);
\filldraw[fill=cancan, draw=black] (4.75, -0.50) rectangle (5.00, -0.75);
\filldraw[fill=cancan, draw=black] (5.00, -0.50) rectangle (5.25, -0.75);
\filldraw[fill=bermuda, draw=black] (5.25, -0.50) rectangle (5.50, -0.75);
\filldraw[fill=bermuda, draw=black] (5.50, -0.50) rectangle (5.75, -0.75);
\filldraw[fill=bermuda, draw=black] (5.75, -0.50) rectangle (6.00, -0.75);
\filldraw[fill=bermuda, draw=black] (6.00, -0.50) rectangle (6.25, -0.75);
\filldraw[fill=bermuda, draw=black] (6.25, -0.50) rectangle (6.50, -0.75);
\filldraw[fill=cancan, draw=black] (6.50, -0.50) rectangle (6.75, -0.75);
\filldraw[fill=bermuda, draw=black] (6.75, -0.50) rectangle (7.00, -0.75);
\filldraw[fill=bermuda, draw=black] (7.00, -0.50) rectangle (7.25, -0.75);
\filldraw[fill=bermuda, draw=black] (7.25, -0.50) rectangle (7.50, -0.75);
\filldraw[fill=cancan, draw=black] (7.50, -0.50) rectangle (7.75, -0.75);
\filldraw[fill=cancan, draw=black] (7.75, -0.50) rectangle (8.00, -0.75);
\filldraw[fill=cancan, draw=black] (8.00, -0.50) rectangle (8.25, -0.75);
\filldraw[fill=bermuda, draw=black] (8.25, -0.50) rectangle (8.50, -0.75);
\filldraw[fill=bermuda, draw=black] (8.50, -0.50) rectangle (8.75, -0.75);
\filldraw[fill=bermuda, draw=black] (8.75, -0.50) rectangle (9.00, -0.75);
\filldraw[fill=cancan, draw=black] (9.00, -0.50) rectangle (9.25, -0.75);
\filldraw[fill=cancan, draw=black] (9.25, -0.50) rectangle (9.50, -0.75);
\filldraw[fill=cancan, draw=black] (9.50, -0.50) rectangle (9.75, -0.75);
\filldraw[fill=cancan, draw=black] (9.75, -0.50) rectangle (10.00, -0.75);
\filldraw[fill=bermuda, draw=black] (10.00, -0.50) rectangle (10.25, -0.75);
\filldraw[fill=bermuda, draw=black] (10.25, -0.50) rectangle (10.50, -0.75);
\filldraw[fill=cancan, draw=black] (10.50, -0.50) rectangle (10.75, -0.75);
\filldraw[fill=bermuda, draw=black] (10.75, -0.50) rectangle (11.00, -0.75);
\filldraw[fill=cancan, draw=black] (11.00, -0.50) rectangle (11.25, -0.75);
\filldraw[fill=cancan, draw=black] (11.25, -0.50) rectangle (11.50, -0.75);
\filldraw[fill=cancan, draw=black] (11.50, -0.50) rectangle (11.75, -0.75);
\filldraw[fill=bermuda, draw=black] (11.75, -0.50) rectangle (12.00, -0.75);
\filldraw[fill=cancan, draw=black] (12.00, -0.50) rectangle (12.25, -0.75);
\filldraw[fill=cancan, draw=black] (12.25, -0.50) rectangle (12.50, -0.75);
\filldraw[fill=cancan, draw=black] (12.50, -0.50) rectangle (12.75, -0.75);
\filldraw[fill=cancan, draw=black] (12.75, -0.50) rectangle (13.00, -0.75);
\filldraw[fill=cancan, draw=black] (13.00, -0.50) rectangle (13.25, -0.75);
\filldraw[fill=cancan, draw=black] (13.25, -0.50) rectangle (13.50, -0.75);
\filldraw[fill=cancan, draw=black] (13.50, -0.50) rectangle (13.75, -0.75);
\filldraw[fill=cancan, draw=black] (13.75, -0.50) rectangle (14.00, -0.75);
\filldraw[fill=cancan, draw=black] (14.00, -0.50) rectangle (14.25, -0.75);
\filldraw[fill=bermuda, draw=black] (14.25, -0.50) rectangle (14.50, -0.75);
\filldraw[fill=bermuda, draw=black] (14.50, -0.50) rectangle (14.75, -0.75);
\filldraw[fill=bermuda, draw=black] (14.75, -0.50) rectangle (15.00, -0.75);
\filldraw[fill=bermuda, draw=black] (0.00, -0.75) rectangle (0.25, -1.00);
\filldraw[fill=bermuda, draw=black] (0.25, -0.75) rectangle (0.50, -1.00);
\filldraw[fill=cancan, draw=black] (0.50, -0.75) rectangle (0.75, -1.00);
\filldraw[fill=bermuda, draw=black] (0.75, -0.75) rectangle (1.00, -1.00);
\filldraw[fill=cancan, draw=black] (1.00, -0.75) rectangle (1.25, -1.00);
\filldraw[fill=cancan, draw=black] (1.25, -0.75) rectangle (1.50, -1.00);
\filldraw[fill=bermuda, draw=black] (1.50, -0.75) rectangle (1.75, -1.00);
\filldraw[fill=bermuda, draw=black] (1.75, -0.75) rectangle (2.00, -1.00);
\filldraw[fill=cancan, draw=black] (2.00, -0.75) rectangle (2.25, -1.00);
\filldraw[fill=bermuda, draw=black] (2.25, -0.75) rectangle (2.50, -1.00);
\filldraw[fill=cancan, draw=black] (2.50, -0.75) rectangle (2.75, -1.00);
\filldraw[fill=cancan, draw=black] (2.75, -0.75) rectangle (3.00, -1.00);
\filldraw[fill=bermuda, draw=black] (3.00, -0.75) rectangle (3.25, -1.00);
\filldraw[fill=bermuda, draw=black] (3.25, -0.75) rectangle (3.50, -1.00);
\filldraw[fill=cancan, draw=black] (3.50, -0.75) rectangle (3.75, -1.00);
\filldraw[fill=bermuda, draw=black] (3.75, -0.75) rectangle (4.00, -1.00);
\filldraw[fill=cancan, draw=black] (4.00, -0.75) rectangle (4.25, -1.00);
\filldraw[fill=cancan, draw=black] (4.25, -0.75) rectangle (4.50, -1.00);
\filldraw[fill=bermuda, draw=black] (4.50, -0.75) rectangle (4.75, -1.00);
\filldraw[fill=bermuda, draw=black] (4.75, -0.75) rectangle (5.00, -1.00);
\filldraw[fill=cancan, draw=black] (5.00, -0.75) rectangle (5.25, -1.00);
\filldraw[fill=bermuda, draw=black] (5.25, -0.75) rectangle (5.50, -1.00);
\filldraw[fill=cancan, draw=black] (5.50, -0.75) rectangle (5.75, -1.00);
\filldraw[fill=bermuda, draw=black] (5.75, -0.75) rectangle (6.00, -1.00);
\filldraw[fill=bermuda, draw=black] (6.00, -0.75) rectangle (6.25, -1.00);
\filldraw[fill=bermuda, draw=black] (6.25, -0.75) rectangle (6.50, -1.00);
\filldraw[fill=cancan, draw=black] (6.50, -0.75) rectangle (6.75, -1.00);
\filldraw[fill=cancan, draw=black] (6.75, -0.75) rectangle (7.00, -1.00);
\filldraw[fill=bermuda, draw=black] (7.00, -0.75) rectangle (7.25, -1.00);
\filldraw[fill=bermuda, draw=black] (7.25, -0.75) rectangle (7.50, -1.00);
\filldraw[fill=cancan, draw=black] (7.50, -0.75) rectangle (7.75, -1.00);
\filldraw[fill=cancan, draw=black] (7.75, -0.75) rectangle (8.00, -1.00);
\filldraw[fill=cancan, draw=black] (8.00, -0.75) rectangle (8.25, -1.00);
\filldraw[fill=cancan, draw=black] (8.25, -0.75) rectangle (8.50, -1.00);
\filldraw[fill=cancan, draw=black] (8.50, -0.75) rectangle (8.75, -1.00);
\filldraw[fill=bermuda, draw=black] (8.75, -0.75) rectangle (9.00, -1.00);
\filldraw[fill=cancan, draw=black] (9.00, -0.75) rectangle (9.25, -1.00);
\filldraw[fill=bermuda, draw=black] (9.25, -0.75) rectangle (9.50, -1.00);
\filldraw[fill=bermuda, draw=black] (9.50, -0.75) rectangle (9.75, -1.00);
\filldraw[fill=bermuda, draw=black] (9.75, -0.75) rectangle (10.00, -1.00);
\filldraw[fill=cancan, draw=black] (10.00, -0.75) rectangle (10.25, -1.00);
\filldraw[fill=bermuda, draw=black] (10.25, -0.75) rectangle (10.50, -1.00);
\filldraw[fill=bermuda, draw=black] (10.50, -0.75) rectangle (10.75, -1.00);
\filldraw[fill=bermuda, draw=black] (10.75, -0.75) rectangle (11.00, -1.00);
\filldraw[fill=bermuda, draw=black] (11.00, -0.75) rectangle (11.25, -1.00);
\filldraw[fill=bermuda, draw=black] (11.25, -0.75) rectangle (11.50, -1.00);
\filldraw[fill=cancan, draw=black] (11.50, -0.75) rectangle (11.75, -1.00);
\filldraw[fill=bermuda, draw=black] (11.75, -0.75) rectangle (12.00, -1.00);
\filldraw[fill=bermuda, draw=black] (12.00, -0.75) rectangle (12.25, -1.00);
\filldraw[fill=bermuda, draw=black] (12.25, -0.75) rectangle (12.50, -1.00);
\filldraw[fill=cancan, draw=black] (12.50, -0.75) rectangle (12.75, -1.00);
\filldraw[fill=cancan, draw=black] (12.75, -0.75) rectangle (13.00, -1.00);
\filldraw[fill=cancan, draw=black] (13.00, -0.75) rectangle (13.25, -1.00);
\filldraw[fill=bermuda, draw=black] (13.25, -0.75) rectangle (13.50, -1.00);
\filldraw[fill=bermuda, draw=black] (13.50, -0.75) rectangle (13.75, -1.00);
\filldraw[fill=bermuda, draw=black] (13.75, -0.75) rectangle (14.00, -1.00);
\filldraw[fill=cancan, draw=black] (14.00, -0.75) rectangle (14.25, -1.00);
\filldraw[fill=bermuda, draw=black] (14.25, -0.75) rectangle (14.50, -1.00);
\filldraw[fill=bermuda, draw=black] (14.50, -0.75) rectangle (14.75, -1.00);
\filldraw[fill=bermuda, draw=black] (14.75, -0.75) rectangle (15.00, -1.00);
\filldraw[fill=cancan, draw=black] (0.00, -1.00) rectangle (0.25, -1.25);
\filldraw[fill=cancan, draw=black] (0.25, -1.00) rectangle (0.50, -1.25);
\filldraw[fill=cancan, draw=black] (0.50, -1.00) rectangle (0.75, -1.25);
\filldraw[fill=cancan, draw=black] (0.75, -1.00) rectangle (1.00, -1.25);
\filldraw[fill=cancan, draw=black] (1.00, -1.00) rectangle (1.25, -1.25);
\filldraw[fill=bermuda, draw=black] (1.25, -1.00) rectangle (1.50, -1.25);
\filldraw[fill=bermuda, draw=black] (1.50, -1.00) rectangle (1.75, -1.25);
\filldraw[fill=bermuda, draw=black] (1.75, -1.00) rectangle (2.00, -1.25);
\filldraw[fill=cancan, draw=black] (2.00, -1.00) rectangle (2.25, -1.25);
\filldraw[fill=bermuda, draw=black] (2.25, -1.00) rectangle (2.50, -1.25);
\filldraw[fill=bermuda, draw=black] (2.50, -1.00) rectangle (2.75, -1.25);
\filldraw[fill=bermuda, draw=black] (2.75, -1.00) rectangle (3.00, -1.25);
\filldraw[fill=bermuda, draw=black] (3.00, -1.00) rectangle (3.25, -1.25);
\filldraw[fill=bermuda, draw=black] (3.25, -1.00) rectangle (3.50, -1.25);
\filldraw[fill=cancan, draw=black] (3.50, -1.00) rectangle (3.75, -1.25);
\filldraw[fill=bermuda, draw=black] (3.75, -1.00) rectangle (4.00, -1.25);
\filldraw[fill=bermuda, draw=black] (4.00, -1.00) rectangle (4.25, -1.25);
\filldraw[fill=bermuda, draw=black] (4.25, -1.00) rectangle (4.50, -1.25);
\filldraw[fill=bermuda, draw=black] (4.50, -1.00) rectangle (4.75, -1.25);
\filldraw[fill=bermuda, draw=black] (4.75, -1.00) rectangle (5.00, -1.25);
\filldraw[fill=cancan, draw=black] (5.00, -1.00) rectangle (5.25, -1.25);
\filldraw[fill=cancan, draw=black] (5.25, -1.00) rectangle (5.50, -1.25);
\filldraw[fill=cancan, draw=black] (5.50, -1.00) rectangle (5.75, -1.25);
\filldraw[fill=cancan, draw=black] (5.75, -1.00) rectangle (6.00, -1.25);
\filldraw[fill=cancan, draw=black] (6.00, -1.00) rectangle (6.25, -1.25);
\filldraw[fill=cancan, draw=black] (6.25, -1.00) rectangle (6.50, -1.25);
\filldraw[fill=cancan, draw=black] (6.50, -1.00) rectangle (6.75, -1.25);
\filldraw[fill=cancan, draw=black] (6.75, -1.00) rectangle (7.00, -1.25);
\filldraw[fill=cancan, draw=black] (7.00, -1.00) rectangle (7.25, -1.25);
\filldraw[fill=cancan, draw=black] (7.25, -1.00) rectangle (7.50, -1.25);
\filldraw[fill=cancan, draw=black] (7.50, -1.00) rectangle (7.75, -1.25);
\filldraw[fill=bermuda, draw=black] (7.75, -1.00) rectangle (8.00, -1.25);
\filldraw[fill=bermuda, draw=black] (8.00, -1.00) rectangle (8.25, -1.25);
\filldraw[fill=bermuda, draw=black] (8.25, -1.00) rectangle (8.50, -1.25);
\filldraw[fill=cancan, draw=black] (8.50, -1.00) rectangle (8.75, -1.25);
\filldraw[fill=cancan, draw=black] (8.75, -1.00) rectangle (9.00, -1.25);
\filldraw[fill=bermuda, draw=black] (9.00, -1.00) rectangle (9.25, -1.25);
\filldraw[fill=bermuda, draw=black] (9.25, -1.00) rectangle (9.50, -1.25);
\filldraw[fill=cancan, draw=black] (9.50, -1.00) rectangle (9.75, -1.25);
\filldraw[fill=cancan, draw=black] (9.75, -1.00) rectangle (10.00, -1.25);
\filldraw[fill=cancan, draw=black] (10.00, -1.00) rectangle (10.25, -1.25);
\filldraw[fill=cancan, draw=black] (10.25, -1.00) rectangle (10.50, -1.25);
\filldraw[fill=bermuda, draw=black] (10.50, -1.00) rectangle (10.75, -1.25);
\filldraw[fill=bermuda, draw=black] (10.75, -1.00) rectangle (11.00, -1.25);
\filldraw[fill=cancan, draw=black] (11.00, -1.00) rectangle (11.25, -1.25);
\filldraw[fill=bermuda, draw=black] (11.25, -1.00) rectangle (11.50, -1.25);
\filldraw[fill=bermuda, draw=black] (11.50, -1.00) rectangle (11.75, -1.25);
\filldraw[fill=bermuda, draw=black] (11.75, -1.00) rectangle (12.00, -1.25);
\filldraw[fill=cancan, draw=black] (12.00, -1.00) rectangle (12.25, -1.25);
\filldraw[fill=cancan, draw=black] (12.25, -1.00) rectangle (12.50, -1.25);
\filldraw[fill=cancan, draw=black] (12.50, -1.00) rectangle (12.75, -1.25);
\filldraw[fill=cancan, draw=black] (12.75, -1.00) rectangle (13.00, -1.25);
\filldraw[fill=cancan, draw=black] (13.00, -1.00) rectangle (13.25, -1.25);
\filldraw[fill=bermuda, draw=black] (13.25, -1.00) rectangle (13.50, -1.25);
\filldraw[fill=cancan, draw=black] (13.50, -1.00) rectangle (13.75, -1.25);
\filldraw[fill=bermuda, draw=black] (13.75, -1.00) rectangle (14.00, -1.25);
\filldraw[fill=bermuda, draw=black] (14.00, -1.00) rectangle (14.25, -1.25);
\filldraw[fill=bermuda, draw=black] (14.25, -1.00) rectangle (14.50, -1.25);
\filldraw[fill=cancan, draw=black] (14.50, -1.00) rectangle (14.75, -1.25);
\filldraw[fill=cancan, draw=black] (14.75, -1.00) rectangle (15.00, -1.25);
\filldraw[fill=cancan, draw=black] (0.00, -1.25) rectangle (0.25, -1.50);
\filldraw[fill=cancan, draw=black] (0.25, -1.25) rectangle (0.50, -1.50);
\filldraw[fill=cancan, draw=black] (0.50, -1.25) rectangle (0.75, -1.50);
\filldraw[fill=cancan, draw=black] (0.75, -1.25) rectangle (1.00, -1.50);
\filldraw[fill=bermuda, draw=black] (1.00, -1.25) rectangle (1.25, -1.50);
\filldraw[fill=bermuda, draw=black] (1.25, -1.25) rectangle (1.50, -1.50);
\filldraw[fill=cancan, draw=black] (1.50, -1.25) rectangle (1.75, -1.50);
\filldraw[fill=cancan, draw=black] (1.75, -1.25) rectangle (2.00, -1.50);
\filldraw[fill=cancan, draw=black] (2.00, -1.25) rectangle (2.25, -1.50);
\filldraw[fill=bermuda, draw=black] (2.25, -1.25) rectangle (2.50, -1.50);
\filldraw[fill=bermuda, draw=black] (2.50, -1.25) rectangle (2.75, -1.50);
\filldraw[fill=bermuda, draw=black] (2.75, -1.25) rectangle (3.00, -1.50);
\filldraw[fill=bermuda, draw=black] (3.00, -1.25) rectangle (3.25, -1.50);
\filldraw[fill=bermuda, draw=black] (3.25, -1.25) rectangle (3.50, -1.50);
\filldraw[fill=cancan, draw=black] (3.50, -1.25) rectangle (3.75, -1.50);
\filldraw[fill=bermuda, draw=black] (3.75, -1.25) rectangle (4.00, -1.50);
\filldraw[fill=cancan, draw=black] (4.00, -1.25) rectangle (4.25, -1.50);
\filldraw[fill=cancan, draw=black] (4.25, -1.25) rectangle (4.50, -1.50);
\filldraw[fill=bermuda, draw=black] (4.50, -1.25) rectangle (4.75, -1.50);
\filldraw[fill=bermuda, draw=black] (4.75, -1.25) rectangle (5.00, -1.50);
\filldraw[fill=cancan, draw=black] (5.00, -1.25) rectangle (5.25, -1.50);
\filldraw[fill=bermuda, draw=black] (5.25, -1.25) rectangle (5.50, -1.50);
\filldraw[fill=cancan, draw=black] (5.50, -1.25) rectangle (5.75, -1.50);
\filldraw[fill=bermuda, draw=black] (5.75, -1.25) rectangle (6.00, -1.50);
\filldraw[fill=bermuda, draw=black] (6.00, -1.25) rectangle (6.25, -1.50);
\filldraw[fill=bermuda, draw=black] (6.25, -1.25) rectangle (6.50, -1.50);
\filldraw[fill=cancan, draw=black] (6.50, -1.25) rectangle (6.75, -1.50);
\filldraw[fill=bermuda, draw=black] (6.75, -1.25) rectangle (7.00, -1.50);
\filldraw[fill=bermuda, draw=black] (7.00, -1.25) rectangle (7.25, -1.50);
\filldraw[fill=bermuda, draw=black] (7.25, -1.25) rectangle (7.50, -1.50);
\filldraw[fill=cancan, draw=black] (7.50, -1.25) rectangle (7.75, -1.50);
\filldraw[fill=bermuda, draw=black] (7.75, -1.25) rectangle (8.00, -1.50);
\filldraw[fill=bermuda, draw=black] (8.00, -1.25) rectangle (8.25, -1.50);
\filldraw[fill=bermuda, draw=black] (8.25, -1.25) rectangle (8.50, -1.50);
\filldraw[fill=cancan, draw=black] (8.50, -1.25) rectangle (8.75, -1.50);
\filldraw[fill=cancan, draw=black] (8.75, -1.25) rectangle (9.00, -1.50);
\filldraw[fill=cancan, draw=black] (9.00, -1.25) rectangle (9.25, -1.50);
\filldraw[fill=bermuda, draw=black] (9.25, -1.25) rectangle (9.50, -1.50);
\filldraw[fill=bermuda, draw=black] (9.50, -1.25) rectangle (9.75, -1.50);
\filldraw[fill=bermuda, draw=black] (9.75, -1.25) rectangle (10.00, -1.50);
\filldraw[fill=bermuda, draw=black] (10.00, -1.25) rectangle (10.25, -1.50);
\filldraw[fill=bermuda, draw=black] (10.25, -1.25) rectangle (10.50, -1.50);
\filldraw[fill=cancan, draw=black] (10.50, -1.25) rectangle (10.75, -1.50);
\filldraw[fill=cancan, draw=black] (10.75, -1.25) rectangle (11.00, -1.50);
\filldraw[fill=cancan, draw=black] (11.00, -1.25) rectangle (11.25, -1.50);
\filldraw[fill=bermuda, draw=black] (11.25, -1.25) rectangle (11.50, -1.50);
\filldraw[fill=bermuda, draw=black] (11.50, -1.25) rectangle (11.75, -1.50);
\filldraw[fill=bermuda, draw=black] (11.75, -1.25) rectangle (12.00, -1.50);
\filldraw[fill=cancan, draw=black] (12.00, -1.25) rectangle (12.25, -1.50);
\filldraw[fill=bermuda, draw=black] (12.25, -1.25) rectangle (12.50, -1.50);
\filldraw[fill=bermuda, draw=black] (12.50, -1.25) rectangle (12.75, -1.50);
\filldraw[fill=bermuda, draw=black] (12.75, -1.25) rectangle (13.00, -1.50);
\filldraw[fill=bermuda, draw=black] (13.00, -1.25) rectangle (13.25, -1.50);
\filldraw[fill=bermuda, draw=black] (13.25, -1.25) rectangle (13.50, -1.50);
\filldraw[fill=cancan, draw=black] (13.50, -1.25) rectangle (13.75, -1.50);
\filldraw[fill=bermuda, draw=black] (13.75, -1.25) rectangle (14.00, -1.50);
\filldraw[fill=cancan, draw=black] (14.00, -1.25) rectangle (14.25, -1.50);
\filldraw[fill=cancan, draw=black] (14.25, -1.25) rectangle (14.50, -1.50);
\filldraw[fill=bermuda, draw=black] (14.50, -1.25) rectangle (14.75, -1.50);
\filldraw[fill=bermuda, draw=black] (14.75, -1.25) rectangle (15.00, -1.50);
\filldraw[fill=cancan, draw=black] (0.00, -1.50) rectangle (0.25, -1.75);
\filldraw[fill=bermuda, draw=black] (0.25, -1.50) rectangle (0.50, -1.75);
\filldraw[fill=cancan, draw=black] (0.50, -1.50) rectangle (0.75, -1.75);
\filldraw[fill=bermuda, draw=black] (0.75, -1.50) rectangle (1.00, -1.75);
\filldraw[fill=cancan, draw=black] (1.00, -1.50) rectangle (1.25, -1.75);
\filldraw[fill=cancan, draw=black] (1.25, -1.50) rectangle (1.50, -1.75);
\filldraw[fill=cancan, draw=black] (1.50, -1.50) rectangle (1.75, -1.75);
\filldraw[fill=bermuda, draw=black] (1.75, -1.50) rectangle (2.00, -1.75);
\filldraw[fill=bermuda, draw=black] (2.00, -1.50) rectangle (2.25, -1.75);
\filldraw[fill=bermuda, draw=black] (2.25, -1.50) rectangle (2.50, -1.75);
\filldraw[fill=cancan, draw=black] (2.50, -1.50) rectangle (2.75, -1.75);
\filldraw[fill=cancan, draw=black] (2.75, -1.50) rectangle (3.00, -1.75);
\filldraw[fill=cancan, draw=black] (3.00, -1.50) rectangle (3.25, -1.75);
\filldraw[fill=bermuda, draw=black] (3.25, -1.50) rectangle (3.50, -1.75);
\filldraw[fill=bermuda, draw=black] (3.50, -1.50) rectangle (3.75, -1.75);
\filldraw[fill=bermuda, draw=black] (3.75, -1.50) rectangle (4.00, -1.75);
\filldraw[fill=cancan, draw=black] (4.00, -1.50) rectangle (4.25, -1.75);
\filldraw[fill=bermuda, draw=black] (4.25, -1.50) rectangle (4.50, -1.75);
\filldraw[fill=cancan, draw=black] (4.50, -1.50) rectangle (4.75, -1.75);
\filldraw[fill=cancan, draw=black] (4.75, -1.50) rectangle (5.00, -1.75);
\filldraw[fill=cancan, draw=black] (5.00, -1.50) rectangle (5.25, -1.75);
\filldraw[fill=cancan, draw=black] (5.25, -1.50) rectangle (5.50, -1.75);
\filldraw[fill=cancan, draw=black] (5.50, -1.50) rectangle (5.75, -1.75);
\filldraw[fill=cancan, draw=black] (5.75, -1.50) rectangle (6.00, -1.75);
\filldraw[fill=cancan, draw=black] (6.00, -1.50) rectangle (6.25, -1.75);
\filldraw[fill=cancan, draw=black] (6.25, -1.50) rectangle (6.50, -1.75);
\filldraw[fill=cancan, draw=black] (6.50, -1.50) rectangle (6.75, -1.75);
\filldraw[fill=bermuda, draw=black] (6.75, -1.50) rectangle (7.00, -1.75);
\filldraw[fill=bermuda, draw=black] (7.00, -1.50) rectangle (7.25, -1.75);
\filldraw[fill=bermuda, draw=black] (7.25, -1.50) rectangle (7.50, -1.75);
\filldraw[fill=cancan, draw=black] (7.50, -1.50) rectangle (7.75, -1.75);
\filldraw[fill=cancan, draw=black] (7.75, -1.50) rectangle (8.00, -1.75);
\filldraw[fill=cancan, draw=black] (8.00, -1.50) rectangle (8.25, -1.75);
\filldraw[fill=bermuda, draw=black] (8.25, -1.50) rectangle (8.50, -1.75);
\filldraw[fill=bermuda, draw=black] (8.50, -1.50) rectangle (8.75, -1.75);
\filldraw[fill=bermuda, draw=black] (8.75, -1.50) rectangle (9.00, -1.75);
\filldraw[fill=cancan, draw=black] (9.00, -1.50) rectangle (9.25, -1.75);
\filldraw[fill=bermuda, draw=black] (9.25, -1.50) rectangle (9.50, -1.75);
\filldraw[fill=bermuda, draw=black] (9.50, -1.50) rectangle (9.75, -1.75);
\filldraw[fill=bermuda, draw=black] (9.75, -1.50) rectangle (10.00, -1.75);
\filldraw[fill=cancan, draw=black] (10.00, -1.50) rectangle (10.25, -1.75);
\filldraw[fill=cancan, draw=black] (10.25, -1.50) rectangle (10.50, -1.75);
\filldraw[fill=cancan, draw=black] (10.50, -1.50) rectangle (10.75, -1.75);
\filldraw[fill=bermuda, draw=black] (10.75, -1.50) rectangle (11.00, -1.75);
\filldraw[fill=bermuda, draw=black] (11.00, -1.50) rectangle (11.25, -1.75);
\filldraw[fill=bermuda, draw=black] (11.25, -1.50) rectangle (11.50, -1.75);
\filldraw[fill=cancan, draw=black] (11.50, -1.50) rectangle (11.75, -1.75);
\filldraw[fill=cancan, draw=black] (11.75, -1.50) rectangle (12.00, -1.75);
\filldraw[fill=cancan, draw=black] (12.00, -1.50) rectangle (12.25, -1.75);
\filldraw[fill=bermuda, draw=black] (12.25, -1.50) rectangle (12.50, -1.75);
\filldraw[fill=bermuda, draw=black] (12.50, -1.50) rectangle (12.75, -1.75);
\filldraw[fill=bermuda, draw=black] (12.75, -1.50) rectangle (13.00, -1.75);
\filldraw[fill=cancan, draw=black] (13.00, -1.50) rectangle (13.25, -1.75);
\filldraw[fill=cancan, draw=black] (13.25, -1.50) rectangle (13.50, -1.75);
\filldraw[fill=cancan, draw=black] (13.50, -1.50) rectangle (13.75, -1.75);
\filldraw[fill=bermuda, draw=black] (13.75, -1.50) rectangle (14.00, -1.75);
\filldraw[fill=bermuda, draw=black] (14.00, -1.50) rectangle (14.25, -1.75);
\filldraw[fill=bermuda, draw=black] (14.25, -1.50) rectangle (14.50, -1.75);
\filldraw[fill=bermuda, draw=black] (14.50, -1.50) rectangle (14.75, -1.75);
\filldraw[fill=bermuda, draw=black] (14.75, -1.50) rectangle (15.00, -1.75);
\filldraw[fill=bermuda, draw=black] (0.00, -1.75) rectangle (0.25, -2.00);
\filldraw[fill=bermuda, draw=black] (0.25, -1.75) rectangle (0.50, -2.00);
\filldraw[fill=cancan, draw=black] (0.50, -1.75) rectangle (0.75, -2.00);
\filldraw[fill=bermuda, draw=black] (0.75, -1.75) rectangle (1.00, -2.00);
\filldraw[fill=cancan, draw=black] (1.00, -1.75) rectangle (1.25, -2.00);
\filldraw[fill=cancan, draw=black] (1.25, -1.75) rectangle (1.50, -2.00);
\filldraw[fill=bermuda, draw=black] (1.50, -1.75) rectangle (1.75, -2.00);
\filldraw[fill=bermuda, draw=black] (1.75, -1.75) rectangle (2.00, -2.00);
\filldraw[fill=cancan, draw=black] (2.00, -1.75) rectangle (2.25, -2.00);
\filldraw[fill=bermuda, draw=black] (2.25, -1.75) rectangle (2.50, -2.00);
\filldraw[fill=cancan, draw=black] (2.50, -1.75) rectangle (2.75, -2.00);
\filldraw[fill=cancan, draw=black] (2.75, -1.75) rectangle (3.00, -2.00);
\filldraw[fill=bermuda, draw=black] (3.00, -1.75) rectangle (3.25, -2.00);
\filldraw[fill=bermuda, draw=black] (3.25, -1.75) rectangle (3.50, -2.00);
\filldraw[fill=cancan, draw=black] (3.50, -1.75) rectangle (3.75, -2.00);
\filldraw[fill=bermuda, draw=black] (3.75, -1.75) rectangle (4.00, -2.00);
\filldraw[fill=cancan, draw=black] (4.00, -1.75) rectangle (4.25, -2.00);
\filldraw[fill=cancan, draw=black] (4.25, -1.75) rectangle (4.50, -2.00);
\filldraw[fill=bermuda, draw=black] (4.50, -1.75) rectangle (4.75, -2.00);
\filldraw[fill=bermuda, draw=black] (4.75, -1.75) rectangle (5.00, -2.00);
\filldraw[fill=cancan, draw=black] (5.00, -1.75) rectangle (5.25, -2.00);
\filldraw[fill=bermuda, draw=black] (5.25, -1.75) rectangle (5.50, -2.00);
\filldraw[fill=cancan, draw=black] (5.50, -1.75) rectangle (5.75, -2.00);
\filldraw[fill=bermuda, draw=black] (5.75, -1.75) rectangle (6.00, -2.00);
\filldraw[fill=bermuda, draw=black] (6.00, -1.75) rectangle (6.25, -2.00);
\filldraw[fill=bermuda, draw=black] (6.25, -1.75) rectangle (6.50, -2.00);
\filldraw[fill=cancan, draw=black] (6.50, -1.75) rectangle (6.75, -2.00);
\filldraw[fill=bermuda, draw=black] (6.75, -1.75) rectangle (7.00, -2.00);
\filldraw[fill=bermuda, draw=black] (7.00, -1.75) rectangle (7.25, -2.00);
\filldraw[fill=bermuda, draw=black] (7.25, -1.75) rectangle (7.50, -2.00);
\filldraw[fill=cancan, draw=black] (7.50, -1.75) rectangle (7.75, -2.00);
\filldraw[fill=bermuda, draw=black] (7.75, -1.75) rectangle (8.00, -2.00);
\filldraw[fill=bermuda, draw=black] (8.00, -1.75) rectangle (8.25, -2.00);
\filldraw[fill=bermuda, draw=black] (8.25, -1.75) rectangle (8.50, -2.00);
\filldraw[fill=cancan, draw=black] (8.50, -1.75) rectangle (8.75, -2.00);
\filldraw[fill=bermuda, draw=black] (8.75, -1.75) rectangle (9.00, -2.00);
\filldraw[fill=bermuda, draw=black] (9.00, -1.75) rectangle (9.25, -2.00);
\filldraw[fill=bermuda, draw=black] (9.25, -1.75) rectangle (9.50, -2.00);
\filldraw[fill=bermuda, draw=black] (9.50, -1.75) rectangle (9.75, -2.00);
\filldraw[fill=bermuda, draw=black] (9.75, -1.75) rectangle (10.00, -2.00);
\filldraw[fill=cancan, draw=black] (10.00, -1.75) rectangle (10.25, -2.00);
\filldraw[fill=bermuda, draw=black] (10.25, -1.75) rectangle (10.50, -2.00);
\filldraw[fill=bermuda, draw=black] (10.50, -1.75) rectangle (10.75, -2.00);
\filldraw[fill=bermuda, draw=black] (10.75, -1.75) rectangle (11.00, -2.00);
\filldraw[fill=bermuda, draw=black] (11.00, -1.75) rectangle (11.25, -2.00);
\filldraw[fill=bermuda, draw=black] (11.25, -1.75) rectangle (11.50, -2.00);
\filldraw[fill=cancan, draw=black] (11.50, -1.75) rectangle (11.75, -2.00);
\filldraw[fill=bermuda, draw=black] (11.75, -1.75) rectangle (12.00, -2.00);
\filldraw[fill=bermuda, draw=black] (12.00, -1.75) rectangle (12.25, -2.00);
\filldraw[fill=bermuda, draw=black] (12.25, -1.75) rectangle (12.50, -2.00);
\filldraw[fill=bermuda, draw=black] (12.50, -1.75) rectangle (12.75, -2.00);
\filldraw[fill=bermuda, draw=black] (12.75, -1.75) rectangle (13.00, -2.00);
\filldraw[fill=cancan, draw=black] (13.00, -1.75) rectangle (13.25, -2.00);
\filldraw[fill=bermuda, draw=black] (13.25, -1.75) rectangle (13.50, -2.00);
\filldraw[fill=bermuda, draw=black] (13.50, -1.75) rectangle (13.75, -2.00);
\filldraw[fill=bermuda, draw=black] (13.75, -1.75) rectangle (14.00, -2.00);
\filldraw[fill=bermuda, draw=black] (14.00, -1.75) rectangle (14.25, -2.00);
\filldraw[fill=bermuda, draw=black] (14.25, -1.75) rectangle (14.50, -2.00);
\filldraw[fill=cancan, draw=black] (14.50, -1.75) rectangle (14.75, -2.00);
\filldraw[fill=bermuda, draw=black] (14.75, -1.75) rectangle (15.00, -2.00);
\filldraw[fill=bermuda, draw=black] (0.00, -2.00) rectangle (0.25, -2.25);
\filldraw[fill=bermuda, draw=black] (0.25, -2.00) rectangle (0.50, -2.25);
\filldraw[fill=bermuda, draw=black] (0.50, -2.00) rectangle (0.75, -2.25);
\filldraw[fill=bermuda, draw=black] (0.75, -2.00) rectangle (1.00, -2.25);
\filldraw[fill=cancan, draw=black] (1.00, -2.00) rectangle (1.25, -2.25);
\filldraw[fill=bermuda, draw=black] (1.25, -2.00) rectangle (1.50, -2.25);
\filldraw[fill=bermuda, draw=black] (1.50, -2.00) rectangle (1.75, -2.25);
\filldraw[fill=bermuda, draw=black] (1.75, -2.00) rectangle (2.00, -2.25);
\filldraw[fill=bermuda, draw=black] (2.00, -2.00) rectangle (2.25, -2.25);
\filldraw[fill=bermuda, draw=black] (2.25, -2.00) rectangle (2.50, -2.25);
\filldraw[fill=bermuda, draw=black] (2.50, -2.00) rectangle (2.75, -2.25);
\filldraw[fill=bermuda, draw=black] (2.75, -2.00) rectangle (3.00, -2.25);
\filldraw[fill=bermuda, draw=black] (3.00, -2.00) rectangle (3.25, -2.25);
\filldraw[fill=bermuda, draw=black] (3.25, -2.00) rectangle (3.50, -2.25);
\filldraw[fill=cancan, draw=black] (3.50, -2.00) rectangle (3.75, -2.25);
\filldraw[fill=bermuda, draw=black] (3.75, -2.00) rectangle (4.00, -2.25);
\filldraw[fill=bermuda, draw=black] (4.00, -2.00) rectangle (4.25, -2.25);
\filldraw[fill=bermuda, draw=black] (4.25, -2.00) rectangle (4.50, -2.25);
\filldraw[fill=cancan, draw=black] (4.50, -2.00) rectangle (4.75, -2.25);
\filldraw[fill=bermuda, draw=black] (4.75, -2.00) rectangle (5.00, -2.25);
\filldraw[fill=bermuda, draw=black] (5.00, -2.00) rectangle (5.25, -2.25);
\filldraw[fill=bermuda, draw=black] (5.25, -2.00) rectangle (5.50, -2.25);
\filldraw[fill=cancan, draw=black] (5.50, -2.00) rectangle (5.75, -2.25);
\filldraw[fill=cancan, draw=black] (5.75, -2.00) rectangle (6.00, -2.25);
\filldraw[fill=cancan, draw=black] (6.00, -2.00) rectangle (6.25, -2.25);
\filldraw[fill=bermuda, draw=black] (6.25, -2.00) rectangle (6.50, -2.25);
\filldraw[fill=bermuda, draw=black] (6.50, -2.00) rectangle (6.75, -2.25);
\filldraw[fill=bermuda, draw=black] (6.75, -2.00) rectangle (7.00, -2.25);
\filldraw[fill=cancan, draw=black] (7.00, -2.00) rectangle (7.25, -2.25);
\filldraw[fill=bermuda, draw=black] (7.25, -2.00) rectangle (7.50, -2.25);
\filldraw[fill=cancan, draw=black] (7.50, -2.00) rectangle (7.75, -2.25);
\filldraw[fill=cancan, draw=black] (7.75, -2.00) rectangle (8.00, -2.25);
\filldraw[fill=cancan, draw=black] (8.00, -2.00) rectangle (8.25, -2.25);
\filldraw[fill=cancan, draw=black] (8.25, -2.00) rectangle (8.50, -2.25);
\filldraw[fill=cancan, draw=black] (8.50, -2.00) rectangle (8.75, -2.25);
\filldraw[fill=cancan, draw=black] (8.75, -2.00) rectangle (9.00, -2.25);
\filldraw[fill=cancan, draw=black] (9.00, -2.00) rectangle (9.25, -2.25);
\filldraw[fill=bermuda, draw=black] (9.25, -2.00) rectangle (9.50, -2.25);
\filldraw[fill=cancan, draw=black] (9.50, -2.00) rectangle (9.75, -2.25);
\filldraw[fill=cancan, draw=black] (9.75, -2.00) rectangle (10.00, -2.25);
\filldraw[fill=cancan, draw=black] (10.00, -2.00) rectangle (10.25, -2.25);
\filldraw[fill=bermuda, draw=black] (10.25, -2.00) rectangle (10.50, -2.25);
\filldraw[fill=bermuda, draw=black] (10.50, -2.00) rectangle (10.75, -2.25);
\filldraw[fill=bermuda, draw=black] (10.75, -2.00) rectangle (11.00, -2.25);
\filldraw[fill=cancan, draw=black] (11.00, -2.00) rectangle (11.25, -2.25);
\filldraw[fill=cancan, draw=black] (11.25, -2.00) rectangle (11.50, -2.25);
\filldraw[fill=cancan, draw=black] (11.50, -2.00) rectangle (11.75, -2.25);
\filldraw[fill=bermuda, draw=black] (11.75, -2.00) rectangle (12.00, -2.25);
\filldraw[fill=bermuda, draw=black] (12.00, -2.00) rectangle (12.25, -2.25);
\filldraw[fill=bermuda, draw=black] (12.25, -2.00) rectangle (12.50, -2.25);
\filldraw[fill=cancan, draw=black] (12.50, -2.00) rectangle (12.75, -2.25);
\filldraw[fill=bermuda, draw=black] (12.75, -2.00) rectangle (13.00, -2.25);
\filldraw[fill=bermuda, draw=black] (13.00, -2.00) rectangle (13.25, -2.25);
\filldraw[fill=bermuda, draw=black] (13.25, -2.00) rectangle (13.50, -2.25);
\filldraw[fill=bermuda, draw=black] (13.50, -2.00) rectangle (13.75, -2.25);
\filldraw[fill=bermuda, draw=black] (13.75, -2.00) rectangle (14.00, -2.25);
\filldraw[fill=bermuda, draw=black] (14.00, -2.00) rectangle (14.25, -2.25);
\filldraw[fill=bermuda, draw=black] (14.25, -2.00) rectangle (14.50, -2.25);
\filldraw[fill=bermuda, draw=black] (14.50, -2.00) rectangle (14.75, -2.25);
\filldraw[fill=bermuda, draw=black] (14.75, -2.00) rectangle (15.00, -2.25);
\filldraw[fill=bermuda, draw=black] (0.00, -2.25) rectangle (0.25, -2.50);
\filldraw[fill=bermuda, draw=black] (0.25, -2.25) rectangle (0.50, -2.50);
\filldraw[fill=cancan, draw=black] (0.50, -2.25) rectangle (0.75, -2.50);
\filldraw[fill=bermuda, draw=black] (0.75, -2.25) rectangle (1.00, -2.50);
\filldraw[fill=cancan, draw=black] (1.00, -2.25) rectangle (1.25, -2.50);
\filldraw[fill=cancan, draw=black] (1.25, -2.25) rectangle (1.50, -2.50);
\filldraw[fill=cancan, draw=black] (1.50, -2.25) rectangle (1.75, -2.50);
\filldraw[fill=cancan, draw=black] (1.75, -2.25) rectangle (2.00, -2.50);
\filldraw[fill=cancan, draw=black] (2.00, -2.25) rectangle (2.25, -2.50);
\filldraw[fill=cancan, draw=black] (2.25, -2.25) rectangle (2.50, -2.50);
\filldraw[fill=cancan, draw=black] (2.50, -2.25) rectangle (2.75, -2.50);
\filldraw[fill=bermuda, draw=black] (2.75, -2.25) rectangle (3.00, -2.50);
\filldraw[fill=cancan, draw=black] (3.00, -2.25) rectangle (3.25, -2.50);
\filldraw[fill=bermuda, draw=black] (3.25, -2.25) rectangle (3.50, -2.50);
\filldraw[fill=cancan, draw=black] (3.50, -2.25) rectangle (3.75, -2.50);
\filldraw[fill=cancan, draw=black] (3.75, -2.25) rectangle (4.00, -2.50);
\filldraw[fill=bermuda, draw=black] (4.00, -2.25) rectangle (4.25, -2.50);
\filldraw[fill=bermuda, draw=black] (4.25, -2.25) rectangle (4.50, -2.50);
\filldraw[fill=cancan, draw=black] (4.50, -2.25) rectangle (4.75, -2.50);
\filldraw[fill=bermuda, draw=black] (4.75, -2.25) rectangle (5.00, -2.50);
\filldraw[fill=bermuda, draw=black] (5.00, -2.25) rectangle (5.25, -2.50);
\filldraw[fill=bermuda, draw=black] (5.25, -2.25) rectangle (5.50, -2.50);
\filldraw[fill=cancan, draw=black] (5.50, -2.25) rectangle (5.75, -2.50);
\filldraw[fill=cancan, draw=black] (5.75, -2.25) rectangle (6.00, -2.50);
\filldraw[fill=cancan, draw=black] (6.00, -2.25) rectangle (6.25, -2.50);
\filldraw[fill=bermuda, draw=black] (6.25, -2.25) rectangle (6.50, -2.50);
\filldraw[fill=bermuda, draw=black] (6.50, -2.25) rectangle (6.75, -2.50);
\filldraw[fill=bermuda, draw=black] (6.75, -2.25) rectangle (7.00, -2.50);
\filldraw[fill=cancan, draw=black] (7.00, -2.25) rectangle (7.25, -2.50);
\filldraw[fill=bermuda, draw=black] (7.25, -2.25) rectangle (7.50, -2.50);
\filldraw[fill=cancan, draw=black] (7.50, -2.25) rectangle (7.75, -2.50);
\filldraw[fill=bermuda, draw=black] (7.75, -2.25) rectangle (8.00, -2.50);
\filldraw[fill=cancan, draw=black] (8.00, -2.25) rectangle (8.25, -2.50);
\filldraw[fill=cancan, draw=black] (8.25, -2.25) rectangle (8.50, -2.50);
\filldraw[fill=cancan, draw=black] (8.50, -2.25) rectangle (8.75, -2.50);
\filldraw[fill=cancan, draw=black] (8.75, -2.25) rectangle (9.00, -2.50);
\filldraw[fill=cancan, draw=black] (9.00, -2.25) rectangle (9.25, -2.50);
\filldraw[fill=cancan, draw=black] (9.25, -2.25) rectangle (9.50, -2.50);
\filldraw[fill=bermuda, draw=black] (9.50, -2.25) rectangle (9.75, -2.50);
\filldraw[fill=bermuda, draw=black] (9.75, -2.25) rectangle (10.00, -2.50);
\filldraw[fill=cancan, draw=black] (10.00, -2.25) rectangle (10.25, -2.50);
\filldraw[fill=bermuda, draw=black] (10.25, -2.25) rectangle (10.50, -2.50);
\filldraw[fill=cancan, draw=black] (10.50, -2.25) rectangle (10.75, -2.50);
\filldraw[fill=cancan, draw=black] (10.75, -2.25) rectangle (11.00, -2.50);
\filldraw[fill=bermuda, draw=black] (11.00, -2.25) rectangle (11.25, -2.50);
\filldraw[fill=bermuda, draw=black] (11.25, -2.25) rectangle (11.50, -2.50);
\filldraw[fill=cancan, draw=black] (11.50, -2.25) rectangle (11.75, -2.50);
\filldraw[fill=bermuda, draw=black] (11.75, -2.25) rectangle (12.00, -2.50);
\filldraw[fill=cancan, draw=black] (12.00, -2.25) rectangle (12.25, -2.50);
\filldraw[fill=bermuda, draw=black] (12.25, -2.25) rectangle (12.50, -2.50);
\filldraw[fill=bermuda, draw=black] (12.50, -2.25) rectangle (12.75, -2.50);
\filldraw[fill=bermuda, draw=black] (12.75, -2.25) rectangle (13.00, -2.50);
\filldraw[fill=cancan, draw=black] (13.00, -2.25) rectangle (13.25, -2.50);
\filldraw[fill=cancan, draw=black] (13.25, -2.25) rectangle (13.50, -2.50);
\filldraw[fill=cancan, draw=black] (13.50, -2.25) rectangle (13.75, -2.50);
\filldraw[fill=cancan, draw=black] (13.75, -2.25) rectangle (14.00, -2.50);
\filldraw[fill=cancan, draw=black] (14.00, -2.25) rectangle (14.25, -2.50);
\filldraw[fill=cancan, draw=black] (14.25, -2.25) rectangle (14.50, -2.50);
\filldraw[fill=cancan, draw=black] (14.50, -2.25) rectangle (14.75, -2.50);
\filldraw[fill=bermuda, draw=black] (14.75, -2.25) rectangle (15.00, -2.50);
\filldraw[fill=bermuda, draw=black] (0.00, -2.50) rectangle (0.25, -2.75);
\filldraw[fill=bermuda, draw=black] (0.25, -2.50) rectangle (0.50, -2.75);
\filldraw[fill=cancan, draw=black] (0.50, -2.50) rectangle (0.75, -2.75);
\filldraw[fill=cancan, draw=black] (0.75, -2.50) rectangle (1.00, -2.75);
\filldraw[fill=cancan, draw=black] (1.00, -2.50) rectangle (1.25, -2.75);
\filldraw[fill=cancan, draw=black] (1.25, -2.50) rectangle (1.50, -2.75);
\filldraw[fill=cancan, draw=black] (1.50, -2.50) rectangle (1.75, -2.75);
\filldraw[fill=bermuda, draw=black] (1.75, -2.50) rectangle (2.00, -2.75);
\filldraw[fill=bermuda, draw=black] (2.00, -2.50) rectangle (2.25, -2.75);
\filldraw[fill=bermuda, draw=black] (2.25, -2.50) rectangle (2.50, -2.75);
\filldraw[fill=cancan, draw=black] (2.50, -2.50) rectangle (2.75, -2.75);
\filldraw[fill=cancan, draw=black] (2.75, -2.50) rectangle (3.00, -2.75);
\filldraw[fill=bermuda, draw=black] (3.00, -2.50) rectangle (3.25, -2.75);
\filldraw[fill=bermuda, draw=black] (3.25, -2.50) rectangle (3.50, -2.75);
\filldraw[fill=cancan, draw=black] (3.50, -2.50) rectangle (3.75, -2.75);
\filldraw[fill=cancan, draw=black] (3.75, -2.50) rectangle (4.00, -2.75);
\filldraw[fill=cancan, draw=black] (4.00, -2.50) rectangle (4.25, -2.75);
\filldraw[fill=cancan, draw=black] (4.25, -2.50) rectangle (4.50, -2.75);
\filldraw[fill=bermuda, draw=black] (4.50, -2.50) rectangle (4.75, -2.75);
\filldraw[fill=bermuda, draw=black] (4.75, -2.50) rectangle (5.00, -2.75);
\filldraw[fill=cancan, draw=black] (5.00, -2.50) rectangle (5.25, -2.75);
\filldraw[fill=bermuda, draw=black] (5.25, -2.50) rectangle (5.50, -2.75);
\filldraw[fill=cancan, draw=black] (5.50, -2.50) rectangle (5.75, -2.75);
\filldraw[fill=cancan, draw=black] (5.75, -2.50) rectangle (6.00, -2.75);
\filldraw[fill=cancan, draw=black] (6.00, -2.50) rectangle (6.25, -2.75);
\filldraw[fill=bermuda, draw=black] (6.25, -2.50) rectangle (6.50, -2.75);
\filldraw[fill=bermuda, draw=black] (6.50, -2.50) rectangle (6.75, -2.75);
\filldraw[fill=bermuda, draw=black] (6.75, -2.50) rectangle (7.00, -2.75);
\filldraw[fill=bermuda, draw=black] (7.00, -2.50) rectangle (7.25, -2.75);
\filldraw[fill=bermuda, draw=black] (7.25, -2.50) rectangle (7.50, -2.75);
\filldraw[fill=cancan, draw=black] (7.50, -2.50) rectangle (7.75, -2.75);
\filldraw[fill=bermuda, draw=black] (7.75, -2.50) rectangle (8.00, -2.75);
\filldraw[fill=bermuda, draw=black] (8.00, -2.50) rectangle (8.25, -2.75);
\filldraw[fill=bermuda, draw=black] (8.25, -2.50) rectangle (8.50, -2.75);
\filldraw[fill=cancan, draw=black] (8.50, -2.50) rectangle (8.75, -2.75);
\filldraw[fill=cancan, draw=black] (8.75, -2.50) rectangle (9.00, -2.75);
\filldraw[fill=cancan, draw=black] (9.00, -2.50) rectangle (9.25, -2.75);
\filldraw[fill=cancan, draw=black] (9.25, -2.50) rectangle (9.50, -2.75);
\filldraw[fill=cancan, draw=black] (9.50, -2.50) rectangle (9.75, -2.75);
\filldraw[fill=bermuda, draw=black] (9.75, -2.50) rectangle (10.00, -2.75);
\filldraw[fill=cancan, draw=black] (10.00, -2.50) rectangle (10.25, -2.75);
\filldraw[fill=bermuda, draw=black] (10.25, -2.50) rectangle (10.50, -2.75);
\filldraw[fill=cancan, draw=black] (10.50, -2.50) rectangle (10.75, -2.75);
\filldraw[fill=cancan, draw=black] (10.75, -2.50) rectangle (11.00, -2.75);
\filldraw[fill=bermuda, draw=black] (11.00, -2.50) rectangle (11.25, -2.75);
\filldraw[fill=bermuda, draw=black] (11.25, -2.50) rectangle (11.50, -2.75);
\filldraw[fill=cancan, draw=black] (11.50, -2.50) rectangle (11.75, -2.75);
\filldraw[fill=bermuda, draw=black] (11.75, -2.50) rectangle (12.00, -2.75);
\filldraw[fill=cancan, draw=black] (12.00, -2.50) rectangle (12.25, -2.75);
\filldraw[fill=bermuda, draw=black] (12.25, -2.50) rectangle (12.50, -2.75);
\filldraw[fill=cancan, draw=black] (12.50, -2.50) rectangle (12.75, -2.75);
\filldraw[fill=cancan, draw=black] (12.75, -2.50) rectangle (13.00, -2.75);
\filldraw[fill=cancan, draw=black] (13.00, -2.50) rectangle (13.25, -2.75);
\filldraw[fill=cancan, draw=black] (13.25, -2.50) rectangle (13.50, -2.75);
\filldraw[fill=bermuda, draw=black] (13.50, -2.50) rectangle (13.75, -2.75);
\filldraw[fill=bermuda, draw=black] (13.75, -2.50) rectangle (14.00, -2.75);
\filldraw[fill=cancan, draw=black] (14.00, -2.50) rectangle (14.25, -2.75);
\filldraw[fill=cancan, draw=black] (14.25, -2.50) rectangle (14.50, -2.75);
\filldraw[fill=cancan, draw=black] (14.50, -2.50) rectangle (14.75, -2.75);
\filldraw[fill=cancan, draw=black] (14.75, -2.50) rectangle (15.00, -2.75);
\filldraw[fill=bermuda, draw=black] (0.00, -2.75) rectangle (0.25, -3.00);
\filldraw[fill=bermuda, draw=black] (0.25, -2.75) rectangle (0.50, -3.00);
\filldraw[fill=cancan, draw=black] (0.50, -2.75) rectangle (0.75, -3.00);
\filldraw[fill=cancan, draw=black] (0.75, -2.75) rectangle (1.00, -3.00);
\filldraw[fill=cancan, draw=black] (1.00, -2.75) rectangle (1.25, -3.00);
\filldraw[fill=cancan, draw=black] (1.25, -2.75) rectangle (1.50, -3.00);
\filldraw[fill=bermuda, draw=black] (1.50, -2.75) rectangle (1.75, -3.00);
\filldraw[fill=bermuda, draw=black] (1.75, -2.75) rectangle (2.00, -3.00);
\filldraw[fill=cancan, draw=black] (2.00, -2.75) rectangle (2.25, -3.00);
\filldraw[fill=cancan, draw=black] (2.25, -2.75) rectangle (2.50, -3.00);
\filldraw[fill=cancan, draw=black] (2.50, -2.75) rectangle (2.75, -3.00);
\filldraw[fill=bermuda, draw=black] (2.75, -2.75) rectangle (3.00, -3.00);
\filldraw[fill=bermuda, draw=black] (3.00, -2.75) rectangle (3.25, -3.00);
\filldraw[fill=bermuda, draw=black] (3.25, -2.75) rectangle (3.50, -3.00);
\filldraw[fill=bermuda, draw=black] (3.50, -2.75) rectangle (3.75, -3.00);
\filldraw[fill=bermuda, draw=black] (3.75, -2.75) rectangle (4.00, -3.00);
\filldraw[fill=cancan, draw=black] (4.00, -2.75) rectangle (4.25, -3.00);
\filldraw[fill=bermuda, draw=black] (4.25, -2.75) rectangle (4.50, -3.00);
\filldraw[fill=bermuda, draw=black] (4.50, -2.75) rectangle (4.75, -3.00);
\filldraw[fill=bermuda, draw=black] (4.75, -2.75) rectangle (5.00, -3.00);
\filldraw[fill=cancan, draw=black] (5.00, -2.75) rectangle (5.25, -3.00);
\filldraw[fill=cancan, draw=black] (5.25, -2.75) rectangle (5.50, -3.00);
\filldraw[fill=cancan, draw=black] (5.50, -2.75) rectangle (5.75, -3.00);
\filldraw[fill=bermuda, draw=black] (5.75, -2.75) rectangle (6.00, -3.00);
\filldraw[fill=bermuda, draw=black] (6.00, -2.75) rectangle (6.25, -3.00);
\filldraw[fill=bermuda, draw=black] (6.25, -2.75) rectangle (6.50, -3.00);
\filldraw[fill=cancan, draw=black] (6.50, -2.75) rectangle (6.75, -3.00);
\filldraw[fill=bermuda, draw=black] (6.75, -2.75) rectangle (7.00, -3.00);
\filldraw[fill=cancan, draw=black] (7.00, -2.75) rectangle (7.25, -3.00);
\filldraw[fill=bermuda, draw=black] (7.25, -2.75) rectangle (7.50, -3.00);
\filldraw[fill=bermuda, draw=black] (7.50, -2.75) rectangle (7.75, -3.00);
\filldraw[fill=bermuda, draw=black] (7.75, -2.75) rectangle (8.00, -3.00);
\filldraw[fill=cancan, draw=black] (8.00, -2.75) rectangle (8.25, -3.00);
\filldraw[fill=bermuda, draw=black] (8.25, -2.75) rectangle (8.50, -3.00);
\filldraw[fill=bermuda, draw=black] (8.50, -2.75) rectangle (8.75, -3.00);
\filldraw[fill=bermuda, draw=black] (8.75, -2.75) rectangle (9.00, -3.00);
\filldraw[fill=cancan, draw=black] (9.00, -2.75) rectangle (9.25, -3.00);
\filldraw[fill=cancan, draw=black] (9.25, -2.75) rectangle (9.50, -3.00);
\filldraw[fill=cancan, draw=black] (9.50, -2.75) rectangle (9.75, -3.00);
\filldraw[fill=cancan, draw=black] (9.75, -2.75) rectangle (10.00, -3.00);
\filldraw[fill=cancan, draw=black] (10.00, -2.75) rectangle (10.25, -3.00);
\filldraw[fill=cancan, draw=black] (10.25, -2.75) rectangle (10.50, -3.00);
\filldraw[fill=bermuda, draw=black] (10.50, -2.75) rectangle (10.75, -3.00);
\filldraw[fill=bermuda, draw=black] (10.75, -2.75) rectangle (11.00, -3.00);
\filldraw[fill=cancan, draw=black] (11.00, -2.75) rectangle (11.25, -3.00);
\filldraw[fill=cancan, draw=black] (11.25, -2.75) rectangle (11.50, -3.00);
\filldraw[fill=cancan, draw=black] (11.50, -2.75) rectangle (11.75, -3.00);
\filldraw[fill=bermuda, draw=black] (11.75, -2.75) rectangle (12.00, -3.00);
\filldraw[fill=bermuda, draw=black] (12.00, -2.75) rectangle (12.25, -3.00);
\filldraw[fill=bermuda, draw=black] (12.25, -2.75) rectangle (12.50, -3.00);
\filldraw[fill=cancan, draw=black] (12.50, -2.75) rectangle (12.75, -3.00);
\filldraw[fill=cancan, draw=black] (12.75, -2.75) rectangle (13.00, -3.00);
\filldraw[fill=bermuda, draw=black] (13.00, -2.75) rectangle (13.25, -3.00);
\filldraw[fill=bermuda, draw=black] (13.25, -2.75) rectangle (13.50, -3.00);
\filldraw[fill=cancan, draw=black] (13.50, -2.75) rectangle (13.75, -3.00);
\filldraw[fill=cancan, draw=black] (13.75, -2.75) rectangle (14.00, -3.00);
\filldraw[fill=cancan, draw=black] (14.00, -2.75) rectangle (14.25, -3.00);
\filldraw[fill=bermuda, draw=black] (14.25, -2.75) rectangle (14.50, -3.00);
\filldraw[fill=bermuda, draw=black] (14.50, -2.75) rectangle (14.75, -3.00);
\filldraw[fill=bermuda, draw=black] (14.75, -2.75) rectangle (15.00, -3.00);
\filldraw[fill=cancan, draw=black] (0.00, -3.00) rectangle (0.25, -3.25);
\filldraw[fill=cancan, draw=black] (0.25, -3.00) rectangle (0.50, -3.25);
\filldraw[fill=cancan, draw=black] (0.50, -3.00) rectangle (0.75, -3.25);
\filldraw[fill=bermuda, draw=black] (0.75, -3.00) rectangle (1.00, -3.25);
\filldraw[fill=bermuda, draw=black] (1.00, -3.00) rectangle (1.25, -3.25);
\filldraw[fill=bermuda, draw=black] (1.25, -3.00) rectangle (1.50, -3.25);
\filldraw[fill=cancan, draw=black] (1.50, -3.00) rectangle (1.75, -3.25);
\filldraw[fill=bermuda, draw=black] (1.75, -3.00) rectangle (2.00, -3.25);
\filldraw[fill=bermuda, draw=black] (2.00, -3.00) rectangle (2.25, -3.25);
\filldraw[fill=bermuda, draw=black] (2.25, -3.00) rectangle (2.50, -3.25);
\filldraw[fill=cancan, draw=black] (2.50, -3.00) rectangle (2.75, -3.25);
\filldraw[fill=cancan, draw=black] (2.75, -3.00) rectangle (3.00, -3.25);
\filldraw[fill=cancan, draw=black] (3.00, -3.00) rectangle (3.25, -3.25);
\filldraw[fill=cancan, draw=black] (3.25, -3.00) rectangle (3.50, -3.25);
\filldraw[fill=bermuda, draw=black] (3.50, -3.00) rectangle (3.75, -3.25);
\filldraw[fill=bermuda, draw=black] (3.75, -3.00) rectangle (4.00, -3.25);
\filldraw[fill=cancan, draw=black] (4.00, -3.00) rectangle (4.25, -3.25);
\filldraw[fill=cancan, draw=black] (4.25, -3.00) rectangle (4.50, -3.25);
\filldraw[fill=cancan, draw=black] (4.50, -3.00) rectangle (4.75, -3.25);
\filldraw[fill=cancan, draw=black] (4.75, -3.00) rectangle (5.00, -3.25);
\filldraw[fill=bermuda, draw=black] (5.00, -3.00) rectangle (5.25, -3.25);
\filldraw[fill=bermuda, draw=black] (5.25, -3.00) rectangle (5.50, -3.25);
\filldraw[fill=cancan, draw=black] (5.50, -3.00) rectangle (5.75, -3.25);
\filldraw[fill=cancan, draw=black] (5.75, -3.00) rectangle (6.00, -3.25);
\filldraw[fill=cancan, draw=black] (6.00, -3.00) rectangle (6.25, -3.25);
\filldraw[fill=cancan, draw=black] (6.25, -3.00) rectangle (6.50, -3.25);
\filldraw[fill=bermuda, draw=black] (6.50, -3.00) rectangle (6.75, -3.25);
\filldraw[fill=bermuda, draw=black] (6.75, -3.00) rectangle (7.00, -3.25);
\filldraw[fill=cancan, draw=black] (7.00, -3.00) rectangle (7.25, -3.25);
\filldraw[fill=cancan, draw=black] (7.25, -3.00) rectangle (7.50, -3.25);
\filldraw[fill=cancan, draw=black] (7.50, -3.00) rectangle (7.75, -3.25);
\filldraw[fill=cancan, draw=black] (7.75, -3.00) rectangle (8.00, -3.25);
\filldraw[fill=cancan, draw=black] (8.00, -3.00) rectangle (8.25, -3.25);
\filldraw[fill=cancan, draw=black] (8.25, -3.00) rectangle (8.50, -3.25);
\filldraw[fill=cancan, draw=black] (8.50, -3.00) rectangle (8.75, -3.25);
\filldraw[fill=bermuda, draw=black] (8.75, -3.00) rectangle (9.00, -3.25);
\filldraw[fill=bermuda, draw=black] (9.00, -3.00) rectangle (9.25, -3.25);
\filldraw[fill=bermuda, draw=black] (9.25, -3.00) rectangle (9.50, -3.25);
\filldraw[fill=cancan, draw=black] (9.50, -3.00) rectangle (9.75, -3.25);
\filldraw[fill=cancan, draw=black] (9.75, -3.00) rectangle (10.00, -3.25);
\filldraw[fill=cancan, draw=black] (10.00, -3.00) rectangle (10.25, -3.25);
\filldraw[fill=bermuda, draw=black] (10.25, -3.00) rectangle (10.50, -3.25);
\filldraw[fill=bermuda, draw=black] (10.50, -3.00) rectangle (10.75, -3.25);
\filldraw[fill=bermuda, draw=black] (10.75, -3.00) rectangle (11.00, -3.25);
\filldraw[fill=cancan, draw=black] (11.00, -3.00) rectangle (11.25, -3.25);
\filldraw[fill=bermuda, draw=black] (11.25, -3.00) rectangle (11.50, -3.25);
\filldraw[fill=bermuda, draw=black] (11.50, -3.00) rectangle (11.75, -3.25);
\filldraw[fill=bermuda, draw=black] (11.75, -3.00) rectangle (12.00, -3.25);
\filldraw[fill=cancan, draw=black] (12.00, -3.00) rectangle (12.25, -3.25);
\filldraw[fill=cancan, draw=black] (12.25, -3.00) rectangle (12.50, -3.25);
\filldraw[fill=cancan, draw=black] (12.50, -3.00) rectangle (12.75, -3.25);
\filldraw[fill=cancan, draw=black] (12.75, -3.00) rectangle (13.00, -3.25);
\filldraw[fill=cancan, draw=black] (13.00, -3.00) rectangle (13.25, -3.25);
\filldraw[fill=cancan, draw=black] (13.25, -3.00) rectangle (13.50, -3.25);
\filldraw[fill=cancan, draw=black] (13.50, -3.00) rectangle (13.75, -3.25);
\filldraw[fill=bermuda, draw=black] (13.75, -3.00) rectangle (14.00, -3.25);
\filldraw[fill=bermuda, draw=black] (14.00, -3.00) rectangle (14.25, -3.25);
\filldraw[fill=bermuda, draw=black] (14.25, -3.00) rectangle (14.50, -3.25);
\filldraw[fill=cancan, draw=black] (14.50, -3.00) rectangle (14.75, -3.25);
\filldraw[fill=cancan, draw=black] (14.75, -3.00) rectangle (15.00, -3.25);
\filldraw[fill=cancan, draw=black] (0.00, -3.25) rectangle (0.25, -3.50);
\filldraw[fill=bermuda, draw=black] (0.25, -3.25) rectangle (0.50, -3.50);
\filldraw[fill=bermuda, draw=black] (0.50, -3.25) rectangle (0.75, -3.50);
\filldraw[fill=bermuda, draw=black] (0.75, -3.25) rectangle (1.00, -3.50);
\filldraw[fill=cancan, draw=black] (1.00, -3.25) rectangle (1.25, -3.50);
\filldraw[fill=cancan, draw=black] (1.25, -3.25) rectangle (1.50, -3.50);
\filldraw[fill=cancan, draw=black] (1.50, -3.25) rectangle (1.75, -3.50);
\filldraw[fill=bermuda, draw=black] (1.75, -3.25) rectangle (2.00, -3.50);
\filldraw[fill=bermuda, draw=black] (2.00, -3.25) rectangle (2.25, -3.50);
\filldraw[fill=bermuda, draw=black] (2.25, -3.25) rectangle (2.50, -3.50);
\filldraw[fill=cancan, draw=black] (2.50, -3.25) rectangle (2.75, -3.50);
} } }\end{equation*}
\begin{equation*}
\hspace{4.6pt} b_{8} = \vcenter{\hbox{ \tikz{
\filldraw[fill=bermuda, draw=black] (0.00, 0.00) rectangle (0.25, -0.25);
\filldraw[fill=bermuda, draw=black] (0.25, 0.00) rectangle (0.50, -0.25);
\filldraw[fill=bermuda, draw=black] (0.50, 0.00) rectangle (0.75, -0.25);
\filldraw[fill=bermuda, draw=black] (0.75, 0.00) rectangle (1.00, -0.25);
\filldraw[fill=bermuda, draw=black] (1.00, 0.00) rectangle (1.25, -0.25);
\filldraw[fill=cancan, draw=black] (1.25, 0.00) rectangle (1.50, -0.25);
\filldraw[fill=bermuda, draw=black] (1.50, 0.00) rectangle (1.75, -0.25);
\filldraw[fill=cancan, draw=black] (1.75, 0.00) rectangle (2.00, -0.25);
\filldraw[fill=cancan, draw=black] (2.00, 0.00) rectangle (2.25, -0.25);
\filldraw[fill=cancan, draw=black] (2.25, 0.00) rectangle (2.50, -0.25);
\filldraw[fill=bermuda, draw=black] (2.50, 0.00) rectangle (2.75, -0.25);
\filldraw[fill=cancan, draw=black] (2.75, 0.00) rectangle (3.00, -0.25);
\filldraw[fill=cancan, draw=black] (3.00, 0.00) rectangle (3.25, -0.25);
\filldraw[fill=cancan, draw=black] (3.25, 0.00) rectangle (3.50, -0.25);
\filldraw[fill=bermuda, draw=black] (3.50, 0.00) rectangle (3.75, -0.25);
\filldraw[fill=bermuda, draw=black] (3.75, 0.00) rectangle (4.00, -0.25);
\filldraw[fill=bermuda, draw=black] (4.00, 0.00) rectangle (4.25, -0.25);
\filldraw[fill=cancan, draw=black] (4.25, 0.00) rectangle (4.50, -0.25);
\filldraw[fill=bermuda, draw=black] (4.50, 0.00) rectangle (4.75, -0.25);
\filldraw[fill=cancan, draw=black] (4.75, 0.00) rectangle (5.00, -0.25);
\filldraw[fill=cancan, draw=black] (5.00, 0.00) rectangle (5.25, -0.25);
\filldraw[fill=cancan, draw=black] (5.25, 0.00) rectangle (5.50, -0.25);
\filldraw[fill=cancan, draw=black] (5.50, 0.00) rectangle (5.75, -0.25);
\filldraw[fill=cancan, draw=black] (5.75, 0.00) rectangle (6.00, -0.25);
\filldraw[fill=bermuda, draw=black] (6.00, 0.00) rectangle (6.25, -0.25);
\filldraw[fill=cancan, draw=black] (6.25, 0.00) rectangle (6.50, -0.25);
\filldraw[fill=bermuda, draw=black] (6.50, 0.00) rectangle (6.75, -0.25);
\filldraw[fill=cancan, draw=black] (6.75, 0.00) rectangle (7.00, -0.25);
\filldraw[fill=cancan, draw=black] (7.00, 0.00) rectangle (7.25, -0.25);
\filldraw[fill=cancan, draw=black] (7.25, 0.00) rectangle (7.50, -0.25);
\filldraw[fill=bermuda, draw=black] (7.50, 0.00) rectangle (7.75, -0.25);
\filldraw[fill=bermuda, draw=black] (7.75, 0.00) rectangle (8.00, -0.25);
\filldraw[fill=bermuda, draw=black] (8.00, 0.00) rectangle (8.25, -0.25);
\filldraw[fill=cancan, draw=black] (8.25, 0.00) rectangle (8.50, -0.25);
\filldraw[fill=cancan, draw=black] (8.50, 0.00) rectangle (8.75, -0.25);
\filldraw[fill=cancan, draw=black] (8.75, 0.00) rectangle (9.00, -0.25);
\filldraw[fill=cancan, draw=black] (9.00, 0.00) rectangle (9.25, -0.25);
\filldraw[fill=bermuda, draw=black] (9.25, 0.00) rectangle (9.50, -0.25);
\filldraw[fill=bermuda, draw=black] (9.50, 0.00) rectangle (9.75, -0.25);
\filldraw[fill=cancan, draw=black] (9.75, 0.00) rectangle (10.00, -0.25);
\filldraw[fill=bermuda, draw=black] (10.00, 0.00) rectangle (10.25, -0.25);
\filldraw[fill=cancan, draw=black] (10.25, 0.00) rectangle (10.50, -0.25);
\filldraw[fill=cancan, draw=black] (10.50, 0.00) rectangle (10.75, -0.25);
\filldraw[fill=bermuda, draw=black] (10.75, 0.00) rectangle (11.00, -0.25);
\filldraw[fill=bermuda, draw=black] (11.00, 0.00) rectangle (11.25, -0.25);
\filldraw[fill=cancan, draw=black] (11.25, 0.00) rectangle (11.50, -0.25);
\filldraw[fill=bermuda, draw=black] (11.50, 0.00) rectangle (11.75, -0.25);
\filldraw[fill=cancan, draw=black] (11.75, 0.00) rectangle (12.00, -0.25);
\filldraw[fill=bermuda, draw=black] (12.00, 0.00) rectangle (12.25, -0.25);
\filldraw[fill=bermuda, draw=black] (12.25, 0.00) rectangle (12.50, -0.25);
\filldraw[fill=bermuda, draw=black] (12.50, 0.00) rectangle (12.75, -0.25);
\filldraw[fill=cancan, draw=black] (12.75, 0.00) rectangle (13.00, -0.25);
\filldraw[fill=cancan, draw=black] (13.00, 0.00) rectangle (13.25, -0.25);
\filldraw[fill=cancan, draw=black] (13.25, 0.00) rectangle (13.50, -0.25);
\filldraw[fill=bermuda, draw=black] (13.50, 0.00) rectangle (13.75, -0.25);
\filldraw[fill=bermuda, draw=black] (13.75, 0.00) rectangle (14.00, -0.25);
\filldraw[fill=bermuda, draw=black] (14.00, 0.00) rectangle (14.25, -0.25);
\filldraw[fill=cancan, draw=black] (14.25, 0.00) rectangle (14.50, -0.25);
\filldraw[fill=cancan, draw=black] (14.50, 0.00) rectangle (14.75, -0.25);
\filldraw[fill=cancan, draw=black] (14.75, 0.00) rectangle (15.00, -0.25);
\filldraw[fill=bermuda, draw=black] (0.00, -0.25) rectangle (0.25, -0.50);
\filldraw[fill=bermuda, draw=black] (0.25, -0.25) rectangle (0.50, -0.50);
\filldraw[fill=bermuda, draw=black] (0.50, -0.25) rectangle (0.75, -0.50);
\filldraw[fill=cancan, draw=black] (0.75, -0.25) rectangle (1.00, -0.50);
\filldraw[fill=cancan, draw=black] (1.00, -0.25) rectangle (1.25, -0.50);
\filldraw[fill=cancan, draw=black] (1.25, -0.25) rectangle (1.50, -0.50);
\filldraw[fill=bermuda, draw=black] (1.50, -0.25) rectangle (1.75, -0.50);
\filldraw[fill=bermuda, draw=black] (1.75, -0.25) rectangle (2.00, -0.50);
\filldraw[fill=bermuda, draw=black] (2.00, -0.25) rectangle (2.25, -0.50);
\filldraw[fill=cancan, draw=black] (2.25, -0.25) rectangle (2.50, -0.50);
\filldraw[fill=cancan, draw=black] (2.50, -0.25) rectangle (2.75, -0.50);
\filldraw[fill=cancan, draw=black] (2.75, -0.25) rectangle (3.00, -0.50);
\filldraw[fill=cancan, draw=black] (3.00, -0.25) rectangle (3.25, -0.50);
\filldraw[fill=cancan, draw=black] (3.25, -0.25) rectangle (3.50, -0.50);
\filldraw[fill=bermuda, draw=black] (3.50, -0.25) rectangle (3.75, -0.50);
\filldraw[fill=cancan, draw=black] (3.75, -0.25) rectangle (4.00, -0.50);
\filldraw[fill=cancan, draw=black] (4.00, -0.25) rectangle (4.25, -0.50);
\filldraw[fill=cancan, draw=black] (4.25, -0.25) rectangle (4.50, -0.50);
\filldraw[fill=bermuda, draw=black] (4.50, -0.25) rectangle (4.75, -0.50);
\filldraw[fill=bermuda, draw=black] (4.75, -0.25) rectangle (5.00, -0.50);
\filldraw[fill=bermuda, draw=black] (5.00, -0.25) rectangle (5.25, -0.50);
\filldraw[fill=bermuda, draw=black] (5.25, -0.25) rectangle (5.50, -0.50);
\filldraw[fill=bermuda, draw=black] (5.50, -0.25) rectangle (5.75, -0.50);
\filldraw[fill=cancan, draw=black] (5.75, -0.25) rectangle (6.00, -0.50);
\filldraw[fill=bermuda, draw=black] (6.00, -0.25) rectangle (6.25, -0.50);
\filldraw[fill=bermuda, draw=black] (6.25, -0.25) rectangle (6.50, -0.50);
\filldraw[fill=bermuda, draw=black] (6.50, -0.25) rectangle (6.75, -0.50);
\filldraw[fill=cancan, draw=black] (6.75, -0.25) rectangle (7.00, -0.50);
\filldraw[fill=cancan, draw=black] (7.00, -0.25) rectangle (7.25, -0.50);
\filldraw[fill=cancan, draw=black] (7.25, -0.25) rectangle (7.50, -0.50);
\filldraw[fill=bermuda, draw=black] (7.50, -0.25) rectangle (7.75, -0.50);
\filldraw[fill=bermuda, draw=black] (7.75, -0.25) rectangle (8.00, -0.50);
\filldraw[fill=bermuda, draw=black] (8.00, -0.25) rectangle (8.25, -0.50);
\filldraw[fill=cancan, draw=black] (8.25, -0.25) rectangle (8.50, -0.50);
\filldraw[fill=bermuda, draw=black] (8.50, -0.25) rectangle (8.75, -0.50);
\filldraw[fill=cancan, draw=black] (8.75, -0.25) rectangle (9.00, -0.50);
\filldraw[fill=cancan, draw=black] (9.00, -0.25) rectangle (9.25, -0.50);
\filldraw[fill=cancan, draw=black] (9.25, -0.25) rectangle (9.50, -0.50);
\filldraw[fill=bermuda, draw=black] (9.50, -0.25) rectangle (9.75, -0.50);
\filldraw[fill=cancan, draw=black] (9.75, -0.25) rectangle (10.00, -0.50);
\filldraw[fill=cancan, draw=black] (10.00, -0.25) rectangle (10.25, -0.50);
\filldraw[fill=cancan, draw=black] (10.25, -0.25) rectangle (10.50, -0.50);
\filldraw[fill=cancan, draw=black] (10.50, -0.25) rectangle (10.75, -0.50);
\filldraw[fill=cancan, draw=black] (10.75, -0.25) rectangle (11.00, -0.50);
\filldraw[fill=bermuda, draw=black] (11.00, -0.25) rectangle (11.25, -0.50);
\filldraw[fill=bermuda, draw=black] (11.25, -0.25) rectangle (11.50, -0.50);
\filldraw[fill=bermuda, draw=black] (11.50, -0.25) rectangle (11.75, -0.50);
\filldraw[fill=cancan, draw=black] (11.75, -0.25) rectangle (12.00, -0.50);
\filldraw[fill=cancan, draw=black] (12.00, -0.25) rectangle (12.25, -0.50);
\filldraw[fill=cancan, draw=black] (12.25, -0.25) rectangle (12.50, -0.50);
\filldraw[fill=bermuda, draw=black] (12.50, -0.25) rectangle (12.75, -0.50);
\filldraw[fill=bermuda, draw=black] (12.75, -0.25) rectangle (13.00, -0.50);
\filldraw[fill=bermuda, draw=black] (13.00, -0.25) rectangle (13.25, -0.50);
\filldraw[fill=cancan, draw=black] (13.25, -0.25) rectangle (13.50, -0.50);
\filldraw[fill=cancan, draw=black] (13.50, -0.25) rectangle (13.75, -0.50);
\filldraw[fill=cancan, draw=black] (13.75, -0.25) rectangle (14.00, -0.50);
\filldraw[fill=bermuda, draw=black] (14.00, -0.25) rectangle (14.25, -0.50);
\filldraw[fill=bermuda, draw=black] (14.25, -0.25) rectangle (14.50, -0.50);
\filldraw[fill=bermuda, draw=black] (14.50, -0.25) rectangle (14.75, -0.50);
\filldraw[fill=cancan, draw=black] (14.75, -0.25) rectangle (15.00, -0.50);
\filldraw[fill=bermuda, draw=black] (0.00, -0.50) rectangle (0.25, -0.75);
\filldraw[fill=bermuda, draw=black] (0.25, -0.50) rectangle (0.50, -0.75);
\filldraw[fill=bermuda, draw=black] (0.50, -0.50) rectangle (0.75, -0.75);
\filldraw[fill=bermuda, draw=black] (0.75, -0.50) rectangle (1.00, -0.75);
\filldraw[fill=bermuda, draw=black] (1.00, -0.50) rectangle (1.25, -0.75);
\filldraw[fill=cancan, draw=black] (1.25, -0.50) rectangle (1.50, -0.75);
\filldraw[fill=bermuda, draw=black] (1.50, -0.50) rectangle (1.75, -0.75);
\filldraw[fill=cancan, draw=black] (1.75, -0.50) rectangle (2.00, -0.75);
\filldraw[fill=cancan, draw=black] (2.00, -0.50) rectangle (2.25, -0.75);
\filldraw[fill=cancan, draw=black] (2.25, -0.50) rectangle (2.50, -0.75);
\filldraw[fill=bermuda, draw=black] (2.50, -0.50) rectangle (2.75, -0.75);
\filldraw[fill=bermuda, draw=black] (2.75, -0.50) rectangle (3.00, -0.75);
\filldraw[fill=bermuda, draw=black] (3.00, -0.50) rectangle (3.25, -0.75);
\filldraw[fill=cancan, draw=black] (3.25, -0.50) rectangle (3.50, -0.75);
\filldraw[fill=cancan, draw=black] (3.50, -0.50) rectangle (3.75, -0.75);
\filldraw[fill=bermuda, draw=black] (3.75, -0.50) rectangle (4.00, -0.75);
\filldraw[fill=bermuda, draw=black] (4.00, -0.50) rectangle (4.25, -0.75);
\filldraw[fill=cancan, draw=black] (4.25, -0.50) rectangle (4.50, -0.75);
\filldraw[fill=cancan, draw=black] (4.50, -0.50) rectangle (4.75, -0.75);
\filldraw[fill=cancan, draw=black] (4.75, -0.50) rectangle (5.00, -0.75);
\filldraw[fill=bermuda, draw=black] (5.00, -0.50) rectangle (5.25, -0.75);
\filldraw[fill=bermuda, draw=black] (5.25, -0.50) rectangle (5.50, -0.75);
\filldraw[fill=bermuda, draw=black] (5.50, -0.50) rectangle (5.75, -0.75);
\filldraw[fill=cancan, draw=black] (5.75, -0.50) rectangle (6.00, -0.75);
\filldraw[fill=bermuda, draw=black] (6.00, -0.50) rectangle (6.25, -0.75);
\filldraw[fill=bermuda, draw=black] (6.25, -0.50) rectangle (6.50, -0.75);
\filldraw[fill=bermuda, draw=black] (6.50, -0.50) rectangle (6.75, -0.75);
\filldraw[fill=bermuda, draw=black] (6.75, -0.50) rectangle (7.00, -0.75);
\filldraw[fill=bermuda, draw=black] (7.00, -0.50) rectangle (7.25, -0.75);
\filldraw[fill=cancan, draw=black] (7.25, -0.50) rectangle (7.50, -0.75);
\filldraw[fill=bermuda, draw=black] (7.50, -0.50) rectangle (7.75, -0.75);
\filldraw[fill=cancan, draw=black] (7.75, -0.50) rectangle (8.00, -0.75);
\filldraw[fill=cancan, draw=black] (8.00, -0.50) rectangle (8.25, -0.75);
\filldraw[fill=bermuda, draw=black] (8.25, -0.50) rectangle (8.50, -0.75);
\filldraw[fill=bermuda, draw=black] (8.50, -0.50) rectangle (8.75, -0.75);
\filldraw[fill=cancan, draw=black] (8.75, -0.50) rectangle (9.00, -0.75);
\filldraw[fill=bermuda, draw=black] (9.00, -0.50) rectangle (9.25, -0.75);
\filldraw[fill=cancan, draw=black] (9.25, -0.50) rectangle (9.50, -0.75);
\filldraw[fill=cancan, draw=black] (9.50, -0.50) rectangle (9.75, -0.75);
\filldraw[fill=cancan, draw=black] (9.75, -0.50) rectangle (10.00, -0.75);
\filldraw[fill=bermuda, draw=black] (10.00, -0.50) rectangle (10.25, -0.75);
\filldraw[fill=bermuda, draw=black] (10.25, -0.50) rectangle (10.50, -0.75);
\filldraw[fill=bermuda, draw=black] (10.50, -0.50) rectangle (10.75, -0.75);
\filldraw[fill=bermuda, draw=black] (10.75, -0.50) rectangle (11.00, -0.75);
\filldraw[fill=bermuda, draw=black] (11.00, -0.50) rectangle (11.25, -0.75);
\filldraw[fill=cancan, draw=black] (11.25, -0.50) rectangle (11.50, -0.75);
\filldraw[fill=cancan, draw=black] (11.50, -0.50) rectangle (11.75, -0.75);
\filldraw[fill=cancan, draw=black] (11.75, -0.50) rectangle (12.00, -0.75);
\filldraw[fill=cancan, draw=black] (12.00, -0.50) rectangle (12.25, -0.75);
\filldraw[fill=cancan, draw=black] (12.25, -0.50) rectangle (12.50, -0.75);
\filldraw[fill=cancan, draw=black] (12.50, -0.50) rectangle (12.75, -0.75);
\filldraw[fill=cancan, draw=black] (12.75, -0.50) rectangle (13.00, -0.75);
\filldraw[fill=cancan, draw=black] (13.00, -0.50) rectangle (13.25, -0.75);
\filldraw[fill=cancan, draw=black] (13.25, -0.50) rectangle (13.50, -0.75);
\filldraw[fill=bermuda, draw=black] (13.50, -0.50) rectangle (13.75, -0.75);
\filldraw[fill=bermuda, draw=black] (13.75, -0.50) rectangle (14.00, -0.75);
\filldraw[fill=bermuda, draw=black] (14.00, -0.50) rectangle (14.25, -0.75);
\filldraw[fill=cancan, draw=black] (14.25, -0.50) rectangle (14.50, -0.75);
\filldraw[fill=bermuda, draw=black] (14.50, -0.50) rectangle (14.75, -0.75);
\filldraw[fill=bermuda, draw=black] (14.75, -0.50) rectangle (15.00, -0.75);
\filldraw[fill=bermuda, draw=black] (0.00, -0.75) rectangle (0.25, -1.00);
\filldraw[fill=bermuda, draw=black] (0.25, -0.75) rectangle (0.50, -1.00);
\filldraw[fill=bermuda, draw=black] (0.50, -0.75) rectangle (0.75, -1.00);
\filldraw[fill=cancan, draw=black] (0.75, -0.75) rectangle (1.00, -1.00);
\filldraw[fill=bermuda, draw=black] (1.00, -0.75) rectangle (1.25, -1.00);
\filldraw[fill=bermuda, draw=black] (1.25, -0.75) rectangle (1.50, -1.00);
\filldraw[fill=bermuda, draw=black] (1.50, -0.75) rectangle (1.75, -1.00);
\filldraw[fill=bermuda, draw=black] (1.75, -0.75) rectangle (2.00, -1.00);
\filldraw[fill=bermuda, draw=black] (2.00, -0.75) rectangle (2.25, -1.00);
\filldraw[fill=cancan, draw=black] (2.25, -0.75) rectangle (2.50, -1.00);
\filldraw[fill=bermuda, draw=black] (2.50, -0.75) rectangle (2.75, -1.00);
\filldraw[fill=bermuda, draw=black] (2.75, -0.75) rectangle (3.00, -1.00);
\filldraw[fill=bermuda, draw=black] (3.00, -0.75) rectangle (3.25, -1.00);
\filldraw[fill=bermuda, draw=black] (3.25, -0.75) rectangle (3.50, -1.00);
\filldraw[fill=bermuda, draw=black] (3.50, -0.75) rectangle (3.75, -1.00);
\filldraw[fill=cancan, draw=black] (3.75, -0.75) rectangle (4.00, -1.00);
\filldraw[fill=bermuda, draw=black] (4.00, -0.75) rectangle (4.25, -1.00);
\filldraw[fill=bermuda, draw=black] (4.25, -0.75) rectangle (4.50, -1.00);
\filldraw[fill=bermuda, draw=black] (4.50, -0.75) rectangle (4.75, -1.00);
\filldraw[fill=bermuda, draw=black] (4.75, -0.75) rectangle (5.00, -1.00);
\filldraw[fill=bermuda, draw=black] (5.00, -0.75) rectangle (5.25, -1.00);
\filldraw[fill=bermuda, draw=black] (5.25, -0.75) rectangle (5.50, -1.00);
\filldraw[fill=bermuda, draw=black] (5.50, -0.75) rectangle (5.75, -1.00);
\filldraw[fill=cancan, draw=black] (5.75, -0.75) rectangle (6.00, -1.00);
\filldraw[fill=bermuda, draw=black] (6.00, -0.75) rectangle (6.25, -1.00);
\filldraw[fill=cancan, draw=black] (6.25, -0.75) rectangle (6.50, -1.00);
\filldraw[fill=cancan, draw=black] (6.50, -0.75) rectangle (6.75, -1.00);
\filldraw[fill=cancan, draw=black] (6.75, -0.75) rectangle (7.00, -1.00);
\filldraw[fill=bermuda, draw=black] (7.00, -0.75) rectangle (7.25, -1.00);
\filldraw[fill=cancan, draw=black] (7.25, -0.75) rectangle (7.50, -1.00);
\filldraw[fill=cancan, draw=black] (7.50, -0.75) rectangle (7.75, -1.00);
\filldraw[fill=cancan, draw=black] (7.75, -0.75) rectangle (8.00, -1.00);
\filldraw[fill=cancan, draw=black] (8.00, -0.75) rectangle (8.25, -1.00);
\filldraw[fill=cancan, draw=black] (8.25, -0.75) rectangle (8.50, -1.00);
\filldraw[fill=cancan, draw=black] (8.50, -0.75) rectangle (8.75, -1.00);
\filldraw[fill=cancan, draw=black] (8.75, -0.75) rectangle (9.00, -1.00);
\filldraw[fill=bermuda, draw=black] (9.00, -0.75) rectangle (9.25, -1.00);
\filldraw[fill=bermuda, draw=black] (9.25, -0.75) rectangle (9.50, -1.00);
\filldraw[fill=bermuda, draw=black] (9.50, -0.75) rectangle (9.75, -1.00);
\filldraw[fill=bermuda, draw=black] (9.75, -0.75) rectangle (10.00, -1.00);
\filldraw[fill=bermuda, draw=black] (10.00, -0.75) rectangle (10.25, -1.00);
\filldraw[fill=cancan, draw=black] (10.25, -0.75) rectangle (10.50, -1.00);
\filldraw[fill=bermuda, draw=black] (10.50, -0.75) rectangle (10.75, -1.00);
\filldraw[fill=bermuda, draw=black] (10.75, -0.75) rectangle (11.00, -1.00);
\filldraw[fill=bermuda, draw=black] (11.00, -0.75) rectangle (11.25, -1.00);
\filldraw[fill=cancan, draw=black] (11.25, -0.75) rectangle (11.50, -1.00);
\filldraw[fill=cancan, draw=black] (11.50, -0.75) rectangle (11.75, -1.00);
\filldraw[fill=cancan, draw=black] (11.75, -0.75) rectangle (12.00, -1.00);
\filldraw[fill=bermuda, draw=black] (12.00, -0.75) rectangle (12.25, -1.00);
\filldraw[fill=bermuda, draw=black] (12.25, -0.75) rectangle (12.50, -1.00);
\filldraw[fill=bermuda, draw=black] (12.50, -0.75) rectangle (12.75, -1.00);
\filldraw[fill=cancan, draw=black] (12.75, -0.75) rectangle (13.00, -1.00);
\filldraw[fill=cancan, draw=black] (13.00, -0.75) rectangle (13.25, -1.00);
\filldraw[fill=cancan, draw=black] (13.25, -0.75) rectangle (13.50, -1.00);
\filldraw[fill=bermuda, draw=black] (13.50, -0.75) rectangle (13.75, -1.00);
\filldraw[fill=bermuda, draw=black] (13.75, -0.75) rectangle (14.00, -1.00);
\filldraw[fill=bermuda, draw=black] (14.00, -0.75) rectangle (14.25, -1.00);
\filldraw[fill=bermuda, draw=black] (14.25, -0.75) rectangle (14.50, -1.00);
\filldraw[fill=bermuda, draw=black] (14.50, -0.75) rectangle (14.75, -1.00);
\filldraw[fill=cancan, draw=black] (14.75, -0.75) rectangle (15.00, -1.00);
\filldraw[fill=bermuda, draw=black] (0.00, -1.00) rectangle (0.25, -1.25);
\filldraw[fill=cancan, draw=black] (0.25, -1.00) rectangle (0.50, -1.25);
\filldraw[fill=cancan, draw=black] (0.50, -1.00) rectangle (0.75, -1.25);
\filldraw[fill=cancan, draw=black] (0.75, -1.00) rectangle (1.00, -1.25);
\filldraw[fill=cancan, draw=black] (1.00, -1.00) rectangle (1.25, -1.25);
\filldraw[fill=cancan, draw=black] (1.25, -1.00) rectangle (1.50, -1.25);
\filldraw[fill=bermuda, draw=black] (1.50, -1.00) rectangle (1.75, -1.25);
\filldraw[fill=bermuda, draw=black] (1.75, -1.00) rectangle (2.00, -1.25);
\filldraw[fill=bermuda, draw=black] (2.00, -1.00) rectangle (2.25, -1.25);
\filldraw[fill=bermuda, draw=black] (2.25, -1.00) rectangle (2.50, -1.25);
\filldraw[fill=bermuda, draw=black] (2.50, -1.00) rectangle (2.75, -1.25);
\filldraw[fill=cancan, draw=black] (2.75, -1.00) rectangle (3.00, -1.25);
\filldraw[fill=bermuda, draw=black] (3.00, -1.00) rectangle (3.25, -1.25);
\filldraw[fill=bermuda, draw=black] (3.25, -1.00) rectangle (3.50, -1.25);
\filldraw[fill=bermuda, draw=black] (3.50, -1.00) rectangle (3.75, -1.25);
\filldraw[fill=cancan, draw=black] (3.75, -1.00) rectangle (4.00, -1.25);
\filldraw[fill=cancan, draw=black] (4.00, -1.00) rectangle (4.25, -1.25);
\filldraw[fill=cancan, draw=black] (4.25, -1.00) rectangle (4.50, -1.25);
\filldraw[fill=bermuda, draw=black] (4.50, -1.00) rectangle (4.75, -1.25);
\filldraw[fill=bermuda, draw=black] (4.75, -1.00) rectangle (5.00, -1.25);
\filldraw[fill=bermuda, draw=black] (5.00, -1.00) rectangle (5.25, -1.25);
\filldraw[fill=cancan, draw=black] (5.25, -1.00) rectangle (5.50, -1.25);
\filldraw[fill=bermuda, draw=black] (5.50, -1.00) rectangle (5.75, -1.25);
\filldraw[fill=cancan, draw=black] (5.75, -1.00) rectangle (6.00, -1.25);
\filldraw[fill=cancan, draw=black] (6.00, -1.00) rectangle (6.25, -1.25);
\filldraw[fill=cancan, draw=black] (6.25, -1.00) rectangle (6.50, -1.25);
\filldraw[fill=cancan, draw=black] (6.50, -1.00) rectangle (6.75, -1.25);
\filldraw[fill=cancan, draw=black] (6.75, -1.00) rectangle (7.00, -1.25);
\filldraw[fill=cancan, draw=black] (7.00, -1.00) rectangle (7.25, -1.25);
\filldraw[fill=cancan, draw=black] (7.25, -1.00) rectangle (7.50, -1.25);
\filldraw[fill=cancan, draw=black] (7.50, -1.00) rectangle (7.75, -1.25);
\filldraw[fill=cancan, draw=black] (7.75, -1.00) rectangle (8.00, -1.25);
\filldraw[fill=bermuda, draw=black] (8.00, -1.00) rectangle (8.25, -1.25);
\filldraw[fill=bermuda, draw=black] (8.25, -1.00) rectangle (8.50, -1.25);
\filldraw[fill=bermuda, draw=black] (8.50, -1.00) rectangle (8.75, -1.25);
\filldraw[fill=cancan, draw=black] (8.75, -1.00) rectangle (9.00, -1.25);
\filldraw[fill=bermuda, draw=black] (9.00, -1.00) rectangle (9.25, -1.25);
\filldraw[fill=bermuda, draw=black] (9.25, -1.00) rectangle (9.50, -1.25);
\filldraw[fill=bermuda, draw=black] (9.50, -1.00) rectangle (9.75, -1.25);
\filldraw[fill=cancan, draw=black] (9.75, -1.00) rectangle (10.00, -1.25);
\filldraw[fill=cancan, draw=black] (10.00, -1.00) rectangle (10.25, -1.25);
\filldraw[fill=bermuda, draw=black] (10.25, -1.00) rectangle (10.50, -1.25);
\filldraw[fill=bermuda, draw=black] (10.50, -1.00) rectangle (10.75, -1.25);
\filldraw[fill=cancan, draw=black] (10.75, -1.00) rectangle (11.00, -1.25);
\filldraw[fill=cancan, draw=black] (11.00, -1.00) rectangle (11.25, -1.25);
\filldraw[fill=cancan, draw=black] (11.25, -1.00) rectangle (11.50, -1.25);
\filldraw[fill=bermuda, draw=black] (11.50, -1.00) rectangle (11.75, -1.25);
\filldraw[fill=bermuda, draw=black] (11.75, -1.00) rectangle (12.00, -1.25);
\filldraw[fill=bermuda, draw=black] (12.00, -1.00) rectangle (12.25, -1.25);
\filldraw[fill=cancan, draw=black] (12.25, -1.00) rectangle (12.50, -1.25);
\filldraw[fill=cancan, draw=black] (12.50, -1.00) rectangle (12.75, -1.25);
\filldraw[fill=cancan, draw=black] (12.75, -1.00) rectangle (13.00, -1.25);
\filldraw[fill=bermuda, draw=black] (13.00, -1.00) rectangle (13.25, -1.25);
\filldraw[fill=cancan, draw=black] (13.25, -1.00) rectangle (13.50, -1.25);
\filldraw[fill=bermuda, draw=black] (13.50, -1.00) rectangle (13.75, -1.25);
\filldraw[fill=bermuda, draw=black] (13.75, -1.00) rectangle (14.00, -1.25);
\filldraw[fill=bermuda, draw=black] (14.00, -1.00) rectangle (14.25, -1.25);
\filldraw[fill=cancan, draw=black] (14.25, -1.00) rectangle (14.50, -1.25);
\filldraw[fill=cancan, draw=black] (14.50, -1.00) rectangle (14.75, -1.25);
\filldraw[fill=cancan, draw=black] (14.75, -1.00) rectangle (15.00, -1.25);
\filldraw[fill=cancan, draw=black] (0.00, -1.25) rectangle (0.25, -1.50);
\filldraw[fill=cancan, draw=black] (0.25, -1.25) rectangle (0.50, -1.50);
\filldraw[fill=bermuda, draw=black] (0.50, -1.25) rectangle (0.75, -1.50);
\filldraw[fill=bermuda, draw=black] (0.75, -1.25) rectangle (1.00, -1.50);
\filldraw[fill=bermuda, draw=black] (1.00, -1.25) rectangle (1.25, -1.50);
\filldraw[fill=cancan, draw=black] (1.25, -1.25) rectangle (1.50, -1.50);
\filldraw[fill=cancan, draw=black] (1.50, -1.25) rectangle (1.75, -1.50);
\filldraw[fill=cancan, draw=black] (1.75, -1.25) rectangle (2.00, -1.50);
\filldraw[fill=bermuda, draw=black] (2.00, -1.25) rectangle (2.25, -1.50);
\filldraw[fill=bermuda, draw=black] (2.25, -1.25) rectangle (2.50, -1.50);
\filldraw[fill=bermuda, draw=black] (2.50, -1.25) rectangle (2.75, -1.50);
\filldraw[fill=cancan, draw=black] (2.75, -1.25) rectangle (3.00, -1.50);
\filldraw[fill=bermuda, draw=black] (3.00, -1.25) rectangle (3.25, -1.50);
\filldraw[fill=cancan, draw=black] (3.25, -1.25) rectangle (3.50, -1.50);
\filldraw[fill=bermuda, draw=black] (3.50, -1.25) rectangle (3.75, -1.50);
\filldraw[fill=cancan, draw=black] (3.75, -1.25) rectangle (4.00, -1.50);
\filldraw[fill=cancan, draw=black] (4.00, -1.25) rectangle (4.25, -1.50);
\filldraw[fill=cancan, draw=black] (4.25, -1.25) rectangle (4.50, -1.50);
\filldraw[fill=bermuda, draw=black] (4.50, -1.25) rectangle (4.75, -1.50);
\filldraw[fill=cancan, draw=black] (4.75, -1.25) rectangle (5.00, -1.50);
\filldraw[fill=bermuda, draw=black] (5.00, -1.25) rectangle (5.25, -1.50);
\filldraw[fill=cancan, draw=black] (5.25, -1.25) rectangle (5.50, -1.50);
\filldraw[fill=bermuda, draw=black] (5.50, -1.25) rectangle (5.75, -1.50);
\filldraw[fill=bermuda, draw=black] (5.75, -1.25) rectangle (6.00, -1.50);
\filldraw[fill=bermuda, draw=black] (6.00, -1.25) rectangle (6.25, -1.50);
\filldraw[fill=cancan, draw=black] (6.25, -1.25) rectangle (6.50, -1.50);
\filldraw[fill=cancan, draw=black] (6.50, -1.25) rectangle (6.75, -1.50);
\filldraw[fill=cancan, draw=black] (6.75, -1.25) rectangle (7.00, -1.50);
\filldraw[fill=bermuda, draw=black] (7.00, -1.25) rectangle (7.25, -1.50);
\filldraw[fill=bermuda, draw=black] (7.25, -1.25) rectangle (7.50, -1.50);
\filldraw[fill=bermuda, draw=black] (7.50, -1.25) rectangle (7.75, -1.50);
\filldraw[fill=bermuda, draw=black] (7.75, -1.25) rectangle (8.00, -1.50);
\filldraw[fill=bermuda, draw=black] (8.00, -1.25) rectangle (8.25, -1.50);
\filldraw[fill=cancan, draw=black] (8.25, -1.25) rectangle (8.50, -1.50);
\filldraw[fill=bermuda, draw=black] (8.50, -1.25) rectangle (8.75, -1.50);
\filldraw[fill=cancan, draw=black] (8.75, -1.25) rectangle (9.00, -1.50);
\filldraw[fill=cancan, draw=black] (9.00, -1.25) rectangle (9.25, -1.50);
\filldraw[fill=bermuda, draw=black] (9.25, -1.25) rectangle (9.50, -1.50);
\filldraw[fill=bermuda, draw=black] (9.50, -1.25) rectangle (9.75, -1.50);
\filldraw[fill=cancan, draw=black] (9.75, -1.25) rectangle (10.00, -1.50);
\filldraw[fill=bermuda, draw=black] (10.00, -1.25) rectangle (10.25, -1.50);
\filldraw[fill=bermuda, draw=black] (10.25, -1.25) rectangle (10.50, -1.50);
\filldraw[fill=bermuda, draw=black] (10.50, -1.25) rectangle (10.75, -1.50);
\filldraw[fill=cancan, draw=black] (10.75, -1.25) rectangle (11.00, -1.50);
\filldraw[fill=cancan, draw=black] (11.00, -1.25) rectangle (11.25, -1.50);
\filldraw[fill=cancan, draw=black] (11.25, -1.25) rectangle (11.50, -1.50);
\filldraw[fill=bermuda, draw=black] (11.50, -1.25) rectangle (11.75, -1.50);
\filldraw[fill=bermuda, draw=black] (11.75, -1.25) rectangle (12.00, -1.50);
\filldraw[fill=bermuda, draw=black] (12.00, -1.25) rectangle (12.25, -1.50);
\filldraw[fill=cancan, draw=black] (12.25, -1.25) rectangle (12.50, -1.50);
\filldraw[fill=cancan, draw=black] (12.50, -1.25) rectangle (12.75, -1.50);
\filldraw[fill=cancan, draw=black] (12.75, -1.25) rectangle (13.00, -1.50);
\filldraw[fill=bermuda, draw=black] (13.00, -1.25) rectangle (13.25, -1.50);
\filldraw[fill=bermuda, draw=black] (13.25, -1.25) rectangle (13.50, -1.50);
\filldraw[fill=bermuda, draw=black] (13.50, -1.25) rectangle (13.75, -1.50);
\filldraw[fill=cancan, draw=black] (13.75, -1.25) rectangle (14.00, -1.50);
\filldraw[fill=bermuda, draw=black] (14.00, -1.25) rectangle (14.25, -1.50);
\filldraw[fill=cancan, draw=black] (14.25, -1.25) rectangle (14.50, -1.50);
\filldraw[fill=bermuda, draw=black] (14.50, -1.25) rectangle (14.75, -1.50);
\filldraw[fill=cancan, draw=black] (14.75, -1.25) rectangle (15.00, -1.50);
\filldraw[fill=cancan, draw=black] (0.00, -1.50) rectangle (0.25, -1.75);
\filldraw[fill=cancan, draw=black] (0.25, -1.50) rectangle (0.50, -1.75);
\filldraw[fill=bermuda, draw=black] (0.50, -1.50) rectangle (0.75, -1.75);
\filldraw[fill=cancan, draw=black] (0.75, -1.50) rectangle (1.00, -1.75);
\filldraw[fill=bermuda, draw=black] (1.00, -1.50) rectangle (1.25, -1.75);
\filldraw[fill=cancan, draw=black] (1.25, -1.50) rectangle (1.50, -1.75);
\filldraw[fill=bermuda, draw=black] (1.50, -1.50) rectangle (1.75, -1.75);
\filldraw[fill=cancan, draw=black] (1.75, -1.50) rectangle (2.00, -1.75);
\filldraw[fill=cancan, draw=black] (2.00, -1.50) rectangle (2.25, -1.75);
\filldraw[fill=bermuda, draw=black] (2.25, -1.50) rectangle (2.50, -1.75);
\filldraw[fill=bermuda, draw=black] (2.50, -1.50) rectangle (2.75, -1.75);
\filldraw[fill=cancan, draw=black] (2.75, -1.50) rectangle (3.00, -1.75);
\filldraw[fill=bermuda, draw=black] (3.00, -1.50) rectangle (3.25, -1.75);
\filldraw[fill=cancan, draw=black] (3.25, -1.50) rectangle (3.50, -1.75);
\filldraw[fill=cancan, draw=black] (3.50, -1.50) rectangle (3.75, -1.75);
\filldraw[fill=bermuda, draw=black] (3.75, -1.50) rectangle (4.00, -1.75);
\filldraw[fill=bermuda, draw=black] (4.00, -1.50) rectangle (4.25, -1.75);
\filldraw[fill=cancan, draw=black] (4.25, -1.50) rectangle (4.50, -1.75);
\filldraw[fill=bermuda, draw=black] (4.50, -1.50) rectangle (4.75, -1.75);
\filldraw[fill=cancan, draw=black] (4.75, -1.50) rectangle (5.00, -1.75);
\filldraw[fill=bermuda, draw=black] (5.00, -1.50) rectangle (5.25, -1.75);
\filldraw[fill=cancan, draw=black] (5.25, -1.50) rectangle (5.50, -1.75);
\filldraw[fill=cancan, draw=black] (5.50, -1.50) rectangle (5.75, -1.75);
\filldraw[fill=cancan, draw=black] (5.75, -1.50) rectangle (6.00, -1.75);
\filldraw[fill=cancan, draw=black] (6.00, -1.50) rectangle (6.25, -1.75);
\filldraw[fill=cancan, draw=black] (6.25, -1.50) rectangle (6.50, -1.75);
\filldraw[fill=cancan, draw=black] (6.50, -1.50) rectangle (6.75, -1.75);
\filldraw[fill=bermuda, draw=black] (6.75, -1.50) rectangle (7.00, -1.75);
\filldraw[fill=bermuda, draw=black] (7.00, -1.50) rectangle (7.25, -1.75);
\filldraw[fill=cancan, draw=black] (7.25, -1.50) rectangle (7.50, -1.75);
\filldraw[fill=bermuda, draw=black] (7.50, -1.50) rectangle (7.75, -1.75);
\filldraw[fill=cancan, draw=black] (7.75, -1.50) rectangle (8.00, -1.75);
\filldraw[fill=cancan, draw=black] (8.00, -1.50) rectangle (8.25, -1.75);
\filldraw[fill=bermuda, draw=black] (8.25, -1.50) rectangle (8.50, -1.75);
\filldraw[fill=bermuda, draw=black] (8.50, -1.50) rectangle (8.75, -1.75);
\filldraw[fill=cancan, draw=black] (8.75, -1.50) rectangle (9.00, -1.75);
\filldraw[fill=bermuda, draw=black] (9.00, -1.50) rectangle (9.25, -1.75);
\filldraw[fill=cancan, draw=black] (9.25, -1.50) rectangle (9.50, -1.75);
\filldraw[fill=bermuda, draw=black] (9.50, -1.50) rectangle (9.75, -1.75);
\filldraw[fill=bermuda, draw=black] (9.75, -1.50) rectangle (10.00, -1.75);
\filldraw[fill=bermuda, draw=black] (10.00, -1.50) rectangle (10.25, -1.75);
\filldraw[fill=cancan, draw=black] (10.25, -1.50) rectangle (10.50, -1.75);
\filldraw[fill=cancan, draw=black] (10.50, -1.50) rectangle (10.75, -1.75);
\filldraw[fill=cancan, draw=black] (10.75, -1.50) rectangle (11.00, -1.75);
\filldraw[fill=bermuda, draw=black] (11.00, -1.50) rectangle (11.25, -1.75);
\filldraw[fill=bermuda, draw=black] (11.25, -1.50) rectangle (11.50, -1.75);
\filldraw[fill=bermuda, draw=black] (11.50, -1.50) rectangle (11.75, -1.75);
\filldraw[fill=cancan, draw=black] (11.75, -1.50) rectangle (12.00, -1.75);
\filldraw[fill=cancan, draw=black] (12.00, -1.50) rectangle (12.25, -1.75);
\filldraw[fill=cancan, draw=black] (12.25, -1.50) rectangle (12.50, -1.75);
\filldraw[fill=bermuda, draw=black] (12.50, -1.50) rectangle (12.75, -1.75);
\filldraw[fill=bermuda, draw=black] (12.75, -1.50) rectangle (13.00, -1.75);
\filldraw[fill=bermuda, draw=black] (13.00, -1.50) rectangle (13.25, -1.75);
\filldraw[fill=cancan, draw=black] (13.25, -1.50) rectangle (13.50, -1.75);
\filldraw[fill=cancan, draw=black] (13.50, -1.50) rectangle (13.75, -1.75);
\filldraw[fill=cancan, draw=black] (13.75, -1.50) rectangle (14.00, -1.75);
\filldraw[fill=bermuda, draw=black] (14.00, -1.50) rectangle (14.25, -1.75);
\filldraw[fill=bermuda, draw=black] (14.25, -1.50) rectangle (14.50, -1.75);
\filldraw[fill=bermuda, draw=black] (14.50, -1.50) rectangle (14.75, -1.75);
\filldraw[fill=cancan, draw=black] (14.75, -1.50) rectangle (15.00, -1.75);
\filldraw[fill=bermuda, draw=black] (0.00, -1.75) rectangle (0.25, -2.00);
\filldraw[fill=bermuda, draw=black] (0.25, -1.75) rectangle (0.50, -2.00);
\filldraw[fill=bermuda, draw=black] (0.50, -1.75) rectangle (0.75, -2.00);
\filldraw[fill=cancan, draw=black] (0.75, -1.75) rectangle (1.00, -2.00);
\filldraw[fill=bermuda, draw=black] (1.00, -1.75) rectangle (1.25, -2.00);
\filldraw[fill=cancan, draw=black] (1.25, -1.75) rectangle (1.50, -2.00);
\filldraw[fill=bermuda, draw=black] (1.50, -1.75) rectangle (1.75, -2.00);
\filldraw[fill=cancan, draw=black] (1.75, -1.75) rectangle (2.00, -2.00);
\filldraw[fill=cancan, draw=black] (2.00, -1.75) rectangle (2.25, -2.00);
\filldraw[fill=cancan, draw=black] (2.25, -1.75) rectangle (2.50, -2.00);
\filldraw[fill=bermuda, draw=black] (2.50, -1.75) rectangle (2.75, -2.00);
\filldraw[fill=cancan, draw=black] (2.75, -1.75) rectangle (3.00, -2.00);
\filldraw[fill=bermuda, draw=black] (3.00, -1.75) rectangle (3.25, -2.00);
\filldraw[fill=cancan, draw=black] (3.25, -1.75) rectangle (3.50, -2.00);
\filldraw[fill=cancan, draw=black] (3.50, -1.75) rectangle (3.75, -2.00);
\filldraw[fill=cancan, draw=black] (3.75, -1.75) rectangle (4.00, -2.00);
\filldraw[fill=bermuda, draw=black] (4.00, -1.75) rectangle (4.25, -2.00);
\filldraw[fill=cancan, draw=black] (4.25, -1.75) rectangle (4.50, -2.00);
\filldraw[fill=bermuda, draw=black] (4.50, -1.75) rectangle (4.75, -2.00);
\filldraw[fill=cancan, draw=black] (4.75, -1.75) rectangle (5.00, -2.00);
\filldraw[fill=cancan, draw=black] (5.00, -1.75) rectangle (5.25, -2.00);
\filldraw[fill=cancan, draw=black] (5.25, -1.75) rectangle (5.50, -2.00);
\filldraw[fill=bermuda, draw=black] (5.50, -1.75) rectangle (5.75, -2.00);
\filldraw[fill=cancan, draw=black] (5.75, -1.75) rectangle (6.00, -2.00);
\filldraw[fill=bermuda, draw=black] (6.00, -1.75) rectangle (6.25, -2.00);
\filldraw[fill=cancan, draw=black] (6.25, -1.75) rectangle (6.50, -2.00);
\filldraw[fill=bermuda, draw=black] (6.50, -1.75) rectangle (6.75, -2.00);
\filldraw[fill=bermuda, draw=black] (6.75, -1.75) rectangle (7.00, -2.00);
\filldraw[fill=bermuda, draw=black] (7.00, -1.75) rectangle (7.25, -2.00);
\filldraw[fill=cancan, draw=black] (7.25, -1.75) rectangle (7.50, -2.00);
\filldraw[fill=cancan, draw=black] (7.50, -1.75) rectangle (7.75, -2.00);
\filldraw[fill=cancan, draw=black] (7.75, -1.75) rectangle (8.00, -2.00);
\filldraw[fill=bermuda, draw=black] (8.00, -1.75) rectangle (8.25, -2.00);
\filldraw[fill=bermuda, draw=black] (8.25, -1.75) rectangle (8.50, -2.00);
\filldraw[fill=bermuda, draw=black] (8.50, -1.75) rectangle (8.75, -2.00);
\filldraw[fill=bermuda, draw=black] (8.75, -1.75) rectangle (9.00, -2.00);
\filldraw[fill=bermuda, draw=black] (9.00, -1.75) rectangle (9.25, -2.00);
\filldraw[fill=cancan, draw=black] (9.25, -1.75) rectangle (9.50, -2.00);
\filldraw[fill=bermuda, draw=black] (9.50, -1.75) rectangle (9.75, -2.00);
\filldraw[fill=cancan, draw=black] (9.75, -1.75) rectangle (10.00, -2.00);
\filldraw[fill=bermuda, draw=black] (10.00, -1.75) rectangle (10.25, -2.00);
\filldraw[fill=bermuda, draw=black] (10.25, -1.75) rectangle (10.50, -2.00);
\filldraw[fill=bermuda, draw=black] (10.50, -1.75) rectangle (10.75, -2.00);
\filldraw[fill=cancan, draw=black] (10.75, -1.75) rectangle (11.00, -2.00);
\filldraw[fill=bermuda, draw=black] (11.00, -1.75) rectangle (11.25, -2.00);
\filldraw[fill=bermuda, draw=black] (11.25, -1.75) rectangle (11.50, -2.00);
\filldraw[fill=bermuda, draw=black] (11.50, -1.75) rectangle (11.75, -2.00);
\filldraw[fill=bermuda, draw=black] (11.75, -1.75) rectangle (12.00, -2.00);
\filldraw[fill=bermuda, draw=black] (12.00, -1.75) rectangle (12.25, -2.00);
\filldraw[fill=cancan, draw=black] (12.25, -1.75) rectangle (12.50, -2.00);
\filldraw[fill=bermuda, draw=black] (12.50, -1.75) rectangle (12.75, -2.00);
\filldraw[fill=bermuda, draw=black] (12.75, -1.75) rectangle (13.00, -2.00);
\filldraw[fill=bermuda, draw=black] (13.00, -1.75) rectangle (13.25, -2.00);
\filldraw[fill=cancan, draw=black] (13.25, -1.75) rectangle (13.50, -2.00);
\filldraw[fill=cancan, draw=black] (13.50, -1.75) rectangle (13.75, -2.00);
\filldraw[fill=cancan, draw=black] (13.75, -1.75) rectangle (14.00, -2.00);
\filldraw[fill=bermuda, draw=black] (14.00, -1.75) rectangle (14.25, -2.00);
\filldraw[fill=bermuda, draw=black] (14.25, -1.75) rectangle (14.50, -2.00);
\filldraw[fill=bermuda, draw=black] (14.50, -1.75) rectangle (14.75, -2.00);
\filldraw[fill=cancan, draw=black] (14.75, -1.75) rectangle (15.00, -2.00);
\filldraw[fill=bermuda, draw=black] (0.00, -2.00) rectangle (0.25, -2.25);
\filldraw[fill=cancan, draw=black] (0.25, -2.00) rectangle (0.50, -2.25);
\filldraw[fill=bermuda, draw=black] (0.50, -2.00) rectangle (0.75, -2.25);
\filldraw[fill=bermuda, draw=black] (0.75, -2.00) rectangle (1.00, -2.25);
\filldraw[fill=bermuda, draw=black] (1.00, -2.00) rectangle (1.25, -2.25);
\filldraw[fill=cancan, draw=black] (1.25, -2.00) rectangle (1.50, -2.25);
\filldraw[fill=bermuda, draw=black] (1.50, -2.00) rectangle (1.75, -2.25);
\filldraw[fill=bermuda, draw=black] (1.75, -2.00) rectangle (2.00, -2.25);
\filldraw[fill=bermuda, draw=black] (2.00, -2.00) rectangle (2.25, -2.25);
\filldraw[fill=bermuda, draw=black] (2.25, -2.00) rectangle (2.50, -2.25);
\filldraw[fill=bermuda, draw=black] (2.50, -2.00) rectangle (2.75, -2.25);
\filldraw[fill=cancan, draw=black] (2.75, -2.00) rectangle (3.00, -2.25);
\filldraw[fill=bermuda, draw=black] (3.00, -2.00) rectangle (3.25, -2.25);
\filldraw[fill=bermuda, draw=black] (3.25, -2.00) rectangle (3.50, -2.25);
\filldraw[fill=bermuda, draw=black] (3.50, -2.00) rectangle (3.75, -2.25);
\filldraw[fill=cancan, draw=black] (3.75, -2.00) rectangle (4.00, -2.25);
\filldraw[fill=cancan, draw=black] (4.00, -2.00) rectangle (4.25, -2.25);
\filldraw[fill=cancan, draw=black] (4.25, -2.00) rectangle (4.50, -2.25);
\filldraw[fill=bermuda, draw=black] (4.50, -2.00) rectangle (4.75, -2.25);
\filldraw[fill=bermuda, draw=black] (4.75, -2.00) rectangle (5.00, -2.25);
\filldraw[fill=bermuda, draw=black] (5.00, -2.00) rectangle (5.25, -2.25);
\filldraw[fill=cancan, draw=black] (5.25, -2.00) rectangle (5.50, -2.25);
\filldraw[fill=bermuda, draw=black] (5.50, -2.00) rectangle (5.75, -2.25);
\filldraw[fill=bermuda, draw=black] (5.75, -2.00) rectangle (6.00, -2.25);
\filldraw[fill=bermuda, draw=black] (6.00, -2.00) rectangle (6.25, -2.25);
\filldraw[fill=cancan, draw=black] (6.25, -2.00) rectangle (6.50, -2.25);
\filldraw[fill=bermuda, draw=black] (6.50, -2.00) rectangle (6.75, -2.25);
\filldraw[fill=bermuda, draw=black] (6.75, -2.00) rectangle (7.00, -2.25);
\filldraw[fill=bermuda, draw=black] (7.00, -2.00) rectangle (7.25, -2.25);
\filldraw[fill=bermuda, draw=black] (7.25, -2.00) rectangle (7.50, -2.25);
\filldraw[fill=bermuda, draw=black] (7.50, -2.00) rectangle (7.75, -2.25);
\filldraw[fill=cancan, draw=black] (7.75, -2.00) rectangle (8.00, -2.25);
\filldraw[fill=bermuda, draw=black] (8.00, -2.00) rectangle (8.25, -2.25);
\filldraw[fill=cancan, draw=black] (8.25, -2.00) rectangle (8.50, -2.25);
\filldraw[fill=bermuda, draw=black] (8.50, -2.00) rectangle (8.75, -2.25);
\filldraw[fill=bermuda, draw=black] (8.75, -2.00) rectangle (9.00, -2.25);
\filldraw[fill=bermuda, draw=black] (9.00, -2.00) rectangle (9.25, -2.25);
\filldraw[fill=cancan, draw=black] (9.25, -2.00) rectangle (9.50, -2.25);
\filldraw[fill=cancan, draw=black] (9.50, -2.00) rectangle (9.75, -2.25);
\filldraw[fill=cancan, draw=black] (9.75, -2.00) rectangle (10.00, -2.25);
\filldraw[fill=bermuda, draw=black] (10.00, -2.00) rectangle (10.25, -2.25);
\filldraw[fill=bermuda, draw=black] (10.25, -2.00) rectangle (10.50, -2.25);
\filldraw[fill=bermuda, draw=black] (10.50, -2.00) rectangle (10.75, -2.25);
\filldraw[fill=cancan, draw=black] (10.75, -2.00) rectangle (11.00, -2.25);
\filldraw[fill=cancan, draw=black] (11.00, -2.00) rectangle (11.25, -2.25);
\filldraw[fill=cancan, draw=black] (11.25, -2.00) rectangle (11.50, -2.25);
\filldraw[fill=cancan, draw=black] (11.50, -2.00) rectangle (11.75, -2.25);
\filldraw[fill=cancan, draw=black] (11.75, -2.00) rectangle (12.00, -2.25);
\filldraw[fill=cancan, draw=black] (12.00, -2.00) rectangle (12.25, -2.25);
\filldraw[fill=cancan, draw=black] (12.25, -2.00) rectangle (12.50, -2.25);
\filldraw[fill=cancan, draw=black] (12.50, -2.00) rectangle (12.75, -2.25);
\filldraw[fill=cancan, draw=black] (12.75, -2.00) rectangle (13.00, -2.25);
\filldraw[fill=cancan, draw=black] (13.00, -2.00) rectangle (13.25, -2.25);
\filldraw[fill=cancan, draw=black] (13.25, -2.00) rectangle (13.50, -2.25);
\filldraw[fill=bermuda, draw=black] (13.50, -2.00) rectangle (13.75, -2.25);
\filldraw[fill=bermuda, draw=black] (13.75, -2.00) rectangle (14.00, -2.25);
\filldraw[fill=bermuda, draw=black] (14.00, -2.00) rectangle (14.25, -2.25);
\filldraw[fill=cancan, draw=black] (14.25, -2.00) rectangle (14.50, -2.25);
\filldraw[fill=cancan, draw=black] (14.50, -2.00) rectangle (14.75, -2.25);
\filldraw[fill=cancan, draw=black] (14.75, -2.00) rectangle (15.00, -2.25);
\filldraw[fill=bermuda, draw=black] (0.00, -2.25) rectangle (0.25, -2.50);
\filldraw[fill=bermuda, draw=black] (0.25, -2.25) rectangle (0.50, -2.50);
\filldraw[fill=bermuda, draw=black] (0.50, -2.25) rectangle (0.75, -2.50);
\filldraw[fill=cancan, draw=black] (0.75, -2.25) rectangle (1.00, -2.50);
\filldraw[fill=cancan, draw=black] (1.00, -2.25) rectangle (1.25, -2.50);
\filldraw[fill=cancan, draw=black] (1.25, -2.25) rectangle (1.50, -2.50);
\filldraw[fill=bermuda, draw=black] (1.50, -2.25) rectangle (1.75, -2.50);
\filldraw[fill=bermuda, draw=black] (1.75, -2.25) rectangle (2.00, -2.50);
\filldraw[fill=bermuda, draw=black] (2.00, -2.25) rectangle (2.25, -2.50);
\filldraw[fill=cancan, draw=black] (2.25, -2.25) rectangle (2.50, -2.50);
\filldraw[fill=bermuda, draw=black] (2.50, -2.25) rectangle (2.75, -2.50);
\filldraw[fill=bermuda, draw=black] (2.75, -2.25) rectangle (3.00, -2.50);
\filldraw[fill=bermuda, draw=black] (3.00, -2.25) rectangle (3.25, -2.50);
\filldraw[fill=cancan, draw=black] (3.25, -2.25) rectangle (3.50, -2.50);
\filldraw[fill=bermuda, draw=black] (3.50, -2.25) rectangle (3.75, -2.50);
\filldraw[fill=bermuda, draw=black] (3.75, -2.25) rectangle (4.00, -2.50);
\filldraw[fill=bermuda, draw=black] (4.00, -2.25) rectangle (4.25, -2.50);
\filldraw[fill=cancan, draw=black] (4.25, -2.25) rectangle (4.50, -2.50);
\filldraw[fill=bermuda, draw=black] (4.50, -2.25) rectangle (4.75, -2.50);
\filldraw[fill=cancan, draw=black] (4.75, -2.25) rectangle (5.00, -2.50);
\filldraw[fill=bermuda, draw=black] (5.00, -2.25) rectangle (5.25, -2.50);
\filldraw[fill=cancan, draw=black] (5.25, -2.25) rectangle (5.50, -2.50);
\filldraw[fill=cancan, draw=black] (5.50, -2.25) rectangle (5.75, -2.50);
\filldraw[fill=cancan, draw=black] (5.75, -2.25) rectangle (6.00, -2.50);
\filldraw[fill=cancan, draw=black] (6.00, -2.25) rectangle (6.25, -2.50);
\filldraw[fill=cancan, draw=black] (6.25, -2.25) rectangle (6.50, -2.50);
\filldraw[fill=bermuda, draw=black] (6.50, -2.25) rectangle (6.75, -2.50);
\filldraw[fill=cancan, draw=black] (6.75, -2.25) rectangle (7.00, -2.50);
\filldraw[fill=bermuda, draw=black] (7.00, -2.25) rectangle (7.25, -2.50);
\filldraw[fill=cancan, draw=black] (7.25, -2.25) rectangle (7.50, -2.50);
\filldraw[fill=cancan, draw=black] (7.50, -2.25) rectangle (7.75, -2.50);
\filldraw[fill=cancan, draw=black] (7.75, -2.25) rectangle (8.00, -2.50);
\filldraw[fill=bermuda, draw=black] (8.00, -2.25) rectangle (8.25, -2.50);
\filldraw[fill=cancan, draw=black] (8.25, -2.25) rectangle (8.50, -2.50);
\filldraw[fill=bermuda, draw=black] (8.50, -2.25) rectangle (8.75, -2.50);
\filldraw[fill=bermuda, draw=black] (8.75, -2.25) rectangle (9.00, -2.50);
\filldraw[fill=bermuda, draw=black] (9.00, -2.25) rectangle (9.25, -2.50);
\filldraw[fill=cancan, draw=black] (9.25, -2.25) rectangle (9.50, -2.50);
\filldraw[fill=cancan, draw=black] (9.50, -2.25) rectangle (9.75, -2.50);
\filldraw[fill=cancan, draw=black] (9.75, -2.25) rectangle (10.00, -2.50);
\filldraw[fill=bermuda, draw=black] (10.00, -2.25) rectangle (10.25, -2.50);
\filldraw[fill=bermuda, draw=black] (10.25, -2.25) rectangle (10.50, -2.50);
\filldraw[fill=bermuda, draw=black] (10.50, -2.25) rectangle (10.75, -2.50);
\filldraw[fill=cancan, draw=black] (10.75, -2.25) rectangle (11.00, -2.50);
\filldraw[fill=cancan, draw=black] (11.00, -2.25) rectangle (11.25, -2.50);
\filldraw[fill=cancan, draw=black] (11.25, -2.25) rectangle (11.50, -2.50);
\filldraw[fill=cancan, draw=black] (11.50, -2.25) rectangle (11.75, -2.50);
\filldraw[fill=cancan, draw=black] (11.75, -2.25) rectangle (12.00, -2.50);
\filldraw[fill=cancan, draw=black] (12.00, -2.25) rectangle (12.25, -2.50);
\filldraw[fill=cancan, draw=black] (12.25, -2.25) rectangle (12.50, -2.50);
\filldraw[fill=cancan, draw=black] (12.50, -2.25) rectangle (12.75, -2.50);
\filldraw[fill=cancan, draw=black] (12.75, -2.25) rectangle (13.00, -2.50);
\filldraw[fill=bermuda, draw=black] (13.00, -2.25) rectangle (13.25, -2.50);
\filldraw[fill=bermuda, draw=black] (13.25, -2.25) rectangle (13.50, -2.50);
\filldraw[fill=bermuda, draw=black] (13.50, -2.25) rectangle (13.75, -2.50);
\filldraw[fill=bermuda, draw=black] (13.75, -2.25) rectangle (14.00, -2.50);
\filldraw[fill=bermuda, draw=black] (14.00, -2.25) rectangle (14.25, -2.50);
\filldraw[fill=cancan, draw=black] (14.25, -2.25) rectangle (14.50, -2.50);
\filldraw[fill=bermuda, draw=black] (14.50, -2.25) rectangle (14.75, -2.50);
\filldraw[fill=bermuda, draw=black] (14.75, -2.25) rectangle (15.00, -2.50);
\filldraw[fill=bermuda, draw=black] (0.00, -2.50) rectangle (0.25, -2.75);
\filldraw[fill=bermuda, draw=black] (0.25, -2.50) rectangle (0.50, -2.75);
\filldraw[fill=bermuda, draw=black] (0.50, -2.50) rectangle (0.75, -2.75);
\filldraw[fill=bermuda, draw=black] (0.75, -2.50) rectangle (1.00, -2.75);
\filldraw[fill=bermuda, draw=black] (1.00, -2.50) rectangle (1.25, -2.75);
\filldraw[fill=cancan, draw=black] (1.25, -2.50) rectangle (1.50, -2.75);
\filldraw[fill=bermuda, draw=black] (1.50, -2.50) rectangle (1.75, -2.75);
\filldraw[fill=cancan, draw=black] (1.75, -2.50) rectangle (2.00, -2.75);
\filldraw[fill=cancan, draw=black] (2.00, -2.50) rectangle (2.25, -2.75);
\filldraw[fill=cancan, draw=black] (2.25, -2.50) rectangle (2.50, -2.75);
\filldraw[fill=cancan, draw=black] (2.50, -2.50) rectangle (2.75, -2.75);
\filldraw[fill=cancan, draw=black] (2.75, -2.50) rectangle (3.00, -2.75);
\filldraw[fill=bermuda, draw=black] (3.00, -2.50) rectangle (3.25, -2.75);
\filldraw[fill=bermuda, draw=black] (3.25, -2.50) rectangle (3.50, -2.75);
\filldraw[fill=bermuda, draw=black] (3.50, -2.50) rectangle (3.75, -2.75);
\filldraw[fill=cancan, draw=black] (3.75, -2.50) rectangle (4.00, -2.75);
\filldraw[fill=cancan, draw=black] (4.00, -2.50) rectangle (4.25, -2.75);
\filldraw[fill=cancan, draw=black] (4.25, -2.50) rectangle (4.50, -2.75);
\filldraw[fill=cancan, draw=black] (4.50, -2.50) rectangle (4.75, -2.75);
\filldraw[fill=bermuda, draw=black] (4.75, -2.50) rectangle (5.00, -2.75);
\filldraw[fill=bermuda, draw=black] (5.00, -2.50) rectangle (5.25, -2.75);
\filldraw[fill=cancan, draw=black] (5.25, -2.50) rectangle (5.50, -2.75);
\filldraw[fill=bermuda, draw=black] (5.50, -2.50) rectangle (5.75, -2.75);
\filldraw[fill=cancan, draw=black] (5.75, -2.50) rectangle (6.00, -2.75);
\filldraw[fill=cancan, draw=black] (6.00, -2.50) rectangle (6.25, -2.75);
\filldraw[fill=cancan, draw=black] (6.25, -2.50) rectangle (6.50, -2.75);
\filldraw[fill=bermuda, draw=black] (6.50, -2.50) rectangle (6.75, -2.75);
\filldraw[fill=cancan, draw=black] (6.75, -2.50) rectangle (7.00, -2.75);
\filldraw[fill=cancan, draw=black] (7.00, -2.50) rectangle (7.25, -2.75);
\filldraw[fill=cancan, draw=black] (7.25, -2.50) rectangle (7.50, -2.75);
\filldraw[fill=bermuda, draw=black] (7.50, -2.50) rectangle (7.75, -2.75);
\filldraw[fill=bermuda, draw=black] (7.75, -2.50) rectangle (8.00, -2.75);
\filldraw[fill=bermuda, draw=black] (8.00, -2.50) rectangle (8.25, -2.75);
\filldraw[fill=bermuda, draw=black] (8.25, -2.50) rectangle (8.50, -2.75);
\filldraw[fill=bermuda, draw=black] (8.50, -2.50) rectangle (8.75, -2.75);
\filldraw[fill=cancan, draw=black] (8.75, -2.50) rectangle (9.00, -2.75);
\filldraw[fill=cancan, draw=black] (9.00, -2.50) rectangle (9.25, -2.75);
\filldraw[fill=cancan, draw=black] (9.25, -2.50) rectangle (9.50, -2.75);
\filldraw[fill=bermuda, draw=black] (9.50, -2.50) rectangle (9.75, -2.75);
\filldraw[fill=bermuda, draw=black] (9.75, -2.50) rectangle (10.00, -2.75);
\filldraw[fill=bermuda, draw=black] (10.00, -2.50) rectangle (10.25, -2.75);
\filldraw[fill=cancan, draw=black] (10.25, -2.50) rectangle (10.50, -2.75);
\filldraw[fill=bermuda, draw=black] (10.50, -2.50) rectangle (10.75, -2.75);
\filldraw[fill=cancan, draw=black] (10.75, -2.50) rectangle (11.00, -2.75);
\filldraw[fill=bermuda, draw=black] (11.00, -2.50) rectangle (11.25, -2.75);
\filldraw[fill=bermuda, draw=black] (11.25, -2.50) rectangle (11.50, -2.75);
\filldraw[fill=bermuda, draw=black] (11.50, -2.50) rectangle (11.75, -2.75);
\filldraw[fill=cancan, draw=black] (11.75, -2.50) rectangle (12.00, -2.75);
\filldraw[fill=cancan, draw=black] (12.00, -2.50) rectangle (12.25, -2.75);
\filldraw[fill=cancan, draw=black] (12.25, -2.50) rectangle (12.50, -2.75);
\filldraw[fill=cancan, draw=black] (12.50, -2.50) rectangle (12.75, -2.75);
\filldraw[fill=bermuda, draw=black] (12.75, -2.50) rectangle (13.00, -2.75);
\filldraw[fill=bermuda, draw=black] (13.00, -2.50) rectangle (13.25, -2.75);
\filldraw[fill=bermuda, draw=black] (13.25, -2.50) rectangle (13.50, -2.75);
\filldraw[fill=bermuda, draw=black] (13.50, -2.50) rectangle (13.75, -2.75);
\filldraw[fill=cancan, draw=black] (13.75, -2.50) rectangle (14.00, -2.75);
\filldraw[fill=bermuda, draw=black] (14.00, -2.50) rectangle (14.25, -2.75);
\filldraw[fill=bermuda, draw=black] (14.25, -2.50) rectangle (14.50, -2.75);
\filldraw[fill=bermuda, draw=black] (14.50, -2.50) rectangle (14.75, -2.75);
\filldraw[fill=bermuda, draw=black] (14.75, -2.50) rectangle (15.00, -2.75);
\filldraw[fill=bermuda, draw=black] (0.00, -2.75) rectangle (0.25, -3.00);
\filldraw[fill=bermuda, draw=black] (0.25, -2.75) rectangle (0.50, -3.00);
\filldraw[fill=bermuda, draw=black] (0.50, -2.75) rectangle (0.75, -3.00);
\filldraw[fill=cancan, draw=black] (0.75, -2.75) rectangle (1.00, -3.00);
\filldraw[fill=cancan, draw=black] (1.00, -2.75) rectangle (1.25, -3.00);
\filldraw[fill=bermuda, draw=black] (1.25, -2.75) rectangle (1.50, -3.00);
\filldraw[fill=bermuda, draw=black] (1.50, -2.75) rectangle (1.75, -3.00);
\filldraw[fill=cancan, draw=black] (1.75, -2.75) rectangle (2.00, -3.00);
\filldraw[fill=cancan, draw=black] (2.00, -2.75) rectangle (2.25, -3.00);
\filldraw[fill=cancan, draw=black] (2.25, -2.75) rectangle (2.50, -3.00);
\filldraw[fill=cancan, draw=black] (2.50, -2.75) rectangle (2.75, -3.00);
\filldraw[fill=cancan, draw=black] (2.75, -2.75) rectangle (3.00, -3.00);
\filldraw[fill=bermuda, draw=black] (3.00, -2.75) rectangle (3.25, -3.00);
\filldraw[fill=cancan, draw=black] (3.25, -2.75) rectangle (3.50, -3.00);
\filldraw[fill=cancan, draw=black] (3.50, -2.75) rectangle (3.75, -3.00);
\filldraw[fill=cancan, draw=black] (3.75, -2.75) rectangle (4.00, -3.00);
\filldraw[fill=bermuda, draw=black] (4.00, -2.75) rectangle (4.25, -3.00);
\filldraw[fill=bermuda, draw=black] (4.25, -2.75) rectangle (4.50, -3.00);
\filldraw[fill=bermuda, draw=black] (4.50, -2.75) rectangle (4.75, -3.00);
\filldraw[fill=cancan, draw=black] (4.75, -2.75) rectangle (5.00, -3.00);
\filldraw[fill=cancan, draw=black] (5.00, -2.75) rectangle (5.25, -3.00);
\filldraw[fill=cancan, draw=black] (5.25, -2.75) rectangle (5.50, -3.00);
\filldraw[fill=bermuda, draw=black] (5.50, -2.75) rectangle (5.75, -3.00);
\filldraw[fill=bermuda, draw=black] (5.75, -2.75) rectangle (6.00, -3.00);
\filldraw[fill=bermuda, draw=black] (6.00, -2.75) rectangle (6.25, -3.00);
\filldraw[fill=cancan, draw=black] (6.25, -2.75) rectangle (6.50, -3.00);
\filldraw[fill=bermuda, draw=black] (6.50, -2.75) rectangle (6.75, -3.00);
\filldraw[fill=cancan, draw=black] (6.75, -2.75) rectangle (7.00, -3.00);
\filldraw[fill=bermuda, draw=black] (7.00, -2.75) rectangle (7.25, -3.00);
\filldraw[fill=bermuda, draw=black] (7.25, -2.75) rectangle (7.50, -3.00);
\filldraw[fill=bermuda, draw=black] (7.50, -2.75) rectangle (7.75, -3.00);
\filldraw[fill=cancan, draw=black] (7.75, -2.75) rectangle (8.00, -3.00);
\filldraw[fill=cancan, draw=black] (8.00, -2.75) rectangle (8.25, -3.00);
\filldraw[fill=cancan, draw=black] (8.25, -2.75) rectangle (8.50, -3.00);
\filldraw[fill=bermuda, draw=black] (8.50, -2.75) rectangle (8.75, -3.00);
\filldraw[fill=bermuda, draw=black] (8.75, -2.75) rectangle (9.00, -3.00);
\filldraw[fill=bermuda, draw=black] (9.00, -2.75) rectangle (9.25, -3.00);
\filldraw[fill=cancan, draw=black] (9.25, -2.75) rectangle (9.50, -3.00);
\filldraw[fill=cancan, draw=black] (9.50, -2.75) rectangle (9.75, -3.00);
\filldraw[fill=cancan, draw=black] (9.75, -2.75) rectangle (10.00, -3.00);
\filldraw[fill=cancan, draw=black] (10.00, -2.75) rectangle (10.25, -3.00);
\filldraw[fill=cancan, draw=black] (10.25, -2.75) rectangle (10.50, -3.00);
\filldraw[fill=bermuda, draw=black] (10.50, -2.75) rectangle (10.75, -3.00);
\filldraw[fill=bermuda, draw=black] (10.75, -2.75) rectangle (11.00, -3.00);
\filldraw[fill=bermuda, draw=black] (11.00, -2.75) rectangle (11.25, -3.00);
\filldraw[fill=bermuda, draw=black] (11.25, -2.75) rectangle (11.50, -3.00);
\filldraw[fill=bermuda, draw=black] (11.50, -2.75) rectangle (11.75, -3.00);
\filldraw[fill=cancan, draw=black] (11.75, -2.75) rectangle (12.00, -3.00);
\filldraw[fill=bermuda, draw=black] (12.00, -2.75) rectangle (12.25, -3.00);
\filldraw[fill=cancan, draw=black] (12.25, -2.75) rectangle (12.50, -3.00);
\filldraw[fill=cancan, draw=black] (12.50, -2.75) rectangle (12.75, -3.00);
\filldraw[fill=cancan, draw=black] (12.75, -2.75) rectangle (13.00, -3.00);
\filldraw[fill=cancan, draw=black] (13.00, -2.75) rectangle (13.25, -3.00);
\filldraw[fill=bermuda, draw=black] (13.25, -2.75) rectangle (13.50, -3.00);
\filldraw[fill=bermuda, draw=black] (13.50, -2.75) rectangle (13.75, -3.00);
\filldraw[fill=cancan, draw=black] (13.75, -2.75) rectangle (14.00, -3.00);
\filldraw[fill=cancan, draw=black] (14.00, -2.75) rectangle (14.25, -3.00);
\filldraw[fill=cancan, draw=black] (14.25, -2.75) rectangle (14.50, -3.00);
\filldraw[fill=bermuda, draw=black] (14.50, -2.75) rectangle (14.75, -3.00);
\filldraw[fill=bermuda, draw=black] (14.75, -2.75) rectangle (15.00, -3.00);
\filldraw[fill=bermuda, draw=black] (0.00, -3.00) rectangle (0.25, -3.25);
\filldraw[fill=cancan, draw=black] (0.25, -3.00) rectangle (0.50, -3.25);
\filldraw[fill=cancan, draw=black] (0.50, -3.00) rectangle (0.75, -3.25);
\filldraw[fill=bermuda, draw=black] (0.75, -3.00) rectangle (1.00, -3.25);
\filldraw[fill=bermuda, draw=black] (1.00, -3.00) rectangle (1.25, -3.25);
\filldraw[fill=cancan, draw=black] (1.25, -3.00) rectangle (1.50, -3.25);
\filldraw[fill=cancan, draw=black] (1.50, -3.00) rectangle (1.75, -3.25);
\filldraw[fill=cancan, draw=black] (1.75, -3.00) rectangle (2.00, -3.25);
\filldraw[fill=bermuda, draw=black] (2.00, -3.00) rectangle (2.25, -3.25);
\filldraw[fill=bermuda, draw=black] (2.25, -3.00) rectangle (2.50, -3.25);
\filldraw[fill=bermuda, draw=black] (2.50, -3.00) rectangle (2.75, -3.25);
\filldraw[fill=cancan, draw=black] (2.75, -3.00) rectangle (3.00, -3.25);
\filldraw[fill=cancan, draw=black] (3.00, -3.00) rectangle (3.25, -3.25);
\filldraw[fill=cancan, draw=black] (3.25, -3.00) rectangle (3.50, -3.25);
\filldraw[fill=bermuda, draw=black] (3.50, -3.00) rectangle (3.75, -3.25);
\filldraw[fill=bermuda, draw=black] (3.75, -3.00) rectangle (4.00, -3.25);
\filldraw[fill=bermuda, draw=black] (4.00, -3.00) rectangle (4.25, -3.25);
\filldraw[fill=cancan, draw=black] (4.25, -3.00) rectangle (4.50, -3.25);
\filldraw[fill=cancan, draw=black] (4.50, -3.00) rectangle (4.75, -3.25);
\filldraw[fill=cancan, draw=black] (4.75, -3.00) rectangle (5.00, -3.25);
\filldraw[fill=bermuda, draw=black] (5.00, -3.00) rectangle (5.25, -3.25);
\filldraw[fill=bermuda, draw=black] (5.25, -3.00) rectangle (5.50, -3.25);
\filldraw[fill=bermuda, draw=black] (5.50, -3.00) rectangle (5.75, -3.25);
\filldraw[fill=cancan, draw=black] (5.75, -3.00) rectangle (6.00, -3.25);
\filldraw[fill=cancan, draw=black] (6.00, -3.00) rectangle (6.25, -3.25);
\filldraw[fill=bermuda, draw=black] (6.25, -3.00) rectangle (6.50, -3.25);
\filldraw[fill=bermuda, draw=black] (6.50, -3.00) rectangle (6.75, -3.25);
\filldraw[fill=cancan, draw=black] (6.75, -3.00) rectangle (7.00, -3.25);
\filldraw[fill=cancan, draw=black] (7.00, -3.00) rectangle (7.25, -3.25);
\filldraw[fill=cancan, draw=black] (7.25, -3.00) rectangle (7.50, -3.25);
\filldraw[fill=cancan, draw=black] (7.50, -3.00) rectangle (7.75, -3.25);
\filldraw[fill=cancan, draw=black] (7.75, -3.00) rectangle (8.00, -3.25);
\filldraw[fill=bermuda, draw=black] (8.00, -3.00) rectangle (8.25, -3.25);
\filldraw[fill=cancan, draw=black] (8.25, -3.00) rectangle (8.50, -3.25);
\filldraw[fill=cancan, draw=black] (8.50, -3.00) rectangle (8.75, -3.25);
\filldraw[fill=cancan, draw=black] (8.75, -3.00) rectangle (9.00, -3.25);
\filldraw[fill=bermuda, draw=black] (9.00, -3.00) rectangle (9.25, -3.25);
\filldraw[fill=bermuda, draw=black] (9.25, -3.00) rectangle (9.50, -3.25);
\filldraw[fill=bermuda, draw=black] (9.50, -3.00) rectangle (9.75, -3.25);
\filldraw[fill=cancan, draw=black] (9.75, -3.00) rectangle (10.00, -3.25);
\filldraw[fill=cancan, draw=black] (10.00, -3.00) rectangle (10.25, -3.25);
\filldraw[fill=cancan, draw=black] (10.25, -3.00) rectangle (10.50, -3.25);
\filldraw[fill=cancan, draw=black] (10.50, -3.00) rectangle (10.75, -3.25);
\filldraw[fill=bermuda, draw=black] (10.75, -3.00) rectangle (11.00, -3.25);
\filldraw[fill=bermuda, draw=black] (11.00, -3.00) rectangle (11.25, -3.25);
\filldraw[fill=cancan, draw=black] (11.25, -3.00) rectangle (11.50, -3.25);
\filldraw[fill=bermuda, draw=black] (11.50, -3.00) rectangle (11.75, -3.25);
\filldraw[fill=cancan, draw=black] (11.75, -3.00) rectangle (12.00, -3.25);
\filldraw[fill=cancan, draw=black] (12.00, -3.00) rectangle (12.25, -3.25);
\filldraw[fill=bermuda, draw=black] (12.25, -3.00) rectangle (12.50, -3.25);
\filldraw[fill=bermuda, draw=black] (12.50, -3.00) rectangle (12.75, -3.25);
\filldraw[fill=cancan, draw=black] (12.75, -3.00) rectangle (13.00, -3.25);
\filldraw[fill=bermuda, draw=black] (13.00, -3.00) rectangle (13.25, -3.25);
\filldraw[fill=cancan, draw=black] (13.25, -3.00) rectangle (13.50, -3.25);
\filldraw[fill=bermuda, draw=black] (13.50, -3.00) rectangle (13.75, -3.25);
\filldraw[fill=bermuda, draw=black] (13.75, -3.00) rectangle (14.00, -3.25);
\filldraw[fill=bermuda, draw=black] (14.00, -3.00) rectangle (14.25, -3.25);
\filldraw[fill=cancan, draw=black] (14.25, -3.00) rectangle (14.50, -3.25);
\filldraw[fill=cancan, draw=black] (14.50, -3.00) rectangle (14.75, -3.25);
\filldraw[fill=cancan, draw=black] (14.75, -3.00) rectangle (15.00, -3.25);
\filldraw[fill=cancan, draw=black] (0.00, -3.25) rectangle (0.25, -3.50);
\filldraw[fill=cancan, draw=black] (0.25, -3.25) rectangle (0.50, -3.50);
\filldraw[fill=cancan, draw=black] (0.50, -3.25) rectangle (0.75, -3.50);
\filldraw[fill=cancan, draw=black] (0.75, -3.25) rectangle (1.00, -3.50);
\filldraw[fill=bermuda, draw=black] (1.00, -3.25) rectangle (1.25, -3.50);
\filldraw[fill=bermuda, draw=black] (1.25, -3.25) rectangle (1.50, -3.50);
\filldraw[fill=bermuda, draw=black] (1.50, -3.25) rectangle (1.75, -3.50);
\filldraw[fill=cancan, draw=black] (1.75, -3.25) rectangle (2.00, -3.50);
\filldraw[fill=cancan, draw=black] (2.00, -3.25) rectangle (2.25, -3.50);
\filldraw[fill=cancan, draw=black] (2.25, -3.25) rectangle (2.50, -3.50);
\filldraw[fill=bermuda, draw=black] (2.50, -3.25) rectangle (2.75, -3.50);
\filldraw[fill=bermuda, draw=black] (2.75, -3.25) rectangle (3.00, -3.50);
\filldraw[fill=bermuda, draw=black] (3.00, -3.25) rectangle (3.25, -3.50);
\filldraw[fill=cancan, draw=black] (3.25, -3.25) rectangle (3.50, -3.50);
\filldraw[fill=cancan, draw=black] (3.50, -3.25) rectangle (3.75, -3.50);
\filldraw[fill=cancan, draw=black] (3.75, -3.25) rectangle (4.00, -3.50);
\filldraw[fill=bermuda, draw=black] (4.00, -3.25) rectangle (4.25, -3.50);
\filldraw[fill=bermuda, draw=black] (4.25, -3.25) rectangle (4.50, -3.50);
\filldraw[fill=bermuda, draw=black] (4.50, -3.25) rectangle (4.75, -3.50);
} } }\end{equation*}
\begin{equation*}
\hspace{4.6pt} b_{9} = \vcenter{\hbox{ \tikz{
\filldraw[fill=bermuda, draw=black] (0.00, 0.00) rectangle (0.25, -0.25);
\filldraw[fill=bermuda, draw=black] (0.25, 0.00) rectangle (0.50, -0.25);
\filldraw[fill=cancan, draw=black] (0.50, 0.00) rectangle (0.75, -0.25);
\filldraw[fill=bermuda, draw=black] (0.75, 0.00) rectangle (1.00, -0.25);
\filldraw[fill=bermuda, draw=black] (1.00, 0.00) rectangle (1.25, -0.25);
\filldraw[fill=bermuda, draw=black] (1.25, 0.00) rectangle (1.50, -0.25);
\filldraw[fill=cancan, draw=black] (1.50, 0.00) rectangle (1.75, -0.25);
\filldraw[fill=cancan, draw=black] (1.75, 0.00) rectangle (2.00, -0.25);
\filldraw[fill=cancan, draw=black] (2.00, 0.00) rectangle (2.25, -0.25);
\filldraw[fill=cancan, draw=black] (2.25, 0.00) rectangle (2.50, -0.25);
\filldraw[fill=cancan, draw=black] (2.50, 0.00) rectangle (2.75, -0.25);
\filldraw[fill=bermuda, draw=black] (2.75, 0.00) rectangle (3.00, -0.25);
\filldraw[fill=cancan, draw=black] (3.00, 0.00) rectangle (3.25, -0.25);
\filldraw[fill=cancan, draw=black] (3.25, 0.00) rectangle (3.50, -0.25);
\filldraw[fill=bermuda, draw=black] (3.50, 0.00) rectangle (3.75, -0.25);
\filldraw[fill=bermuda, draw=black] (3.75, 0.00) rectangle (4.00, -0.25);
\filldraw[fill=cancan, draw=black] (4.00, 0.00) rectangle (4.25, -0.25);
\filldraw[fill=bermuda, draw=black] (4.25, 0.00) rectangle (4.50, -0.25);
\filldraw[fill=cancan, draw=black] (4.50, 0.00) rectangle (4.75, -0.25);
\filldraw[fill=bermuda, draw=black] (4.75, 0.00) rectangle (5.00, -0.25);
\filldraw[fill=cancan, draw=black] (5.00, 0.00) rectangle (5.25, -0.25);
\filldraw[fill=cancan, draw=black] (5.25, 0.00) rectangle (5.50, -0.25);
\filldraw[fill=cancan, draw=black] (5.50, 0.00) rectangle (5.75, -0.25);
\filldraw[fill=cancan, draw=black] (5.75, 0.00) rectangle (6.00, -0.25);
\filldraw[fill=cancan, draw=black] (6.00, 0.00) rectangle (6.25, -0.25);
\filldraw[fill=cancan, draw=black] (6.25, 0.00) rectangle (6.50, -0.25);
\filldraw[fill=cancan, draw=black] (6.50, 0.00) rectangle (6.75, -0.25);
\filldraw[fill=cancan, draw=black] (6.75, 0.00) rectangle (7.00, -0.25);
\filldraw[fill=cancan, draw=black] (7.00, 0.00) rectangle (7.25, -0.25);
\filldraw[fill=bermuda, draw=black] (7.25, 0.00) rectangle (7.50, -0.25);
\filldraw[fill=bermuda, draw=black] (7.50, 0.00) rectangle (7.75, -0.25);
\filldraw[fill=bermuda, draw=black] (7.75, 0.00) rectangle (8.00, -0.25);
\filldraw[fill=cancan, draw=black] (8.00, 0.00) rectangle (8.25, -0.25);
\filldraw[fill=cancan, draw=black] (8.25, 0.00) rectangle (8.50, -0.25);
\filldraw[fill=cancan, draw=black] (8.50, 0.00) rectangle (8.75, -0.25);
\filldraw[fill=cancan, draw=black] (8.75, 0.00) rectangle (9.00, -0.25);
\filldraw[fill=cancan, draw=black] (9.00, 0.00) rectangle (9.25, -0.25);
\filldraw[fill=bermuda, draw=black] (9.25, 0.00) rectangle (9.50, -0.25);
\filldraw[fill=cancan, draw=black] (9.50, 0.00) rectangle (9.75, -0.25);
\filldraw[fill=bermuda, draw=black] (9.75, 0.00) rectangle (10.00, -0.25);
\filldraw[fill=cancan, draw=black] (10.00, 0.00) rectangle (10.25, -0.25);
\filldraw[fill=cancan, draw=black] (10.25, 0.00) rectangle (10.50, -0.25);
\filldraw[fill=cancan, draw=black] (10.50, 0.00) rectangle (10.75, -0.25);
\filldraw[fill=bermuda, draw=black] (10.75, 0.00) rectangle (11.00, -0.25);
\filldraw[fill=cancan, draw=black] (11.00, 0.00) rectangle (11.25, -0.25);
\filldraw[fill=bermuda, draw=black] (11.25, 0.00) rectangle (11.50, -0.25);
\filldraw[fill=cancan, draw=black] (11.50, 0.00) rectangle (11.75, -0.25);
\filldraw[fill=bermuda, draw=black] (11.75, 0.00) rectangle (12.00, -0.25);
\filldraw[fill=bermuda, draw=black] (12.00, 0.00) rectangle (12.25, -0.25);
\filldraw[fill=bermuda, draw=black] (12.25, 0.00) rectangle (12.50, -0.25);
\filldraw[fill=cancan, draw=black] (12.50, 0.00) rectangle (12.75, -0.25);
\filldraw[fill=cancan, draw=black] (12.75, 0.00) rectangle (13.00, -0.25);
\filldraw[fill=cancan, draw=black] (13.00, 0.00) rectangle (13.25, -0.25);
\filldraw[fill=bermuda, draw=black] (13.25, 0.00) rectangle (13.50, -0.25);
\filldraw[fill=bermuda, draw=black] (13.50, 0.00) rectangle (13.75, -0.25);
\filldraw[fill=bermuda, draw=black] (13.75, 0.00) rectangle (14.00, -0.25);
\filldraw[fill=cancan, draw=black] (14.00, 0.00) rectangle (14.25, -0.25);
\filldraw[fill=cancan, draw=black] (14.25, 0.00) rectangle (14.50, -0.25);
\filldraw[fill=cancan, draw=black] (14.50, 0.00) rectangle (14.75, -0.25);
\filldraw[fill=bermuda, draw=black] (14.75, 0.00) rectangle (15.00, -0.25);
\filldraw[fill=bermuda, draw=black] (0.00, -0.25) rectangle (0.25, -0.50);
\filldraw[fill=bermuda, draw=black] (0.25, -0.25) rectangle (0.50, -0.50);
\filldraw[fill=cancan, draw=black] (0.50, -0.25) rectangle (0.75, -0.50);
\filldraw[fill=cancan, draw=black] (0.75, -0.25) rectangle (1.00, -0.50);
\filldraw[fill=cancan, draw=black] (1.00, -0.25) rectangle (1.25, -0.50);
\filldraw[fill=bermuda, draw=black] (1.25, -0.25) rectangle (1.50, -0.50);
\filldraw[fill=bermuda, draw=black] (1.50, -0.25) rectangle (1.75, -0.50);
\filldraw[fill=bermuda, draw=black] (1.75, -0.25) rectangle (2.00, -0.50);
\filldraw[fill=cancan, draw=black] (2.00, -0.25) rectangle (2.25, -0.50);
\filldraw[fill=cancan, draw=black] (2.25, -0.25) rectangle (2.50, -0.50);
\filldraw[fill=cancan, draw=black] (2.50, -0.25) rectangle (2.75, -0.50);
\filldraw[fill=cancan, draw=black] (2.75, -0.25) rectangle (3.00, -0.50);
\filldraw[fill=bermuda, draw=black] (3.00, -0.25) rectangle (3.25, -0.50);
\filldraw[fill=bermuda, draw=black] (3.25, -0.25) rectangle (3.50, -0.50);
\filldraw[fill=cancan, draw=black] (3.50, -0.25) rectangle (3.75, -0.50);
\filldraw[fill=cancan, draw=black] (3.75, -0.25) rectangle (4.00, -0.50);
\filldraw[fill=cancan, draw=black] (4.00, -0.25) rectangle (4.25, -0.50);
\filldraw[fill=bermuda, draw=black] (4.25, -0.25) rectangle (4.50, -0.50);
\filldraw[fill=bermuda, draw=black] (4.50, -0.25) rectangle (4.75, -0.50);
\filldraw[fill=bermuda, draw=black] (4.75, -0.25) rectangle (5.00, -0.50);
\filldraw[fill=cancan, draw=black] (5.00, -0.25) rectangle (5.25, -0.50);
\filldraw[fill=bermuda, draw=black] (5.25, -0.25) rectangle (5.50, -0.50);
\filldraw[fill=cancan, draw=black] (5.50, -0.25) rectangle (5.75, -0.50);
\filldraw[fill=bermuda, draw=black] (5.75, -0.25) rectangle (6.00, -0.50);
\filldraw[fill=bermuda, draw=black] (6.00, -0.25) rectangle (6.25, -0.50);
\filldraw[fill=bermuda, draw=black] (6.25, -0.25) rectangle (6.50, -0.50);
\filldraw[fill=cancan, draw=black] (6.50, -0.25) rectangle (6.75, -0.50);
\filldraw[fill=cancan, draw=black] (6.75, -0.25) rectangle (7.00, -0.50);
\filldraw[fill=cancan, draw=black] (7.00, -0.25) rectangle (7.25, -0.50);
\filldraw[fill=bermuda, draw=black] (7.25, -0.25) rectangle (7.50, -0.50);
\filldraw[fill=bermuda, draw=black] (7.50, -0.25) rectangle (7.75, -0.50);
\filldraw[fill=bermuda, draw=black] (7.75, -0.25) rectangle (8.00, -0.50);
\filldraw[fill=cancan, draw=black] (8.00, -0.25) rectangle (8.25, -0.50);
\filldraw[fill=cancan, draw=black] (8.25, -0.25) rectangle (8.50, -0.50);
\filldraw[fill=cancan, draw=black] (8.50, -0.25) rectangle (8.75, -0.50);
\filldraw[fill=cancan, draw=black] (8.75, -0.25) rectangle (9.00, -0.50);
\filldraw[fill=cancan, draw=black] (9.00, -0.25) rectangle (9.25, -0.50);
\filldraw[fill=bermuda, draw=black] (9.25, -0.25) rectangle (9.50, -0.50);
\filldraw[fill=cancan, draw=black] (9.50, -0.25) rectangle (9.75, -0.50);
\filldraw[fill=cancan, draw=black] (9.75, -0.25) rectangle (10.00, -0.50);
\filldraw[fill=cancan, draw=black] (10.00, -0.25) rectangle (10.25, -0.50);
\filldraw[fill=cancan, draw=black] (10.25, -0.25) rectangle (10.50, -0.50);
\filldraw[fill=cancan, draw=black] (10.50, -0.25) rectangle (10.75, -0.50);
\filldraw[fill=bermuda, draw=black] (10.75, -0.25) rectangle (11.00, -0.50);
\filldraw[fill=bermuda, draw=black] (11.00, -0.25) rectangle (11.25, -0.50);
\filldraw[fill=bermuda, draw=black] (11.25, -0.25) rectangle (11.50, -0.50);
\filldraw[fill=cancan, draw=black] (11.50, -0.25) rectangle (11.75, -0.50);
\filldraw[fill=cancan, draw=black] (11.75, -0.25) rectangle (12.00, -0.50);
\filldraw[fill=cancan, draw=black] (12.00, -0.25) rectangle (12.25, -0.50);
\filldraw[fill=bermuda, draw=black] (12.25, -0.25) rectangle (12.50, -0.50);
\filldraw[fill=bermuda, draw=black] (12.50, -0.25) rectangle (12.75, -0.50);
\filldraw[fill=bermuda, draw=black] (12.75, -0.25) rectangle (13.00, -0.50);
\filldraw[fill=cancan, draw=black] (13.00, -0.25) rectangle (13.25, -0.50);
\filldraw[fill=cancan, draw=black] (13.25, -0.25) rectangle (13.50, -0.50);
\filldraw[fill=cancan, draw=black] (13.50, -0.25) rectangle (13.75, -0.50);
\filldraw[fill=bermuda, draw=black] (13.75, -0.25) rectangle (14.00, -0.50);
\filldraw[fill=bermuda, draw=black] (14.00, -0.25) rectangle (14.25, -0.50);
\filldraw[fill=bermuda, draw=black] (14.25, -0.25) rectangle (14.50, -0.50);
\filldraw[fill=bermuda, draw=black] (14.50, -0.25) rectangle (14.75, -0.50);
\filldraw[fill=bermuda, draw=black] (14.75, -0.25) rectangle (15.00, -0.50);
\filldraw[fill=cancan, draw=black] (0.00, -0.50) rectangle (0.25, -0.75);
\filldraw[fill=bermuda, draw=black] (0.25, -0.50) rectangle (0.50, -0.75);
\filldraw[fill=bermuda, draw=black] (0.50, -0.50) rectangle (0.75, -0.75);
\filldraw[fill=bermuda, draw=black] (0.75, -0.50) rectangle (1.00, -0.75);
\filldraw[fill=cancan, draw=black] (1.00, -0.50) rectangle (1.25, -0.75);
\filldraw[fill=cancan, draw=black] (1.25, -0.50) rectangle (1.50, -0.75);
\filldraw[fill=cancan, draw=black] (1.50, -0.50) rectangle (1.75, -0.75);
\filldraw[fill=cancan, draw=black] (1.75, -0.50) rectangle (2.00, -0.75);
\filldraw[fill=cancan, draw=black] (2.00, -0.50) rectangle (2.25, -0.75);
\filldraw[fill=bermuda, draw=black] (2.25, -0.50) rectangle (2.50, -0.75);
\filldraw[fill=bermuda, draw=black] (2.50, -0.50) rectangle (2.75, -0.75);
\filldraw[fill=bermuda, draw=black] (2.75, -0.50) rectangle (3.00, -0.75);
\filldraw[fill=cancan, draw=black] (3.00, -0.50) rectangle (3.25, -0.75);
\filldraw[fill=cancan, draw=black] (3.25, -0.50) rectangle (3.50, -0.75);
\filldraw[fill=bermuda, draw=black] (3.50, -0.50) rectangle (3.75, -0.75);
\filldraw[fill=bermuda, draw=black] (3.75, -0.50) rectangle (4.00, -0.75);
\filldraw[fill=cancan, draw=black] (4.00, -0.50) rectangle (4.25, -0.75);
\filldraw[fill=cancan, draw=black] (4.25, -0.50) rectangle (4.50, -0.75);
\filldraw[fill=cancan, draw=black] (4.50, -0.50) rectangle (4.75, -0.75);
\filldraw[fill=bermuda, draw=black] (4.75, -0.50) rectangle (5.00, -0.75);
\filldraw[fill=bermuda, draw=black] (5.00, -0.50) rectangle (5.25, -0.75);
\filldraw[fill=bermuda, draw=black] (5.25, -0.50) rectangle (5.50, -0.75);
\filldraw[fill=cancan, draw=black] (5.50, -0.50) rectangle (5.75, -0.75);
\filldraw[fill=cancan, draw=black] (5.75, -0.50) rectangle (6.00, -0.75);
\filldraw[fill=cancan, draw=black] (6.00, -0.50) rectangle (6.25, -0.75);
\filldraw[fill=bermuda, draw=black] (6.25, -0.50) rectangle (6.50, -0.75);
\filldraw[fill=bermuda, draw=black] (6.50, -0.50) rectangle (6.75, -0.75);
\filldraw[fill=bermuda, draw=black] (6.75, -0.50) rectangle (7.00, -0.75);
\filldraw[fill=cancan, draw=black] (7.00, -0.50) rectangle (7.25, -0.75);
\filldraw[fill=bermuda, draw=black] (7.25, -0.50) rectangle (7.50, -0.75);
\filldraw[fill=cancan, draw=black] (7.50, -0.50) rectangle (7.75, -0.75);
\filldraw[fill=bermuda, draw=black] (7.75, -0.50) rectangle (8.00, -0.75);
\filldraw[fill=cancan, draw=black] (8.00, -0.50) rectangle (8.25, -0.75);
\filldraw[fill=cancan, draw=black] (8.25, -0.50) rectangle (8.50, -0.75);
\filldraw[fill=cancan, draw=black] (8.50, -0.50) rectangle (8.75, -0.75);
\filldraw[fill=bermuda, draw=black] (8.75, -0.50) rectangle (9.00, -0.75);
\filldraw[fill=cancan, draw=black] (9.00, -0.50) rectangle (9.25, -0.75);
\filldraw[fill=bermuda, draw=black] (9.25, -0.50) rectangle (9.50, -0.75);
\filldraw[fill=cancan, draw=black] (9.50, -0.50) rectangle (9.75, -0.75);
\filldraw[fill=cancan, draw=black] (9.75, -0.50) rectangle (10.00, -0.75);
\filldraw[fill=bermuda, draw=black] (10.00, -0.50) rectangle (10.25, -0.75);
\filldraw[fill=bermuda, draw=black] (10.25, -0.50) rectangle (10.50, -0.75);
\filldraw[fill=cancan, draw=black] (10.50, -0.50) rectangle (10.75, -0.75);
\filldraw[fill=bermuda, draw=black] (10.75, -0.50) rectangle (11.00, -0.75);
\filldraw[fill=bermuda, draw=black] (11.00, -0.50) rectangle (11.25, -0.75);
\filldraw[fill=bermuda, draw=black] (11.25, -0.50) rectangle (11.50, -0.75);
\filldraw[fill=cancan, draw=black] (11.50, -0.50) rectangle (11.75, -0.75);
\filldraw[fill=cancan, draw=black] (11.75, -0.50) rectangle (12.00, -0.75);
\filldraw[fill=cancan, draw=black] (12.00, -0.50) rectangle (12.25, -0.75);
\filldraw[fill=cancan, draw=black] (12.25, -0.50) rectangle (12.50, -0.75);
\filldraw[fill=cancan, draw=black] (12.50, -0.50) rectangle (12.75, -0.75);
\filldraw[fill=cancan, draw=black] (12.75, -0.50) rectangle (13.00, -0.75);
\filldraw[fill=cancan, draw=black] (13.00, -0.50) rectangle (13.25, -0.75);
\filldraw[fill=bermuda, draw=black] (13.25, -0.50) rectangle (13.50, -0.75);
\filldraw[fill=bermuda, draw=black] (13.50, -0.50) rectangle (13.75, -0.75);
\filldraw[fill=bermuda, draw=black] (13.75, -0.50) rectangle (14.00, -0.75);
\filldraw[fill=cancan, draw=black] (14.00, -0.50) rectangle (14.25, -0.75);
\filldraw[fill=cancan, draw=black] (14.25, -0.50) rectangle (14.50, -0.75);
\filldraw[fill=cancan, draw=black] (14.50, -0.50) rectangle (14.75, -0.75);
\filldraw[fill=bermuda, draw=black] (14.75, -0.50) rectangle (15.00, -0.75);
\filldraw[fill=bermuda, draw=black] (0.00, -0.75) rectangle (0.25, -1.00);
\filldraw[fill=bermuda, draw=black] (0.25, -0.75) rectangle (0.50, -1.00);
\filldraw[fill=cancan, draw=black] (0.50, -0.75) rectangle (0.75, -1.00);
\filldraw[fill=bermuda, draw=black] (0.75, -0.75) rectangle (1.00, -1.00);
\filldraw[fill=cancan, draw=black] (1.00, -0.75) rectangle (1.25, -1.00);
\filldraw[fill=bermuda, draw=black] (1.25, -0.75) rectangle (1.50, -1.00);
\filldraw[fill=bermuda, draw=black] (1.50, -0.75) rectangle (1.75, -1.00);
\filldraw[fill=bermuda, draw=black] (1.75, -0.75) rectangle (2.00, -1.00);
\filldraw[fill=cancan, draw=black] (2.00, -0.75) rectangle (2.25, -1.00);
\filldraw[fill=bermuda, draw=black] (2.25, -0.75) rectangle (2.50, -1.00);
\filldraw[fill=bermuda, draw=black] (2.50, -0.75) rectangle (2.75, -1.00);
\filldraw[fill=bermuda, draw=black] (2.75, -0.75) rectangle (3.00, -1.00);
\filldraw[fill=bermuda, draw=black] (3.00, -0.75) rectangle (3.25, -1.00);
\filldraw[fill=bermuda, draw=black] (3.25, -0.75) rectangle (3.50, -1.00);
\filldraw[fill=cancan, draw=black] (3.50, -0.75) rectangle (3.75, -1.00);
\filldraw[fill=bermuda, draw=black] (3.75, -0.75) rectangle (4.00, -1.00);
\filldraw[fill=bermuda, draw=black] (4.00, -0.75) rectangle (4.25, -1.00);
\filldraw[fill=bermuda, draw=black] (4.25, -0.75) rectangle (4.50, -1.00);
\filldraw[fill=cancan, draw=black] (4.50, -0.75) rectangle (4.75, -1.00);
\filldraw[fill=cancan, draw=black] (4.75, -0.75) rectangle (5.00, -1.00);
\filldraw[fill=cancan, draw=black] (5.00, -0.75) rectangle (5.25, -1.00);
\filldraw[fill=bermuda, draw=black] (5.25, -0.75) rectangle (5.50, -1.00);
\filldraw[fill=bermuda, draw=black] (5.50, -0.75) rectangle (5.75, -1.00);
\filldraw[fill=bermuda, draw=black] (5.75, -0.75) rectangle (6.00, -1.00);
\filldraw[fill=cancan, draw=black] (6.00, -0.75) rectangle (6.25, -1.00);
\filldraw[fill=bermuda, draw=black] (6.25, -0.75) rectangle (6.50, -1.00);
\filldraw[fill=bermuda, draw=black] (6.50, -0.75) rectangle (6.75, -1.00);
\filldraw[fill=bermuda, draw=black] (6.75, -0.75) rectangle (7.00, -1.00);
\filldraw[fill=cancan, draw=black] (7.00, -0.75) rectangle (7.25, -1.00);
\filldraw[fill=cancan, draw=black] (7.25, -0.75) rectangle (7.50, -1.00);
\filldraw[fill=cancan, draw=black] (7.50, -0.75) rectangle (7.75, -1.00);
\filldraw[fill=cancan, draw=black] (7.75, -0.75) rectangle (8.00, -1.00);
\filldraw[fill=cancan, draw=black] (8.00, -0.75) rectangle (8.25, -1.00);
\filldraw[fill=bermuda, draw=black] (8.25, -0.75) rectangle (8.50, -1.00);
\filldraw[fill=cancan, draw=black] (8.50, -0.75) rectangle (8.75, -1.00);
\filldraw[fill=cancan, draw=black] (8.75, -0.75) rectangle (9.00, -1.00);
\filldraw[fill=cancan, draw=black] (9.00, -0.75) rectangle (9.25, -1.00);
\filldraw[fill=cancan, draw=black] (9.25, -0.75) rectangle (9.50, -1.00);
\filldraw[fill=cancan, draw=black] (9.50, -0.75) rectangle (9.75, -1.00);
\filldraw[fill=bermuda, draw=black] (9.75, -0.75) rectangle (10.00, -1.00);
\filldraw[fill=bermuda, draw=black] (10.00, -0.75) rectangle (10.25, -1.00);
\filldraw[fill=bermuda, draw=black] (10.25, -0.75) rectangle (10.50, -1.00);
\filldraw[fill=cancan, draw=black] (10.50, -0.75) rectangle (10.75, -1.00);
\filldraw[fill=bermuda, draw=black] (10.75, -0.75) rectangle (11.00, -1.00);
\filldraw[fill=bermuda, draw=black] (11.00, -0.75) rectangle (11.25, -1.00);
\filldraw[fill=bermuda, draw=black] (11.25, -0.75) rectangle (11.50, -1.00);
\filldraw[fill=bermuda, draw=black] (11.50, -0.75) rectangle (11.75, -1.00);
\filldraw[fill=bermuda, draw=black] (11.75, -0.75) rectangle (12.00, -1.00);
\filldraw[fill=cancan, draw=black] (12.00, -0.75) rectangle (12.25, -1.00);
\filldraw[fill=bermuda, draw=black] (12.25, -0.75) rectangle (12.50, -1.00);
\filldraw[fill=cancan, draw=black] (12.50, -0.75) rectangle (12.75, -1.00);
\filldraw[fill=cancan, draw=black] (12.75, -0.75) rectangle (13.00, -1.00);
\filldraw[fill=bermuda, draw=black] (13.00, -0.75) rectangle (13.25, -1.00);
\filldraw[fill=bermuda, draw=black] (13.25, -0.75) rectangle (13.50, -1.00);
\filldraw[fill=cancan, draw=black] (13.50, -0.75) rectangle (13.75, -1.00);
\filldraw[fill=bermuda, draw=black] (13.75, -0.75) rectangle (14.00, -1.00);
\filldraw[fill=cancan, draw=black] (14.00, -0.75) rectangle (14.25, -1.00);
\filldraw[fill=cancan, draw=black] (14.25, -0.75) rectangle (14.50, -1.00);
\filldraw[fill=bermuda, draw=black] (14.50, -0.75) rectangle (14.75, -1.00);
\filldraw[fill=bermuda, draw=black] (14.75, -0.75) rectangle (15.00, -1.00);
\filldraw[fill=cancan, draw=black] (0.00, -1.00) rectangle (0.25, -1.25);
\filldraw[fill=bermuda, draw=black] (0.25, -1.00) rectangle (0.50, -1.25);
\filldraw[fill=bermuda, draw=black] (0.50, -1.00) rectangle (0.75, -1.25);
\filldraw[fill=bermuda, draw=black] (0.75, -1.00) rectangle (1.00, -1.25);
\filldraw[fill=cancan, draw=black] (1.00, -1.00) rectangle (1.25, -1.25);
\filldraw[fill=cancan, draw=black] (1.25, -1.00) rectangle (1.50, -1.25);
\filldraw[fill=cancan, draw=black] (1.50, -1.00) rectangle (1.75, -1.25);
\filldraw[fill=cancan, draw=black] (1.75, -1.00) rectangle (2.00, -1.25);
\filldraw[fill=cancan, draw=black] (2.00, -1.00) rectangle (2.25, -1.25);
\filldraw[fill=cancan, draw=black] (2.25, -1.00) rectangle (2.50, -1.25);
\filldraw[fill=bermuda, draw=black] (2.50, -1.00) rectangle (2.75, -1.25);
\filldraw[fill=bermuda, draw=black] (2.75, -1.00) rectangle (3.00, -1.25);
\filldraw[fill=cancan, draw=black] (3.00, -1.00) rectangle (3.25, -1.25);
\filldraw[fill=bermuda, draw=black] (3.25, -1.00) rectangle (3.50, -1.25);
\filldraw[fill=bermuda, draw=black] (3.50, -1.00) rectangle (3.75, -1.25);
\filldraw[fill=bermuda, draw=black] (3.75, -1.00) rectangle (4.00, -1.25);
\filldraw[fill=cancan, draw=black] (4.00, -1.00) rectangle (4.25, -1.25);
\filldraw[fill=cancan, draw=black] (4.25, -1.00) rectangle (4.50, -1.25);
\filldraw[fill=cancan, draw=black] (4.50, -1.00) rectangle (4.75, -1.25);
\filldraw[fill=bermuda, draw=black] (4.75, -1.00) rectangle (5.00, -1.25);
\filldraw[fill=bermuda, draw=black] (5.00, -1.00) rectangle (5.25, -1.25);
\filldraw[fill=bermuda, draw=black] (5.25, -1.00) rectangle (5.50, -1.25);
\filldraw[fill=cancan, draw=black] (5.50, -1.00) rectangle (5.75, -1.25);
\filldraw[fill=cancan, draw=black] (5.75, -1.00) rectangle (6.00, -1.25);
\filldraw[fill=cancan, draw=black] (6.00, -1.00) rectangle (6.25, -1.25);
\filldraw[fill=bermuda, draw=black] (6.25, -1.00) rectangle (6.50, -1.25);
\filldraw[fill=bermuda, draw=black] (6.50, -1.00) rectangle (6.75, -1.25);
\filldraw[fill=bermuda, draw=black] (6.75, -1.00) rectangle (7.00, -1.25);
\filldraw[fill=bermuda, draw=black] (7.00, -1.00) rectangle (7.25, -1.25);
\filldraw[fill=bermuda, draw=black] (7.25, -1.00) rectangle (7.50, -1.25);
\filldraw[fill=cancan, draw=black] (7.50, -1.00) rectangle (7.75, -1.25);
\filldraw[fill=cancan, draw=black] (7.75, -1.00) rectangle (8.00, -1.25);
\filldraw[fill=cancan, draw=black] (8.00, -1.00) rectangle (8.25, -1.25);
\filldraw[fill=cancan, draw=black] (8.25, -1.00) rectangle (8.50, -1.25);
\filldraw[fill=cancan, draw=black] (8.50, -1.00) rectangle (8.75, -1.25);
\filldraw[fill=cancan, draw=black] (8.75, -1.00) rectangle (9.00, -1.25);
\filldraw[fill=cancan, draw=black] (9.00, -1.00) rectangle (9.25, -1.25);
\filldraw[fill=bermuda, draw=black] (9.25, -1.00) rectangle (9.50, -1.25);
\filldraw[fill=bermuda, draw=black] (9.50, -1.00) rectangle (9.75, -1.25);
\filldraw[fill=bermuda, draw=black] (9.75, -1.00) rectangle (10.00, -1.25);
\filldraw[fill=bermuda, draw=black] (10.00, -1.00) rectangle (10.25, -1.25);
\filldraw[fill=bermuda, draw=black] (10.25, -1.00) rectangle (10.50, -1.25);
\filldraw[fill=cancan, draw=black] (10.50, -1.00) rectangle (10.75, -1.25);
\filldraw[fill=bermuda, draw=black] (10.75, -1.00) rectangle (11.00, -1.25);
\filldraw[fill=cancan, draw=black] (11.00, -1.00) rectangle (11.25, -1.25);
\filldraw[fill=cancan, draw=black] (11.25, -1.00) rectangle (11.50, -1.25);
\filldraw[fill=cancan, draw=black] (11.50, -1.00) rectangle (11.75, -1.25);
\filldraw[fill=bermuda, draw=black] (11.75, -1.00) rectangle (12.00, -1.25);
\filldraw[fill=cancan, draw=black] (12.00, -1.00) rectangle (12.25, -1.25);
\filldraw[fill=cancan, draw=black] (12.25, -1.00) rectangle (12.50, -1.25);
\filldraw[fill=bermuda, draw=black] (12.50, -1.00) rectangle (12.75, -1.25);
\filldraw[fill=bermuda, draw=black] (12.75, -1.00) rectangle (13.00, -1.25);
\filldraw[fill=cancan, draw=black] (13.00, -1.00) rectangle (13.25, -1.25);
\filldraw[fill=bermuda, draw=black] (13.25, -1.00) rectangle (13.50, -1.25);
\filldraw[fill=cancan, draw=black] (13.50, -1.00) rectangle (13.75, -1.25);
\filldraw[fill=cancan, draw=black] (13.75, -1.00) rectangle (14.00, -1.25);
\filldraw[fill=bermuda, draw=black] (14.00, -1.00) rectangle (14.25, -1.25);
\filldraw[fill=bermuda, draw=black] (14.25, -1.00) rectangle (14.50, -1.25);
\filldraw[fill=bermuda, draw=black] (14.50, -1.00) rectangle (14.75, -1.25);
\filldraw[fill=bermuda, draw=black] (14.75, -1.00) rectangle (15.00, -1.25);
\filldraw[fill=cancan, draw=black] (0.00, -1.25) rectangle (0.25, -1.50);
\filldraw[fill=bermuda, draw=black] (0.25, -1.25) rectangle (0.50, -1.50);
\filldraw[fill=cancan, draw=black] (0.50, -1.25) rectangle (0.75, -1.50);
\filldraw[fill=cancan, draw=black] (0.75, -1.25) rectangle (1.00, -1.50);
\filldraw[fill=cancan, draw=black] (1.00, -1.25) rectangle (1.25, -1.50);
\filldraw[fill=cancan, draw=black] (1.25, -1.25) rectangle (1.50, -1.50);
\filldraw[fill=cancan, draw=black] (1.50, -1.25) rectangle (1.75, -1.50);
\filldraw[fill=bermuda, draw=black] (1.75, -1.25) rectangle (2.00, -1.50);
\filldraw[fill=bermuda, draw=black] (2.00, -1.25) rectangle (2.25, -1.50);
\filldraw[fill=bermuda, draw=black] (2.25, -1.25) rectangle (2.50, -1.50);
\filldraw[fill=cancan, draw=black] (2.50, -1.25) rectangle (2.75, -1.50);
\filldraw[fill=cancan, draw=black] (2.75, -1.25) rectangle (3.00, -1.50);
\filldraw[fill=cancan, draw=black] (3.00, -1.25) rectangle (3.25, -1.50);
\filldraw[fill=bermuda, draw=black] (3.25, -1.25) rectangle (3.50, -1.50);
\filldraw[fill=bermuda, draw=black] (3.50, -1.25) rectangle (3.75, -1.50);
\filldraw[fill=bermuda, draw=black] (3.75, -1.25) rectangle (4.00, -1.50);
\filldraw[fill=bermuda, draw=black] (4.00, -1.25) rectangle (4.25, -1.50);
\filldraw[fill=bermuda, draw=black] (4.25, -1.25) rectangle (4.50, -1.50);
\filldraw[fill=bermuda, draw=black] (4.50, -1.25) rectangle (4.75, -1.50);
\filldraw[fill=bermuda, draw=black] (4.75, -1.25) rectangle (5.00, -1.50);
\filldraw[fill=cancan, draw=black] (5.00, -1.25) rectangle (5.25, -1.50);
\filldraw[fill=cancan, draw=black] (5.25, -1.25) rectangle (5.50, -1.50);
\filldraw[fill=cancan, draw=black] (5.50, -1.25) rectangle (5.75, -1.50);
\filldraw[fill=cancan, draw=black] (5.75, -1.25) rectangle (6.00, -1.50);
\filldraw[fill=cancan, draw=black] (6.00, -1.25) rectangle (6.25, -1.50);
\filldraw[fill=cancan, draw=black] (6.25, -1.25) rectangle (6.50, -1.50);
\filldraw[fill=cancan, draw=black] (6.50, -1.25) rectangle (6.75, -1.50);
\filldraw[fill=bermuda, draw=black] (6.75, -1.25) rectangle (7.00, -1.50);
\filldraw[fill=bermuda, draw=black] (7.00, -1.25) rectangle (7.25, -1.50);
\filldraw[fill=bermuda, draw=black] (7.25, -1.25) rectangle (7.50, -1.50);
\filldraw[fill=cancan, draw=black] (7.50, -1.25) rectangle (7.75, -1.50);
\filldraw[fill=cancan, draw=black] (7.75, -1.25) rectangle (8.00, -1.50);
\filldraw[fill=cancan, draw=black] (8.00, -1.25) rectangle (8.25, -1.50);
\filldraw[fill=bermuda, draw=black] (8.25, -1.25) rectangle (8.50, -1.50);
\filldraw[fill=bermuda, draw=black] (8.50, -1.25) rectangle (8.75, -1.50);
\filldraw[fill=bermuda, draw=black] (8.75, -1.25) rectangle (9.00, -1.50);
\filldraw[fill=cancan, draw=black] (9.00, -1.25) rectangle (9.25, -1.50);
\filldraw[fill=bermuda, draw=black] (9.25, -1.25) rectangle (9.50, -1.50);
\filldraw[fill=cancan, draw=black] (9.50, -1.25) rectangle (9.75, -1.50);
\filldraw[fill=bermuda, draw=black] (9.75, -1.25) rectangle (10.00, -1.50);
\filldraw[fill=cancan, draw=black] (10.00, -1.25) rectangle (10.25, -1.50);
\filldraw[fill=cancan, draw=black] (10.25, -1.25) rectangle (10.50, -1.50);
\filldraw[fill=cancan, draw=black] (10.50, -1.25) rectangle (10.75, -1.50);
\filldraw[fill=bermuda, draw=black] (10.75, -1.25) rectangle (11.00, -1.50);
\filldraw[fill=cancan, draw=black] (11.00, -1.25) rectangle (11.25, -1.50);
\filldraw[fill=bermuda, draw=black] (11.25, -1.25) rectangle (11.50, -1.50);
\filldraw[fill=bermuda, draw=black] (11.50, -1.25) rectangle (11.75, -1.50);
\filldraw[fill=bermuda, draw=black] (11.75, -1.25) rectangle (12.00, -1.50);
\filldraw[fill=cancan, draw=black] (12.00, -1.25) rectangle (12.25, -1.50);
\filldraw[fill=cancan, draw=black] (12.25, -1.25) rectangle (12.50, -1.50);
\filldraw[fill=cancan, draw=black] (12.50, -1.25) rectangle (12.75, -1.50);
\filldraw[fill=bermuda, draw=black] (12.75, -1.25) rectangle (13.00, -1.50);
\filldraw[fill=bermuda, draw=black] (13.00, -1.25) rectangle (13.25, -1.50);
\filldraw[fill=bermuda, draw=black] (13.25, -1.25) rectangle (13.50, -1.50);
\filldraw[fill=cancan, draw=black] (13.50, -1.25) rectangle (13.75, -1.50);
\filldraw[fill=cancan, draw=black] (13.75, -1.25) rectangle (14.00, -1.50);
\filldraw[fill=cancan, draw=black] (14.00, -1.25) rectangle (14.25, -1.50);
\filldraw[fill=bermuda, draw=black] (14.25, -1.25) rectangle (14.50, -1.50);
\filldraw[fill=bermuda, draw=black] (14.50, -1.25) rectangle (14.75, -1.50);
\filldraw[fill=bermuda, draw=black] (14.75, -1.25) rectangle (15.00, -1.50);
\filldraw[fill=cancan, draw=black] (0.00, -1.50) rectangle (0.25, -1.75);
\filldraw[fill=cancan, draw=black] (0.25, -1.50) rectangle (0.50, -1.75);
\filldraw[fill=cancan, draw=black] (0.50, -1.50) rectangle (0.75, -1.75);
\filldraw[fill=cancan, draw=black] (0.75, -1.50) rectangle (1.00, -1.75);
\filldraw[fill=cancan, draw=black] (1.00, -1.50) rectangle (1.25, -1.75);
\filldraw[fill=cancan, draw=black] (1.25, -1.50) rectangle (1.50, -1.75);
\filldraw[fill=cancan, draw=black] (1.50, -1.50) rectangle (1.75, -1.75);
\filldraw[fill=bermuda, draw=black] (1.75, -1.50) rectangle (2.00, -1.75);
\filldraw[fill=cancan, draw=black] (2.00, -1.50) rectangle (2.25, -1.75);
\filldraw[fill=bermuda, draw=black] (2.25, -1.50) rectangle (2.50, -1.75);
\filldraw[fill=cancan, draw=black] (2.50, -1.50) rectangle (2.75, -1.75);
\filldraw[fill=bermuda, draw=black] (2.75, -1.50) rectangle (3.00, -1.75);
\filldraw[fill=cancan, draw=black] (3.00, -1.50) rectangle (3.25, -1.75);
\filldraw[fill=cancan, draw=black] (3.25, -1.50) rectangle (3.50, -1.75);
\filldraw[fill=cancan, draw=black] (3.50, -1.50) rectangle (3.75, -1.75);
\filldraw[fill=bermuda, draw=black] (3.75, -1.50) rectangle (4.00, -1.75);
\filldraw[fill=cancan, draw=black] (4.00, -1.50) rectangle (4.25, -1.75);
\filldraw[fill=bermuda, draw=black] (4.25, -1.50) rectangle (4.50, -1.75);
\filldraw[fill=cancan, draw=black] (4.50, -1.50) rectangle (4.75, -1.75);
\filldraw[fill=cancan, draw=black] (4.75, -1.50) rectangle (5.00, -1.75);
\filldraw[fill=cancan, draw=black] (5.00, -1.50) rectangle (5.25, -1.75);
\filldraw[fill=bermuda, draw=black] (5.25, -1.50) rectangle (5.50, -1.75);
\filldraw[fill=cancan, draw=black] (5.50, -1.50) rectangle (5.75, -1.75);
\filldraw[fill=bermuda, draw=black] (5.75, -1.50) rectangle (6.00, -1.75);
\filldraw[fill=cancan, draw=black] (6.00, -1.50) rectangle (6.25, -1.75);
\filldraw[fill=bermuda, draw=black] (6.25, -1.50) rectangle (6.50, -1.75);
\filldraw[fill=cancan, draw=black] (6.50, -1.50) rectangle (6.75, -1.75);
\filldraw[fill=cancan, draw=black] (6.75, -1.50) rectangle (7.00, -1.75);
\filldraw[fill=cancan, draw=black] (7.00, -1.50) rectangle (7.25, -1.75);
\filldraw[fill=cancan, draw=black] (7.25, -1.50) rectangle (7.50, -1.75);
\filldraw[fill=bermuda, draw=black] (7.50, -1.50) rectangle (7.75, -1.75);
\filldraw[fill=bermuda, draw=black] (7.75, -1.50) rectangle (8.00, -1.75);
\filldraw[fill=cancan, draw=black] (8.00, -1.50) rectangle (8.25, -1.75);
\filldraw[fill=cancan, draw=black] (8.25, -1.50) rectangle (8.50, -1.75);
\filldraw[fill=cancan, draw=black] (8.50, -1.50) rectangle (8.75, -1.75);
\filldraw[fill=cancan, draw=black] (8.75, -1.50) rectangle (9.00, -1.75);
\filldraw[fill=bermuda, draw=black] (9.00, -1.50) rectangle (9.25, -1.75);
\filldraw[fill=bermuda, draw=black] (9.25, -1.50) rectangle (9.50, -1.75);
\filldraw[fill=cancan, draw=black] (9.50, -1.50) rectangle (9.75, -1.75);
\filldraw[fill=cancan, draw=black] (9.75, -1.50) rectangle (10.00, -1.75);
\filldraw[fill=cancan, draw=black] (10.00, -1.50) rectangle (10.25, -1.75);
\filldraw[fill=cancan, draw=black] (10.25, -1.50) rectangle (10.50, -1.75);
\filldraw[fill=cancan, draw=black] (10.50, -1.50) rectangle (10.75, -1.75);
\filldraw[fill=bermuda, draw=black] (10.75, -1.50) rectangle (11.00, -1.75);
\filldraw[fill=bermuda, draw=black] (11.00, -1.50) rectangle (11.25, -1.75);
\filldraw[fill=bermuda, draw=black] (11.25, -1.50) rectangle (11.50, -1.75);
\filldraw[fill=cancan, draw=black] (11.50, -1.50) rectangle (11.75, -1.75);
\filldraw[fill=cancan, draw=black] (11.75, -1.50) rectangle (12.00, -1.75);
\filldraw[fill=cancan, draw=black] (12.00, -1.50) rectangle (12.25, -1.75);
\filldraw[fill=bermuda, draw=black] (12.25, -1.50) rectangle (12.50, -1.75);
\filldraw[fill=bermuda, draw=black] (12.50, -1.50) rectangle (12.75, -1.75);
\filldraw[fill=bermuda, draw=black] (12.75, -1.50) rectangle (13.00, -1.75);
\filldraw[fill=cancan, draw=black] (13.00, -1.50) rectangle (13.25, -1.75);
\filldraw[fill=cancan, draw=black] (13.25, -1.50) rectangle (13.50, -1.75);
\filldraw[fill=cancan, draw=black] (13.50, -1.50) rectangle (13.75, -1.75);
\filldraw[fill=bermuda, draw=black] (13.75, -1.50) rectangle (14.00, -1.75);
\filldraw[fill=bermuda, draw=black] (14.00, -1.50) rectangle (14.25, -1.75);
\filldraw[fill=bermuda, draw=black] (14.25, -1.50) rectangle (14.50, -1.75);
\filldraw[fill=cancan, draw=black] (14.50, -1.50) rectangle (14.75, -1.75);
\filldraw[fill=cancan, draw=black] (14.75, -1.50) rectangle (15.00, -1.75);
\filldraw[fill=cancan, draw=black] (0.00, -1.75) rectangle (0.25, -2.00);
\filldraw[fill=bermuda, draw=black] (0.25, -1.75) rectangle (0.50, -2.00);
\filldraw[fill=bermuda, draw=black] (0.50, -1.75) rectangle (0.75, -2.00);
\filldraw[fill=bermuda, draw=black] (0.75, -1.75) rectangle (1.00, -2.00);
\filldraw[fill=bermuda, draw=black] (1.00, -1.75) rectangle (1.25, -2.00);
\filldraw[fill=bermuda, draw=black] (1.25, -1.75) rectangle (1.50, -2.00);
\filldraw[fill=cancan, draw=black] (1.50, -1.75) rectangle (1.75, -2.00);
\filldraw[fill=bermuda, draw=black] (1.75, -1.75) rectangle (2.00, -2.00);
\filldraw[fill=cancan, draw=black] (2.00, -1.75) rectangle (2.25, -2.00);
\filldraw[fill=bermuda, draw=black] (2.25, -1.75) rectangle (2.50, -2.00);
\filldraw[fill=cancan, draw=black] (2.50, -1.75) rectangle (2.75, -2.00);
\filldraw[fill=bermuda, draw=black] (2.75, -1.75) rectangle (3.00, -2.00);
\filldraw[fill=cancan, draw=black] (3.00, -1.75) rectangle (3.25, -2.00);
\filldraw[fill=cancan, draw=black] (3.25, -1.75) rectangle (3.50, -2.00);
\filldraw[fill=bermuda, draw=black] (3.50, -1.75) rectangle (3.75, -2.00);
\filldraw[fill=bermuda, draw=black] (3.75, -1.75) rectangle (4.00, -2.00);
\filldraw[fill=bermuda, draw=black] (4.00, -1.75) rectangle (4.25, -2.00);
\filldraw[fill=bermuda, draw=black] (4.25, -1.75) rectangle (4.50, -2.00);
\filldraw[fill=cancan, draw=black] (4.50, -1.75) rectangle (4.75, -2.00);
\filldraw[fill=cancan, draw=black] (4.75, -1.75) rectangle (5.00, -2.00);
\filldraw[fill=cancan, draw=black] (5.00, -1.75) rectangle (5.25, -2.00);
\filldraw[fill=cancan, draw=black] (5.25, -1.75) rectangle (5.50, -2.00);
\filldraw[fill=cancan, draw=black] (5.50, -1.75) rectangle (5.75, -2.00);
\filldraw[fill=cancan, draw=black] (5.75, -1.75) rectangle (6.00, -2.00);
\filldraw[fill=cancan, draw=black] (6.00, -1.75) rectangle (6.25, -2.00);
\filldraw[fill=bermuda, draw=black] (6.25, -1.75) rectangle (6.50, -2.00);
\filldraw[fill=cancan, draw=black] (6.50, -1.75) rectangle (6.75, -2.00);
\filldraw[fill=bermuda, draw=black] (6.75, -1.75) rectangle (7.00, -2.00);
\filldraw[fill=cancan, draw=black] (7.00, -1.75) rectangle (7.25, -2.00);
\filldraw[fill=bermuda, draw=black] (7.25, -1.75) rectangle (7.50, -2.00);
\filldraw[fill=bermuda, draw=black] (7.50, -1.75) rectangle (7.75, -2.00);
\filldraw[fill=bermuda, draw=black] (7.75, -1.75) rectangle (8.00, -2.00);
\filldraw[fill=cancan, draw=black] (8.00, -1.75) rectangle (8.25, -2.00);
\filldraw[fill=cancan, draw=black] (8.25, -1.75) rectangle (8.50, -2.00);
\filldraw[fill=cancan, draw=black] (8.50, -1.75) rectangle (8.75, -2.00);
\filldraw[fill=bermuda, draw=black] (8.75, -1.75) rectangle (9.00, -2.00);
\filldraw[fill=bermuda, draw=black] (9.00, -1.75) rectangle (9.25, -2.00);
\filldraw[fill=bermuda, draw=black] (9.25, -1.75) rectangle (9.50, -2.00);
\filldraw[fill=cancan, draw=black] (9.50, -1.75) rectangle (9.75, -2.00);
\filldraw[fill=bermuda, draw=black] (9.75, -1.75) rectangle (10.00, -2.00);
\filldraw[fill=bermuda, draw=black] (10.00, -1.75) rectangle (10.25, -2.00);
\filldraw[fill=bermuda, draw=black] (10.25, -1.75) rectangle (10.50, -2.00);
\filldraw[fill=bermuda, draw=black] (10.50, -1.75) rectangle (10.75, -2.00);
\filldraw[fill=bermuda, draw=black] (10.75, -1.75) rectangle (11.00, -2.00);
\filldraw[fill=bermuda, draw=black] (11.00, -1.75) rectangle (11.25, -2.00);
\filldraw[fill=bermuda, draw=black] (11.25, -1.75) rectangle (11.50, -2.00);
\filldraw[fill=cancan, draw=black] (11.50, -1.75) rectangle (11.75, -2.00);
\filldraw[fill=bermuda, draw=black] (11.75, -1.75) rectangle (12.00, -2.00);
\filldraw[fill=cancan, draw=black] (12.00, -1.75) rectangle (12.25, -2.00);
\filldraw[fill=bermuda, draw=black] (12.25, -1.75) rectangle (12.50, -2.00);
\filldraw[fill=bermuda, draw=black] (12.50, -1.75) rectangle (12.75, -2.00);
\filldraw[fill=bermuda, draw=black] (12.75, -1.75) rectangle (13.00, -2.00);
\filldraw[fill=cancan, draw=black] (13.00, -1.75) rectangle (13.25, -2.00);
\filldraw[fill=bermuda, draw=black] (13.25, -1.75) rectangle (13.50, -2.00);
\filldraw[fill=bermuda, draw=black] (13.50, -1.75) rectangle (13.75, -2.00);
\filldraw[fill=bermuda, draw=black] (13.75, -1.75) rectangle (14.00, -2.00);
\filldraw[fill=bermuda, draw=black] (14.00, -1.75) rectangle (14.25, -2.00);
\filldraw[fill=bermuda, draw=black] (14.25, -1.75) rectangle (14.50, -2.00);
\filldraw[fill=cancan, draw=black] (14.50, -1.75) rectangle (14.75, -2.00);
\filldraw[fill=bermuda, draw=black] (14.75, -1.75) rectangle (15.00, -2.00);
\filldraw[fill=cancan, draw=black] (0.00, -2.00) rectangle (0.25, -2.25);
\filldraw[fill=cancan, draw=black] (0.25, -2.00) rectangle (0.50, -2.25);
\filldraw[fill=bermuda, draw=black] (0.50, -2.00) rectangle (0.75, -2.25);
\filldraw[fill=bermuda, draw=black] (0.75, -2.00) rectangle (1.00, -2.25);
\filldraw[fill=cancan, draw=black] (1.00, -2.00) rectangle (1.25, -2.25);
\filldraw[fill=bermuda, draw=black] (1.25, -2.00) rectangle (1.50, -2.25);
\filldraw[fill=cancan, draw=black] (1.50, -2.00) rectangle (1.75, -2.25);
\filldraw[fill=bermuda, draw=black] (1.75, -2.00) rectangle (2.00, -2.25);
\filldraw[fill=cancan, draw=black] (2.00, -2.00) rectangle (2.25, -2.25);
\filldraw[fill=bermuda, draw=black] (2.25, -2.00) rectangle (2.50, -2.25);
\filldraw[fill=bermuda, draw=black] (2.50, -2.00) rectangle (2.75, -2.25);
\filldraw[fill=bermuda, draw=black] (2.75, -2.00) rectangle (3.00, -2.25);
\filldraw[fill=cancan, draw=black] (3.00, -2.00) rectangle (3.25, -2.25);
\filldraw[fill=cancan, draw=black] (3.25, -2.00) rectangle (3.50, -2.25);
\filldraw[fill=cancan, draw=black] (3.50, -2.00) rectangle (3.75, -2.25);
\filldraw[fill=bermuda, draw=black] (3.75, -2.00) rectangle (4.00, -2.25);
\filldraw[fill=bermuda, draw=black] (4.00, -2.00) rectangle (4.25, -2.25);
\filldraw[fill=bermuda, draw=black] (4.25, -2.00) rectangle (4.50, -2.25);
\filldraw[fill=cancan, draw=black] (4.50, -2.00) rectangle (4.75, -2.25);
\filldraw[fill=bermuda, draw=black] (4.75, -2.00) rectangle (5.00, -2.25);
\filldraw[fill=cancan, draw=black] (5.00, -2.00) rectangle (5.25, -2.25);
\filldraw[fill=bermuda, draw=black] (5.25, -2.00) rectangle (5.50, -2.25);
\filldraw[fill=bermuda, draw=black] (5.50, -2.00) rectangle (5.75, -2.25);
\filldraw[fill=bermuda, draw=black] (5.75, -2.00) rectangle (6.00, -2.25);
\filldraw[fill=cancan, draw=black] (6.00, -2.00) rectangle (6.25, -2.25);
\filldraw[fill=cancan, draw=black] (6.25, -2.00) rectangle (6.50, -2.25);
\filldraw[fill=cancan, draw=black] (6.50, -2.00) rectangle (6.75, -2.25);
\filldraw[fill=bermuda, draw=black] (6.75, -2.00) rectangle (7.00, -2.25);
\filldraw[fill=bermuda, draw=black] (7.00, -2.00) rectangle (7.25, -2.25);
\filldraw[fill=bermuda, draw=black] (7.25, -2.00) rectangle (7.50, -2.25);
\filldraw[fill=cancan, draw=black] (7.50, -2.00) rectangle (7.75, -2.25);
\filldraw[fill=cancan, draw=black] (7.75, -2.00) rectangle (8.00, -2.25);
\filldraw[fill=cancan, draw=black] (8.00, -2.00) rectangle (8.25, -2.25);
\filldraw[fill=bermuda, draw=black] (8.25, -2.00) rectangle (8.50, -2.25);
\filldraw[fill=bermuda, draw=black] (8.50, -2.00) rectangle (8.75, -2.25);
\filldraw[fill=bermuda, draw=black] (8.75, -2.00) rectangle (9.00, -2.25);
\filldraw[fill=bermuda, draw=black] (9.00, -2.00) rectangle (9.25, -2.25);
\filldraw[fill=bermuda, draw=black] (9.25, -2.00) rectangle (9.50, -2.25);
\filldraw[fill=cancan, draw=black] (9.50, -2.00) rectangle (9.75, -2.25);
\filldraw[fill=bermuda, draw=black] (9.75, -2.00) rectangle (10.00, -2.25);
\filldraw[fill=bermuda, draw=black] (10.00, -2.00) rectangle (10.25, -2.25);
\filldraw[fill=bermuda, draw=black] (10.25, -2.00) rectangle (10.50, -2.25);
\filldraw[fill=cancan, draw=black] (10.50, -2.00) rectangle (10.75, -2.25);
\filldraw[fill=cancan, draw=black] (10.75, -2.00) rectangle (11.00, -2.25);
\filldraw[fill=cancan, draw=black] (11.00, -2.00) rectangle (11.25, -2.25);
\filldraw[fill=cancan, draw=black] (11.25, -2.00) rectangle (11.50, -2.25);
\filldraw[fill=cancan, draw=black] (11.50, -2.00) rectangle (11.75, -2.25);
\filldraw[fill=bermuda, draw=black] (11.75, -2.00) rectangle (12.00, -2.25);
\filldraw[fill=bermuda, draw=black] (12.00, -2.00) rectangle (12.25, -2.25);
\filldraw[fill=bermuda, draw=black] (12.25, -2.00) rectangle (12.50, -2.25);
\filldraw[fill=cancan, draw=black] (12.50, -2.00) rectangle (12.75, -2.25);
\filldraw[fill=cancan, draw=black] (12.75, -2.00) rectangle (13.00, -2.25);
\filldraw[fill=cancan, draw=black] (13.00, -2.00) rectangle (13.25, -2.25);
\filldraw[fill=bermuda, draw=black] (13.25, -2.00) rectangle (13.50, -2.25);
\filldraw[fill=bermuda, draw=black] (13.50, -2.00) rectangle (13.75, -2.25);
\filldraw[fill=bermuda, draw=black] (13.75, -2.00) rectangle (14.00, -2.25);
\filldraw[fill=cancan, draw=black] (14.00, -2.00) rectangle (14.25, -2.25);
\filldraw[fill=cancan, draw=black] (14.25, -2.00) rectangle (14.50, -2.25);
\filldraw[fill=cancan, draw=black] (14.50, -2.00) rectangle (14.75, -2.25);
\filldraw[fill=cancan, draw=black] (14.75, -2.00) rectangle (15.00, -2.25);
\filldraw[fill=cancan, draw=black] (0.00, -2.25) rectangle (0.25, -2.50);
\filldraw[fill=cancan, draw=black] (0.25, -2.25) rectangle (0.50, -2.50);
\filldraw[fill=cancan, draw=black] (0.50, -2.25) rectangle (0.75, -2.50);
\filldraw[fill=bermuda, draw=black] (0.75, -2.25) rectangle (1.00, -2.50);
\filldraw[fill=bermuda, draw=black] (1.00, -2.25) rectangle (1.25, -2.50);
\filldraw[fill=bermuda, draw=black] (1.25, -2.25) rectangle (1.50, -2.50);
\filldraw[fill=cancan, draw=black] (1.50, -2.25) rectangle (1.75, -2.50);
\filldraw[fill=cancan, draw=black] (1.75, -2.25) rectangle (2.00, -2.50);
\filldraw[fill=cancan, draw=black] (2.00, -2.25) rectangle (2.25, -2.50);
\filldraw[fill=bermuda, draw=black] (2.25, -2.25) rectangle (2.50, -2.50);
\filldraw[fill=bermuda, draw=black] (2.50, -2.25) rectangle (2.75, -2.50);
\filldraw[fill=bermuda, draw=black] (2.75, -2.25) rectangle (3.00, -2.50);
\filldraw[fill=cancan, draw=black] (3.00, -2.25) rectangle (3.25, -2.50);
\filldraw[fill=cancan, draw=black] (3.25, -2.25) rectangle (3.50, -2.50);
\filldraw[fill=cancan, draw=black] (3.50, -2.25) rectangle (3.75, -2.50);
\filldraw[fill=bermuda, draw=black] (3.75, -2.25) rectangle (4.00, -2.50);
\filldraw[fill=bermuda, draw=black] (4.00, -2.25) rectangle (4.25, -2.50);
\filldraw[fill=bermuda, draw=black] (4.25, -2.25) rectangle (4.50, -2.50);
\filldraw[fill=cancan, draw=black] (4.50, -2.25) rectangle (4.75, -2.50);
\filldraw[fill=cancan, draw=black] (4.75, -2.25) rectangle (5.00, -2.50);
\filldraw[fill=cancan, draw=black] (5.00, -2.25) rectangle (5.25, -2.50);
\filldraw[fill=bermuda, draw=black] (5.25, -2.25) rectangle (5.50, -2.50);
\filldraw[fill=bermuda, draw=black] (5.50, -2.25) rectangle (5.75, -2.50);
\filldraw[fill=bermuda, draw=black] (5.75, -2.25) rectangle (6.00, -2.50);
\filldraw[fill=bermuda, draw=black] (6.00, -2.25) rectangle (6.25, -2.50);
\filldraw[fill=bermuda, draw=black] (6.25, -2.25) rectangle (6.50, -2.50);
\filldraw[fill=cancan, draw=black] (6.50, -2.25) rectangle (6.75, -2.50);
\filldraw[fill=bermuda, draw=black] (6.75, -2.25) rectangle (7.00, -2.50);
\filldraw[fill=cancan, draw=black] (7.00, -2.25) rectangle (7.25, -2.50);
\filldraw[fill=bermuda, draw=black] (7.25, -2.25) rectangle (7.50, -2.50);
\filldraw[fill=cancan, draw=black] (7.50, -2.25) rectangle (7.75, -2.50);
\filldraw[fill=bermuda, draw=black] (7.75, -2.25) rectangle (8.00, -2.50);
\filldraw[fill=cancan, draw=black] (8.00, -2.25) rectangle (8.25, -2.50);
\filldraw[fill=cancan, draw=black] (8.25, -2.25) rectangle (8.50, -2.50);
\filldraw[fill=cancan, draw=black] (8.50, -2.25) rectangle (8.75, -2.50);
\filldraw[fill=cancan, draw=black] (8.75, -2.25) rectangle (9.00, -2.50);
\filldraw[fill=cancan, draw=black] (9.00, -2.25) rectangle (9.25, -2.50);
\filldraw[fill=cancan, draw=black] (9.25, -2.25) rectangle (9.50, -2.50);
\filldraw[fill=cancan, draw=black] (9.50, -2.25) rectangle (9.75, -2.50);
\filldraw[fill=cancan, draw=black] (9.75, -2.25) rectangle (10.00, -2.50);
\filldraw[fill=cancan, draw=black] (10.00, -2.25) rectangle (10.25, -2.50);
\filldraw[fill=bermuda, draw=black] (10.25, -2.25) rectangle (10.50, -2.50);
\filldraw[fill=cancan, draw=black] (10.50, -2.25) rectangle (10.75, -2.50);
\filldraw[fill=bermuda, draw=black] (10.75, -2.25) rectangle (11.00, -2.50);
\filldraw[fill=bermuda, draw=black] (11.00, -2.25) rectangle (11.25, -2.50);
\filldraw[fill=bermuda, draw=black] (11.25, -2.25) rectangle (11.50, -2.50);
\filldraw[fill=cancan, draw=black] (11.50, -2.25) rectangle (11.75, -2.50);
\filldraw[fill=cancan, draw=black] (11.75, -2.25) rectangle (12.00, -2.50);
\filldraw[fill=cancan, draw=black] (12.00, -2.25) rectangle (12.25, -2.50);
\filldraw[fill=bermuda, draw=black] (12.25, -2.25) rectangle (12.50, -2.50);
\filldraw[fill=bermuda, draw=black] (12.50, -2.25) rectangle (12.75, -2.50);
\filldraw[fill=bermuda, draw=black] (12.75, -2.25) rectangle (13.00, -2.50);
\filldraw[fill=cancan, draw=black] (13.00, -2.25) rectangle (13.25, -2.50);
\filldraw[fill=cancan, draw=black] (13.25, -2.25) rectangle (13.50, -2.50);
\filldraw[fill=cancan, draw=black] (13.50, -2.25) rectangle (13.75, -2.50);
\filldraw[fill=cancan, draw=black] (13.75, -2.25) rectangle (14.00, -2.50);
\filldraw[fill=cancan, draw=black] (14.00, -2.25) rectangle (14.25, -2.50);
\filldraw[fill=cancan, draw=black] (14.25, -2.25) rectangle (14.50, -2.50);
\filldraw[fill=cancan, draw=black] (14.50, -2.25) rectangle (14.75, -2.50);
\filldraw[fill=bermuda, draw=black] (14.75, -2.25) rectangle (15.00, -2.50);
\filldraw[fill=bermuda, draw=black] (0.00, -2.50) rectangle (0.25, -2.75);
\filldraw[fill=bermuda, draw=black] (0.25, -2.50) rectangle (0.50, -2.75);
\filldraw[fill=cancan, draw=black] (0.50, -2.50) rectangle (0.75, -2.75);
\filldraw[fill=bermuda, draw=black] (0.75, -2.50) rectangle (1.00, -2.75);
\filldraw[fill=bermuda, draw=black] (1.00, -2.50) rectangle (1.25, -2.75);
\filldraw[fill=bermuda, draw=black] (1.25, -2.50) rectangle (1.50, -2.75);
\filldraw[fill=bermuda, draw=black] (1.50, -2.50) rectangle (1.75, -2.75);
\filldraw[fill=bermuda, draw=black] (1.75, -2.50) rectangle (2.00, -2.75);
\filldraw[fill=cancan, draw=black] (2.00, -2.50) rectangle (2.25, -2.75);
\filldraw[fill=bermuda, draw=black] (2.25, -2.50) rectangle (2.50, -2.75);
\filldraw[fill=bermuda, draw=black] (2.50, -2.50) rectangle (2.75, -2.75);
\filldraw[fill=bermuda, draw=black] (2.75, -2.50) rectangle (3.00, -2.75);
\filldraw[fill=cancan, draw=black] (3.00, -2.50) rectangle (3.25, -2.75);
\filldraw[fill=bermuda, draw=black] (3.25, -2.50) rectangle (3.50, -2.75);
\filldraw[fill=bermuda, draw=black] (3.50, -2.50) rectangle (3.75, -2.75);
\filldraw[fill=bermuda, draw=black] (3.75, -2.50) rectangle (4.00, -2.75);
\filldraw[fill=bermuda, draw=black] (4.00, -2.50) rectangle (4.25, -2.75);
\filldraw[fill=bermuda, draw=black] (4.25, -2.50) rectangle (4.50, -2.75);
\filldraw[fill=cancan, draw=black] (4.50, -2.50) rectangle (4.75, -2.75);
\filldraw[fill=cancan, draw=black] (4.75, -2.50) rectangle (5.00, -2.75);
\filldraw[fill=cancan, draw=black] (5.00, -2.50) rectangle (5.25, -2.75);
\filldraw[fill=bermuda, draw=black] (5.25, -2.50) rectangle (5.50, -2.75);
\filldraw[fill=bermuda, draw=black] (5.50, -2.50) rectangle (5.75, -2.75);
\filldraw[fill=bermuda, draw=black] (5.75, -2.50) rectangle (6.00, -2.75);
\filldraw[fill=cancan, draw=black] (6.00, -2.50) rectangle (6.25, -2.75);
\filldraw[fill=cancan, draw=black] (6.25, -2.50) rectangle (6.50, -2.75);
\filldraw[fill=cancan, draw=black] (6.50, -2.50) rectangle (6.75, -2.75);
\filldraw[fill=cancan, draw=black] (6.75, -2.50) rectangle (7.00, -2.75);
\filldraw[fill=cancan, draw=black] (7.00, -2.50) rectangle (7.25, -2.75);
\filldraw[fill=bermuda, draw=black] (7.25, -2.50) rectangle (7.50, -2.75);
\filldraw[fill=cancan, draw=black] (7.50, -2.50) rectangle (7.75, -2.75);
\filldraw[fill=bermuda, draw=black] (7.75, -2.50) rectangle (8.00, -2.75);
\filldraw[fill=cancan, draw=black] (8.00, -2.50) rectangle (8.25, -2.75);
\filldraw[fill=cancan, draw=black] (8.25, -2.50) rectangle (8.50, -2.75);
\filldraw[fill=bermuda, draw=black] (8.50, -2.50) rectangle (8.75, -2.75);
\filldraw[fill=bermuda, draw=black] (8.75, -2.50) rectangle (9.00, -2.75);
\filldraw[fill=cancan, draw=black] (9.00, -2.50) rectangle (9.25, -2.75);
\filldraw[fill=cancan, draw=black] (9.25, -2.50) rectangle (9.50, -2.75);
\filldraw[fill=cancan, draw=black] (9.50, -2.50) rectangle (9.75, -2.75);
\filldraw[fill=bermuda, draw=black] (9.75, -2.50) rectangle (10.00, -2.75);
\filldraw[fill=bermuda, draw=black] (10.00, -2.50) rectangle (10.25, -2.75);
\filldraw[fill=bermuda, draw=black] (10.25, -2.50) rectangle (10.50, -2.75);
\filldraw[fill=cancan, draw=black] (10.50, -2.50) rectangle (10.75, -2.75);
\filldraw[fill=bermuda, draw=black] (10.75, -2.50) rectangle (11.00, -2.75);
\filldraw[fill=cancan, draw=black] (11.00, -2.50) rectangle (11.25, -2.75);
\filldraw[fill=cancan, draw=black] (11.25, -2.50) rectangle (11.50, -2.75);
\filldraw[fill=bermuda, draw=black] (11.50, -2.50) rectangle (11.75, -2.75);
\filldraw[fill=bermuda, draw=black] (11.75, -2.50) rectangle (12.00, -2.75);
\filldraw[fill=cancan, draw=black] (12.00, -2.50) rectangle (12.25, -2.75);
\filldraw[fill=bermuda, draw=black] (12.25, -2.50) rectangle (12.50, -2.75);
\filldraw[fill=cancan, draw=black] (12.50, -2.50) rectangle (12.75, -2.75);
\filldraw[fill=bermuda, draw=black] (12.75, -2.50) rectangle (13.00, -2.75);
\filldraw[fill=cancan, draw=black] (13.00, -2.50) rectangle (13.25, -2.75);
\filldraw[fill=bermuda, draw=black] (13.25, -2.50) rectangle (13.50, -2.75);
\filldraw[fill=bermuda, draw=black] (13.50, -2.50) rectangle (13.75, -2.75);
\filldraw[fill=bermuda, draw=black] (13.75, -2.50) rectangle (14.00, -2.75);
\filldraw[fill=cancan, draw=black] (14.00, -2.50) rectangle (14.25, -2.75);
\filldraw[fill=cancan, draw=black] (14.25, -2.50) rectangle (14.50, -2.75);
\filldraw[fill=cancan, draw=black] (14.50, -2.50) rectangle (14.75, -2.75);
\filldraw[fill=cancan, draw=black] (14.75, -2.50) rectangle (15.00, -2.75);
\filldraw[fill=cancan, draw=black] (0.00, -2.75) rectangle (0.25, -3.00);
\filldraw[fill=bermuda, draw=black] (0.25, -2.75) rectangle (0.50, -3.00);
\filldraw[fill=bermuda, draw=black] (0.50, -2.75) rectangle (0.75, -3.00);
\filldraw[fill=bermuda, draw=black] (0.75, -2.75) rectangle (1.00, -3.00);
\filldraw[fill=cancan, draw=black] (1.00, -2.75) rectangle (1.25, -3.00);
\filldraw[fill=cancan, draw=black] (1.25, -2.75) rectangle (1.50, -3.00);
\filldraw[fill=cancan, draw=black] (1.50, -2.75) rectangle (1.75, -3.00);
\filldraw[fill=bermuda, draw=black] (1.75, -2.75) rectangle (2.00, -3.00);
\filldraw[fill=bermuda, draw=black] (2.00, -2.75) rectangle (2.25, -3.00);
\filldraw[fill=bermuda, draw=black] (2.25, -2.75) rectangle (2.50, -3.00);
\filldraw[fill=cancan, draw=black] (2.50, -2.75) rectangle (2.75, -3.00);
\filldraw[fill=bermuda, draw=black] (2.75, -2.75) rectangle (3.00, -3.00);
\filldraw[fill=bermuda, draw=black] (3.00, -2.75) rectangle (3.25, -3.00);
\filldraw[fill=bermuda, draw=black] (3.25, -2.75) rectangle (3.50, -3.00);
\filldraw[fill=cancan, draw=black] (3.50, -2.75) rectangle (3.75, -3.00);
\filldraw[fill=cancan, draw=black] (3.75, -2.75) rectangle (4.00, -3.00);
\filldraw[fill=bermuda, draw=black] (4.00, -2.75) rectangle (4.25, -3.00);
\filldraw[fill=bermuda, draw=black] (4.25, -2.75) rectangle (4.50, -3.00);
\filldraw[fill=cancan, draw=black] (4.50, -2.75) rectangle (4.75, -3.00);
\filldraw[fill=cancan, draw=black] (4.75, -2.75) rectangle (5.00, -3.00);
\filldraw[fill=cancan, draw=black] (5.00, -2.75) rectangle (5.25, -3.00);
\filldraw[fill=cancan, draw=black] (5.25, -2.75) rectangle (5.50, -3.00);
\filldraw[fill=bermuda, draw=black] (5.50, -2.75) rectangle (5.75, -3.00);
\filldraw[fill=bermuda, draw=black] (5.75, -2.75) rectangle (6.00, -3.00);
\filldraw[fill=cancan, draw=black] (6.00, -2.75) rectangle (6.25, -3.00);
\filldraw[fill=cancan, draw=black] (6.25, -2.75) rectangle (6.50, -3.00);
\filldraw[fill=cancan, draw=black] (6.50, -2.75) rectangle (6.75, -3.00);
\filldraw[fill=bermuda, draw=black] (6.75, -2.75) rectangle (7.00, -3.00);
\filldraw[fill=bermuda, draw=black] (7.00, -2.75) rectangle (7.25, -3.00);
\filldraw[fill=bermuda, draw=black] (7.25, -2.75) rectangle (7.50, -3.00);
\filldraw[fill=cancan, draw=black] (7.50, -2.75) rectangle (7.75, -3.00);
\filldraw[fill=cancan, draw=black] (7.75, -2.75) rectangle (8.00, -3.00);
\filldraw[fill=cancan, draw=black] (8.00, -2.75) rectangle (8.25, -3.00);
\filldraw[fill=bermuda, draw=black] (8.25, -2.75) rectangle (8.50, -3.00);
\filldraw[fill=bermuda, draw=black] (8.50, -2.75) rectangle (8.75, -3.00);
\filldraw[fill=bermuda, draw=black] (8.75, -2.75) rectangle (9.00, -3.00);
\filldraw[fill=bermuda, draw=black] (9.00, -2.75) rectangle (9.25, -3.00);
\filldraw[fill=bermuda, draw=black] (9.25, -2.75) rectangle (9.50, -3.00);
\filldraw[fill=bermuda, draw=black] (9.50, -2.75) rectangle (9.75, -3.00);
\filldraw[fill=bermuda, draw=black] (9.75, -2.75) rectangle (10.00, -3.00);
\filldraw[fill=cancan, draw=black] (10.00, -2.75) rectangle (10.25, -3.00);
\filldraw[fill=bermuda, draw=black] (10.25, -2.75) rectangle (10.50, -3.00);
\filldraw[fill=cancan, draw=black] (10.50, -2.75) rectangle (10.75, -3.00);
\filldraw[fill=cancan, draw=black] (10.75, -2.75) rectangle (11.00, -3.00);
\filldraw[fill=bermuda, draw=black] (11.00, -2.75) rectangle (11.25, -3.00);
\filldraw[fill=bermuda, draw=black] (11.25, -2.75) rectangle (11.50, -3.00);
\filldraw[fill=cancan, draw=black] (11.50, -2.75) rectangle (11.75, -3.00);
\filldraw[fill=bermuda, draw=black] (11.75, -2.75) rectangle (12.00, -3.00);
\filldraw[fill=cancan, draw=black] (12.00, -2.75) rectangle (12.25, -3.00);
\filldraw[fill=cancan, draw=black] (12.25, -2.75) rectangle (12.50, -3.00);
\filldraw[fill=cancan, draw=black] (12.50, -2.75) rectangle (12.75, -3.00);
\filldraw[fill=cancan, draw=black] (12.75, -2.75) rectangle (13.00, -3.00);
\filldraw[fill=cancan, draw=black] (13.00, -2.75) rectangle (13.25, -3.00);
\filldraw[fill=bermuda, draw=black] (13.25, -2.75) rectangle (13.50, -3.00);
\filldraw[fill=bermuda, draw=black] (13.50, -2.75) rectangle (13.75, -3.00);
\filldraw[fill=bermuda, draw=black] (13.75, -2.75) rectangle (14.00, -3.00);
\filldraw[fill=cancan, draw=black] (14.00, -2.75) rectangle (14.25, -3.00);
\filldraw[fill=bermuda, draw=black] (14.25, -2.75) rectangle (14.50, -3.00);
\filldraw[fill=cancan, draw=black] (14.50, -2.75) rectangle (14.75, -3.00);
\filldraw[fill=bermuda, draw=black] (14.75, -2.75) rectangle (15.00, -3.00);
\filldraw[fill=cancan, draw=black] (0.00, -3.00) rectangle (0.25, -3.25);
\filldraw[fill=cancan, draw=black] (0.25, -3.00) rectangle (0.50, -3.25);
\filldraw[fill=cancan, draw=black] (0.50, -3.00) rectangle (0.75, -3.25);
\filldraw[fill=bermuda, draw=black] (0.75, -3.00) rectangle (1.00, -3.25);
\filldraw[fill=bermuda, draw=black] (1.00, -3.00) rectangle (1.25, -3.25);
\filldraw[fill=bermuda, draw=black] (1.25, -3.00) rectangle (1.50, -3.25);
\filldraw[fill=cancan, draw=black] (1.50, -3.00) rectangle (1.75, -3.25);
\filldraw[fill=cancan, draw=black] (1.75, -3.00) rectangle (2.00, -3.25);
\filldraw[fill=cancan, draw=black] (2.00, -3.00) rectangle (2.25, -3.25);
\filldraw[fill=bermuda, draw=black] (2.25, -3.00) rectangle (2.50, -3.25);
\filldraw[fill=bermuda, draw=black] (2.50, -3.00) rectangle (2.75, -3.25);
\filldraw[fill=bermuda, draw=black] (2.75, -3.00) rectangle (3.00, -3.25);
\filldraw[fill=cancan, draw=black] (3.00, -3.00) rectangle (3.25, -3.25);
\filldraw[fill=bermuda, draw=black] (3.25, -3.00) rectangle (3.50, -3.25);
\filldraw[fill=bermuda, draw=black] (3.50, -3.00) rectangle (3.75, -3.25);
\filldraw[fill=bermuda, draw=black] (3.75, -3.00) rectangle (4.00, -3.25);
\filldraw[fill=cancan, draw=black] (4.00, -3.00) rectangle (4.25, -3.25);
\filldraw[fill=cancan, draw=black] (4.25, -3.00) rectangle (4.50, -3.25);
\filldraw[fill=cancan, draw=black] (4.50, -3.00) rectangle (4.75, -3.25);
\filldraw[fill=bermuda, draw=black] (4.75, -3.00) rectangle (5.00, -3.25);
\filldraw[fill=bermuda, draw=black] (5.00, -3.00) rectangle (5.25, -3.25);
\filldraw[fill=bermuda, draw=black] (5.25, -3.00) rectangle (5.50, -3.25);
\filldraw[fill=cancan, draw=black] (5.50, -3.00) rectangle (5.75, -3.25);
\filldraw[fill=cancan, draw=black] (5.75, -3.00) rectangle (6.00, -3.25);
\filldraw[fill=cancan, draw=black] (6.00, -3.00) rectangle (6.25, -3.25);
\filldraw[fill=bermuda, draw=black] (6.25, -3.00) rectangle (6.50, -3.25);
\filldraw[fill=bermuda, draw=black] (6.50, -3.00) rectangle (6.75, -3.25);
\filldraw[fill=bermuda, draw=black] (6.75, -3.00) rectangle (7.00, -3.25);
\filldraw[fill=cancan, draw=black] (7.00, -3.00) rectangle (7.25, -3.25);
\filldraw[fill=cancan, draw=black] (7.25, -3.00) rectangle (7.50, -3.25);
\filldraw[fill=cancan, draw=black] (7.50, -3.00) rectangle (7.75, -3.25);
\filldraw[fill=bermuda, draw=black] (7.75, -3.00) rectangle (8.00, -3.25);
\filldraw[fill=bermuda, draw=black] (8.00, -3.00) rectangle (8.25, -3.25);
\filldraw[fill=bermuda, draw=black] (8.25, -3.00) rectangle (8.50, -3.25);
\filldraw[fill=cancan, draw=black] (8.50, -3.00) rectangle (8.75, -3.25);
\filldraw[fill=bermuda, draw=black] (8.75, -3.00) rectangle (9.00, -3.25);
\filldraw[fill=bermuda, draw=black] (9.00, -3.00) rectangle (9.25, -3.25);
\filldraw[fill=bermuda, draw=black] (9.25, -3.00) rectangle (9.50, -3.25);
\filldraw[fill=cancan, draw=black] (9.50, -3.00) rectangle (9.75, -3.25);
\filldraw[fill=cancan, draw=black] (9.75, -3.00) rectangle (10.00, -3.25);
\filldraw[fill=cancan, draw=black] (10.00, -3.00) rectangle (10.25, -3.25);
\filldraw[fill=bermuda, draw=black] (10.25, -3.00) rectangle (10.50, -3.25);
\filldraw[fill=cancan, draw=black] (10.50, -3.00) rectangle (10.75, -3.25);
\filldraw[fill=cancan, draw=black] (10.75, -3.00) rectangle (11.00, -3.25);
\filldraw[fill=bermuda, draw=black] (11.00, -3.00) rectangle (11.25, -3.25);
\filldraw[fill=bermuda, draw=black] (11.25, -3.00) rectangle (11.50, -3.25);
\filldraw[fill=cancan, draw=black] (11.50, -3.00) rectangle (11.75, -3.25);
\filldraw[fill=bermuda, draw=black] (11.75, -3.00) rectangle (12.00, -3.25);
\filldraw[fill=cancan, draw=black] (12.00, -3.00) rectangle (12.25, -3.25);
\filldraw[fill=cancan, draw=black] (12.25, -3.00) rectangle (12.50, -3.25);
\filldraw[fill=cancan, draw=black] (12.50, -3.00) rectangle (12.75, -3.25);
\filldraw[fill=bermuda, draw=black] (12.75, -3.00) rectangle (13.00, -3.25);
\filldraw[fill=bermuda, draw=black] (13.00, -3.00) rectangle (13.25, -3.25);
\filldraw[fill=bermuda, draw=black] (13.25, -3.00) rectangle (13.50, -3.25);
\filldraw[fill=bermuda, draw=black] (13.50, -3.00) rectangle (13.75, -3.25);
\filldraw[fill=bermuda, draw=black] (13.75, -3.00) rectangle (14.00, -3.25);
\filldraw[fill=cancan, draw=black] (14.00, -3.00) rectangle (14.25, -3.25);
\filldraw[fill=bermuda, draw=black] (14.25, -3.00) rectangle (14.50, -3.25);
\filldraw[fill=bermuda, draw=black] (14.50, -3.00) rectangle (14.75, -3.25);
\filldraw[fill=bermuda, draw=black] (14.75, -3.00) rectangle (15.00, -3.25);
\filldraw[fill=bermuda, draw=black] (0.00, -3.25) rectangle (0.25, -3.50);
\filldraw[fill=bermuda, draw=black] (0.25, -3.25) rectangle (0.50, -3.50);
\filldraw[fill=bermuda, draw=black] (0.50, -3.25) rectangle (0.75, -3.50);
\filldraw[fill=bermuda, draw=black] (0.75, -3.25) rectangle (1.00, -3.50);
\filldraw[fill=cancan, draw=black] (1.00, -3.25) rectangle (1.25, -3.50);
\filldraw[fill=bermuda, draw=black] (1.25, -3.25) rectangle (1.50, -3.50);
\filldraw[fill=cancan, draw=black] (1.50, -3.25) rectangle (1.75, -3.50);
\filldraw[fill=cancan, draw=black] (1.75, -3.25) rectangle (2.00, -3.50);
\filldraw[fill=cancan, draw=black] (2.00, -3.25) rectangle (2.25, -3.50);
\filldraw[fill=cancan, draw=black] (2.25, -3.25) rectangle (2.50, -3.50);
\filldraw[fill=cancan, draw=black] (2.50, -3.25) rectangle (2.75, -3.50);
\filldraw[fill=bermuda, draw=black] (2.75, -3.25) rectangle (3.00, -3.50);
\filldraw[fill=bermuda, draw=black] (3.00, -3.25) rectangle (3.25, -3.50);
\filldraw[fill=bermuda, draw=black] (3.25, -3.25) rectangle (3.50, -3.50);
\filldraw[fill=cancan, draw=black] (3.50, -3.25) rectangle (3.75, -3.50);
\filldraw[fill=cancan, draw=black] (3.75, -3.25) rectangle (4.00, -3.50);
\filldraw[fill=cancan, draw=black] (4.00, -3.25) rectangle (4.25, -3.50);
\filldraw[fill=bermuda, draw=black] (4.25, -3.25) rectangle (4.50, -3.50);
\filldraw[fill=bermuda, draw=black] (4.50, -3.25) rectangle (4.75, -3.50);
\filldraw[fill=bermuda, draw=black] (4.75, -3.25) rectangle (5.00, -3.50);
\filldraw[fill=cancan, draw=black] (5.00, -3.25) rectangle (5.25, -3.50);
\filldraw[fill=cancan, draw=black] (5.25, -3.25) rectangle (5.50, -3.50);
\filldraw[fill=cancan, draw=black] (5.50, -3.25) rectangle (5.75, -3.50);
\filldraw[fill=bermuda, draw=black] (5.75, -3.25) rectangle (6.00, -3.50);
\filldraw[fill=bermuda, draw=black] (6.00, -3.25) rectangle (6.25, -3.50);
\filldraw[fill=bermuda, draw=black] (6.25, -3.25) rectangle (6.50, -3.50);
\filldraw[fill=cancan, draw=black] (6.50, -3.25) rectangle (6.75, -3.50);
} } }\end{equation*}
\begin{equation*}
\hspace{0.3pt} b_{10} = \vcenter{\hbox{ \tikz{
\filldraw[fill=bermuda, draw=black] (0.00, 0.00) rectangle (0.25, -0.25);
\filldraw[fill=bermuda, draw=black] (0.25, 0.00) rectangle (0.50, -0.25);
\filldraw[fill=bermuda, draw=black] (0.50, 0.00) rectangle (0.75, -0.25);
\filldraw[fill=bermuda, draw=black] (0.75, 0.00) rectangle (1.00, -0.25);
\filldraw[fill=bermuda, draw=black] (1.00, 0.00) rectangle (1.25, -0.25);
\filldraw[fill=cancan, draw=black] (1.25, 0.00) rectangle (1.50, -0.25);
\filldraw[fill=bermuda, draw=black] (1.50, 0.00) rectangle (1.75, -0.25);
\filldraw[fill=cancan, draw=black] (1.75, 0.00) rectangle (2.00, -0.25);
\filldraw[fill=cancan, draw=black] (2.00, 0.00) rectangle (2.25, -0.25);
\filldraw[fill=cancan, draw=black] (2.25, 0.00) rectangle (2.50, -0.25);
\filldraw[fill=cancan, draw=black] (2.50, 0.00) rectangle (2.75, -0.25);
\filldraw[fill=cancan, draw=black] (2.75, 0.00) rectangle (3.00, -0.25);
\filldraw[fill=cancan, draw=black] (3.00, 0.00) rectangle (3.25, -0.25);
\filldraw[fill=bermuda, draw=black] (3.25, 0.00) rectangle (3.50, -0.25);
\filldraw[fill=bermuda, draw=black] (3.50, 0.00) rectangle (3.75, -0.25);
\filldraw[fill=cancan, draw=black] (3.75, 0.00) rectangle (4.00, -0.25);
\filldraw[fill=cancan, draw=black] (4.00, 0.00) rectangle (4.25, -0.25);
\filldraw[fill=cancan, draw=black] (4.25, 0.00) rectangle (4.50, -0.25);
\filldraw[fill=cancan, draw=black] (4.50, 0.00) rectangle (4.75, -0.25);
\filldraw[fill=cancan, draw=black] (4.75, 0.00) rectangle (5.00, -0.25);
\filldraw[fill=cancan, draw=black] (5.00, 0.00) rectangle (5.25, -0.25);
\filldraw[fill=cancan, draw=black] (5.25, 0.00) rectangle (5.50, -0.25);
\filldraw[fill=cancan, draw=black] (5.50, 0.00) rectangle (5.75, -0.25);
\filldraw[fill=cancan, draw=black] (5.75, 0.00) rectangle (6.00, -0.25);
\filldraw[fill=cancan, draw=black] (6.00, 0.00) rectangle (6.25, -0.25);
\filldraw[fill=cancan, draw=black] (6.25, 0.00) rectangle (6.50, -0.25);
\filldraw[fill=bermuda, draw=black] (6.50, 0.00) rectangle (6.75, -0.25);
\filldraw[fill=bermuda, draw=black] (6.75, 0.00) rectangle (7.00, -0.25);
\filldraw[fill=bermuda, draw=black] (7.00, 0.00) rectangle (7.25, -0.25);
\filldraw[fill=cancan, draw=black] (7.25, 0.00) rectangle (7.50, -0.25);
\filldraw[fill=cancan, draw=black] (7.50, 0.00) rectangle (7.75, -0.25);
\filldraw[fill=cancan, draw=black] (7.75, 0.00) rectangle (8.00, -0.25);
\filldraw[fill=cancan, draw=black] (8.00, 0.00) rectangle (8.25, -0.25);
\filldraw[fill=bermuda, draw=black] (8.25, 0.00) rectangle (8.50, -0.25);
\filldraw[fill=bermuda, draw=black] (8.50, 0.00) rectangle (8.75, -0.25);
\filldraw[fill=bermuda, draw=black] (8.75, 0.00) rectangle (9.00, -0.25);
\filldraw[fill=bermuda, draw=black] (9.00, 0.00) rectangle (9.25, -0.25);
\filldraw[fill=cancan, draw=black] (9.25, 0.00) rectangle (9.50, -0.25);
\filldraw[fill=cancan, draw=black] (9.50, 0.00) rectangle (9.75, -0.25);
\filldraw[fill=cancan, draw=black] (9.75, 0.00) rectangle (10.00, -0.25);
\filldraw[fill=cancan, draw=black] (10.00, 0.00) rectangle (10.25, -0.25);
\filldraw[fill=cancan, draw=black] (10.25, 0.00) rectangle (10.50, -0.25);
\filldraw[fill=cancan, draw=black] (10.50, 0.00) rectangle (10.75, -0.25);
\filldraw[fill=cancan, draw=black] (10.75, 0.00) rectangle (11.00, -0.25);
\filldraw[fill=bermuda, draw=black] (11.00, 0.00) rectangle (11.25, -0.25);
\filldraw[fill=bermuda, draw=black] (11.25, 0.00) rectangle (11.50, -0.25);
\filldraw[fill=bermuda, draw=black] (11.50, 0.00) rectangle (11.75, -0.25);
\filldraw[fill=cancan, draw=black] (11.75, 0.00) rectangle (12.00, -0.25);
\filldraw[fill=cancan, draw=black] (12.00, 0.00) rectangle (12.25, -0.25);
\filldraw[fill=cancan, draw=black] (12.25, 0.00) rectangle (12.50, -0.25);
\filldraw[fill=bermuda, draw=black] (12.50, 0.00) rectangle (12.75, -0.25);
\filldraw[fill=bermuda, draw=black] (12.75, 0.00) rectangle (13.00, -0.25);
\filldraw[fill=bermuda, draw=black] (13.00, 0.00) rectangle (13.25, -0.25);
\filldraw[fill=cancan, draw=black] (13.25, 0.00) rectangle (13.50, -0.25);
\filldraw[fill=cancan, draw=black] (13.50, 0.00) rectangle (13.75, -0.25);
\filldraw[fill=cancan, draw=black] (13.75, 0.00) rectangle (14.00, -0.25);
\filldraw[fill=bermuda, draw=black] (14.00, 0.00) rectangle (14.25, -0.25);
\filldraw[fill=bermuda, draw=black] (14.25, 0.00) rectangle (14.50, -0.25);
\filldraw[fill=bermuda, draw=black] (14.50, 0.00) rectangle (14.75, -0.25);
\filldraw[fill=cancan, draw=black] (14.75, 0.00) rectangle (15.00, -0.25);
\filldraw[fill=cancan, draw=black] (0.00, -0.25) rectangle (0.25, -0.50);
\filldraw[fill=cancan, draw=black] (0.25, -0.25) rectangle (0.50, -0.50);
\filldraw[fill=bermuda, draw=black] (0.50, -0.25) rectangle (0.75, -0.50);
\filldraw[fill=bermuda, draw=black] (0.75, -0.25) rectangle (1.00, -0.50);
\filldraw[fill=bermuda, draw=black] (1.00, -0.25) rectangle (1.25, -0.50);
\filldraw[fill=cancan, draw=black] (1.25, -0.25) rectangle (1.50, -0.50);
\filldraw[fill=cancan, draw=black] (1.50, -0.25) rectangle (1.75, -0.50);
\filldraw[fill=bermuda, draw=black] (1.75, -0.25) rectangle (2.00, -0.50);
\filldraw[fill=bermuda, draw=black] (2.00, -0.25) rectangle (2.25, -0.50);
\filldraw[fill=cancan, draw=black] (2.25, -0.25) rectangle (2.50, -0.50);
\filldraw[fill=cancan, draw=black] (2.50, -0.25) rectangle (2.75, -0.50);
\filldraw[fill=cancan, draw=black] (2.75, -0.25) rectangle (3.00, -0.50);
\filldraw[fill=bermuda, draw=black] (3.00, -0.25) rectangle (3.25, -0.50);
\filldraw[fill=bermuda, draw=black] (3.25, -0.25) rectangle (3.50, -0.50);
\filldraw[fill=bermuda, draw=black] (3.50, -0.25) rectangle (3.75, -0.50);
\filldraw[fill=cancan, draw=black] (3.75, -0.25) rectangle (4.00, -0.50);
\filldraw[fill=cancan, draw=black] (4.00, -0.25) rectangle (4.25, -0.50);
\filldraw[fill=cancan, draw=black] (4.25, -0.25) rectangle (4.50, -0.50);
\filldraw[fill=cancan, draw=black] (4.50, -0.25) rectangle (4.75, -0.50);
\filldraw[fill=cancan, draw=black] (4.75, -0.25) rectangle (5.00, -0.50);
\filldraw[fill=bermuda, draw=black] (5.00, -0.25) rectangle (5.25, -0.50);
\filldraw[fill=bermuda, draw=black] (5.25, -0.25) rectangle (5.50, -0.50);
\filldraw[fill=bermuda, draw=black] (5.50, -0.25) rectangle (5.75, -0.50);
\filldraw[fill=cancan, draw=black] (5.75, -0.25) rectangle (6.00, -0.50);
\filldraw[fill=cancan, draw=black] (6.00, -0.25) rectangle (6.25, -0.50);
\filldraw[fill=cancan, draw=black] (6.25, -0.25) rectangle (6.50, -0.50);
\filldraw[fill=bermuda, draw=black] (6.50, -0.25) rectangle (6.75, -0.50);
\filldraw[fill=bermuda, draw=black] (6.75, -0.25) rectangle (7.00, -0.50);
\filldraw[fill=bermuda, draw=black] (7.00, -0.25) rectangle (7.25, -0.50);
\filldraw[fill=cancan, draw=black] (7.25, -0.25) rectangle (7.50, -0.50);
\filldraw[fill=cancan, draw=black] (7.50, -0.25) rectangle (7.75, -0.50);
\filldraw[fill=cancan, draw=black] (7.75, -0.25) rectangle (8.00, -0.50);
\filldraw[fill=bermuda, draw=black] (8.00, -0.25) rectangle (8.25, -0.50);
\filldraw[fill=cancan, draw=black] (8.25, -0.25) rectangle (8.50, -0.50);
\filldraw[fill=cancan, draw=black] (8.50, -0.25) rectangle (8.75, -0.50);
\filldraw[fill=cancan, draw=black] (8.75, -0.25) rectangle (9.00, -0.50);
\filldraw[fill=cancan, draw=black] (9.00, -0.25) rectangle (9.25, -0.50);
\filldraw[fill=cancan, draw=black] (9.25, -0.25) rectangle (9.50, -0.50);
\filldraw[fill=bermuda, draw=black] (9.50, -0.25) rectangle (9.75, -0.50);
\filldraw[fill=bermuda, draw=black] (9.75, -0.25) rectangle (10.00, -0.50);
\filldraw[fill=bermuda, draw=black] (10.00, -0.25) rectangle (10.25, -0.50);
\filldraw[fill=cancan, draw=black] (10.25, -0.25) rectangle (10.50, -0.50);
\filldraw[fill=cancan, draw=black] (10.50, -0.25) rectangle (10.75, -0.50);
\filldraw[fill=cancan, draw=black] (10.75, -0.25) rectangle (11.00, -0.50);
\filldraw[fill=bermuda, draw=black] (11.00, -0.25) rectangle (11.25, -0.50);
\filldraw[fill=bermuda, draw=black] (11.25, -0.25) rectangle (11.50, -0.50);
\filldraw[fill=bermuda, draw=black] (11.50, -0.25) rectangle (11.75, -0.50);
\filldraw[fill=cancan, draw=black] (11.75, -0.25) rectangle (12.00, -0.50);
\filldraw[fill=cancan, draw=black] (12.00, -0.25) rectangle (12.25, -0.50);
\filldraw[fill=cancan, draw=black] (12.25, -0.25) rectangle (12.50, -0.50);
\filldraw[fill=bermuda, draw=black] (12.50, -0.25) rectangle (12.75, -0.50);
\filldraw[fill=bermuda, draw=black] (12.75, -0.25) rectangle (13.00, -0.50);
\filldraw[fill=bermuda, draw=black] (13.00, -0.25) rectangle (13.25, -0.50);
\filldraw[fill=cancan, draw=black] (13.25, -0.25) rectangle (13.50, -0.50);
\filldraw[fill=bermuda, draw=black] (13.50, -0.25) rectangle (13.75, -0.50);
\filldraw[fill=cancan, draw=black] (13.75, -0.25) rectangle (14.00, -0.50);
\filldraw[fill=bermuda, draw=black] (14.00, -0.25) rectangle (14.25, -0.50);
\filldraw[fill=bermuda, draw=black] (14.25, -0.25) rectangle (14.50, -0.50);
\filldraw[fill=bermuda, draw=black] (14.50, -0.25) rectangle (14.75, -0.50);
\filldraw[fill=cancan, draw=black] (14.75, -0.25) rectangle (15.00, -0.50);
\filldraw[fill=cancan, draw=black] (0.00, -0.50) rectangle (0.25, -0.75);
\filldraw[fill=cancan, draw=black] (0.25, -0.50) rectangle (0.50, -0.75);
\filldraw[fill=cancan, draw=black] (0.50, -0.50) rectangle (0.75, -0.75);
\filldraw[fill=bermuda, draw=black] (0.75, -0.50) rectangle (1.00, -0.75);
\filldraw[fill=bermuda, draw=black] (1.00, -0.50) rectangle (1.25, -0.75);
\filldraw[fill=cancan, draw=black] (1.25, -0.50) rectangle (1.50, -0.75);
\filldraw[fill=bermuda, draw=black] (1.50, -0.50) rectangle (1.75, -0.75);
\filldraw[fill=cancan, draw=black] (1.75, -0.50) rectangle (2.00, -0.75);
\filldraw[fill=cancan, draw=black] (2.00, -0.50) rectangle (2.25, -0.75);
\filldraw[fill=cancan, draw=black] (2.25, -0.50) rectangle (2.50, -0.75);
\filldraw[fill=bermuda, draw=black] (2.50, -0.50) rectangle (2.75, -0.75);
\filldraw[fill=cancan, draw=black] (2.75, -0.50) rectangle (3.00, -0.75);
\filldraw[fill=cancan, draw=black] (3.00, -0.50) rectangle (3.25, -0.75);
\filldraw[fill=bermuda, draw=black] (3.25, -0.50) rectangle (3.50, -0.75);
\filldraw[fill=bermuda, draw=black] (3.50, -0.50) rectangle (3.75, -0.75);
\filldraw[fill=cancan, draw=black] (3.75, -0.50) rectangle (4.00, -0.75);
\filldraw[fill=bermuda, draw=black] (4.00, -0.50) rectangle (4.25, -0.75);
\filldraw[fill=cancan, draw=black] (4.25, -0.50) rectangle (4.50, -0.75);
\filldraw[fill=cancan, draw=black] (4.50, -0.50) rectangle (4.75, -0.75);
\filldraw[fill=bermuda, draw=black] (4.75, -0.50) rectangle (5.00, -0.75);
\filldraw[fill=bermuda, draw=black] (5.00, -0.50) rectangle (5.25, -0.75);
\filldraw[fill=cancan, draw=black] (5.25, -0.50) rectangle (5.50, -0.75);
\filldraw[fill=bermuda, draw=black] (5.50, -0.50) rectangle (5.75, -0.75);
\filldraw[fill=cancan, draw=black] (5.75, -0.50) rectangle (6.00, -0.75);
\filldraw[fill=bermuda, draw=black] (6.00, -0.50) rectangle (6.25, -0.75);
\filldraw[fill=cancan, draw=black] (6.25, -0.50) rectangle (6.50, -0.75);
\filldraw[fill=bermuda, draw=black] (6.50, -0.50) rectangle (6.75, -0.75);
\filldraw[fill=cancan, draw=black] (6.75, -0.50) rectangle (7.00, -0.75);
\filldraw[fill=cancan, draw=black] (7.00, -0.50) rectangle (7.25, -0.75);
\filldraw[fill=bermuda, draw=black] (7.25, -0.50) rectangle (7.50, -0.75);
\filldraw[fill=bermuda, draw=black] (7.50, -0.50) rectangle (7.75, -0.75);
\filldraw[fill=bermuda, draw=black] (7.75, -0.50) rectangle (8.00, -0.75);
\filldraw[fill=bermuda, draw=black] (8.00, -0.50) rectangle (8.25, -0.75);
\filldraw[fill=cancan, draw=black] (8.25, -0.50) rectangle (8.50, -0.75);
\filldraw[fill=bermuda, draw=black] (8.50, -0.50) rectangle (8.75, -0.75);
\filldraw[fill=bermuda, draw=black] (8.75, -0.50) rectangle (9.00, -0.75);
\filldraw[fill=bermuda, draw=black] (9.00, -0.50) rectangle (9.25, -0.75);
\filldraw[fill=bermuda, draw=black] (9.25, -0.50) rectangle (9.50, -0.75);
\filldraw[fill=bermuda, draw=black] (9.50, -0.50) rectangle (9.75, -0.75);
\filldraw[fill=cancan, draw=black] (9.75, -0.50) rectangle (10.00, -0.75);
\filldraw[fill=bermuda, draw=black] (10.00, -0.50) rectangle (10.25, -0.75);
\filldraw[fill=cancan, draw=black] (10.25, -0.50) rectangle (10.50, -0.75);
\filldraw[fill=cancan, draw=black] (10.50, -0.50) rectangle (10.75, -0.75);
\filldraw[fill=cancan, draw=black] (10.75, -0.50) rectangle (11.00, -0.75);
\filldraw[fill=cancan, draw=black] (11.00, -0.50) rectangle (11.25, -0.75);
\filldraw[fill=cancan, draw=black] (11.25, -0.50) rectangle (11.50, -0.75);
\filldraw[fill=bermuda, draw=black] (11.50, -0.50) rectangle (11.75, -0.75);
\filldraw[fill=bermuda, draw=black] (11.75, -0.50) rectangle (12.00, -0.75);
\filldraw[fill=bermuda, draw=black] (12.00, -0.50) rectangle (12.25, -0.75);
\filldraw[fill=cancan, draw=black] (12.25, -0.50) rectangle (12.50, -0.75);
\filldraw[fill=cancan, draw=black] (12.50, -0.50) rectangle (12.75, -0.75);
\filldraw[fill=cancan, draw=black] (12.75, -0.50) rectangle (13.00, -0.75);
\filldraw[fill=bermuda, draw=black] (13.00, -0.50) rectangle (13.25, -0.75);
\filldraw[fill=bermuda, draw=black] (13.25, -0.50) rectangle (13.50, -0.75);
\filldraw[fill=bermuda, draw=black] (13.50, -0.50) rectangle (13.75, -0.75);
\filldraw[fill=cancan, draw=black] (13.75, -0.50) rectangle (14.00, -0.75);
\filldraw[fill=cancan, draw=black] (14.00, -0.50) rectangle (14.25, -0.75);
\filldraw[fill=cancan, draw=black] (14.25, -0.50) rectangle (14.50, -0.75);
\filldraw[fill=cancan, draw=black] (14.50, -0.50) rectangle (14.75, -0.75);
\filldraw[fill=cancan, draw=black] (14.75, -0.50) rectangle (15.00, -0.75);
\filldraw[fill=bermuda, draw=black] (0.00, -0.75) rectangle (0.25, -1.00);
\filldraw[fill=bermuda, draw=black] (0.25, -0.75) rectangle (0.50, -1.00);
\filldraw[fill=bermuda, draw=black] (0.50, -0.75) rectangle (0.75, -1.00);
\filldraw[fill=bermuda, draw=black] (0.75, -0.75) rectangle (1.00, -1.00);
\filldraw[fill=bermuda, draw=black] (1.00, -0.75) rectangle (1.25, -1.00);
\filldraw[fill=cancan, draw=black] (1.25, -0.75) rectangle (1.50, -1.00);
\filldraw[fill=bermuda, draw=black] (1.50, -0.75) rectangle (1.75, -1.00);
\filldraw[fill=bermuda, draw=black] (1.75, -0.75) rectangle (2.00, -1.00);
\filldraw[fill=bermuda, draw=black] (2.00, -0.75) rectangle (2.25, -1.00);
\filldraw[fill=cancan, draw=black] (2.25, -0.75) rectangle (2.50, -1.00);
\filldraw[fill=cancan, draw=black] (2.50, -0.75) rectangle (2.75, -1.00);
\filldraw[fill=cancan, draw=black] (2.75, -0.75) rectangle (3.00, -1.00);
\filldraw[fill=bermuda, draw=black] (3.00, -0.75) rectangle (3.25, -1.00);
\filldraw[fill=bermuda, draw=black] (3.25, -0.75) rectangle (3.50, -1.00);
\filldraw[fill=bermuda, draw=black] (3.50, -0.75) rectangle (3.75, -1.00);
\filldraw[fill=cancan, draw=black] (3.75, -0.75) rectangle (4.00, -1.00);
\filldraw[fill=cancan, draw=black] (4.00, -0.75) rectangle (4.25, -1.00);
\filldraw[fill=cancan, draw=black] (4.25, -0.75) rectangle (4.50, -1.00);
\filldraw[fill=bermuda, draw=black] (4.50, -0.75) rectangle (4.75, -1.00);
\filldraw[fill=bermuda, draw=black] (4.75, -0.75) rectangle (5.00, -1.00);
\filldraw[fill=bermuda, draw=black] (5.00, -0.75) rectangle (5.25, -1.00);
\filldraw[fill=bermuda, draw=black] (5.25, -0.75) rectangle (5.50, -1.00);
\filldraw[fill=bermuda, draw=black] (5.50, -0.75) rectangle (5.75, -1.00);
\filldraw[fill=cancan, draw=black] (5.75, -0.75) rectangle (6.00, -1.00);
\filldraw[fill=bermuda, draw=black] (6.00, -0.75) rectangle (6.25, -1.00);
\filldraw[fill=cancan, draw=black] (6.25, -0.75) rectangle (6.50, -1.00);
\filldraw[fill=cancan, draw=black] (6.50, -0.75) rectangle (6.75, -1.00);
\filldraw[fill=cancan, draw=black] (6.75, -0.75) rectangle (7.00, -1.00);
\filldraw[fill=cancan, draw=black] (7.00, -0.75) rectangle (7.25, -1.00);
\filldraw[fill=cancan, draw=black] (7.25, -0.75) rectangle (7.50, -1.00);
\filldraw[fill=cancan, draw=black] (7.50, -0.75) rectangle (7.75, -1.00);
\filldraw[fill=cancan, draw=black] (7.75, -0.75) rectangle (8.00, -1.00);
\filldraw[fill=cancan, draw=black] (8.00, -0.75) rectangle (8.25, -1.00);
\filldraw[fill=bermuda, draw=black] (8.25, -0.75) rectangle (8.50, -1.00);
\filldraw[fill=bermuda, draw=black] (8.50, -0.75) rectangle (8.75, -1.00);
\filldraw[fill=cancan, draw=black] (8.75, -0.75) rectangle (9.00, -1.00);
\filldraw[fill=bermuda, draw=black] (9.00, -0.75) rectangle (9.25, -1.00);
\filldraw[fill=cancan, draw=black] (9.25, -0.75) rectangle (9.50, -1.00);
\filldraw[fill=bermuda, draw=black] (9.50, -0.75) rectangle (9.75, -1.00);
\filldraw[fill=bermuda, draw=black] (9.75, -0.75) rectangle (10.00, -1.00);
\filldraw[fill=bermuda, draw=black] (10.00, -0.75) rectangle (10.25, -1.00);
\filldraw[fill=cancan, draw=black] (10.25, -0.75) rectangle (10.50, -1.00);
\filldraw[fill=bermuda, draw=black] (10.50, -0.75) rectangle (10.75, -1.00);
\filldraw[fill=bermuda, draw=black] (10.75, -0.75) rectangle (11.00, -1.00);
\filldraw[fill=bermuda, draw=black] (11.00, -0.75) rectangle (11.25, -1.00);
\filldraw[fill=bermuda, draw=black] (11.25, -0.75) rectangle (11.50, -1.00);
\filldraw[fill=bermuda, draw=black] (11.50, -0.75) rectangle (11.75, -1.00);
\filldraw[fill=cancan, draw=black] (11.75, -0.75) rectangle (12.00, -1.00);
\filldraw[fill=bermuda, draw=black] (12.00, -0.75) rectangle (12.25, -1.00);
\filldraw[fill=bermuda, draw=black] (12.25, -0.75) rectangle (12.50, -1.00);
\filldraw[fill=bermuda, draw=black] (12.50, -0.75) rectangle (12.75, -1.00);
\filldraw[fill=bermuda, draw=black] (12.75, -0.75) rectangle (13.00, -1.00);
\filldraw[fill=bermuda, draw=black] (13.00, -0.75) rectangle (13.25, -1.00);
\filldraw[fill=cancan, draw=black] (13.25, -0.75) rectangle (13.50, -1.00);
\filldraw[fill=bermuda, draw=black] (13.50, -0.75) rectangle (13.75, -1.00);
\filldraw[fill=bermuda, draw=black] (13.75, -0.75) rectangle (14.00, -1.00);
\filldraw[fill=bermuda, draw=black] (14.00, -0.75) rectangle (14.25, -1.00);
\filldraw[fill=bermuda, draw=black] (14.25, -0.75) rectangle (14.50, -1.00);
\filldraw[fill=bermuda, draw=black] (14.50, -0.75) rectangle (14.75, -1.00);
\filldraw[fill=cancan, draw=black] (14.75, -0.75) rectangle (15.00, -1.00);
\filldraw[fill=bermuda, draw=black] (0.00, -1.00) rectangle (0.25, -1.25);
\filldraw[fill=cancan, draw=black] (0.25, -1.00) rectangle (0.50, -1.25);
\filldraw[fill=cancan, draw=black] (0.50, -1.00) rectangle (0.75, -1.25);
\filldraw[fill=cancan, draw=black] (0.75, -1.00) rectangle (1.00, -1.25);
\filldraw[fill=cancan, draw=black] (1.00, -1.00) rectangle (1.25, -1.25);
\filldraw[fill=bermuda, draw=black] (1.25, -1.00) rectangle (1.50, -1.25);
\filldraw[fill=bermuda, draw=black] (1.50, -1.00) rectangle (1.75, -1.25);
\filldraw[fill=cancan, draw=black] (1.75, -1.00) rectangle (2.00, -1.25);
\filldraw[fill=cancan, draw=black] (2.00, -1.00) rectangle (2.25, -1.25);
\filldraw[fill=cancan, draw=black] (2.25, -1.00) rectangle (2.50, -1.25);
\filldraw[fill=bermuda, draw=black] (2.50, -1.00) rectangle (2.75, -1.25);
\filldraw[fill=bermuda, draw=black] (2.75, -1.00) rectangle (3.00, -1.25);
\filldraw[fill=bermuda, draw=black] (3.00, -1.00) rectangle (3.25, -1.25);
\filldraw[fill=cancan, draw=black] (3.25, -1.00) rectangle (3.50, -1.25);
\filldraw[fill=cancan, draw=black] (3.50, -1.00) rectangle (3.75, -1.25);
\filldraw[fill=cancan, draw=black] (3.75, -1.00) rectangle (4.00, -1.25);
\filldraw[fill=bermuda, draw=black] (4.00, -1.00) rectangle (4.25, -1.25);
\filldraw[fill=bermuda, draw=black] (4.25, -1.00) rectangle (4.50, -1.25);
\filldraw[fill=bermuda, draw=black] (4.50, -1.00) rectangle (4.75, -1.25);
\filldraw[fill=cancan, draw=black] (4.75, -1.00) rectangle (5.00, -1.25);
\filldraw[fill=cancan, draw=black] (5.00, -1.00) rectangle (5.25, -1.25);
\filldraw[fill=cancan, draw=black] (5.25, -1.00) rectangle (5.50, -1.25);
\filldraw[fill=bermuda, draw=black] (5.50, -1.00) rectangle (5.75, -1.25);
\filldraw[fill=bermuda, draw=black] (5.75, -1.00) rectangle (6.00, -1.25);
\filldraw[fill=bermuda, draw=black] (6.00, -1.00) rectangle (6.25, -1.25);
\filldraw[fill=cancan, draw=black] (6.25, -1.00) rectangle (6.50, -1.25);
\filldraw[fill=bermuda, draw=black] (6.50, -1.00) rectangle (6.75, -1.25);
\filldraw[fill=cancan, draw=black] (6.75, -1.00) rectangle (7.00, -1.25);
\filldraw[fill=cancan, draw=black] (7.00, -1.00) rectangle (7.25, -1.25);
\filldraw[fill=cancan, draw=black] (7.25, -1.00) rectangle (7.50, -1.25);
\filldraw[fill=cancan, draw=black] (7.50, -1.00) rectangle (7.75, -1.25);
\filldraw[fill=cancan, draw=black] (7.75, -1.00) rectangle (8.00, -1.25);
\filldraw[fill=bermuda, draw=black] (8.00, -1.00) rectangle (8.25, -1.25);
\filldraw[fill=bermuda, draw=black] (8.25, -1.00) rectangle (8.50, -1.25);
\filldraw[fill=bermuda, draw=black] (8.50, -1.00) rectangle (8.75, -1.25);
\filldraw[fill=cancan, draw=black] (8.75, -1.00) rectangle (9.00, -1.25);
\filldraw[fill=bermuda, draw=black] (9.00, -1.00) rectangle (9.25, -1.25);
\filldraw[fill=bermuda, draw=black] (9.25, -1.00) rectangle (9.50, -1.25);
\filldraw[fill=bermuda, draw=black] (9.50, -1.00) rectangle (9.75, -1.25);
\filldraw[fill=bermuda, draw=black] (9.75, -1.00) rectangle (10.00, -1.25);
\filldraw[fill=bermuda, draw=black] (10.00, -1.00) rectangle (10.25, -1.25);
\filldraw[fill=cancan, draw=black] (10.25, -1.00) rectangle (10.50, -1.25);
\filldraw[fill=cancan, draw=black] (10.50, -1.00) rectangle (10.75, -1.25);
\filldraw[fill=cancan, draw=black] (10.75, -1.00) rectangle (11.00, -1.25);
\filldraw[fill=cancan, draw=black] (11.00, -1.00) rectangle (11.25, -1.25);
\filldraw[fill=bermuda, draw=black] (11.25, -1.00) rectangle (11.50, -1.25);
\filldraw[fill=bermuda, draw=black] (11.50, -1.00) rectangle (11.75, -1.25);
\filldraw[fill=cancan, draw=black] (11.75, -1.00) rectangle (12.00, -1.25);
\filldraw[fill=cancan, draw=black] (12.00, -1.00) rectangle (12.25, -1.25);
\filldraw[fill=cancan, draw=black] (12.25, -1.00) rectangle (12.50, -1.25);
\filldraw[fill=cancan, draw=black] (12.50, -1.00) rectangle (12.75, -1.25);
\filldraw[fill=bermuda, draw=black] (12.75, -1.00) rectangle (13.00, -1.25);
\filldraw[fill=bermuda, draw=black] (13.00, -1.00) rectangle (13.25, -1.25);
\filldraw[fill=cancan, draw=black] (13.25, -1.00) rectangle (13.50, -1.25);
\filldraw[fill=bermuda, draw=black] (13.50, -1.00) rectangle (13.75, -1.25);
\filldraw[fill=bermuda, draw=black] (13.75, -1.00) rectangle (14.00, -1.25);
\filldraw[fill=bermuda, draw=black] (14.00, -1.00) rectangle (14.25, -1.25);
\filldraw[fill=bermuda, draw=black] (14.25, -1.00) rectangle (14.50, -1.25);
\filldraw[fill=bermuda, draw=black] (14.50, -1.00) rectangle (14.75, -1.25);
\filldraw[fill=cancan, draw=black] (14.75, -1.00) rectangle (15.00, -1.25);
\filldraw[fill=cancan, draw=black] (0.00, -1.25) rectangle (0.25, -1.50);
\filldraw[fill=cancan, draw=black] (0.25, -1.25) rectangle (0.50, -1.50);
\filldraw[fill=bermuda, draw=black] (0.50, -1.25) rectangle (0.75, -1.50);
\filldraw[fill=bermuda, draw=black] (0.75, -1.25) rectangle (1.00, -1.50);
\filldraw[fill=bermuda, draw=black] (1.00, -1.25) rectangle (1.25, -1.50);
\filldraw[fill=cancan, draw=black] (1.25, -1.25) rectangle (1.50, -1.50);
\filldraw[fill=cancan, draw=black] (1.50, -1.25) rectangle (1.75, -1.50);
\filldraw[fill=cancan, draw=black] (1.75, -1.25) rectangle (2.00, -1.50);
\filldraw[fill=bermuda, draw=black] (2.00, -1.25) rectangle (2.25, -1.50);
\filldraw[fill=bermuda, draw=black] (2.25, -1.25) rectangle (2.50, -1.50);
\filldraw[fill=bermuda, draw=black] (2.50, -1.25) rectangle (2.75, -1.50);
\filldraw[fill=cancan, draw=black] (2.75, -1.25) rectangle (3.00, -1.50);
\filldraw[fill=bermuda, draw=black] (3.00, -1.25) rectangle (3.25, -1.50);
\filldraw[fill=bermuda, draw=black] (3.25, -1.25) rectangle (3.50, -1.50);
\filldraw[fill=bermuda, draw=black] (3.50, -1.25) rectangle (3.75, -1.50);
\filldraw[fill=cancan, draw=black] (3.75, -1.25) rectangle (4.00, -1.50);
\filldraw[fill=bermuda, draw=black] (4.00, -1.25) rectangle (4.25, -1.50);
\filldraw[fill=cancan, draw=black] (4.25, -1.25) rectangle (4.50, -1.50);
\filldraw[fill=cancan, draw=black] (4.50, -1.25) rectangle (4.75, -1.50);
\filldraw[fill=cancan, draw=black] (4.75, -1.25) rectangle (5.00, -1.50);
\filldraw[fill=cancan, draw=black] (5.00, -1.25) rectangle (5.25, -1.50);
\filldraw[fill=cancan, draw=black] (5.25, -1.25) rectangle (5.50, -1.50);
\filldraw[fill=bermuda, draw=black] (5.50, -1.25) rectangle (5.75, -1.50);
\filldraw[fill=bermuda, draw=black] (5.75, -1.25) rectangle (6.00, -1.50);
\filldraw[fill=bermuda, draw=black] (6.00, -1.25) rectangle (6.25, -1.50);
\filldraw[fill=cancan, draw=black] (6.25, -1.25) rectangle (6.50, -1.50);
\filldraw[fill=cancan, draw=black] (6.50, -1.25) rectangle (6.75, -1.50);
\filldraw[fill=cancan, draw=black] (6.75, -1.25) rectangle (7.00, -1.50);
\filldraw[fill=bermuda, draw=black] (7.00, -1.25) rectangle (7.25, -1.50);
\filldraw[fill=bermuda, draw=black] (7.25, -1.25) rectangle (7.50, -1.50);
\filldraw[fill=bermuda, draw=black] (7.50, -1.25) rectangle (7.75, -1.50);
\filldraw[fill=cancan, draw=black] (7.75, -1.25) rectangle (8.00, -1.50);
\filldraw[fill=cancan, draw=black] (8.00, -1.25) rectangle (8.25, -1.50);
\filldraw[fill=cancan, draw=black] (8.25, -1.25) rectangle (8.50, -1.50);
\filldraw[fill=cancan, draw=black] (8.50, -1.25) rectangle (8.75, -1.50);
\filldraw[fill=cancan, draw=black] (8.75, -1.25) rectangle (9.00, -1.50);
\filldraw[fill=cancan, draw=black] (9.00, -1.25) rectangle (9.25, -1.50);
\filldraw[fill=cancan, draw=black] (9.25, -1.25) rectangle (9.50, -1.50);
\filldraw[fill=bermuda, draw=black] (9.50, -1.25) rectangle (9.75, -1.50);
\filldraw[fill=cancan, draw=black] (9.75, -1.25) rectangle (10.00, -1.50);
\filldraw[fill=bermuda, draw=black] (10.00, -1.25) rectangle (10.25, -1.50);
\filldraw[fill=bermuda, draw=black] (10.25, -1.25) rectangle (10.50, -1.50);
\filldraw[fill=bermuda, draw=black] (10.50, -1.25) rectangle (10.75, -1.50);
\filldraw[fill=cancan, draw=black] (10.75, -1.25) rectangle (11.00, -1.50);
\filldraw[fill=cancan, draw=black] (11.00, -1.25) rectangle (11.25, -1.50);
\filldraw[fill=cancan, draw=black] (11.25, -1.25) rectangle (11.50, -1.50);
\filldraw[fill=bermuda, draw=black] (11.50, -1.25) rectangle (11.75, -1.50);
\filldraw[fill=bermuda, draw=black] (11.75, -1.25) rectangle (12.00, -1.50);
\filldraw[fill=bermuda, draw=black] (12.00, -1.25) rectangle (12.25, -1.50);
\filldraw[fill=cancan, draw=black] (12.25, -1.25) rectangle (12.50, -1.50);
\filldraw[fill=cancan, draw=black] (12.50, -1.25) rectangle (12.75, -1.50);
\filldraw[fill=cancan, draw=black] (12.75, -1.25) rectangle (13.00, -1.50);
\filldraw[fill=bermuda, draw=black] (13.00, -1.25) rectangle (13.25, -1.50);
\filldraw[fill=bermuda, draw=black] (13.25, -1.25) rectangle (13.50, -1.50);
\filldraw[fill=bermuda, draw=black] (13.50, -1.25) rectangle (13.75, -1.50);
\filldraw[fill=cancan, draw=black] (13.75, -1.25) rectangle (14.00, -1.50);
\filldraw[fill=cancan, draw=black] (14.00, -1.25) rectangle (14.25, -1.50);
\filldraw[fill=cancan, draw=black] (14.25, -1.25) rectangle (14.50, -1.50);
\filldraw[fill=cancan, draw=black] (14.50, -1.25) rectangle (14.75, -1.50);
\filldraw[fill=cancan, draw=black] (14.75, -1.25) rectangle (15.00, -1.50);
\filldraw[fill=cancan, draw=black] (0.00, -1.50) rectangle (0.25, -1.75);
\filldraw[fill=cancan, draw=black] (0.25, -1.50) rectangle (0.50, -1.75);
\filldraw[fill=cancan, draw=black] (0.50, -1.50) rectangle (0.75, -1.75);
\filldraw[fill=cancan, draw=black] (0.75, -1.50) rectangle (1.00, -1.75);
\filldraw[fill=cancan, draw=black] (1.00, -1.50) rectangle (1.25, -1.75);
\filldraw[fill=cancan, draw=black] (1.25, -1.50) rectangle (1.50, -1.75);
\filldraw[fill=cancan, draw=black] (1.50, -1.50) rectangle (1.75, -1.75);
\filldraw[fill=cancan, draw=black] (1.75, -1.50) rectangle (2.00, -1.75);
\filldraw[fill=bermuda, draw=black] (2.00, -1.50) rectangle (2.25, -1.75);
\filldraw[fill=cancan, draw=black] (2.25, -1.50) rectangle (2.50, -1.75);
\filldraw[fill=bermuda, draw=black] (2.50, -1.50) rectangle (2.75, -1.75);
\filldraw[fill=cancan, draw=black] (2.75, -1.50) rectangle (3.00, -1.75);
\filldraw[fill=cancan, draw=black] (3.00, -1.50) rectangle (3.25, -1.75);
\filldraw[fill=bermuda, draw=black] (3.25, -1.50) rectangle (3.50, -1.75);
\filldraw[fill=bermuda, draw=black] (3.50, -1.50) rectangle (3.75, -1.75);
\filldraw[fill=bermuda, draw=black] (3.75, -1.50) rectangle (4.00, -1.75);
\filldraw[fill=bermuda, draw=black] (4.00, -1.50) rectangle (4.25, -1.75);
\filldraw[fill=bermuda, draw=black] (4.25, -1.50) rectangle (4.50, -1.75);
\filldraw[fill=bermuda, draw=black] (4.50, -1.50) rectangle (4.75, -1.75);
\filldraw[fill=cancan, draw=black] (4.75, -1.50) rectangle (5.00, -1.75);
\filldraw[fill=cancan, draw=black] (5.00, -1.50) rectangle (5.25, -1.75);
\filldraw[fill=bermuda, draw=black] (5.25, -1.50) rectangle (5.50, -1.75);
\filldraw[fill=bermuda, draw=black] (5.50, -1.50) rectangle (5.75, -1.75);
\filldraw[fill=cancan, draw=black] (5.75, -1.50) rectangle (6.00, -1.75);
\filldraw[fill=cancan, draw=black] (6.00, -1.50) rectangle (6.25, -1.75);
\filldraw[fill=cancan, draw=black] (6.25, -1.50) rectangle (6.50, -1.75);
\filldraw[fill=cancan, draw=black] (6.50, -1.50) rectangle (6.75, -1.75);
\filldraw[fill=cancan, draw=black] (6.75, -1.50) rectangle (7.00, -1.75);
\filldraw[fill=bermuda, draw=black] (7.00, -1.50) rectangle (7.25, -1.75);
\filldraw[fill=cancan, draw=black] (7.25, -1.50) rectangle (7.50, -1.75);
\filldraw[fill=cancan, draw=black] (7.50, -1.50) rectangle (7.75, -1.75);
\filldraw[fill=cancan, draw=black] (7.75, -1.50) rectangle (8.00, -1.75);
\filldraw[fill=cancan, draw=black] (8.00, -1.50) rectangle (8.25, -1.75);
\filldraw[fill=cancan, draw=black] (8.25, -1.50) rectangle (8.50, -1.75);
\filldraw[fill=bermuda, draw=black] (8.50, -1.50) rectangle (8.75, -1.75);
\filldraw[fill=bermuda, draw=black] (8.75, -1.50) rectangle (9.00, -1.75);
\filldraw[fill=bermuda, draw=black] (9.00, -1.50) rectangle (9.25, -1.75);
\filldraw[fill=cancan, draw=black] (9.25, -1.50) rectangle (9.50, -1.75);
\filldraw[fill=cancan, draw=black] (9.50, -1.50) rectangle (9.75, -1.75);
\filldraw[fill=cancan, draw=black] (9.75, -1.50) rectangle (10.00, -1.75);
\filldraw[fill=bermuda, draw=black] (10.00, -1.50) rectangle (10.25, -1.75);
\filldraw[fill=bermuda, draw=black] (10.25, -1.50) rectangle (10.50, -1.75);
\filldraw[fill=bermuda, draw=black] (10.50, -1.50) rectangle (10.75, -1.75);
\filldraw[fill=cancan, draw=black] (10.75, -1.50) rectangle (11.00, -1.75);
\filldraw[fill=cancan, draw=black] (11.00, -1.50) rectangle (11.25, -1.75);
\filldraw[fill=cancan, draw=black] (11.25, -1.50) rectangle (11.50, -1.75);
\filldraw[fill=bermuda, draw=black] (11.50, -1.50) rectangle (11.75, -1.75);
\filldraw[fill=bermuda, draw=black] (11.75, -1.50) rectangle (12.00, -1.75);
\filldraw[fill=bermuda, draw=black] (12.00, -1.50) rectangle (12.25, -1.75);
\filldraw[fill=cancan, draw=black] (12.25, -1.50) rectangle (12.50, -1.75);
\filldraw[fill=cancan, draw=black] (12.50, -1.50) rectangle (12.75, -1.75);
\filldraw[fill=cancan, draw=black] (12.75, -1.50) rectangle (13.00, -1.75);
\filldraw[fill=bermuda, draw=black] (13.00, -1.50) rectangle (13.25, -1.75);
\filldraw[fill=bermuda, draw=black] (13.25, -1.50) rectangle (13.50, -1.75);
\filldraw[fill=bermuda, draw=black] (13.50, -1.50) rectangle (13.75, -1.75);
\filldraw[fill=cancan, draw=black] (13.75, -1.50) rectangle (14.00, -1.75);
\filldraw[fill=bermuda, draw=black] (14.00, -1.50) rectangle (14.25, -1.75);
\filldraw[fill=cancan, draw=black] (14.25, -1.50) rectangle (14.50, -1.75);
\filldraw[fill=bermuda, draw=black] (14.50, -1.50) rectangle (14.75, -1.75);
\filldraw[fill=cancan, draw=black] (14.75, -1.50) rectangle (15.00, -1.75);
\filldraw[fill=bermuda, draw=black] (0.00, -1.75) rectangle (0.25, -2.00);
\filldraw[fill=cancan, draw=black] (0.25, -1.75) rectangle (0.50, -2.00);
\filldraw[fill=bermuda, draw=black] (0.50, -1.75) rectangle (0.75, -2.00);
\filldraw[fill=cancan, draw=black] (0.75, -1.75) rectangle (1.00, -2.00);
\filldraw[fill=cancan, draw=black] (1.00, -1.75) rectangle (1.25, -2.00);
\filldraw[fill=cancan, draw=black] (1.25, -1.75) rectangle (1.50, -2.00);
\filldraw[fill=bermuda, draw=black] (1.50, -1.75) rectangle (1.75, -2.00);
\filldraw[fill=bermuda, draw=black] (1.75, -1.75) rectangle (2.00, -2.00);
\filldraw[fill=bermuda, draw=black] (2.00, -1.75) rectangle (2.25, -2.00);
\filldraw[fill=cancan, draw=black] (2.25, -1.75) rectangle (2.50, -2.00);
\filldraw[fill=cancan, draw=black] (2.50, -1.75) rectangle (2.75, -2.00);
\filldraw[fill=cancan, draw=black] (2.75, -1.75) rectangle (3.00, -2.00);
\filldraw[fill=cancan, draw=black] (3.00, -1.75) rectangle (3.25, -2.00);
\filldraw[fill=cancan, draw=black] (3.25, -1.75) rectangle (3.50, -2.00);
\filldraw[fill=cancan, draw=black] (3.50, -1.75) rectangle (3.75, -2.00);
\filldraw[fill=cancan, draw=black] (3.75, -1.75) rectangle (4.00, -2.00);
\filldraw[fill=cancan, draw=black] (4.00, -1.75) rectangle (4.25, -2.00);
\filldraw[fill=cancan, draw=black] (4.25, -1.75) rectangle (4.50, -2.00);
\filldraw[fill=cancan, draw=black] (4.50, -1.75) rectangle (4.75, -2.00);
\filldraw[fill=cancan, draw=black] (4.75, -1.75) rectangle (5.00, -2.00);
\filldraw[fill=bermuda, draw=black] (5.00, -1.75) rectangle (5.25, -2.00);
\filldraw[fill=bermuda, draw=black] (5.25, -1.75) rectangle (5.50, -2.00);
\filldraw[fill=bermuda, draw=black] (5.50, -1.75) rectangle (5.75, -2.00);
\filldraw[fill=cancan, draw=black] (5.75, -1.75) rectangle (6.00, -2.00);
\filldraw[fill=cancan, draw=black] (6.00, -1.75) rectangle (6.25, -2.00);
\filldraw[fill=cancan, draw=black] (6.25, -1.75) rectangle (6.50, -2.00);
\filldraw[fill=bermuda, draw=black] (6.50, -1.75) rectangle (6.75, -2.00);
\filldraw[fill=bermuda, draw=black] (6.75, -1.75) rectangle (7.00, -2.00);
\filldraw[fill=bermuda, draw=black] (7.00, -1.75) rectangle (7.25, -2.00);
\filldraw[fill=bermuda, draw=black] (7.25, -1.75) rectangle (7.50, -2.00);
\filldraw[fill=bermuda, draw=black] (7.50, -1.75) rectangle (7.75, -2.00);
\filldraw[fill=cancan, draw=black] (7.75, -1.75) rectangle (8.00, -2.00);
\filldraw[fill=bermuda, draw=black] (8.00, -1.75) rectangle (8.25, -2.00);
\filldraw[fill=bermuda, draw=black] (8.25, -1.75) rectangle (8.50, -2.00);
\filldraw[fill=bermuda, draw=black] (8.50, -1.75) rectangle (8.75, -2.00);
\filldraw[fill=cancan, draw=black] (8.75, -1.75) rectangle (9.00, -2.00);
\filldraw[fill=bermuda, draw=black] (9.00, -1.75) rectangle (9.25, -2.00);
\filldraw[fill=bermuda, draw=black] (9.25, -1.75) rectangle (9.50, -2.00);
\filldraw[fill=bermuda, draw=black] (9.50, -1.75) rectangle (9.75, -2.00);
\filldraw[fill=bermuda, draw=black] (9.75, -1.75) rectangle (10.00, -2.00);
\filldraw[fill=bermuda, draw=black] (10.00, -1.75) rectangle (10.25, -2.00);
\filldraw[fill=bermuda, draw=black] (10.25, -1.75) rectangle (10.50, -2.00);
\filldraw[fill=bermuda, draw=black] (10.50, -1.75) rectangle (10.75, -2.00);
\filldraw[fill=cancan, draw=black] (10.75, -1.75) rectangle (11.00, -2.00);
\filldraw[fill=bermuda, draw=black] (11.00, -1.75) rectangle (11.25, -2.00);
\filldraw[fill=cancan, draw=black] (11.25, -1.75) rectangle (11.50, -2.00);
\filldraw[fill=bermuda, draw=black] (11.50, -1.75) rectangle (11.75, -2.00);
\filldraw[fill=bermuda, draw=black] (11.75, -1.75) rectangle (12.00, -2.00);
\filldraw[fill=bermuda, draw=black] (12.00, -1.75) rectangle (12.25, -2.00);
\filldraw[fill=cancan, draw=black] (12.25, -1.75) rectangle (12.50, -2.00);
\filldraw[fill=bermuda, draw=black] (12.50, -1.75) rectangle (12.75, -2.00);
\filldraw[fill=bermuda, draw=black] (12.75, -1.75) rectangle (13.00, -2.00);
\filldraw[fill=bermuda, draw=black] (13.00, -1.75) rectangle (13.25, -2.00);
\filldraw[fill=bermuda, draw=black] (13.25, -1.75) rectangle (13.50, -2.00);
\filldraw[fill=bermuda, draw=black] (13.50, -1.75) rectangle (13.75, -2.00);
\filldraw[fill=cancan, draw=black] (13.75, -1.75) rectangle (14.00, -2.00);
\filldraw[fill=bermuda, draw=black] (14.00, -1.75) rectangle (14.25, -2.00);
\filldraw[fill=bermuda, draw=black] (14.25, -1.75) rectangle (14.50, -2.00);
\filldraw[fill=bermuda, draw=black] (14.50, -1.75) rectangle (14.75, -2.00);
\filldraw[fill=bermuda, draw=black] (14.75, -1.75) rectangle (15.00, -2.00);
\filldraw[fill=bermuda, draw=black] (0.00, -2.00) rectangle (0.25, -2.25);
\filldraw[fill=bermuda, draw=black] (0.25, -2.00) rectangle (0.50, -2.25);
\filldraw[fill=bermuda, draw=black] (0.50, -2.00) rectangle (0.75, -2.25);
\filldraw[fill=bermuda, draw=black] (0.75, -2.00) rectangle (1.00, -2.25);
\filldraw[fill=bermuda, draw=black] (1.00, -2.00) rectangle (1.25, -2.25);
\filldraw[fill=cancan, draw=black] (1.25, -2.00) rectangle (1.50, -2.25);
\filldraw[fill=bermuda, draw=black] (1.50, -2.00) rectangle (1.75, -2.25);
\filldraw[fill=cancan, draw=black] (1.75, -2.00) rectangle (2.00, -2.25);
\filldraw[fill=cancan, draw=black] (2.00, -2.00) rectangle (2.25, -2.25);
\filldraw[fill=bermuda, draw=black] (2.25, -2.00) rectangle (2.50, -2.25);
\filldraw[fill=bermuda, draw=black] (2.50, -2.00) rectangle (2.75, -2.25);
\filldraw[fill=cancan, draw=black] (2.75, -2.00) rectangle (3.00, -2.25);
\filldraw[fill=bermuda, draw=black] (3.00, -2.00) rectangle (3.25, -2.25);
\filldraw[fill=cancan, draw=black] (3.25, -2.00) rectangle (3.50, -2.25);
\filldraw[fill=bermuda, draw=black] (3.50, -2.00) rectangle (3.75, -2.25);
\filldraw[fill=cancan, draw=black] (3.75, -2.00) rectangle (4.00, -2.25);
\filldraw[fill=bermuda, draw=black] (4.00, -2.00) rectangle (4.25, -2.25);
\filldraw[fill=bermuda, draw=black] (4.25, -2.00) rectangle (4.50, -2.25);
\filldraw[fill=bermuda, draw=black] (4.50, -2.00) rectangle (4.75, -2.25);
\filldraw[fill=cancan, draw=black] (4.75, -2.00) rectangle (5.00, -2.25);
\filldraw[fill=cancan, draw=black] (5.00, -2.00) rectangle (5.25, -2.25);
\filldraw[fill=cancan, draw=black] (5.25, -2.00) rectangle (5.50, -2.25);
\filldraw[fill=bermuda, draw=black] (5.50, -2.00) rectangle (5.75, -2.25);
\filldraw[fill=bermuda, draw=black] (5.75, -2.00) rectangle (6.00, -2.25);
\filldraw[fill=bermuda, draw=black] (6.00, -2.00) rectangle (6.25, -2.25);
\filldraw[fill=cancan, draw=black] (6.25, -2.00) rectangle (6.50, -2.25);
\filldraw[fill=cancan, draw=black] (6.50, -2.00) rectangle (6.75, -2.25);
\filldraw[fill=cancan, draw=black] (6.75, -2.00) rectangle (7.00, -2.25);
\filldraw[fill=bermuda, draw=black] (7.00, -2.00) rectangle (7.25, -2.25);
\filldraw[fill=bermuda, draw=black] (7.25, -2.00) rectangle (7.50, -2.25);
\filldraw[fill=bermuda, draw=black] (7.50, -2.00) rectangle (7.75, -2.25);
\filldraw[fill=cancan, draw=black] (7.75, -2.00) rectangle (8.00, -2.25);
\filldraw[fill=bermuda, draw=black] (8.00, -2.00) rectangle (8.25, -2.25);
\filldraw[fill=bermuda, draw=black] (8.25, -2.00) rectangle (8.50, -2.25);
\filldraw[fill=bermuda, draw=black] (8.50, -2.00) rectangle (8.75, -2.25);
\filldraw[fill=bermuda, draw=black] (8.75, -2.00) rectangle (9.00, -2.25);
\filldraw[fill=bermuda, draw=black] (9.00, -2.00) rectangle (9.25, -2.25);
\filldraw[fill=cancan, draw=black] (9.25, -2.00) rectangle (9.50, -2.25);
\filldraw[fill=bermuda, draw=black] (9.50, -2.00) rectangle (9.75, -2.25);
\filldraw[fill=cancan, draw=black] (9.75, -2.00) rectangle (10.00, -2.25);
\filldraw[fill=cancan, draw=black] (10.00, -2.00) rectangle (10.25, -2.25);
\filldraw[fill=cancan, draw=black] (10.25, -2.00) rectangle (10.50, -2.25);
\filldraw[fill=cancan, draw=black] (10.50, -2.00) rectangle (10.75, -2.25);
\filldraw[fill=cancan, draw=black] (10.75, -2.00) rectangle (11.00, -2.25);
\filldraw[fill=bermuda, draw=black] (11.00, -2.00) rectangle (11.25, -2.25);
\filldraw[fill=bermuda, draw=black] (11.25, -2.00) rectangle (11.50, -2.25);
\filldraw[fill=bermuda, draw=black] (11.50, -2.00) rectangle (11.75, -2.25);
\filldraw[fill=cancan, draw=black] (11.75, -2.00) rectangle (12.00, -2.25);
\filldraw[fill=cancan, draw=black] (12.00, -2.00) rectangle (12.25, -2.25);
\filldraw[fill=cancan, draw=black] (12.25, -2.00) rectangle (12.50, -2.25);
\filldraw[fill=bermuda, draw=black] (12.50, -2.00) rectangle (12.75, -2.25);
\filldraw[fill=bermuda, draw=black] (12.75, -2.00) rectangle (13.00, -2.25);
\filldraw[fill=bermuda, draw=black] (13.00, -2.00) rectangle (13.25, -2.25);
\filldraw[fill=cancan, draw=black] (13.25, -2.00) rectangle (13.50, -2.25);
\filldraw[fill=cancan, draw=black] (13.50, -2.00) rectangle (13.75, -2.25);
\filldraw[fill=cancan, draw=black] (13.75, -2.00) rectangle (14.00, -2.25);
\filldraw[fill=cancan, draw=black] (14.00, -2.00) rectangle (14.25, -2.25);
\filldraw[fill=bermuda, draw=black] (14.25, -2.00) rectangle (14.50, -2.25);
\filldraw[fill=bermuda, draw=black] (14.50, -2.00) rectangle (14.75, -2.25);
\filldraw[fill=cancan, draw=black] (14.75, -2.00) rectangle (15.00, -2.25);
\filldraw[fill=bermuda, draw=black] (0.00, -2.25) rectangle (0.25, -2.50);
\filldraw[fill=cancan, draw=black] (0.25, -2.25) rectangle (0.50, -2.50);
\filldraw[fill=cancan, draw=black] (0.50, -2.25) rectangle (0.75, -2.50);
\filldraw[fill=bermuda, draw=black] (0.75, -2.25) rectangle (1.00, -2.50);
\filldraw[fill=bermuda, draw=black] (1.00, -2.25) rectangle (1.25, -2.50);
\filldraw[fill=cancan, draw=black] (1.25, -2.25) rectangle (1.50, -2.50);
\filldraw[fill=bermuda, draw=black] (1.50, -2.25) rectangle (1.75, -2.50);
\filldraw[fill=cancan, draw=black] (1.75, -2.25) rectangle (2.00, -2.50);
\filldraw[fill=cancan, draw=black] (2.00, -2.25) rectangle (2.25, -2.50);
\filldraw[fill=bermuda, draw=black] (2.25, -2.25) rectangle (2.50, -2.50);
\filldraw[fill=bermuda, draw=black] (2.50, -2.25) rectangle (2.75, -2.50);
\filldraw[fill=cancan, draw=black] (2.75, -2.25) rectangle (3.00, -2.50);
\filldraw[fill=bermuda, draw=black] (3.00, -2.25) rectangle (3.25, -2.50);
\filldraw[fill=cancan, draw=black] (3.25, -2.25) rectangle (3.50, -2.50);
\filldraw[fill=cancan, draw=black] (3.50, -2.25) rectangle (3.75, -2.50);
\filldraw[fill=bermuda, draw=black] (3.75, -2.25) rectangle (4.00, -2.50);
\filldraw[fill=bermuda, draw=black] (4.00, -2.25) rectangle (4.25, -2.50);
\filldraw[fill=cancan, draw=black] (4.25, -2.25) rectangle (4.50, -2.50);
\filldraw[fill=bermuda, draw=black] (4.50, -2.25) rectangle (4.75, -2.50);
\filldraw[fill=bermuda, draw=black] (4.75, -2.25) rectangle (5.00, -2.50);
\filldraw[fill=bermuda, draw=black] (5.00, -2.25) rectangle (5.25, -2.50);
\filldraw[fill=cancan, draw=black] (5.25, -2.25) rectangle (5.50, -2.50);
\filldraw[fill=cancan, draw=black] (5.50, -2.25) rectangle (5.75, -2.50);
\filldraw[fill=cancan, draw=black] (5.75, -2.25) rectangle (6.00, -2.50);
\filldraw[fill=cancan, draw=black] (6.00, -2.25) rectangle (6.25, -2.50);
\filldraw[fill=cancan, draw=black] (6.25, -2.25) rectangle (6.50, -2.50);
\filldraw[fill=cancan, draw=black] (6.50, -2.25) rectangle (6.75, -2.50);
\filldraw[fill=cancan, draw=black] (6.75, -2.25) rectangle (7.00, -2.50);
\filldraw[fill=cancan, draw=black] (7.00, -2.25) rectangle (7.25, -2.50);
\filldraw[fill=cancan, draw=black] (7.25, -2.25) rectangle (7.50, -2.50);
\filldraw[fill=cancan, draw=black] (7.50, -2.25) rectangle (7.75, -2.50);
\filldraw[fill=cancan, draw=black] (7.75, -2.25) rectangle (8.00, -2.50);
\filldraw[fill=cancan, draw=black] (8.00, -2.25) rectangle (8.25, -2.50);
\filldraw[fill=cancan, draw=black] (8.25, -2.25) rectangle (8.50, -2.50);
\filldraw[fill=bermuda, draw=black] (8.50, -2.25) rectangle (8.75, -2.50);
\filldraw[fill=cancan, draw=black] (8.75, -2.25) rectangle (9.00, -2.50);
\filldraw[fill=bermuda, draw=black] (9.00, -2.25) rectangle (9.25, -2.50);
\filldraw[fill=bermuda, draw=black] (9.25, -2.25) rectangle (9.50, -2.50);
\filldraw[fill=bermuda, draw=black] (9.50, -2.25) rectangle (9.75, -2.50);
\filldraw[fill=cancan, draw=black] (9.75, -2.25) rectangle (10.00, -2.50);
\filldraw[fill=cancan, draw=black] (10.00, -2.25) rectangle (10.25, -2.50);
\filldraw[fill=cancan, draw=black] (10.25, -2.25) rectangle (10.50, -2.50);
\filldraw[fill=bermuda, draw=black] (10.50, -2.25) rectangle (10.75, -2.50);
\filldraw[fill=bermuda, draw=black] (10.75, -2.25) rectangle (11.00, -2.50);
\filldraw[fill=bermuda, draw=black] (11.00, -2.25) rectangle (11.25, -2.50);
\filldraw[fill=cancan, draw=black] (11.25, -2.25) rectangle (11.50, -2.50);
\filldraw[fill=cancan, draw=black] (11.50, -2.25) rectangle (11.75, -2.50);
\filldraw[fill=cancan, draw=black] (11.75, -2.25) rectangle (12.00, -2.50);
\filldraw[fill=cancan, draw=black] (12.00, -2.25) rectangle (12.25, -2.50);
\filldraw[fill=cancan, draw=black] (12.25, -2.25) rectangle (12.50, -2.50);
\filldraw[fill=cancan, draw=black] (12.50, -2.25) rectangle (12.75, -2.50);
\filldraw[fill=cancan, draw=black] (12.75, -2.25) rectangle (13.00, -2.50);
\filldraw[fill=bermuda, draw=black] (13.00, -2.25) rectangle (13.25, -2.50);
\filldraw[fill=bermuda, draw=black] (13.25, -2.25) rectangle (13.50, -2.50);
\filldraw[fill=bermuda, draw=black] (13.50, -2.25) rectangle (13.75, -2.50);
\filldraw[fill=bermuda, draw=black] (13.75, -2.25) rectangle (14.00, -2.50);
\filldraw[fill=bermuda, draw=black] (14.00, -2.25) rectangle (14.25, -2.50);
\filldraw[fill=cancan, draw=black] (14.25, -2.25) rectangle (14.50, -2.50);
\filldraw[fill=bermuda, draw=black] (14.50, -2.25) rectangle (14.75, -2.50);
\filldraw[fill=bermuda, draw=black] (14.75, -2.25) rectangle (15.00, -2.50);
\filldraw[fill=bermuda, draw=black] (0.00, -2.50) rectangle (0.25, -2.75);
\filldraw[fill=cancan, draw=black] (0.25, -2.50) rectangle (0.50, -2.75);
\filldraw[fill=cancan, draw=black] (0.50, -2.50) rectangle (0.75, -2.75);
\filldraw[fill=cancan, draw=black] (0.75, -2.50) rectangle (1.00, -2.75);
\filldraw[fill=bermuda, draw=black] (1.00, -2.50) rectangle (1.25, -2.75);
\filldraw[fill=bermuda, draw=black] (1.25, -2.50) rectangle (1.50, -2.75);
\filldraw[fill=bermuda, draw=black] (1.50, -2.50) rectangle (1.75, -2.75);
\filldraw[fill=bermuda, draw=black] (1.75, -2.50) rectangle (2.00, -2.75);
\filldraw[fill=bermuda, draw=black] (2.00, -2.50) rectangle (2.25, -2.75);
\filldraw[fill=cancan, draw=black] (2.25, -2.50) rectangle (2.50, -2.75);
\filldraw[fill=bermuda, draw=black] (2.50, -2.50) rectangle (2.75, -2.75);
\filldraw[fill=bermuda, draw=black] (2.75, -2.50) rectangle (3.00, -2.75);
\filldraw[fill=bermuda, draw=black] (3.00, -2.50) rectangle (3.25, -2.75);
\filldraw[fill=cancan, draw=black] (3.25, -2.50) rectangle (3.50, -2.75);
\filldraw[fill=cancan, draw=black] (3.50, -2.50) rectangle (3.75, -2.75);
\filldraw[fill=cancan, draw=black] (3.75, -2.50) rectangle (4.00, -2.75);
\filldraw[fill=bermuda, draw=black] (4.00, -2.50) rectangle (4.25, -2.75);
\filldraw[fill=bermuda, draw=black] (4.25, -2.50) rectangle (4.50, -2.75);
\filldraw[fill=bermuda, draw=black] (4.50, -2.50) rectangle (4.75, -2.75);
\filldraw[fill=cancan, draw=black] (4.75, -2.50) rectangle (5.00, -2.75);
\filldraw[fill=cancan, draw=black] (5.00, -2.50) rectangle (5.25, -2.75);
\filldraw[fill=cancan, draw=black] (5.25, -2.50) rectangle (5.50, -2.75);
\filldraw[fill=cancan, draw=black] (5.50, -2.50) rectangle (5.75, -2.75);
\filldraw[fill=bermuda, draw=black] (5.75, -2.50) rectangle (6.00, -2.75);
\filldraw[fill=bermuda, draw=black] (6.00, -2.50) rectangle (6.25, -2.75);
\filldraw[fill=bermuda, draw=black] (6.25, -2.50) rectangle (6.50, -2.75);
\filldraw[fill=bermuda, draw=black] (6.50, -2.50) rectangle (6.75, -2.75);
\filldraw[fill=cancan, draw=black] (6.75, -2.50) rectangle (7.00, -2.75);
\filldraw[fill=bermuda, draw=black] (7.00, -2.50) rectangle (7.25, -2.75);
\filldraw[fill=bermuda, draw=black] (7.25, -2.50) rectangle (7.50, -2.75);
\filldraw[fill=bermuda, draw=black] (7.50, -2.50) rectangle (7.75, -2.75);
\filldraw[fill=cancan, draw=black] (7.75, -2.50) rectangle (8.00, -2.75);
\filldraw[fill=cancan, draw=black] (8.00, -2.50) rectangle (8.25, -2.75);
\filldraw[fill=cancan, draw=black] (8.25, -2.50) rectangle (8.50, -2.75);
\filldraw[fill=bermuda, draw=black] (8.50, -2.50) rectangle (8.75, -2.75);
\filldraw[fill=bermuda, draw=black] (8.75, -2.50) rectangle (9.00, -2.75);
\filldraw[fill=bermuda, draw=black] (9.00, -2.50) rectangle (9.25, -2.75);
\filldraw[fill=bermuda, draw=black] (9.25, -2.50) rectangle (9.50, -2.75);
\filldraw[fill=bermuda, draw=black] (9.50, -2.50) rectangle (9.75, -2.75);
\filldraw[fill=cancan, draw=black] (9.75, -2.50) rectangle (10.00, -2.75);
\filldraw[fill=bermuda, draw=black] (10.00, -2.50) rectangle (10.25, -2.75);
\filldraw[fill=bermuda, draw=black] (10.25, -2.50) rectangle (10.50, -2.75);
\filldraw[fill=bermuda, draw=black] (10.50, -2.50) rectangle (10.75, -2.75);
\filldraw[fill=bermuda, draw=black] (10.75, -2.50) rectangle (11.00, -2.75);
\filldraw[fill=bermuda, draw=black] (11.00, -2.50) rectangle (11.25, -2.75);
\filldraw[fill=bermuda, draw=black] (11.25, -2.50) rectangle (11.50, -2.75);
\filldraw[fill=bermuda, draw=black] (11.50, -2.50) rectangle (11.75, -2.75);
\filldraw[fill=bermuda, draw=black] (11.75, -2.50) rectangle (12.00, -2.75);
\filldraw[fill=bermuda, draw=black] (12.00, -2.50) rectangle (12.25, -2.75);
\filldraw[fill=cancan, draw=black] (12.25, -2.50) rectangle (12.50, -2.75);
\filldraw[fill=bermuda, draw=black] (12.50, -2.50) rectangle (12.75, -2.75);
\filldraw[fill=cancan, draw=black] (12.75, -2.50) rectangle (13.00, -2.75);
\filldraw[fill=cancan, draw=black] (13.00, -2.50) rectangle (13.25, -2.75);
\filldraw[fill=cancan, draw=black] (13.25, -2.50) rectangle (13.50, -2.75);
\filldraw[fill=cancan, draw=black] (13.50, -2.50) rectangle (13.75, -2.75);
\filldraw[fill=cancan, draw=black] (13.75, -2.50) rectangle (14.00, -2.75);
\filldraw[fill=bermuda, draw=black] (14.00, -2.50) rectangle (14.25, -2.75);
\filldraw[fill=bermuda, draw=black] (14.25, -2.50) rectangle (14.50, -2.75);
\filldraw[fill=bermuda, draw=black] (14.50, -2.50) rectangle (14.75, -2.75);
\filldraw[fill=cancan, draw=black] (14.75, -2.50) rectangle (15.00, -2.75);
\filldraw[fill=cancan, draw=black] (0.00, -2.75) rectangle (0.25, -3.00);
\filldraw[fill=cancan, draw=black] (0.25, -2.75) rectangle (0.50, -3.00);
\filldraw[fill=bermuda, draw=black] (0.50, -2.75) rectangle (0.75, -3.00);
\filldraw[fill=bermuda, draw=black] (0.75, -2.75) rectangle (1.00, -3.00);
\filldraw[fill=bermuda, draw=black] (1.00, -2.75) rectangle (1.25, -3.00);
\filldraw[fill=bermuda, draw=black] (1.25, -2.75) rectangle (1.50, -3.00);
\filldraw[fill=bermuda, draw=black] (1.50, -2.75) rectangle (1.75, -3.00);
\filldraw[fill=cancan, draw=black] (1.75, -2.75) rectangle (2.00, -3.00);
\filldraw[fill=bermuda, draw=black] (2.00, -2.75) rectangle (2.25, -3.00);
\filldraw[fill=cancan, draw=black] (2.25, -2.75) rectangle (2.50, -3.00);
\filldraw[fill=cancan, draw=black] (2.50, -2.75) rectangle (2.75, -3.00);
\filldraw[fill=cancan, draw=black] (2.75, -2.75) rectangle (3.00, -3.00);
\filldraw[fill=bermuda, draw=black] (3.00, -2.75) rectangle (3.25, -3.00);
\filldraw[fill=cancan, draw=black] (3.25, -2.75) rectangle (3.50, -3.00);
\filldraw[fill=cancan, draw=black] (3.50, -2.75) rectangle (3.75, -3.00);
\filldraw[fill=cancan, draw=black] (3.75, -2.75) rectangle (4.00, -3.00);
\filldraw[fill=bermuda, draw=black] (4.00, -2.75) rectangle (4.25, -3.00);
\filldraw[fill=bermuda, draw=black] (4.25, -2.75) rectangle (4.50, -3.00);
\filldraw[fill=bermuda, draw=black] (4.50, -2.75) rectangle (4.75, -3.00);
\filldraw[fill=cancan, draw=black] (4.75, -2.75) rectangle (5.00, -3.00);
\filldraw[fill=cancan, draw=black] (5.00, -2.75) rectangle (5.25, -3.00);
\filldraw[fill=cancan, draw=black] (5.25, -2.75) rectangle (5.50, -3.00);
\filldraw[fill=bermuda, draw=black] (5.50, -2.75) rectangle (5.75, -3.00);
\filldraw[fill=bermuda, draw=black] (5.75, -2.75) rectangle (6.00, -3.00);
\filldraw[fill=bermuda, draw=black] (6.00, -2.75) rectangle (6.25, -3.00);
\filldraw[fill=cancan, draw=black] (6.25, -2.75) rectangle (6.50, -3.00);
\filldraw[fill=cancan, draw=black] (6.50, -2.75) rectangle (6.75, -3.00);
\filldraw[fill=cancan, draw=black] (6.75, -2.75) rectangle (7.00, -3.00);
\filldraw[fill=bermuda, draw=black] (7.00, -2.75) rectangle (7.25, -3.00);
\filldraw[fill=bermuda, draw=black] (7.25, -2.75) rectangle (7.50, -3.00);
\filldraw[fill=bermuda, draw=black] (7.50, -2.75) rectangle (7.75, -3.00);
\filldraw[fill=cancan, draw=black] (7.75, -2.75) rectangle (8.00, -3.00);
\filldraw[fill=bermuda, draw=black] (8.00, -2.75) rectangle (8.25, -3.00);
\filldraw[fill=bermuda, draw=black] (8.25, -2.75) rectangle (8.50, -3.00);
\filldraw[fill=bermuda, draw=black] (8.50, -2.75) rectangle (8.75, -3.00);
\filldraw[fill=cancan, draw=black] (8.75, -2.75) rectangle (9.00, -3.00);
\filldraw[fill=cancan, draw=black] (9.00, -2.75) rectangle (9.25, -3.00);
\filldraw[fill=cancan, draw=black] (9.25, -2.75) rectangle (9.50, -3.00);
\filldraw[fill=cancan, draw=black] (9.50, -2.75) rectangle (9.75, -3.00);
\filldraw[fill=bermuda, draw=black] (9.75, -2.75) rectangle (10.00, -3.00);
\filldraw[fill=bermuda, draw=black] (10.00, -2.75) rectangle (10.25, -3.00);
\filldraw[fill=cancan, draw=black] (10.25, -2.75) rectangle (10.50, -3.00);
\filldraw[fill=cancan, draw=black] (10.50, -2.75) rectangle (10.75, -3.00);
\filldraw[fill=cancan, draw=black] (10.75, -2.75) rectangle (11.00, -3.00);
\filldraw[fill=cancan, draw=black] (11.00, -2.75) rectangle (11.25, -3.00);
\filldraw[fill=cancan, draw=black] (11.25, -2.75) rectangle (11.50, -3.00);
\filldraw[fill=cancan, draw=black] (11.50, -2.75) rectangle (11.75, -3.00);
\filldraw[fill=bermuda, draw=black] (11.75, -2.75) rectangle (12.00, -3.00);
\filldraw[fill=bermuda, draw=black] (12.00, -2.75) rectangle (12.25, -3.00);
\filldraw[fill=cancan, draw=black] (12.25, -2.75) rectangle (12.50, -3.00);
\filldraw[fill=bermuda, draw=black] (12.50, -2.75) rectangle (12.75, -3.00);
\filldraw[fill=cancan, draw=black] (12.75, -2.75) rectangle (13.00, -3.00);
\filldraw[fill=bermuda, draw=black] (13.00, -2.75) rectangle (13.25, -3.00);
\filldraw[fill=cancan, draw=black] (13.25, -2.75) rectangle (13.50, -3.00);
\filldraw[fill=bermuda, draw=black] (13.50, -2.75) rectangle (13.75, -3.00);
\filldraw[fill=cancan, draw=black] (13.75, -2.75) rectangle (14.00, -3.00);
\filldraw[fill=cancan, draw=black] (14.00, -2.75) rectangle (14.25, -3.00);
\filldraw[fill=bermuda, draw=black] (14.25, -2.75) rectangle (14.50, -3.00);
\filldraw[fill=bermuda, draw=black] (14.50, -2.75) rectangle (14.75, -3.00);
\filldraw[fill=cancan, draw=black] (14.75, -2.75) rectangle (15.00, -3.00);
\filldraw[fill=bermuda, draw=black] (0.00, -3.00) rectangle (0.25, -3.25);
\filldraw[fill=cancan, draw=black] (0.25, -3.00) rectangle (0.50, -3.25);
\filldraw[fill=cancan, draw=black] (0.50, -3.00) rectangle (0.75, -3.25);
\filldraw[fill=bermuda, draw=black] (0.75, -3.00) rectangle (1.00, -3.25);
\filldraw[fill=bermuda, draw=black] (1.00, -3.00) rectangle (1.25, -3.25);
\filldraw[fill=cancan, draw=black] (1.25, -3.00) rectangle (1.50, -3.25);
\filldraw[fill=bermuda, draw=black] (1.50, -3.00) rectangle (1.75, -3.25);
\filldraw[fill=cancan, draw=black] (1.75, -3.00) rectangle (2.00, -3.25);
\filldraw[fill=bermuda, draw=black] (2.00, -3.00) rectangle (2.25, -3.25);
\filldraw[fill=bermuda, draw=black] (2.25, -3.00) rectangle (2.50, -3.25);
\filldraw[fill=bermuda, draw=black] (2.50, -3.00) rectangle (2.75, -3.25);
\filldraw[fill=cancan, draw=black] (2.75, -3.00) rectangle (3.00, -3.25);
\filldraw[fill=cancan, draw=black] (3.00, -3.00) rectangle (3.25, -3.25);
\filldraw[fill=cancan, draw=black] (3.25, -3.00) rectangle (3.50, -3.25);
\filldraw[fill=bermuda, draw=black] (3.50, -3.00) rectangle (3.75, -3.25);
\filldraw[fill=bermuda, draw=black] (3.75, -3.00) rectangle (4.00, -3.25);
\filldraw[fill=bermuda, draw=black] (4.00, -3.00) rectangle (4.25, -3.25);
\filldraw[fill=cancan, draw=black] (4.25, -3.00) rectangle (4.50, -3.25);
\filldraw[fill=cancan, draw=black] (4.50, -3.00) rectangle (4.75, -3.25);
\filldraw[fill=cancan, draw=black] (4.75, -3.00) rectangle (5.00, -3.25);
\filldraw[fill=bermuda, draw=black] (5.00, -3.00) rectangle (5.25, -3.25);
\filldraw[fill=bermuda, draw=black] (5.25, -3.00) rectangle (5.50, -3.25);
\filldraw[fill=bermuda, draw=black] (5.50, -3.00) rectangle (5.75, -3.25);
\filldraw[fill=cancan, draw=black] (5.75, -3.00) rectangle (6.00, -3.25);
\filldraw[fill=cancan, draw=black] (6.00, -3.00) rectangle (6.25, -3.25);
\filldraw[fill=cancan, draw=black] (6.25, -3.00) rectangle (6.50, -3.25);
\filldraw[fill=bermuda, draw=black] (6.50, -3.00) rectangle (6.75, -3.25);
\filldraw[fill=bermuda, draw=black] (6.75, -3.00) rectangle (7.00, -3.25);
\filldraw[fill=bermuda, draw=black] (7.00, -3.00) rectangle (7.25, -3.25);
\filldraw[fill=cancan, draw=black] (7.25, -3.00) rectangle (7.50, -3.25);
\filldraw[fill=cancan, draw=black] (7.50, -3.00) rectangle (7.75, -3.25);
\filldraw[fill=cancan, draw=black] (7.75, -3.00) rectangle (8.00, -3.25);
\filldraw[fill=bermuda, draw=black] (8.00, -3.00) rectangle (8.25, -3.25);
\filldraw[fill=bermuda, draw=black] (8.25, -3.00) rectangle (8.50, -3.25);
\filldraw[fill=bermuda, draw=black] (8.50, -3.00) rectangle (8.75, -3.25);
\filldraw[fill=cancan, draw=black] (8.75, -3.00) rectangle (9.00, -3.25);
\filldraw[fill=cancan, draw=black] (9.00, -3.00) rectangle (9.25, -3.25);
\filldraw[fill=cancan, draw=black] (9.25, -3.00) rectangle (9.50, -3.25);
\filldraw[fill=cancan, draw=black] (9.50, -3.00) rectangle (9.75, -3.25);
\filldraw[fill=bermuda, draw=black] (9.75, -3.00) rectangle (10.00, -3.25);
\filldraw[fill=bermuda, draw=black] (10.00, -3.00) rectangle (10.25, -3.25);
\filldraw[fill=cancan, draw=black] (10.25, -3.00) rectangle (10.50, -3.25);
\filldraw[fill=cancan, draw=black] (10.50, -3.00) rectangle (10.75, -3.25);
\filldraw[fill=cancan, draw=black] (10.75, -3.00) rectangle (11.00, -3.25);
\filldraw[fill=cancan, draw=black] (11.00, -3.00) rectangle (11.25, -3.25);
\filldraw[fill=cancan, draw=black] (11.25, -3.00) rectangle (11.50, -3.25);
\filldraw[fill=bermuda, draw=black] (11.50, -3.00) rectangle (11.75, -3.25);
\filldraw[fill=bermuda, draw=black] (11.75, -3.00) rectangle (12.00, -3.25);
\filldraw[fill=bermuda, draw=black] (12.00, -3.00) rectangle (12.25, -3.25);
\filldraw[fill=cancan, draw=black] (12.25, -3.00) rectangle (12.50, -3.25);
\filldraw[fill=bermuda, draw=black] (12.50, -3.00) rectangle (12.75, -3.25);
\filldraw[fill=bermuda, draw=black] (12.75, -3.00) rectangle (13.00, -3.25);
\filldraw[fill=bermuda, draw=black] (13.00, -3.00) rectangle (13.25, -3.25);
\filldraw[fill=bermuda, draw=black] (13.25, -3.00) rectangle (13.50, -3.25);
\filldraw[fill=bermuda, draw=black] (13.50, -3.00) rectangle (13.75, -3.25);
\filldraw[fill=cancan, draw=black] (13.75, -3.00) rectangle (14.00, -3.25);
\filldraw[fill=bermuda, draw=black] (14.00, -3.00) rectangle (14.25, -3.25);
\filldraw[fill=bermuda, draw=black] (14.25, -3.00) rectangle (14.50, -3.25);
\filldraw[fill=bermuda, draw=black] (14.50, -3.00) rectangle (14.75, -3.25);
\filldraw[fill=cancan, draw=black] (14.75, -3.00) rectangle (15.00, -3.25);
\filldraw[fill=bermuda, draw=black] (0.00, -3.25) rectangle (0.25, -3.50);
\filldraw[fill=bermuda, draw=black] (0.25, -3.25) rectangle (0.50, -3.50);
\filldraw[fill=bermuda, draw=black] (0.50, -3.25) rectangle (0.75, -3.50);
\filldraw[fill=bermuda, draw=black] (0.75, -3.25) rectangle (1.00, -3.50);
\filldraw[fill=bermuda, draw=black] (1.00, -3.25) rectangle (1.25, -3.50);
\filldraw[fill=cancan, draw=black] (1.25, -3.25) rectangle (1.50, -3.50);
\filldraw[fill=cancan, draw=black] (1.50, -3.25) rectangle (1.75, -3.50);
\filldraw[fill=cancan, draw=black] (1.75, -3.25) rectangle (2.00, -3.50);
\filldraw[fill=bermuda, draw=black] (2.00, -3.25) rectangle (2.25, -3.50);
\filldraw[fill=bermuda, draw=black] (2.25, -3.25) rectangle (2.50, -3.50);
\filldraw[fill=bermuda, draw=black] (2.50, -3.25) rectangle (2.75, -3.50);
\filldraw[fill=cancan, draw=black] (2.75, -3.25) rectangle (3.00, -3.50);
\filldraw[fill=cancan, draw=black] (3.00, -3.25) rectangle (3.25, -3.50);
\filldraw[fill=cancan, draw=black] (3.25, -3.25) rectangle (3.50, -3.50);
\filldraw[fill=bermuda, draw=black] (3.50, -3.25) rectangle (3.75, -3.50);
\filldraw[fill=bermuda, draw=black] (3.75, -3.25) rectangle (4.00, -3.50);
\filldraw[fill=bermuda, draw=black] (4.00, -3.25) rectangle (4.25, -3.50);
\filldraw[fill=cancan, draw=black] (4.25, -3.25) rectangle (4.50, -3.50);
\filldraw[fill=cancan, draw=black] (4.50, -3.25) rectangle (4.75, -3.50);
\filldraw[fill=cancan, draw=black] (4.75, -3.25) rectangle (5.00, -3.50);
\filldraw[fill=bermuda, draw=black] (5.00, -3.25) rectangle (5.25, -3.50);
\filldraw[fill=bermuda, draw=black] (5.25, -3.25) rectangle (5.50, -3.50);
\filldraw[fill=bermuda, draw=black] (5.50, -3.25) rectangle (5.75, -3.50);
\filldraw[fill=cancan, draw=black] (5.75, -3.25) rectangle (6.00, -3.50);
\filldraw[fill=cancan, draw=black] (6.00, -3.25) rectangle (6.25, -3.50);
} } }\end{equation*}
\begin{equation*}
\hspace{0.3pt} b_{11} = \vcenter{\hbox{ \tikz{
\filldraw[fill=cancan, draw=black] (0.00, 0.00) rectangle (0.25, -0.25);
\filldraw[fill=bermuda, draw=black] (0.25, 0.00) rectangle (0.50, -0.25);
\filldraw[fill=bermuda, draw=black] (0.50, 0.00) rectangle (0.75, -0.25);
\filldraw[fill=bermuda, draw=black] (0.75, 0.00) rectangle (1.00, -0.25);
\filldraw[fill=cancan, draw=black] (1.00, 0.00) rectangle (1.25, -0.25);
\filldraw[fill=bermuda, draw=black] (1.25, 0.00) rectangle (1.50, -0.25);
\filldraw[fill=cancan, draw=black] (1.50, 0.00) rectangle (1.75, -0.25);
\filldraw[fill=bermuda, draw=black] (1.75, 0.00) rectangle (2.00, -0.25);
\filldraw[fill=cancan, draw=black] (2.00, 0.00) rectangle (2.25, -0.25);
\filldraw[fill=cancan, draw=black] (2.25, 0.00) rectangle (2.50, -0.25);
\filldraw[fill=cancan, draw=black] (2.50, 0.00) rectangle (2.75, -0.25);
\filldraw[fill=cancan, draw=black] (2.75, 0.00) rectangle (3.00, -0.25);
\filldraw[fill=bermuda, draw=black] (3.00, 0.00) rectangle (3.25, -0.25);
\filldraw[fill=bermuda, draw=black] (3.25, 0.00) rectangle (3.50, -0.25);
\filldraw[fill=cancan, draw=black] (3.50, 0.00) rectangle (3.75, -0.25);
\filldraw[fill=cancan, draw=black] (3.75, 0.00) rectangle (4.00, -0.25);
\filldraw[fill=cancan, draw=black] (4.00, 0.00) rectangle (4.25, -0.25);
\filldraw[fill=cancan, draw=black] (4.25, 0.00) rectangle (4.50, -0.25);
\filldraw[fill=cancan, draw=black] (4.50, 0.00) rectangle (4.75, -0.25);
\filldraw[fill=cancan, draw=black] (4.75, 0.00) rectangle (5.00, -0.25);
\filldraw[fill=cancan, draw=black] (5.00, 0.00) rectangle (5.25, -0.25);
\filldraw[fill=cancan, draw=black] (5.25, 0.00) rectangle (5.50, -0.25);
\filldraw[fill=bermuda, draw=black] (5.50, 0.00) rectangle (5.75, -0.25);
\filldraw[fill=bermuda, draw=black] (5.75, 0.00) rectangle (6.00, -0.25);
\filldraw[fill=cancan, draw=black] (6.00, 0.00) rectangle (6.25, -0.25);
\filldraw[fill=bermuda, draw=black] (6.25, 0.00) rectangle (6.50, -0.25);
\filldraw[fill=cancan, draw=black] (6.50, 0.00) rectangle (6.75, -0.25);
\filldraw[fill=cancan, draw=black] (6.75, 0.00) rectangle (7.00, -0.25);
\filldraw[fill=cancan, draw=black] (7.00, 0.00) rectangle (7.25, -0.25);
\filldraw[fill=bermuda, draw=black] (7.25, 0.00) rectangle (7.50, -0.25);
\filldraw[fill=bermuda, draw=black] (7.50, 0.00) rectangle (7.75, -0.25);
\filldraw[fill=bermuda, draw=black] (7.75, 0.00) rectangle (8.00, -0.25);
\filldraw[fill=cancan, draw=black] (8.00, 0.00) rectangle (8.25, -0.25);
\filldraw[fill=bermuda, draw=black] (8.25, 0.00) rectangle (8.50, -0.25);
\filldraw[fill=cancan, draw=black] (8.50, 0.00) rectangle (8.75, -0.25);
\filldraw[fill=cancan, draw=black] (8.75, 0.00) rectangle (9.00, -0.25);
\filldraw[fill=cancan, draw=black] (9.00, 0.00) rectangle (9.25, -0.25);
\filldraw[fill=cancan, draw=black] (9.25, 0.00) rectangle (9.50, -0.25);
\filldraw[fill=cancan, draw=black] (9.50, 0.00) rectangle (9.75, -0.25);
\filldraw[fill=bermuda, draw=black] (9.75, 0.00) rectangle (10.00, -0.25);
\filldraw[fill=bermuda, draw=black] (10.00, 0.00) rectangle (10.25, -0.25);
\filldraw[fill=bermuda, draw=black] (10.25, 0.00) rectangle (10.50, -0.25);
\filldraw[fill=cancan, draw=black] (10.50, 0.00) rectangle (10.75, -0.25);
\filldraw[fill=cancan, draw=black] (10.75, 0.00) rectangle (11.00, -0.25);
\filldraw[fill=cancan, draw=black] (11.00, 0.00) rectangle (11.25, -0.25);
\filldraw[fill=bermuda, draw=black] (11.25, 0.00) rectangle (11.50, -0.25);
\filldraw[fill=bermuda, draw=black] (11.50, 0.00) rectangle (11.75, -0.25);
\filldraw[fill=bermuda, draw=black] (11.75, 0.00) rectangle (12.00, -0.25);
\filldraw[fill=cancan, draw=black] (12.00, 0.00) rectangle (12.25, -0.25);
\filldraw[fill=cancan, draw=black] (12.25, 0.00) rectangle (12.50, -0.25);
\filldraw[fill=cancan, draw=black] (12.50, 0.00) rectangle (12.75, -0.25);
\filldraw[fill=bermuda, draw=black] (12.75, 0.00) rectangle (13.00, -0.25);
\filldraw[fill=bermuda, draw=black] (13.00, 0.00) rectangle (13.25, -0.25);
\filldraw[fill=bermuda, draw=black] (13.25, 0.00) rectangle (13.50, -0.25);
\filldraw[fill=cancan, draw=black] (13.50, 0.00) rectangle (13.75, -0.25);
\filldraw[fill=cancan, draw=black] (13.75, 0.00) rectangle (14.00, -0.25);
\filldraw[fill=cancan, draw=black] (14.00, 0.00) rectangle (14.25, -0.25);
\filldraw[fill=bermuda, draw=black] (14.25, 0.00) rectangle (14.50, -0.25);
\filldraw[fill=bermuda, draw=black] (14.50, 0.00) rectangle (14.75, -0.25);
\filldraw[fill=bermuda, draw=black] (14.75, 0.00) rectangle (15.00, -0.25);
\filldraw[fill=cancan, draw=black] (0.00, -0.25) rectangle (0.25, -0.50);
\filldraw[fill=cancan, draw=black] (0.25, -0.25) rectangle (0.50, -0.50);
\filldraw[fill=bermuda, draw=black] (0.50, -0.25) rectangle (0.75, -0.50);
\filldraw[fill=bermuda, draw=black] (0.75, -0.25) rectangle (1.00, -0.50);
\filldraw[fill=cancan, draw=black] (1.00, -0.25) rectangle (1.25, -0.50);
\filldraw[fill=cancan, draw=black] (1.25, -0.25) rectangle (1.50, -0.50);
\filldraw[fill=cancan, draw=black] (1.50, -0.25) rectangle (1.75, -0.50);
\filldraw[fill=bermuda, draw=black] (1.75, -0.25) rectangle (2.00, -0.50);
\filldraw[fill=bermuda, draw=black] (2.00, -0.25) rectangle (2.25, -0.50);
\filldraw[fill=bermuda, draw=black] (2.25, -0.25) rectangle (2.50, -0.50);
\filldraw[fill=cancan, draw=black] (2.50, -0.25) rectangle (2.75, -0.50);
\filldraw[fill=cancan, draw=black] (2.75, -0.25) rectangle (3.00, -0.50);
\filldraw[fill=cancan, draw=black] (3.00, -0.25) rectangle (3.25, -0.50);
\filldraw[fill=cancan, draw=black] (3.25, -0.25) rectangle (3.50, -0.50);
\filldraw[fill=bermuda, draw=black] (3.50, -0.25) rectangle (3.75, -0.50);
\filldraw[fill=bermuda, draw=black] (3.75, -0.25) rectangle (4.00, -0.50);
\filldraw[fill=cancan, draw=black] (4.00, -0.25) rectangle (4.25, -0.50);
\filldraw[fill=bermuda, draw=black] (4.25, -0.25) rectangle (4.50, -0.50);
\filldraw[fill=cancan, draw=black] (4.50, -0.25) rectangle (4.75, -0.50);
\filldraw[fill=cancan, draw=black] (4.75, -0.25) rectangle (5.00, -0.50);
\filldraw[fill=bermuda, draw=black] (5.00, -0.25) rectangle (5.25, -0.50);
\filldraw[fill=bermuda, draw=black] (5.25, -0.25) rectangle (5.50, -0.50);
\filldraw[fill=cancan, draw=black] (5.50, -0.25) rectangle (5.75, -0.50);
\filldraw[fill=bermuda, draw=black] (5.75, -0.25) rectangle (6.00, -0.50);
\filldraw[fill=cancan, draw=black] (6.00, -0.25) rectangle (6.25, -0.50);
\filldraw[fill=cancan, draw=black] (6.25, -0.25) rectangle (6.50, -0.50);
\filldraw[fill=bermuda, draw=black] (6.50, -0.25) rectangle (6.75, -0.50);
\filldraw[fill=bermuda, draw=black] (6.75, -0.25) rectangle (7.00, -0.50);
\filldraw[fill=cancan, draw=black] (7.00, -0.25) rectangle (7.25, -0.50);
\filldraw[fill=cancan, draw=black] (7.25, -0.25) rectangle (7.50, -0.50);
\filldraw[fill=cancan, draw=black] (7.50, -0.25) rectangle (7.75, -0.50);
\filldraw[fill=bermuda, draw=black] (7.75, -0.25) rectangle (8.00, -0.50);
\filldraw[fill=bermuda, draw=black] (8.00, -0.25) rectangle (8.25, -0.50);
\filldraw[fill=bermuda, draw=black] (8.25, -0.25) rectangle (8.50, -0.50);
\filldraw[fill=cancan, draw=black] (8.50, -0.25) rectangle (8.75, -0.50);
\filldraw[fill=cancan, draw=black] (8.75, -0.25) rectangle (9.00, -0.50);
\filldraw[fill=cancan, draw=black] (9.00, -0.25) rectangle (9.25, -0.50);
\filldraw[fill=bermuda, draw=black] (9.25, -0.25) rectangle (9.50, -0.50);
\filldraw[fill=bermuda, draw=black] (9.50, -0.25) rectangle (9.75, -0.50);
\filldraw[fill=bermuda, draw=black] (9.75, -0.25) rectangle (10.00, -0.50);
\filldraw[fill=cancan, draw=black] (10.00, -0.25) rectangle (10.25, -0.50);
\filldraw[fill=cancan, draw=black] (10.25, -0.25) rectangle (10.50, -0.50);
\filldraw[fill=cancan, draw=black] (10.50, -0.25) rectangle (10.75, -0.50);
\filldraw[fill=bermuda, draw=black] (10.75, -0.25) rectangle (11.00, -0.50);
\filldraw[fill=bermuda, draw=black] (11.00, -0.25) rectangle (11.25, -0.50);
\filldraw[fill=bermuda, draw=black] (11.25, -0.25) rectangle (11.50, -0.50);
\filldraw[fill=cancan, draw=black] (11.50, -0.25) rectangle (11.75, -0.50);
\filldraw[fill=cancan, draw=black] (11.75, -0.25) rectangle (12.00, -0.50);
\filldraw[fill=cancan, draw=black] (12.00, -0.25) rectangle (12.25, -0.50);
\filldraw[fill=cancan, draw=black] (12.25, -0.25) rectangle (12.50, -0.50);
\filldraw[fill=cancan, draw=black] (12.50, -0.25) rectangle (12.75, -0.50);
\filldraw[fill=bermuda, draw=black] (12.75, -0.25) rectangle (13.00, -0.50);
\filldraw[fill=bermuda, draw=black] (13.00, -0.25) rectangle (13.25, -0.50);
\filldraw[fill=bermuda, draw=black] (13.25, -0.25) rectangle (13.50, -0.50);
\filldraw[fill=cancan, draw=black] (13.50, -0.25) rectangle (13.75, -0.50);
\filldraw[fill=cancan, draw=black] (13.75, -0.25) rectangle (14.00, -0.50);
\filldraw[fill=bermuda, draw=black] (14.00, -0.25) rectangle (14.25, -0.50);
\filldraw[fill=bermuda, draw=black] (14.25, -0.25) rectangle (14.50, -0.50);
\filldraw[fill=cancan, draw=black] (14.50, -0.25) rectangle (14.75, -0.50);
\filldraw[fill=cancan, draw=black] (14.75, -0.25) rectangle (15.00, -0.50);
\filldraw[fill=cancan, draw=black] (0.00, -0.50) rectangle (0.25, -0.75);
\filldraw[fill=cancan, draw=black] (0.25, -0.50) rectangle (0.50, -0.75);
\filldraw[fill=cancan, draw=black] (0.50, -0.50) rectangle (0.75, -0.75);
\filldraw[fill=bermuda, draw=black] (0.75, -0.50) rectangle (1.00, -0.75);
\filldraw[fill=cancan, draw=black] (1.00, -0.50) rectangle (1.25, -0.75);
\filldraw[fill=cancan, draw=black] (1.25, -0.50) rectangle (1.50, -0.75);
\filldraw[fill=cancan, draw=black] (1.50, -0.50) rectangle (1.75, -0.75);
\filldraw[fill=bermuda, draw=black] (1.75, -0.50) rectangle (2.00, -0.75);
\filldraw[fill=cancan, draw=black] (2.00, -0.50) rectangle (2.25, -0.75);
\filldraw[fill=bermuda, draw=black] (2.25, -0.50) rectangle (2.50, -0.75);
\filldraw[fill=cancan, draw=black] (2.50, -0.50) rectangle (2.75, -0.75);
\filldraw[fill=cancan, draw=black] (2.75, -0.50) rectangle (3.00, -0.75);
\filldraw[fill=cancan, draw=black] (3.00, -0.50) rectangle (3.25, -0.75);
\filldraw[fill=bermuda, draw=black] (3.25, -0.50) rectangle (3.50, -0.75);
\filldraw[fill=cancan, draw=black] (3.50, -0.50) rectangle (3.75, -0.75);
\filldraw[fill=bermuda, draw=black] (3.75, -0.50) rectangle (4.00, -0.75);
\filldraw[fill=cancan, draw=black] (4.00, -0.50) rectangle (4.25, -0.75);
\filldraw[fill=bermuda, draw=black] (4.25, -0.50) rectangle (4.50, -0.75);
\filldraw[fill=cancan, draw=black] (4.50, -0.50) rectangle (4.75, -0.75);
\filldraw[fill=bermuda, draw=black] (4.75, -0.50) rectangle (5.00, -0.75);
\filldraw[fill=cancan, draw=black] (5.00, -0.50) rectangle (5.25, -0.75);
\filldraw[fill=cancan, draw=black] (5.25, -0.50) rectangle (5.50, -0.75);
\filldraw[fill=cancan, draw=black] (5.50, -0.50) rectangle (5.75, -0.75);
\filldraw[fill=bermuda, draw=black] (5.75, -0.50) rectangle (6.00, -0.75);
\filldraw[fill=bermuda, draw=black] (6.00, -0.50) rectangle (6.25, -0.75);
\filldraw[fill=bermuda, draw=black] (6.25, -0.50) rectangle (6.50, -0.75);
\filldraw[fill=cancan, draw=black] (6.50, -0.50) rectangle (6.75, -0.75);
\filldraw[fill=cancan, draw=black] (6.75, -0.50) rectangle (7.00, -0.75);
\filldraw[fill=cancan, draw=black] (7.00, -0.50) rectangle (7.25, -0.75);
\filldraw[fill=bermuda, draw=black] (7.25, -0.50) rectangle (7.50, -0.75);
\filldraw[fill=bermuda, draw=black] (7.50, -0.50) rectangle (7.75, -0.75);
\filldraw[fill=bermuda, draw=black] (7.75, -0.50) rectangle (8.00, -0.75);
\filldraw[fill=cancan, draw=black] (8.00, -0.50) rectangle (8.25, -0.75);
\filldraw[fill=bermuda, draw=black] (8.25, -0.50) rectangle (8.50, -0.75);
\filldraw[fill=cancan, draw=black] (8.50, -0.50) rectangle (8.75, -0.75);
\filldraw[fill=bermuda, draw=black] (8.75, -0.50) rectangle (9.00, -0.75);
\filldraw[fill=cancan, draw=black] (9.00, -0.50) rectangle (9.25, -0.75);
\filldraw[fill=cancan, draw=black] (9.25, -0.50) rectangle (9.50, -0.75);
\filldraw[fill=cancan, draw=black] (9.50, -0.50) rectangle (9.75, -0.75);
\filldraw[fill=cancan, draw=black] (9.75, -0.50) rectangle (10.00, -0.75);
\filldraw[fill=cancan, draw=black] (10.00, -0.50) rectangle (10.25, -0.75);
\filldraw[fill=bermuda, draw=black] (10.25, -0.50) rectangle (10.50, -0.75);
\filldraw[fill=bermuda, draw=black] (10.50, -0.50) rectangle (10.75, -0.75);
\filldraw[fill=bermuda, draw=black] (10.75, -0.50) rectangle (11.00, -0.75);
\filldraw[fill=cancan, draw=black] (11.00, -0.50) rectangle (11.25, -0.75);
\filldraw[fill=cancan, draw=black] (11.25, -0.50) rectangle (11.50, -0.75);
\filldraw[fill=cancan, draw=black] (11.50, -0.50) rectangle (11.75, -0.75);
\filldraw[fill=bermuda, draw=black] (11.75, -0.50) rectangle (12.00, -0.75);
\filldraw[fill=bermuda, draw=black] (12.00, -0.50) rectangle (12.25, -0.75);
\filldraw[fill=bermuda, draw=black] (12.25, -0.50) rectangle (12.50, -0.75);
\filldraw[fill=cancan, draw=black] (12.50, -0.50) rectangle (12.75, -0.75);
\filldraw[fill=cancan, draw=black] (12.75, -0.50) rectangle (13.00, -0.75);
\filldraw[fill=cancan, draw=black] (13.00, -0.50) rectangle (13.25, -0.75);
\filldraw[fill=bermuda, draw=black] (13.25, -0.50) rectangle (13.50, -0.75);
\filldraw[fill=bermuda, draw=black] (13.50, -0.50) rectangle (13.75, -0.75);
\filldraw[fill=bermuda, draw=black] (13.75, -0.50) rectangle (14.00, -0.75);
\filldraw[fill=cancan, draw=black] (14.00, -0.50) rectangle (14.25, -0.75);
\filldraw[fill=bermuda, draw=black] (14.25, -0.50) rectangle (14.50, -0.75);
\filldraw[fill=bermuda, draw=black] (14.50, -0.50) rectangle (14.75, -0.75);
\filldraw[fill=bermuda, draw=black] (14.75, -0.50) rectangle (15.00, -0.75);
\filldraw[fill=bermuda, draw=black] (0.00, -0.75) rectangle (0.25, -1.00);
\filldraw[fill=bermuda, draw=black] (0.25, -0.75) rectangle (0.50, -1.00);
\filldraw[fill=cancan, draw=black] (0.50, -0.75) rectangle (0.75, -1.00);
\filldraw[fill=bermuda, draw=black] (0.75, -0.75) rectangle (1.00, -1.00);
\filldraw[fill=cancan, draw=black] (1.00, -0.75) rectangle (1.25, -1.00);
\filldraw[fill=cancan, draw=black] (1.25, -0.75) rectangle (1.50, -1.00);
\filldraw[fill=bermuda, draw=black] (1.50, -0.75) rectangle (1.75, -1.00);
\filldraw[fill=bermuda, draw=black] (1.75, -0.75) rectangle (2.00, -1.00);
\filldraw[fill=cancan, draw=black] (2.00, -0.75) rectangle (2.25, -1.00);
\filldraw[fill=bermuda, draw=black] (2.25, -0.75) rectangle (2.50, -1.00);
\filldraw[fill=cancan, draw=black] (2.50, -0.75) rectangle (2.75, -1.00);
\filldraw[fill=cancan, draw=black] (2.75, -0.75) rectangle (3.00, -1.00);
\filldraw[fill=bermuda, draw=black] (3.00, -0.75) rectangle (3.25, -1.00);
\filldraw[fill=bermuda, draw=black] (3.25, -0.75) rectangle (3.50, -1.00);
\filldraw[fill=cancan, draw=black] (3.50, -0.75) rectangle (3.75, -1.00);
\filldraw[fill=bermuda, draw=black] (3.75, -0.75) rectangle (4.00, -1.00);
\filldraw[fill=bermuda, draw=black] (4.00, -0.75) rectangle (4.25, -1.00);
\filldraw[fill=bermuda, draw=black] (4.25, -0.75) rectangle (4.50, -1.00);
\filldraw[fill=cancan, draw=black] (4.50, -0.75) rectangle (4.75, -1.00);
\filldraw[fill=cancan, draw=black] (4.75, -0.75) rectangle (5.00, -1.00);
\filldraw[fill=cancan, draw=black] (5.00, -0.75) rectangle (5.25, -1.00);
\filldraw[fill=cancan, draw=black] (5.25, -0.75) rectangle (5.50, -1.00);
\filldraw[fill=cancan, draw=black] (5.50, -0.75) rectangle (5.75, -1.00);
\filldraw[fill=cancan, draw=black] (5.75, -0.75) rectangle (6.00, -1.00);
\filldraw[fill=cancan, draw=black] (6.00, -0.75) rectangle (6.25, -1.00);
\filldraw[fill=cancan, draw=black] (6.25, -0.75) rectangle (6.50, -1.00);
\filldraw[fill=bermuda, draw=black] (6.50, -0.75) rectangle (6.75, -1.00);
\filldraw[fill=bermuda, draw=black] (6.75, -0.75) rectangle (7.00, -1.00);
\filldraw[fill=cancan, draw=black] (7.00, -0.75) rectangle (7.25, -1.00);
\filldraw[fill=cancan, draw=black] (7.25, -0.75) rectangle (7.50, -1.00);
\filldraw[fill=cancan, draw=black] (7.50, -0.75) rectangle (7.75, -1.00);
\filldraw[fill=cancan, draw=black] (7.75, -0.75) rectangle (8.00, -1.00);
\filldraw[fill=cancan, draw=black] (8.00, -0.75) rectangle (8.25, -1.00);
\filldraw[fill=bermuda, draw=black] (8.25, -0.75) rectangle (8.50, -1.00);
\filldraw[fill=bermuda, draw=black] (8.50, -0.75) rectangle (8.75, -1.00);
\filldraw[fill=bermuda, draw=black] (8.75, -0.75) rectangle (9.00, -1.00);
\filldraw[fill=bermuda, draw=black] (9.00, -0.75) rectangle (9.25, -1.00);
\filldraw[fill=bermuda, draw=black] (9.25, -0.75) rectangle (9.50, -1.00);
\filldraw[fill=cancan, draw=black] (9.50, -0.75) rectangle (9.75, -1.00);
\filldraw[fill=bermuda, draw=black] (9.75, -0.75) rectangle (10.00, -1.00);
\filldraw[fill=bermuda, draw=black] (10.00, -0.75) rectangle (10.25, -1.00);
\filldraw[fill=bermuda, draw=black] (10.25, -0.75) rectangle (10.50, -1.00);
\filldraw[fill=cancan, draw=black] (10.50, -0.75) rectangle (10.75, -1.00);
\filldraw[fill=cancan, draw=black] (10.75, -0.75) rectangle (11.00, -1.00);
\filldraw[fill=cancan, draw=black] (11.00, -0.75) rectangle (11.25, -1.00);
\filldraw[fill=bermuda, draw=black] (11.25, -0.75) rectangle (11.50, -1.00);
\filldraw[fill=bermuda, draw=black] (11.50, -0.75) rectangle (11.75, -1.00);
\filldraw[fill=bermuda, draw=black] (11.75, -0.75) rectangle (12.00, -1.00);
\filldraw[fill=cancan, draw=black] (12.00, -0.75) rectangle (12.25, -1.00);
\filldraw[fill=bermuda, draw=black] (12.25, -0.75) rectangle (12.50, -1.00);
\filldraw[fill=cancan, draw=black] (12.50, -0.75) rectangle (12.75, -1.00);
\filldraw[fill=bermuda, draw=black] (12.75, -0.75) rectangle (13.00, -1.00);
\filldraw[fill=bermuda, draw=black] (13.00, -0.75) rectangle (13.25, -1.00);
\filldraw[fill=bermuda, draw=black] (13.25, -0.75) rectangle (13.50, -1.00);
\filldraw[fill=cancan, draw=black] (13.50, -0.75) rectangle (13.75, -1.00);
\filldraw[fill=bermuda, draw=black] (13.75, -0.75) rectangle (14.00, -1.00);
\filldraw[fill=bermuda, draw=black] (14.00, -0.75) rectangle (14.25, -1.00);
\filldraw[fill=bermuda, draw=black] (14.25, -0.75) rectangle (14.50, -1.00);
\filldraw[fill=bermuda, draw=black] (14.50, -0.75) rectangle (14.75, -1.00);
\filldraw[fill=bermuda, draw=black] (14.75, -0.75) rectangle (15.00, -1.00);
\filldraw[fill=cancan, draw=black] (0.00, -1.00) rectangle (0.25, -1.25);
\filldraw[fill=cancan, draw=black] (0.25, -1.00) rectangle (0.50, -1.25);
\filldraw[fill=bermuda, draw=black] (0.50, -1.00) rectangle (0.75, -1.25);
\filldraw[fill=bermuda, draw=black] (0.75, -1.00) rectangle (1.00, -1.25);
\filldraw[fill=cancan, draw=black] (1.00, -1.00) rectangle (1.25, -1.25);
\filldraw[fill=cancan, draw=black] (1.25, -1.00) rectangle (1.50, -1.25);
\filldraw[fill=cancan, draw=black] (1.50, -1.00) rectangle (1.75, -1.25);
\filldraw[fill=bermuda, draw=black] (1.75, -1.00) rectangle (2.00, -1.25);
\filldraw[fill=bermuda, draw=black] (2.00, -1.00) rectangle (2.25, -1.25);
\filldraw[fill=bermuda, draw=black] (2.25, -1.00) rectangle (2.50, -1.25);
\filldraw[fill=cancan, draw=black] (2.50, -1.00) rectangle (2.75, -1.25);
\filldraw[fill=cancan, draw=black] (2.75, -1.00) rectangle (3.00, -1.25);
\filldraw[fill=cancan, draw=black] (3.00, -1.00) rectangle (3.25, -1.25);
\filldraw[fill=bermuda, draw=black] (3.25, -1.00) rectangle (3.50, -1.25);
\filldraw[fill=bermuda, draw=black] (3.50, -1.00) rectangle (3.75, -1.25);
\filldraw[fill=bermuda, draw=black] (3.75, -1.00) rectangle (4.00, -1.25);
\filldraw[fill=cancan, draw=black] (4.00, -1.00) rectangle (4.25, -1.25);
\filldraw[fill=cancan, draw=black] (4.25, -1.00) rectangle (4.50, -1.25);
\filldraw[fill=cancan, draw=black] (4.50, -1.00) rectangle (4.75, -1.25);
\filldraw[fill=bermuda, draw=black] (4.75, -1.00) rectangle (5.00, -1.25);
\filldraw[fill=bermuda, draw=black] (5.00, -1.00) rectangle (5.25, -1.25);
\filldraw[fill=bermuda, draw=black] (5.25, -1.00) rectangle (5.50, -1.25);
\filldraw[fill=cancan, draw=black] (5.50, -1.00) rectangle (5.75, -1.25);
\filldraw[fill=cancan, draw=black] (5.75, -1.00) rectangle (6.00, -1.25);
\filldraw[fill=cancan, draw=black] (6.00, -1.00) rectangle (6.25, -1.25);
\filldraw[fill=cancan, draw=black] (6.25, -1.00) rectangle (6.50, -1.25);
\filldraw[fill=cancan, draw=black] (6.50, -1.00) rectangle (6.75, -1.25);
\filldraw[fill=cancan, draw=black] (6.75, -1.00) rectangle (7.00, -1.25);
\filldraw[fill=bermuda, draw=black] (7.00, -1.00) rectangle (7.25, -1.25);
\filldraw[fill=bermuda, draw=black] (7.25, -1.00) rectangle (7.50, -1.25);
\filldraw[fill=cancan, draw=black] (7.50, -1.00) rectangle (7.75, -1.25);
\filldraw[fill=bermuda, draw=black] (7.75, -1.00) rectangle (8.00, -1.25);
\filldraw[fill=cancan, draw=black] (8.00, -1.00) rectangle (8.25, -1.25);
\filldraw[fill=bermuda, draw=black] (8.25, -1.00) rectangle (8.50, -1.25);
\filldraw[fill=bermuda, draw=black] (8.50, -1.00) rectangle (8.75, -1.25);
\filldraw[fill=bermuda, draw=black] (8.75, -1.00) rectangle (9.00, -1.25);
\filldraw[fill=cancan, draw=black] (9.00, -1.00) rectangle (9.25, -1.25);
\filldraw[fill=bermuda, draw=black] (9.25, -1.00) rectangle (9.50, -1.25);
\filldraw[fill=bermuda, draw=black] (9.50, -1.00) rectangle (9.75, -1.25);
\filldraw[fill=bermuda, draw=black] (9.75, -1.00) rectangle (10.00, -1.25);
\filldraw[fill=cancan, draw=black] (10.00, -1.00) rectangle (10.25, -1.25);
\filldraw[fill=cancan, draw=black] (10.25, -1.00) rectangle (10.50, -1.25);
\filldraw[fill=bermuda, draw=black] (10.50, -1.00) rectangle (10.75, -1.25);
\filldraw[fill=bermuda, draw=black] (10.75, -1.00) rectangle (11.00, -1.25);
\filldraw[fill=cancan, draw=black] (11.00, -1.00) rectangle (11.25, -1.25);
\filldraw[fill=cancan, draw=black] (11.25, -1.00) rectangle (11.50, -1.25);
\filldraw[fill=cancan, draw=black] (11.50, -1.00) rectangle (11.75, -1.25);
\filldraw[fill=cancan, draw=black] (11.75, -1.00) rectangle (12.00, -1.25);
\filldraw[fill=cancan, draw=black] (12.00, -1.00) rectangle (12.25, -1.25);
\filldraw[fill=bermuda, draw=black] (12.25, -1.00) rectangle (12.50, -1.25);
\filldraw[fill=cancan, draw=black] (12.50, -1.00) rectangle (12.75, -1.25);
\filldraw[fill=bermuda, draw=black] (12.75, -1.00) rectangle (13.00, -1.25);
\filldraw[fill=bermuda, draw=black] (13.00, -1.00) rectangle (13.25, -1.25);
\filldraw[fill=bermuda, draw=black] (13.25, -1.00) rectangle (13.50, -1.25);
\filldraw[fill=cancan, draw=black] (13.50, -1.00) rectangle (13.75, -1.25);
\filldraw[fill=bermuda, draw=black] (13.75, -1.00) rectangle (14.00, -1.25);
\filldraw[fill=bermuda, draw=black] (14.00, -1.00) rectangle (14.25, -1.25);
\filldraw[fill=bermuda, draw=black] (14.25, -1.00) rectangle (14.50, -1.25);
\filldraw[fill=cancan, draw=black] (14.50, -1.00) rectangle (14.75, -1.25);
\filldraw[fill=cancan, draw=black] (14.75, -1.00) rectangle (15.00, -1.25);
\filldraw[fill=cancan, draw=black] (0.00, -1.25) rectangle (0.25, -1.50);
\filldraw[fill=bermuda, draw=black] (0.25, -1.25) rectangle (0.50, -1.50);
\filldraw[fill=bermuda, draw=black] (0.50, -1.25) rectangle (0.75, -1.50);
\filldraw[fill=bermuda, draw=black] (0.75, -1.25) rectangle (1.00, -1.50);
\filldraw[fill=cancan, draw=black] (1.00, -1.25) rectangle (1.25, -1.50);
\filldraw[fill=cancan, draw=black] (1.25, -1.25) rectangle (1.50, -1.50);
\filldraw[fill=cancan, draw=black] (1.50, -1.25) rectangle (1.75, -1.50);
\filldraw[fill=bermuda, draw=black] (1.75, -1.25) rectangle (2.00, -1.50);
\filldraw[fill=bermuda, draw=black] (2.00, -1.25) rectangle (2.25, -1.50);
\filldraw[fill=bermuda, draw=black] (2.25, -1.25) rectangle (2.50, -1.50);
\filldraw[fill=bermuda, draw=black] (2.50, -1.25) rectangle (2.75, -1.50);
\filldraw[fill=bermuda, draw=black] (2.75, -1.25) rectangle (3.00, -1.50);
\filldraw[fill=cancan, draw=black] (3.00, -1.25) rectangle (3.25, -1.50);
\filldraw[fill=bermuda, draw=black] (3.25, -1.25) rectangle (3.50, -1.50);
\filldraw[fill=cancan, draw=black] (3.50, -1.25) rectangle (3.75, -1.50);
\filldraw[fill=bermuda, draw=black] (3.75, -1.25) rectangle (4.00, -1.50);
\filldraw[fill=cancan, draw=black] (4.00, -1.25) rectangle (4.25, -1.50);
\filldraw[fill=cancan, draw=black] (4.25, -1.25) rectangle (4.50, -1.50);
\filldraw[fill=cancan, draw=black] (4.50, -1.25) rectangle (4.75, -1.50);
\filldraw[fill=cancan, draw=black] (4.75, -1.25) rectangle (5.00, -1.50);
\filldraw[fill=cancan, draw=black] (5.00, -1.25) rectangle (5.25, -1.50);
\filldraw[fill=bermuda, draw=black] (5.25, -1.25) rectangle (5.50, -1.50);
\filldraw[fill=bermuda, draw=black] (5.50, -1.25) rectangle (5.75, -1.50);
\filldraw[fill=bermuda, draw=black] (5.75, -1.25) rectangle (6.00, -1.50);
\filldraw[fill=cancan, draw=black] (6.00, -1.25) rectangle (6.25, -1.50);
\filldraw[fill=cancan, draw=black] (6.25, -1.25) rectangle (6.50, -1.50);
\filldraw[fill=cancan, draw=black] (6.50, -1.25) rectangle (6.75, -1.50);
\filldraw[fill=bermuda, draw=black] (6.75, -1.25) rectangle (7.00, -1.50);
\filldraw[fill=bermuda, draw=black] (7.00, -1.25) rectangle (7.25, -1.50);
\filldraw[fill=bermuda, draw=black] (7.25, -1.25) rectangle (7.50, -1.50);
\filldraw[fill=cancan, draw=black] (7.50, -1.25) rectangle (7.75, -1.50);
\filldraw[fill=cancan, draw=black] (7.75, -1.25) rectangle (8.00, -1.50);
\filldraw[fill=cancan, draw=black] (8.00, -1.25) rectangle (8.25, -1.50);
\filldraw[fill=cancan, draw=black] (8.25, -1.25) rectangle (8.50, -1.50);
\filldraw[fill=bermuda, draw=black] (8.50, -1.25) rectangle (8.75, -1.50);
\filldraw[fill=bermuda, draw=black] (8.75, -1.25) rectangle (9.00, -1.50);
\filldraw[fill=bermuda, draw=black] (9.00, -1.25) rectangle (9.25, -1.50);
\filldraw[fill=bermuda, draw=black] (9.25, -1.25) rectangle (9.50, -1.50);
\filldraw[fill=cancan, draw=black] (9.50, -1.25) rectangle (9.75, -1.50);
\filldraw[fill=bermuda, draw=black] (9.75, -1.25) rectangle (10.00, -1.50);
\filldraw[fill=cancan, draw=black] (10.00, -1.25) rectangle (10.25, -1.50);
\filldraw[fill=cancan, draw=black] (10.25, -1.25) rectangle (10.50, -1.50);
\filldraw[fill=bermuda, draw=black] (10.50, -1.25) rectangle (10.75, -1.50);
\filldraw[fill=bermuda, draw=black] (10.75, -1.25) rectangle (11.00, -1.50);
\filldraw[fill=cancan, draw=black] (11.00, -1.25) rectangle (11.25, -1.50);
\filldraw[fill=bermuda, draw=black] (11.25, -1.25) rectangle (11.50, -1.50);
\filldraw[fill=cancan, draw=black] (11.50, -1.25) rectangle (11.75, -1.50);
\filldraw[fill=cancan, draw=black] (11.75, -1.25) rectangle (12.00, -1.50);
\filldraw[fill=bermuda, draw=black] (12.00, -1.25) rectangle (12.25, -1.50);
\filldraw[fill=bermuda, draw=black] (12.25, -1.25) rectangle (12.50, -1.50);
\filldraw[fill=cancan, draw=black] (12.50, -1.25) rectangle (12.75, -1.50);
\filldraw[fill=bermuda, draw=black] (12.75, -1.25) rectangle (13.00, -1.50);
\filldraw[fill=cancan, draw=black] (13.00, -1.25) rectangle (13.25, -1.50);
\filldraw[fill=cancan, draw=black] (13.25, -1.25) rectangle (13.50, -1.50);
\filldraw[fill=cancan, draw=black] (13.50, -1.25) rectangle (13.75, -1.50);
\filldraw[fill=cancan, draw=black] (13.75, -1.25) rectangle (14.00, -1.50);
\filldraw[fill=cancan, draw=black] (14.00, -1.25) rectangle (14.25, -1.50);
\filldraw[fill=cancan, draw=black] (14.25, -1.25) rectangle (14.50, -1.50);
\filldraw[fill=cancan, draw=black] (14.50, -1.25) rectangle (14.75, -1.50);
\filldraw[fill=cancan, draw=black] (14.75, -1.25) rectangle (15.00, -1.50);
\filldraw[fill=cancan, draw=black] (0.00, -1.50) rectangle (0.25, -1.75);
\filldraw[fill=bermuda, draw=black] (0.25, -1.50) rectangle (0.50, -1.75);
\filldraw[fill=cancan, draw=black] (0.50, -1.50) rectangle (0.75, -1.75);
\filldraw[fill=bermuda, draw=black] (0.75, -1.50) rectangle (1.00, -1.75);
\filldraw[fill=cancan, draw=black] (1.00, -1.50) rectangle (1.25, -1.75);
\filldraw[fill=cancan, draw=black] (1.25, -1.50) rectangle (1.50, -1.75);
\filldraw[fill=cancan, draw=black] (1.50, -1.50) rectangle (1.75, -1.75);
\filldraw[fill=bermuda, draw=black] (1.75, -1.50) rectangle (2.00, -1.75);
\filldraw[fill=bermuda, draw=black] (2.00, -1.50) rectangle (2.25, -1.75);
\filldraw[fill=bermuda, draw=black] (2.25, -1.50) rectangle (2.50, -1.75);
\filldraw[fill=cancan, draw=black] (2.50, -1.50) rectangle (2.75, -1.75);
\filldraw[fill=bermuda, draw=black] (2.75, -1.50) rectangle (3.00, -1.75);
\filldraw[fill=bermuda, draw=black] (3.00, -1.50) rectangle (3.25, -1.75);
\filldraw[fill=bermuda, draw=black] (3.25, -1.50) rectangle (3.50, -1.75);
\filldraw[fill=cancan, draw=black] (3.50, -1.50) rectangle (3.75, -1.75);
\filldraw[fill=bermuda, draw=black] (3.75, -1.50) rectangle (4.00, -1.75);
\filldraw[fill=bermuda, draw=black] (4.00, -1.50) rectangle (4.25, -1.75);
\filldraw[fill=bermuda, draw=black] (4.25, -1.50) rectangle (4.50, -1.75);
\filldraw[fill=cancan, draw=black] (4.50, -1.50) rectangle (4.75, -1.75);
\filldraw[fill=cancan, draw=black] (4.75, -1.50) rectangle (5.00, -1.75);
\filldraw[fill=cancan, draw=black] (5.00, -1.50) rectangle (5.25, -1.75);
\filldraw[fill=bermuda, draw=black] (5.25, -1.50) rectangle (5.50, -1.75);
\filldraw[fill=cancan, draw=black] (5.50, -1.50) rectangle (5.75, -1.75);
\filldraw[fill=cancan, draw=black] (5.75, -1.50) rectangle (6.00, -1.75);
\filldraw[fill=cancan, draw=black] (6.00, -1.50) rectangle (6.25, -1.75);
\filldraw[fill=cancan, draw=black] (6.25, -1.50) rectangle (6.50, -1.75);
\filldraw[fill=cancan, draw=black] (6.50, -1.50) rectangle (6.75, -1.75);
\filldraw[fill=bermuda, draw=black] (6.75, -1.50) rectangle (7.00, -1.75);
\filldraw[fill=bermuda, draw=black] (7.00, -1.50) rectangle (7.25, -1.75);
\filldraw[fill=bermuda, draw=black] (7.25, -1.50) rectangle (7.50, -1.75);
\filldraw[fill=cancan, draw=black] (7.50, -1.50) rectangle (7.75, -1.75);
\filldraw[fill=cancan, draw=black] (7.75, -1.50) rectangle (8.00, -1.75);
\filldraw[fill=cancan, draw=black] (8.00, -1.50) rectangle (8.25, -1.75);
\filldraw[fill=bermuda, draw=black] (8.25, -1.50) rectangle (8.50, -1.75);
\filldraw[fill=bermuda, draw=black] (8.50, -1.50) rectangle (8.75, -1.75);
\filldraw[fill=bermuda, draw=black] (8.75, -1.50) rectangle (9.00, -1.75);
\filldraw[fill=cancan, draw=black] (9.00, -1.50) rectangle (9.25, -1.75);
\filldraw[fill=cancan, draw=black] (9.25, -1.50) rectangle (9.50, -1.75);
\filldraw[fill=cancan, draw=black] (9.50, -1.50) rectangle (9.75, -1.75);
\filldraw[fill=bermuda, draw=black] (9.75, -1.50) rectangle (10.00, -1.75);
\filldraw[fill=bermuda, draw=black] (10.00, -1.50) rectangle (10.25, -1.75);
\filldraw[fill=bermuda, draw=black] (10.25, -1.50) rectangle (10.50, -1.75);
\filldraw[fill=cancan, draw=black] (10.50, -1.50) rectangle (10.75, -1.75);
\filldraw[fill=cancan, draw=black] (10.75, -1.50) rectangle (11.00, -1.75);
\filldraw[fill=cancan, draw=black] (11.00, -1.50) rectangle (11.25, -1.75);
\filldraw[fill=bermuda, draw=black] (11.25, -1.50) rectangle (11.50, -1.75);
\filldraw[fill=bermuda, draw=black] (11.50, -1.50) rectangle (11.75, -1.75);
\filldraw[fill=bermuda, draw=black] (11.75, -1.50) rectangle (12.00, -1.75);
\filldraw[fill=bermuda, draw=black] (12.00, -1.50) rectangle (12.25, -1.75);
\filldraw[fill=bermuda, draw=black] (12.25, -1.50) rectangle (12.50, -1.75);
\filldraw[fill=bermuda, draw=black] (12.50, -1.50) rectangle (12.75, -1.75);
\filldraw[fill=bermuda, draw=black] (12.75, -1.50) rectangle (13.00, -1.75);
\filldraw[fill=bermuda, draw=black] (13.00, -1.50) rectangle (13.25, -1.75);
\filldraw[fill=bermuda, draw=black] (13.25, -1.50) rectangle (13.50, -1.75);
\filldraw[fill=bermuda, draw=black] (13.50, -1.50) rectangle (13.75, -1.75);
\filldraw[fill=bermuda, draw=black] (13.75, -1.50) rectangle (14.00, -1.75);
\filldraw[fill=cancan, draw=black] (14.00, -1.50) rectangle (14.25, -1.75);
\filldraw[fill=cancan, draw=black] (14.25, -1.50) rectangle (14.50, -1.75);
\filldraw[fill=cancan, draw=black] (14.50, -1.50) rectangle (14.75, -1.75);
\filldraw[fill=bermuda, draw=black] (14.75, -1.50) rectangle (15.00, -1.75);
\filldraw[fill=bermuda, draw=black] (0.00, -1.75) rectangle (0.25, -2.00);
\filldraw[fill=bermuda, draw=black] (0.25, -1.75) rectangle (0.50, -2.00);
\filldraw[fill=cancan, draw=black] (0.50, -1.75) rectangle (0.75, -2.00);
\filldraw[fill=cancan, draw=black] (0.75, -1.75) rectangle (1.00, -2.00);
\filldraw[fill=cancan, draw=black] (1.00, -1.75) rectangle (1.25, -2.00);
\filldraw[fill=cancan, draw=black] (1.25, -1.75) rectangle (1.50, -2.00);
\filldraw[fill=cancan, draw=black] (1.50, -1.75) rectangle (1.75, -2.00);
\filldraw[fill=cancan, draw=black] (1.75, -1.75) rectangle (2.00, -2.00);
\filldraw[fill=cancan, draw=black] (2.00, -1.75) rectangle (2.25, -2.00);
\filldraw[fill=bermuda, draw=black] (2.25, -1.75) rectangle (2.50, -2.00);
\filldraw[fill=bermuda, draw=black] (2.50, -1.75) rectangle (2.75, -2.00);
\filldraw[fill=bermuda, draw=black] (2.75, -1.75) rectangle (3.00, -2.00);
\filldraw[fill=cancan, draw=black] (3.00, -1.75) rectangle (3.25, -2.00);
\filldraw[fill=cancan, draw=black] (3.25, -1.75) rectangle (3.50, -2.00);
\filldraw[fill=cancan, draw=black] (3.50, -1.75) rectangle (3.75, -2.00);
\filldraw[fill=bermuda, draw=black] (3.75, -1.75) rectangle (4.00, -2.00);
\filldraw[fill=bermuda, draw=black] (4.00, -1.75) rectangle (4.25, -2.00);
\filldraw[fill=bermuda, draw=black] (4.25, -1.75) rectangle (4.50, -2.00);
\filldraw[fill=cancan, draw=black] (4.50, -1.75) rectangle (4.75, -2.00);
\filldraw[fill=bermuda, draw=black] (4.75, -1.75) rectangle (5.00, -2.00);
\filldraw[fill=bermuda, draw=black] (5.00, -1.75) rectangle (5.25, -2.00);
\filldraw[fill=bermuda, draw=black] (5.25, -1.75) rectangle (5.50, -2.00);
\filldraw[fill=bermuda, draw=black] (5.50, -1.75) rectangle (5.75, -2.00);
\filldraw[fill=bermuda, draw=black] (5.75, -1.75) rectangle (6.00, -2.00);
\filldraw[fill=cancan, draw=black] (6.00, -1.75) rectangle (6.25, -2.00);
\filldraw[fill=bermuda, draw=black] (6.25, -1.75) rectangle (6.50, -2.00);
\filldraw[fill=cancan, draw=black] (6.50, -1.75) rectangle (6.75, -2.00);
\filldraw[fill=bermuda, draw=black] (6.75, -1.75) rectangle (7.00, -2.00);
\filldraw[fill=bermuda, draw=black] (7.00, -1.75) rectangle (7.25, -2.00);
\filldraw[fill=bermuda, draw=black] (7.25, -1.75) rectangle (7.50, -2.00);
\filldraw[fill=cancan, draw=black] (7.50, -1.75) rectangle (7.75, -2.00);
\filldraw[fill=bermuda, draw=black] (7.75, -1.75) rectangle (8.00, -2.00);
\filldraw[fill=bermuda, draw=black] (8.00, -1.75) rectangle (8.25, -2.00);
\filldraw[fill=bermuda, draw=black] (8.25, -1.75) rectangle (8.50, -2.00);
\filldraw[fill=cancan, draw=black] (8.50, -1.75) rectangle (8.75, -2.00);
\filldraw[fill=bermuda, draw=black] (8.75, -1.75) rectangle (9.00, -2.00);
\filldraw[fill=cancan, draw=black] (9.00, -1.75) rectangle (9.25, -2.00);
\filldraw[fill=bermuda, draw=black] (9.25, -1.75) rectangle (9.50, -2.00);
\filldraw[fill=cancan, draw=black] (9.50, -1.75) rectangle (9.75, -2.00);
\filldraw[fill=bermuda, draw=black] (9.75, -1.75) rectangle (10.00, -2.00);
\filldraw[fill=bermuda, draw=black] (10.00, -1.75) rectangle (10.25, -2.00);
\filldraw[fill=bermuda, draw=black] (10.25, -1.75) rectangle (10.50, -2.00);
\filldraw[fill=cancan, draw=black] (10.50, -1.75) rectangle (10.75, -2.00);
\filldraw[fill=cancan, draw=black] (10.75, -1.75) rectangle (11.00, -2.00);
\filldraw[fill=cancan, draw=black] (11.00, -1.75) rectangle (11.25, -2.00);
\filldraw[fill=bermuda, draw=black] (11.25, -1.75) rectangle (11.50, -2.00);
\filldraw[fill=bermuda, draw=black] (11.50, -1.75) rectangle (11.75, -2.00);
\filldraw[fill=bermuda, draw=black] (11.75, -1.75) rectangle (12.00, -2.00);
\filldraw[fill=cancan, draw=black] (12.00, -1.75) rectangle (12.25, -2.00);
\filldraw[fill=bermuda, draw=black] (12.25, -1.75) rectangle (12.50, -2.00);
\filldraw[fill=cancan, draw=black] (12.50, -1.75) rectangle (12.75, -2.00);
\filldraw[fill=bermuda, draw=black] (12.75, -1.75) rectangle (13.00, -2.00);
\filldraw[fill=bermuda, draw=black] (13.00, -1.75) rectangle (13.25, -2.00);
\filldraw[fill=bermuda, draw=black] (13.25, -1.75) rectangle (13.50, -2.00);
\filldraw[fill=cancan, draw=black] (13.50, -1.75) rectangle (13.75, -2.00);
\filldraw[fill=bermuda, draw=black] (13.75, -1.75) rectangle (14.00, -2.00);
\filldraw[fill=bermuda, draw=black] (14.00, -1.75) rectangle (14.25, -2.00);
\filldraw[fill=bermuda, draw=black] (14.25, -1.75) rectangle (14.50, -2.00);
\filldraw[fill=cancan, draw=black] (14.50, -1.75) rectangle (14.75, -2.00);
\filldraw[fill=bermuda, draw=black] (14.75, -1.75) rectangle (15.00, -2.00);
\filldraw[fill=bermuda, draw=black] (0.00, -2.00) rectangle (0.25, -2.25);
\filldraw[fill=bermuda, draw=black] (0.25, -2.00) rectangle (0.50, -2.25);
\filldraw[fill=cancan, draw=black] (0.50, -2.00) rectangle (0.75, -2.25);
\filldraw[fill=cancan, draw=black] (0.75, -2.00) rectangle (1.00, -2.25);
\filldraw[fill=cancan, draw=black] (1.00, -2.00) rectangle (1.25, -2.25);
\filldraw[fill=cancan, draw=black] (1.25, -2.00) rectangle (1.50, -2.25);
\filldraw[fill=bermuda, draw=black] (1.50, -2.00) rectangle (1.75, -2.25);
\filldraw[fill=bermuda, draw=black] (1.75, -2.00) rectangle (2.00, -2.25);
\filldraw[fill=cancan, draw=black] (2.00, -2.00) rectangle (2.25, -2.25);
\filldraw[fill=cancan, draw=black] (2.25, -2.00) rectangle (2.50, -2.25);
\filldraw[fill=cancan, draw=black] (2.50, -2.00) rectangle (2.75, -2.25);
\filldraw[fill=cancan, draw=black] (2.75, -2.00) rectangle (3.00, -2.25);
\filldraw[fill=cancan, draw=black] (3.00, -2.00) rectangle (3.25, -2.25);
\filldraw[fill=cancan, draw=black] (3.25, -2.00) rectangle (3.50, -2.25);
\filldraw[fill=cancan, draw=black] (3.50, -2.00) rectangle (3.75, -2.25);
\filldraw[fill=bermuda, draw=black] (3.75, -2.00) rectangle (4.00, -2.25);
\filldraw[fill=bermuda, draw=black] (4.00, -2.00) rectangle (4.25, -2.25);
\filldraw[fill=bermuda, draw=black] (4.25, -2.00) rectangle (4.50, -2.25);
\filldraw[fill=cancan, draw=black] (4.50, -2.00) rectangle (4.75, -2.25);
\filldraw[fill=cancan, draw=black] (4.75, -2.00) rectangle (5.00, -2.25);
\filldraw[fill=cancan, draw=black] (5.00, -2.00) rectangle (5.25, -2.25);
\filldraw[fill=bermuda, draw=black] (5.25, -2.00) rectangle (5.50, -2.25);
\filldraw[fill=bermuda, draw=black] (5.50, -2.00) rectangle (5.75, -2.25);
\filldraw[fill=bermuda, draw=black] (5.75, -2.00) rectangle (6.00, -2.25);
\filldraw[fill=cancan, draw=black] (6.00, -2.00) rectangle (6.25, -2.25);
\filldraw[fill=cancan, draw=black] (6.25, -2.00) rectangle (6.50, -2.25);
\filldraw[fill=cancan, draw=black] (6.50, -2.00) rectangle (6.75, -2.25);
\filldraw[fill=bermuda, draw=black] (6.75, -2.00) rectangle (7.00, -2.25);
\filldraw[fill=bermuda, draw=black] (7.00, -2.00) rectangle (7.25, -2.25);
\filldraw[fill=bermuda, draw=black] (7.25, -2.00) rectangle (7.50, -2.25);
\filldraw[fill=bermuda, draw=black] (7.50, -2.00) rectangle (7.75, -2.25);
\filldraw[fill=bermuda, draw=black] (7.75, -2.00) rectangle (8.00, -2.25);
\filldraw[fill=cancan, draw=black] (8.00, -2.00) rectangle (8.25, -2.25);
\filldraw[fill=bermuda, draw=black] (8.25, -2.00) rectangle (8.50, -2.25);
\filldraw[fill=bermuda, draw=black] (8.50, -2.00) rectangle (8.75, -2.25);
\filldraw[fill=bermuda, draw=black] (8.75, -2.00) rectangle (9.00, -2.25);
\filldraw[fill=cancan, draw=black] (9.00, -2.00) rectangle (9.25, -2.25);
\filldraw[fill=cancan, draw=black] (9.25, -2.00) rectangle (9.50, -2.25);
\filldraw[fill=cancan, draw=black] (9.50, -2.00) rectangle (9.75, -2.25);
\filldraw[fill=cancan, draw=black] (9.75, -2.00) rectangle (10.00, -2.25);
\filldraw[fill=cancan, draw=black] (10.00, -2.00) rectangle (10.25, -2.25);
\filldraw[fill=cancan, draw=black] (10.25, -2.00) rectangle (10.50, -2.25);
\filldraw[fill=bermuda, draw=black] (10.50, -2.00) rectangle (10.75, -2.25);
\filldraw[fill=bermuda, draw=black] (10.75, -2.00) rectangle (11.00, -2.25);
\filldraw[fill=cancan, draw=black] (11.00, -2.00) rectangle (11.25, -2.25);
\filldraw[fill=bermuda, draw=black] (11.25, -2.00) rectangle (11.50, -2.25);
\filldraw[fill=cancan, draw=black] (11.50, -2.00) rectangle (11.75, -2.25);
\filldraw[fill=cancan, draw=black] (11.75, -2.00) rectangle (12.00, -2.25);
\filldraw[fill=bermuda, draw=black] (12.00, -2.00) rectangle (12.25, -2.25);
\filldraw[fill=bermuda, draw=black] (12.25, -2.00) rectangle (12.50, -2.25);
\filldraw[fill=cancan, draw=black] (12.50, -2.00) rectangle (12.75, -2.25);
\filldraw[fill=bermuda, draw=black] (12.75, -2.00) rectangle (13.00, -2.25);
\filldraw[fill=cancan, draw=black] (13.00, -2.00) rectangle (13.25, -2.25);
\filldraw[fill=cancan, draw=black] (13.25, -2.00) rectangle (13.50, -2.25);
\filldraw[fill=cancan, draw=black] (13.50, -2.00) rectangle (13.75, -2.25);
\filldraw[fill=bermuda, draw=black] (13.75, -2.00) rectangle (14.00, -2.25);
\filldraw[fill=bermuda, draw=black] (14.00, -2.00) rectangle (14.25, -2.25);
\filldraw[fill=bermuda, draw=black] (14.25, -2.00) rectangle (14.50, -2.25);
\filldraw[fill=bermuda, draw=black] (14.50, -2.00) rectangle (14.75, -2.25);
\filldraw[fill=bermuda, draw=black] (14.75, -2.00) rectangle (15.00, -2.25);
\filldraw[fill=cancan, draw=black] (0.00, -2.25) rectangle (0.25, -2.50);
\filldraw[fill=bermuda, draw=black] (0.25, -2.25) rectangle (0.50, -2.50);
\filldraw[fill=bermuda, draw=black] (0.50, -2.25) rectangle (0.75, -2.50);
\filldraw[fill=bermuda, draw=black] (0.75, -2.25) rectangle (1.00, -2.50);
\filldraw[fill=bermuda, draw=black] (1.00, -2.25) rectangle (1.25, -2.50);
\filldraw[fill=bermuda, draw=black] (1.25, -2.25) rectangle (1.50, -2.50);
\filldraw[fill=cancan, draw=black] (1.50, -2.25) rectangle (1.75, -2.50);
\filldraw[fill=bermuda, draw=black] (1.75, -2.25) rectangle (2.00, -2.50);
\filldraw[fill=bermuda, draw=black] (2.00, -2.25) rectangle (2.25, -2.50);
\filldraw[fill=bermuda, draw=black] (2.25, -2.25) rectangle (2.50, -2.50);
\filldraw[fill=bermuda, draw=black] (2.50, -2.25) rectangle (2.75, -2.50);
\filldraw[fill=bermuda, draw=black] (2.75, -2.25) rectangle (3.00, -2.50);
\filldraw[fill=cancan, draw=black] (3.00, -2.25) rectangle (3.25, -2.50);
\filldraw[fill=bermuda, draw=black] (3.25, -2.25) rectangle (3.50, -2.50);
\filldraw[fill=bermuda, draw=black] (3.50, -2.25) rectangle (3.75, -2.50);
\filldraw[fill=bermuda, draw=black] (3.75, -2.25) rectangle (4.00, -2.50);
\filldraw[fill=bermuda, draw=black] (4.00, -2.25) rectangle (4.25, -2.50);
\filldraw[fill=bermuda, draw=black] (4.25, -2.25) rectangle (4.50, -2.50);
\filldraw[fill=cancan, draw=black] (4.50, -2.25) rectangle (4.75, -2.50);
\filldraw[fill=bermuda, draw=black] (4.75, -2.25) rectangle (5.00, -2.50);
\filldraw[fill=cancan, draw=black] (5.00, -2.25) rectangle (5.25, -2.50);
\filldraw[fill=cancan, draw=black] (5.25, -2.25) rectangle (5.50, -2.50);
\filldraw[fill=cancan, draw=black] (5.50, -2.25) rectangle (5.75, -2.50);
\filldraw[fill=cancan, draw=black] (5.75, -2.25) rectangle (6.00, -2.50);
\filldraw[fill=cancan, draw=black] (6.00, -2.25) rectangle (6.25, -2.50);
\filldraw[fill=cancan, draw=black] (6.25, -2.25) rectangle (6.50, -2.50);
\filldraw[fill=cancan, draw=black] (6.50, -2.25) rectangle (6.75, -2.50);
\filldraw[fill=cancan, draw=black] (6.75, -2.25) rectangle (7.00, -2.50);
\filldraw[fill=cancan, draw=black] (7.00, -2.25) rectangle (7.25, -2.50);
\filldraw[fill=bermuda, draw=black] (7.25, -2.25) rectangle (7.50, -2.50);
\filldraw[fill=cancan, draw=black] (7.50, -2.25) rectangle (7.75, -2.50);
\filldraw[fill=bermuda, draw=black] (7.75, -2.25) rectangle (8.00, -2.50);
\filldraw[fill=bermuda, draw=black] (8.00, -2.25) rectangle (8.25, -2.50);
\filldraw[fill=bermuda, draw=black] (8.25, -2.25) rectangle (8.50, -2.50);
\filldraw[fill=cancan, draw=black] (8.50, -2.25) rectangle (8.75, -2.50);
\filldraw[fill=cancan, draw=black] (8.75, -2.25) rectangle (9.00, -2.50);
\filldraw[fill=cancan, draw=black] (9.00, -2.25) rectangle (9.25, -2.50);
\filldraw[fill=bermuda, draw=black] (9.25, -2.25) rectangle (9.50, -2.50);
\filldraw[fill=bermuda, draw=black] (9.50, -2.25) rectangle (9.75, -2.50);
\filldraw[fill=bermuda, draw=black] (9.75, -2.25) rectangle (10.00, -2.50);
\filldraw[fill=cancan, draw=black] (10.00, -2.25) rectangle (10.25, -2.50);
\filldraw[fill=cancan, draw=black] (10.25, -2.25) rectangle (10.50, -2.50);
\filldraw[fill=cancan, draw=black] (10.50, -2.25) rectangle (10.75, -2.50);
\filldraw[fill=cancan, draw=black] (10.75, -2.25) rectangle (11.00, -2.50);
\filldraw[fill=cancan, draw=black] (11.00, -2.25) rectangle (11.25, -2.50);
\filldraw[fill=cancan, draw=black] (11.25, -2.25) rectangle (11.50, -2.50);
\filldraw[fill=cancan, draw=black] (11.50, -2.25) rectangle (11.75, -2.50);
\filldraw[fill=bermuda, draw=black] (11.75, -2.25) rectangle (12.00, -2.50);
\filldraw[fill=bermuda, draw=black] (12.00, -2.25) rectangle (12.25, -2.50);
\filldraw[fill=bermuda, draw=black] (12.25, -2.25) rectangle (12.50, -2.50);
\filldraw[fill=cancan, draw=black] (12.50, -2.25) rectangle (12.75, -2.50);
\filldraw[fill=bermuda, draw=black] (12.75, -2.25) rectangle (13.00, -2.50);
\filldraw[fill=cancan, draw=black] (13.00, -2.25) rectangle (13.25, -2.50);
\filldraw[fill=bermuda, draw=black] (13.25, -2.25) rectangle (13.50, -2.50);
\filldraw[fill=bermuda, draw=black] (13.50, -2.25) rectangle (13.75, -2.50);
\filldraw[fill=bermuda, draw=black] (13.75, -2.25) rectangle (14.00, -2.50);
\filldraw[fill=cancan, draw=black] (14.00, -2.25) rectangle (14.25, -2.50);
\filldraw[fill=cancan, draw=black] (14.25, -2.25) rectangle (14.50, -2.50);
\filldraw[fill=cancan, draw=black] (14.50, -2.25) rectangle (14.75, -2.50);
\filldraw[fill=bermuda, draw=black] (14.75, -2.25) rectangle (15.00, -2.50);
\filldraw[fill=bermuda, draw=black] (0.00, -2.50) rectangle (0.25, -2.75);
\filldraw[fill=bermuda, draw=black] (0.25, -2.50) rectangle (0.50, -2.75);
\filldraw[fill=cancan, draw=black] (0.50, -2.50) rectangle (0.75, -2.75);
\filldraw[fill=bermuda, draw=black] (0.75, -2.50) rectangle (1.00, -2.75);
\filldraw[fill=bermuda, draw=black] (1.00, -2.50) rectangle (1.25, -2.75);
\filldraw[fill=bermuda, draw=black] (1.25, -2.50) rectangle (1.50, -2.75);
\filldraw[fill=bermuda, draw=black] (1.50, -2.50) rectangle (1.75, -2.75);
\filldraw[fill=bermuda, draw=black] (1.75, -2.50) rectangle (2.00, -2.75);
\filldraw[fill=cancan, draw=black] (2.00, -2.50) rectangle (2.25, -2.75);
\filldraw[fill=bermuda, draw=black] (2.25, -2.50) rectangle (2.50, -2.75);
\filldraw[fill=cancan, draw=black] (2.50, -2.50) rectangle (2.75, -2.75);
\filldraw[fill=cancan, draw=black] (2.75, -2.50) rectangle (3.00, -2.75);
\filldraw[fill=bermuda, draw=black] (3.00, -2.50) rectangle (3.25, -2.75);
\filldraw[fill=bermuda, draw=black] (3.25, -2.50) rectangle (3.50, -2.75);
\filldraw[fill=cancan, draw=black] (3.50, -2.50) rectangle (3.75, -2.75);
\filldraw[fill=bermuda, draw=black] (3.75, -2.50) rectangle (4.00, -2.75);
\filldraw[fill=cancan, draw=black] (4.00, -2.50) rectangle (4.25, -2.75);
\filldraw[fill=cancan, draw=black] (4.25, -2.50) rectangle (4.50, -2.75);
\filldraw[fill=cancan, draw=black] (4.50, -2.50) rectangle (4.75, -2.75);
\filldraw[fill=bermuda, draw=black] (4.75, -2.50) rectangle (5.00, -2.75);
\filldraw[fill=bermuda, draw=black] (5.00, -2.50) rectangle (5.25, -2.75);
\filldraw[fill=bermuda, draw=black] (5.25, -2.50) rectangle (5.50, -2.75);
\filldraw[fill=cancan, draw=black] (5.50, -2.50) rectangle (5.75, -2.75);
\filldraw[fill=bermuda, draw=black] (5.75, -2.50) rectangle (6.00, -2.75);
\filldraw[fill=cancan, draw=black] (6.00, -2.50) rectangle (6.25, -2.75);
\filldraw[fill=bermuda, draw=black] (6.25, -2.50) rectangle (6.50, -2.75);
\filldraw[fill=bermuda, draw=black] (6.50, -2.50) rectangle (6.75, -2.75);
\filldraw[fill=bermuda, draw=black] (6.75, -2.50) rectangle (7.00, -2.75);
\filldraw[fill=cancan, draw=black] (7.00, -2.50) rectangle (7.25, -2.75);
\filldraw[fill=cancan, draw=black] (7.25, -2.50) rectangle (7.50, -2.75);
\filldraw[fill=cancan, draw=black] (7.50, -2.50) rectangle (7.75, -2.75);
\filldraw[fill=bermuda, draw=black] (7.75, -2.50) rectangle (8.00, -2.75);
\filldraw[fill=bermuda, draw=black] (8.00, -2.50) rectangle (8.25, -2.75);
\filldraw[fill=bermuda, draw=black] (8.25, -2.50) rectangle (8.50, -2.75);
\filldraw[fill=cancan, draw=black] (8.50, -2.50) rectangle (8.75, -2.75);
\filldraw[fill=bermuda, draw=black] (8.75, -2.50) rectangle (9.00, -2.75);
\filldraw[fill=bermuda, draw=black] (9.00, -2.50) rectangle (9.25, -2.75);
\filldraw[fill=bermuda, draw=black] (9.25, -2.50) rectangle (9.50, -2.75);
\filldraw[fill=bermuda, draw=black] (9.50, -2.50) rectangle (9.75, -2.75);
\filldraw[fill=bermuda, draw=black] (9.75, -2.50) rectangle (10.00, -2.75);
\filldraw[fill=cancan, draw=black] (10.00, -2.50) rectangle (10.25, -2.75);
\filldraw[fill=bermuda, draw=black] (10.25, -2.50) rectangle (10.50, -2.75);
\filldraw[fill=bermuda, draw=black] (10.50, -2.50) rectangle (10.75, -2.75);
\filldraw[fill=bermuda, draw=black] (10.75, -2.50) rectangle (11.00, -2.75);
\filldraw[fill=cancan, draw=black] (11.00, -2.50) rectangle (11.25, -2.75);
\filldraw[fill=bermuda, draw=black] (11.25, -2.50) rectangle (11.50, -2.75);
\filldraw[fill=bermuda, draw=black] (11.50, -2.50) rectangle (11.75, -2.75);
\filldraw[fill=bermuda, draw=black] (11.75, -2.50) rectangle (12.00, -2.75);
\filldraw[fill=cancan, draw=black] (12.00, -2.50) rectangle (12.25, -2.75);
\filldraw[fill=bermuda, draw=black] (12.25, -2.50) rectangle (12.50, -2.75);
\filldraw[fill=cancan, draw=black] (12.50, -2.50) rectangle (12.75, -2.75);
\filldraw[fill=bermuda, draw=black] (12.75, -2.50) rectangle (13.00, -2.75);
\filldraw[fill=cancan, draw=black] (13.00, -2.50) rectangle (13.25, -2.75);
\filldraw[fill=cancan, draw=black] (13.25, -2.50) rectangle (13.50, -2.75);
\filldraw[fill=cancan, draw=black] (13.50, -2.50) rectangle (13.75, -2.75);
\filldraw[fill=cancan, draw=black] (13.75, -2.50) rectangle (14.00, -2.75);
\filldraw[fill=cancan, draw=black] (14.00, -2.50) rectangle (14.25, -2.75);
\filldraw[fill=bermuda, draw=black] (14.25, -2.50) rectangle (14.50, -2.75);
\filldraw[fill=bermuda, draw=black] (14.50, -2.50) rectangle (14.75, -2.75);
\filldraw[fill=bermuda, draw=black] (14.75, -2.50) rectangle (15.00, -2.75);
\filldraw[fill=cancan, draw=black] (0.00, -2.75) rectangle (0.25, -3.00);
\filldraw[fill=cancan, draw=black] (0.25, -2.75) rectangle (0.50, -3.00);
\filldraw[fill=cancan, draw=black] (0.50, -2.75) rectangle (0.75, -3.00);
\filldraw[fill=bermuda, draw=black] (0.75, -2.75) rectangle (1.00, -3.00);
\filldraw[fill=bermuda, draw=black] (1.00, -2.75) rectangle (1.25, -3.00);
\filldraw[fill=bermuda, draw=black] (1.25, -2.75) rectangle (1.50, -3.00);
\filldraw[fill=cancan, draw=black] (1.50, -2.75) rectangle (1.75, -3.00);
\filldraw[fill=bermuda, draw=black] (1.75, -2.75) rectangle (2.00, -3.00);
\filldraw[fill=cancan, draw=black] (2.00, -2.75) rectangle (2.25, -3.00);
\filldraw[fill=bermuda, draw=black] (2.25, -2.75) rectangle (2.50, -3.00);
\filldraw[fill=cancan, draw=black] (2.50, -2.75) rectangle (2.75, -3.00);
\filldraw[fill=cancan, draw=black] (2.75, -2.75) rectangle (3.00, -3.00);
\filldraw[fill=bermuda, draw=black] (3.00, -2.75) rectangle (3.25, -3.00);
\filldraw[fill=bermuda, draw=black] (3.25, -2.75) rectangle (3.50, -3.00);
\filldraw[fill=cancan, draw=black] (3.50, -2.75) rectangle (3.75, -3.00);
\filldraw[fill=cancan, draw=black] (3.75, -2.75) rectangle (4.00, -3.00);
\filldraw[fill=cancan, draw=black] (4.00, -2.75) rectangle (4.25, -3.00);
\filldraw[fill=bermuda, draw=black] (4.25, -2.75) rectangle (4.50, -3.00);
\filldraw[fill=bermuda, draw=black] (4.50, -2.75) rectangle (4.75, -3.00);
\filldraw[fill=bermuda, draw=black] (4.75, -2.75) rectangle (5.00, -3.00);
\filldraw[fill=cancan, draw=black] (5.00, -2.75) rectangle (5.25, -3.00);
\filldraw[fill=cancan, draw=black] (5.25, -2.75) rectangle (5.50, -3.00);
\filldraw[fill=cancan, draw=black] (5.50, -2.75) rectangle (5.75, -3.00);
\filldraw[fill=bermuda, draw=black] (5.75, -2.75) rectangle (6.00, -3.00);
\filldraw[fill=bermuda, draw=black] (6.00, -2.75) rectangle (6.25, -3.00);
\filldraw[fill=bermuda, draw=black] (6.25, -2.75) rectangle (6.50, -3.00);
\filldraw[fill=cancan, draw=black] (6.50, -2.75) rectangle (6.75, -3.00);
\filldraw[fill=cancan, draw=black] (6.75, -2.75) rectangle (7.00, -3.00);
\filldraw[fill=cancan, draw=black] (7.00, -2.75) rectangle (7.25, -3.00);
\filldraw[fill=bermuda, draw=black] (7.25, -2.75) rectangle (7.50, -3.00);
\filldraw[fill=bermuda, draw=black] (7.50, -2.75) rectangle (7.75, -3.00);
\filldraw[fill=bermuda, draw=black] (7.75, -2.75) rectangle (8.00, -3.00);
\filldraw[fill=bermuda, draw=black] (8.00, -2.75) rectangle (8.25, -3.00);
\filldraw[fill=bermuda, draw=black] (8.25, -2.75) rectangle (8.50, -3.00);
\filldraw[fill=cancan, draw=black] (8.50, -2.75) rectangle (8.75, -3.00);
\filldraw[fill=bermuda, draw=black] (8.75, -2.75) rectangle (9.00, -3.00);
\filldraw[fill=cancan, draw=black] (9.00, -2.75) rectangle (9.25, -3.00);
\filldraw[fill=cancan, draw=black] (9.25, -2.75) rectangle (9.50, -3.00);
\filldraw[fill=cancan, draw=black] (9.50, -2.75) rectangle (9.75, -3.00);
\filldraw[fill=bermuda, draw=black] (9.75, -2.75) rectangle (10.00, -3.00);
\filldraw[fill=bermuda, draw=black] (10.00, -2.75) rectangle (10.25, -3.00);
\filldraw[fill=bermuda, draw=black] (10.25, -2.75) rectangle (10.50, -3.00);
\filldraw[fill=cancan, draw=black] (10.50, -2.75) rectangle (10.75, -3.00);
\filldraw[fill=cancan, draw=black] (10.75, -2.75) rectangle (11.00, -3.00);
\filldraw[fill=cancan, draw=black] (11.00, -2.75) rectangle (11.25, -3.00);
\filldraw[fill=cancan, draw=black] (11.25, -2.75) rectangle (11.50, -3.00);
\filldraw[fill=cancan, draw=black] (11.50, -2.75) rectangle (11.75, -3.00);
\filldraw[fill=bermuda, draw=black] (11.75, -2.75) rectangle (12.00, -3.00);
\filldraw[fill=cancan, draw=black] (12.00, -2.75) rectangle (12.25, -3.00);
\filldraw[fill=bermuda, draw=black] (12.25, -2.75) rectangle (12.50, -3.00);
\filldraw[fill=cancan, draw=black] (12.50, -2.75) rectangle (12.75, -3.00);
\filldraw[fill=bermuda, draw=black] (12.75, -2.75) rectangle (13.00, -3.00);
\filldraw[fill=cancan, draw=black] (13.00, -2.75) rectangle (13.25, -3.00);
\filldraw[fill=bermuda, draw=black] (13.25, -2.75) rectangle (13.50, -3.00);
\filldraw[fill=cancan, draw=black] (13.50, -2.75) rectangle (13.75, -3.00);
\filldraw[fill=cancan, draw=black] (13.75, -2.75) rectangle (14.00, -3.00);
\filldraw[fill=cancan, draw=black] (14.00, -2.75) rectangle (14.25, -3.00);
\filldraw[fill=bermuda, draw=black] (14.25, -2.75) rectangle (14.50, -3.00);
\filldraw[fill=cancan, draw=black] (14.50, -2.75) rectangle (14.75, -3.00);
\filldraw[fill=bermuda, draw=black] (14.75, -2.75) rectangle (15.00, -3.00);
\filldraw[fill=cancan, draw=black] (0.00, -3.00) rectangle (0.25, -3.25);
\filldraw[fill=cancan, draw=black] (0.25, -3.00) rectangle (0.50, -3.25);
\filldraw[fill=cancan, draw=black] (0.50, -3.00) rectangle (0.75, -3.25);
\filldraw[fill=bermuda, draw=black] (0.75, -3.00) rectangle (1.00, -3.25);
\filldraw[fill=cancan, draw=black] (1.00, -3.00) rectangle (1.25, -3.25);
\filldraw[fill=bermuda, draw=black] (1.25, -3.00) rectangle (1.50, -3.25);
\filldraw[fill=cancan, draw=black] (1.50, -3.00) rectangle (1.75, -3.25);
\filldraw[fill=bermuda, draw=black] (1.75, -3.00) rectangle (2.00, -3.25);
\filldraw[fill=bermuda, draw=black] (2.00, -3.00) rectangle (2.25, -3.25);
\filldraw[fill=bermuda, draw=black] (2.25, -3.00) rectangle (2.50, -3.25);
\filldraw[fill=cancan, draw=black] (2.50, -3.00) rectangle (2.75, -3.25);
\filldraw[fill=cancan, draw=black] (2.75, -3.00) rectangle (3.00, -3.25);
\filldraw[fill=cancan, draw=black] (3.00, -3.00) rectangle (3.25, -3.25);
\filldraw[fill=bermuda, draw=black] (3.25, -3.00) rectangle (3.50, -3.25);
\filldraw[fill=bermuda, draw=black] (3.50, -3.00) rectangle (3.75, -3.25);
\filldraw[fill=bermuda, draw=black] (3.75, -3.00) rectangle (4.00, -3.25);
\filldraw[fill=cancan, draw=black] (4.00, -3.00) rectangle (4.25, -3.25);
\filldraw[fill=cancan, draw=black] (4.25, -3.00) rectangle (4.50, -3.25);
\filldraw[fill=cancan, draw=black] (4.50, -3.00) rectangle (4.75, -3.25);
\filldraw[fill=bermuda, draw=black] (4.75, -3.00) rectangle (5.00, -3.25);
\filldraw[fill=bermuda, draw=black] (5.00, -3.00) rectangle (5.25, -3.25);
\filldraw[fill=bermuda, draw=black] (5.25, -3.00) rectangle (5.50, -3.25);
\filldraw[fill=cancan, draw=black] (5.50, -3.00) rectangle (5.75, -3.25);
\filldraw[fill=cancan, draw=black] (5.75, -3.00) rectangle (6.00, -3.25);
\filldraw[fill=cancan, draw=black] (6.00, -3.00) rectangle (6.25, -3.25);
\filldraw[fill=bermuda, draw=black] (6.25, -3.00) rectangle (6.50, -3.25);
\filldraw[fill=bermuda, draw=black] (6.50, -3.00) rectangle (6.75, -3.25);
\filldraw[fill=bermuda, draw=black] (6.75, -3.00) rectangle (7.00, -3.25);
\filldraw[fill=cancan, draw=black] (7.00, -3.00) rectangle (7.25, -3.25);
\filldraw[fill=cancan, draw=black] (7.25, -3.00) rectangle (7.50, -3.25);
\filldraw[fill=cancan, draw=black] (7.50, -3.00) rectangle (7.75, -3.25);
\filldraw[fill=bermuda, draw=black] (7.75, -3.00) rectangle (8.00, -3.25);
\filldraw[fill=bermuda, draw=black] (8.00, -3.00) rectangle (8.25, -3.25);
\filldraw[fill=bermuda, draw=black] (8.25, -3.00) rectangle (8.50, -3.25);
\filldraw[fill=cancan, draw=black] (8.50, -3.00) rectangle (8.75, -3.25);
\filldraw[fill=cancan, draw=black] (8.75, -3.00) rectangle (9.00, -3.25);
\filldraw[fill=cancan, draw=black] (9.00, -3.00) rectangle (9.25, -3.25);
\filldraw[fill=cancan, draw=black] (9.25, -3.00) rectangle (9.50, -3.25);
\filldraw[fill=cancan, draw=black] (9.50, -3.00) rectangle (9.75, -3.25);
\filldraw[fill=bermuda, draw=black] (9.75, -3.00) rectangle (10.00, -3.25);
\filldraw[fill=cancan, draw=black] (10.00, -3.00) rectangle (10.25, -3.25);
\filldraw[fill=cancan, draw=black] (10.25, -3.00) rectangle (10.50, -3.25);
\filldraw[fill=cancan, draw=black] (10.50, -3.00) rectangle (10.75, -3.25);
\filldraw[fill=cancan, draw=black] (10.75, -3.00) rectangle (11.00, -3.25);
\filldraw[fill=cancan, draw=black] (11.00, -3.00) rectangle (11.25, -3.25);
\filldraw[fill=bermuda, draw=black] (11.25, -3.00) rectangle (11.50, -3.25);
\filldraw[fill=bermuda, draw=black] (11.50, -3.00) rectangle (11.75, -3.25);
\filldraw[fill=bermuda, draw=black] (11.75, -3.00) rectangle (12.00, -3.25);
\filldraw[fill=bermuda, draw=black] (12.00, -3.00) rectangle (12.25, -3.25);
\filldraw[fill=bermuda, draw=black] (12.25, -3.00) rectangle (12.50, -3.25);
\filldraw[fill=cancan, draw=black] (12.50, -3.00) rectangle (12.75, -3.25);
\filldraw[fill=bermuda, draw=black] (12.75, -3.00) rectangle (13.00, -3.25);
\filldraw[fill=bermuda, draw=black] (13.00, -3.00) rectangle (13.25, -3.25);
\filldraw[fill=bermuda, draw=black] (13.25, -3.00) rectangle (13.50, -3.25);
\filldraw[fill=cancan, draw=black] (13.50, -3.00) rectangle (13.75, -3.25);
\filldraw[fill=cancan, draw=black] (13.75, -3.00) rectangle (14.00, -3.25);
\filldraw[fill=cancan, draw=black] (14.00, -3.00) rectangle (14.25, -3.25);
\filldraw[fill=bermuda, draw=black] (14.25, -3.00) rectangle (14.50, -3.25);
\filldraw[fill=bermuda, draw=black] (14.50, -3.00) rectangle (14.75, -3.25);
\filldraw[fill=bermuda, draw=black] (14.75, -3.00) rectangle (15.00, -3.25);
\filldraw[fill=bermuda, draw=black] (0.00, -3.25) rectangle (0.25, -3.50);
\filldraw[fill=bermuda, draw=black] (0.25, -3.25) rectangle (0.50, -3.50);
\filldraw[fill=cancan, draw=black] (0.50, -3.25) rectangle (0.75, -3.50);
\filldraw[fill=bermuda, draw=black] (0.75, -3.25) rectangle (1.00, -3.50);
\filldraw[fill=bermuda, draw=black] (1.00, -3.25) rectangle (1.25, -3.50);
\filldraw[fill=bermuda, draw=black] (1.25, -3.25) rectangle (1.50, -3.50);
\filldraw[fill=cancan, draw=black] (1.50, -3.25) rectangle (1.75, -3.50);
\filldraw[fill=cancan, draw=black] (1.75, -3.25) rectangle (2.00, -3.50);
\filldraw[fill=cancan, draw=black] (2.00, -3.25) rectangle (2.25, -3.50);
\filldraw[fill=bermuda, draw=black] (2.25, -3.25) rectangle (2.50, -3.50);
\filldraw[fill=bermuda, draw=black] (2.50, -3.25) rectangle (2.75, -3.50);
\filldraw[fill=bermuda, draw=black] (2.75, -3.25) rectangle (3.00, -3.50);
\filldraw[fill=cancan, draw=black] (3.00, -3.25) rectangle (3.25, -3.50);
\filldraw[fill=cancan, draw=black] (3.25, -3.25) rectangle (3.50, -3.50);
\filldraw[fill=cancan, draw=black] (3.50, -3.25) rectangle (3.75, -3.50);
\filldraw[fill=bermuda, draw=black] (3.75, -3.25) rectangle (4.00, -3.50);
\filldraw[fill=bermuda, draw=black] (4.00, -3.25) rectangle (4.25, -3.50);
\filldraw[fill=bermuda, draw=black] (4.25, -3.25) rectangle (4.50, -3.50);
\filldraw[fill=cancan, draw=black] (4.50, -3.25) rectangle (4.75, -3.50);
\filldraw[fill=cancan, draw=black] (4.75, -3.25) rectangle (5.00, -3.50);
\filldraw[fill=cancan, draw=black] (5.00, -3.25) rectangle (5.25, -3.50);
\filldraw[fill=bermuda, draw=black] (5.25, -3.25) rectangle (5.50, -3.50);
\filldraw[fill=bermuda, draw=black] (5.50, -3.25) rectangle (5.75, -3.50);
\filldraw[fill=bermuda, draw=black] (5.75, -3.25) rectangle (6.00, -3.50);
\filldraw[fill=cancan, draw=black] (6.00, -3.25) rectangle (6.25, -3.50);
\filldraw[fill=cancan, draw=black] (6.25, -3.25) rectangle (6.50, -3.50);
} } }\end{equation*}
\begin{equation*}
\hspace{0.3pt} b_{12} = \vcenter{\hbox{ \tikz{
\filldraw[fill=cancan, draw=black] (0.00, 0.00) rectangle (0.25, -0.25);
\filldraw[fill=bermuda, draw=black] (0.25, 0.00) rectangle (0.50, -0.25);
\filldraw[fill=bermuda, draw=black] (0.50, 0.00) rectangle (0.75, -0.25);
\filldraw[fill=bermuda, draw=black] (0.75, 0.00) rectangle (1.00, -0.25);
\filldraw[fill=cancan, draw=black] (1.00, 0.00) rectangle (1.25, -0.25);
\filldraw[fill=cancan, draw=black] (1.25, 0.00) rectangle (1.50, -0.25);
\filldraw[fill=cancan, draw=black] (1.50, 0.00) rectangle (1.75, -0.25);
\filldraw[fill=cancan, draw=black] (1.75, 0.00) rectangle (2.00, -0.25);
\filldraw[fill=cancan, draw=black] (2.00, 0.00) rectangle (2.25, -0.25);
\filldraw[fill=cancan, draw=black] (2.25, 0.00) rectangle (2.50, -0.25);
\filldraw[fill=cancan, draw=black] (2.50, 0.00) rectangle (2.75, -0.25);
\filldraw[fill=cancan, draw=black] (2.75, 0.00) rectangle (3.00, -0.25);
\filldraw[fill=cancan, draw=black] (3.00, 0.00) rectangle (3.25, -0.25);
\filldraw[fill=bermuda, draw=black] (3.25, 0.00) rectangle (3.50, -0.25);
\filldraw[fill=cancan, draw=black] (3.50, 0.00) rectangle (3.75, -0.25);
\filldraw[fill=cancan, draw=black] (3.75, 0.00) rectangle (4.00, -0.25);
\filldraw[fill=cancan, draw=black] (4.00, 0.00) rectangle (4.25, -0.25);
\filldraw[fill=cancan, draw=black] (4.25, 0.00) rectangle (4.50, -0.25);
\filldraw[fill=cancan, draw=black] (4.50, 0.00) rectangle (4.75, -0.25);
\filldraw[fill=cancan, draw=black] (4.75, 0.00) rectangle (5.00, -0.25);
\filldraw[fill=cancan, draw=black] (5.00, 0.00) rectangle (5.25, -0.25);
\filldraw[fill=bermuda, draw=black] (5.25, 0.00) rectangle (5.50, -0.25);
\filldraw[fill=cancan, draw=black] (5.50, 0.00) rectangle (5.75, -0.25);
\filldraw[fill=bermuda, draw=black] (5.75, 0.00) rectangle (6.00, -0.25);
\filldraw[fill=cancan, draw=black] (6.00, 0.00) rectangle (6.25, -0.25);
\filldraw[fill=cancan, draw=black] (6.25, 0.00) rectangle (6.50, -0.25);
\filldraw[fill=bermuda, draw=black] (6.50, 0.00) rectangle (6.75, -0.25);
\filldraw[fill=bermuda, draw=black] (6.75, 0.00) rectangle (7.00, -0.25);
\filldraw[fill=cancan, draw=black] (7.00, 0.00) rectangle (7.25, -0.25);
\filldraw[fill=bermuda, draw=black] (7.25, 0.00) rectangle (7.50, -0.25);
\filldraw[fill=cancan, draw=black] (7.50, 0.00) rectangle (7.75, -0.25);
\filldraw[fill=bermuda, draw=black] (7.75, 0.00) rectangle (8.00, -0.25);
\filldraw[fill=cancan, draw=black] (8.00, 0.00) rectangle (8.25, -0.25);
\filldraw[fill=cancan, draw=black] (8.25, 0.00) rectangle (8.50, -0.25);
\filldraw[fill=cancan, draw=black] (8.50, 0.00) rectangle (8.75, -0.25);
\filldraw[fill=cancan, draw=black] (8.75, 0.00) rectangle (9.00, -0.25);
\filldraw[fill=cancan, draw=black] (9.00, 0.00) rectangle (9.25, -0.25);
\filldraw[fill=bermuda, draw=black] (9.25, 0.00) rectangle (9.50, -0.25);
\filldraw[fill=bermuda, draw=black] (9.50, 0.00) rectangle (9.75, -0.25);
\filldraw[fill=bermuda, draw=black] (9.75, 0.00) rectangle (10.00, -0.25);
\filldraw[fill=cancan, draw=black] (10.00, 0.00) rectangle (10.25, -0.25);
\filldraw[fill=cancan, draw=black] (10.25, 0.00) rectangle (10.50, -0.25);
\filldraw[fill=cancan, draw=black] (10.50, 0.00) rectangle (10.75, -0.25);
\filldraw[fill=bermuda, draw=black] (10.75, 0.00) rectangle (11.00, -0.25);
\filldraw[fill=bermuda, draw=black] (11.00, 0.00) rectangle (11.25, -0.25);
\filldraw[fill=bermuda, draw=black] (11.25, 0.00) rectangle (11.50, -0.25);
\filldraw[fill=cancan, draw=black] (11.50, 0.00) rectangle (11.75, -0.25);
\filldraw[fill=cancan, draw=black] (11.75, 0.00) rectangle (12.00, -0.25);
\filldraw[fill=cancan, draw=black] (12.00, 0.00) rectangle (12.25, -0.25);
\filldraw[fill=bermuda, draw=black] (12.25, 0.00) rectangle (12.50, -0.25);
\filldraw[fill=bermuda, draw=black] (12.50, 0.00) rectangle (12.75, -0.25);
\filldraw[fill=bermuda, draw=black] (12.75, 0.00) rectangle (13.00, -0.25);
\filldraw[fill=cancan, draw=black] (13.00, 0.00) rectangle (13.25, -0.25);
\filldraw[fill=cancan, draw=black] (13.25, 0.00) rectangle (13.50, -0.25);
\filldraw[fill=cancan, draw=black] (13.50, 0.00) rectangle (13.75, -0.25);
\filldraw[fill=bermuda, draw=black] (13.75, 0.00) rectangle (14.00, -0.25);
\filldraw[fill=bermuda, draw=black] (14.00, 0.00) rectangle (14.25, -0.25);
\filldraw[fill=bermuda, draw=black] (14.25, 0.00) rectangle (14.50, -0.25);
\filldraw[fill=cancan, draw=black] (14.50, 0.00) rectangle (14.75, -0.25);
\filldraw[fill=bermuda, draw=black] (14.75, 0.00) rectangle (15.00, -0.25);
\filldraw[fill=bermuda, draw=black] (0.00, -0.25) rectangle (0.25, -0.50);
\filldraw[fill=bermuda, draw=black] (0.25, -0.25) rectangle (0.50, -0.50);
\filldraw[fill=cancan, draw=black] (0.50, -0.25) rectangle (0.75, -0.50);
\filldraw[fill=cancan, draw=black] (0.75, -0.25) rectangle (1.00, -0.50);
\filldraw[fill=cancan, draw=black] (1.00, -0.25) rectangle (1.25, -0.50);
\filldraw[fill=bermuda, draw=black] (1.25, -0.25) rectangle (1.50, -0.50);
\filldraw[fill=bermuda, draw=black] (1.50, -0.25) rectangle (1.75, -0.50);
\filldraw[fill=bermuda, draw=black] (1.75, -0.25) rectangle (2.00, -0.50);
\filldraw[fill=cancan, draw=black] (2.00, -0.25) rectangle (2.25, -0.50);
\filldraw[fill=cancan, draw=black] (2.25, -0.25) rectangle (2.50, -0.50);
\filldraw[fill=bermuda, draw=black] (2.50, -0.25) rectangle (2.75, -0.50);
\filldraw[fill=bermuda, draw=black] (2.75, -0.25) rectangle (3.00, -0.50);
\filldraw[fill=cancan, draw=black] (3.00, -0.25) rectangle (3.25, -0.50);
\filldraw[fill=cancan, draw=black] (3.25, -0.25) rectangle (3.50, -0.50);
\filldraw[fill=cancan, draw=black] (3.50, -0.25) rectangle (3.75, -0.50);
\filldraw[fill=cancan, draw=black] (3.75, -0.25) rectangle (4.00, -0.50);
\filldraw[fill=bermuda, draw=black] (4.00, -0.25) rectangle (4.25, -0.50);
\filldraw[fill=bermuda, draw=black] (4.25, -0.25) rectangle (4.50, -0.50);
\filldraw[fill=cancan, draw=black] (4.50, -0.25) rectangle (4.75, -0.50);
\filldraw[fill=cancan, draw=black] (4.75, -0.25) rectangle (5.00, -0.50);
\filldraw[fill=cancan, draw=black] (5.00, -0.25) rectangle (5.25, -0.50);
\filldraw[fill=cancan, draw=black] (5.25, -0.25) rectangle (5.50, -0.50);
\filldraw[fill=bermuda, draw=black] (5.50, -0.25) rectangle (5.75, -0.50);
\filldraw[fill=bermuda, draw=black] (5.75, -0.25) rectangle (6.00, -0.50);
\filldraw[fill=cancan, draw=black] (6.00, -0.25) rectangle (6.25, -0.50);
\filldraw[fill=cancan, draw=black] (6.25, -0.25) rectangle (6.50, -0.50);
\filldraw[fill=cancan, draw=black] (6.50, -0.25) rectangle (6.75, -0.50);
\filldraw[fill=bermuda, draw=black] (6.75, -0.25) rectangle (7.00, -0.50);
\filldraw[fill=bermuda, draw=black] (7.00, -0.25) rectangle (7.25, -0.50);
\filldraw[fill=bermuda, draw=black] (7.25, -0.25) rectangle (7.50, -0.50);
\filldraw[fill=cancan, draw=black] (7.50, -0.25) rectangle (7.75, -0.50);
\filldraw[fill=cancan, draw=black] (7.75, -0.25) rectangle (8.00, -0.50);
\filldraw[fill=cancan, draw=black] (8.00, -0.25) rectangle (8.25, -0.50);
\filldraw[fill=bermuda, draw=black] (8.25, -0.25) rectangle (8.50, -0.50);
\filldraw[fill=bermuda, draw=black] (8.50, -0.25) rectangle (8.75, -0.50);
\filldraw[fill=bermuda, draw=black] (8.75, -0.25) rectangle (9.00, -0.50);
\filldraw[fill=cancan, draw=black] (9.00, -0.25) rectangle (9.25, -0.50);
\filldraw[fill=cancan, draw=black] (9.25, -0.25) rectangle (9.50, -0.50);
\filldraw[fill=cancan, draw=black] (9.50, -0.25) rectangle (9.75, -0.50);
\filldraw[fill=bermuda, draw=black] (9.75, -0.25) rectangle (10.00, -0.50);
\filldraw[fill=bermuda, draw=black] (10.00, -0.25) rectangle (10.25, -0.50);
\filldraw[fill=bermuda, draw=black] (10.25, -0.25) rectangle (10.50, -0.50);
\filldraw[fill=cancan, draw=black] (10.50, -0.25) rectangle (10.75, -0.50);
\filldraw[fill=cancan, draw=black] (10.75, -0.25) rectangle (11.00, -0.50);
\filldraw[fill=cancan, draw=black] (11.00, -0.25) rectangle (11.25, -0.50);
\filldraw[fill=cancan, draw=black] (11.25, -0.25) rectangle (11.50, -0.50);
\filldraw[fill=bermuda, draw=black] (11.50, -0.25) rectangle (11.75, -0.50);
\filldraw[fill=bermuda, draw=black] (11.75, -0.25) rectangle (12.00, -0.50);
\filldraw[fill=cancan, draw=black] (12.00, -0.25) rectangle (12.25, -0.50);
\filldraw[fill=bermuda, draw=black] (12.25, -0.25) rectangle (12.50, -0.50);
\filldraw[fill=cancan, draw=black] (12.50, -0.25) rectangle (12.75, -0.50);
\filldraw[fill=cancan, draw=black] (12.75, -0.25) rectangle (13.00, -0.50);
\filldraw[fill=cancan, draw=black] (13.00, -0.25) rectangle (13.25, -0.50);
\filldraw[fill=bermuda, draw=black] (13.25, -0.25) rectangle (13.50, -0.50);
\filldraw[fill=cancan, draw=black] (13.50, -0.25) rectangle (13.75, -0.50);
\filldraw[fill=cancan, draw=black] (13.75, -0.25) rectangle (14.00, -0.50);
\filldraw[fill=cancan, draw=black] (14.00, -0.25) rectangle (14.25, -0.50);
\filldraw[fill=cancan, draw=black] (14.25, -0.25) rectangle (14.50, -0.50);
\filldraw[fill=cancan, draw=black] (14.50, -0.25) rectangle (14.75, -0.50);
\filldraw[fill=cancan, draw=black] (14.75, -0.25) rectangle (15.00, -0.50);
\filldraw[fill=cancan, draw=black] (0.00, -0.50) rectangle (0.25, -0.75);
\filldraw[fill=bermuda, draw=black] (0.25, -0.50) rectangle (0.50, -0.75);
\filldraw[fill=cancan, draw=black] (0.50, -0.50) rectangle (0.75, -0.75);
\filldraw[fill=bermuda, draw=black] (0.75, -0.50) rectangle (1.00, -0.75);
\filldraw[fill=cancan, draw=black] (1.00, -0.50) rectangle (1.25, -0.75);
\filldraw[fill=cancan, draw=black] (1.25, -0.50) rectangle (1.50, -0.75);
\filldraw[fill=bermuda, draw=black] (1.50, -0.50) rectangle (1.75, -0.75);
\filldraw[fill=bermuda, draw=black] (1.75, -0.50) rectangle (2.00, -0.75);
\filldraw[fill=bermuda, draw=black] (2.00, -0.50) rectangle (2.25, -0.75);
\filldraw[fill=bermuda, draw=black] (2.25, -0.50) rectangle (2.50, -0.75);
\filldraw[fill=bermuda, draw=black] (2.50, -0.50) rectangle (2.75, -0.75);
\filldraw[fill=bermuda, draw=black] (2.75, -0.50) rectangle (3.00, -0.75);
\filldraw[fill=bermuda, draw=black] (3.00, -0.50) rectangle (3.25, -0.75);
\filldraw[fill=bermuda, draw=black] (3.25, -0.50) rectangle (3.50, -0.75);
\filldraw[fill=cancan, draw=black] (3.50, -0.50) rectangle (3.75, -0.75);
\filldraw[fill=cancan, draw=black] (3.75, -0.50) rectangle (4.00, -0.75);
\filldraw[fill=cancan, draw=black] (4.00, -0.50) rectangle (4.25, -0.75);
\filldraw[fill=bermuda, draw=black] (4.25, -0.50) rectangle (4.50, -0.75);
\filldraw[fill=bermuda, draw=black] (4.50, -0.50) rectangle (4.75, -0.75);
\filldraw[fill=bermuda, draw=black] (4.75, -0.50) rectangle (5.00, -0.75);
\filldraw[fill=cancan, draw=black] (5.00, -0.50) rectangle (5.25, -0.75);
\filldraw[fill=cancan, draw=black] (5.25, -0.50) rectangle (5.50, -0.75);
\filldraw[fill=cancan, draw=black] (5.50, -0.50) rectangle (5.75, -0.75);
\filldraw[fill=bermuda, draw=black] (5.75, -0.50) rectangle (6.00, -0.75);
\filldraw[fill=bermuda, draw=black] (6.00, -0.50) rectangle (6.25, -0.75);
\filldraw[fill=bermuda, draw=black] (6.25, -0.50) rectangle (6.50, -0.75);
\filldraw[fill=bermuda, draw=black] (6.50, -0.50) rectangle (6.75, -0.75);
\filldraw[fill=bermuda, draw=black] (6.75, -0.50) rectangle (7.00, -0.75);
\filldraw[fill=bermuda, draw=black] (7.00, -0.50) rectangle (7.25, -0.75);
\filldraw[fill=bermuda, draw=black] (7.25, -0.50) rectangle (7.50, -0.75);
\filldraw[fill=cancan, draw=black] (7.50, -0.50) rectangle (7.75, -0.75);
\filldraw[fill=cancan, draw=black] (7.75, -0.50) rectangle (8.00, -0.75);
\filldraw[fill=cancan, draw=black] (8.00, -0.50) rectangle (8.25, -0.75);
\filldraw[fill=bermuda, draw=black] (8.25, -0.50) rectangle (8.50, -0.75);
\filldraw[fill=bermuda, draw=black] (8.50, -0.50) rectangle (8.75, -0.75);
\filldraw[fill=bermuda, draw=black] (8.75, -0.50) rectangle (9.00, -0.75);
\filldraw[fill=cancan, draw=black] (9.00, -0.50) rectangle (9.25, -0.75);
\filldraw[fill=cancan, draw=black] (9.25, -0.50) rectangle (9.50, -0.75);
\filldraw[fill=cancan, draw=black] (9.50, -0.50) rectangle (9.75, -0.75);
\filldraw[fill=bermuda, draw=black] (9.75, -0.50) rectangle (10.00, -0.75);
\filldraw[fill=bermuda, draw=black] (10.00, -0.50) rectangle (10.25, -0.75);
\filldraw[fill=bermuda, draw=black] (10.25, -0.50) rectangle (10.50, -0.75);
\filldraw[fill=cancan, draw=black] (10.50, -0.50) rectangle (10.75, -0.75);
\filldraw[fill=cancan, draw=black] (10.75, -0.50) rectangle (11.00, -0.75);
\filldraw[fill=cancan, draw=black] (11.00, -0.50) rectangle (11.25, -0.75);
\filldraw[fill=bermuda, draw=black] (11.25, -0.50) rectangle (11.50, -0.75);
\filldraw[fill=bermuda, draw=black] (11.50, -0.50) rectangle (11.75, -0.75);
\filldraw[fill=bermuda, draw=black] (11.75, -0.50) rectangle (12.00, -0.75);
\filldraw[fill=cancan, draw=black] (12.00, -0.50) rectangle (12.25, -0.75);
\filldraw[fill=cancan, draw=black] (12.25, -0.50) rectangle (12.50, -0.75);
\filldraw[fill=cancan, draw=black] (12.50, -0.50) rectangle (12.75, -0.75);
\filldraw[fill=bermuda, draw=black] (12.75, -0.50) rectangle (13.00, -0.75);
\filldraw[fill=bermuda, draw=black] (13.00, -0.50) rectangle (13.25, -0.75);
\filldraw[fill=bermuda, draw=black] (13.25, -0.50) rectangle (13.50, -0.75);
\filldraw[fill=cancan, draw=black] (13.50, -0.50) rectangle (13.75, -0.75);
\filldraw[fill=bermuda, draw=black] (13.75, -0.50) rectangle (14.00, -0.75);
\filldraw[fill=cancan, draw=black] (14.00, -0.50) rectangle (14.25, -0.75);
\filldraw[fill=bermuda, draw=black] (14.25, -0.50) rectangle (14.50, -0.75);
\filldraw[fill=cancan, draw=black] (14.50, -0.50) rectangle (14.75, -0.75);
\filldraw[fill=cancan, draw=black] (14.75, -0.50) rectangle (15.00, -0.75);
\filldraw[fill=cancan, draw=black] (0.00, -0.75) rectangle (0.25, -1.00);
\filldraw[fill=bermuda, draw=black] (0.25, -0.75) rectangle (0.50, -1.00);
\filldraw[fill=cancan, draw=black] (0.50, -0.75) rectangle (0.75, -1.00);
\filldraw[fill=bermuda, draw=black] (0.75, -0.75) rectangle (1.00, -1.00);
\filldraw[fill=cancan, draw=black] (1.00, -0.75) rectangle (1.25, -1.00);
\filldraw[fill=cancan, draw=black] (1.25, -0.75) rectangle (1.50, -1.00);
\filldraw[fill=cancan, draw=black] (1.50, -0.75) rectangle (1.75, -1.00);
\filldraw[fill=bermuda, draw=black] (1.75, -0.75) rectangle (2.00, -1.00);
\filldraw[fill=cancan, draw=black] (2.00, -0.75) rectangle (2.25, -1.00);
\filldraw[fill=bermuda, draw=black] (2.25, -0.75) rectangle (2.50, -1.00);
\filldraw[fill=bermuda, draw=black] (2.50, -0.75) rectangle (2.75, -1.00);
\filldraw[fill=bermuda, draw=black] (2.75, -0.75) rectangle (3.00, -1.00);
\filldraw[fill=cancan, draw=black] (3.00, -0.75) rectangle (3.25, -1.00);
\filldraw[fill=cancan, draw=black] (3.25, -0.75) rectangle (3.50, -1.00);
\filldraw[fill=cancan, draw=black] (3.50, -0.75) rectangle (3.75, -1.00);
\filldraw[fill=cancan, draw=black] (3.75, -0.75) rectangle (4.00, -1.00);
\filldraw[fill=cancan, draw=black] (4.00, -0.75) rectangle (4.25, -1.00);
\filldraw[fill=cancan, draw=black] (4.25, -0.75) rectangle (4.50, -1.00);
\filldraw[fill=bermuda, draw=black] (4.50, -0.75) rectangle (4.75, -1.00);
\filldraw[fill=bermuda, draw=black] (4.75, -0.75) rectangle (5.00, -1.00);
\filldraw[fill=cancan, draw=black] (5.00, -0.75) rectangle (5.25, -1.00);
\filldraw[fill=cancan, draw=black] (5.25, -0.75) rectangle (5.50, -1.00);
\filldraw[fill=cancan, draw=black] (5.50, -0.75) rectangle (5.75, -1.00);
\filldraw[fill=cancan, draw=black] (5.75, -0.75) rectangle (6.00, -1.00);
\filldraw[fill=bermuda, draw=black] (6.00, -0.75) rectangle (6.25, -1.00);
\filldraw[fill=bermuda, draw=black] (6.25, -0.75) rectangle (6.50, -1.00);
\filldraw[fill=cancan, draw=black] (6.50, -0.75) rectangle (6.75, -1.00);
\filldraw[fill=bermuda, draw=black] (6.75, -0.75) rectangle (7.00, -1.00);
\filldraw[fill=bermuda, draw=black] (7.00, -0.75) rectangle (7.25, -1.00);
\filldraw[fill=bermuda, draw=black] (7.25, -0.75) rectangle (7.50, -1.00);
\filldraw[fill=cancan, draw=black] (7.50, -0.75) rectangle (7.75, -1.00);
\filldraw[fill=cancan, draw=black] (7.75, -0.75) rectangle (8.00, -1.00);
\filldraw[fill=cancan, draw=black] (8.00, -0.75) rectangle (8.25, -1.00);
\filldraw[fill=bermuda, draw=black] (8.25, -0.75) rectangle (8.50, -1.00);
\filldraw[fill=bermuda, draw=black] (8.50, -0.75) rectangle (8.75, -1.00);
\filldraw[fill=bermuda, draw=black] (8.75, -0.75) rectangle (9.00, -1.00);
\filldraw[fill=cancan, draw=black] (9.00, -0.75) rectangle (9.25, -1.00);
\filldraw[fill=cancan, draw=black] (9.25, -0.75) rectangle (9.50, -1.00);
\filldraw[fill=cancan, draw=black] (9.50, -0.75) rectangle (9.75, -1.00);
\filldraw[fill=bermuda, draw=black] (9.75, -0.75) rectangle (10.00, -1.00);
\filldraw[fill=bermuda, draw=black] (10.00, -0.75) rectangle (10.25, -1.00);
\filldraw[fill=bermuda, draw=black] (10.25, -0.75) rectangle (10.50, -1.00);
\filldraw[fill=bermuda, draw=black] (10.50, -0.75) rectangle (10.75, -1.00);
\filldraw[fill=bermuda, draw=black] (10.75, -0.75) rectangle (11.00, -1.00);
\filldraw[fill=bermuda, draw=black] (11.00, -0.75) rectangle (11.25, -1.00);
\filldraw[fill=bermuda, draw=black] (11.25, -0.75) rectangle (11.50, -1.00);
\filldraw[fill=cancan, draw=black] (11.50, -0.75) rectangle (11.75, -1.00);
\filldraw[fill=bermuda, draw=black] (11.75, -0.75) rectangle (12.00, -1.00);
\filldraw[fill=cancan, draw=black] (12.00, -0.75) rectangle (12.25, -1.00);
\filldraw[fill=bermuda, draw=black] (12.25, -0.75) rectangle (12.50, -1.00);
\filldraw[fill=bermuda, draw=black] (12.50, -0.75) rectangle (12.75, -1.00);
\filldraw[fill=bermuda, draw=black] (12.75, -0.75) rectangle (13.00, -1.00);
\filldraw[fill=cancan, draw=black] (13.00, -0.75) rectangle (13.25, -1.00);
\filldraw[fill=bermuda, draw=black] (13.25, -0.75) rectangle (13.50, -1.00);
\filldraw[fill=bermuda, draw=black] (13.50, -0.75) rectangle (13.75, -1.00);
\filldraw[fill=bermuda, draw=black] (13.75, -0.75) rectangle (14.00, -1.00);
\filldraw[fill=cancan, draw=black] (14.00, -0.75) rectangle (14.25, -1.00);
\filldraw[fill=bermuda, draw=black] (14.25, -0.75) rectangle (14.50, -1.00);
\filldraw[fill=bermuda, draw=black] (14.50, -0.75) rectangle (14.75, -1.00);
\filldraw[fill=bermuda, draw=black] (14.75, -0.75) rectangle (15.00, -1.00);
\filldraw[fill=cancan, draw=black] (0.00, -1.00) rectangle (0.25, -1.25);
\filldraw[fill=cancan, draw=black] (0.25, -1.00) rectangle (0.50, -1.25);
\filldraw[fill=cancan, draw=black] (0.50, -1.00) rectangle (0.75, -1.25);
\filldraw[fill=bermuda, draw=black] (0.75, -1.00) rectangle (1.00, -1.25);
\filldraw[fill=bermuda, draw=black] (1.00, -1.00) rectangle (1.25, -1.25);
\filldraw[fill=bermuda, draw=black] (1.25, -1.00) rectangle (1.50, -1.25);
\filldraw[fill=cancan, draw=black] (1.50, -1.00) rectangle (1.75, -1.25);
\filldraw[fill=cancan, draw=black] (1.75, -1.00) rectangle (2.00, -1.25);
\filldraw[fill=cancan, draw=black] (2.00, -1.00) rectangle (2.25, -1.25);
\filldraw[fill=bermuda, draw=black] (2.25, -1.00) rectangle (2.50, -1.25);
\filldraw[fill=bermuda, draw=black] (2.50, -1.00) rectangle (2.75, -1.25);
\filldraw[fill=bermuda, draw=black] (2.75, -1.00) rectangle (3.00, -1.25);
\filldraw[fill=cancan, draw=black] (3.00, -1.00) rectangle (3.25, -1.25);
\filldraw[fill=cancan, draw=black] (3.25, -1.00) rectangle (3.50, -1.25);
\filldraw[fill=cancan, draw=black] (3.50, -1.00) rectangle (3.75, -1.25);
\filldraw[fill=bermuda, draw=black] (3.75, -1.00) rectangle (4.00, -1.25);
\filldraw[fill=bermuda, draw=black] (4.00, -1.00) rectangle (4.25, -1.25);
\filldraw[fill=bermuda, draw=black] (4.25, -1.00) rectangle (4.50, -1.25);
\filldraw[fill=cancan, draw=black] (4.50, -1.00) rectangle (4.75, -1.25);
\filldraw[fill=cancan, draw=black] (4.75, -1.00) rectangle (5.00, -1.25);
\filldraw[fill=cancan, draw=black] (5.00, -1.00) rectangle (5.25, -1.25);
\filldraw[fill=cancan, draw=black] (5.25, -1.00) rectangle (5.50, -1.25);
\filldraw[fill=cancan, draw=black] (5.50, -1.00) rectangle (5.75, -1.25);
\filldraw[fill=bermuda, draw=black] (5.75, -1.00) rectangle (6.00, -1.25);
\filldraw[fill=cancan, draw=black] (6.00, -1.00) rectangle (6.25, -1.25);
\filldraw[fill=bermuda, draw=black] (6.25, -1.00) rectangle (6.50, -1.25);
\filldraw[fill=cancan, draw=black] (6.50, -1.00) rectangle (6.75, -1.25);
\filldraw[fill=bermuda, draw=black] (6.75, -1.00) rectangle (7.00, -1.25);
\filldraw[fill=bermuda, draw=black] (7.00, -1.00) rectangle (7.25, -1.25);
\filldraw[fill=bermuda, draw=black] (7.25, -1.00) rectangle (7.50, -1.25);
\filldraw[fill=cancan, draw=black] (7.50, -1.00) rectangle (7.75, -1.25);
\filldraw[fill=cancan, draw=black] (7.75, -1.00) rectangle (8.00, -1.25);
\filldraw[fill=cancan, draw=black] (8.00, -1.00) rectangle (8.25, -1.25);
\filldraw[fill=bermuda, draw=black] (8.25, -1.00) rectangle (8.50, -1.25);
\filldraw[fill=bermuda, draw=black] (8.50, -1.00) rectangle (8.75, -1.25);
\filldraw[fill=bermuda, draw=black] (8.75, -1.00) rectangle (9.00, -1.25);
\filldraw[fill=cancan, draw=black] (9.00, -1.00) rectangle (9.25, -1.25);
\filldraw[fill=bermuda, draw=black] (9.25, -1.00) rectangle (9.50, -1.25);
\filldraw[fill=bermuda, draw=black] (9.50, -1.00) rectangle (9.75, -1.25);
\filldraw[fill=bermuda, draw=black] (9.75, -1.00) rectangle (10.00, -1.25);
\filldraw[fill=cancan, draw=black] (10.00, -1.00) rectangle (10.25, -1.25);
\filldraw[fill=cancan, draw=black] (10.25, -1.00) rectangle (10.50, -1.25);
\filldraw[fill=cancan, draw=black] (10.50, -1.00) rectangle (10.75, -1.25);
\filldraw[fill=bermuda, draw=black] (10.75, -1.00) rectangle (11.00, -1.25);
\filldraw[fill=cancan, draw=black] (11.00, -1.00) rectangle (11.25, -1.25);
\filldraw[fill=bermuda, draw=black] (11.25, -1.00) rectangle (11.50, -1.25);
\filldraw[fill=bermuda, draw=black] (11.50, -1.00) rectangle (11.75, -1.25);
\filldraw[fill=bermuda, draw=black] (11.75, -1.00) rectangle (12.00, -1.25);
\filldraw[fill=cancan, draw=black] (12.00, -1.00) rectangle (12.25, -1.25);
\filldraw[fill=cancan, draw=black] (12.25, -1.00) rectangle (12.50, -1.25);
\filldraw[fill=cancan, draw=black] (12.50, -1.00) rectangle (12.75, -1.25);
\filldraw[fill=bermuda, draw=black] (12.75, -1.00) rectangle (13.00, -1.25);
\filldraw[fill=bermuda, draw=black] (13.00, -1.00) rectangle (13.25, -1.25);
\filldraw[fill=bermuda, draw=black] (13.25, -1.00) rectangle (13.50, -1.25);
\filldraw[fill=cancan, draw=black] (13.50, -1.00) rectangle (13.75, -1.25);
\filldraw[fill=cancan, draw=black] (13.75, -1.00) rectangle (14.00, -1.25);
\filldraw[fill=cancan, draw=black] (14.00, -1.00) rectangle (14.25, -1.25);
\filldraw[fill=bermuda, draw=black] (14.25, -1.00) rectangle (14.50, -1.25);
\filldraw[fill=bermuda, draw=black] (14.50, -1.00) rectangle (14.75, -1.25);
\filldraw[fill=bermuda, draw=black] (14.75, -1.00) rectangle (15.00, -1.25);
\filldraw[fill=cancan, draw=black] (0.00, -1.25) rectangle (0.25, -1.50);
\filldraw[fill=cancan, draw=black] (0.25, -1.25) rectangle (0.50, -1.50);
\filldraw[fill=cancan, draw=black] (0.50, -1.25) rectangle (0.75, -1.50);
\filldraw[fill=bermuda, draw=black] (0.75, -1.25) rectangle (1.00, -1.50);
\filldraw[fill=bermuda, draw=black] (1.00, -1.25) rectangle (1.25, -1.50);
\filldraw[fill=bermuda, draw=black] (1.25, -1.25) rectangle (1.50, -1.50);
\filldraw[fill=cancan, draw=black] (1.50, -1.25) rectangle (1.75, -1.50);
\filldraw[fill=bermuda, draw=black] (1.75, -1.25) rectangle (2.00, -1.50);
\filldraw[fill=cancan, draw=black] (2.00, -1.25) rectangle (2.25, -1.50);
\filldraw[fill=bermuda, draw=black] (2.25, -1.25) rectangle (2.50, -1.50);
\filldraw[fill=cancan, draw=black] (2.50, -1.25) rectangle (2.75, -1.50);
\filldraw[fill=bermuda, draw=black] (2.75, -1.25) rectangle (3.00, -1.50);
\filldraw[fill=cancan, draw=black] (3.00, -1.25) rectangle (3.25, -1.50);
\filldraw[fill=cancan, draw=black] (3.25, -1.25) rectangle (3.50, -1.50);
\filldraw[fill=cancan, draw=black] (3.50, -1.25) rectangle (3.75, -1.50);
\filldraw[fill=cancan, draw=black] (3.75, -1.25) rectangle (4.00, -1.50);
\filldraw[fill=cancan, draw=black] (4.00, -1.25) rectangle (4.25, -1.50);
\filldraw[fill=bermuda, draw=black] (4.25, -1.25) rectangle (4.50, -1.50);
\filldraw[fill=bermuda, draw=black] (4.50, -1.25) rectangle (4.75, -1.50);
\filldraw[fill=bermuda, draw=black] (4.75, -1.25) rectangle (5.00, -1.50);
\filldraw[fill=cancan, draw=black] (5.00, -1.25) rectangle (5.25, -1.50);
\filldraw[fill=cancan, draw=black] (5.25, -1.25) rectangle (5.50, -1.50);
\filldraw[fill=cancan, draw=black] (5.50, -1.25) rectangle (5.75, -1.50);
\filldraw[fill=bermuda, draw=black] (5.75, -1.25) rectangle (6.00, -1.50);
\filldraw[fill=bermuda, draw=black] (6.00, -1.25) rectangle (6.25, -1.50);
\filldraw[fill=bermuda, draw=black] (6.25, -1.25) rectangle (6.50, -1.50);
\filldraw[fill=cancan, draw=black] (6.50, -1.25) rectangle (6.75, -1.50);
\filldraw[fill=cancan, draw=black] (6.75, -1.25) rectangle (7.00, -1.50);
\filldraw[fill=bermuda, draw=black] (7.00, -1.25) rectangle (7.25, -1.50);
\filldraw[fill=bermuda, draw=black] (7.25, -1.25) rectangle (7.50, -1.50);
\filldraw[fill=cancan, draw=black] (7.50, -1.25) rectangle (7.75, -1.50);
\filldraw[fill=bermuda, draw=black] (7.75, -1.25) rectangle (8.00, -1.50);
\filldraw[fill=bermuda, draw=black] (8.00, -1.25) rectangle (8.25, -1.50);
\filldraw[fill=bermuda, draw=black] (8.25, -1.25) rectangle (8.50, -1.50);
\filldraw[fill=bermuda, draw=black] (8.50, -1.25) rectangle (8.75, -1.50);
\filldraw[fill=bermuda, draw=black] (8.75, -1.25) rectangle (9.00, -1.50);
\filldraw[fill=cancan, draw=black] (9.00, -1.25) rectangle (9.25, -1.50);
\filldraw[fill=bermuda, draw=black] (9.25, -1.25) rectangle (9.50, -1.50);
\filldraw[fill=bermuda, draw=black] (9.50, -1.25) rectangle (9.75, -1.50);
\filldraw[fill=bermuda, draw=black] (9.75, -1.25) rectangle (10.00, -1.50);
\filldraw[fill=bermuda, draw=black] (10.00, -1.25) rectangle (10.25, -1.50);
\filldraw[fill=bermuda, draw=black] (10.25, -1.25) rectangle (10.50, -1.50);
\filldraw[fill=cancan, draw=black] (10.50, -1.25) rectangle (10.75, -1.50);
\filldraw[fill=bermuda, draw=black] (10.75, -1.25) rectangle (11.00, -1.50);
\filldraw[fill=bermuda, draw=black] (11.00, -1.25) rectangle (11.25, -1.50);
\filldraw[fill=bermuda, draw=black] (11.25, -1.25) rectangle (11.50, -1.50);
\filldraw[fill=bermuda, draw=black] (11.50, -1.25) rectangle (11.75, -1.50);
\filldraw[fill=bermuda, draw=black] (11.75, -1.25) rectangle (12.00, -1.50);
\filldraw[fill=cancan, draw=black] (12.00, -1.25) rectangle (12.25, -1.50);
\filldraw[fill=cancan, draw=black] (12.25, -1.25) rectangle (12.50, -1.50);
\filldraw[fill=cancan, draw=black] (12.50, -1.25) rectangle (12.75, -1.50);
\filldraw[fill=cancan, draw=black] (12.75, -1.25) rectangle (13.00, -1.50);
\filldraw[fill=cancan, draw=black] (13.00, -1.25) rectangle (13.25, -1.50);
\filldraw[fill=cancan, draw=black] (13.25, -1.25) rectangle (13.50, -1.50);
\filldraw[fill=cancan, draw=black] (13.50, -1.25) rectangle (13.75, -1.50);
\filldraw[fill=cancan, draw=black] (13.75, -1.25) rectangle (14.00, -1.50);
\filldraw[fill=cancan, draw=black] (14.00, -1.25) rectangle (14.25, -1.50);
\filldraw[fill=cancan, draw=black] (14.25, -1.25) rectangle (14.50, -1.50);
\filldraw[fill=cancan, draw=black] (14.50, -1.25) rectangle (14.75, -1.50);
\filldraw[fill=bermuda, draw=black] (14.75, -1.25) rectangle (15.00, -1.50);
\filldraw[fill=bermuda, draw=black] (0.00, -1.50) rectangle (0.25, -1.75);
\filldraw[fill=bermuda, draw=black] (0.25, -1.50) rectangle (0.50, -1.75);
\filldraw[fill=cancan, draw=black] (0.50, -1.50) rectangle (0.75, -1.75);
\filldraw[fill=cancan, draw=black] (0.75, -1.50) rectangle (1.00, -1.75);
\filldraw[fill=cancan, draw=black] (1.00, -1.50) rectangle (1.25, -1.75);
\filldraw[fill=bermuda, draw=black] (1.25, -1.50) rectangle (1.50, -1.75);
\filldraw[fill=bermuda, draw=black] (1.50, -1.50) rectangle (1.75, -1.75);
\filldraw[fill=bermuda, draw=black] (1.75, -1.50) rectangle (2.00, -1.75);
\filldraw[fill=bermuda, draw=black] (2.00, -1.50) rectangle (2.25, -1.75);
\filldraw[fill=bermuda, draw=black] (2.25, -1.50) rectangle (2.50, -1.75);
\filldraw[fill=cancan, draw=black] (2.50, -1.50) rectangle (2.75, -1.75);
\filldraw[fill=bermuda, draw=black] (2.75, -1.50) rectangle (3.00, -1.75);
\filldraw[fill=cancan, draw=black] (3.00, -1.50) rectangle (3.25, -1.75);
\filldraw[fill=cancan, draw=black] (3.25, -1.50) rectangle (3.50, -1.75);
\filldraw[fill=bermuda, draw=black] (3.50, -1.50) rectangle (3.75, -1.75);
\filldraw[fill=bermuda, draw=black] (3.75, -1.50) rectangle (4.00, -1.75);
\filldraw[fill=cancan, draw=black] (4.00, -1.50) rectangle (4.25, -1.75);
\filldraw[fill=cancan, draw=black] (4.25, -1.50) rectangle (4.50, -1.75);
\filldraw[fill=cancan, draw=black] (4.50, -1.50) rectangle (4.75, -1.75);
\filldraw[fill=bermuda, draw=black] (4.75, -1.50) rectangle (5.00, -1.75);
\filldraw[fill=bermuda, draw=black] (5.00, -1.50) rectangle (5.25, -1.75);
\filldraw[fill=bermuda, draw=black] (5.25, -1.50) rectangle (5.50, -1.75);
\filldraw[fill=cancan, draw=black] (5.50, -1.50) rectangle (5.75, -1.75);
\filldraw[fill=cancan, draw=black] (5.75, -1.50) rectangle (6.00, -1.75);
\filldraw[fill=cancan, draw=black] (6.00, -1.50) rectangle (6.25, -1.75);
\filldraw[fill=bermuda, draw=black] (6.25, -1.50) rectangle (6.50, -1.75);
\filldraw[fill=bermuda, draw=black] (6.50, -1.50) rectangle (6.75, -1.75);
\filldraw[fill=bermuda, draw=black] (6.75, -1.50) rectangle (7.00, -1.75);
\filldraw[fill=cancan, draw=black] (7.00, -1.50) rectangle (7.25, -1.75);
\filldraw[fill=cancan, draw=black] (7.25, -1.50) rectangle (7.50, -1.75);
\filldraw[fill=cancan, draw=black] (7.50, -1.50) rectangle (7.75, -1.75);
\filldraw[fill=bermuda, draw=black] (7.75, -1.50) rectangle (8.00, -1.75);
\filldraw[fill=bermuda, draw=black] (8.00, -1.50) rectangle (8.25, -1.75);
\filldraw[fill=bermuda, draw=black] (8.25, -1.50) rectangle (8.50, -1.75);
\filldraw[fill=cancan, draw=black] (8.50, -1.50) rectangle (8.75, -1.75);
\filldraw[fill=cancan, draw=black] (8.75, -1.50) rectangle (9.00, -1.75);
\filldraw[fill=cancan, draw=black] (9.00, -1.50) rectangle (9.25, -1.75);
\filldraw[fill=bermuda, draw=black] (9.25, -1.50) rectangle (9.50, -1.75);
\filldraw[fill=bermuda, draw=black] (9.50, -1.50) rectangle (9.75, -1.75);
\filldraw[fill=bermuda, draw=black] (9.75, -1.50) rectangle (10.00, -1.75);
\filldraw[fill=cancan, draw=black] (10.00, -1.50) rectangle (10.25, -1.75);
\filldraw[fill=bermuda, draw=black] (10.25, -1.50) rectangle (10.50, -1.75);
\filldraw[fill=bermuda, draw=black] (10.50, -1.50) rectangle (10.75, -1.75);
\filldraw[fill=bermuda, draw=black] (10.75, -1.50) rectangle (11.00, -1.75);
\filldraw[fill=cancan, draw=black] (11.00, -1.50) rectangle (11.25, -1.75);
\filldraw[fill=bermuda, draw=black] (11.25, -1.50) rectangle (11.50, -1.75);
\filldraw[fill=bermuda, draw=black] (11.50, -1.50) rectangle (11.75, -1.75);
\filldraw[fill=bermuda, draw=black] (11.75, -1.50) rectangle (12.00, -1.75);
\filldraw[fill=cancan, draw=black] (12.00, -1.50) rectangle (12.25, -1.75);
\filldraw[fill=bermuda, draw=black] (12.25, -1.50) rectangle (12.50, -1.75);
\filldraw[fill=bermuda, draw=black] (12.50, -1.50) rectangle (12.75, -1.75);
\filldraw[fill=bermuda, draw=black] (12.75, -1.50) rectangle (13.00, -1.75);
\filldraw[fill=cancan, draw=black] (13.00, -1.50) rectangle (13.25, -1.75);
\filldraw[fill=cancan, draw=black] (13.25, -1.50) rectangle (13.50, -1.75);
\filldraw[fill=cancan, draw=black] (13.50, -1.50) rectangle (13.75, -1.75);
\filldraw[fill=bermuda, draw=black] (13.75, -1.50) rectangle (14.00, -1.75);
\filldraw[fill=bermuda, draw=black] (14.00, -1.50) rectangle (14.25, -1.75);
\filldraw[fill=bermuda, draw=black] (14.25, -1.50) rectangle (14.50, -1.75);
\filldraw[fill=cancan, draw=black] (14.50, -1.50) rectangle (14.75, -1.75);
\filldraw[fill=cancan, draw=black] (14.75, -1.50) rectangle (15.00, -1.75);
\filldraw[fill=cancan, draw=black] (0.00, -1.75) rectangle (0.25, -2.00);
\filldraw[fill=cancan, draw=black] (0.25, -1.75) rectangle (0.50, -2.00);
\filldraw[fill=bermuda, draw=black] (0.50, -1.75) rectangle (0.75, -2.00);
\filldraw[fill=bermuda, draw=black] (0.75, -1.75) rectangle (1.00, -2.00);
\filldraw[fill=cancan, draw=black] (1.00, -1.75) rectangle (1.25, -2.00);
\filldraw[fill=bermuda, draw=black] (1.25, -1.75) rectangle (1.50, -2.00);
\filldraw[fill=cancan, draw=black] (1.50, -1.75) rectangle (1.75, -2.00);
\filldraw[fill=cancan, draw=black] (1.75, -1.75) rectangle (2.00, -2.00);
\filldraw[fill=bermuda, draw=black] (2.00, -1.75) rectangle (2.25, -2.00);
\filldraw[fill=bermuda, draw=black] (2.25, -1.75) rectangle (2.50, -2.00);
\filldraw[fill=cancan, draw=black] (2.50, -1.75) rectangle (2.75, -2.00);
\filldraw[fill=bermuda, draw=black] (2.75, -1.75) rectangle (3.00, -2.00);
\filldraw[fill=cancan, draw=black] (3.00, -1.75) rectangle (3.25, -2.00);
\filldraw[fill=bermuda, draw=black] (3.25, -1.75) rectangle (3.50, -2.00);
\filldraw[fill=bermuda, draw=black] (3.50, -1.75) rectangle (3.75, -2.00);
\filldraw[fill=bermuda, draw=black] (3.75, -1.75) rectangle (4.00, -2.00);
\filldraw[fill=cancan, draw=black] (4.00, -1.75) rectangle (4.25, -2.00);
\filldraw[fill=bermuda, draw=black] (4.25, -1.75) rectangle (4.50, -2.00);
\filldraw[fill=bermuda, draw=black] (4.50, -1.75) rectangle (4.75, -2.00);
\filldraw[fill=bermuda, draw=black] (4.75, -1.75) rectangle (5.00, -2.00);
\filldraw[fill=bermuda, draw=black] (5.00, -1.75) rectangle (5.25, -2.00);
\filldraw[fill=bermuda, draw=black] (5.25, -1.75) rectangle (5.50, -2.00);
\filldraw[fill=bermuda, draw=black] (5.50, -1.75) rectangle (5.75, -2.00);
\filldraw[fill=bermuda, draw=black] (5.75, -1.75) rectangle (6.00, -2.00);
\filldraw[fill=cancan, draw=black] (6.00, -1.75) rectangle (6.25, -2.00);
\filldraw[fill=bermuda, draw=black] (6.25, -1.75) rectangle (6.50, -2.00);
\filldraw[fill=cancan, draw=black] (6.50, -1.75) rectangle (6.75, -2.00);
\filldraw[fill=bermuda, draw=black] (6.75, -1.75) rectangle (7.00, -2.00);
\filldraw[fill=bermuda, draw=black] (7.00, -1.75) rectangle (7.25, -2.00);
\filldraw[fill=bermuda, draw=black] (7.25, -1.75) rectangle (7.50, -2.00);
\filldraw[fill=cancan, draw=black] (7.50, -1.75) rectangle (7.75, -2.00);
\filldraw[fill=cancan, draw=black] (7.75, -1.75) rectangle (8.00, -2.00);
\filldraw[fill=cancan, draw=black] (8.00, -1.75) rectangle (8.25, -2.00);
\filldraw[fill=cancan, draw=black] (8.25, -1.75) rectangle (8.50, -2.00);
\filldraw[fill=cancan, draw=black] (8.50, -1.75) rectangle (8.75, -2.00);
\filldraw[fill=cancan, draw=black] (8.75, -1.75) rectangle (9.00, -2.00);
\filldraw[fill=cancan, draw=black] (9.00, -1.75) rectangle (9.25, -2.00);
\filldraw[fill=bermuda, draw=black] (9.25, -1.75) rectangle (9.50, -2.00);
\filldraw[fill=bermuda, draw=black] (9.50, -1.75) rectangle (9.75, -2.00);
\filldraw[fill=bermuda, draw=black] (9.75, -1.75) rectangle (10.00, -2.00);
\filldraw[fill=cancan, draw=black] (10.00, -1.75) rectangle (10.25, -2.00);
\filldraw[fill=cancan, draw=black] (10.25, -1.75) rectangle (10.50, -2.00);
\filldraw[fill=cancan, draw=black] (10.50, -1.75) rectangle (10.75, -2.00);
\filldraw[fill=bermuda, draw=black] (10.75, -1.75) rectangle (11.00, -2.00);
\filldraw[fill=bermuda, draw=black] (11.00, -1.75) rectangle (11.25, -2.00);
\filldraw[fill=bermuda, draw=black] (11.25, -1.75) rectangle (11.50, -2.00);
\filldraw[fill=bermuda, draw=black] (11.50, -1.75) rectangle (11.75, -2.00);
\filldraw[fill=bermuda, draw=black] (11.75, -1.75) rectangle (12.00, -2.00);
\filldraw[fill=bermuda, draw=black] (12.00, -1.75) rectangle (12.25, -2.00);
\filldraw[fill=bermuda, draw=black] (12.25, -1.75) rectangle (12.50, -2.00);
\filldraw[fill=cancan, draw=black] (12.50, -1.75) rectangle (12.75, -2.00);
\filldraw[fill=bermuda, draw=black] (12.75, -1.75) rectangle (13.00, -2.00);
\filldraw[fill=cancan, draw=black] (13.00, -1.75) rectangle (13.25, -2.00);
\filldraw[fill=bermuda, draw=black] (13.25, -1.75) rectangle (13.50, -2.00);
\filldraw[fill=bermuda, draw=black] (13.50, -1.75) rectangle (13.75, -2.00);
\filldraw[fill=bermuda, draw=black] (13.75, -1.75) rectangle (14.00, -2.00);
\filldraw[fill=cancan, draw=black] (14.00, -1.75) rectangle (14.25, -2.00);
\filldraw[fill=cancan, draw=black] (14.25, -1.75) rectangle (14.50, -2.00);
\filldraw[fill=cancan, draw=black] (14.50, -1.75) rectangle (14.75, -2.00);
\filldraw[fill=bermuda, draw=black] (14.75, -1.75) rectangle (15.00, -2.00);
\filldraw[fill=bermuda, draw=black] (0.00, -2.00) rectangle (0.25, -2.25);
\filldraw[fill=bermuda, draw=black] (0.25, -2.00) rectangle (0.50, -2.25);
\filldraw[fill=cancan, draw=black] (0.50, -2.00) rectangle (0.75, -2.25);
\filldraw[fill=cancan, draw=black] (0.75, -2.00) rectangle (1.00, -2.25);
\filldraw[fill=bermuda, draw=black] (1.00, -2.00) rectangle (1.25, -2.25);
\filldraw[fill=bermuda, draw=black] (1.25, -2.00) rectangle (1.50, -2.25);
\filldraw[fill=cancan, draw=black] (1.50, -2.00) rectangle (1.75, -2.25);
\filldraw[fill=cancan, draw=black] (1.75, -2.00) rectangle (2.00, -2.25);
\filldraw[fill=cancan, draw=black] (2.00, -2.00) rectangle (2.25, -2.25);
\filldraw[fill=cancan, draw=black] (2.25, -2.00) rectangle (2.50, -2.25);
\filldraw[fill=cancan, draw=black] (2.50, -2.00) rectangle (2.75, -2.25);
\filldraw[fill=cancan, draw=black] (2.75, -2.00) rectangle (3.00, -2.25);
\filldraw[fill=cancan, draw=black] (3.00, -2.00) rectangle (3.25, -2.25);
\filldraw[fill=bermuda, draw=black] (3.25, -2.00) rectangle (3.50, -2.25);
\filldraw[fill=bermuda, draw=black] (3.50, -2.00) rectangle (3.75, -2.25);
\filldraw[fill=bermuda, draw=black] (3.75, -2.00) rectangle (4.00, -2.25);
\filldraw[fill=cancan, draw=black] (4.00, -2.00) rectangle (4.25, -2.25);
\filldraw[fill=cancan, draw=black] (4.25, -2.00) rectangle (4.50, -2.25);
\filldraw[fill=cancan, draw=black] (4.50, -2.00) rectangle (4.75, -2.25);
\filldraw[fill=bermuda, draw=black] (4.75, -2.00) rectangle (5.00, -2.25);
\filldraw[fill=bermuda, draw=black] (5.00, -2.00) rectangle (5.25, -2.25);
\filldraw[fill=bermuda, draw=black] (5.25, -2.00) rectangle (5.50, -2.25);
\filldraw[fill=cancan, draw=black] (5.50, -2.00) rectangle (5.75, -2.25);
\filldraw[fill=cancan, draw=black] (5.75, -2.00) rectangle (6.00, -2.25);
\filldraw[fill=cancan, draw=black] (6.00, -2.00) rectangle (6.25, -2.25);
\filldraw[fill=bermuda, draw=black] (6.25, -2.00) rectangle (6.50, -2.25);
\filldraw[fill=bermuda, draw=black] (6.50, -2.00) rectangle (6.75, -2.25);
\filldraw[fill=bermuda, draw=black] (6.75, -2.00) rectangle (7.00, -2.25);
\filldraw[fill=cancan, draw=black] (7.00, -2.00) rectangle (7.25, -2.25);
\filldraw[fill=bermuda, draw=black] (7.25, -2.00) rectangle (7.50, -2.25);
\filldraw[fill=cancan, draw=black] (7.50, -2.00) rectangle (7.75, -2.25);
\filldraw[fill=bermuda, draw=black] (7.75, -2.00) rectangle (8.00, -2.25);
\filldraw[fill=bermuda, draw=black] (8.00, -2.00) rectangle (8.25, -2.25);
\filldraw[fill=bermuda, draw=black] (8.25, -2.00) rectangle (8.50, -2.25);
\filldraw[fill=cancan, draw=black] (8.50, -2.00) rectangle (8.75, -2.25);
\filldraw[fill=cancan, draw=black] (8.75, -2.00) rectangle (9.00, -2.25);
\filldraw[fill=cancan, draw=black] (9.00, -2.00) rectangle (9.25, -2.25);
\filldraw[fill=cancan, draw=black] (9.25, -2.00) rectangle (9.50, -2.25);
\filldraw[fill=cancan, draw=black] (9.50, -2.00) rectangle (9.75, -2.25);
\filldraw[fill=bermuda, draw=black] (9.75, -2.00) rectangle (10.00, -2.25);
\filldraw[fill=bermuda, draw=black] (10.00, -2.00) rectangle (10.25, -2.25);
\filldraw[fill=bermuda, draw=black] (10.25, -2.00) rectangle (10.50, -2.25);
\filldraw[fill=bermuda, draw=black] (10.50, -2.00) rectangle (10.75, -2.25);
\filldraw[fill=bermuda, draw=black] (10.75, -2.00) rectangle (11.00, -2.25);
\filldraw[fill=cancan, draw=black] (11.00, -2.00) rectangle (11.25, -2.25);
\filldraw[fill=bermuda, draw=black] (11.25, -2.00) rectangle (11.50, -2.25);
\filldraw[fill=bermuda, draw=black] (11.50, -2.00) rectangle (11.75, -2.25);
\filldraw[fill=bermuda, draw=black] (11.75, -2.00) rectangle (12.00, -2.25);
\filldraw[fill=bermuda, draw=black] (12.00, -2.00) rectangle (12.25, -2.25);
\filldraw[fill=bermuda, draw=black] (12.25, -2.00) rectangle (12.50, -2.25);
\filldraw[fill=cancan, draw=black] (12.50, -2.00) rectangle (12.75, -2.25);
\filldraw[fill=cancan, draw=black] (12.75, -2.00) rectangle (13.00, -2.25);
\filldraw[fill=cancan, draw=black] (13.00, -2.00) rectangle (13.25, -2.25);
\filldraw[fill=bermuda, draw=black] (13.25, -2.00) rectangle (13.50, -2.25);
\filldraw[fill=bermuda, draw=black] (13.50, -2.00) rectangle (13.75, -2.25);
\filldraw[fill=bermuda, draw=black] (13.75, -2.00) rectangle (14.00, -2.25);
\filldraw[fill=cancan, draw=black] (14.00, -2.00) rectangle (14.25, -2.25);
\filldraw[fill=bermuda, draw=black] (14.25, -2.00) rectangle (14.50, -2.25);
\filldraw[fill=cancan, draw=black] (14.50, -2.00) rectangle (14.75, -2.25);
\filldraw[fill=bermuda, draw=black] (14.75, -2.00) rectangle (15.00, -2.25);
\filldraw[fill=bermuda, draw=black] (0.00, -2.25) rectangle (0.25, -2.50);
\filldraw[fill=bermuda, draw=black] (0.25, -2.25) rectangle (0.50, -2.50);
\filldraw[fill=cancan, draw=black] (0.50, -2.25) rectangle (0.75, -2.50);
\filldraw[fill=bermuda, draw=black] (0.75, -2.25) rectangle (1.00, -2.50);
\filldraw[fill=bermuda, draw=black] (1.00, -2.25) rectangle (1.25, -2.50);
\filldraw[fill=bermuda, draw=black] (1.25, -2.25) rectangle (1.50, -2.50);
\filldraw[fill=bermuda, draw=black] (1.50, -2.25) rectangle (1.75, -2.50);
\filldraw[fill=bermuda, draw=black] (1.75, -2.25) rectangle (2.00, -2.50);
\filldraw[fill=cancan, draw=black] (2.00, -2.25) rectangle (2.25, -2.50);
\filldraw[fill=bermuda, draw=black] (2.25, -2.25) rectangle (2.50, -2.50);
\filldraw[fill=bermuda, draw=black] (2.50, -2.25) rectangle (2.75, -2.50);
\filldraw[fill=bermuda, draw=black] (2.75, -2.25) rectangle (3.00, -2.50);
\filldraw[fill=cancan, draw=black] (3.00, -2.25) rectangle (3.25, -2.50);
\filldraw[fill=cancan, draw=black] (3.25, -2.25) rectangle (3.50, -2.50);
\filldraw[fill=cancan, draw=black] (3.50, -2.25) rectangle (3.75, -2.50);
\filldraw[fill=bermuda, draw=black] (3.75, -2.25) rectangle (4.00, -2.50);
\filldraw[fill=bermuda, draw=black] (4.00, -2.25) rectangle (4.25, -2.50);
\filldraw[fill=bermuda, draw=black] (4.25, -2.25) rectangle (4.50, -2.50);
\filldraw[fill=cancan, draw=black] (4.50, -2.25) rectangle (4.75, -2.50);
\filldraw[fill=bermuda, draw=black] (4.75, -2.25) rectangle (5.00, -2.50);
\filldraw[fill=cancan, draw=black] (5.00, -2.25) rectangle (5.25, -2.50);
\filldraw[fill=bermuda, draw=black] (5.25, -2.25) rectangle (5.50, -2.50);
\filldraw[fill=cancan, draw=black] (5.50, -2.25) rectangle (5.75, -2.50);
\filldraw[fill=cancan, draw=black] (5.75, -2.25) rectangle (6.00, -2.50);
\filldraw[fill=cancan, draw=black] (6.00, -2.25) rectangle (6.25, -2.50);
\filldraw[fill=cancan, draw=black] (6.25, -2.25) rectangle (6.50, -2.50);
\filldraw[fill=cancan, draw=black] (6.50, -2.25) rectangle (6.75, -2.50);
\filldraw[fill=cancan, draw=black] (6.75, -2.25) rectangle (7.00, -2.50);
\filldraw[fill=bermuda, draw=black] (7.00, -2.25) rectangle (7.25, -2.50);
\filldraw[fill=bermuda, draw=black] (7.25, -2.25) rectangle (7.50, -2.50);
\filldraw[fill=bermuda, draw=black] (7.50, -2.25) rectangle (7.75, -2.50);
\filldraw[fill=bermuda, draw=black] (7.75, -2.25) rectangle (8.00, -2.50);
\filldraw[fill=cancan, draw=black] (8.00, -2.25) rectangle (8.25, -2.50);
\filldraw[fill=bermuda, draw=black] (8.25, -2.25) rectangle (8.50, -2.50);
\filldraw[fill=cancan, draw=black] (8.50, -2.25) rectangle (8.75, -2.50);
\filldraw[fill=cancan, draw=black] (8.75, -2.25) rectangle (9.00, -2.50);
\filldraw[fill=bermuda, draw=black] (9.00, -2.25) rectangle (9.25, -2.50);
\filldraw[fill=bermuda, draw=black] (9.25, -2.25) rectangle (9.50, -2.50);
\filldraw[fill=cancan, draw=black] (9.50, -2.25) rectangle (9.75, -2.50);
\filldraw[fill=bermuda, draw=black] (9.75, -2.25) rectangle (10.00, -2.50);
\filldraw[fill=cancan, draw=black] (10.00, -2.25) rectangle (10.25, -2.50);
\filldraw[fill=cancan, draw=black] (10.25, -2.25) rectangle (10.50, -2.50);
\filldraw[fill=cancan, draw=black] (10.50, -2.25) rectangle (10.75, -2.50);
\filldraw[fill=cancan, draw=black] (10.75, -2.25) rectangle (11.00, -2.50);
\filldraw[fill=cancan, draw=black] (11.00, -2.25) rectangle (11.25, -2.50);
\filldraw[fill=bermuda, draw=black] (11.25, -2.25) rectangle (11.50, -2.50);
\filldraw[fill=bermuda, draw=black] (11.50, -2.25) rectangle (11.75, -2.50);
\filldraw[fill=bermuda, draw=black] (11.75, -2.25) rectangle (12.00, -2.50);
\filldraw[fill=cancan, draw=black] (12.00, -2.25) rectangle (12.25, -2.50);
\filldraw[fill=cancan, draw=black] (12.25, -2.25) rectangle (12.50, -2.50);
\filldraw[fill=cancan, draw=black] (12.50, -2.25) rectangle (12.75, -2.50);
\filldraw[fill=cancan, draw=black] (12.75, -2.25) rectangle (13.00, -2.50);
\filldraw[fill=cancan, draw=black] (13.00, -2.25) rectangle (13.25, -2.50);
\filldraw[fill=bermuda, draw=black] (13.25, -2.25) rectangle (13.50, -2.50);
\filldraw[fill=bermuda, draw=black] (13.50, -2.25) rectangle (13.75, -2.50);
\filldraw[fill=bermuda, draw=black] (13.75, -2.25) rectangle (14.00, -2.50);
\filldraw[fill=cancan, draw=black] (14.00, -2.25) rectangle (14.25, -2.50);
\filldraw[fill=cancan, draw=black] (14.25, -2.25) rectangle (14.50, -2.50);
\filldraw[fill=cancan, draw=black] (14.50, -2.25) rectangle (14.75, -2.50);
\filldraw[fill=bermuda, draw=black] (14.75, -2.25) rectangle (15.00, -2.50);
\filldraw[fill=bermuda, draw=black] (0.00, -2.50) rectangle (0.25, -2.75);
\filldraw[fill=bermuda, draw=black] (0.25, -2.50) rectangle (0.50, -2.75);
\filldraw[fill=bermuda, draw=black] (0.50, -2.50) rectangle (0.75, -2.75);
\filldraw[fill=bermuda, draw=black] (0.75, -2.50) rectangle (1.00, -2.75);
\filldraw[fill=cancan, draw=black] (1.00, -2.50) rectangle (1.25, -2.75);
\filldraw[fill=bermuda, draw=black] (1.25, -2.50) rectangle (1.50, -2.75);
\filldraw[fill=bermuda, draw=black] (1.50, -2.50) rectangle (1.75, -2.75);
\filldraw[fill=bermuda, draw=black] (1.75, -2.50) rectangle (2.00, -2.75);
\filldraw[fill=cancan, draw=black] (2.00, -2.50) rectangle (2.25, -2.75);
\filldraw[fill=cancan, draw=black] (2.25, -2.50) rectangle (2.50, -2.75);
\filldraw[fill=cancan, draw=black] (2.50, -2.50) rectangle (2.75, -2.75);
\filldraw[fill=cancan, draw=black] (2.75, -2.50) rectangle (3.00, -2.75);
\filldraw[fill=bermuda, draw=black] (3.00, -2.50) rectangle (3.25, -2.75);
\filldraw[fill=bermuda, draw=black] (3.25, -2.50) rectangle (3.50, -2.75);
\filldraw[fill=cancan, draw=black] (3.50, -2.50) rectangle (3.75, -2.75);
\filldraw[fill=cancan, draw=black] (3.75, -2.50) rectangle (4.00, -2.75);
\filldraw[fill=cancan, draw=black] (4.00, -2.50) rectangle (4.25, -2.75);
\filldraw[fill=cancan, draw=black] (4.25, -2.50) rectangle (4.50, -2.75);
\filldraw[fill=cancan, draw=black] (4.50, -2.50) rectangle (4.75, -2.75);
\filldraw[fill=bermuda, draw=black] (4.75, -2.50) rectangle (5.00, -2.75);
\filldraw[fill=bermuda, draw=black] (5.00, -2.50) rectangle (5.25, -2.75);
\filldraw[fill=bermuda, draw=black] (5.25, -2.50) rectangle (5.50, -2.75);
\filldraw[fill=cancan, draw=black] (5.50, -2.50) rectangle (5.75, -2.75);
\filldraw[fill=cancan, draw=black] (5.75, -2.50) rectangle (6.00, -2.75);
\filldraw[fill=cancan, draw=black] (6.00, -2.50) rectangle (6.25, -2.75);
\filldraw[fill=cancan, draw=black] (6.25, -2.50) rectangle (6.50, -2.75);
\filldraw[fill=cancan, draw=black] (6.50, -2.50) rectangle (6.75, -2.75);
\filldraw[fill=bermuda, draw=black] (6.75, -2.50) rectangle (7.00, -2.75);
\filldraw[fill=bermuda, draw=black] (7.00, -2.50) rectangle (7.25, -2.75);
\filldraw[fill=bermuda, draw=black] (7.25, -2.50) rectangle (7.50, -2.75);
\filldraw[fill=cancan, draw=black] (7.50, -2.50) rectangle (7.75, -2.75);
\filldraw[fill=cancan, draw=black] (7.75, -2.50) rectangle (8.00, -2.75);
\filldraw[fill=cancan, draw=black] (8.00, -2.50) rectangle (8.25, -2.75);
\filldraw[fill=bermuda, draw=black] (8.25, -2.50) rectangle (8.50, -2.75);
\filldraw[fill=bermuda, draw=black] (8.50, -2.50) rectangle (8.75, -2.75);
\filldraw[fill=bermuda, draw=black] (8.75, -2.50) rectangle (9.00, -2.75);
\filldraw[fill=bermuda, draw=black] (9.00, -2.50) rectangle (9.25, -2.75);
\filldraw[fill=bermuda, draw=black] (9.25, -2.50) rectangle (9.50, -2.75);
\filldraw[fill=cancan, draw=black] (9.50, -2.50) rectangle (9.75, -2.75);
\filldraw[fill=bermuda, draw=black] (9.75, -2.50) rectangle (10.00, -2.75);
\filldraw[fill=bermuda, draw=black] (10.00, -2.50) rectangle (10.25, -2.75);
\filldraw[fill=bermuda, draw=black] (10.25, -2.50) rectangle (10.50, -2.75);
\filldraw[fill=cancan, draw=black] (10.50, -2.50) rectangle (10.75, -2.75);
\filldraw[fill=cancan, draw=black] (10.75, -2.50) rectangle (11.00, -2.75);
\filldraw[fill=cancan, draw=black] (11.00, -2.50) rectangle (11.25, -2.75);
\filldraw[fill=bermuda, draw=black] (11.25, -2.50) rectangle (11.50, -2.75);
\filldraw[fill=bermuda, draw=black] (11.50, -2.50) rectangle (11.75, -2.75);
\filldraw[fill=bermuda, draw=black] (11.75, -2.50) rectangle (12.00, -2.75);
\filldraw[fill=bermuda, draw=black] (12.00, -2.50) rectangle (12.25, -2.75);
\filldraw[fill=bermuda, draw=black] (12.25, -2.50) rectangle (12.50, -2.75);
\filldraw[fill=cancan, draw=black] (12.50, -2.50) rectangle (12.75, -2.75);
\filldraw[fill=bermuda, draw=black] (12.75, -2.50) rectangle (13.00, -2.75);
\filldraw[fill=cancan, draw=black] (13.00, -2.50) rectangle (13.25, -2.75);
\filldraw[fill=bermuda, draw=black] (13.25, -2.50) rectangle (13.50, -2.75);
\filldraw[fill=cancan, draw=black] (13.50, -2.50) rectangle (13.75, -2.75);
\filldraw[fill=bermuda, draw=black] (13.75, -2.50) rectangle (14.00, -2.75);
\filldraw[fill=cancan, draw=black] (14.00, -2.50) rectangle (14.25, -2.75);
\filldraw[fill=cancan, draw=black] (14.25, -2.50) rectangle (14.50, -2.75);
\filldraw[fill=cancan, draw=black] (14.50, -2.50) rectangle (14.75, -2.75);
\filldraw[fill=cancan, draw=black] (14.75, -2.50) rectangle (15.00, -2.75);
\filldraw[fill=cancan, draw=black] (0.00, -2.75) rectangle (0.25, -3.00);
\filldraw[fill=bermuda, draw=black] (0.25, -2.75) rectangle (0.50, -3.00);
\filldraw[fill=bermuda, draw=black] (0.50, -2.75) rectangle (0.75, -3.00);
\filldraw[fill=bermuda, draw=black] (0.75, -2.75) rectangle (1.00, -3.00);
\filldraw[fill=cancan, draw=black] (1.00, -2.75) rectangle (1.25, -3.00);
\filldraw[fill=cancan, draw=black] (1.25, -2.75) rectangle (1.50, -3.00);
\filldraw[fill=cancan, draw=black] (1.50, -2.75) rectangle (1.75, -3.00);
\filldraw[fill=bermuda, draw=black] (1.75, -2.75) rectangle (2.00, -3.00);
\filldraw[fill=bermuda, draw=black] (2.00, -2.75) rectangle (2.25, -3.00);
\filldraw[fill=bermuda, draw=black] (2.25, -2.75) rectangle (2.50, -3.00);
\filldraw[fill=bermuda, draw=black] (2.50, -2.75) rectangle (2.75, -3.00);
\filldraw[fill=bermuda, draw=black] (2.75, -2.75) rectangle (3.00, -3.00);
\filldraw[fill=bermuda, draw=black] (3.00, -2.75) rectangle (3.25, -3.00);
\filldraw[fill=bermuda, draw=black] (3.25, -2.75) rectangle (3.50, -3.00);
\filldraw[fill=cancan, draw=black] (3.50, -2.75) rectangle (3.75, -3.00);
\filldraw[fill=bermuda, draw=black] (3.75, -2.75) rectangle (4.00, -3.00);
\filldraw[fill=bermuda, draw=black] (4.00, -2.75) rectangle (4.25, -3.00);
\filldraw[fill=bermuda, draw=black] (4.25, -2.75) rectangle (4.50, -3.00);
\filldraw[fill=cancan, draw=black] (4.50, -2.75) rectangle (4.75, -3.00);
\filldraw[fill=cancan, draw=black] (4.75, -2.75) rectangle (5.00, -3.00);
\filldraw[fill=cancan, draw=black] (5.00, -2.75) rectangle (5.25, -3.00);
\filldraw[fill=bermuda, draw=black] (5.25, -2.75) rectangle (5.50, -3.00);
\filldraw[fill=bermuda, draw=black] (5.50, -2.75) rectangle (5.75, -3.00);
\filldraw[fill=bermuda, draw=black] (5.75, -2.75) rectangle (6.00, -3.00);
\filldraw[fill=cancan, draw=black] (6.00, -2.75) rectangle (6.25, -3.00);
\filldraw[fill=cancan, draw=black] (6.25, -2.75) rectangle (6.50, -3.00);
\filldraw[fill=cancan, draw=black] (6.50, -2.75) rectangle (6.75, -3.00);
\filldraw[fill=bermuda, draw=black] (6.75, -2.75) rectangle (7.00, -3.00);
\filldraw[fill=bermuda, draw=black] (7.00, -2.75) rectangle (7.25, -3.00);
\filldraw[fill=bermuda, draw=black] (7.25, -2.75) rectangle (7.50, -3.00);
\filldraw[fill=cancan, draw=black] (7.50, -2.75) rectangle (7.75, -3.00);
\filldraw[fill=cancan, draw=black] (7.75, -2.75) rectangle (8.00, -3.00);
\filldraw[fill=cancan, draw=black] (8.00, -2.75) rectangle (8.25, -3.00);
\filldraw[fill=bermuda, draw=black] (8.25, -2.75) rectangle (8.50, -3.00);
\filldraw[fill=bermuda, draw=black] (8.50, -2.75) rectangle (8.75, -3.00);
\filldraw[fill=bermuda, draw=black] (8.75, -2.75) rectangle (9.00, -3.00);
\filldraw[fill=cancan, draw=black] (9.00, -2.75) rectangle (9.25, -3.00);
\filldraw[fill=bermuda, draw=black] (9.25, -2.75) rectangle (9.50, -3.00);
\filldraw[fill=cancan, draw=black] (9.50, -2.75) rectangle (9.75, -3.00);
\filldraw[fill=bermuda, draw=black] (9.75, -2.75) rectangle (10.00, -3.00);
\filldraw[fill=cancan, draw=black] (10.00, -2.75) rectangle (10.25, -3.00);
\filldraw[fill=cancan, draw=black] (10.25, -2.75) rectangle (10.50, -3.00);
\filldraw[fill=bermuda, draw=black] (10.50, -2.75) rectangle (10.75, -3.00);
\filldraw[fill=bermuda, draw=black] (10.75, -2.75) rectangle (11.00, -3.00);
\filldraw[fill=cancan, draw=black] (11.00, -2.75) rectangle (11.25, -3.00);
\filldraw[fill=bermuda, draw=black] (11.25, -2.75) rectangle (11.50, -3.00);
\filldraw[fill=cancan, draw=black] (11.50, -2.75) rectangle (11.75, -3.00);
\filldraw[fill=cancan, draw=black] (11.75, -2.75) rectangle (12.00, -3.00);
\filldraw[fill=cancan, draw=black] (12.00, -2.75) rectangle (12.25, -3.00);
\filldraw[fill=cancan, draw=black] (12.25, -2.75) rectangle (12.50, -3.00);
\filldraw[fill=cancan, draw=black] (12.50, -2.75) rectangle (12.75, -3.00);
\filldraw[fill=cancan, draw=black] (12.75, -2.75) rectangle (13.00, -3.00);
\filldraw[fill=cancan, draw=black] (13.00, -2.75) rectangle (13.25, -3.00);
\filldraw[fill=cancan, draw=black] (13.25, -2.75) rectangle (13.50, -3.00);
\filldraw[fill=cancan, draw=black] (13.50, -2.75) rectangle (13.75, -3.00);
\filldraw[fill=cancan, draw=black] (13.75, -2.75) rectangle (14.00, -3.00);
\filldraw[fill=cancan, draw=black] (14.00, -2.75) rectangle (14.25, -3.00);
\filldraw[fill=cancan, draw=black] (14.25, -2.75) rectangle (14.50, -3.00);
\filldraw[fill=cancan, draw=black] (14.50, -2.75) rectangle (14.75, -3.00);
\filldraw[fill=bermuda, draw=black] (14.75, -2.75) rectangle (15.00, -3.00);
\filldraw[fill=cancan, draw=black] (0.00, -3.00) rectangle (0.25, -3.25);
\filldraw[fill=bermuda, draw=black] (0.25, -3.00) rectangle (0.50, -3.25);
\filldraw[fill=cancan, draw=black] (0.50, -3.00) rectangle (0.75, -3.25);
\filldraw[fill=cancan, draw=black] (0.75, -3.00) rectangle (1.00, -3.25);
\filldraw[fill=bermuda, draw=black] (1.00, -3.00) rectangle (1.25, -3.25);
\filldraw[fill=bermuda, draw=black] (1.25, -3.00) rectangle (1.50, -3.25);
\filldraw[fill=bermuda, draw=black] (1.50, -3.00) rectangle (1.75, -3.25);
\filldraw[fill=bermuda, draw=black] (1.75, -3.00) rectangle (2.00, -3.25);
\filldraw[fill=bermuda, draw=black] (2.00, -3.00) rectangle (2.25, -3.25);
\filldraw[fill=bermuda, draw=black] (2.25, -3.00) rectangle (2.50, -3.25);
\filldraw[fill=cancan, draw=black] (2.50, -3.00) rectangle (2.75, -3.25);
\filldraw[fill=bermuda, draw=black] (2.75, -3.00) rectangle (3.00, -3.25);
\filldraw[fill=cancan, draw=black] (3.00, -3.00) rectangle (3.25, -3.25);
\filldraw[fill=cancan, draw=black] (3.25, -3.00) rectangle (3.50, -3.25);
\filldraw[fill=bermuda, draw=black] (3.50, -3.00) rectangle (3.75, -3.25);
\filldraw[fill=bermuda, draw=black] (3.75, -3.00) rectangle (4.00, -3.25);
\filldraw[fill=cancan, draw=black] (4.00, -3.00) rectangle (4.25, -3.25);
\filldraw[fill=bermuda, draw=black] (4.25, -3.00) rectangle (4.50, -3.25);
\filldraw[fill=cancan, draw=black] (4.50, -3.00) rectangle (4.75, -3.25);
\filldraw[fill=cancan, draw=black] (4.75, -3.00) rectangle (5.00, -3.25);
\filldraw[fill=bermuda, draw=black] (5.00, -3.00) rectangle (5.25, -3.25);
\filldraw[fill=bermuda, draw=black] (5.25, -3.00) rectangle (5.50, -3.25);
\filldraw[fill=cancan, draw=black] (5.50, -3.00) rectangle (5.75, -3.25);
\filldraw[fill=bermuda, draw=black] (5.75, -3.00) rectangle (6.00, -3.25);
\filldraw[fill=cancan, draw=black] (6.00, -3.00) rectangle (6.25, -3.25);
\filldraw[fill=cancan, draw=black] (6.25, -3.00) rectangle (6.50, -3.25);
\filldraw[fill=bermuda, draw=black] (6.50, -3.00) rectangle (6.75, -3.25);
\filldraw[fill=bermuda, draw=black] (6.75, -3.00) rectangle (7.00, -3.25);
\filldraw[fill=cancan, draw=black] (7.00, -3.00) rectangle (7.25, -3.25);
\filldraw[fill=bermuda, draw=black] (7.25, -3.00) rectangle (7.50, -3.25);
\filldraw[fill=cancan, draw=black] (7.50, -3.00) rectangle (7.75, -3.25);
\filldraw[fill=cancan, draw=black] (7.75, -3.00) rectangle (8.00, -3.25);
\filldraw[fill=bermuda, draw=black] (8.00, -3.00) rectangle (8.25, -3.25);
\filldraw[fill=bermuda, draw=black] (8.25, -3.00) rectangle (8.50, -3.25);
\filldraw[fill=cancan, draw=black] (8.50, -3.00) rectangle (8.75, -3.25);
\filldraw[fill=bermuda, draw=black] (8.75, -3.00) rectangle (9.00, -3.25);
\filldraw[fill=cancan, draw=black] (9.00, -3.00) rectangle (9.25, -3.25);
\filldraw[fill=cancan, draw=black] (9.25, -3.00) rectangle (9.50, -3.25);
\filldraw[fill=cancan, draw=black] (9.50, -3.00) rectangle (9.75, -3.25);
\filldraw[fill=cancan, draw=black] (9.75, -3.00) rectangle (10.00, -3.25);
\filldraw[fill=cancan, draw=black] (10.00, -3.00) rectangle (10.25, -3.25);
\filldraw[fill=cancan, draw=black] (10.25, -3.00) rectangle (10.50, -3.25);
\filldraw[fill=cancan, draw=black] (10.50, -3.00) rectangle (10.75, -3.25);
\filldraw[fill=cancan, draw=black] (10.75, -3.00) rectangle (11.00, -3.25);
\filldraw[fill=bermuda, draw=black] (11.00, -3.00) rectangle (11.25, -3.25);
\filldraw[fill=bermuda, draw=black] (11.25, -3.00) rectangle (11.50, -3.25);
\filldraw[fill=cancan, draw=black] (11.50, -3.00) rectangle (11.75, -3.25);
\filldraw[fill=bermuda, draw=black] (11.75, -3.00) rectangle (12.00, -3.25);
\filldraw[fill=bermuda, draw=black] (12.00, -3.00) rectangle (12.25, -3.25);
\filldraw[fill=bermuda, draw=black] (12.25, -3.00) rectangle (12.50, -3.25);
\filldraw[fill=cancan, draw=black] (12.50, -3.00) rectangle (12.75, -3.25);
\filldraw[fill=cancan, draw=black] (12.75, -3.00) rectangle (13.00, -3.25);
\filldraw[fill=cancan, draw=black] (13.00, -3.00) rectangle (13.25, -3.25);
\filldraw[fill=bermuda, draw=black] (13.25, -3.00) rectangle (13.50, -3.25);
\filldraw[fill=bermuda, draw=black] (13.50, -3.00) rectangle (13.75, -3.25);
\filldraw[fill=bermuda, draw=black] (13.75, -3.00) rectangle (14.00, -3.25);
\filldraw[fill=cancan, draw=black] (14.00, -3.00) rectangle (14.25, -3.25);
\filldraw[fill=cancan, draw=black] (14.25, -3.00) rectangle (14.50, -3.25);
\filldraw[fill=cancan, draw=black] (14.50, -3.00) rectangle (14.75, -3.25);
\filldraw[fill=bermuda, draw=black] (14.75, -3.00) rectangle (15.00, -3.25);
\filldraw[fill=bermuda, draw=black] (0.00, -3.25) rectangle (0.25, -3.50);
\filldraw[fill=bermuda, draw=black] (0.25, -3.25) rectangle (0.50, -3.50);
\filldraw[fill=cancan, draw=black] (0.50, -3.25) rectangle (0.75, -3.50);
\filldraw[fill=bermuda, draw=black] (0.75, -3.25) rectangle (1.00, -3.50);
\filldraw[fill=cancan, draw=black] (1.00, -3.25) rectangle (1.25, -3.50);
\filldraw[fill=bermuda, draw=black] (1.25, -3.25) rectangle (1.50, -3.50);
\filldraw[fill=bermuda, draw=black] (1.50, -3.25) rectangle (1.75, -3.50);
\filldraw[fill=bermuda, draw=black] (1.75, -3.25) rectangle (2.00, -3.50);
\filldraw[fill=cancan, draw=black] (2.00, -3.25) rectangle (2.25, -3.50);
\filldraw[fill=cancan, draw=black] (2.25, -3.25) rectangle (2.50, -3.50);
\filldraw[fill=cancan, draw=black] (2.50, -3.25) rectangle (2.75, -3.50);
\filldraw[fill=bermuda, draw=black] (2.75, -3.25) rectangle (3.00, -3.50);
\filldraw[fill=bermuda, draw=black] (3.00, -3.25) rectangle (3.25, -3.50);
\filldraw[fill=bermuda, draw=black] (3.25, -3.25) rectangle (3.50, -3.50);
\filldraw[fill=cancan, draw=black] (3.50, -3.25) rectangle (3.75, -3.50);
\filldraw[fill=cancan, draw=black] (3.75, -3.25) rectangle (4.00, -3.50);
\filldraw[fill=cancan, draw=black] (4.00, -3.25) rectangle (4.25, -3.50);
\filldraw[fill=bermuda, draw=black] (4.25, -3.25) rectangle (4.50, -3.50);
\filldraw[fill=bermuda, draw=black] (4.50, -3.25) rectangle (4.75, -3.50);
\filldraw[fill=bermuda, draw=black] (4.75, -3.25) rectangle (5.00, -3.50);
\filldraw[fill=cancan, draw=black] (5.00, -3.25) rectangle (5.25, -3.50);
\filldraw[fill=cancan, draw=black] (5.25, -3.25) rectangle (5.50, -3.50);
\filldraw[fill=cancan, draw=black] (5.50, -3.25) rectangle (5.75, -3.50);
\filldraw[fill=bermuda, draw=black] (5.75, -3.25) rectangle (6.00, -3.50);
\filldraw[fill=bermuda, draw=black] (6.00, -3.25) rectangle (6.25, -3.50);
\filldraw[fill=bermuda, draw=black] (6.25, -3.25) rectangle (6.50, -3.50);
\filldraw[fill=cancan, draw=black] (6.50, -3.25) rectangle (6.75, -3.50);
\filldraw[fill=cancan, draw=black] (6.75, -3.25) rectangle (7.00, -3.50);
} } }\end{equation*}
\begin{equation*}
\hspace{0.3pt} b_{13} = \vcenter{\hbox{ \tikz{
\filldraw[fill=cancan, draw=black] (0.00, 0.00) rectangle (0.25, -0.25);
\filldraw[fill=bermuda, draw=black] (0.25, 0.00) rectangle (0.50, -0.25);
\filldraw[fill=bermuda, draw=black] (0.50, 0.00) rectangle (0.75, -0.25);
\filldraw[fill=bermuda, draw=black] (0.75, 0.00) rectangle (1.00, -0.25);
\filldraw[fill=cancan, draw=black] (1.00, 0.00) rectangle (1.25, -0.25);
\filldraw[fill=cancan, draw=black] (1.25, 0.00) rectangle (1.50, -0.25);
\filldraw[fill=cancan, draw=black] (1.50, 0.00) rectangle (1.75, -0.25);
\filldraw[fill=cancan, draw=black] (1.75, 0.00) rectangle (2.00, -0.25);
\filldraw[fill=cancan, draw=black] (2.00, 0.00) rectangle (2.25, -0.25);
\filldraw[fill=cancan, draw=black] (2.25, 0.00) rectangle (2.50, -0.25);
\filldraw[fill=cancan, draw=black] (2.50, 0.00) rectangle (2.75, -0.25);
\filldraw[fill=bermuda, draw=black] (2.75, 0.00) rectangle (3.00, -0.25);
\filldraw[fill=cancan, draw=black] (3.00, 0.00) rectangle (3.25, -0.25);
\filldraw[fill=cancan, draw=black] (3.25, 0.00) rectangle (3.50, -0.25);
\filldraw[fill=cancan, draw=black] (3.50, 0.00) rectangle (3.75, -0.25);
\filldraw[fill=cancan, draw=black] (3.75, 0.00) rectangle (4.00, -0.25);
\filldraw[fill=cancan, draw=black] (4.00, 0.00) rectangle (4.25, -0.25);
\filldraw[fill=bermuda, draw=black] (4.25, 0.00) rectangle (4.50, -0.25);
\filldraw[fill=cancan, draw=black] (4.50, 0.00) rectangle (4.75, -0.25);
\filldraw[fill=bermuda, draw=black] (4.75, 0.00) rectangle (5.00, -0.25);
\filldraw[fill=cancan, draw=black] (5.00, 0.00) rectangle (5.25, -0.25);
\filldraw[fill=cancan, draw=black] (5.25, 0.00) rectangle (5.50, -0.25);
\filldraw[fill=cancan, draw=black] (5.50, 0.00) rectangle (5.75, -0.25);
\filldraw[fill=bermuda, draw=black] (5.75, 0.00) rectangle (6.00, -0.25);
\filldraw[fill=cancan, draw=black] (6.00, 0.00) rectangle (6.25, -0.25);
\filldraw[fill=bermuda, draw=black] (6.25, 0.00) rectangle (6.50, -0.25);
\filldraw[fill=cancan, draw=black] (6.50, 0.00) rectangle (6.75, -0.25);
\filldraw[fill=bermuda, draw=black] (6.75, 0.00) rectangle (7.00, -0.25);
\filldraw[fill=cancan, draw=black] (7.00, 0.00) rectangle (7.25, -0.25);
\filldraw[fill=cancan, draw=black] (7.25, 0.00) rectangle (7.50, -0.25);
\filldraw[fill=cancan, draw=black] (7.50, 0.00) rectangle (7.75, -0.25);
\filldraw[fill=cancan, draw=black] (7.75, 0.00) rectangle (8.00, -0.25);
\filldraw[fill=cancan, draw=black] (8.00, 0.00) rectangle (8.25, -0.25);
\filldraw[fill=bermuda, draw=black] (8.25, 0.00) rectangle (8.50, -0.25);
\filldraw[fill=bermuda, draw=black] (8.50, 0.00) rectangle (8.75, -0.25);
\filldraw[fill=bermuda, draw=black] (8.75, 0.00) rectangle (9.00, -0.25);
\filldraw[fill=cancan, draw=black] (9.00, 0.00) rectangle (9.25, -0.25);
\filldraw[fill=cancan, draw=black] (9.25, 0.00) rectangle (9.50, -0.25);
\filldraw[fill=cancan, draw=black] (9.50, 0.00) rectangle (9.75, -0.25);
\filldraw[fill=bermuda, draw=black] (9.75, 0.00) rectangle (10.00, -0.25);
\filldraw[fill=bermuda, draw=black] (10.00, 0.00) rectangle (10.25, -0.25);
\filldraw[fill=bermuda, draw=black] (10.25, 0.00) rectangle (10.50, -0.25);
\filldraw[fill=cancan, draw=black] (10.50, 0.00) rectangle (10.75, -0.25);
\filldraw[fill=cancan, draw=black] (10.75, 0.00) rectangle (11.00, -0.25);
\filldraw[fill=cancan, draw=black] (11.00, 0.00) rectangle (11.25, -0.25);
\filldraw[fill=bermuda, draw=black] (11.25, 0.00) rectangle (11.50, -0.25);
\filldraw[fill=bermuda, draw=black] (11.50, 0.00) rectangle (11.75, -0.25);
\filldraw[fill=bermuda, draw=black] (11.75, 0.00) rectangle (12.00, -0.25);
\filldraw[fill=cancan, draw=black] (12.00, 0.00) rectangle (12.25, -0.25);
\filldraw[fill=cancan, draw=black] (12.25, 0.00) rectangle (12.50, -0.25);
\filldraw[fill=cancan, draw=black] (12.50, 0.00) rectangle (12.75, -0.25);
\filldraw[fill=bermuda, draw=black] (12.75, 0.00) rectangle (13.00, -0.25);
\filldraw[fill=bermuda, draw=black] (13.00, 0.00) rectangle (13.25, -0.25);
\filldraw[fill=bermuda, draw=black] (13.25, 0.00) rectangle (13.50, -0.25);
\filldraw[fill=bermuda, draw=black] (13.50, 0.00) rectangle (13.75, -0.25);
\filldraw[fill=bermuda, draw=black] (13.75, 0.00) rectangle (14.00, -0.25);
\filldraw[fill=cancan, draw=black] (14.00, 0.00) rectangle (14.25, -0.25);
\filldraw[fill=bermuda, draw=black] (14.25, 0.00) rectangle (14.50, -0.25);
\filldraw[fill=cancan, draw=black] (14.50, 0.00) rectangle (14.75, -0.25);
\filldraw[fill=cancan, draw=black] (14.75, 0.00) rectangle (15.00, -0.25);
\filldraw[fill=bermuda, draw=black] (0.00, -0.25) rectangle (0.25, -0.50);
\filldraw[fill=bermuda, draw=black] (0.25, -0.25) rectangle (0.50, -0.50);
\filldraw[fill=cancan, draw=black] (0.50, -0.25) rectangle (0.75, -0.50);
\filldraw[fill=bermuda, draw=black] (0.75, -0.25) rectangle (1.00, -0.50);
\filldraw[fill=cancan, draw=black] (1.00, -0.25) rectangle (1.25, -0.50);
\filldraw[fill=cancan, draw=black] (1.25, -0.25) rectangle (1.50, -0.50);
\filldraw[fill=cancan, draw=black] (1.50, -0.25) rectangle (1.75, -0.50);
\filldraw[fill=bermuda, draw=black] (1.75, -0.25) rectangle (2.00, -0.50);
\filldraw[fill=cancan, draw=black] (2.00, -0.25) rectangle (2.25, -0.50);
\filldraw[fill=cancan, draw=black] (2.25, -0.25) rectangle (2.50, -0.50);
\filldraw[fill=cancan, draw=black] (2.50, -0.25) rectangle (2.75, -0.50);
\filldraw[fill=bermuda, draw=black] (2.75, -0.25) rectangle (3.00, -0.50);
\filldraw[fill=bermuda, draw=black] (3.00, -0.25) rectangle (3.25, -0.50);
\filldraw[fill=bermuda, draw=black] (3.25, -0.25) rectangle (3.50, -0.50);
\filldraw[fill=cancan, draw=black] (3.50, -0.25) rectangle (3.75, -0.50);
\filldraw[fill=cancan, draw=black] (3.75, -0.25) rectangle (4.00, -0.50);
\filldraw[fill=bermuda, draw=black] (4.00, -0.25) rectangle (4.25, -0.50);
\filldraw[fill=bermuda, draw=black] (4.25, -0.25) rectangle (4.50, -0.50);
\filldraw[fill=cancan, draw=black] (4.50, -0.25) rectangle (4.75, -0.50);
\filldraw[fill=cancan, draw=black] (4.75, -0.25) rectangle (5.00, -0.50);
\filldraw[fill=cancan, draw=black] (5.00, -0.25) rectangle (5.25, -0.50);
\filldraw[fill=bermuda, draw=black] (5.25, -0.25) rectangle (5.50, -0.50);
\filldraw[fill=bermuda, draw=black] (5.50, -0.25) rectangle (5.75, -0.50);
\filldraw[fill=bermuda, draw=black] (5.75, -0.25) rectangle (6.00, -0.50);
\filldraw[fill=cancan, draw=black] (6.00, -0.25) rectangle (6.25, -0.50);
\filldraw[fill=cancan, draw=black] (6.25, -0.25) rectangle (6.50, -0.50);
\filldraw[fill=cancan, draw=black] (6.50, -0.25) rectangle (6.75, -0.50);
\filldraw[fill=bermuda, draw=black] (6.75, -0.25) rectangle (7.00, -0.50);
\filldraw[fill=bermuda, draw=black] (7.00, -0.25) rectangle (7.25, -0.50);
\filldraw[fill=bermuda, draw=black] (7.25, -0.25) rectangle (7.50, -0.50);
\filldraw[fill=cancan, draw=black] (7.50, -0.25) rectangle (7.75, -0.50);
\filldraw[fill=cancan, draw=black] (7.75, -0.25) rectangle (8.00, -0.50);
\filldraw[fill=cancan, draw=black] (8.00, -0.25) rectangle (8.25, -0.50);
\filldraw[fill=bermuda, draw=black] (8.25, -0.25) rectangle (8.50, -0.50);
\filldraw[fill=bermuda, draw=black] (8.50, -0.25) rectangle (8.75, -0.50);
\filldraw[fill=bermuda, draw=black] (8.75, -0.25) rectangle (9.00, -0.50);
\filldraw[fill=cancan, draw=black] (9.00, -0.25) rectangle (9.25, -0.50);
\filldraw[fill=cancan, draw=black] (9.25, -0.25) rectangle (9.50, -0.50);
\filldraw[fill=cancan, draw=black] (9.50, -0.25) rectangle (9.75, -0.50);
\filldraw[fill=cancan, draw=black] (9.75, -0.25) rectangle (10.00, -0.50);
\filldraw[fill=cancan, draw=black] (10.00, -0.25) rectangle (10.25, -0.50);
\filldraw[fill=bermuda, draw=black] (10.25, -0.25) rectangle (10.50, -0.50);
\filldraw[fill=cancan, draw=black] (10.50, -0.25) rectangle (10.75, -0.50);
\filldraw[fill=bermuda, draw=black] (10.75, -0.25) rectangle (11.00, -0.50);
\filldraw[fill=cancan, draw=black] (11.00, -0.25) rectangle (11.25, -0.50);
\filldraw[fill=cancan, draw=black] (11.25, -0.25) rectangle (11.50, -0.50);
\filldraw[fill=bermuda, draw=black] (11.50, -0.25) rectangle (11.75, -0.50);
\filldraw[fill=bermuda, draw=black] (11.75, -0.25) rectangle (12.00, -0.50);
\filldraw[fill=cancan, draw=black] (12.00, -0.25) rectangle (12.25, -0.50);
\filldraw[fill=cancan, draw=black] (12.25, -0.25) rectangle (12.50, -0.50);
\filldraw[fill=cancan, draw=black] (12.50, -0.25) rectangle (12.75, -0.50);
\filldraw[fill=cancan, draw=black] (12.75, -0.25) rectangle (13.00, -0.50);
\filldraw[fill=bermuda, draw=black] (13.00, -0.25) rectangle (13.25, -0.50);
\filldraw[fill=bermuda, draw=black] (13.25, -0.25) rectangle (13.50, -0.50);
\filldraw[fill=bermuda, draw=black] (13.50, -0.25) rectangle (13.75, -0.50);
\filldraw[fill=bermuda, draw=black] (13.75, -0.25) rectangle (14.00, -0.50);
\filldraw[fill=cancan, draw=black] (14.00, -0.25) rectangle (14.25, -0.50);
\filldraw[fill=bermuda, draw=black] (14.25, -0.25) rectangle (14.50, -0.50);
\filldraw[fill=bermuda, draw=black] (14.50, -0.25) rectangle (14.75, -0.50);
\filldraw[fill=bermuda, draw=black] (14.75, -0.25) rectangle (15.00, -0.50);
\filldraw[fill=cancan, draw=black] (0.00, -0.50) rectangle (0.25, -0.75);
\filldraw[fill=bermuda, draw=black] (0.25, -0.50) rectangle (0.50, -0.75);
\filldraw[fill=bermuda, draw=black] (0.50, -0.50) rectangle (0.75, -0.75);
\filldraw[fill=bermuda, draw=black] (0.75, -0.50) rectangle (1.00, -0.75);
\filldraw[fill=cancan, draw=black] (1.00, -0.50) rectangle (1.25, -0.75);
\filldraw[fill=bermuda, draw=black] (1.25, -0.50) rectangle (1.50, -0.75);
\filldraw[fill=bermuda, draw=black] (1.50, -0.50) rectangle (1.75, -0.75);
\filldraw[fill=bermuda, draw=black] (1.75, -0.50) rectangle (2.00, -0.75);
\filldraw[fill=cancan, draw=black] (2.00, -0.50) rectangle (2.25, -0.75);
\filldraw[fill=cancan, draw=black] (2.25, -0.50) rectangle (2.50, -0.75);
\filldraw[fill=cancan, draw=black] (2.50, -0.50) rectangle (2.75, -0.75);
\filldraw[fill=bermuda, draw=black] (2.75, -0.50) rectangle (3.00, -0.75);
\filldraw[fill=bermuda, draw=black] (3.00, -0.50) rectangle (3.25, -0.75);
\filldraw[fill=bermuda, draw=black] (3.25, -0.50) rectangle (3.50, -0.75);
\filldraw[fill=cancan, draw=black] (3.50, -0.50) rectangle (3.75, -0.75);
\filldraw[fill=cancan, draw=black] (3.75, -0.50) rectangle (4.00, -0.75);
\filldraw[fill=cancan, draw=black] (4.00, -0.50) rectangle (4.25, -0.75);
\filldraw[fill=bermuda, draw=black] (4.25, -0.50) rectangle (4.50, -0.75);
\filldraw[fill=bermuda, draw=black] (4.50, -0.50) rectangle (4.75, -0.75);
\filldraw[fill=bermuda, draw=black] (4.75, -0.50) rectangle (5.00, -0.75);
\filldraw[fill=cancan, draw=black] (5.00, -0.50) rectangle (5.25, -0.75);
\filldraw[fill=bermuda, draw=black] (5.25, -0.50) rectangle (5.50, -0.75);
\filldraw[fill=bermuda, draw=black] (5.50, -0.50) rectangle (5.75, -0.75);
\filldraw[fill=bermuda, draw=black] (5.75, -0.50) rectangle (6.00, -0.75);
\filldraw[fill=cancan, draw=black] (6.00, -0.50) rectangle (6.25, -0.75);
\filldraw[fill=cancan, draw=black] (6.25, -0.50) rectangle (6.50, -0.75);
\filldraw[fill=cancan, draw=black] (6.50, -0.50) rectangle (6.75, -0.75);
\filldraw[fill=bermuda, draw=black] (6.75, -0.50) rectangle (7.00, -0.75);
\filldraw[fill=bermuda, draw=black] (7.00, -0.50) rectangle (7.25, -0.75);
\filldraw[fill=bermuda, draw=black] (7.25, -0.50) rectangle (7.50, -0.75);
\filldraw[fill=cancan, draw=black] (7.50, -0.50) rectangle (7.75, -0.75);
\filldraw[fill=cancan, draw=black] (7.75, -0.50) rectangle (8.00, -0.75);
\filldraw[fill=cancan, draw=black] (8.00, -0.50) rectangle (8.25, -0.75);
\filldraw[fill=bermuda, draw=black] (8.25, -0.50) rectangle (8.50, -0.75);
\filldraw[fill=bermuda, draw=black] (8.50, -0.50) rectangle (8.75, -0.75);
\filldraw[fill=bermuda, draw=black] (8.75, -0.50) rectangle (9.00, -0.75);
\filldraw[fill=cancan, draw=black] (9.00, -0.50) rectangle (9.25, -0.75);
\filldraw[fill=cancan, draw=black] (9.25, -0.50) rectangle (9.50, -0.75);
\filldraw[fill=cancan, draw=black] (9.50, -0.50) rectangle (9.75, -0.75);
\filldraw[fill=bermuda, draw=black] (9.75, -0.50) rectangle (10.00, -0.75);
\filldraw[fill=bermuda, draw=black] (10.00, -0.50) rectangle (10.25, -0.75);
\filldraw[fill=bermuda, draw=black] (10.25, -0.50) rectangle (10.50, -0.75);
\filldraw[fill=cancan, draw=black] (10.50, -0.50) rectangle (10.75, -0.75);
\filldraw[fill=cancan, draw=black] (10.75, -0.50) rectangle (11.00, -0.75);
\filldraw[fill=cancan, draw=black] (11.00, -0.50) rectangle (11.25, -0.75);
\filldraw[fill=bermuda, draw=black] (11.25, -0.50) rectangle (11.50, -0.75);
\filldraw[fill=bermuda, draw=black] (11.50, -0.50) rectangle (11.75, -0.75);
\filldraw[fill=bermuda, draw=black] (11.75, -0.50) rectangle (12.00, -0.75);
\filldraw[fill=cancan, draw=black] (12.00, -0.50) rectangle (12.25, -0.75);
\filldraw[fill=cancan, draw=black] (12.25, -0.50) rectangle (12.50, -0.75);
\filldraw[fill=cancan, draw=black] (12.50, -0.50) rectangle (12.75, -0.75);
\filldraw[fill=cancan, draw=black] (12.75, -0.50) rectangle (13.00, -0.75);
\filldraw[fill=cancan, draw=black] (13.00, -0.50) rectangle (13.25, -0.75);
\filldraw[fill=cancan, draw=black] (13.25, -0.50) rectangle (13.50, -0.75);
\filldraw[fill=cancan, draw=black] (13.50, -0.50) rectangle (13.75, -0.75);
\filldraw[fill=bermuda, draw=black] (13.75, -0.50) rectangle (14.00, -0.75);
\filldraw[fill=cancan, draw=black] (14.00, -0.50) rectangle (14.25, -0.75);
\filldraw[fill=bermuda, draw=black] (14.25, -0.50) rectangle (14.50, -0.75);
\filldraw[fill=cancan, draw=black] (14.50, -0.50) rectangle (14.75, -0.75);
\filldraw[fill=cancan, draw=black] (14.75, -0.50) rectangle (15.00, -0.75);
\filldraw[fill=bermuda, draw=black] (0.00, -0.75) rectangle (0.25, -1.00);
\filldraw[fill=bermuda, draw=black] (0.25, -0.75) rectangle (0.50, -1.00);
\filldraw[fill=bermuda, draw=black] (0.50, -0.75) rectangle (0.75, -1.00);
\filldraw[fill=bermuda, draw=black] (0.75, -0.75) rectangle (1.00, -1.00);
\filldraw[fill=cancan, draw=black] (1.00, -0.75) rectangle (1.25, -1.00);
\filldraw[fill=bermuda, draw=black] (1.25, -0.75) rectangle (1.50, -1.00);
\filldraw[fill=cancan, draw=black] (1.50, -0.75) rectangle (1.75, -1.00);
\filldraw[fill=cancan, draw=black] (1.75, -0.75) rectangle (2.00, -1.00);
\filldraw[fill=cancan, draw=black] (2.00, -0.75) rectangle (2.25, -1.00);
\filldraw[fill=cancan, draw=black] (2.25, -0.75) rectangle (2.50, -1.00);
\filldraw[fill=bermuda, draw=black] (2.50, -0.75) rectangle (2.75, -1.00);
\filldraw[fill=bermuda, draw=black] (2.75, -0.75) rectangle (3.00, -1.00);
\filldraw[fill=cancan, draw=black] (3.00, -0.75) rectangle (3.25, -1.00);
\filldraw[fill=cancan, draw=black] (3.25, -0.75) rectangle (3.50, -1.00);
\filldraw[fill=cancan, draw=black] (3.50, -0.75) rectangle (3.75, -1.00);
\filldraw[fill=cancan, draw=black] (3.75, -0.75) rectangle (4.00, -1.00);
\filldraw[fill=cancan, draw=black] (4.00, -0.75) rectangle (4.25, -1.00);
\filldraw[fill=bermuda, draw=black] (4.25, -0.75) rectangle (4.50, -1.00);
\filldraw[fill=cancan, draw=black] (4.50, -0.75) rectangle (4.75, -1.00);
\filldraw[fill=bermuda, draw=black] (4.75, -0.75) rectangle (5.00, -1.00);
\filldraw[fill=bermuda, draw=black] (5.00, -0.75) rectangle (5.25, -1.00);
\filldraw[fill=bermuda, draw=black] (5.25, -0.75) rectangle (5.50, -1.00);
\filldraw[fill=cancan, draw=black] (5.50, -0.75) rectangle (5.75, -1.00);
\filldraw[fill=cancan, draw=black] (5.75, -0.75) rectangle (6.00, -1.00);
\filldraw[fill=cancan, draw=black] (6.00, -0.75) rectangle (6.25, -1.00);
\filldraw[fill=bermuda, draw=black] (6.25, -0.75) rectangle (6.50, -1.00);
\filldraw[fill=bermuda, draw=black] (6.50, -0.75) rectangle (6.75, -1.00);
\filldraw[fill=bermuda, draw=black] (6.75, -0.75) rectangle (7.00, -1.00);
\filldraw[fill=cancan, draw=black] (7.00, -0.75) rectangle (7.25, -1.00);
\filldraw[fill=cancan, draw=black] (7.25, -0.75) rectangle (7.50, -1.00);
\filldraw[fill=cancan, draw=black] (7.50, -0.75) rectangle (7.75, -1.00);
\filldraw[fill=bermuda, draw=black] (7.75, -0.75) rectangle (8.00, -1.00);
\filldraw[fill=bermuda, draw=black] (8.00, -0.75) rectangle (8.25, -1.00);
\filldraw[fill=bermuda, draw=black] (8.25, -0.75) rectangle (8.50, -1.00);
\filldraw[fill=cancan, draw=black] (8.50, -0.75) rectangle (8.75, -1.00);
\filldraw[fill=bermuda, draw=black] (8.75, -0.75) rectangle (9.00, -1.00);
\filldraw[fill=bermuda, draw=black] (9.00, -0.75) rectangle (9.25, -1.00);
\filldraw[fill=bermuda, draw=black] (9.25, -0.75) rectangle (9.50, -1.00);
\filldraw[fill=cancan, draw=black] (9.50, -0.75) rectangle (9.75, -1.00);
\filldraw[fill=bermuda, draw=black] (9.75, -0.75) rectangle (10.00, -1.00);
\filldraw[fill=cancan, draw=black] (10.00, -0.75) rectangle (10.25, -1.00);
\filldraw[fill=bermuda, draw=black] (10.25, -0.75) rectangle (10.50, -1.00);
\filldraw[fill=cancan, draw=black] (10.50, -0.75) rectangle (10.75, -1.00);
\filldraw[fill=bermuda, draw=black] (10.75, -0.75) rectangle (11.00, -1.00);
\filldraw[fill=bermuda, draw=black] (11.00, -0.75) rectangle (11.25, -1.00);
\filldraw[fill=bermuda, draw=black] (11.25, -0.75) rectangle (11.50, -1.00);
\filldraw[fill=cancan, draw=black] (11.50, -0.75) rectangle (11.75, -1.00);
\filldraw[fill=cancan, draw=black] (11.75, -0.75) rectangle (12.00, -1.00);
\filldraw[fill=cancan, draw=black] (12.00, -0.75) rectangle (12.25, -1.00);
\filldraw[fill=bermuda, draw=black] (12.25, -0.75) rectangle (12.50, -1.00);
\filldraw[fill=bermuda, draw=black] (12.50, -0.75) rectangle (12.75, -1.00);
\filldraw[fill=bermuda, draw=black] (12.75, -0.75) rectangle (13.00, -1.00);
\filldraw[fill=bermuda, draw=black] (13.00, -0.75) rectangle (13.25, -1.00);
\filldraw[fill=bermuda, draw=black] (13.25, -0.75) rectangle (13.50, -1.00);
\filldraw[fill=cancan, draw=black] (13.50, -0.75) rectangle (13.75, -1.00);
\filldraw[fill=bermuda, draw=black] (13.75, -0.75) rectangle (14.00, -1.00);
\filldraw[fill=cancan, draw=black] (14.00, -0.75) rectangle (14.25, -1.00);
\filldraw[fill=cancan, draw=black] (14.25, -0.75) rectangle (14.50, -1.00);
\filldraw[fill=bermuda, draw=black] (14.50, -0.75) rectangle (14.75, -1.00);
\filldraw[fill=bermuda, draw=black] (14.75, -0.75) rectangle (15.00, -1.00);
\filldraw[fill=cancan, draw=black] (0.00, -1.00) rectangle (0.25, -1.25);
\filldraw[fill=bermuda, draw=black] (0.25, -1.00) rectangle (0.50, -1.25);
\filldraw[fill=cancan, draw=black] (0.50, -1.00) rectangle (0.75, -1.25);
\filldraw[fill=cancan, draw=black] (0.75, -1.00) rectangle (1.00, -1.25);
\filldraw[fill=bermuda, draw=black] (1.00, -1.00) rectangle (1.25, -1.25);
\filldraw[fill=bermuda, draw=black] (1.25, -1.00) rectangle (1.50, -1.25);
\filldraw[fill=cancan, draw=black] (1.50, -1.00) rectangle (1.75, -1.25);
\filldraw[fill=bermuda, draw=black] (1.75, -1.00) rectangle (2.00, -1.25);
\filldraw[fill=cancan, draw=black] (2.00, -1.00) rectangle (2.25, -1.25);
\filldraw[fill=cancan, draw=black] (2.25, -1.00) rectangle (2.50, -1.25);
\filldraw[fill=bermuda, draw=black] (2.50, -1.00) rectangle (2.75, -1.25);
\filldraw[fill=bermuda, draw=black] (2.75, -1.00) rectangle (3.00, -1.25);
\filldraw[fill=cancan, draw=black] (3.00, -1.00) rectangle (3.25, -1.25);
\filldraw[fill=bermuda, draw=black] (3.25, -1.00) rectangle (3.50, -1.25);
\filldraw[fill=cancan, draw=black] (3.50, -1.00) rectangle (3.75, -1.25);
\filldraw[fill=cancan, draw=black] (3.75, -1.00) rectangle (4.00, -1.25);
\filldraw[fill=cancan, draw=black] (4.00, -1.00) rectangle (4.25, -1.25);
\filldraw[fill=cancan, draw=black] (4.25, -1.00) rectangle (4.50, -1.25);
\filldraw[fill=cancan, draw=black] (4.50, -1.00) rectangle (4.75, -1.25);
\filldraw[fill=cancan, draw=black] (4.75, -1.00) rectangle (5.00, -1.25);
\filldraw[fill=cancan, draw=black] (5.00, -1.00) rectangle (5.25, -1.25);
\filldraw[fill=cancan, draw=black] (5.25, -1.00) rectangle (5.50, -1.25);
\filldraw[fill=cancan, draw=black] (5.50, -1.00) rectangle (5.75, -1.25);
\filldraw[fill=bermuda, draw=black] (5.75, -1.00) rectangle (6.00, -1.25);
\filldraw[fill=bermuda, draw=black] (6.00, -1.00) rectangle (6.25, -1.25);
\filldraw[fill=bermuda, draw=black] (6.25, -1.00) rectangle (6.50, -1.25);
\filldraw[fill=cancan, draw=black] (6.50, -1.00) rectangle (6.75, -1.25);
\filldraw[fill=cancan, draw=black] (6.75, -1.00) rectangle (7.00, -1.25);
\filldraw[fill=cancan, draw=black] (7.00, -1.00) rectangle (7.25, -1.25);
\filldraw[fill=bermuda, draw=black] (7.25, -1.00) rectangle (7.50, -1.25);
\filldraw[fill=bermuda, draw=black] (7.50, -1.00) rectangle (7.75, -1.25);
\filldraw[fill=bermuda, draw=black] (7.75, -1.00) rectangle (8.00, -1.25);
\filldraw[fill=bermuda, draw=black] (8.00, -1.00) rectangle (8.25, -1.25);
\filldraw[fill=bermuda, draw=black] (8.25, -1.00) rectangle (8.50, -1.25);
\filldraw[fill=cancan, draw=black] (8.50, -1.00) rectangle (8.75, -1.25);
\filldraw[fill=bermuda, draw=black] (8.75, -1.00) rectangle (9.00, -1.25);
\filldraw[fill=cancan, draw=black] (9.00, -1.00) rectangle (9.25, -1.25);
\filldraw[fill=cancan, draw=black] (9.25, -1.00) rectangle (9.50, -1.25);
\filldraw[fill=bermuda, draw=black] (9.50, -1.00) rectangle (9.75, -1.25);
\filldraw[fill=bermuda, draw=black] (9.75, -1.00) rectangle (10.00, -1.25);
\filldraw[fill=bermuda, draw=black] (10.00, -1.00) rectangle (10.25, -1.25);
\filldraw[fill=bermuda, draw=black] (10.25, -1.00) rectangle (10.50, -1.25);
\filldraw[fill=cancan, draw=black] (10.50, -1.00) rectangle (10.75, -1.25);
\filldraw[fill=bermuda, draw=black] (10.75, -1.00) rectangle (11.00, -1.25);
\filldraw[fill=cancan, draw=black] (11.00, -1.00) rectangle (11.25, -1.25);
\filldraw[fill=cancan, draw=black] (11.25, -1.00) rectangle (11.50, -1.25);
\filldraw[fill=bermuda, draw=black] (11.50, -1.00) rectangle (11.75, -1.25);
\filldraw[fill=bermuda, draw=black] (11.75, -1.00) rectangle (12.00, -1.25);
\filldraw[fill=cancan, draw=black] (12.00, -1.00) rectangle (12.25, -1.25);
\filldraw[fill=bermuda, draw=black] (12.25, -1.00) rectangle (12.50, -1.25);
\filldraw[fill=cancan, draw=black] (12.50, -1.00) rectangle (12.75, -1.25);
\filldraw[fill=cancan, draw=black] (12.75, -1.00) rectangle (13.00, -1.25);
\filldraw[fill=bermuda, draw=black] (13.00, -1.00) rectangle (13.25, -1.25);
\filldraw[fill=bermuda, draw=black] (13.25, -1.00) rectangle (13.50, -1.25);
\filldraw[fill=cancan, draw=black] (13.50, -1.00) rectangle (13.75, -1.25);
\filldraw[fill=bermuda, draw=black] (13.75, -1.00) rectangle (14.00, -1.25);
\filldraw[fill=cancan, draw=black] (14.00, -1.00) rectangle (14.25, -1.25);
\filldraw[fill=cancan, draw=black] (14.25, -1.00) rectangle (14.50, -1.25);
\filldraw[fill=bermuda, draw=black] (14.50, -1.00) rectangle (14.75, -1.25);
\filldraw[fill=bermuda, draw=black] (14.75, -1.00) rectangle (15.00, -1.25);
\filldraw[fill=cancan, draw=black] (0.00, -1.25) rectangle (0.25, -1.50);
\filldraw[fill=bermuda, draw=black] (0.25, -1.25) rectangle (0.50, -1.50);
\filldraw[fill=cancan, draw=black] (0.50, -1.25) rectangle (0.75, -1.50);
\filldraw[fill=bermuda, draw=black] (0.75, -1.25) rectangle (1.00, -1.50);
\filldraw[fill=cancan, draw=black] (1.00, -1.25) rectangle (1.25, -1.50);
\filldraw[fill=bermuda, draw=black] (1.25, -1.25) rectangle (1.50, -1.50);
\filldraw[fill=cancan, draw=black] (1.50, -1.25) rectangle (1.75, -1.50);
\filldraw[fill=bermuda, draw=black] (1.75, -1.25) rectangle (2.00, -1.50);
\filldraw[fill=cancan, draw=black] (2.00, -1.25) rectangle (2.25, -1.50);
\filldraw[fill=cancan, draw=black] (2.25, -1.25) rectangle (2.50, -1.50);
\filldraw[fill=cancan, draw=black] (2.50, -1.25) rectangle (2.75, -1.50);
\filldraw[fill=cancan, draw=black] (2.75, -1.25) rectangle (3.00, -1.50);
\filldraw[fill=cancan, draw=black] (3.00, -1.25) rectangle (3.25, -1.50);
\filldraw[fill=bermuda, draw=black] (3.25, -1.25) rectangle (3.50, -1.50);
\filldraw[fill=bermuda, draw=black] (3.50, -1.25) rectangle (3.75, -1.50);
\filldraw[fill=bermuda, draw=black] (3.75, -1.25) rectangle (4.00, -1.50);
\filldraw[fill=cancan, draw=black] (4.00, -1.25) rectangle (4.25, -1.50);
\filldraw[fill=cancan, draw=black] (4.25, -1.25) rectangle (4.50, -1.50);
\filldraw[fill=cancan, draw=black] (4.50, -1.25) rectangle (4.75, -1.50);
\filldraw[fill=bermuda, draw=black] (4.75, -1.25) rectangle (5.00, -1.50);
\filldraw[fill=bermuda, draw=black] (5.00, -1.25) rectangle (5.25, -1.50);
\filldraw[fill=bermuda, draw=black] (5.25, -1.25) rectangle (5.50, -1.50);
\filldraw[fill=cancan, draw=black] (5.50, -1.25) rectangle (5.75, -1.50);
\filldraw[fill=bermuda, draw=black] (5.75, -1.25) rectangle (6.00, -1.50);
\filldraw[fill=bermuda, draw=black] (6.00, -1.25) rectangle (6.25, -1.50);
\filldraw[fill=bermuda, draw=black] (6.25, -1.25) rectangle (6.50, -1.50);
\filldraw[fill=bermuda, draw=black] (6.50, -1.25) rectangle (6.75, -1.50);
\filldraw[fill=bermuda, draw=black] (6.75, -1.25) rectangle (7.00, -1.50);
\filldraw[fill=cancan, draw=black] (7.00, -1.25) rectangle (7.25, -1.50);
\filldraw[fill=bermuda, draw=black] (7.25, -1.25) rectangle (7.50, -1.50);
\filldraw[fill=bermuda, draw=black] (7.50, -1.25) rectangle (7.75, -1.50);
\filldraw[fill=bermuda, draw=black] (7.75, -1.25) rectangle (8.00, -1.50);
\filldraw[fill=cancan, draw=black] (8.00, -1.25) rectangle (8.25, -1.50);
\filldraw[fill=cancan, draw=black] (8.25, -1.25) rectangle (8.50, -1.50);
\filldraw[fill=cancan, draw=black] (8.50, -1.25) rectangle (8.75, -1.50);
\filldraw[fill=bermuda, draw=black] (8.75, -1.25) rectangle (9.00, -1.50);
\filldraw[fill=bermuda, draw=black] (9.00, -1.25) rectangle (9.25, -1.50);
\filldraw[fill=bermuda, draw=black] (9.25, -1.25) rectangle (9.50, -1.50);
\filldraw[fill=cancan, draw=black] (9.50, -1.25) rectangle (9.75, -1.50);
\filldraw[fill=bermuda, draw=black] (9.75, -1.25) rectangle (10.00, -1.50);
\filldraw[fill=cancan, draw=black] (10.00, -1.25) rectangle (10.25, -1.50);
\filldraw[fill=bermuda, draw=black] (10.25, -1.25) rectangle (10.50, -1.50);
\filldraw[fill=bermuda, draw=black] (10.50, -1.25) rectangle (10.75, -1.50);
\filldraw[fill=bermuda, draw=black] (10.75, -1.25) rectangle (11.00, -1.50);
\filldraw[fill=cancan, draw=black] (11.00, -1.25) rectangle (11.25, -1.50);
\filldraw[fill=bermuda, draw=black] (11.25, -1.25) rectangle (11.50, -1.50);
\filldraw[fill=bermuda, draw=black] (11.50, -1.25) rectangle (11.75, -1.50);
\filldraw[fill=bermuda, draw=black] (11.75, -1.25) rectangle (12.00, -1.50);
\filldraw[fill=cancan, draw=black] (12.00, -1.25) rectangle (12.25, -1.50);
\filldraw[fill=cancan, draw=black] (12.25, -1.25) rectangle (12.50, -1.50);
\filldraw[fill=cancan, draw=black] (12.50, -1.25) rectangle (12.75, -1.50);
\filldraw[fill=cancan, draw=black] (12.75, -1.25) rectangle (13.00, -1.50);
\filldraw[fill=cancan, draw=black] (13.00, -1.25) rectangle (13.25, -1.50);
\filldraw[fill=cancan, draw=black] (13.25, -1.25) rectangle (13.50, -1.50);
\filldraw[fill=cancan, draw=black] (13.50, -1.25) rectangle (13.75, -1.50);
\filldraw[fill=bermuda, draw=black] (13.75, -1.25) rectangle (14.00, -1.50);
\filldraw[fill=bermuda, draw=black] (14.00, -1.25) rectangle (14.25, -1.50);
\filldraw[fill=bermuda, draw=black] (14.25, -1.25) rectangle (14.50, -1.50);
\filldraw[fill=cancan, draw=black] (14.50, -1.25) rectangle (14.75, -1.50);
\filldraw[fill=cancan, draw=black] (14.75, -1.25) rectangle (15.00, -1.50);
\filldraw[fill=cancan, draw=black] (0.00, -1.50) rectangle (0.25, -1.75);
\filldraw[fill=bermuda, draw=black] (0.25, -1.50) rectangle (0.50, -1.75);
\filldraw[fill=bermuda, draw=black] (0.50, -1.50) rectangle (0.75, -1.75);
\filldraw[fill=bermuda, draw=black] (0.75, -1.50) rectangle (1.00, -1.75);
\filldraw[fill=cancan, draw=black] (1.00, -1.50) rectangle (1.25, -1.75);
\filldraw[fill=bermuda, draw=black] (1.25, -1.50) rectangle (1.50, -1.75);
\filldraw[fill=bermuda, draw=black] (1.50, -1.50) rectangle (1.75, -1.75);
\filldraw[fill=bermuda, draw=black] (1.75, -1.50) rectangle (2.00, -1.75);
\filldraw[fill=bermuda, draw=black] (2.00, -1.50) rectangle (2.25, -1.75);
\filldraw[fill=bermuda, draw=black] (2.25, -1.50) rectangle (2.50, -1.75);
\filldraw[fill=cancan, draw=black] (2.50, -1.50) rectangle (2.75, -1.75);
\filldraw[fill=bermuda, draw=black] (2.75, -1.50) rectangle (3.00, -1.75);
\filldraw[fill=bermuda, draw=black] (3.00, -1.50) rectangle (3.25, -1.75);
\filldraw[fill=bermuda, draw=black] (3.25, -1.50) rectangle (3.50, -1.75);
\filldraw[fill=cancan, draw=black] (3.50, -1.50) rectangle (3.75, -1.75);
\filldraw[fill=cancan, draw=black] (3.75, -1.50) rectangle (4.00, -1.75);
\filldraw[fill=cancan, draw=black] (4.00, -1.50) rectangle (4.25, -1.75);
\filldraw[fill=bermuda, draw=black] (4.25, -1.50) rectangle (4.50, -1.75);
\filldraw[fill=bermuda, draw=black] (4.50, -1.50) rectangle (4.75, -1.75);
\filldraw[fill=bermuda, draw=black] (4.75, -1.50) rectangle (5.00, -1.75);
\filldraw[fill=cancan, draw=black] (5.00, -1.50) rectangle (5.25, -1.75);
\filldraw[fill=cancan, draw=black] (5.25, -1.50) rectangle (5.50, -1.75);
\filldraw[fill=cancan, draw=black] (5.50, -1.50) rectangle (5.75, -1.75);
\filldraw[fill=bermuda, draw=black] (5.75, -1.50) rectangle (6.00, -1.75);
\filldraw[fill=bermuda, draw=black] (6.00, -1.50) rectangle (6.25, -1.75);
\filldraw[fill=bermuda, draw=black] (6.25, -1.50) rectangle (6.50, -1.75);
\filldraw[fill=cancan, draw=black] (6.50, -1.50) rectangle (6.75, -1.75);
\filldraw[fill=cancan, draw=black] (6.75, -1.50) rectangle (7.00, -1.75);
\filldraw[fill=cancan, draw=black] (7.00, -1.50) rectangle (7.25, -1.75);
\filldraw[fill=bermuda, draw=black] (7.25, -1.50) rectangle (7.50, -1.75);
\filldraw[fill=bermuda, draw=black] (7.50, -1.50) rectangle (7.75, -1.75);
\filldraw[fill=bermuda, draw=black] (7.75, -1.50) rectangle (8.00, -1.75);
\filldraw[fill=cancan, draw=black] (8.00, -1.50) rectangle (8.25, -1.75);
\filldraw[fill=cancan, draw=black] (8.25, -1.50) rectangle (8.50, -1.75);
\filldraw[fill=cancan, draw=black] (8.50, -1.50) rectangle (8.75, -1.75);
\filldraw[fill=bermuda, draw=black] (8.75, -1.50) rectangle (9.00, -1.75);
\filldraw[fill=bermuda, draw=black] (9.00, -1.50) rectangle (9.25, -1.75);
\filldraw[fill=bermuda, draw=black] (9.25, -1.50) rectangle (9.50, -1.75);
\filldraw[fill=bermuda, draw=black] (9.50, -1.50) rectangle (9.75, -1.75);
\filldraw[fill=bermuda, draw=black] (9.75, -1.50) rectangle (10.00, -1.75);
\filldraw[fill=cancan, draw=black] (10.00, -1.50) rectangle (10.25, -1.75);
\filldraw[fill=bermuda, draw=black] (10.25, -1.50) rectangle (10.50, -1.75);
\filldraw[fill=cancan, draw=black] (10.50, -1.50) rectangle (10.75, -1.75);
\filldraw[fill=bermuda, draw=black] (10.75, -1.50) rectangle (11.00, -1.75);
\filldraw[fill=bermuda, draw=black] (11.00, -1.50) rectangle (11.25, -1.75);
\filldraw[fill=bermuda, draw=black] (11.25, -1.50) rectangle (11.50, -1.75);
\filldraw[fill=cancan, draw=black] (11.50, -1.50) rectangle (11.75, -1.75);
\filldraw[fill=cancan, draw=black] (11.75, -1.50) rectangle (12.00, -1.75);
\filldraw[fill=cancan, draw=black] (12.00, -1.50) rectangle (12.25, -1.75);
\filldraw[fill=bermuda, draw=black] (12.25, -1.50) rectangle (12.50, -1.75);
\filldraw[fill=bermuda, draw=black] (12.50, -1.50) rectangle (12.75, -1.75);
\filldraw[fill=bermuda, draw=black] (12.75, -1.50) rectangle (13.00, -1.75);
\filldraw[fill=cancan, draw=black] (13.00, -1.50) rectangle (13.25, -1.75);
\filldraw[fill=cancan, draw=black] (13.25, -1.50) rectangle (13.50, -1.75);
\filldraw[fill=cancan, draw=black] (13.50, -1.50) rectangle (13.75, -1.75);
\filldraw[fill=bermuda, draw=black] (13.75, -1.50) rectangle (14.00, -1.75);
\filldraw[fill=bermuda, draw=black] (14.00, -1.50) rectangle (14.25, -1.75);
\filldraw[fill=bermuda, draw=black] (14.25, -1.50) rectangle (14.50, -1.75);
\filldraw[fill=cancan, draw=black] (14.50, -1.50) rectangle (14.75, -1.75);
\filldraw[fill=cancan, draw=black] (14.75, -1.50) rectangle (15.00, -1.75);
\filldraw[fill=bermuda, draw=black] (0.00, -1.75) rectangle (0.25, -2.00);
\filldraw[fill=bermuda, draw=black] (0.25, -1.75) rectangle (0.50, -2.00);
\filldraw[fill=cancan, draw=black] (0.50, -1.75) rectangle (0.75, -2.00);
\filldraw[fill=cancan, draw=black] (0.75, -1.75) rectangle (1.00, -2.00);
\filldraw[fill=cancan, draw=black] (1.00, -1.75) rectangle (1.25, -2.00);
\filldraw[fill=cancan, draw=black] (1.25, -1.75) rectangle (1.50, -2.00);
\filldraw[fill=bermuda, draw=black] (1.50, -1.75) rectangle (1.75, -2.00);
\filldraw[fill=bermuda, draw=black] (1.75, -1.75) rectangle (2.00, -2.00);
\filldraw[fill=cancan, draw=black] (2.00, -1.75) rectangle (2.25, -2.00);
\filldraw[fill=cancan, draw=black] (2.25, -1.75) rectangle (2.50, -2.00);
\filldraw[fill=cancan, draw=black] (2.50, -1.75) rectangle (2.75, -2.00);
\filldraw[fill=cancan, draw=black] (2.75, -1.75) rectangle (3.00, -2.00);
\filldraw[fill=cancan, draw=black] (3.00, -1.75) rectangle (3.25, -2.00);
\filldraw[fill=bermuda, draw=black] (3.25, -1.75) rectangle (3.50, -2.00);
\filldraw[fill=bermuda, draw=black] (3.50, -1.75) rectangle (3.75, -2.00);
\filldraw[fill=bermuda, draw=black] (3.75, -1.75) rectangle (4.00, -2.00);
\filldraw[fill=bermuda, draw=black] (4.00, -1.75) rectangle (4.25, -2.00);
\filldraw[fill=bermuda, draw=black] (4.25, -1.75) rectangle (4.50, -2.00);
\filldraw[fill=cancan, draw=black] (4.50, -1.75) rectangle (4.75, -2.00);
\filldraw[fill=bermuda, draw=black] (4.75, -1.75) rectangle (5.00, -2.00);
\filldraw[fill=bermuda, draw=black] (5.00, -1.75) rectangle (5.25, -2.00);
\filldraw[fill=bermuda, draw=black] (5.25, -1.75) rectangle (5.50, -2.00);
\filldraw[fill=cancan, draw=black] (5.50, -1.75) rectangle (5.75, -2.00);
\filldraw[fill=bermuda, draw=black] (5.75, -1.75) rectangle (6.00, -2.00);
\filldraw[fill=bermuda, draw=black] (6.00, -1.75) rectangle (6.25, -2.00);
\filldraw[fill=bermuda, draw=black] (6.25, -1.75) rectangle (6.50, -2.00);
\filldraw[fill=bermuda, draw=black] (6.50, -1.75) rectangle (6.75, -2.00);
\filldraw[fill=bermuda, draw=black] (6.75, -1.75) rectangle (7.00, -2.00);
\filldraw[fill=bermuda, draw=black] (7.00, -1.75) rectangle (7.25, -2.00);
\filldraw[fill=bermuda, draw=black] (7.25, -1.75) rectangle (7.50, -2.00);
\filldraw[fill=cancan, draw=black] (7.50, -1.75) rectangle (7.75, -2.00);
\filldraw[fill=bermuda, draw=black] (7.75, -1.75) rectangle (8.00, -2.00);
\filldraw[fill=cancan, draw=black] (8.00, -1.75) rectangle (8.25, -2.00);
\filldraw[fill=cancan, draw=black] (8.25, -1.75) rectangle (8.50, -2.00);
\filldraw[fill=cancan, draw=black] (8.50, -1.75) rectangle (8.75, -2.00);
\filldraw[fill=cancan, draw=black] (8.75, -1.75) rectangle (9.00, -2.00);
\filldraw[fill=cancan, draw=black] (9.00, -1.75) rectangle (9.25, -2.00);
\filldraw[fill=bermuda, draw=black] (9.25, -1.75) rectangle (9.50, -2.00);
\filldraw[fill=bermuda, draw=black] (9.50, -1.75) rectangle (9.75, -2.00);
\filldraw[fill=bermuda, draw=black] (9.75, -1.75) rectangle (10.00, -2.00);
\filldraw[fill=cancan, draw=black] (10.00, -1.75) rectangle (10.25, -2.00);
\filldraw[fill=cancan, draw=black] (10.25, -1.75) rectangle (10.50, -2.00);
\filldraw[fill=cancan, draw=black] (10.50, -1.75) rectangle (10.75, -2.00);
\filldraw[fill=bermuda, draw=black] (10.75, -1.75) rectangle (11.00, -2.00);
\filldraw[fill=bermuda, draw=black] (11.00, -1.75) rectangle (11.25, -2.00);
\filldraw[fill=bermuda, draw=black] (11.25, -1.75) rectangle (11.50, -2.00);
\filldraw[fill=cancan, draw=black] (11.50, -1.75) rectangle (11.75, -2.00);
\filldraw[fill=bermuda, draw=black] (11.75, -1.75) rectangle (12.00, -2.00);
\filldraw[fill=bermuda, draw=black] (12.00, -1.75) rectangle (12.25, -2.00);
\filldraw[fill=bermuda, draw=black] (12.25, -1.75) rectangle (12.50, -2.00);
\filldraw[fill=cancan, draw=black] (12.50, -1.75) rectangle (12.75, -2.00);
\filldraw[fill=bermuda, draw=black] (12.75, -1.75) rectangle (13.00, -2.00);
\filldraw[fill=cancan, draw=black] (13.00, -1.75) rectangle (13.25, -2.00);
\filldraw[fill=bermuda, draw=black] (13.25, -1.75) rectangle (13.50, -2.00);
\filldraw[fill=cancan, draw=black] (13.50, -1.75) rectangle (13.75, -2.00);
\filldraw[fill=bermuda, draw=black] (13.75, -1.75) rectangle (14.00, -2.00);
\filldraw[fill=bermuda, draw=black] (14.00, -1.75) rectangle (14.25, -2.00);
\filldraw[fill=bermuda, draw=black] (14.25, -1.75) rectangle (14.50, -2.00);
\filldraw[fill=cancan, draw=black] (14.50, -1.75) rectangle (14.75, -2.00);
\filldraw[fill=cancan, draw=black] (14.75, -1.75) rectangle (15.00, -2.00);
\filldraw[fill=cancan, draw=black] (0.00, -2.00) rectangle (0.25, -2.25);
\filldraw[fill=bermuda, draw=black] (0.25, -2.00) rectangle (0.50, -2.25);
\filldraw[fill=bermuda, draw=black] (0.50, -2.00) rectangle (0.75, -2.25);
\filldraw[fill=bermuda, draw=black] (0.75, -2.00) rectangle (1.00, -2.25);
\filldraw[fill=cancan, draw=black] (1.00, -2.00) rectangle (1.25, -2.25);
\filldraw[fill=cancan, draw=black] (1.25, -2.00) rectangle (1.50, -2.25);
\filldraw[fill=bermuda, draw=black] (1.50, -2.00) rectangle (1.75, -2.25);
\filldraw[fill=bermuda, draw=black] (1.75, -2.00) rectangle (2.00, -2.25);
\filldraw[fill=cancan, draw=black] (2.00, -2.00) rectangle (2.25, -2.25);
\filldraw[fill=cancan, draw=black] (2.25, -2.00) rectangle (2.50, -2.25);
\filldraw[fill=cancan, draw=black] (2.50, -2.00) rectangle (2.75, -2.25);
\filldraw[fill=cancan, draw=black] (2.75, -2.00) rectangle (3.00, -2.25);
\filldraw[fill=cancan, draw=black] (3.00, -2.00) rectangle (3.25, -2.25);
\filldraw[fill=cancan, draw=black] (3.25, -2.00) rectangle (3.50, -2.25);
\filldraw[fill=cancan, draw=black] (3.50, -2.00) rectangle (3.75, -2.25);
\filldraw[fill=bermuda, draw=black] (3.75, -2.00) rectangle (4.00, -2.25);
\filldraw[fill=bermuda, draw=black] (4.00, -2.00) rectangle (4.25, -2.25);
\filldraw[fill=bermuda, draw=black] (4.25, -2.00) rectangle (4.50, -2.25);
\filldraw[fill=cancan, draw=black] (4.50, -2.00) rectangle (4.75, -2.25);
\filldraw[fill=cancan, draw=black] (4.75, -2.00) rectangle (5.00, -2.25);
\filldraw[fill=cancan, draw=black] (5.00, -2.00) rectangle (5.25, -2.25);
\filldraw[fill=bermuda, draw=black] (5.25, -2.00) rectangle (5.50, -2.25);
\filldraw[fill=bermuda, draw=black] (5.50, -2.00) rectangle (5.75, -2.25);
\filldraw[fill=bermuda, draw=black] (5.75, -2.00) rectangle (6.00, -2.25);
\filldraw[fill=cancan, draw=black] (6.00, -2.00) rectangle (6.25, -2.25);
\filldraw[fill=cancan, draw=black] (6.25, -2.00) rectangle (6.50, -2.25);
\filldraw[fill=cancan, draw=black] (6.50, -2.00) rectangle (6.75, -2.25);
\filldraw[fill=bermuda, draw=black] (6.75, -2.00) rectangle (7.00, -2.25);
\filldraw[fill=bermuda, draw=black] (7.00, -2.00) rectangle (7.25, -2.25);
\filldraw[fill=bermuda, draw=black] (7.25, -2.00) rectangle (7.50, -2.25);
\filldraw[fill=bermuda, draw=black] (7.50, -2.00) rectangle (7.75, -2.25);
\filldraw[fill=bermuda, draw=black] (7.75, -2.00) rectangle (8.00, -2.25);
\filldraw[fill=bermuda, draw=black] (8.00, -2.00) rectangle (8.25, -2.25);
\filldraw[fill=bermuda, draw=black] (8.25, -2.00) rectangle (8.50, -2.25);
\filldraw[fill=cancan, draw=black] (8.50, -2.00) rectangle (8.75, -2.25);
\filldraw[fill=bermuda, draw=black] (8.75, -2.00) rectangle (9.00, -2.25);
\filldraw[fill=cancan, draw=black] (9.00, -2.00) rectangle (9.25, -2.25);
\filldraw[fill=cancan, draw=black] (9.25, -2.00) rectangle (9.50, -2.25);
\filldraw[fill=cancan, draw=black] (9.50, -2.00) rectangle (9.75, -2.25);
\filldraw[fill=cancan, draw=black] (9.75, -2.00) rectangle (10.00, -2.25);
\filldraw[fill=cancan, draw=black] (10.00, -2.00) rectangle (10.25, -2.25);
\filldraw[fill=bermuda, draw=black] (10.25, -2.00) rectangle (10.50, -2.25);
\filldraw[fill=bermuda, draw=black] (10.50, -2.00) rectangle (10.75, -2.25);
\filldraw[fill=bermuda, draw=black] (10.75, -2.00) rectangle (11.00, -2.25);
\filldraw[fill=cancan, draw=black] (11.00, -2.00) rectangle (11.25, -2.25);
\filldraw[fill=bermuda, draw=black] (11.25, -2.00) rectangle (11.50, -2.25);
\filldraw[fill=cancan, draw=black] (11.50, -2.00) rectangle (11.75, -2.25);
\filldraw[fill=bermuda, draw=black] (11.75, -2.00) rectangle (12.00, -2.25);
\filldraw[fill=bermuda, draw=black] (12.00, -2.00) rectangle (12.25, -2.25);
\filldraw[fill=bermuda, draw=black] (12.25, -2.00) rectangle (12.50, -2.25);
\filldraw[fill=cancan, draw=black] (12.50, -2.00) rectangle (12.75, -2.25);
\filldraw[fill=bermuda, draw=black] (12.75, -2.00) rectangle (13.00, -2.25);
\filldraw[fill=bermuda, draw=black] (13.00, -2.00) rectangle (13.25, -2.25);
\filldraw[fill=bermuda, draw=black] (13.25, -2.00) rectangle (13.50, -2.25);
\filldraw[fill=cancan, draw=black] (13.50, -2.00) rectangle (13.75, -2.25);
\filldraw[fill=cancan, draw=black] (13.75, -2.00) rectangle (14.00, -2.25);
\filldraw[fill=cancan, draw=black] (14.00, -2.00) rectangle (14.25, -2.25);
\filldraw[fill=bermuda, draw=black] (14.25, -2.00) rectangle (14.50, -2.25);
\filldraw[fill=bermuda, draw=black] (14.50, -2.00) rectangle (14.75, -2.25);
\filldraw[fill=bermuda, draw=black] (14.75, -2.00) rectangle (15.00, -2.25);
\filldraw[fill=bermuda, draw=black] (0.00, -2.25) rectangle (0.25, -2.50);
\filldraw[fill=bermuda, draw=black] (0.25, -2.25) rectangle (0.50, -2.50);
\filldraw[fill=bermuda, draw=black] (0.50, -2.25) rectangle (0.75, -2.50);
\filldraw[fill=bermuda, draw=black] (0.75, -2.25) rectangle (1.00, -2.50);
\filldraw[fill=cancan, draw=black] (1.00, -2.25) rectangle (1.25, -2.50);
\filldraw[fill=bermuda, draw=black] (1.25, -2.25) rectangle (1.50, -2.50);
\filldraw[fill=cancan, draw=black] (1.50, -2.25) rectangle (1.75, -2.50);
\filldraw[fill=bermuda, draw=black] (1.75, -2.25) rectangle (2.00, -2.50);
\filldraw[fill=bermuda, draw=black] (2.00, -2.25) rectangle (2.25, -2.50);
\filldraw[fill=bermuda, draw=black] (2.25, -2.25) rectangle (2.50, -2.50);
\filldraw[fill=cancan, draw=black] (2.50, -2.25) rectangle (2.75, -2.50);
\filldraw[fill=bermuda, draw=black] (2.75, -2.25) rectangle (3.00, -2.50);
\filldraw[fill=bermuda, draw=black] (3.00, -2.25) rectangle (3.25, -2.50);
\filldraw[fill=bermuda, draw=black] (3.25, -2.25) rectangle (3.50, -2.50);
\filldraw[fill=bermuda, draw=black] (3.50, -2.25) rectangle (3.75, -2.50);
\filldraw[fill=bermuda, draw=black] (3.75, -2.25) rectangle (4.00, -2.50);
\filldraw[fill=cancan, draw=black] (4.00, -2.25) rectangle (4.25, -2.50);
\filldraw[fill=bermuda, draw=black] (4.25, -2.25) rectangle (4.50, -2.50);
\filldraw[fill=cancan, draw=black] (4.50, -2.25) rectangle (4.75, -2.50);
\filldraw[fill=cancan, draw=black] (4.75, -2.25) rectangle (5.00, -2.50);
\filldraw[fill=bermuda, draw=black] (5.00, -2.25) rectangle (5.25, -2.50);
\filldraw[fill=bermuda, draw=black] (5.25, -2.25) rectangle (5.50, -2.50);
\filldraw[fill=cancan, draw=black] (5.50, -2.25) rectangle (5.75, -2.50);
\filldraw[fill=bermuda, draw=black] (5.75, -2.25) rectangle (6.00, -2.50);
\filldraw[fill=cancan, draw=black] (6.00, -2.25) rectangle (6.25, -2.50);
\filldraw[fill=bermuda, draw=black] (6.25, -2.25) rectangle (6.50, -2.50);
\filldraw[fill=cancan, draw=black] (6.50, -2.25) rectangle (6.75, -2.50);
\filldraw[fill=bermuda, draw=black] (6.75, -2.25) rectangle (7.00, -2.50);
\filldraw[fill=cancan, draw=black] (7.00, -2.25) rectangle (7.25, -2.50);
\filldraw[fill=cancan, draw=black] (7.25, -2.25) rectangle (7.50, -2.50);
\filldraw[fill=cancan, draw=black] (7.50, -2.25) rectangle (7.75, -2.50);
\filldraw[fill=cancan, draw=black] (7.75, -2.25) rectangle (8.00, -2.50);
\filldraw[fill=bermuda, draw=black] (8.00, -2.25) rectangle (8.25, -2.50);
\filldraw[fill=bermuda, draw=black] (8.25, -2.25) rectangle (8.50, -2.50);
\filldraw[fill=cancan, draw=black] (8.50, -2.25) rectangle (8.75, -2.50);
\filldraw[fill=bermuda, draw=black] (8.75, -2.25) rectangle (9.00, -2.50);
\filldraw[fill=bermuda, draw=black] (9.00, -2.25) rectangle (9.25, -2.50);
\filldraw[fill=bermuda, draw=black] (9.25, -2.25) rectangle (9.50, -2.50);
\filldraw[fill=bermuda, draw=black] (9.50, -2.25) rectangle (9.75, -2.50);
\filldraw[fill=bermuda, draw=black] (9.75, -2.25) rectangle (10.00, -2.50);
\filldraw[fill=cancan, draw=black] (10.00, -2.25) rectangle (10.25, -2.50);
\filldraw[fill=bermuda, draw=black] (10.25, -2.25) rectangle (10.50, -2.50);
\filldraw[fill=bermuda, draw=black] (10.50, -2.25) rectangle (10.75, -2.50);
\filldraw[fill=bermuda, draw=black] (10.75, -2.25) rectangle (11.00, -2.50);
\filldraw[fill=bermuda, draw=black] (11.00, -2.25) rectangle (11.25, -2.50);
\filldraw[fill=bermuda, draw=black] (11.25, -2.25) rectangle (11.50, -2.50);
\filldraw[fill=cancan, draw=black] (11.50, -2.25) rectangle (11.75, -2.50);
\filldraw[fill=cancan, draw=black] (11.75, -2.25) rectangle (12.00, -2.50);
\filldraw[fill=cancan, draw=black] (12.00, -2.25) rectangle (12.25, -2.50);
\filldraw[fill=bermuda, draw=black] (12.25, -2.25) rectangle (12.50, -2.50);
\filldraw[fill=bermuda, draw=black] (12.50, -2.25) rectangle (12.75, -2.50);
\filldraw[fill=bermuda, draw=black] (12.75, -2.25) rectangle (13.00, -2.50);
\filldraw[fill=cancan, draw=black] (13.00, -2.25) rectangle (13.25, -2.50);
\filldraw[fill=cancan, draw=black] (13.25, -2.25) rectangle (13.50, -2.50);
\filldraw[fill=cancan, draw=black] (13.50, -2.25) rectangle (13.75, -2.50);
\filldraw[fill=cancan, draw=black] (13.75, -2.25) rectangle (14.00, -2.50);
\filldraw[fill=bermuda, draw=black] (14.00, -2.25) rectangle (14.25, -2.50);
\filldraw[fill=bermuda, draw=black] (14.25, -2.25) rectangle (14.50, -2.50);
\filldraw[fill=cancan, draw=black] (14.50, -2.25) rectangle (14.75, -2.50);
\filldraw[fill=bermuda, draw=black] (14.75, -2.25) rectangle (15.00, -2.50);
\filldraw[fill=cancan, draw=black] (0.00, -2.50) rectangle (0.25, -2.75);
\filldraw[fill=cancan, draw=black] (0.25, -2.50) rectangle (0.50, -2.75);
\filldraw[fill=bermuda, draw=black] (0.50, -2.50) rectangle (0.75, -2.75);
\filldraw[fill=bermuda, draw=black] (0.75, -2.50) rectangle (1.00, -2.75);
\filldraw[fill=cancan, draw=black] (1.00, -2.50) rectangle (1.25, -2.75);
\filldraw[fill=bermuda, draw=black] (1.25, -2.50) rectangle (1.50, -2.75);
\filldraw[fill=bermuda, draw=black] (1.50, -2.50) rectangle (1.75, -2.75);
\filldraw[fill=bermuda, draw=black] (1.75, -2.50) rectangle (2.00, -2.75);
\filldraw[fill=cancan, draw=black] (2.00, -2.50) rectangle (2.25, -2.75);
\filldraw[fill=cancan, draw=black] (2.25, -2.50) rectangle (2.50, -2.75);
\filldraw[fill=cancan, draw=black] (2.50, -2.50) rectangle (2.75, -2.75);
\filldraw[fill=bermuda, draw=black] (2.75, -2.50) rectangle (3.00, -2.75);
\filldraw[fill=bermuda, draw=black] (3.00, -2.50) rectangle (3.25, -2.75);
\filldraw[fill=bermuda, draw=black] (3.25, -2.50) rectangle (3.50, -2.75);
\filldraw[fill=cancan, draw=black] (3.50, -2.50) rectangle (3.75, -2.75);
\filldraw[fill=cancan, draw=black] (3.75, -2.50) rectangle (4.00, -2.75);
\filldraw[fill=bermuda, draw=black] (4.00, -2.50) rectangle (4.25, -2.75);
\filldraw[fill=bermuda, draw=black] (4.25, -2.50) rectangle (4.50, -2.75);
\filldraw[fill=cancan, draw=black] (4.50, -2.50) rectangle (4.75, -2.75);
\filldraw[fill=cancan, draw=black] (4.75, -2.50) rectangle (5.00, -2.75);
\filldraw[fill=cancan, draw=black] (5.00, -2.50) rectangle (5.25, -2.75);
\filldraw[fill=cancan, draw=black] (5.25, -2.50) rectangle (5.50, -2.75);
\filldraw[fill=bermuda, draw=black] (5.50, -2.50) rectangle (5.75, -2.75);
\filldraw[fill=bermuda, draw=black] (5.75, -2.50) rectangle (6.00, -2.75);
\filldraw[fill=cancan, draw=black] (6.00, -2.50) rectangle (6.25, -2.75);
\filldraw[fill=bermuda, draw=black] (6.25, -2.50) rectangle (6.50, -2.75);
\filldraw[fill=cancan, draw=black] (6.50, -2.50) rectangle (6.75, -2.75);
\filldraw[fill=cancan, draw=black] (6.75, -2.50) rectangle (7.00, -2.75);
\filldraw[fill=cancan, draw=black] (7.00, -2.50) rectangle (7.25, -2.75);
\filldraw[fill=cancan, draw=black] (7.25, -2.50) rectangle (7.50, -2.75);
\filldraw[fill=cancan, draw=black] (7.50, -2.50) rectangle (7.75, -2.75);
\filldraw[fill=bermuda, draw=black] (7.75, -2.50) rectangle (8.00, -2.75);
\filldraw[fill=bermuda, draw=black] (8.00, -2.50) rectangle (8.25, -2.75);
\filldraw[fill=bermuda, draw=black] (8.25, -2.50) rectangle (8.50, -2.75);
\filldraw[fill=cancan, draw=black] (8.50, -2.50) rectangle (8.75, -2.75);
\filldraw[fill=cancan, draw=black] (8.75, -2.50) rectangle (9.00, -2.75);
\filldraw[fill=cancan, draw=black] (9.00, -2.50) rectangle (9.25, -2.75);
\filldraw[fill=bermuda, draw=black] (9.25, -2.50) rectangle (9.50, -2.75);
\filldraw[fill=bermuda, draw=black] (9.50, -2.50) rectangle (9.75, -2.75);
\filldraw[fill=bermuda, draw=black] (9.75, -2.50) rectangle (10.00, -2.75);
\filldraw[fill=cancan, draw=black] (10.00, -2.50) rectangle (10.25, -2.75);
\filldraw[fill=bermuda, draw=black] (10.25, -2.50) rectangle (10.50, -2.75);
\filldraw[fill=cancan, draw=black] (10.50, -2.50) rectangle (10.75, -2.75);
\filldraw[fill=bermuda, draw=black] (10.75, -2.50) rectangle (11.00, -2.75);
\filldraw[fill=bermuda, draw=black] (11.00, -2.50) rectangle (11.25, -2.75);
\filldraw[fill=bermuda, draw=black] (11.25, -2.50) rectangle (11.50, -2.75);
\filldraw[fill=cancan, draw=black] (11.50, -2.50) rectangle (11.75, -2.75);
\filldraw[fill=cancan, draw=black] (11.75, -2.50) rectangle (12.00, -2.75);
\filldraw[fill=cancan, draw=black] (12.00, -2.50) rectangle (12.25, -2.75);
\filldraw[fill=bermuda, draw=black] (12.25, -2.50) rectangle (12.50, -2.75);
\filldraw[fill=bermuda, draw=black] (12.50, -2.50) rectangle (12.75, -2.75);
\filldraw[fill=bermuda, draw=black] (12.75, -2.50) rectangle (13.00, -2.75);
\filldraw[fill=cancan, draw=black] (13.00, -2.50) rectangle (13.25, -2.75);
\filldraw[fill=bermuda, draw=black] (13.25, -2.50) rectangle (13.50, -2.75);
\filldraw[fill=bermuda, draw=black] (13.50, -2.50) rectangle (13.75, -2.75);
\filldraw[fill=bermuda, draw=black] (13.75, -2.50) rectangle (14.00, -2.75);
\filldraw[fill=bermuda, draw=black] (14.00, -2.50) rectangle (14.25, -2.75);
\filldraw[fill=bermuda, draw=black] (14.25, -2.50) rectangle (14.50, -2.75);
\filldraw[fill=bermuda, draw=black] (14.50, -2.50) rectangle (14.75, -2.75);
\filldraw[fill=bermuda, draw=black] (14.75, -2.50) rectangle (15.00, -2.75);
\filldraw[fill=bermuda, draw=black] (0.00, -2.75) rectangle (0.25, -3.00);
\filldraw[fill=bermuda, draw=black] (0.25, -2.75) rectangle (0.50, -3.00);
\filldraw[fill=cancan, draw=black] (0.50, -2.75) rectangle (0.75, -3.00);
\filldraw[fill=cancan, draw=black] (0.75, -2.75) rectangle (1.00, -3.00);
\filldraw[fill=cancan, draw=black] (1.00, -2.75) rectangle (1.25, -3.00);
\filldraw[fill=bermuda, draw=black] (1.25, -2.75) rectangle (1.50, -3.00);
\filldraw[fill=bermuda, draw=black] (1.50, -2.75) rectangle (1.75, -3.00);
\filldraw[fill=bermuda, draw=black] (1.75, -2.75) rectangle (2.00, -3.00);
\filldraw[fill=cancan, draw=black] (2.00, -2.75) rectangle (2.25, -3.00);
\filldraw[fill=cancan, draw=black] (2.25, -2.75) rectangle (2.50, -3.00);
\filldraw[fill=cancan, draw=black] (2.50, -2.75) rectangle (2.75, -3.00);
\filldraw[fill=bermuda, draw=black] (2.75, -2.75) rectangle (3.00, -3.00);
\filldraw[fill=bermuda, draw=black] (3.00, -2.75) rectangle (3.25, -3.00);
\filldraw[fill=bermuda, draw=black] (3.25, -2.75) rectangle (3.50, -3.00);
\filldraw[fill=cancan, draw=black] (3.50, -2.75) rectangle (3.75, -3.00);
\filldraw[fill=bermuda, draw=black] (3.75, -2.75) rectangle (4.00, -3.00);
\filldraw[fill=bermuda, draw=black] (4.00, -2.75) rectangle (4.25, -3.00);
\filldraw[fill=bermuda, draw=black] (4.25, -2.75) rectangle (4.50, -3.00);
\filldraw[fill=cancan, draw=black] (4.50, -2.75) rectangle (4.75, -3.00);
\filldraw[fill=bermuda, draw=black] (4.75, -2.75) rectangle (5.00, -3.00);
\filldraw[fill=cancan, draw=black] (5.00, -2.75) rectangle (5.25, -3.00);
\filldraw[fill=bermuda, draw=black] (5.25, -2.75) rectangle (5.50, -3.00);
\filldraw[fill=bermuda, draw=black] (5.50, -2.75) rectangle (5.75, -3.00);
\filldraw[fill=bermuda, draw=black] (5.75, -2.75) rectangle (6.00, -3.00);
\filldraw[fill=cancan, draw=black] (6.00, -2.75) rectangle (6.25, -3.00);
\filldraw[fill=cancan, draw=black] (6.25, -2.75) rectangle (6.50, -3.00);
\filldraw[fill=cancan, draw=black] (6.50, -2.75) rectangle (6.75, -3.00);
\filldraw[fill=bermuda, draw=black] (6.75, -2.75) rectangle (7.00, -3.00);
\filldraw[fill=bermuda, draw=black] (7.00, -2.75) rectangle (7.25, -3.00);
\filldraw[fill=bermuda, draw=black] (7.25, -2.75) rectangle (7.50, -3.00);
\filldraw[fill=cancan, draw=black] (7.50, -2.75) rectangle (7.75, -3.00);
\filldraw[fill=cancan, draw=black] (7.75, -2.75) rectangle (8.00, -3.00);
\filldraw[fill=cancan, draw=black] (8.00, -2.75) rectangle (8.25, -3.00);
\filldraw[fill=bermuda, draw=black] (8.25, -2.75) rectangle (8.50, -3.00);
\filldraw[fill=bermuda, draw=black] (8.50, -2.75) rectangle (8.75, -3.00);
\filldraw[fill=bermuda, draw=black] (8.75, -2.75) rectangle (9.00, -3.00);
\filldraw[fill=cancan, draw=black] (9.00, -2.75) rectangle (9.25, -3.00);
\filldraw[fill=cancan, draw=black] (9.25, -2.75) rectangle (9.50, -3.00);
\filldraw[fill=cancan, draw=black] (9.50, -2.75) rectangle (9.75, -3.00);
\filldraw[fill=bermuda, draw=black] (9.75, -2.75) rectangle (10.00, -3.00);
\filldraw[fill=bermuda, draw=black] (10.00, -2.75) rectangle (10.25, -3.00);
\filldraw[fill=bermuda, draw=black] (10.25, -2.75) rectangle (10.50, -3.00);
\filldraw[fill=cancan, draw=black] (10.50, -2.75) rectangle (10.75, -3.00);
\filldraw[fill=cancan, draw=black] (10.75, -2.75) rectangle (11.00, -3.00);
\filldraw[fill=cancan, draw=black] (11.00, -2.75) rectangle (11.25, -3.00);
\filldraw[fill=cancan, draw=black] (11.25, -2.75) rectangle (11.50, -3.00);
\filldraw[fill=cancan, draw=black] (11.50, -2.75) rectangle (11.75, -3.00);
\filldraw[fill=cancan, draw=black] (11.75, -2.75) rectangle (12.00, -3.00);
\filldraw[fill=bermuda, draw=black] (12.00, -2.75) rectangle (12.25, -3.00);
\filldraw[fill=bermuda, draw=black] (12.25, -2.75) rectangle (12.50, -3.00);
\filldraw[fill=cancan, draw=black] (12.50, -2.75) rectangle (12.75, -3.00);
\filldraw[fill=cancan, draw=black] (12.75, -2.75) rectangle (13.00, -3.00);
\filldraw[fill=cancan, draw=black] (13.00, -2.75) rectangle (13.25, -3.00);
\filldraw[fill=cancan, draw=black] (13.25, -2.75) rectangle (13.50, -3.00);
\filldraw[fill=cancan, draw=black] (13.50, -2.75) rectangle (13.75, -3.00);
\filldraw[fill=cancan, draw=black] (13.75, -2.75) rectangle (14.00, -3.00);
\filldraw[fill=cancan, draw=black] (14.00, -2.75) rectangle (14.25, -3.00);
\filldraw[fill=cancan, draw=black] (14.25, -2.75) rectangle (14.50, -3.00);
\filldraw[fill=cancan, draw=black] (14.50, -2.75) rectangle (14.75, -3.00);
\filldraw[fill=cancan, draw=black] (14.75, -2.75) rectangle (15.00, -3.00);
\filldraw[fill=cancan, draw=black] (0.00, -3.00) rectangle (0.25, -3.25);
\filldraw[fill=cancan, draw=black] (0.25, -3.00) rectangle (0.50, -3.25);
\filldraw[fill=cancan, draw=black] (0.50, -3.00) rectangle (0.75, -3.25);
\filldraw[fill=cancan, draw=black] (0.75, -3.00) rectangle (1.00, -3.25);
\filldraw[fill=cancan, draw=black] (1.00, -3.00) rectangle (1.25, -3.25);
\filldraw[fill=cancan, draw=black] (1.25, -3.00) rectangle (1.50, -3.25);
\filldraw[fill=bermuda, draw=black] (1.50, -3.00) rectangle (1.75, -3.25);
\filldraw[fill=bermuda, draw=black] (1.75, -3.00) rectangle (2.00, -3.25);
\filldraw[fill=cancan, draw=black] (2.00, -3.00) rectangle (2.25, -3.25);
\filldraw[fill=bermuda, draw=black] (2.25, -3.00) rectangle (2.50, -3.25);
\filldraw[fill=bermuda, draw=black] (2.50, -3.00) rectangle (2.75, -3.25);
\filldraw[fill=bermuda, draw=black] (2.75, -3.00) rectangle (3.00, -3.25);
\filldraw[fill=cancan, draw=black] (3.00, -3.00) rectangle (3.25, -3.25);
\filldraw[fill=bermuda, draw=black] (3.25, -3.00) rectangle (3.50, -3.25);
\filldraw[fill=cancan, draw=black] (3.50, -3.00) rectangle (3.75, -3.25);
\filldraw[fill=bermuda, draw=black] (3.75, -3.00) rectangle (4.00, -3.25);
\filldraw[fill=cancan, draw=black] (4.00, -3.00) rectangle (4.25, -3.25);
\filldraw[fill=cancan, draw=black] (4.25, -3.00) rectangle (4.50, -3.25);
\filldraw[fill=cancan, draw=black] (4.50, -3.00) rectangle (4.75, -3.25);
\filldraw[fill=bermuda, draw=black] (4.75, -3.00) rectangle (5.00, -3.25);
\filldraw[fill=cancan, draw=black] (5.00, -3.00) rectangle (5.25, -3.25);
\filldraw[fill=bermuda, draw=black] (5.25, -3.00) rectangle (5.50, -3.25);
\filldraw[fill=cancan, draw=black] (5.50, -3.00) rectangle (5.75, -3.25);
\filldraw[fill=cancan, draw=black] (5.75, -3.00) rectangle (6.00, -3.25);
\filldraw[fill=cancan, draw=black] (6.00, -3.00) rectangle (6.25, -3.25);
\filldraw[fill=bermuda, draw=black] (6.25, -3.00) rectangle (6.50, -3.25);
\filldraw[fill=cancan, draw=black] (6.50, -3.00) rectangle (6.75, -3.25);
\filldraw[fill=bermuda, draw=black] (6.75, -3.00) rectangle (7.00, -3.25);
\filldraw[fill=cancan, draw=black] (7.00, -3.00) rectangle (7.25, -3.25);
\filldraw[fill=cancan, draw=black] (7.25, -3.00) rectangle (7.50, -3.25);
\filldraw[fill=cancan, draw=black] (7.50, -3.00) rectangle (7.75, -3.25);
\filldraw[fill=bermuda, draw=black] (7.75, -3.00) rectangle (8.00, -3.25);
\filldraw[fill=cancan, draw=black] (8.00, -3.00) rectangle (8.25, -3.25);
\filldraw[fill=bermuda, draw=black] (8.25, -3.00) rectangle (8.50, -3.25);
\filldraw[fill=cancan, draw=black] (8.50, -3.00) rectangle (8.75, -3.25);
\filldraw[fill=cancan, draw=black] (8.75, -3.00) rectangle (9.00, -3.25);
\filldraw[fill=cancan, draw=black] (9.00, -3.00) rectangle (9.25, -3.25);
\filldraw[fill=bermuda, draw=black] (9.25, -3.00) rectangle (9.50, -3.25);
\filldraw[fill=cancan, draw=black] (9.50, -3.00) rectangle (9.75, -3.25);
\filldraw[fill=bermuda, draw=black] (9.75, -3.00) rectangle (10.00, -3.25);
\filldraw[fill=cancan, draw=black] (10.00, -3.00) rectangle (10.25, -3.25);
\filldraw[fill=cancan, draw=black] (10.25, -3.00) rectangle (10.50, -3.25);
\filldraw[fill=cancan, draw=black] (10.50, -3.00) rectangle (10.75, -3.25);
\filldraw[fill=cancan, draw=black] (10.75, -3.00) rectangle (11.00, -3.25);
\filldraw[fill=cancan, draw=black] (11.00, -3.00) rectangle (11.25, -3.25);
\filldraw[fill=cancan, draw=black] (11.25, -3.00) rectangle (11.50, -3.25);
\filldraw[fill=cancan, draw=black] (11.50, -3.00) rectangle (11.75, -3.25);
\filldraw[fill=bermuda, draw=black] (11.75, -3.00) rectangle (12.00, -3.25);
\filldraw[fill=cancan, draw=black] (12.00, -3.00) rectangle (12.25, -3.25);
\filldraw[fill=bermuda, draw=black] (12.25, -3.00) rectangle (12.50, -3.25);
\filldraw[fill=bermuda, draw=black] (12.50, -3.00) rectangle (12.75, -3.25);
\filldraw[fill=bermuda, draw=black] (12.75, -3.00) rectangle (13.00, -3.25);
\filldraw[fill=cancan, draw=black] (13.00, -3.00) rectangle (13.25, -3.25);
\filldraw[fill=cancan, draw=black] (13.25, -3.00) rectangle (13.50, -3.25);
\filldraw[fill=cancan, draw=black] (13.50, -3.00) rectangle (13.75, -3.25);
\filldraw[fill=bermuda, draw=black] (13.75, -3.00) rectangle (14.00, -3.25);
\filldraw[fill=bermuda, draw=black] (14.00, -3.00) rectangle (14.25, -3.25);
\filldraw[fill=bermuda, draw=black] (14.25, -3.00) rectangle (14.50, -3.25);
\filldraw[fill=cancan, draw=black] (14.50, -3.00) rectangle (14.75, -3.25);
\filldraw[fill=cancan, draw=black] (14.75, -3.00) rectangle (15.00, -3.25);
\filldraw[fill=cancan, draw=black] (0.00, -3.25) rectangle (0.25, -3.50);
\filldraw[fill=bermuda, draw=black] (0.25, -3.25) rectangle (0.50, -3.50);
\filldraw[fill=bermuda, draw=black] (0.50, -3.25) rectangle (0.75, -3.50);
\filldraw[fill=bermuda, draw=black] (0.75, -3.25) rectangle (1.00, -3.50);
\filldraw[fill=cancan, draw=black] (1.00, -3.25) rectangle (1.25, -3.50);
\filldraw[fill=cancan, draw=black] (1.25, -3.25) rectangle (1.50, -3.50);
\filldraw[fill=cancan, draw=black] (1.50, -3.25) rectangle (1.75, -3.50);
\filldraw[fill=cancan, draw=black] (1.75, -3.25) rectangle (2.00, -3.50);
\filldraw[fill=cancan, draw=black] (2.00, -3.25) rectangle (2.25, -3.50);
\filldraw[fill=bermuda, draw=black] (2.25, -3.25) rectangle (2.50, -3.50);
\filldraw[fill=bermuda, draw=black] (2.50, -3.25) rectangle (2.75, -3.50);
\filldraw[fill=bermuda, draw=black] (2.75, -3.25) rectangle (3.00, -3.50);
\filldraw[fill=cancan, draw=black] (3.00, -3.25) rectangle (3.25, -3.50);
\filldraw[fill=cancan, draw=black] (3.25, -3.25) rectangle (3.50, -3.50);
\filldraw[fill=cancan, draw=black] (3.50, -3.25) rectangle (3.75, -3.50);
\filldraw[fill=bermuda, draw=black] (3.75, -3.25) rectangle (4.00, -3.50);
\filldraw[fill=bermuda, draw=black] (4.00, -3.25) rectangle (4.25, -3.50);
\filldraw[fill=bermuda, draw=black] (4.25, -3.25) rectangle (4.50, -3.50);
\filldraw[fill=cancan, draw=black] (4.50, -3.25) rectangle (4.75, -3.50);
\filldraw[fill=cancan, draw=black] (4.75, -3.25) rectangle (5.00, -3.50);
\filldraw[fill=cancan, draw=black] (5.00, -3.25) rectangle (5.25, -3.50);
\filldraw[fill=bermuda, draw=black] (5.25, -3.25) rectangle (5.50, -3.50);
\filldraw[fill=bermuda, draw=black] (5.50, -3.25) rectangle (5.75, -3.50);
\filldraw[fill=bermuda, draw=black] (5.75, -3.25) rectangle (6.00, -3.50);
\filldraw[fill=cancan, draw=black] (6.00, -3.25) rectangle (6.25, -3.50);
\filldraw[fill=cancan, draw=black] (6.25, -3.25) rectangle (6.50, -3.50);
\filldraw[fill=cancan, draw=black] (6.50, -3.25) rectangle (6.75, -3.50);
\filldraw[fill=bermuda, draw=black] (6.75, -3.25) rectangle (7.00, -3.50);
\filldraw[fill=bermuda, draw=black] (7.00, -3.25) rectangle (7.25, -3.50);
\filldraw[fill=bermuda, draw=black] (7.25, -3.25) rectangle (7.50, -3.50);
\filldraw[fill=cancan, draw=black] (7.50, -3.25) rectangle (7.75, -3.50);
} } }\end{equation*}
\begin{equation*}
\hspace{0.3pt} b_{14} = \vcenter{\hbox{ \tikz{
\filldraw[fill=cancan, draw=black] (0.00, 0.00) rectangle (0.25, -0.25);
\filldraw[fill=cancan, draw=black] (0.25, 0.00) rectangle (0.50, -0.25);
\filldraw[fill=bermuda, draw=black] (0.50, 0.00) rectangle (0.75, -0.25);
\filldraw[fill=bermuda, draw=black] (0.75, 0.00) rectangle (1.00, -0.25);
\filldraw[fill=bermuda, draw=black] (1.00, 0.00) rectangle (1.25, -0.25);
\filldraw[fill=cancan, draw=black] (1.25, 0.00) rectangle (1.50, -0.25);
\filldraw[fill=cancan, draw=black] (1.50, 0.00) rectangle (1.75, -0.25);
\filldraw[fill=cancan, draw=black] (1.75, 0.00) rectangle (2.00, -0.25);
\filldraw[fill=cancan, draw=black] (2.00, 0.00) rectangle (2.25, -0.25);
\filldraw[fill=bermuda, draw=black] (2.25, 0.00) rectangle (2.50, -0.25);
\filldraw[fill=bermuda, draw=black] (2.50, 0.00) rectangle (2.75, -0.25);
\filldraw[fill=cancan, draw=black] (2.75, 0.00) rectangle (3.00, -0.25);
\filldraw[fill=cancan, draw=black] (3.00, 0.00) rectangle (3.25, -0.25);
\filldraw[fill=cancan, draw=black] (3.25, 0.00) rectangle (3.50, -0.25);
\filldraw[fill=bermuda, draw=black] (3.50, 0.00) rectangle (3.75, -0.25);
\filldraw[fill=cancan, draw=black] (3.75, 0.00) rectangle (4.00, -0.25);
\filldraw[fill=bermuda, draw=black] (4.00, 0.00) rectangle (4.25, -0.25);
\filldraw[fill=cancan, draw=black] (4.25, 0.00) rectangle (4.50, -0.25);
\filldraw[fill=cancan, draw=black] (4.50, 0.00) rectangle (4.75, -0.25);
\filldraw[fill=bermuda, draw=black] (4.75, 0.00) rectangle (5.00, -0.25);
\filldraw[fill=bermuda, draw=black] (5.00, 0.00) rectangle (5.25, -0.25);
\filldraw[fill=bermuda, draw=black] (5.25, 0.00) rectangle (5.50, -0.25);
\filldraw[fill=bermuda, draw=black] (5.50, 0.00) rectangle (5.75, -0.25);
\filldraw[fill=bermuda, draw=black] (5.75, 0.00) rectangle (6.00, -0.25);
\filldraw[fill=bermuda, draw=black] (6.00, 0.00) rectangle (6.25, -0.25);
\filldraw[fill=cancan, draw=black] (6.25, 0.00) rectangle (6.50, -0.25);
\filldraw[fill=cancan, draw=black] (6.50, 0.00) rectangle (6.75, -0.25);
\filldraw[fill=cancan, draw=black] (6.75, 0.00) rectangle (7.00, -0.25);
\filldraw[fill=bermuda, draw=black] (7.00, 0.00) rectangle (7.25, -0.25);
\filldraw[fill=bermuda, draw=black] (7.25, 0.00) rectangle (7.50, -0.25);
\filldraw[fill=bermuda, draw=black] (7.50, 0.00) rectangle (7.75, -0.25);
\filldraw[fill=cancan, draw=black] (7.75, 0.00) rectangle (8.00, -0.25);
\filldraw[fill=cancan, draw=black] (8.00, 0.00) rectangle (8.25, -0.25);
\filldraw[fill=cancan, draw=black] (8.25, 0.00) rectangle (8.50, -0.25);
\filldraw[fill=bermuda, draw=black] (8.50, 0.00) rectangle (8.75, -0.25);
\filldraw[fill=bermuda, draw=black] (8.75, 0.00) rectangle (9.00, -0.25);
\filldraw[fill=bermuda, draw=black] (9.00, 0.00) rectangle (9.25, -0.25);
\filldraw[fill=cancan, draw=black] (9.25, 0.00) rectangle (9.50, -0.25);
\filldraw[fill=cancan, draw=black] (9.50, 0.00) rectangle (9.75, -0.25);
\filldraw[fill=cancan, draw=black] (9.75, 0.00) rectangle (10.00, -0.25);
\filldraw[fill=bermuda, draw=black] (10.00, 0.00) rectangle (10.25, -0.25);
\filldraw[fill=bermuda, draw=black] (10.25, 0.00) rectangle (10.50, -0.25);
\filldraw[fill=bermuda, draw=black] (10.50, 0.00) rectangle (10.75, -0.25);
\filldraw[fill=cancan, draw=black] (10.75, 0.00) rectangle (11.00, -0.25);
\filldraw[fill=cancan, draw=black] (11.00, 0.00) rectangle (11.25, -0.25);
\filldraw[fill=cancan, draw=black] (11.25, 0.00) rectangle (11.50, -0.25);
\filldraw[fill=bermuda, draw=black] (11.50, 0.00) rectangle (11.75, -0.25);
\filldraw[fill=bermuda, draw=black] (11.75, 0.00) rectangle (12.00, -0.25);
\filldraw[fill=bermuda, draw=black] (12.00, 0.00) rectangle (12.25, -0.25);
\filldraw[fill=cancan, draw=black] (12.25, 0.00) rectangle (12.50, -0.25);
\filldraw[fill=bermuda, draw=black] (12.50, 0.00) rectangle (12.75, -0.25);
\filldraw[fill=bermuda, draw=black] (12.75, 0.00) rectangle (13.00, -0.25);
\filldraw[fill=bermuda, draw=black] (13.00, 0.00) rectangle (13.25, -0.25);
\filldraw[fill=bermuda, draw=black] (13.25, 0.00) rectangle (13.50, -0.25);
\filldraw[fill=bermuda, draw=black] (13.50, 0.00) rectangle (13.75, -0.25);
\filldraw[fill=cancan, draw=black] (13.75, 0.00) rectangle (14.00, -0.25);
\filldraw[fill=bermuda, draw=black] (14.00, 0.00) rectangle (14.25, -0.25);
\filldraw[fill=bermuda, draw=black] (14.25, 0.00) rectangle (14.50, -0.25);
\filldraw[fill=bermuda, draw=black] (14.50, 0.00) rectangle (14.75, -0.25);
\filldraw[fill=bermuda, draw=black] (14.75, 0.00) rectangle (15.00, -0.25);
\filldraw[fill=bermuda, draw=black] (0.00, -0.25) rectangle (0.25, -0.50);
\filldraw[fill=cancan, draw=black] (0.25, -0.25) rectangle (0.50, -0.50);
\filldraw[fill=cancan, draw=black] (0.50, -0.25) rectangle (0.75, -0.50);
\filldraw[fill=cancan, draw=black] (0.75, -0.25) rectangle (1.00, -0.50);
\filldraw[fill=cancan, draw=black] (1.00, -0.25) rectangle (1.25, -0.50);
\filldraw[fill=cancan, draw=black] (1.25, -0.25) rectangle (1.50, -0.50);
\filldraw[fill=bermuda, draw=black] (1.50, -0.25) rectangle (1.75, -0.50);
\filldraw[fill=bermuda, draw=black] (1.75, -0.25) rectangle (2.00, -0.50);
\filldraw[fill=bermuda, draw=black] (2.00, -0.25) rectangle (2.25, -0.50);
\filldraw[fill=cancan, draw=black] (2.25, -0.25) rectangle (2.50, -0.50);
\filldraw[fill=cancan, draw=black] (2.50, -0.25) rectangle (2.75, -0.50);
\filldraw[fill=bermuda, draw=black] (2.75, -0.25) rectangle (3.00, -0.50);
\filldraw[fill=bermuda, draw=black] (3.00, -0.25) rectangle (3.25, -0.50);
\filldraw[fill=cancan, draw=black] (3.25, -0.25) rectangle (3.50, -0.50);
\filldraw[fill=cancan, draw=black] (3.50, -0.25) rectangle (3.75, -0.50);
\filldraw[fill=cancan, draw=black] (3.75, -0.25) rectangle (4.00, -0.50);
\filldraw[fill=bermuda, draw=black] (4.00, -0.25) rectangle (4.25, -0.50);
\filldraw[fill=bermuda, draw=black] (4.25, -0.25) rectangle (4.50, -0.50);
\filldraw[fill=bermuda, draw=black] (4.50, -0.25) rectangle (4.75, -0.50);
\filldraw[fill=cancan, draw=black] (4.75, -0.25) rectangle (5.00, -0.50);
\filldraw[fill=cancan, draw=black] (5.00, -0.25) rectangle (5.25, -0.50);
\filldraw[fill=cancan, draw=black] (5.25, -0.25) rectangle (5.50, -0.50);
\filldraw[fill=bermuda, draw=black] (5.50, -0.25) rectangle (5.75, -0.50);
\filldraw[fill=bermuda, draw=black] (5.75, -0.25) rectangle (6.00, -0.50);
\filldraw[fill=bermuda, draw=black] (6.00, -0.25) rectangle (6.25, -0.50);
\filldraw[fill=cancan, draw=black] (6.25, -0.25) rectangle (6.50, -0.50);
\filldraw[fill=cancan, draw=black] (6.50, -0.25) rectangle (6.75, -0.50);
\filldraw[fill=cancan, draw=black] (6.75, -0.25) rectangle (7.00, -0.50);
\filldraw[fill=bermuda, draw=black] (7.00, -0.25) rectangle (7.25, -0.50);
\filldraw[fill=bermuda, draw=black] (7.25, -0.25) rectangle (7.50, -0.50);
\filldraw[fill=bermuda, draw=black] (7.50, -0.25) rectangle (7.75, -0.50);
\filldraw[fill=cancan, draw=black] (7.75, -0.25) rectangle (8.00, -0.50);
\filldraw[fill=cancan, draw=black] (8.00, -0.25) rectangle (8.25, -0.50);
\filldraw[fill=cancan, draw=black] (8.25, -0.25) rectangle (8.50, -0.50);
\filldraw[fill=cancan, draw=black] (8.50, -0.25) rectangle (8.75, -0.50);
\filldraw[fill=bermuda, draw=black] (8.75, -0.25) rectangle (9.00, -0.50);
\filldraw[fill=bermuda, draw=black] (9.00, -0.25) rectangle (9.25, -0.50);
\filldraw[fill=bermuda, draw=black] (9.25, -0.25) rectangle (9.50, -0.50);
\filldraw[fill=bermuda, draw=black] (9.50, -0.25) rectangle (9.75, -0.50);
\filldraw[fill=cancan, draw=black] (9.75, -0.25) rectangle (10.00, -0.50);
\filldraw[fill=bermuda, draw=black] (10.00, -0.25) rectangle (10.25, -0.50);
\filldraw[fill=bermuda, draw=black] (10.25, -0.25) rectangle (10.50, -0.50);
\filldraw[fill=bermuda, draw=black] (10.50, -0.25) rectangle (10.75, -0.50);
\filldraw[fill=cancan, draw=black] (10.75, -0.25) rectangle (11.00, -0.50);
\filldraw[fill=cancan, draw=black] (11.00, -0.25) rectangle (11.25, -0.50);
\filldraw[fill=bermuda, draw=black] (11.25, -0.25) rectangle (11.50, -0.50);
\filldraw[fill=bermuda, draw=black] (11.50, -0.25) rectangle (11.75, -0.50);
\filldraw[fill=cancan, draw=black] (11.75, -0.25) rectangle (12.00, -0.50);
\filldraw[fill=bermuda, draw=black] (12.00, -0.25) rectangle (12.25, -0.50);
\filldraw[fill=bermuda, draw=black] (12.25, -0.25) rectangle (12.50, -0.50);
\filldraw[fill=bermuda, draw=black] (12.50, -0.25) rectangle (12.75, -0.50);
\filldraw[fill=bermuda, draw=black] (12.75, -0.25) rectangle (13.00, -0.50);
\filldraw[fill=bermuda, draw=black] (13.00, -0.25) rectangle (13.25, -0.50);
\filldraw[fill=cancan, draw=black] (13.25, -0.25) rectangle (13.50, -0.50);
\filldraw[fill=bermuda, draw=black] (13.50, -0.25) rectangle (13.75, -0.50);
\filldraw[fill=cancan, draw=black] (13.75, -0.25) rectangle (14.00, -0.50);
\filldraw[fill=bermuda, draw=black] (14.00, -0.25) rectangle (14.25, -0.50);
\filldraw[fill=bermuda, draw=black] (14.25, -0.25) rectangle (14.50, -0.50);
\filldraw[fill=bermuda, draw=black] (14.50, -0.25) rectangle (14.75, -0.50);
\filldraw[fill=cancan, draw=black] (14.75, -0.25) rectangle (15.00, -0.50);
\filldraw[fill=cancan, draw=black] (0.00, -0.50) rectangle (0.25, -0.75);
\filldraw[fill=cancan, draw=black] (0.25, -0.50) rectangle (0.50, -0.75);
\filldraw[fill=bermuda, draw=black] (0.50, -0.50) rectangle (0.75, -0.75);
\filldraw[fill=bermuda, draw=black] (0.75, -0.50) rectangle (1.00, -0.75);
\filldraw[fill=bermuda, draw=black] (1.00, -0.50) rectangle (1.25, -0.75);
\filldraw[fill=cancan, draw=black] (1.25, -0.50) rectangle (1.50, -0.75);
\filldraw[fill=cancan, draw=black] (1.50, -0.50) rectangle (1.75, -0.75);
\filldraw[fill=cancan, draw=black] (1.75, -0.50) rectangle (2.00, -0.75);
\filldraw[fill=bermuda, draw=black] (2.00, -0.50) rectangle (2.25, -0.75);
\filldraw[fill=bermuda, draw=black] (2.25, -0.50) rectangle (2.50, -0.75);
\filldraw[fill=bermuda, draw=black] (2.50, -0.50) rectangle (2.75, -0.75);
\filldraw[fill=cancan, draw=black] (2.75, -0.50) rectangle (3.00, -0.75);
\filldraw[fill=cancan, draw=black] (3.00, -0.50) rectangle (3.25, -0.75);
\filldraw[fill=cancan, draw=black] (3.25, -0.50) rectangle (3.50, -0.75);
\filldraw[fill=bermuda, draw=black] (3.50, -0.50) rectangle (3.75, -0.75);
\filldraw[fill=bermuda, draw=black] (3.75, -0.50) rectangle (4.00, -0.75);
\filldraw[fill=bermuda, draw=black] (4.00, -0.50) rectangle (4.25, -0.75);
\filldraw[fill=bermuda, draw=black] (4.25, -0.50) rectangle (4.50, -0.75);
\filldraw[fill=bermuda, draw=black] (4.50, -0.50) rectangle (4.75, -0.75);
\filldraw[fill=cancan, draw=black] (4.75, -0.50) rectangle (5.00, -0.75);
\filldraw[fill=bermuda, draw=black] (5.00, -0.50) rectangle (5.25, -0.75);
\filldraw[fill=cancan, draw=black] (5.25, -0.50) rectangle (5.50, -0.75);
\filldraw[fill=cancan, draw=black] (5.50, -0.50) rectangle (5.75, -0.75);
\filldraw[fill=bermuda, draw=black] (5.75, -0.50) rectangle (6.00, -0.75);
\filldraw[fill=bermuda, draw=black] (6.00, -0.50) rectangle (6.25, -0.75);
\filldraw[fill=cancan, draw=black] (6.25, -0.50) rectangle (6.50, -0.75);
\filldraw[fill=bermuda, draw=black] (6.50, -0.50) rectangle (6.75, -0.75);
\filldraw[fill=cancan, draw=black] (6.75, -0.50) rectangle (7.00, -0.75);
\filldraw[fill=cancan, draw=black] (7.00, -0.50) rectangle (7.25, -0.75);
\filldraw[fill=bermuda, draw=black] (7.25, -0.50) rectangle (7.50, -0.75);
\filldraw[fill=bermuda, draw=black] (7.50, -0.50) rectangle (7.75, -0.75);
\filldraw[fill=cancan, draw=black] (7.75, -0.50) rectangle (8.00, -0.75);
\filldraw[fill=bermuda, draw=black] (8.00, -0.50) rectangle (8.25, -0.75);
\filldraw[fill=cancan, draw=black] (8.25, -0.50) rectangle (8.50, -0.75);
\filldraw[fill=cancan, draw=black] (8.50, -0.50) rectangle (8.75, -0.75);
\filldraw[fill=bermuda, draw=black] (8.75, -0.50) rectangle (9.00, -0.75);
\filldraw[fill=bermuda, draw=black] (9.00, -0.50) rectangle (9.25, -0.75);
\filldraw[fill=cancan, draw=black] (9.25, -0.50) rectangle (9.50, -0.75);
\filldraw[fill=bermuda, draw=black] (9.50, -0.50) rectangle (9.75, -0.75);
\filldraw[fill=cancan, draw=black] (9.75, -0.50) rectangle (10.00, -0.75);
\filldraw[fill=cancan, draw=black] (10.00, -0.50) rectangle (10.25, -0.75);
\filldraw[fill=bermuda, draw=black] (10.25, -0.50) rectangle (10.50, -0.75);
\filldraw[fill=bermuda, draw=black] (10.50, -0.50) rectangle (10.75, -0.75);
\filldraw[fill=cancan, draw=black] (10.75, -0.50) rectangle (11.00, -0.75);
\filldraw[fill=bermuda, draw=black] (11.00, -0.50) rectangle (11.25, -0.75);
\filldraw[fill=cancan, draw=black] (11.25, -0.50) rectangle (11.50, -0.75);
\filldraw[fill=cancan, draw=black] (11.50, -0.50) rectangle (11.75, -0.75);
\filldraw[fill=cancan, draw=black] (11.75, -0.50) rectangle (12.00, -0.75);
\filldraw[fill=cancan, draw=black] (12.00, -0.50) rectangle (12.25, -0.75);
\filldraw[fill=cancan, draw=black] (12.25, -0.50) rectangle (12.50, -0.75);
\filldraw[fill=bermuda, draw=black] (12.50, -0.50) rectangle (12.75, -0.75);
\filldraw[fill=cancan, draw=black] (12.75, -0.50) rectangle (13.00, -0.75);
\filldraw[fill=bermuda, draw=black] (13.00, -0.50) rectangle (13.25, -0.75);
\filldraw[fill=cancan, draw=black] (13.25, -0.50) rectangle (13.50, -0.75);
\filldraw[fill=cancan, draw=black] (13.50, -0.50) rectangle (13.75, -0.75);
\filldraw[fill=cancan, draw=black] (13.75, -0.50) rectangle (14.00, -0.75);
\filldraw[fill=bermuda, draw=black] (14.00, -0.50) rectangle (14.25, -0.75);
\filldraw[fill=bermuda, draw=black] (14.25, -0.50) rectangle (14.50, -0.75);
\filldraw[fill=bermuda, draw=black] (14.50, -0.50) rectangle (14.75, -0.75);
\filldraw[fill=cancan, draw=black] (14.75, -0.50) rectangle (15.00, -0.75);
\filldraw[fill=cancan, draw=black] (0.00, -0.75) rectangle (0.25, -1.00);
\filldraw[fill=cancan, draw=black] (0.25, -0.75) rectangle (0.50, -1.00);
\filldraw[fill=cancan, draw=black] (0.50, -0.75) rectangle (0.75, -1.00);
\filldraw[fill=cancan, draw=black] (0.75, -0.75) rectangle (1.00, -1.00);
\filldraw[fill=cancan, draw=black] (1.00, -0.75) rectangle (1.25, -1.00);
\filldraw[fill=cancan, draw=black] (1.25, -0.75) rectangle (1.50, -1.00);
\filldraw[fill=bermuda, draw=black] (1.50, -0.75) rectangle (1.75, -1.00);
\filldraw[fill=cancan, draw=black] (1.75, -0.75) rectangle (2.00, -1.00);
\filldraw[fill=cancan, draw=black] (2.00, -0.75) rectangle (2.25, -1.00);
\filldraw[fill=cancan, draw=black] (2.25, -0.75) rectangle (2.50, -1.00);
\filldraw[fill=cancan, draw=black] (2.50, -0.75) rectangle (2.75, -1.00);
\filldraw[fill=cancan, draw=black] (2.75, -0.75) rectangle (3.00, -1.00);
\filldraw[fill=cancan, draw=black] (3.00, -0.75) rectangle (3.25, -1.00);
\filldraw[fill=cancan, draw=black] (3.25, -0.75) rectangle (3.50, -1.00);
\filldraw[fill=bermuda, draw=black] (3.50, -0.75) rectangle (3.75, -1.00);
\filldraw[fill=bermuda, draw=black] (3.75, -0.75) rectangle (4.00, -1.00);
\filldraw[fill=bermuda, draw=black] (4.00, -0.75) rectangle (4.25, -1.00);
\filldraw[fill=cancan, draw=black] (4.25, -0.75) rectangle (4.50, -1.00);
\filldraw[fill=cancan, draw=black] (4.50, -0.75) rectangle (4.75, -1.00);
\filldraw[fill=cancan, draw=black] (4.75, -0.75) rectangle (5.00, -1.00);
\filldraw[fill=bermuda, draw=black] (5.00, -0.75) rectangle (5.25, -1.00);
\filldraw[fill=bermuda, draw=black] (5.25, -0.75) rectangle (5.50, -1.00);
\filldraw[fill=bermuda, draw=black] (5.50, -0.75) rectangle (5.75, -1.00);
\filldraw[fill=cancan, draw=black] (5.75, -0.75) rectangle (6.00, -1.00);
\filldraw[fill=cancan, draw=black] (6.00, -0.75) rectangle (6.25, -1.00);
\filldraw[fill=cancan, draw=black] (6.25, -0.75) rectangle (6.50, -1.00);
\filldraw[fill=bermuda, draw=black] (6.50, -0.75) rectangle (6.75, -1.00);
\filldraw[fill=bermuda, draw=black] (6.75, -0.75) rectangle (7.00, -1.00);
\filldraw[fill=bermuda, draw=black] (7.00, -0.75) rectangle (7.25, -1.00);
\filldraw[fill=bermuda, draw=black] (7.25, -0.75) rectangle (7.50, -1.00);
\filldraw[fill=bermuda, draw=black] (7.50, -0.75) rectangle (7.75, -1.00);
\filldraw[fill=cancan, draw=black] (7.75, -0.75) rectangle (8.00, -1.00);
\filldraw[fill=bermuda, draw=black] (8.00, -0.75) rectangle (8.25, -1.00);
\filldraw[fill=cancan, draw=black] (8.25, -0.75) rectangle (8.50, -1.00);
\filldraw[fill=bermuda, draw=black] (8.50, -0.75) rectangle (8.75, -1.00);
\filldraw[fill=cancan, draw=black] (8.75, -0.75) rectangle (9.00, -1.00);
\filldraw[fill=bermuda, draw=black] (9.00, -0.75) rectangle (9.25, -1.00);
\filldraw[fill=cancan, draw=black] (9.25, -0.75) rectangle (9.50, -1.00);
\filldraw[fill=bermuda, draw=black] (9.50, -0.75) rectangle (9.75, -1.00);
\filldraw[fill=bermuda, draw=black] (9.75, -0.75) rectangle (10.00, -1.00);
\filldraw[fill=bermuda, draw=black] (10.00, -0.75) rectangle (10.25, -1.00);
\filldraw[fill=cancan, draw=black] (10.25, -0.75) rectangle (10.50, -1.00);
\filldraw[fill=cancan, draw=black] (10.50, -0.75) rectangle (10.75, -1.00);
\filldraw[fill=cancan, draw=black] (10.75, -0.75) rectangle (11.00, -1.00);
\filldraw[fill=bermuda, draw=black] (11.00, -0.75) rectangle (11.25, -1.00);
\filldraw[fill=bermuda, draw=black] (11.25, -0.75) rectangle (11.50, -1.00);
\filldraw[fill=bermuda, draw=black] (11.50, -0.75) rectangle (11.75, -1.00);
\filldraw[fill=cancan, draw=black] (11.75, -0.75) rectangle (12.00, -1.00);
\filldraw[fill=bermuda, draw=black] (12.00, -0.75) rectangle (12.25, -1.00);
\filldraw[fill=bermuda, draw=black] (12.25, -0.75) rectangle (12.50, -1.00);
\filldraw[fill=bermuda, draw=black] (12.50, -0.75) rectangle (12.75, -1.00);
\filldraw[fill=bermuda, draw=black] (12.75, -0.75) rectangle (13.00, -1.00);
\filldraw[fill=bermuda, draw=black] (13.00, -0.75) rectangle (13.25, -1.00);
\filldraw[fill=cancan, draw=black] (13.25, -0.75) rectangle (13.50, -1.00);
\filldraw[fill=bermuda, draw=black] (13.50, -0.75) rectangle (13.75, -1.00);
\filldraw[fill=bermuda, draw=black] (13.75, -0.75) rectangle (14.00, -1.00);
\filldraw[fill=bermuda, draw=black] (14.00, -0.75) rectangle (14.25, -1.00);
\filldraw[fill=bermuda, draw=black] (14.25, -0.75) rectangle (14.50, -1.00);
\filldraw[fill=bermuda, draw=black] (14.50, -0.75) rectangle (14.75, -1.00);
\filldraw[fill=cancan, draw=black] (14.75, -0.75) rectangle (15.00, -1.00);
\filldraw[fill=bermuda, draw=black] (0.00, -1.00) rectangle (0.25, -1.25);
\filldraw[fill=bermuda, draw=black] (0.25, -1.00) rectangle (0.50, -1.25);
\filldraw[fill=bermuda, draw=black] (0.50, -1.00) rectangle (0.75, -1.25);
\filldraw[fill=bermuda, draw=black] (0.75, -1.00) rectangle (1.00, -1.25);
\filldraw[fill=bermuda, draw=black] (1.00, -1.00) rectangle (1.25, -1.25);
\filldraw[fill=cancan, draw=black] (1.25, -1.00) rectangle (1.50, -1.25);
\filldraw[fill=bermuda, draw=black] (1.50, -1.00) rectangle (1.75, -1.25);
\filldraw[fill=bermuda, draw=black] (1.75, -1.00) rectangle (2.00, -1.25);
\filldraw[fill=bermuda, draw=black] (2.00, -1.00) rectangle (2.25, -1.25);
\filldraw[fill=bermuda, draw=black] (2.25, -1.00) rectangle (2.50, -1.25);
\filldraw[fill=bermuda, draw=black] (2.50, -1.00) rectangle (2.75, -1.25);
\filldraw[fill=cancan, draw=black] (2.75, -1.00) rectangle (3.00, -1.25);
\filldraw[fill=cancan, draw=black] (3.00, -1.00) rectangle (3.25, -1.25);
\filldraw[fill=cancan, draw=black] (3.25, -1.00) rectangle (3.50, -1.25);
\filldraw[fill=cancan, draw=black] (3.50, -1.00) rectangle (3.75, -1.25);
\filldraw[fill=cancan, draw=black] (3.75, -1.00) rectangle (4.00, -1.25);
\filldraw[fill=cancan, draw=black] (4.00, -1.00) rectangle (4.25, -1.25);
\filldraw[fill=cancan, draw=black] (4.25, -1.00) rectangle (4.50, -1.25);
\filldraw[fill=bermuda, draw=black] (4.50, -1.00) rectangle (4.75, -1.25);
\filldraw[fill=bermuda, draw=black] (4.75, -1.00) rectangle (5.00, -1.25);
\filldraw[fill=bermuda, draw=black] (5.00, -1.00) rectangle (5.25, -1.25);
\filldraw[fill=cancan, draw=black] (5.25, -1.00) rectangle (5.50, -1.25);
\filldraw[fill=cancan, draw=black] (5.50, -1.00) rectangle (5.75, -1.25);
\filldraw[fill=cancan, draw=black] (5.75, -1.00) rectangle (6.00, -1.25);
\filldraw[fill=bermuda, draw=black] (6.00, -1.00) rectangle (6.25, -1.25);
\filldraw[fill=bermuda, draw=black] (6.25, -1.00) rectangle (6.50, -1.25);
\filldraw[fill=bermuda, draw=black] (6.50, -1.00) rectangle (6.75, -1.25);
\filldraw[fill=cancan, draw=black] (6.75, -1.00) rectangle (7.00, -1.25);
\filldraw[fill=bermuda, draw=black] (7.00, -1.00) rectangle (7.25, -1.25);
\filldraw[fill=cancan, draw=black] (7.25, -1.00) rectangle (7.50, -1.25);
\filldraw[fill=bermuda, draw=black] (7.50, -1.00) rectangle (7.75, -1.25);
\filldraw[fill=cancan, draw=black] (7.75, -1.00) rectangle (8.00, -1.25);
\filldraw[fill=cancan, draw=black] (8.00, -1.00) rectangle (8.25, -1.25);
\filldraw[fill=cancan, draw=black] (8.25, -1.00) rectangle (8.50, -1.25);
\filldraw[fill=bermuda, draw=black] (8.50, -1.00) rectangle (8.75, -1.25);
\filldraw[fill=bermuda, draw=black] (8.75, -1.00) rectangle (9.00, -1.25);
\filldraw[fill=bermuda, draw=black] (9.00, -1.00) rectangle (9.25, -1.25);
\filldraw[fill=cancan, draw=black] (9.25, -1.00) rectangle (9.50, -1.25);
\filldraw[fill=cancan, draw=black] (9.50, -1.00) rectangle (9.75, -1.25);
\filldraw[fill=cancan, draw=black] (9.75, -1.00) rectangle (10.00, -1.25);
\filldraw[fill=cancan, draw=black] (10.00, -1.00) rectangle (10.25, -1.25);
\filldraw[fill=bermuda, draw=black] (10.25, -1.00) rectangle (10.50, -1.25);
\filldraw[fill=bermuda, draw=black] (10.50, -1.00) rectangle (10.75, -1.25);
\filldraw[fill=cancan, draw=black] (10.75, -1.00) rectangle (11.00, -1.25);
\filldraw[fill=cancan, draw=black] (11.00, -1.00) rectangle (11.25, -1.25);
\filldraw[fill=cancan, draw=black] (11.25, -1.00) rectangle (11.50, -1.25);
\filldraw[fill=cancan, draw=black] (11.50, -1.00) rectangle (11.75, -1.25);
\filldraw[fill=bermuda, draw=black] (11.75, -1.00) rectangle (12.00, -1.25);
\filldraw[fill=bermuda, draw=black] (12.00, -1.00) rectangle (12.25, -1.25);
\filldraw[fill=cancan, draw=black] (12.25, -1.00) rectangle (12.50, -1.25);
\filldraw[fill=cancan, draw=black] (12.50, -1.00) rectangle (12.75, -1.25);
\filldraw[fill=cancan, draw=black] (12.75, -1.00) rectangle (13.00, -1.25);
\filldraw[fill=cancan, draw=black] (13.00, -1.00) rectangle (13.25, -1.25);
\filldraw[fill=bermuda, draw=black] (13.25, -1.00) rectangle (13.50, -1.25);
\filldraw[fill=bermuda, draw=black] (13.50, -1.00) rectangle (13.75, -1.25);
\filldraw[fill=cancan, draw=black] (13.75, -1.00) rectangle (14.00, -1.25);
\filldraw[fill=cancan, draw=black] (14.00, -1.00) rectangle (14.25, -1.25);
\filldraw[fill=cancan, draw=black] (14.25, -1.00) rectangle (14.50, -1.25);
\filldraw[fill=cancan, draw=black] (14.50, -1.00) rectangle (14.75, -1.25);
\filldraw[fill=cancan, draw=black] (14.75, -1.00) rectangle (15.00, -1.25);
\filldraw[fill=cancan, draw=black] (0.00, -1.25) rectangle (0.25, -1.50);
\filldraw[fill=cancan, draw=black] (0.25, -1.25) rectangle (0.50, -1.50);
\filldraw[fill=cancan, draw=black] (0.50, -1.25) rectangle (0.75, -1.50);
\filldraw[fill=cancan, draw=black] (0.75, -1.25) rectangle (1.00, -1.50);
\filldraw[fill=cancan, draw=black] (1.00, -1.25) rectangle (1.25, -1.50);
\filldraw[fill=cancan, draw=black] (1.25, -1.25) rectangle (1.50, -1.50);
\filldraw[fill=cancan, draw=black] (1.50, -1.25) rectangle (1.75, -1.50);
\filldraw[fill=bermuda, draw=black] (1.75, -1.25) rectangle (2.00, -1.50);
\filldraw[fill=bermuda, draw=black] (2.00, -1.25) rectangle (2.25, -1.50);
\filldraw[fill=cancan, draw=black] (2.25, -1.25) rectangle (2.50, -1.50);
\filldraw[fill=bermuda, draw=black] (2.50, -1.25) rectangle (2.75, -1.50);
\filldraw[fill=cancan, draw=black] (2.75, -1.25) rectangle (3.00, -1.50);
\filldraw[fill=cancan, draw=black] (3.00, -1.25) rectangle (3.25, -1.50);
\filldraw[fill=bermuda, draw=black] (3.25, -1.25) rectangle (3.50, -1.50);
\filldraw[fill=bermuda, draw=black] (3.50, -1.25) rectangle (3.75, -1.50);
\filldraw[fill=cancan, draw=black] (3.75, -1.25) rectangle (4.00, -1.50);
\filldraw[fill=bermuda, draw=black] (4.00, -1.25) rectangle (4.25, -1.50);
\filldraw[fill=cancan, draw=black] (4.25, -1.25) rectangle (4.50, -1.50);
\filldraw[fill=bermuda, draw=black] (4.50, -1.25) rectangle (4.75, -1.50);
\filldraw[fill=bermuda, draw=black] (4.75, -1.25) rectangle (5.00, -1.50);
\filldraw[fill=bermuda, draw=black] (5.00, -1.25) rectangle (5.25, -1.50);
\filldraw[fill=cancan, draw=black] (5.25, -1.25) rectangle (5.50, -1.50);
\filldraw[fill=bermuda, draw=black] (5.50, -1.25) rectangle (5.75, -1.50);
\filldraw[fill=bermuda, draw=black] (5.75, -1.25) rectangle (6.00, -1.50);
\filldraw[fill=bermuda, draw=black] (6.00, -1.25) rectangle (6.25, -1.50);
\filldraw[fill=bermuda, draw=black] (6.25, -1.25) rectangle (6.50, -1.50);
\filldraw[fill=bermuda, draw=black] (6.50, -1.25) rectangle (6.75, -1.50);
\filldraw[fill=cancan, draw=black] (6.75, -1.25) rectangle (7.00, -1.50);
\filldraw[fill=bermuda, draw=black] (7.00, -1.25) rectangle (7.25, -1.50);
\filldraw[fill=cancan, draw=black] (7.25, -1.25) rectangle (7.50, -1.50);
\filldraw[fill=cancan, draw=black] (7.50, -1.25) rectangle (7.75, -1.50);
\filldraw[fill=bermuda, draw=black] (7.75, -1.25) rectangle (8.00, -1.50);
\filldraw[fill=bermuda, draw=black] (8.00, -1.25) rectangle (8.25, -1.50);
\filldraw[fill=cancan, draw=black] (8.25, -1.25) rectangle (8.50, -1.50);
\filldraw[fill=bermuda, draw=black] (8.50, -1.25) rectangle (8.75, -1.50);
\filldraw[fill=cancan, draw=black] (8.75, -1.25) rectangle (9.00, -1.50);
\filldraw[fill=bermuda, draw=black] (9.00, -1.25) rectangle (9.25, -1.50);
\filldraw[fill=cancan, draw=black] (9.25, -1.25) rectangle (9.50, -1.50);
\filldraw[fill=bermuda, draw=black] (9.50, -1.25) rectangle (9.75, -1.50);
\filldraw[fill=bermuda, draw=black] (9.75, -1.25) rectangle (10.00, -1.50);
\filldraw[fill=bermuda, draw=black] (10.00, -1.25) rectangle (10.25, -1.50);
\filldraw[fill=cancan, draw=black] (10.25, -1.25) rectangle (10.50, -1.50);
\filldraw[fill=cancan, draw=black] (10.50, -1.25) rectangle (10.75, -1.50);
\filldraw[fill=cancan, draw=black] (10.75, -1.25) rectangle (11.00, -1.50);
\filldraw[fill=bermuda, draw=black] (11.00, -1.25) rectangle (11.25, -1.50);
\filldraw[fill=bermuda, draw=black] (11.25, -1.25) rectangle (11.50, -1.50);
\filldraw[fill=bermuda, draw=black] (11.50, -1.25) rectangle (11.75, -1.50);
\filldraw[fill=cancan, draw=black] (11.75, -1.25) rectangle (12.00, -1.50);
\filldraw[fill=cancan, draw=black] (12.00, -1.25) rectangle (12.25, -1.50);
\filldraw[fill=cancan, draw=black] (12.25, -1.25) rectangle (12.50, -1.50);
\filldraw[fill=cancan, draw=black] (12.50, -1.25) rectangle (12.75, -1.50);
\filldraw[fill=bermuda, draw=black] (12.75, -1.25) rectangle (13.00, -1.50);
\filldraw[fill=bermuda, draw=black] (13.00, -1.25) rectangle (13.25, -1.50);
\filldraw[fill=cancan, draw=black] (13.25, -1.25) rectangle (13.50, -1.50);
\filldraw[fill=bermuda, draw=black] (13.50, -1.25) rectangle (13.75, -1.50);
\filldraw[fill=cancan, draw=black] (13.75, -1.25) rectangle (14.00, -1.50);
\filldraw[fill=cancan, draw=black] (14.00, -1.25) rectangle (14.25, -1.50);
\filldraw[fill=bermuda, draw=black] (14.25, -1.25) rectangle (14.50, -1.50);
\filldraw[fill=bermuda, draw=black] (14.50, -1.25) rectangle (14.75, -1.50);
\filldraw[fill=cancan, draw=black] (14.75, -1.25) rectangle (15.00, -1.50);
\filldraw[fill=bermuda, draw=black] (0.00, -1.50) rectangle (0.25, -1.75);
\filldraw[fill=cancan, draw=black] (0.25, -1.50) rectangle (0.50, -1.75);
\filldraw[fill=bermuda, draw=black] (0.50, -1.50) rectangle (0.75, -1.75);
\filldraw[fill=bermuda, draw=black] (0.75, -1.50) rectangle (1.00, -1.75);
\filldraw[fill=bermuda, draw=black] (1.00, -1.50) rectangle (1.25, -1.75);
\filldraw[fill=cancan, draw=black] (1.25, -1.50) rectangle (1.50, -1.75);
\filldraw[fill=bermuda, draw=black] (1.50, -1.50) rectangle (1.75, -1.75);
\filldraw[fill=bermuda, draw=black] (1.75, -1.50) rectangle (2.00, -1.75);
\filldraw[fill=bermuda, draw=black] (2.00, -1.50) rectangle (2.25, -1.75);
\filldraw[fill=bermuda, draw=black] (2.25, -1.50) rectangle (2.50, -1.75);
\filldraw[fill=bermuda, draw=black] (2.50, -1.50) rectangle (2.75, -1.75);
\filldraw[fill=cancan, draw=black] (2.75, -1.50) rectangle (3.00, -1.75);
\filldraw[fill=bermuda, draw=black] (3.00, -1.50) rectangle (3.25, -1.75);
\filldraw[fill=cancan, draw=black] (3.25, -1.50) rectangle (3.50, -1.75);
\filldraw[fill=cancan, draw=black] (3.50, -1.50) rectangle (3.75, -1.75);
\filldraw[fill=bermuda, draw=black] (3.75, -1.50) rectangle (4.00, -1.75);
\filldraw[fill=bermuda, draw=black] (4.00, -1.50) rectangle (4.25, -1.75);
\filldraw[fill=cancan, draw=black] (4.25, -1.50) rectangle (4.50, -1.75);
\filldraw[fill=bermuda, draw=black] (4.50, -1.50) rectangle (4.75, -1.75);
\filldraw[fill=cancan, draw=black] (4.75, -1.50) rectangle (5.00, -1.75);
\filldraw[fill=cancan, draw=black] (5.00, -1.50) rectangle (5.25, -1.75);
\filldraw[fill=bermuda, draw=black] (5.25, -1.50) rectangle (5.50, -1.75);
\filldraw[fill=bermuda, draw=black] (5.50, -1.50) rectangle (5.75, -1.75);
\filldraw[fill=cancan, draw=black] (5.75, -1.50) rectangle (6.00, -1.75);
\filldraw[fill=bermuda, draw=black] (6.00, -1.50) rectangle (6.25, -1.75);
\filldraw[fill=cancan, draw=black] (6.25, -1.50) rectangle (6.50, -1.75);
\filldraw[fill=cancan, draw=black] (6.50, -1.50) rectangle (6.75, -1.75);
\filldraw[fill=bermuda, draw=black] (6.75, -1.50) rectangle (7.00, -1.75);
\filldraw[fill=bermuda, draw=black] (7.00, -1.50) rectangle (7.25, -1.75);
\filldraw[fill=cancan, draw=black] (7.25, -1.50) rectangle (7.50, -1.75);
\filldraw[fill=bermuda, draw=black] (7.50, -1.50) rectangle (7.75, -1.75);
\filldraw[fill=cancan, draw=black] (7.75, -1.50) rectangle (8.00, -1.75);
\filldraw[fill=cancan, draw=black] (8.00, -1.50) rectangle (8.25, -1.75);
\filldraw[fill=bermuda, draw=black] (8.25, -1.50) rectangle (8.50, -1.75);
\filldraw[fill=bermuda, draw=black] (8.50, -1.50) rectangle (8.75, -1.75);
\filldraw[fill=cancan, draw=black] (8.75, -1.50) rectangle (9.00, -1.75);
\filldraw[fill=bermuda, draw=black] (9.00, -1.50) rectangle (9.25, -1.75);
\filldraw[fill=bermuda, draw=black] (9.25, -1.50) rectangle (9.50, -1.75);
\filldraw[fill=bermuda, draw=black] (9.50, -1.50) rectangle (9.75, -1.75);
\filldraw[fill=cancan, draw=black] (9.75, -1.50) rectangle (10.00, -1.75);
\filldraw[fill=cancan, draw=black] (10.00, -1.50) rectangle (10.25, -1.75);
\filldraw[fill=cancan, draw=black] (10.25, -1.50) rectangle (10.50, -1.75);
\filldraw[fill=cancan, draw=black] (10.50, -1.50) rectangle (10.75, -1.75);
\filldraw[fill=cancan, draw=black] (10.75, -1.50) rectangle (11.00, -1.75);
\filldraw[fill=bermuda, draw=black] (11.00, -1.50) rectangle (11.25, -1.75);
\filldraw[fill=bermuda, draw=black] (11.25, -1.50) rectangle (11.50, -1.75);
\filldraw[fill=bermuda, draw=black] (11.50, -1.50) rectangle (11.75, -1.75);
\filldraw[fill=cancan, draw=black] (11.75, -1.50) rectangle (12.00, -1.75);
\filldraw[fill=cancan, draw=black] (12.00, -1.50) rectangle (12.25, -1.75);
\filldraw[fill=cancan, draw=black] (12.25, -1.50) rectangle (12.50, -1.75);
\filldraw[fill=bermuda, draw=black] (12.50, -1.50) rectangle (12.75, -1.75);
\filldraw[fill=bermuda, draw=black] (12.75, -1.50) rectangle (13.00, -1.75);
\filldraw[fill=bermuda, draw=black] (13.00, -1.50) rectangle (13.25, -1.75);
\filldraw[fill=cancan, draw=black] (13.25, -1.50) rectangle (13.50, -1.75);
\filldraw[fill=cancan, draw=black] (13.50, -1.50) rectangle (13.75, -1.75);
\filldraw[fill=cancan, draw=black] (13.75, -1.50) rectangle (14.00, -1.75);
\filldraw[fill=bermuda, draw=black] (14.00, -1.50) rectangle (14.25, -1.75);
\filldraw[fill=bermuda, draw=black] (14.25, -1.50) rectangle (14.50, -1.75);
\filldraw[fill=bermuda, draw=black] (14.50, -1.50) rectangle (14.75, -1.75);
\filldraw[fill=cancan, draw=black] (14.75, -1.50) rectangle (15.00, -1.75);
\filldraw[fill=bermuda, draw=black] (0.00, -1.75) rectangle (0.25, -2.00);
\filldraw[fill=bermuda, draw=black] (0.25, -1.75) rectangle (0.50, -2.00);
\filldraw[fill=bermuda, draw=black] (0.50, -1.75) rectangle (0.75, -2.00);
\filldraw[fill=cancan, draw=black] (0.75, -1.75) rectangle (1.00, -2.00);
\filldraw[fill=cancan, draw=black] (1.00, -1.75) rectangle (1.25, -2.00);
\filldraw[fill=bermuda, draw=black] (1.25, -1.75) rectangle (1.50, -2.00);
\filldraw[fill=bermuda, draw=black] (1.50, -1.75) rectangle (1.75, -2.00);
\filldraw[fill=cancan, draw=black] (1.75, -1.75) rectangle (2.00, -2.00);
\filldraw[fill=cancan, draw=black] (2.00, -1.75) rectangle (2.25, -2.00);
\filldraw[fill=cancan, draw=black] (2.25, -1.75) rectangle (2.50, -2.00);
\filldraw[fill=cancan, draw=black] (2.50, -1.75) rectangle (2.75, -2.00);
\filldraw[fill=bermuda, draw=black] (2.75, -1.75) rectangle (3.00, -2.00);
\filldraw[fill=bermuda, draw=black] (3.00, -1.75) rectangle (3.25, -2.00);
\filldraw[fill=cancan, draw=black] (3.25, -1.75) rectangle (3.50, -2.00);
\filldraw[fill=bermuda, draw=black] (3.50, -1.75) rectangle (3.75, -2.00);
\filldraw[fill=bermuda, draw=black] (3.75, -1.75) rectangle (4.00, -2.00);
\filldraw[fill=bermuda, draw=black] (4.00, -1.75) rectangle (4.25, -2.00);
\filldraw[fill=cancan, draw=black] (4.25, -1.75) rectangle (4.50, -2.00);
\filldraw[fill=cancan, draw=black] (4.50, -1.75) rectangle (4.75, -2.00);
\filldraw[fill=cancan, draw=black] (4.75, -1.75) rectangle (5.00, -2.00);
\filldraw[fill=bermuda, draw=black] (5.00, -1.75) rectangle (5.25, -2.00);
\filldraw[fill=bermuda, draw=black] (5.25, -1.75) rectangle (5.50, -2.00);
\filldraw[fill=bermuda, draw=black] (5.50, -1.75) rectangle (5.75, -2.00);
\filldraw[fill=bermuda, draw=black] (5.75, -1.75) rectangle (6.00, -2.00);
\filldraw[fill=bermuda, draw=black] (6.00, -1.75) rectangle (6.25, -2.00);
\filldraw[fill=cancan, draw=black] (6.25, -1.75) rectangle (6.50, -2.00);
\filldraw[fill=bermuda, draw=black] (6.50, -1.75) rectangle (6.75, -2.00);
\filldraw[fill=bermuda, draw=black] (6.75, -1.75) rectangle (7.00, -2.00);
\filldraw[fill=bermuda, draw=black] (7.00, -1.75) rectangle (7.25, -2.00);
\filldraw[fill=cancan, draw=black] (7.25, -1.75) rectangle (7.50, -2.00);
\filldraw[fill=bermuda, draw=black] (7.50, -1.75) rectangle (7.75, -2.00);
\filldraw[fill=bermuda, draw=black] (7.75, -1.75) rectangle (8.00, -2.00);
\filldraw[fill=bermuda, draw=black] (8.00, -1.75) rectangle (8.25, -2.00);
\filldraw[fill=bermuda, draw=black] (8.25, -1.75) rectangle (8.50, -2.00);
\filldraw[fill=bermuda, draw=black] (8.50, -1.75) rectangle (8.75, -2.00);
\filldraw[fill=cancan, draw=black] (8.75, -1.75) rectangle (9.00, -2.00);
\filldraw[fill=cancan, draw=black] (9.00, -1.75) rectangle (9.25, -2.00);
\filldraw[fill=cancan, draw=black] (9.25, -1.75) rectangle (9.50, -2.00);
\filldraw[fill=bermuda, draw=black] (9.50, -1.75) rectangle (9.75, -2.00);
\filldraw[fill=bermuda, draw=black] (9.75, -1.75) rectangle (10.00, -2.00);
\filldraw[fill=bermuda, draw=black] (10.00, -1.75) rectangle (10.25, -2.00);
\filldraw[fill=cancan, draw=black] (10.25, -1.75) rectangle (10.50, -2.00);
\filldraw[fill=cancan, draw=black] (10.50, -1.75) rectangle (10.75, -2.00);
\filldraw[fill=cancan, draw=black] (10.75, -1.75) rectangle (11.00, -2.00);
\filldraw[fill=bermuda, draw=black] (11.00, -1.75) rectangle (11.25, -2.00);
\filldraw[fill=bermuda, draw=black] (11.25, -1.75) rectangle (11.50, -2.00);
\filldraw[fill=bermuda, draw=black] (11.50, -1.75) rectangle (11.75, -2.00);
\filldraw[fill=cancan, draw=black] (11.75, -1.75) rectangle (12.00, -2.00);
\filldraw[fill=cancan, draw=black] (12.00, -1.75) rectangle (12.25, -2.00);
\filldraw[fill=cancan, draw=black] (12.25, -1.75) rectangle (12.50, -2.00);
\filldraw[fill=bermuda, draw=black] (12.50, -1.75) rectangle (12.75, -2.00);
\filldraw[fill=bermuda, draw=black] (12.75, -1.75) rectangle (13.00, -2.00);
\filldraw[fill=bermuda, draw=black] (13.00, -1.75) rectangle (13.25, -2.00);
\filldraw[fill=bermuda, draw=black] (13.25, -1.75) rectangle (13.50, -2.00);
\filldraw[fill=bermuda, draw=black] (13.50, -1.75) rectangle (13.75, -2.00);
\filldraw[fill=bermuda, draw=black] (13.75, -1.75) rectangle (14.00, -2.00);
\filldraw[fill=bermuda, draw=black] (14.00, -1.75) rectangle (14.25, -2.00);
\filldraw[fill=bermuda, draw=black] (14.25, -1.75) rectangle (14.50, -2.00);
\filldraw[fill=bermuda, draw=black] (14.50, -1.75) rectangle (14.75, -2.00);
\filldraw[fill=cancan, draw=black] (14.75, -1.75) rectangle (15.00, -2.00);
\filldraw[fill=bermuda, draw=black] (0.00, -2.00) rectangle (0.25, -2.25);
\filldraw[fill=cancan, draw=black] (0.25, -2.00) rectangle (0.50, -2.25);
\filldraw[fill=cancan, draw=black] (0.50, -2.00) rectangle (0.75, -2.25);
\filldraw[fill=bermuda, draw=black] (0.75, -2.00) rectangle (1.00, -2.25);
\filldraw[fill=bermuda, draw=black] (1.00, -2.00) rectangle (1.25, -2.25);
\filldraw[fill=cancan, draw=black] (1.25, -2.00) rectangle (1.50, -2.25);
\filldraw[fill=bermuda, draw=black] (1.50, -2.00) rectangle (1.75, -2.25);
\filldraw[fill=cancan, draw=black] (1.75, -2.00) rectangle (2.00, -2.25);
\filldraw[fill=cancan, draw=black] (2.00, -2.00) rectangle (2.25, -2.25);
\filldraw[fill=cancan, draw=black] (2.25, -2.00) rectangle (2.50, -2.25);
\filldraw[fill=bermuda, draw=black] (2.50, -2.00) rectangle (2.75, -2.25);
\filldraw[fill=cancan, draw=black] (2.75, -2.00) rectangle (3.00, -2.25);
\filldraw[fill=cancan, draw=black] (3.00, -2.00) rectangle (3.25, -2.25);
\filldraw[fill=cancan, draw=black] (3.25, -2.00) rectangle (3.50, -2.25);
\filldraw[fill=cancan, draw=black] (3.50, -2.00) rectangle (3.75, -2.25);
\filldraw[fill=cancan, draw=black] (3.75, -2.00) rectangle (4.00, -2.25);
\filldraw[fill=bermuda, draw=black] (4.00, -2.00) rectangle (4.25, -2.25);
\filldraw[fill=bermuda, draw=black] (4.25, -2.00) rectangle (4.50, -2.25);
\filldraw[fill=bermuda, draw=black] (4.50, -2.00) rectangle (4.75, -2.25);
\filldraw[fill=cancan, draw=black] (4.75, -2.00) rectangle (5.00, -2.25);
\filldraw[fill=cancan, draw=black] (5.00, -2.00) rectangle (5.25, -2.25);
\filldraw[fill=cancan, draw=black] (5.25, -2.00) rectangle (5.50, -2.25);
\filldraw[fill=bermuda, draw=black] (5.50, -2.00) rectangle (5.75, -2.25);
\filldraw[fill=bermuda, draw=black] (5.75, -2.00) rectangle (6.00, -2.25);
\filldraw[fill=bermuda, draw=black] (6.00, -2.00) rectangle (6.25, -2.25);
\filldraw[fill=cancan, draw=black] (6.25, -2.00) rectangle (6.50, -2.25);
\filldraw[fill=cancan, draw=black] (6.50, -2.00) rectangle (6.75, -2.25);
\filldraw[fill=cancan, draw=black] (6.75, -2.00) rectangle (7.00, -2.25);
\filldraw[fill=bermuda, draw=black] (7.00, -2.00) rectangle (7.25, -2.25);
\filldraw[fill=bermuda, draw=black] (7.25, -2.00) rectangle (7.50, -2.25);
\filldraw[fill=bermuda, draw=black] (7.50, -2.00) rectangle (7.75, -2.25);
\filldraw[fill=cancan, draw=black] (7.75, -2.00) rectangle (8.00, -2.25);
\filldraw[fill=bermuda, draw=black] (8.00, -2.00) rectangle (8.25, -2.25);
\filldraw[fill=bermuda, draw=black] (8.25, -2.00) rectangle (8.50, -2.25);
\filldraw[fill=bermuda, draw=black] (8.50, -2.00) rectangle (8.75, -2.25);
\filldraw[fill=cancan, draw=black] (8.75, -2.00) rectangle (9.00, -2.25);
\filldraw[fill=bermuda, draw=black] (9.00, -2.00) rectangle (9.25, -2.25);
\filldraw[fill=cancan, draw=black] (9.25, -2.00) rectangle (9.50, -2.25);
\filldraw[fill=bermuda, draw=black] (9.50, -2.00) rectangle (9.75, -2.25);
\filldraw[fill=cancan, draw=black] (9.75, -2.00) rectangle (10.00, -2.25);
\filldraw[fill=cancan, draw=black] (10.00, -2.00) rectangle (10.25, -2.25);
\filldraw[fill=cancan, draw=black] (10.25, -2.00) rectangle (10.50, -2.25);
\filldraw[fill=cancan, draw=black] (10.50, -2.00) rectangle (10.75, -2.25);
\filldraw[fill=cancan, draw=black] (10.75, -2.00) rectangle (11.00, -2.25);
\filldraw[fill=bermuda, draw=black] (11.00, -2.00) rectangle (11.25, -2.25);
\filldraw[fill=bermuda, draw=black] (11.25, -2.00) rectangle (11.50, -2.25);
\filldraw[fill=bermuda, draw=black] (11.50, -2.00) rectangle (11.75, -2.25);
\filldraw[fill=bermuda, draw=black] (11.75, -2.00) rectangle (12.00, -2.25);
\filldraw[fill=bermuda, draw=black] (12.00, -2.00) rectangle (12.25, -2.25);
\filldraw[fill=bermuda, draw=black] (12.25, -2.00) rectangle (12.50, -2.25);
\filldraw[fill=bermuda, draw=black] (12.50, -2.00) rectangle (12.75, -2.25);
\filldraw[fill=cancan, draw=black] (12.75, -2.00) rectangle (13.00, -2.25);
\filldraw[fill=bermuda, draw=black] (13.00, -2.00) rectangle (13.25, -2.25);
\filldraw[fill=cancan, draw=black] (13.25, -2.00) rectangle (13.50, -2.25);
\filldraw[fill=bermuda, draw=black] (13.50, -2.00) rectangle (13.75, -2.25);
\filldraw[fill=bermuda, draw=black] (13.75, -2.00) rectangle (14.00, -2.25);
\filldraw[fill=bermuda, draw=black] (14.00, -2.00) rectangle (14.25, -2.25);
\filldraw[fill=cancan, draw=black] (14.25, -2.00) rectangle (14.50, -2.25);
\filldraw[fill=cancan, draw=black] (14.50, -2.00) rectangle (14.75, -2.25);
\filldraw[fill=cancan, draw=black] (14.75, -2.00) rectangle (15.00, -2.25);
\filldraw[fill=bermuda, draw=black] (0.00, -2.25) rectangle (0.25, -2.50);
\filldraw[fill=bermuda, draw=black] (0.25, -2.25) rectangle (0.50, -2.50);
\filldraw[fill=bermuda, draw=black] (0.50, -2.25) rectangle (0.75, -2.50);
\filldraw[fill=cancan, draw=black] (0.75, -2.25) rectangle (1.00, -2.50);
\filldraw[fill=bermuda, draw=black] (1.00, -2.25) rectangle (1.25, -2.50);
\filldraw[fill=bermuda, draw=black] (1.25, -2.25) rectangle (1.50, -2.50);
\filldraw[fill=bermuda, draw=black] (1.50, -2.25) rectangle (1.75, -2.50);
\filldraw[fill=cancan, draw=black] (1.75, -2.25) rectangle (2.00, -2.50);
\filldraw[fill=bermuda, draw=black] (2.00, -2.25) rectangle (2.25, -2.50);
\filldraw[fill=cancan, draw=black] (2.25, -2.25) rectangle (2.50, -2.50);
\filldraw[fill=bermuda, draw=black] (2.50, -2.25) rectangle (2.75, -2.50);
\filldraw[fill=cancan, draw=black] (2.75, -2.25) rectangle (3.00, -2.50);
\filldraw[fill=bermuda, draw=black] (3.00, -2.25) rectangle (3.25, -2.50);
\filldraw[fill=bermuda, draw=black] (3.25, -2.25) rectangle (3.50, -2.50);
\filldraw[fill=bermuda, draw=black] (3.50, -2.25) rectangle (3.75, -2.50);
\filldraw[fill=cancan, draw=black] (3.75, -2.25) rectangle (4.00, -2.50);
\filldraw[fill=cancan, draw=black] (4.00, -2.25) rectangle (4.25, -2.50);
\filldraw[fill=cancan, draw=black] (4.25, -2.25) rectangle (4.50, -2.50);
\filldraw[fill=bermuda, draw=black] (4.50, -2.25) rectangle (4.75, -2.50);
\filldraw[fill=bermuda, draw=black] (4.75, -2.25) rectangle (5.00, -2.50);
\filldraw[fill=bermuda, draw=black] (5.00, -2.25) rectangle (5.25, -2.50);
\filldraw[fill=cancan, draw=black] (5.25, -2.25) rectangle (5.50, -2.50);
\filldraw[fill=bermuda, draw=black] (5.50, -2.25) rectangle (5.75, -2.50);
\filldraw[fill=cancan, draw=black] (5.75, -2.25) rectangle (6.00, -2.50);
\filldraw[fill=bermuda, draw=black] (6.00, -2.25) rectangle (6.25, -2.50);
\filldraw[fill=cancan, draw=black] (6.25, -2.25) rectangle (6.50, -2.50);
\filldraw[fill=cancan, draw=black] (6.50, -2.25) rectangle (6.75, -2.50);
\filldraw[fill=cancan, draw=black] (6.75, -2.25) rectangle (7.00, -2.50);
\filldraw[fill=bermuda, draw=black] (7.00, -2.25) rectangle (7.25, -2.50);
\filldraw[fill=cancan, draw=black] (7.25, -2.25) rectangle (7.50, -2.50);
\filldraw[fill=bermuda, draw=black] (7.50, -2.25) rectangle (7.75, -2.50);
\filldraw[fill=cancan, draw=black] (7.75, -2.25) rectangle (8.00, -2.50);
\filldraw[fill=bermuda, draw=black] (8.00, -2.25) rectangle (8.25, -2.50);
\filldraw[fill=cancan, draw=black] (8.25, -2.25) rectangle (8.50, -2.50);
\filldraw[fill=bermuda, draw=black] (8.50, -2.25) rectangle (8.75, -2.50);
\filldraw[fill=cancan, draw=black] (8.75, -2.25) rectangle (9.00, -2.50);
\filldraw[fill=cancan, draw=black] (9.00, -2.25) rectangle (9.25, -2.50);
\filldraw[fill=cancan, draw=black] (9.25, -2.25) rectangle (9.50, -2.50);
\filldraw[fill=bermuda, draw=black] (9.50, -2.25) rectangle (9.75, -2.50);
\filldraw[fill=bermuda, draw=black] (9.75, -2.25) rectangle (10.00, -2.50);
\filldraw[fill=bermuda, draw=black] (10.00, -2.25) rectangle (10.25, -2.50);
\filldraw[fill=bermuda, draw=black] (10.25, -2.25) rectangle (10.50, -2.50);
\filldraw[fill=bermuda, draw=black] (10.50, -2.25) rectangle (10.75, -2.50);
\filldraw[fill=cancan, draw=black] (10.75, -2.25) rectangle (11.00, -2.50);
\filldraw[fill=bermuda, draw=black] (11.00, -2.25) rectangle (11.25, -2.50);
\filldraw[fill=bermuda, draw=black] (11.25, -2.25) rectangle (11.50, -2.50);
\filldraw[fill=bermuda, draw=black] (11.50, -2.25) rectangle (11.75, -2.50);
\filldraw[fill=cancan, draw=black] (11.75, -2.25) rectangle (12.00, -2.50);
\filldraw[fill=cancan, draw=black] (12.00, -2.25) rectangle (12.25, -2.50);
\filldraw[fill=cancan, draw=black] (12.25, -2.25) rectangle (12.50, -2.50);
\filldraw[fill=bermuda, draw=black] (12.50, -2.25) rectangle (12.75, -2.50);
\filldraw[fill=bermuda, draw=black] (12.75, -2.25) rectangle (13.00, -2.50);
\filldraw[fill=bermuda, draw=black] (13.00, -2.25) rectangle (13.25, -2.50);
\filldraw[fill=cancan, draw=black] (13.25, -2.25) rectangle (13.50, -2.50);
\filldraw[fill=bermuda, draw=black] (13.50, -2.25) rectangle (13.75, -2.50);
\filldraw[fill=cancan, draw=black] (13.75, -2.25) rectangle (14.00, -2.50);
\filldraw[fill=cancan, draw=black] (14.00, -2.25) rectangle (14.25, -2.50);
\filldraw[fill=bermuda, draw=black] (14.25, -2.25) rectangle (14.50, -2.50);
\filldraw[fill=bermuda, draw=black] (14.50, -2.25) rectangle (14.75, -2.50);
\filldraw[fill=cancan, draw=black] (14.75, -2.25) rectangle (15.00, -2.50);
\filldraw[fill=bermuda, draw=black] (0.00, -2.50) rectangle (0.25, -2.75);
\filldraw[fill=cancan, draw=black] (0.25, -2.50) rectangle (0.50, -2.75);
\filldraw[fill=cancan, draw=black] (0.50, -2.50) rectangle (0.75, -2.75);
\filldraw[fill=cancan, draw=black] (0.75, -2.50) rectangle (1.00, -2.75);
\filldraw[fill=bermuda, draw=black] (1.00, -2.50) rectangle (1.25, -2.75);
\filldraw[fill=bermuda, draw=black] (1.25, -2.50) rectangle (1.50, -2.75);
\filldraw[fill=bermuda, draw=black] (1.50, -2.50) rectangle (1.75, -2.75);
\filldraw[fill=bermuda, draw=black] (1.75, -2.50) rectangle (2.00, -2.75);
\filldraw[fill=bermuda, draw=black] (2.00, -2.50) rectangle (2.25, -2.75);
\filldraw[fill=cancan, draw=black] (2.25, -2.50) rectangle (2.50, -2.75);
\filldraw[fill=bermuda, draw=black] (2.50, -2.50) rectangle (2.75, -2.75);
\filldraw[fill=bermuda, draw=black] (2.75, -2.50) rectangle (3.00, -2.75);
\filldraw[fill=bermuda, draw=black] (3.00, -2.50) rectangle (3.25, -2.75);
\filldraw[fill=bermuda, draw=black] (3.25, -2.50) rectangle (3.50, -2.75);
\filldraw[fill=bermuda, draw=black] (3.50, -2.50) rectangle (3.75, -2.75);
\filldraw[fill=cancan, draw=black] (3.75, -2.50) rectangle (4.00, -2.75);
\filldraw[fill=bermuda, draw=black] (4.00, -2.50) rectangle (4.25, -2.75);
\filldraw[fill=cancan, draw=black] (4.25, -2.50) rectangle (4.50, -2.75);
\filldraw[fill=cancan, draw=black] (4.50, -2.50) rectangle (4.75, -2.75);
\filldraw[fill=bermuda, draw=black] (4.75, -2.50) rectangle (5.00, -2.75);
\filldraw[fill=bermuda, draw=black] (5.00, -2.50) rectangle (5.25, -2.75);
\filldraw[fill=cancan, draw=black] (5.25, -2.50) rectangle (5.50, -2.75);
\filldraw[fill=bermuda, draw=black] (5.50, -2.50) rectangle (5.75, -2.75);
\filldraw[fill=cancan, draw=black] (5.75, -2.50) rectangle (6.00, -2.75);
\filldraw[fill=cancan, draw=black] (6.00, -2.50) rectangle (6.25, -2.75);
\filldraw[fill=cancan, draw=black] (6.25, -2.50) rectangle (6.50, -2.75);
\filldraw[fill=bermuda, draw=black] (6.50, -2.50) rectangle (6.75, -2.75);
\filldraw[fill=cancan, draw=black] (6.75, -2.50) rectangle (7.00, -2.75);
\filldraw[fill=cancan, draw=black] (7.00, -2.50) rectangle (7.25, -2.75);
\filldraw[fill=cancan, draw=black] (7.25, -2.50) rectangle (7.50, -2.75);
\filldraw[fill=bermuda, draw=black] (7.50, -2.50) rectangle (7.75, -2.75);
\filldraw[fill=bermuda, draw=black] (7.75, -2.50) rectangle (8.00, -2.75);
\filldraw[fill=bermuda, draw=black] (8.00, -2.50) rectangle (8.25, -2.75);
\filldraw[fill=bermuda, draw=black] (8.25, -2.50) rectangle (8.50, -2.75);
\filldraw[fill=bermuda, draw=black] (8.50, -2.50) rectangle (8.75, -2.75);
\filldraw[fill=cancan, draw=black] (8.75, -2.50) rectangle (9.00, -2.75);
\filldraw[fill=cancan, draw=black] (9.00, -2.50) rectangle (9.25, -2.75);
\filldraw[fill=cancan, draw=black] (9.25, -2.50) rectangle (9.50, -2.75);
\filldraw[fill=bermuda, draw=black] (9.50, -2.50) rectangle (9.75, -2.75);
\filldraw[fill=bermuda, draw=black] (9.75, -2.50) rectangle (10.00, -2.75);
\filldraw[fill=bermuda, draw=black] (10.00, -2.50) rectangle (10.25, -2.75);
\filldraw[fill=cancan, draw=black] (10.25, -2.50) rectangle (10.50, -2.75);
\filldraw[fill=cancan, draw=black] (10.50, -2.50) rectangle (10.75, -2.75);
\filldraw[fill=cancan, draw=black] (10.75, -2.50) rectangle (11.00, -2.75);
\filldraw[fill=bermuda, draw=black] (11.00, -2.50) rectangle (11.25, -2.75);
\filldraw[fill=bermuda, draw=black] (11.25, -2.50) rectangle (11.50, -2.75);
\filldraw[fill=bermuda, draw=black] (11.50, -2.50) rectangle (11.75, -2.75);
\filldraw[fill=cancan, draw=black] (11.75, -2.50) rectangle (12.00, -2.75);
\filldraw[fill=cancan, draw=black] (12.00, -2.50) rectangle (12.25, -2.75);
\filldraw[fill=cancan, draw=black] (12.25, -2.50) rectangle (12.50, -2.75);
\filldraw[fill=cancan, draw=black] (12.50, -2.50) rectangle (12.75, -2.75);
\filldraw[fill=cancan, draw=black] (12.75, -2.50) rectangle (13.00, -2.75);
\filldraw[fill=bermuda, draw=black] (13.00, -2.50) rectangle (13.25, -2.75);
\filldraw[fill=bermuda, draw=black] (13.25, -2.50) rectangle (13.50, -2.75);
\filldraw[fill=bermuda, draw=black] (13.50, -2.50) rectangle (13.75, -2.75);
\filldraw[fill=cancan, draw=black] (13.75, -2.50) rectangle (14.00, -2.75);
\filldraw[fill=cancan, draw=black] (14.00, -2.50) rectangle (14.25, -2.75);
\filldraw[fill=cancan, draw=black] (14.25, -2.50) rectangle (14.50, -2.75);
\filldraw[fill=bermuda, draw=black] (14.50, -2.50) rectangle (14.75, -2.75);
\filldraw[fill=bermuda, draw=black] (14.75, -2.50) rectangle (15.00, -2.75);
\filldraw[fill=bermuda, draw=black] (0.00, -2.75) rectangle (0.25, -3.00);
\filldraw[fill=bermuda, draw=black] (0.25, -2.75) rectangle (0.50, -3.00);
\filldraw[fill=bermuda, draw=black] (0.50, -2.75) rectangle (0.75, -3.00);
\filldraw[fill=cancan, draw=black] (0.75, -2.75) rectangle (1.00, -3.00);
\filldraw[fill=bermuda, draw=black] (1.00, -2.75) rectangle (1.25, -3.00);
\filldraw[fill=bermuda, draw=black] (1.25, -2.75) rectangle (1.50, -3.00);
\filldraw[fill=bermuda, draw=black] (1.50, -2.75) rectangle (1.75, -3.00);
\filldraw[fill=cancan, draw=black] (1.75, -2.75) rectangle (2.00, -3.00);
\filldraw[fill=bermuda, draw=black] (2.00, -2.75) rectangle (2.25, -3.00);
\filldraw[fill=bermuda, draw=black] (2.25, -2.75) rectangle (2.50, -3.00);
\filldraw[fill=bermuda, draw=black] (2.50, -2.75) rectangle (2.75, -3.00);
\filldraw[fill=cancan, draw=black] (2.75, -2.75) rectangle (3.00, -3.00);
\filldraw[fill=bermuda, draw=black] (3.00, -2.75) rectangle (3.25, -3.00);
\filldraw[fill=cancan, draw=black] (3.25, -2.75) rectangle (3.50, -3.00);
\filldraw[fill=cancan, draw=black] (3.50, -2.75) rectangle (3.75, -3.00);
\filldraw[fill=bermuda, draw=black] (3.75, -2.75) rectangle (4.00, -3.00);
\filldraw[fill=bermuda, draw=black] (4.00, -2.75) rectangle (4.25, -3.00);
\filldraw[fill=cancan, draw=black] (4.25, -2.75) rectangle (4.50, -3.00);
\filldraw[fill=bermuda, draw=black] (4.50, -2.75) rectangle (4.75, -3.00);
\filldraw[fill=cancan, draw=black] (4.75, -2.75) rectangle (5.00, -3.00);
\filldraw[fill=cancan, draw=black] (5.00, -2.75) rectangle (5.25, -3.00);
\filldraw[fill=bermuda, draw=black] (5.25, -2.75) rectangle (5.50, -3.00);
\filldraw[fill=bermuda, draw=black] (5.50, -2.75) rectangle (5.75, -3.00);
\filldraw[fill=cancan, draw=black] (5.75, -2.75) rectangle (6.00, -3.00);
\filldraw[fill=bermuda, draw=black] (6.00, -2.75) rectangle (6.25, -3.00);
\filldraw[fill=cancan, draw=black] (6.25, -2.75) rectangle (6.50, -3.00);
\filldraw[fill=bermuda, draw=black] (6.50, -2.75) rectangle (6.75, -3.00);
\filldraw[fill=bermuda, draw=black] (6.75, -2.75) rectangle (7.00, -3.00);
\filldraw[fill=bermuda, draw=black] (7.00, -2.75) rectangle (7.25, -3.00);
\filldraw[fill=cancan, draw=black] (7.25, -2.75) rectangle (7.50, -3.00);
\filldraw[fill=cancan, draw=black] (7.50, -2.75) rectangle (7.75, -3.00);
\filldraw[fill=cancan, draw=black] (7.75, -2.75) rectangle (8.00, -3.00);
\filldraw[fill=cancan, draw=black] (8.00, -2.75) rectangle (8.25, -3.00);
\filldraw[fill=cancan, draw=black] (8.25, -2.75) rectangle (8.50, -3.00);
\filldraw[fill=bermuda, draw=black] (8.50, -2.75) rectangle (8.75, -3.00);
\filldraw[fill=bermuda, draw=black] (8.75, -2.75) rectangle (9.00, -3.00);
\filldraw[fill=bermuda, draw=black] (9.00, -2.75) rectangle (9.25, -3.00);
\filldraw[fill=cancan, draw=black] (9.25, -2.75) rectangle (9.50, -3.00);
\filldraw[fill=cancan, draw=black] (9.50, -2.75) rectangle (9.75, -3.00);
\filldraw[fill=cancan, draw=black] (9.75, -2.75) rectangle (10.00, -3.00);
\filldraw[fill=bermuda, draw=black] (10.00, -2.75) rectangle (10.25, -3.00);
\filldraw[fill=bermuda, draw=black] (10.25, -2.75) rectangle (10.50, -3.00);
\filldraw[fill=bermuda, draw=black] (10.50, -2.75) rectangle (10.75, -3.00);
\filldraw[fill=cancan, draw=black] (10.75, -2.75) rectangle (11.00, -3.00);
\filldraw[fill=cancan, draw=black] (11.00, -2.75) rectangle (11.25, -3.00);
\filldraw[fill=cancan, draw=black] (11.25, -2.75) rectangle (11.50, -3.00);
\filldraw[fill=bermuda, draw=black] (11.50, -2.75) rectangle (11.75, -3.00);
\filldraw[fill=bermuda, draw=black] (11.75, -2.75) rectangle (12.00, -3.00);
\filldraw[fill=bermuda, draw=black] (12.00, -2.75) rectangle (12.25, -3.00);
\filldraw[fill=cancan, draw=black] (12.25, -2.75) rectangle (12.50, -3.00);
\filldraw[fill=cancan, draw=black] (12.50, -2.75) rectangle (12.75, -3.00);
\filldraw[fill=cancan, draw=black] (12.75, -2.75) rectangle (13.00, -3.00);
\filldraw[fill=bermuda, draw=black] (13.00, -2.75) rectangle (13.25, -3.00);
\filldraw[fill=bermuda, draw=black] (13.25, -2.75) rectangle (13.50, -3.00);
\filldraw[fill=bermuda, draw=black] (13.50, -2.75) rectangle (13.75, -3.00);
\filldraw[fill=cancan, draw=black] (13.75, -2.75) rectangle (14.00, -3.00);
\filldraw[fill=cancan, draw=black] (14.00, -2.75) rectangle (14.25, -3.00);
\filldraw[fill=cancan, draw=black] (14.25, -2.75) rectangle (14.50, -3.00);
\filldraw[fill=cancan, draw=black] (14.50, -2.75) rectangle (14.75, -3.00);
\filldraw[fill=cancan, draw=black] (14.75, -2.75) rectangle (15.00, -3.00);
\filldraw[fill=bermuda, draw=black] (0.00, -3.00) rectangle (0.25, -3.25);
\filldraw[fill=cancan, draw=black] (0.25, -3.00) rectangle (0.50, -3.25);
\filldraw[fill=cancan, draw=black] (0.50, -3.00) rectangle (0.75, -3.25);
\filldraw[fill=cancan, draw=black] (0.75, -3.00) rectangle (1.00, -3.25);
\filldraw[fill=cancan, draw=black] (1.00, -3.00) rectangle (1.25, -3.25);
\filldraw[fill=cancan, draw=black] (1.25, -3.00) rectangle (1.50, -3.25);
\filldraw[fill=cancan, draw=black] (1.50, -3.00) rectangle (1.75, -3.25);
\filldraw[fill=cancan, draw=black] (1.75, -3.00) rectangle (2.00, -3.25);
\filldraw[fill=cancan, draw=black] (2.00, -3.00) rectangle (2.25, -3.25);
\filldraw[fill=cancan, draw=black] (2.25, -3.00) rectangle (2.50, -3.25);
\filldraw[fill=cancan, draw=black] (2.50, -3.00) rectangle (2.75, -3.25);
\filldraw[fill=cancan, draw=black] (2.75, -3.00) rectangle (3.00, -3.25);
\filldraw[fill=bermuda, draw=black] (3.00, -3.00) rectangle (3.25, -3.25);
\filldraw[fill=bermuda, draw=black] (3.25, -3.00) rectangle (3.50, -3.25);
\filldraw[fill=bermuda, draw=black] (3.50, -3.00) rectangle (3.75, -3.25);
\filldraw[fill=bermuda, draw=black] (3.75, -3.00) rectangle (4.00, -3.25);
\filldraw[fill=bermuda, draw=black] (4.00, -3.00) rectangle (4.25, -3.25);
\filldraw[fill=cancan, draw=black] (4.25, -3.00) rectangle (4.50, -3.25);
\filldraw[fill=bermuda, draw=black] (4.50, -3.00) rectangle (4.75, -3.25);
\filldraw[fill=cancan, draw=black] (4.75, -3.00) rectangle (5.00, -3.25);
\filldraw[fill=bermuda, draw=black] (5.00, -3.00) rectangle (5.25, -3.25);
\filldraw[fill=cancan, draw=black] (5.25, -3.00) rectangle (5.50, -3.25);
\filldraw[fill=bermuda, draw=black] (5.50, -3.00) rectangle (5.75, -3.25);
\filldraw[fill=cancan, draw=black] (5.75, -3.00) rectangle (6.00, -3.25);
\filldraw[fill=cancan, draw=black] (6.00, -3.00) rectangle (6.25, -3.25);
\filldraw[fill=bermuda, draw=black] (6.25, -3.00) rectangle (6.50, -3.25);
\filldraw[fill=bermuda, draw=black] (6.50, -3.00) rectangle (6.75, -3.25);
\filldraw[fill=bermuda, draw=black] (6.75, -3.00) rectangle (7.00, -3.25);
\filldraw[fill=bermuda, draw=black] (7.00, -3.00) rectangle (7.25, -3.25);
\filldraw[fill=cancan, draw=black] (7.25, -3.00) rectangle (7.50, -3.25);
\filldraw[fill=cancan, draw=black] (7.50, -3.00) rectangle (7.75, -3.25);
\filldraw[fill=cancan, draw=black] (7.75, -3.00) rectangle (8.00, -3.25);
\filldraw[fill=cancan, draw=black] (8.00, -3.00) rectangle (8.25, -3.25);
\filldraw[fill=cancan, draw=black] (8.25, -3.00) rectangle (8.50, -3.25);
\filldraw[fill=cancan, draw=black] (8.50, -3.00) rectangle (8.75, -3.25);
\filldraw[fill=cancan, draw=black] (8.75, -3.00) rectangle (9.00, -3.25);
\filldraw[fill=bermuda, draw=black] (9.00, -3.00) rectangle (9.25, -3.25);
\filldraw[fill=cancan, draw=black] (9.25, -3.00) rectangle (9.50, -3.25);
\filldraw[fill=bermuda, draw=black] (9.50, -3.00) rectangle (9.75, -3.25);
\filldraw[fill=cancan, draw=black] (9.75, -3.00) rectangle (10.00, -3.25);
\filldraw[fill=cancan, draw=black] (10.00, -3.00) rectangle (10.25, -3.25);
\filldraw[fill=bermuda, draw=black] (10.25, -3.00) rectangle (10.50, -3.25);
\filldraw[fill=bermuda, draw=black] (10.50, -3.00) rectangle (10.75, -3.25);
\filldraw[fill=bermuda, draw=black] (10.75, -3.00) rectangle (11.00, -3.25);
\filldraw[fill=bermuda, draw=black] (11.00, -3.00) rectangle (11.25, -3.25);
\filldraw[fill=cancan, draw=black] (11.25, -3.00) rectangle (11.50, -3.25);
\filldraw[fill=cancan, draw=black] (11.50, -3.00) rectangle (11.75, -3.25);
\filldraw[fill=cancan, draw=black] (11.75, -3.00) rectangle (12.00, -3.25);
\filldraw[fill=cancan, draw=black] (12.00, -3.00) rectangle (12.25, -3.25);
\filldraw[fill=bermuda, draw=black] (12.25, -3.00) rectangle (12.50, -3.25);
\filldraw[fill=bermuda, draw=black] (12.50, -3.00) rectangle (12.75, -3.25);
\filldraw[fill=bermuda, draw=black] (12.75, -3.00) rectangle (13.00, -3.25);
\filldraw[fill=bermuda, draw=black] (13.00, -3.00) rectangle (13.25, -3.25);
\filldraw[fill=cancan, draw=black] (13.25, -3.00) rectangle (13.50, -3.25);
\filldraw[fill=bermuda, draw=black] (13.50, -3.00) rectangle (13.75, -3.25);
\filldraw[fill=cancan, draw=black] (13.75, -3.00) rectangle (14.00, -3.25);
\filldraw[fill=cancan, draw=black] (14.00, -3.00) rectangle (14.25, -3.25);
\filldraw[fill=bermuda, draw=black] (14.25, -3.00) rectangle (14.50, -3.25);
\filldraw[fill=bermuda, draw=black] (14.50, -3.00) rectangle (14.75, -3.25);
\filldraw[fill=cancan, draw=black] (14.75, -3.00) rectangle (15.00, -3.25);
\filldraw[fill=bermuda, draw=black] (0.00, -3.25) rectangle (0.25, -3.50);
\filldraw[fill=cancan, draw=black] (0.25, -3.25) rectangle (0.50, -3.50);
\filldraw[fill=cancan, draw=black] (0.50, -3.25) rectangle (0.75, -3.50);
\filldraw[fill=bermuda, draw=black] (0.75, -3.25) rectangle (1.00, -3.50);
\filldraw[fill=bermuda, draw=black] (1.00, -3.25) rectangle (1.25, -3.50);
\filldraw[fill=cancan, draw=black] (1.25, -3.25) rectangle (1.50, -3.50);
\filldraw[fill=bermuda, draw=black] (1.50, -3.25) rectangle (1.75, -3.50);
\filldraw[fill=cancan, draw=black] (1.75, -3.25) rectangle (2.00, -3.50);
\filldraw[fill=cancan, draw=black] (2.00, -3.25) rectangle (2.25, -3.50);
\filldraw[fill=cancan, draw=black] (2.25, -3.25) rectangle (2.50, -3.50);
\filldraw[fill=cancan, draw=black] (2.50, -3.25) rectangle (2.75, -3.50);
\filldraw[fill=cancan, draw=black] (2.75, -3.25) rectangle (3.00, -3.50);
\filldraw[fill=bermuda, draw=black] (3.00, -3.25) rectangle (3.25, -3.50);
\filldraw[fill=bermuda, draw=black] (3.25, -3.25) rectangle (3.50, -3.50);
\filldraw[fill=bermuda, draw=black] (3.50, -3.25) rectangle (3.75, -3.50);
\filldraw[fill=cancan, draw=black] (3.75, -3.25) rectangle (4.00, -3.50);
\filldraw[fill=cancan, draw=black] (4.00, -3.25) rectangle (4.25, -3.50);
\filldraw[fill=cancan, draw=black] (4.25, -3.25) rectangle (4.50, -3.50);
\filldraw[fill=bermuda, draw=black] (4.50, -3.25) rectangle (4.75, -3.50);
\filldraw[fill=bermuda, draw=black] (4.75, -3.25) rectangle (5.00, -3.50);
\filldraw[fill=bermuda, draw=black] (5.00, -3.25) rectangle (5.25, -3.50);
\filldraw[fill=cancan, draw=black] (5.25, -3.25) rectangle (5.50, -3.50);
\filldraw[fill=cancan, draw=black] (5.50, -3.25) rectangle (5.75, -3.50);
\filldraw[fill=cancan, draw=black] (5.75, -3.25) rectangle (6.00, -3.50);
\filldraw[fill=bermuda, draw=black] (6.00, -3.25) rectangle (6.25, -3.50);
\filldraw[fill=bermuda, draw=black] (6.25, -3.25) rectangle (6.50, -3.50);
\filldraw[fill=bermuda, draw=black] (6.50, -3.25) rectangle (6.75, -3.50);
\filldraw[fill=cancan, draw=black] (6.75, -3.25) rectangle (7.00, -3.50);
\filldraw[fill=cancan, draw=black] (7.00, -3.25) rectangle (7.25, -3.50);
\filldraw[fill=cancan, draw=black] (7.25, -3.25) rectangle (7.50, -3.50);
\filldraw[fill=bermuda, draw=black] (7.50, -3.25) rectangle (7.75, -3.50);
\filldraw[fill=bermuda, draw=black] (7.75, -3.25) rectangle (8.00, -3.50);
\filldraw[fill=bermuda, draw=black] (8.00, -3.25) rectangle (8.25, -3.50);
} } }\end{equation*}

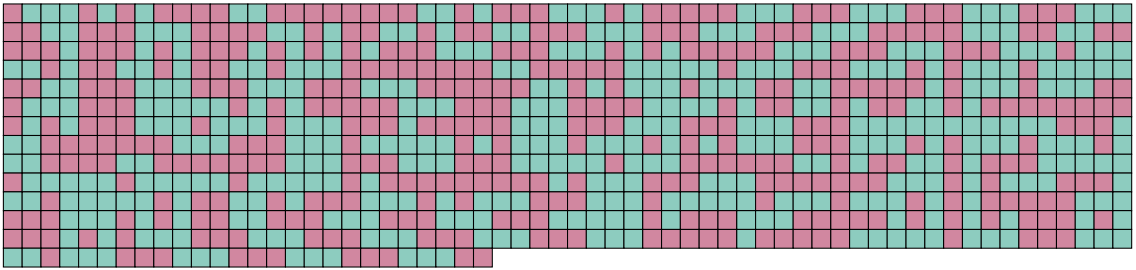
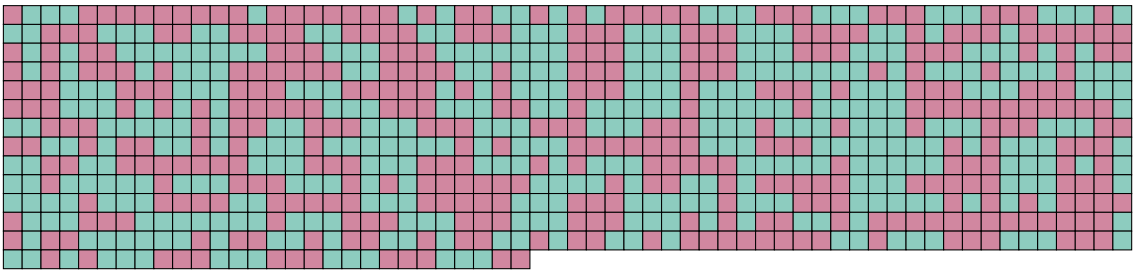
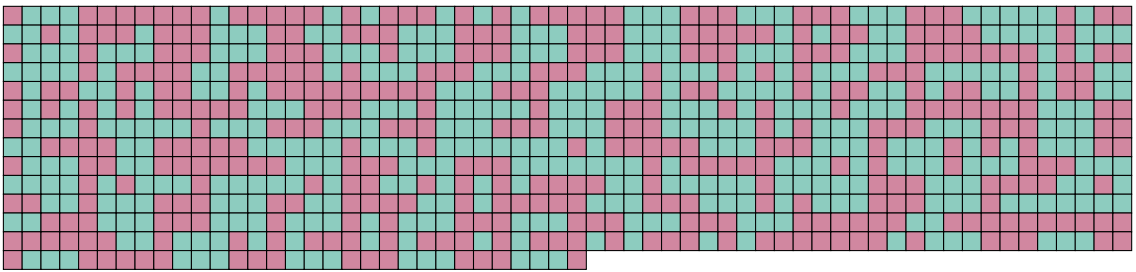
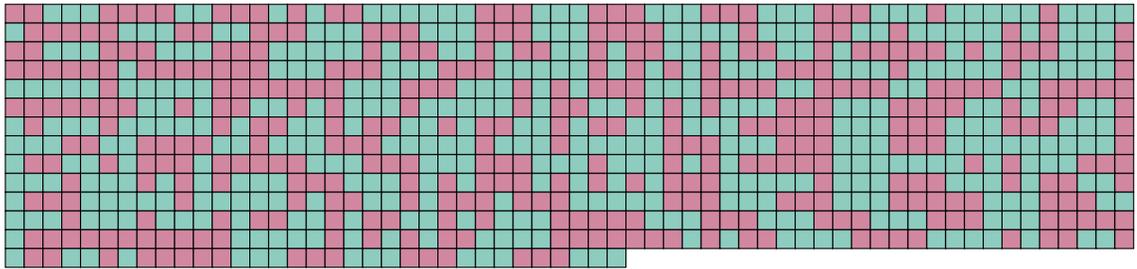
\end{document}